\documentclass[12pt,openany]{cmuthesis}

\usepackage{mathptmx}

\usepackage[binary-units=true,parse-numbers=false,per-mode=symbol]{siunitx}
\DeclareSIUnit\year{year}
\usepackage{xstring}
\usepackage{pifont}
\usepackage{makecell}
\usepackage{colortbl}
\usepackage{hhline}
\usepackage{algorithm}
\usepackage{placeins}
\usepackage{rotating}
\usepackage{pdflscape}

\usepackage{fullpage}
\usepackage{graphicx}
\usepackage{amsmath}
\usepackage{enumitem}
\usepackage{csquotes}
\usepackage[numbers,sort]{natbib}
\usepackage{subcaption}
\usepackage{booktabs}
\usepackage{multirow}
\usepackage{array}
\usepackage[backref,pageanchor=true,plainpages=false, pdfpagelabels, bookmarks,bookmarksnumbered,hidelinks]{hyperref}

\makeatletter

\AtBeginDocument{%
  \patchcmd{\@starttoc}{\begingroup}{\begingroup\@fileswfalse}{}{}%
  \patchcmd{\@starttoc}{\if@filesw}{\endgroup\begingroup\if@filesw}{}{}%
}
\makeatother

\usepackage{datetime}
\newdateformat{monthyeardate}{%
  \monthname[\THEMONTH] \THEYEAR}

\usepackage{dblfloatfix}
\newcolumntype{L}[1]{>{\raggedright\let\newline\\\arraybackslash\hspace{0pt}}m{#1}}
\newcolumntype{C}[1]{>{\centering\let\newline\\\arraybackslash\hspace{0pt}}m{#1}}
\newcolumntype{R}[1]{>{\raggedleft\let\newline\\\arraybackslash\hspace{0pt}}m{#1}}

\newcommand*\circled[1]{\raisebox{.2pt}{\textcircled{\raisebox{-.3pt} {\footnotesize #1}}}}

\newif\ifcameraready%
\camerareadytrue%

\newif\ifeps%
\epsfalse%

\newcommand{\paratitle}[1]{\vspace{8pt}\textbf{#1.}}
\newcommand{\ignore}[1]{}
\ifcameraready
  \newcommand{\todo}[1][]{}
  \newcommand{\chI}[0]{}
  \newcommand{\chII}[0]{}
  \newcommand{\chIII}[0]{}
  \newcommand{\chIV}[0]{}
  \newcommand{\chV}[0]{}
  \newcommand{\chVI}[0]{}
  \newcommand{\chVII}[0]{}
  \newcommand{\chVIII}[0]{}
  \newcommand{\chIX}[0]{}
  \newcommand{\chX}[0]{}
  \newcommand{\chXI}[0]{}
  \newcommand{\chXII}[0]{}
  \newcommand{\yixin}[0]{}
  \newcommand{\sg}[0]{}
  \newcommand{\sph}[0]{}
\else
  \newcommand{\todo}[1][]{\textbf{\scriptsize \fcolorbox{black}{red}{\color{white}{TODO}}} \underline{$\overline{\hbox{\emph{#1}}}$}}
  \newcommand{\chI}[0]{}
  \newcommand{\chII}[0]{}
  \newcommand{\chIII}[0]{}
  \newcommand{\chIV}[0]{}
  \newcommand{\chV}[0]{}
  \newcommand{\chVI}[0]{}
  \newcommand{\chVII}[0]{}
  \newcommand{\chVIII}[0]{}
  \newcommand{\chIX}[0]{}
  \newcommand{\chX}[0]{}
  \newcommand{\chXI}[0]{}
  \newcommand{\chXII}[0]{}
  \newcommand{\yixin}[0]{}
  \newcommand{\sg}[0]{}
  \newcommand{\sph}[0]{}
\fi

\newif\iftwocolumn%
\twocolumnfalse%

\iftwocolumn
  \newcommand{\figscale}{1.}
\else
  \newcommand{\figscale}{.7}
\fi

\newcommand{\incircle}[1]{%
    \IfEqCase{#1}{%
        {1}{\ding{182}}%
        {2}{\ding{183}}%
        {3}{\ding{184}}%
        {4}{\ding{185}}%
        {5}{\ding{186}}%
        {6}{\ding{187}}%
        {7}{\ding{188}}%
        {8}{\ding{189}}%
    }[\PackageError{incircle}{Undefined option to incircle: #1}{}]%
}%

\usepackage[letterpaper,twoside,vscale=.8,hscale=.75,nomarginpar,hmarginratio=1:1]{geometry}

\begin{document}
\frontmatter

\pagestyle{empty}

\title{
{\bf Architectural Techniques for Improving NAND Flash Memory Reliability}}
\author{Yixin Luo}
\date{\monthyeardate\today}
\Year{2018}
\trnumber{CMU-CS-18-101}

\committee{
Onur Mutlu, Chair \\
Phillip B. Gibbons \\
James C. Hoe \\
Yu Cai, SK Hynix \\
Erich F. Haratsch, Seagate Technology
}





\maketitle

\pagestyle{plain} 



\begin{abstract}
\addcontentsline{toc}{chapter}{Abstract}

Over the past decade, NAND flash memory has rapidly grown in popularity within
modern computing systems, thanks to its short random access latency, high
internal parallelism, low static power consumption, and \chII{small form
factor. Today, NAND flash memory is widely used as the primary storage medium
for smartphones, personal laptops, and data center servers.} This growth
\chII{in flash memory popularity} has been sustained by the decreasing cost per
bit of NAND flash memory devices over each \chIII{technology} generation\chV{, which is} due to the increased
storage density. The \chII{higher} density, however, \chII{comes} at a cost of
\chII{reduced} storage reliability. Unreliable primary data storage could lead
to permanent loss of \chII{valuable} data. Thus, we must improve NAND flash
memory reliability to prevent data loss.

Existing techniques to keep NAND flash memory reliable are costly. For example,
a strong error correcting code (ECC), such as \chII{low-density parity-check
(LDPC) code}, is typically used to tolerate up to a relatively high raw bit
error rate  (e.g., between $10^{-3}$ and $10^{-2}$) from the flash memory.
However, such ECC requires \chII{significant redundancy, high latency, and high
area overhead in today's designs.} To tolerate more errors in future \chV{generations of}
NAND flash memories, \chII{a stronger ECC is needed, which requires an even
larger amount of data redundancy, latency and area overhead than the ECC used
in today's flash memory.} \chIII{Such ECC is not only very costly, but it \chV{also may}
not be as effective as other novel techniques that are specialized for
different error types. Our} goal in this
dissertation is to \chIII{greatly} improve
flash \chIII{memory} reliability at low cost.

We identify \chII{three opportunities to improve the cost-efficiency} \chIII{of
flash} reliability enhancement techniques. First, we can adapt
the flash controller to various NAND flash memory error characteristics, or
even to the error characteristics of each individual flash \chIII{chip}. Second, we
can adapt the flash controller to how the host uses the NAND flash memory,
e.g., application access patterns and \chII{environmental} temperature. Third, \chII{the
flash chips} are typically managed by a powerful \chV{controller}
within the Solid-State Drive (SSD). \chIII{This} powerful computing resource is
underutilized when the SSD is idle or when the workload \chIII{has low access
intensity}. \chIII{We}
\chII{can use the flash controller \chIII{to optimize} flash reliability in the
background without capacity or performance loss.}

To exploit these opportunities in improving flash memory reliability, the \chII
{main thesis} of our approach is to specialize the flash controller algorithms
to the device \chIII{and workload} characteristics, rather than using powerful but
expensive generic error tolerance techniques such as ECC\@. In this
dissertation, we (1)~develop a new understanding of the NAND flash memory error
characteristics and the workload behavior through \chII{rigorous} experimental
characterization, and (2)~design smart flash controller algorithms based on this
\chII{new} understanding to improve flash reliability at low cost. \chII{To
this end, we make four major contributions.}

First, we propose a new technique called \emph{WARM} \chII{to improve flash
memory lifetime}. The key idea is to identify and exploit the write-hotness of
the workload in the flash controller \chV{in order} to improve flash reliability. We show that
existing write-hotness agnostic techniques lead to redundant \chII{refresh
operations} that degrade flash lifetime. WARM manages write-hot data and
write-cold data differently\chIII{,} and effectively improves flash lifetime with low hardware
and performance overhead.

Second, we propose a new framework to learn an online flash channel model that
predicts the underlying threshold voltage distribution of each flash chip.
The threshold voltage distribution decides the error characteristics of the
flash chip and \chIII{changes} over time due to flash memory wearout. We show that
existing analytical threshold voltage distribution models are unsuitable for
online flash channel modeling\chIII{,} as they are either inaccurate or expensive to
compute. We show that Student's t-distribution and \chIII{the power law} distribution can
be used to model the static and dynamic (changing) behavior of the threshold
voltage distribution with high \chII{accuracy} and low latency. We also show that
\chII{a} variety of existing techniques can be tuned using this online model to
improve flash memory lifetime.

Third, we perform the first detailed, comprehensive characterization and
analysis of 3D NAND flash memory errors. Through this analysis, we identify
three new error characteristics in 3D NAND flash memory due to its unique
structure and cell design. We develop models for two of the new error
characteristics that are significant in \chIII{current-generation} 3D NAND flash chips.
We develop four new mechanisms within the flash controller to mitigate the
three new error types, \chIII{and thus greatly reduce the error rate,}
at low cost.

Fourth, \chIII{we} perform
\chII{the first} experimental characterization of the self-recovery effect on
3D NAND \chIII{flash memory} and show that dwell time, i.e., the idle time between write cycles,
and temperature significantly impact retention loss speed and program
accuracy. We develop a new unified model of these effects, called \chIII{the
\underline{U}nified self-\underline{R}ecovery and
\underline{T}emperature} \chII{model (URT)}. Using this model,
we propose a new technique called \emph{HeatWatch} \chII{to mitigate
errors due to early retention loss in 3D NAND flash memory}.
HeatWatch reduces \chIII{the} raw bit error rate by tuning the read reference voltages to
the dwell time of the workload and the operating temperature of the flash
memory. We show that \chIII{HeatWatch efficiently} tracks the temperature and
dwell time of NAND flash memory and \chIII{greatly} mitigates retention errors in 3D
\chIII{NAND \chIII{flash memory} using} this information.

\chII{Overall, this dissertation (1)~deepens the understanding of the error
characteristics of \chIII{\emph{both}} \chIII{planar and} 3D NAND flash memory through rigorous
experimental characterization \chIII{and}, (2)~develops new flash controller algorithms
that \chIII{improve \chIII{NAND} flash memory} reliability \chIII{(both lifetime and
error rate)} at low cost by taking advantage of the
flash device and workload characteristics that we find \chIII{based on our
new understandings}.}

\end{abstract}

\begin{acknowledgments}
\addcontentsline{toc}{chapter}{Acknowledgments}

First of all, I would like to thank my adviser, Onur Mutlu, for always
trusting me to do good work, encouraging me whenever a paper gets rejected,
giving me enough resources and opportunities to do great research, and pushing
me to keep improving my presentation and writing skills.

I am grateful to Saugata Ghose, Yu Cai, and Erich Haratsch for being both my
mentors and collaborators. I am grateful to the members of my PhD committee:
Phillip Gibbons and James Hoe for their valuable feedback and for making the
final steps towards my PhD very smooth. I am grateful to Deb Cavlovich who
allowed me to focus on my research by magically solving all other problems.

I am grateful to SAFARI group members that were more than just lab mates.
Saugata Ghose was not only a great mentor and collaborator to me who helped me
improve in all aspects, but also a great friend who was always available for
help. Hongyi Xin was like a big brother to me who was always supportive and
gave me kind advice that kept me on track during my early struggles, and was
never disappointing to have a fun discussion with about bioinformatics or history.
Rachata Ausavarungnirun was a fantastic cook and was always kind to share
his delicious food that reminded me of my grandmother's cooking. Donghyuk Lee
encouraged me when it was most needed, and taught me his highest work ethic and
kindness by example. Justin Meza helped me improve my writing and presentation
skills during my early years in a very friendly manner. Chris Fallin was kind
and patient enough to share his wisdom and time for insightful research
discussions, which helped to improve my research skills. Kevin Chang was a
perfect role model for me to follow, who also happened to share similar hobbies
with me. Yoongu Kim pushed me to work harder and keep improving my research,
and set a high standard for the group in terms of quality of research,
writing, and presentation. Vivek Seshadri, Gennady Pekhimenko, Samira Khan
were always eager to share valuable research and career advice. Kevin Hsieh
was a great roommate at the PDL Retreat, and was always friendly to chat with. Amirali
Boroumand was a \chV{nice buddy until he} cold-bloodedly left us for nicer weather
in California. I also thank other members of the SAFARI group for their
assistance and support: Yang Li, Nandita Vijaykumar, Jamie Liu, HanBin Yoon,
Ben Jaiyen, Damla Senol, Arash Tavakkol, Minesh Patel, and Jeremie Kim.

During my time at Carnegie Mellon, I met a lot of wonderful people: Haoxian
Chen, Yan Gu, Yihan Sun, Yu Zhao, Yong He, Yanzhe Yang, Junchen Jiang, Xuezhi
Wang, Abutalib Aghayev, Jin Kyu Kim, Joy Arulraj, Lin Ma, Michael Zhang,
Huanchen Zhang, Jinliang Wei, Dominic Chen, Guangshuo Liu, Junyan Zhu, Tianshi
Li, Jiyuan Zhang, and many others who helped and supported me in many
different ways. I am also grateful to people in the PDL and CALCM groups, for
accepting me into their communities and for all the feedback and comments that
helped improve my work significantly.

I am grateful to my internship mentors for giving me the opportunities and
resources to do research that is not only practical enough to benefit the
company but also forward-looking enough to lead to this thesis. At Microsoft
Research, I had the privilege to closely work with Jie Liu, Sriram Govindan,
Bikash Sharma, Mark Santaniello, and Aman Kansal. At Seagate, I had the
privilege to closely work with Erich Haratsch, Ludovic Danjean, Thuy Nguyen,
Hakim Alhussien, Sundararajan Sankaranarayanan, Lei Chen, Hongmei Xie, and many others.

I would like to acknowledge the enormous love and support that I received from
my friends and family. Firstly, I would like to thank my girlfriend, Fan Yang,
for all her understanding and support that made me brave and focused enough to
finish my PhD. I would like to thank my friends all over the world: Kaiyu
Shen, Zixiao Chen, Siyuan Sun, Haishan Zhu, Yifei Huang, Yuqin Mu, Pengyao
Chen, Xuanxuan and her lovely family, and many others who really helped me
through my hardest days and brought me a lot of joy whenever I visited. Lastly, I
would like to thank my parents, who raised me to be the person I am today. I thank
my mother, Yi, for her encouragement, support, love, and sacrifice. I thank my
father, Wenping, for setting a high standard for success.

\end{acknowledgments}

\tableofcontents
\listoffigures
\listoftables

\mainmatter%


%
%
%
%
%


\chapter{Introduction}
\label{sec:introduction}

\section{The Problem: The Cost of Flash Reliability}
\label{sec:introduction:problem}

\begin{figure}[h]
  \centering
  \begin{subfigure}[t]{0.48\linewidth}
    \centering
    \includegraphics[trim=10 240 370 0, clip, width=\linewidth]
    {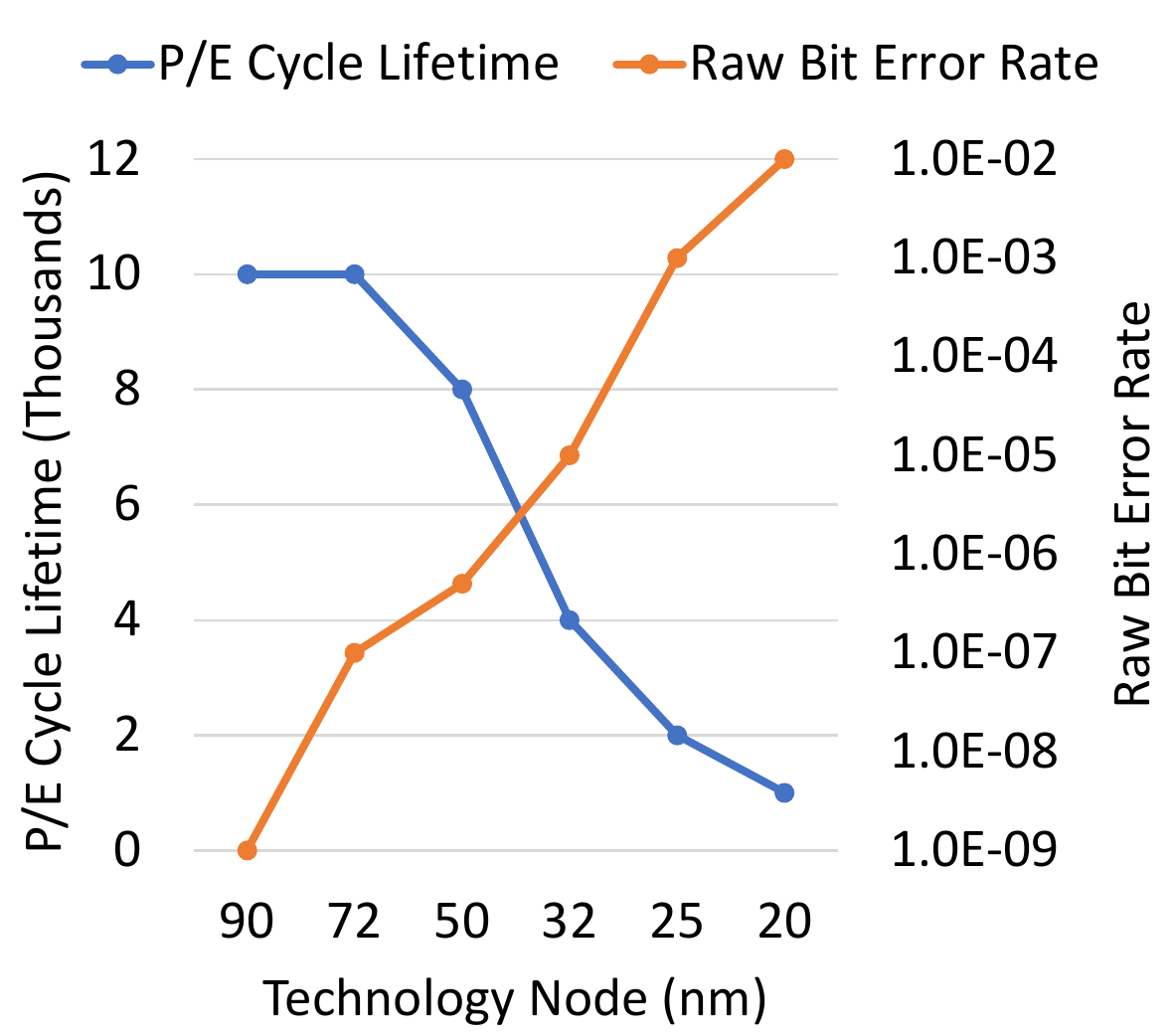}
    \caption{Decreasing flash reliability}
    \label{fig:technology-trend-1}%
  \end{subfigure}%
  ~\hspace{1em}
  \begin{subfigure}[t]{0.48\linewidth}
    \centering
    \includegraphics[trim=370 240 10 0, clip, width=\linewidth]
    {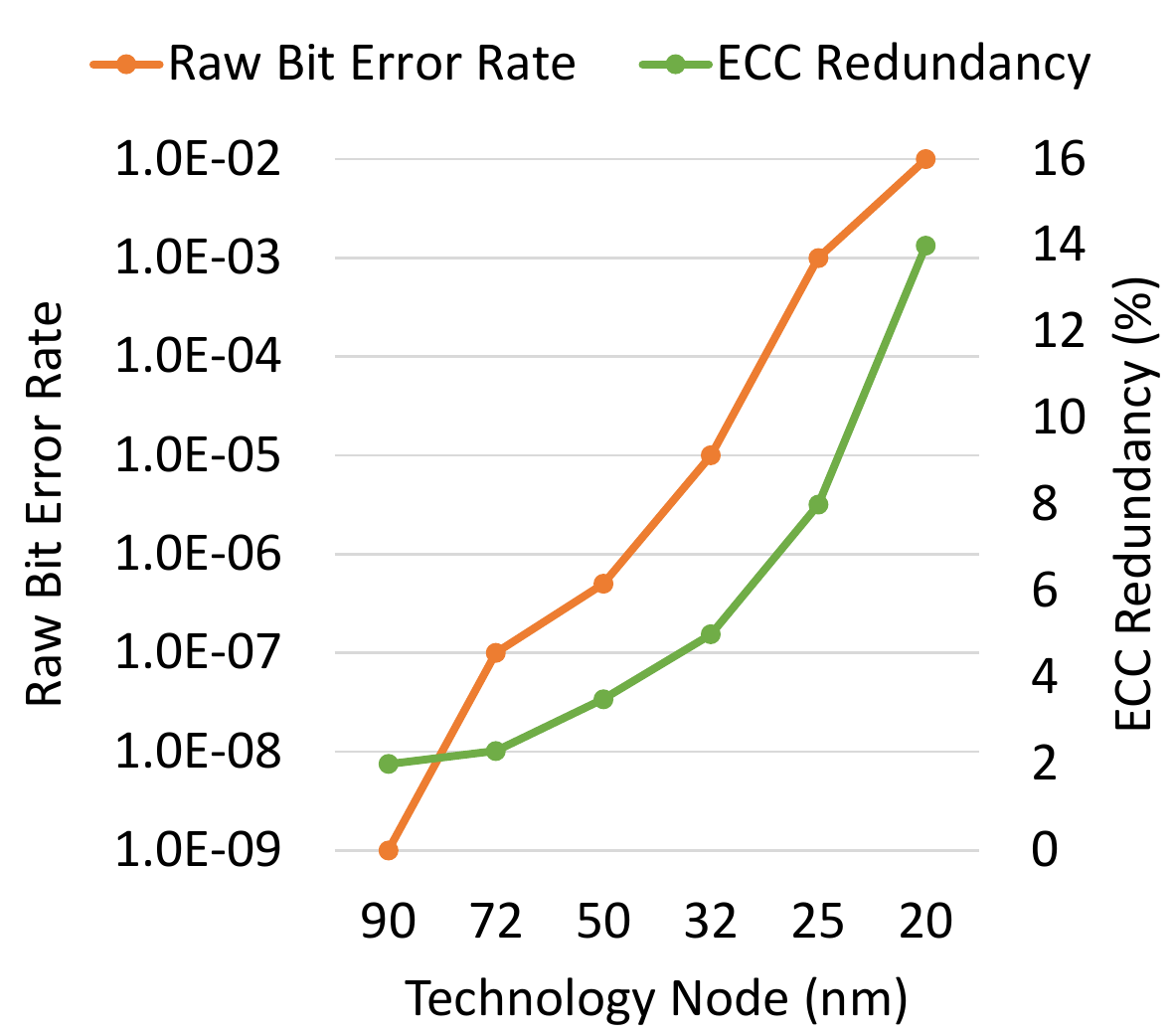}
    \caption{Increasing ECC Redundancy}
    \label{fig:technology-trend-2}%
  \end{subfigure}
  \caption[(a)~Flash reliability (i.e., P/E cycle lifetime and raw bit error rate)
  and (b)~ECC redundancy for each NAND flash memory technology node \chIII{
  (for single-level cell, i.e., SLC devices for \SIrange{90}{72}{\nano\meter}
  technology nodes; for multi-level cell, i.e., MLC devices for
  \SI{50}{\nano\meter} technology node; for triple-level cell, i.e., TLC
  devices for \SIrange{32}{20}{\nano\meter} technology nodes)}.]
  {(a)~Flash reliability (i.e., P/E cycle lifetime and raw bit error rate)
  and (b)~ECC redundancy for each NAND flash memory technology node \chIII{
  (for single-level cell, i.e., SLC devices for \SIrange{90}{72}{\nano\meter}
  technology nodes; for multi-level cell, i.e., MLC devices for
  \SI{50}{\nano\meter} technology node; for triple-level cell, i.e., TLC
  devices for \SIrange{32}{20}{\nano\meter} technology nodes)}.
  Reproduced from~\cite{nazarian.edn13}.}
  \label{fig:technology-trend}%
\end{figure}

In many modern servers and mobile devices, NAND flash memory \chII{(i.e.,
``flash'' or ``NAND flash'')} is used as the primary persistent storage
\chII{device} due to its lower access latency compared to a magnetic disk
drive. As we generate more data at an increasingly faster speed, the need for
higher density NAND flash memory also increases. In the past decade, flash
density has increased by more than $1000\times$ through aggressive process
technology \chII{scaling} from \SI{90}{\nano\meter} to \SI{15}{\nano\meter}.
This rapid increase in flash density, however, has come at the cost of
severely degraded flash memory reliability and lifetime, as is shown in
Figure~\ref{fig:technology-trend-1}. For example, the number of times that a
cell can be reliably programmed and erased before wearing out (i.e., \emph{P/E
cycle lifetime}, \chII{shown as the blue curve in
Figure~\ref{fig:technology-trend-1}}) has dropped from 10,000 times for \SI{72}
{\nano\meter} NAND flash \chIII{(for a single-level cell, i.e., SLC device)}
to only 1,000 times for \SI{20}{\nano\meter} NAND
flash \chIII{(for a triple-level cell, i.e., TLC device)}. In the meantime,
data reliability, measured by the number of bit errors
in a unit of data stored within the flash memory (i.e., \emph{raw bit error
rate}, \chII{shown as the \chIII{orange} curve in Figure~\ref{fig:technology-trend-1}}),
\chIII{increased} by seven orders of magnitude from technology node \SI{90}
{\nano\meter} to \SI{20}{\nano\meter}. Since the data stored on primary
storage is often the only copy, unreliable data storage could lead to
permanent loss of \chII{valuable} data. Thus, we must improve NAND flash memory
reliability to prevent data loss.

\subsection{Flash Reliability Problems}

\begin{figure}[h]
\centering
\includegraphics[width=.8\linewidth, trim=0 385 0 10, clip]
{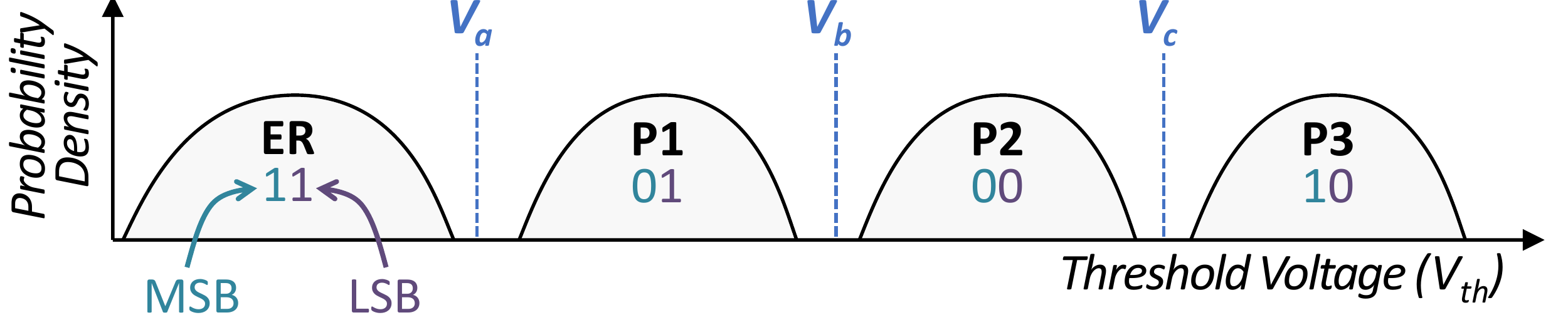}%
\caption{Threshold voltage distribution for MLC
NAND flash memory.}%
\label{fig:intro-vth}%
\end{figure}

\chII{First, we briefly introduce the basics of NAND flash memory to better
understand the flash reliability problems. More detailed background on modern
flash-memory-based solid-state drives (SSDs) and NAND flash memory can be found
in Chapter~\ref{sec:background}. In NAND flash memory, each \emph{flash cell}
consists of a transistor that can store charge. A flash cell represents a
certain data value based on the \emph{threshold voltage} ($V_{th}$) of its
transistor. In \emph{multi-level cell} (MLC) flash memory, each cell stores two
bits of data. A threshold voltage window (i.e., a \emph{state}) is assigned for
each possible two-bit value. Figure~\ref{fig:intro-vth} shows the four possible
states (i.e., ER, P1, P2, P3) in MLC flash memory, along with their
corresponding two-bit values (i.e., \chIII{the} most significant bit, MSB, or,
\chIII{the} least
significant bit, LSB). The state of each flash cell can be \emph{read} by
applying one of three \emph{read reference voltages} (i.e., $V_a$, $V_b$, and
$V_c$) to the cell. Before each flash cell can be \chIII{\emph{programmed}
to} a new state, \chIII{i.e., be written,} the flash cell needs to be 
\emph{erased} to the ER
state. Due to the combination of \emph{program variation}, i.e., the variation
in program operations, and \emph{manufacturing process variation}, the
threshold voltage of cells programmed to the same state follow a Gaussian-like
distribution across the voltage window of the state~\cite{cai.date13,
parnell.globecom14, luo.jsac16}, depicted as \chIII{the curve for each state} in
Figure~\ref{fig:intro-vth}.}

\begin{figure}[h]
\centering
\includegraphics[width=.8\linewidth, trim=0 380 0 10, clip]
{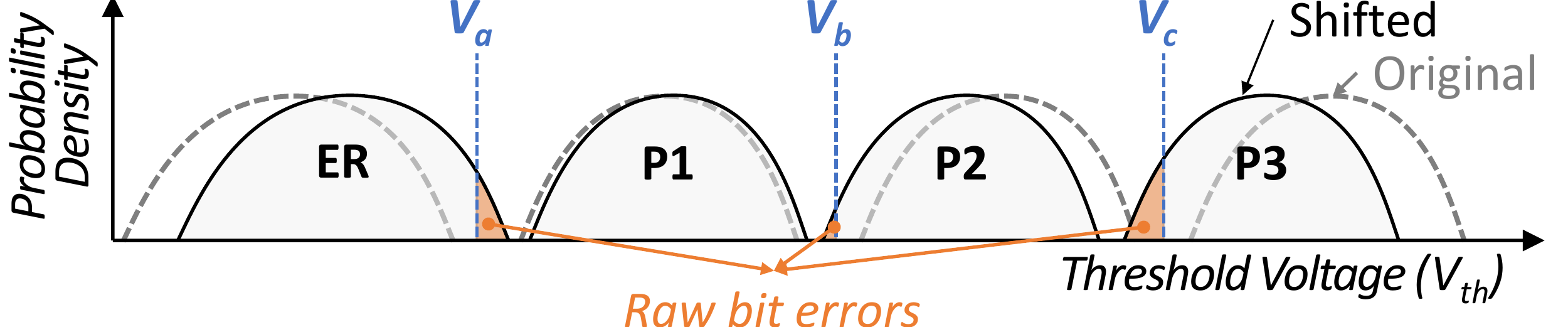}%
\caption[Shifted threshold voltage distribution for MLC
NAND flash memory.]
{Shifted threshold voltage distribution for MLC
NAND flash memory. Reproduced from~\cite{luo.hpca18}.}%
\label{fig:intro-vth-shift}%
\end{figure}

\chII{The reliability \chII{problems} in NAND flash memory \chII{are} caused by
various types of \chIII{noise} that arise from writing, reading, or idling of the NAND
flash memory, or from process variation~\cite{cai.procieee17, mielke.irps08,
cai.date12, cai.date13, cai.sigmetrics14, cai.iccd12, cai.iccd13, cai.hpca15,
cai.dsn15, papandreou.glsvlsi14}. \chIII{Such noise shifts} the threshold voltage
distribution across the read reference voltages, as we show by comparing the
relative position of the original and the shifted distributions in
Figure~\ref{fig:intro-vth-shift}. As a consequence of this shift, some of the
cells
are misread as \chIII{being in} a different state \chIII{than the state they
were programmed to}. \chIII{This phenomenon
leads to} a number of \emph{raw bit errors}.
More detailed background on \chIII{the} various types of NAND flash
memory errors can be found in Section~\ref{sec:errors}. These errors include
P/E cycling errors~\cite{cai.date13, parnell.globecom14, luo.jsac16},
cell-to-cell program interference errors~\cite{cai.iccd13, cai.sigmetrics14},
program errors~\cite{parnell.globecom14, luo.jsac16, cai.hpca17}, read disturb
errors~\cite{parnell.globecom14, cai.dsn15}, retention errors~\cite{cai.iccd12,
cai.hpca15}, and process variation errors~\cite{prabhu.trust11, cai.date12}.}
For instance, \chII{a flash cell} wears out \chII{each} \chV{time
we} write data to it \chII{via} program or erase operations. \chII{Thus,
\chV{the} \emph{P/E cycling error} rate increases over multiple program/erase operation}
cycles, or \emph{P/E cycles}. As another example, \chII{the \chIII{electrical charge}
stored in a \chII{flash cell} \chIII{leaks} over time. Thus, \chV{the} \emph{retention
error} rate increases during the idle time after the data is programmed to the
cell, or \emph{retention time}.} To limit the raw bit error rate to a tolerable
level, flash vendors guarantee reliable operation \chIII{only} for a limited
number of \emph{P/E cycle lifetime} and a limited amount of \emph{retention
time}\chIII{~\cite{jesd218.jedec10}}.

\subsection{The Cost of Improving Flash Reliability}


To \chII{improve} the reliable operation of NAND flash memory \chII{in} the
presence of raw bit errors, \chII{the common \chIII{modern} approach is to}
trade-off storage
density or performance for higher reliability. \chII{We call this the ``naive
approach''.} For example, a strong error correcting code
(ECC) or a redundant array of independent disk (RAID) \chV{techniques use} \chIII{\emph{data
redundancy}} to detect and correct raw bit errors, trading off storage density
\chII{for better reliability}. \chIII{As}
Figure~\ref{fig:technology-trend-2} shows, \chV{in order} to tolerate the \chII{significant
increase in} raw bit error rate \chII{that \chIII{happens} when going from a \SI{90}
{\nano\meter} technology node to a \SI{20}{\nano\meter} technology node}, 
\chII{the amount of redundant storage reserved for ECC (i.e., \emph{ECC
redundancy}, shown as the green curve in Figure~\ref{fig:technology-trend-2})
has to} \chV{increase exponentially} from
$\sim$2\% to $\sim$14\%. Today's NAND flash memory-based solid-state drives
(SSDs) use low-density parity-check (LDPC) codes. \chII{The ECC redundancy
required by LDPC codes for a certain \emph{error correction capability}, i.e.,
the ability to correct a certain number of errors,} is already approaching the
theoretical limit\chIII{,} known as the \emph{Shannon limit}~\cite{shannon.book49,
mackay.letters96, mackay.letters97}. Further improving the error correction
capability \chII{of the ECC} requires either higher ECC redundancy or larger
coding granularity. The former \chIII{greatly degrades} storage density. The
latter \chIII{greatly degrades} performance \chIII{and also} increases the hardware
overhead of the ECC
decoder/encoder~\cite{cai.iccd12, cai.itj13}. Another naive way to improve
NAND flash memory reliability is to increase the precision of program and
erase operations to mitigate write-induced errors. However, this method
significantly increases write latency, \chIII{degrading} SSD performance.


Compared to the naive approach of providing better ECC \chIII{at high cost
and high performance overhead}, we believe it is
more \chIII{appealing} to \chII{develop new techniques} to achieve better flash
reliability \emph{without} trading off storage density or \chIII{performance.} \chII
{These new techniques \chIII{would} enable flash vendors to scale NAND flash memory
density more aggressively and \chIII{would} make NAND flash technology appealing for
even more use cases \chIII{and applications in future computing systems}.}
These new techniques \chIII{would} also enable us to increase the
storage density or performance by tuning down flash reliability when it is not
required. For example, in geo-replicated data centers, data reliability is
already handled by application-level error tolerance techniques. Thus, in
these data centers, single-device reliability can be lowered to improve
storage density by using quadruple-level cell (QLC) technology, \chIII{which
is very unreliable}. \chIII{If we could accomplish this without degrading
data loss guarantees, we could buy} back
the data redundancy caused by data center geo-replication, reducing the total
cost of ownership (TCO) of the data center. \chIII{Our goal in this
dissertation is to build a fundamental understanding of flash devices and
application behavior to enable such techniques.}

\section{Related Work}

\chII{Several prior works characterize NAND flash memory errors and develop
mechanisms to improve flash reliability. In this section, we discuss some
closely related prior approaches. We group prior works based on their
high-level approach and discuss their shortcomings.}

\subsection{NAND Flash Memory Error Characterization}

\chII{Prior work has characterized all types of raw bit errors in planar (or 2D) NAND
flash memory, including P/E cycling errors~\cite{mielke.irps08, cai.procieee17,
cai.date13, parnell.globecom14, luo.jsac16}, program errors~\cite{cai.hpca17,
parnell.globecom14, luo.jsac16}, cell-to-cell program interference
errors~\cite{cai.procieee17, cai.iccd13, cai.sigmetrics14}, retention
errors~\cite{mielke.irps08, cai.procieee17, cai.iccd12, cai.hpca15}, read
disturb errors~\cite{mielke.irps08, cai.procieee17, cai.dsn15,
papandreou.glsvlsi14}, and process variation errors~\cite{prabhu.trust11,
cai.date12}. The findings from these characterizations can be found in
Section~\ref{sec:errors}.}

\chIII{While these works altogether have comprehensively studied the error
characteristics of planar NAND flash memory}, \chII{the \chIII{raw} data from
these \chIII{past characterization works}
are unavailable to the public for further analysis. \chIII{These past
characterizations \chV{were} also done on the devices that are old by today's
standards.} Our \chIII{new and more detailed} characterization of \chIII{much
newer}
planar NAND flash memory in \chIII{Chapter}~\ref{sec:vthmodel} allow us to develop new,
\chIII{and more accurate\chV{,} online}
models and mitigation techniques for raw bit errors \chIII{in more modern
flash memory chips}. Our characterization in
\chIII{Chapter}~\ref{sec:3derror} covers all these errors in the new 3D NAND flash
memory (or \emph{3D NAND}) devices and \chIII{compares} the error characteristics of 3D
NAND with the results in prior work to \chIII{uncover} the differences between 3D NAND
and planar NAND flash memories.}

\subsection{Improving Flash Reliability with Device Awareness}

\chII{\chIII{Flash memory} device characteristics (i.e., NAND
flash memory error characteristics) significantly affect flash
reliability\chIII{~\cite{mielke.irps08, cai.date12, cai.hpca15, cai.dsn15,
papandreou.glsvlsi14, cai.date13, cai.arxiv17, cai.procieee17}}. Many
prior works propose mechanisms to take advantage of these device
characteristics to improve flash reliability. For example, prior work proposes
mechanisms that periodically learn and adapt the read reference voltage to the
threshold voltage distribution shift to mitigate P/E cycling errors, retention
errors, and read disturb errors~\cite{papandreou.glsvlsi14, cai.hpca15,
cai.dsn15}. To mitigate cell-to-cell program interference errors, prior work
proposes a technique that reads the values stored in neighboring cells to
assist correcting errors when a normal read operation
fails~\cite{cai.sigmetrics14, cai.iccd13}. To \chIII{reduce}
program errors in modern MLC NAND flash memory, recent work proposes
to buffer the data stored in \chIII{the LSB page} within the flash
chip~\cite{cai.hpca17}. To
mitigate retention errors, prior work proposes to use various flash refresh
techniques that periodically \chIII{rewrite} all data with high retention
age~\cite{cai.iccd12, cai.itj13}. To mitigate read disturb errors, prior work
proposes to opportunistically reduce the pass-through voltage, i.e., the
highest possible read reference \chIII{voltage,} when the retention age is
low~\cite{cai.dsn15}.}

\chII{While these works demonstrate that improving flash reliability with device
awareness has \chIII{many} benefits in planar NAND, none of these works \chIII
{exploit an online model of flash device characteristics or} design
mechanisms customized for 3D NAND\@. \chIII{In Chapter~\ref{sec:vthmodel},}
we propose new techniques that further \chIII{increase
flash device characteristics} awareness to improve flash reliability by
\chIII{learning an} online model of the flash cells \chIII{within} the flash
controller \chIII{and adapting flash controller policies to this online
model}.} The \chIII{mechanisms we} propose in
Chapters~\ref{sec:3derror} and~\ref{sec:heatwatch} are designed for the unique
error characteristics that we find in 3D NAND\@.

\subsection{Improving Flash Reliability with Workload Awareness}

\chII{Flash reliability varies significantly depending on the workload \chIII
{data access} pattern~\cite{cai.iccd12, cai.dsn15, cai.iccd13, cai.sigmetrics14,
cai.arxiv17, cai.procieee17}. Thus,
to improve flash reliability and lifetime, prior \chIII{work proposes} better
flash management techniques to ensure \chV{a} friendly \chIII{data access} pattern by
optimizing
flash controller algorithms. For example, to reduce unnecessary erase
operations, prior work optimizes \chIII{the flash} page allocation policy to
achieve higher
spatial locality of write operations~\cite{ma.csur14, gupta.asplos09,
park.sigmetrics10, ma.sigmod11}. Prior work also proposes techniques to
minimize P/E cycles consumed by metadata within the flash
controller~\cite{dayan.sigmod16, park.dac16}. To mitigate program interference
errors, prior work proposes to use certain program sequence, instead of
allowing random writes, also managed by the controller~\cite{cai.iccd13,
cai.sigmetrics14, park.dac16}. To mitigate read disturb errors, prior work
proposes to redistribute read-hot pages across different flash
blocks~\cite{liu.systor15}.}

\chIII{While these works demonstrate several effective examples of improving flash
reliability with workload awareness, none of these works design mechanisms to
mitigate retention errors.} \chII{The technique we propose in
Chapter~\ref{sec:warm} partitions
data with different write-hotness and \chIII{applies} the most suitable flash
management policy for each partition to mitigate retention errors.} \chIII
{The techniques we propose in Chapter~\ref{sec:heatwatch} exploits workload
write-intensity awareness and environment temperature awareness to mitigate
retention errors in 3D NAND.}

\subsection{Summary}

\chII{Demonstrated by these prior \chIII{works}, exploiting both device characteristics and
application behavior in the flash controller often leads to significant
reliability improvements with low overhead. This dissertation further \chIII{improves,
expands, or complements} these techniques by \chIII{exploiting \emph{newly-discovered}
(1)~modern} device characteristics and \chIII{(2)~}application behavior \chIII
{characteristics, to greatly improve both bit error rate and lifetime of
the state-of-the-art flash memory devices}.}

\section{Thesis Statement and Overview}
\label{sec:introduction:overview}

Our goal in this dissertation is to improve flash reliability at low cost and
with low performance overhead.  As we can see, \chII{the naive approach of
providing better ECC trades off storage density or performance for improving
flash reliability, and thus} \chV{does} \emph{not} meet our goal of low cost and low
performance overhead. To this end, our thesis statement is that,

\begin{displayquote}

\textit{
NAND flash memory reliability can be improved at low cost and with low
performance overhead by deploying various architectural techniques that are
aware of higher-level application behavior and underlying flash device
characteristics.}

\end{displayquote}


Our approach is to understand flash \chIII{memory device} error
characteristics and workload behavior
through \chII{rigorous experimental} characterization, and to design
\chIII{intelligent and efficient}
flash controller algorithms that utilize this understanding to improve flash
reliability. \chII{This approach is based on \emph{three} observations.} First,
we can take advantage of higher-level application behavior such as write
frequency and locality, and develop the \chII{most suitable and customized}
flash reliability techniques for different \chIII{types of} data stored in the
NAND flash
memory. Second, we can take advantage of underlying device characteristics,
such as variations in temperature, retention time, or error rate, to develop
more efficient device-aware reliability techniques. Third, we can take
advantage of the unused computing resources in the flash controller during idle
time to enable more effective flash reliability techniques. In this
dissertation, we
investigate four directions to exploit \chII{the above three observations} and
efficiently improve flash reliability.

\subsection{WARM---Write-hotness Aware Retention Management}
\label{sec:introduction:overview:warm}

\chIII{Due to charge leakage from the flash cells, data retention errors increase
\chV{over time} after the data has been programmed onto NAND flash memory, i.e.,
the \emph{retention time}. To} \chII{
mitigate retention errors, we limit the amount of retention time in NAND flash
memory by providing a limited amount of guaranteed \emph{internal retention
time}, the duration for which the flash memory correctly holds data within its
P/E cycle lifetime. Flash lifetime can be extended by relaxing this 
\emph{internal retention time}. However,} such relaxation cannot be exposed
externally \chII{to the workload} to avoid altering the expected \emph{data
integrity} property of a flash memory device, \chII{or the
\emph{non-volatility} property expected from a storage device.} Reliability
mechanisms, most prominently \emph{refresh}, restore the duration of data
integrity, but greatly reduce the lifetime \chIII{improvement} from retention time
relaxation by performing a large number of write operations. We find that
retention time relaxation can be achieved more efficiently by exploiting
heterogeneity in \emph{write-hotness}, i.e., the frequency at which each page
is written.

We propose WARM, a write-hotness aware retention management policy for flash
memory. \emph{This is an example of our approach that exploits application-level
\emph{write-hotness} and device level \emph{retention} characteristics to improve
flash lifetime.} The key idea of WARM is to identify and to physically group
together write-hot data within the flash device, allowing the flash controller
to selectively perform retention time relaxation with little cost. When applied
alone, WARM improves overall flash lifetime by an average of 3.24$\times$ over a
\chIII{baseline that \chV{does} not relax the internal retention time}, across a variety of real I/O
workload traces. When WARM is applied together with an adaptive refresh
mechanism, the average lifetime improves by 12.9$\times$ \chIII{over the
baseline}. More
details are in \chIII{Chapter}~\ref{sec:warm} and our MSST
2015 paper~\cite{luo.msst15}.

\subsection{Online Flash Channel Modeling \chII{and Its Applications}}
\label{sec:introduction:overview:vthmodel}

NAND flash memory can be treated as a noisy channel. Each \chII{flash cell}
stores data as the \emph{threshold voltage} of a floating gate transistor. The
threshold voltage can shift as a result of various types of circuit-level
noise, introducing errors when data \chIII{is} read from the channel and ultimately
reducing flash lifetime. An accurate model of the threshold voltage
distribution across flash cells can enable mechanisms within the flash
controller that improve channel reliability and device lifetime.
Unfortunately, existing threshold voltage distribution models are either \chIII{\emph{not}}
accurate enough or have \chIII{\emph{high}} computational complexity, which makes them
unsuitable for \chIII{\emph{online}} implementation within the controller.

We propose a new, low-complexity flash memory \chII{channel} model, built upon a
modified version of the Student's t-distribution and the power law, which
captures the threshold voltage distribution and predicts future distribution
shifts as wear increases.  \emph{This is an example of our approach that
exploits the unused computing resources in the flash controller and enables
greater device-awareness.}  Using our experimental characterization of the
state-of-the-art \SI{1X}{\nano\meter} (i.e., 15--\SI{19}{\nano\meter})
multi-level cell planar NAND flash chips, we show that our
model is highly accurate (with an average modeling error of 0.68\%), and also
simple to compute within the flash controller (requiring 4.41 times less
computation time than the most accurate prior model, with negligible decrease
in accuracy). Our model also predicts future threshold voltage distribution
shifts with a 2.72\% \chIII{average} modeling error.

We demonstrate several example applications of our \chIII{new} model in the
flash
controller, which improve flash channel reliability significantly, including a
new mechanism to predict the remaining lifetime of a flash device. Our
evaluations for two of these applications show that our model: (1)~helps
improve flash memory lifetime by 48.9\% and/or (2)~enables the flash device to
safely sustain 69.9\% more write operations than manufacturer specifications.
More details are in \chIII{Chapter}~\ref{sec:vthmodel} and our \chII{IEEE} JSAC 2016
paper~\cite{luo.jsac16}.

\subsection{3D NAND Flash Memory Error \chII{Characterization and} Mitigation}
\label{sec:introduction:overview:3derror}

Compared to planar NAND flash memory, 3D NAND flash memory uses a new flash
cell design, and vertically stacks dozens of silicon layers in a single chip.
This allows 3D NAND flash memory to increase storage density using a much less
aggressive manufacturing process technology than planar NAND flash memory. The
circuit-level and structural changes in 3D NAND flash memory significantly
alter how different error sources affect the reliability of the memory.

Through experimental characterization of real, state-of-the-art 3D NAND flash
memory chips, we find that 3D NAND flash memory exhibits \emph{three} new
error sources that were \chIII{\emph{not}} previously observed in planar NAND flash memory:
(1)~\emph{layer-to-layer process variation}, where the error rate of each
layer \chIII{of memory in the 3D stack} is very different; (2)~\emph{early
retention loss}, where charge leaks
quickly out of a flash cell soon after the cell is programmed; and
(3)~\emph{retention interference}, where the charge leakage speed of a flash
cell depends on the value stored in the neighboring cell.

Based on our experimental results, we develop new analytical models for
layer-to-layer process variation and retention loss effects in 3D NAND flash
memory. Motivated by our new findings and models, we develop four new
techniques to mitigate process variation and early retention loss in 3D NAND
flash memory.
\emph{These techniques are examples of our approach that exploits device
awareness in the flash controller.}  Our first technique, \chII{layer-to-layer
variation aware reading} (LaVAR), reduces the effect of layer-to-layer process
variation by tuning the read reference voltage for each layer. Our second
technique, \chII{layer-interleaved RAID} (LI-RAID), reorganizes how pages from
different layers are paired together for RAID to reduce the overall error
count. Our third technique, \chII{retention model aware reading} (ReMAR), uses
our retention model to adapt the read reference voltage \emph{to} each cell
based on the \chIII{cell's} retention age. Our fourth technique, \chII{neighbor-cell
assisted
retention interference correction} (NARIC), mitigates retention interference
by predicting and adapting \chIII{the read reference voltages} to the amount of
interference during each read
operation. Compared with similar state-of-the-art error mitigation techniques
developed for planar NAND flash memory, LaVAR and ReMAR reduce the average raw
bit error rate by 51.9\% and 43.3\%, respectively, while LI-RAID reduces the
\chIII{worst-case} RBER by 66.9\%. We conclude that our newly-proposed
techniques successfully mitigate the new error patterns that we discover in 3D
NAND flash memory. More details are in Chapter~\ref{sec:3derror} and our paper
under submission~\cite{luo.sigmetrics18, luo.pomacs18}.

\subsection{HeatWatch: \chII{Self-Recovery and Temperature Aware Retention Error Mitigation}}
\label{sec:introduction:overview:heatwatch}

NAND flash memory wearout can be partially repaired on its own during the idle
time between program or erase operations \chII{(known as the \emph{dwell
time})}, via a phenomenon known as the \emph{self-recovery
effect}\chIII{~\cite{mohan.hotstorage10, wu.hotstorage11}}. We can
exploit the self-recovery effect to \emph{significantly} improve flash
lifetime, by applying \chIII{\emph{high temperature}} to the flash memory
\chIII{during P/E cycling to amplify the self-recovery effect}. As NAND flash
memory \chII{lifetime continues to reduce due to reduced} reliability, the
self-recovery effect provides an appealing opportunity to mitigate \chII{the
poor lifetime}.

While \chIII{flash} self-recovery has been studied for planar 2D NAND flash
memory in the
past\chIII{~\cite{mohan.hotstorage10, wu.hotstorage11}}, the self-recovery
effect in 3D NAND flash memory is \chIII{\emph{not}} well known,
despite the rapidly-growing commercial popularity of 3D NAND flash memory. We
close this gap by characterizing the effects of self-recovery and temperature
on \emph{real, state-of-the-art 3D NAND devices}. We show that these effects
significantly change the \emph{program accuracy} (i.e., the robustness of
flash program operations) and the retention loss speed (i.e., the speed at
which a flash cell leaks charge).  We demonstrate that self-recovery and
temperature affect 3D NAND flash \chIII{\emph{quite differently} from how}
they affect planar
NAND flash, rendering prior models of self-recovery and temperature
ineffective for 3D NAND flash. Using our characterization results, we develop
\chII{a new unified model for 3D NAND self-recovery and temperature effects called
\emph{URT}, \underline{U}nified self-\underline{R}ecovery and
\underline{T}emperature.} \chIII{URT provides a model for raw
bit error rate and threshold voltage distribution based on wearout, retention
time, dwell time, and temperature.}

Based on our \chIII{new} findings and our \chIII{new} model, we propose 
\emph{HeatWatch}, a mechanism
that aims to improve 3D NAND flash reliability and lifetime. \emph{This is an
example of our approach that improves flash reliability by exploiting
device-level behavior.} The key idea of HeatWatch is to optimize read operations
by adapting to the dwell time of the workload and the current operating
temperature. HeatWatch first efficiently tracks flash temperature and the time
of each operation online. Then, HeatWatch uses this information to apply URT
in order to optimize the \chIII{read reference voltages}. Our detailed
experimental evaluations show that HeatWatch reduces the raw bit error rate by
93.5\% and improves flash lifetime by $3.85\times$ over a baseline using a
fixed read voltage, averaged across 28 real workload traces. More details are
in \chIII{Chapter}~\ref{sec:heatwatch} and our HPCA 2018 paper~\cite{luo.hpca18}.

\section{Thesis Outline}
\label{sec:introduction:outline}

In Chapter~\ref{sec:background}, we introduce the basics of modern SSDs and
NAND flash memory. In Chapter~\ref{sec:related}, we provide additional
background and discuss related work on NAND flash memory reliability. A
significant fraction of the material in Chapter~\ref{sec:background} and
Chapter~\ref{sec:related} is borrowed from the author's co-authored work that
\chIII{appears} as an invited paper in the Proceedings of the
IEEE~\cite{cai.procieee17} and placed on arxiv.org~\cite{cai.arxiv17}. In
Chapter~\ref{sec:warm}, we introduce WARM, which \chIII{is} published in
MSST~\cite{luo.msst15}. In Chapter~\ref{sec:vthmodel}, we introduce online
flash channel modeling, which \chIII{is} published in JSAC~\cite{luo.jsac16}. In
Chapter~\ref{sec:3derror}, we introduce our characterization, modeling, and
mitigation techniques for 3D NAND flash memory, which is currently under
\chIII{submission~\cite{luo.sigmetrics18, luo.pomacs18}}. In Chapter~\ref{sec:heatwatch}, we
introduce HeatWatch, which \chIII{is published} in HPCA~\cite{luo.hpca18}.

\section{Contributions}
\label{sec:introduction:contributions}

To our knowledge, this dissertation is the first to propose various \chIII{\emph{device-}} and
\chIII{\emph{workload-}}aware techniques in the flash controller that significantly improve
\chIII{\emph{both}} planar NAND and 3D NAND reliability at low cost. This dissertation makes
the following major contributions to the field:

\begin{enumerate}

\item We propose a new technique called \emph{WARM} that exploits the
write-hotness of the workload in the flash controller to eliminate redundant
flash \chIII{refresh operations}. Chapter~\ref{sec:warm} describes WARM in detail.

\begin{enumerate}[label*=\arabic*.]

\item We propose \emph{\underline{W}rite-hotness \underline{A}ware
\underline{R}etention \underline{M}anagement} (WARM), a
heterogeneous retention management policy for NAND flash memory that
physically partitions write-hot data and write-cold data into two separate
groups of flash blocks, so that we can relax the retention time for only the
write-hot data \chIII{\emph{without}} the need for refreshing such data. We show that doing
so improves flash lifetime by 3.24$\times$ on average across a variety of
server and system workloads, over a write-hotness-oblivious baseline that does
not perform any refresh. WARM can also be combined with an adaptive refresh
mechanism~\cite{cai.iccd12, cai.itj13} to further improve flash lifetime by
12.9$\times$ over the baseline.

\item We propose a mechanism that combines write-hotness-aware retention
management with an adaptive refresh mechanism~\cite{cai.iccd12, cai.itj13}. By
using WARM and applying refresh to write-cold data, we can further
improve flash lifetime due to the benefits of both techniques. We show that
the combined approach improves flash lifetime by 1.21$\times$ over using
adaptive refresh alone homogeneously across the entire flash memory.

\item We propose a simple, yet effective, window-based online algorithm to
identify frequently-written pages. This mechanism can dynamically adapt to
workload behavior and correctly size the identified subset of write-hot pages.
This \chIII{mechanism} fully exploits the existing data structures in the flash
controller to keep track of the write-hotness to reduce \chIII{the tracking}
overhead.
We believe this mechanism can also be used for purposes other than lifetime
management, such as cache performance and energy management.

\end{enumerate}

\item We propose a new framework that \chIII{effectively} learns an online
\chIII{model for the flash channel, while the controller and memory are
under operation}.
Our new model can be learned with low performance overhead and can accurately
emulate the error characteristics of each flash chip.
Chapter~\ref{sec:vthmodel} describes our framework and its applications.

\begin{enumerate}[label*=\arabic*.]

\item We provide an experimental characterization of the threshold voltage
distribution, and how the distribution changes with wear, for state-of-the-art
\chIII{1X-nm MLC} planar NAND flash memory chips. Like prior work, we
find that program errors can cause the tail of the distribution to fatten
significantly, but, \chIII{\emph{unlike}} prior work, we observe that this fat tail can show
up \chIII{\emph{much earlier}} in the lifetime of the flash device than previously thought.

\item We propose a new, simple\chV{,} and accurate static model for the
threshold voltage distribution of MLC NAND flash memory at a particular P/E
cycle count, based upon our modified version of the Student's t-distribution.
The model is capable of accurately capturing the threshold voltage
distribution, with a 0.68\% average modeling error, while requiring little
computation in the flash controller.

\item We propose a new model to dynamically estimate how the threshold voltage
distribution shifts as a function of P/E cycles. This model works in
conjunction with our proposed static model, and it accurately predicts how the
threshold voltage distribution changes in the future, with an average modeling
error of 2.72\%.

\item We demonstrate several \chIII{\emph{practical}} uses of our online threshold voltage
distribution model in a flash controller, which allows the flash controller to
dynamically adapt \chIII{its policies} to threshold voltage shifts and thereby
better improve flash
memory reliability. We propose a new mechanism to estimate the actual
remaining flash lifetime, based on the expected growth in bit error rate. Our
mechanisms improve flash memory lifetime by 48.9\% and/or enable the flash
device to safely endure 69.9\% more P/E cycles than the manufacturer
specification.

\end{enumerate}

\item We perform a detailed, comprehensive characterization and analysis of 3D
NAND flash memory errors using state-of-the-art, real 3D NAND flash memory
chips. Through this analysis, we observe three new error
characteristics in 3D NAND flash memory due to its unique structure and cell
design. To mitigate these errors, we develop four new mechanisms within the
flash controller, called \emph{LaVAR}, \emph{LI-RAID}, \emph{ReMAR}, and 
\emph{NARIC}. Chapter~\ref{sec:3derror} describes our characterization and
analysis, and the proposed mechanisms in detail.

\begin{enumerate}[label*=\arabic*.]

\item We perform the \emph{first comprehensive experimental characterization}
of 3D NAND flash memory errors using real, state-of-the-art MLC 3D NAND flash
memory chips. Based on this characterization, we present an in-depth
comparison and analysis of the different error characteristics of the
well-known error types between 3D NAND and planar NAND flash memories.

\item We present the \emph{first in-depth analysis} of layer-to-layer process
variation, early retention loss, and retention interference, \emph{three
new error characteristics} inherent to 3D NAND flash memory.

\item We develop \emph{new analytical models} for (1)~layer-to-layer process
variation and (2)~early retention loss in 3D NAND flash memory.


\item We propose a new mechanism called \emph{Layer Variation Aware Reading}
(LaVAR) to mitigate the effect of layer-to-layer process variation. LaVAR uses
our layer-to-layer process variation model to fine-tune the read reference
voltage independently for each \chIII{3D stack memory} layer. On average,
LaVAR reduces raw bit error
rate by 43.3\% over a variation-agnostic baseline.

\item We propose a new RAID \chIII{(i.e., Redundant Array of Independent Disks)}
scheme called \emph{Layer-Interleaved RAID}
(LI-RAID) to mitigate the error rate variation between pages within each block
caused by \chIII{manufacturing} process variation and MSB-LSB variation.
LI-RAID eliminates the page
with the worst-case error rate within a block by pairing up pages from
different layers and from different bits within a cell. By doing this, LI-RAID
reduces 99th-percentile tail raw bit error rate by 63.8\% over a baseline
using LaVAR\@.

\item We propose a new mechanism called \emph{Retention Model Aware Reading}
(ReMAR) to mitigate early retention errors in 3D NAND\@. ReMAR tracks the
retention age of the data by recording the \chIII{programming time of} each block and
uses our retention model to adapt the read reference voltage to the retention
age of the data. On average, ReMAR reduces raw bit error rate by 51.9\% over a
retention-agnostic baseline.

\item We propose \emph{Neighbor-cell Assisted Retention Interference Correction}
(NARIC) to mitigate retention interference in 3D NAND devices. NARIC
improves upon Neighbor-cell Assisted Correction (NAC)\chIII{~\cite{cai.sigmetrics14}},
previously proposed to
\chIII{mitigate only} cell-to-cell program interference. In addition to NAC, NARIC
also predicts and adapts \chIII{the read reference voltages} to the amount of
retention interference based on the
threshold voltage of \chIII{the cells on adjacent wordlines}.

\end{enumerate}

\item We propose a new technique called \emph{HeatWatch}, \chIII{a new
mechanism to improve 3D NAND flash memory reliability}. We observe that, due
to self-recovery, data retention in 3D NAND is significantly affected by
temperature and dwell time. HeatWatch significantly improves flash reliability
in 3D NAND by adapting the flash controller algorithms to self-recovery and
temperature.
Chapter~\ref{sec:heatwatch} describes our characterization of self-recovery
effect and HeatWatch in detail.

\begin{enumerate}[label*=\arabic*.]

\item \chIII{Using real, state-of-the-art 3D charge trap NAND flash chips from a
major vendor, we experimentally characterize the effects of self-recovery and
temperature on retention loss speed and program variation. We show that 3D
NAND flash memory exhibits different self-recovery and temperature effects
than planar NAND flash memory.}

\item \chIII{Based on our experimental characterization data, we construct URT, a
unified model for retention loss, wearout, self-recovery, and temperature in
3D NAND flash memory. Our model quantifies these four effects to accurately
predict the raw bit error rate and threshold voltage shift.}

\item \chIII{We propose HeatWatch, a mechanism for 3D NAND flash memory that improves
flash reliability and lifetime. HeatWatch (1)~tracks the temperature, dwell
time, and retention time online, and (2)~sends this information to URT to
accurately predict the optimal read reference voltage. By using the predicted
optimal read reference voltage for flash read operations, HeatWatch reduces
the raw bit error rate by 93.5\%, and improves flash lifetime by $3.85\times$,
over a baseline that uses a fixed read reference voltage.}

\end{enumerate}

\end{enumerate}


\chapter{Basics of Modern SSDs and NAND Flash Memory}
\label{sec:background}

In this chapter, we introduce the basic background of how modern SSDs work
internally (Section~\ref{sec:ssdarch}), and how NAND flash memory operates to
store data (Section~\ref{sec:flash}). This background knowledge helps us
understand the root cause of reliability issues for flash-memory-based SSDs
as well as the \chIII{state-of-the-art} techniques to mitigate these issues, which are introduced in
Chapter~\ref{sec:related}, \chIII{and the new techniques we propose in
Chapters~\ref{sec:warm}--\ref{sec:heatwatch}}.


\section{State-of-the-Art SSD Architecture}
\label{sec:ssdarch}

\begin{figure}[h]
  \centering
  \includegraphics[width=0.75\columnwidth]{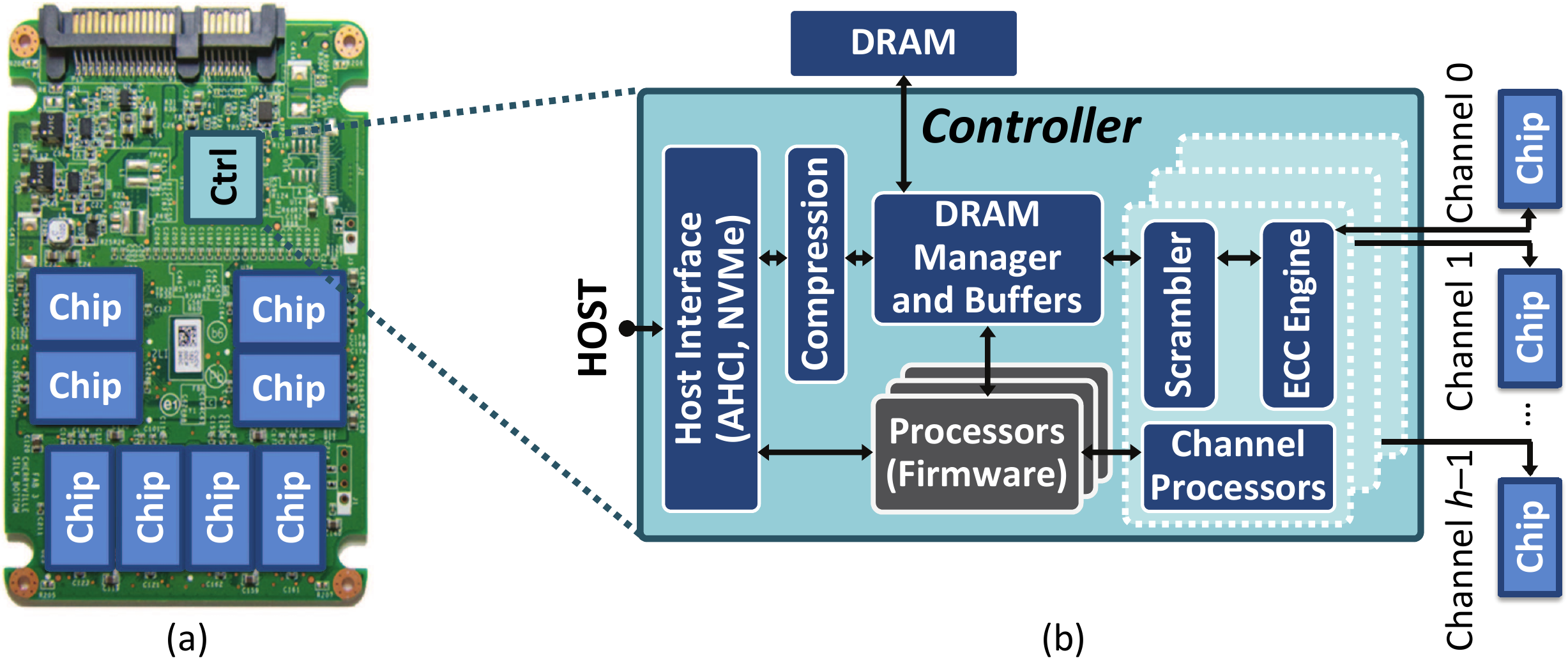}%
  \caption[(a)~SSD system architecture, showing controller (Ctrl)
and chips. (b)~Detailed view of connections between controller
components and chips.]
  {(a)~SSD system architecture, showing controller (Ctrl)
and chips. (b)~Detailed view of connections between controller
components and chips. \chI{Adapted from~\cite{cai.arxiv17}.}}%
  \label{fig:F1}%
\end{figure}

In order to understand the root causes of reliability issues
within SSDs, we first provide an overview of the system architecture
of a state-of-the-art SSD. The SSD consists of a group of
NAND flash memories (or \emph{chips}) and a \emph{controller}, as shown in
Figure~\ref{fig:F1}. A host computer communicates with the SSD through a
high-speed host interface (e.g., \chI{AHCI, NVMe}; \chI{see Section~\ref{sec:ssdarch:ctrl:scheduling}}), which
connects to the SSD controller. The controller is then connected
to each of the NAND flash chips via memory \emph{channels}.

\FloatBarrier

\subsection{Flash Memory Organization}
\label{sec:ssdarch:flash}

\begin{figure}[h]
  \centering
  \includegraphics[width=0.65\columnwidth]{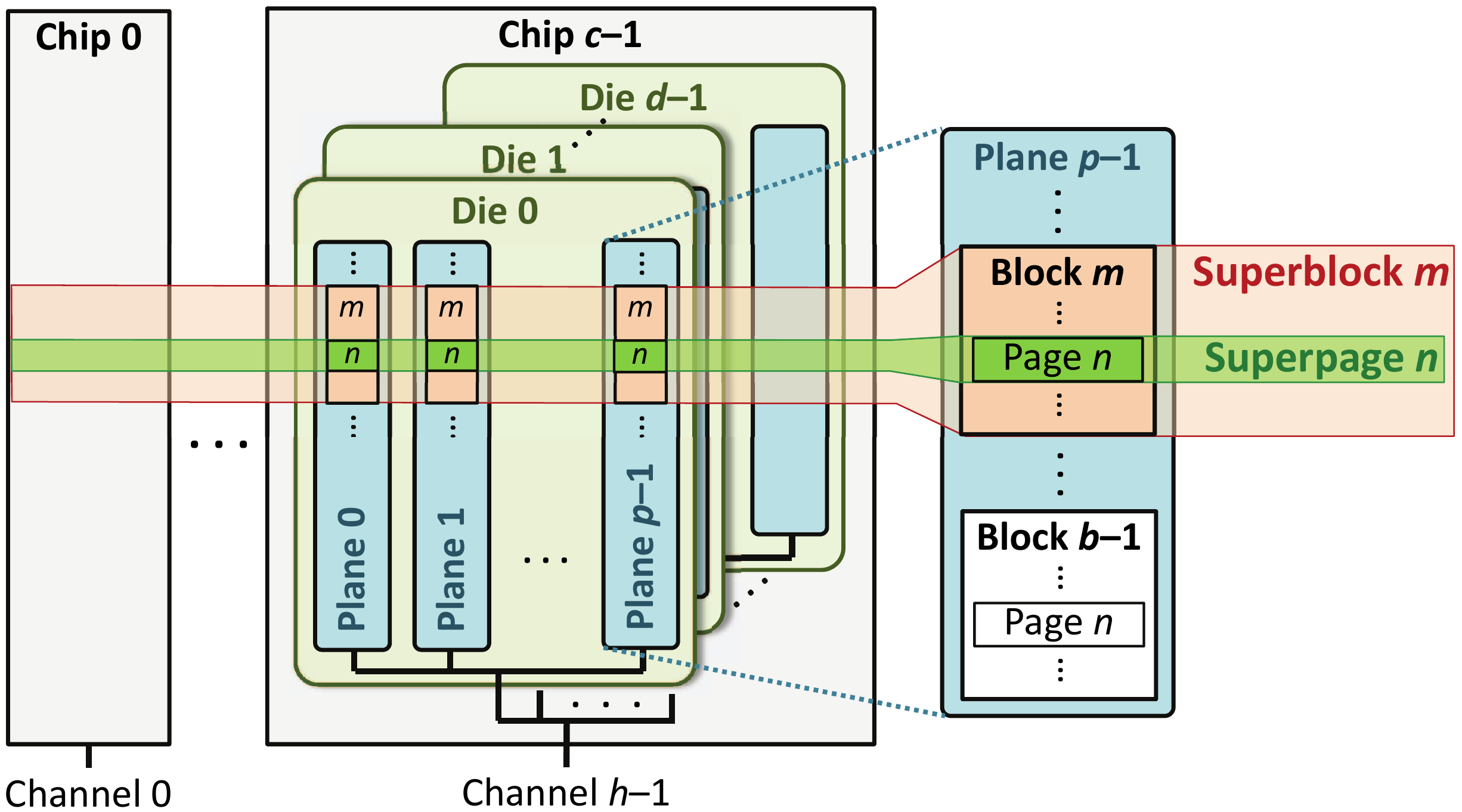}%
  \caption[Flash memory organization.]
  {Flash memory organization. \chI{Reproduced from~\cite{cai.arxiv17}.}}%
  \label{fig:F2}%
\end{figure}

Figure~\ref{fig:F2} shows an example of how NAND flash memory is
organized within an SSD. The flash memory is spread across
multiple flash chips, where each chip contains one or more
flash \emph{dies}, which are individual pieces of silicon wafer that
are connected together to the pins of the chip. Contemporary
SSDs typically have 4--16 chips per SSD, and can have as many
as 16 dies per chip. Each chip is connected to one or more
physical memory channels, and these memory channels are
not shared across chips. A flash die operates independently
of other flash dies, and contains between one and four \emph{planes}.
Each plane contains hundreds to thousands of flash \emph{blocks}.
Each block is a 2D array that contains hundreds of rows of
flash cells (typically 256--1024 rows) where the rows store
contiguous pieces of data. Much like banks in a multi-bank
memory (e.g., DRAM banks~\cite{kim.isca12, lee.hpca13, kim.ch14, moscibroda.security07, mutlu.isca08, mutlu.micro07, chang.hpca16, lee.taco16, lee.pact15, lee.micro09}), 
the planes can execute flash operations
in parallel, but the planes within a die share a single set of data
and control buses~\cite{agrawal.atc08}. Hence, an operation can be started in
a different plane in the same die in a pipelined manner, every
cycle. Figure~\ref{fig:F2} shows how blocks are organized within chips
across multiple channels. In the rest of this chapter, without
loss of generality, we assume that a chip contains a single die.

\FloatBarrier

Data in a block is written at the unit of a \emph{page}, which is
typically between 8 and \SI{16}{\kilo\byte} in size in NAND flash memory.
All read and write operations are performed at the granularity
of a page. Each block typically contains hundreds of pages.
Blocks in each plane are numbered with an ID that is unique
within the plane, but is shared across multiple planes. Within
the block, each page is numbered in sequence. The controller
firmware groups blocks with the same ID number across
multiple chips and planes together into a \emph{superblock}. Within
each superblock, the pages with the same page number are
considered a \emph{superpage}. The controller \emph{opens} one superblock
(i.e., an empty superblock is selected for write operations) at a
time, and typically writes data to the NAND flash memory one
superpage at a time to improve sequential read/write performance
and make error correction efficient, since some parity
information is kept at superpage granularity (see \chI{Section~\ref{sec:ssdarch:ctrl:parity}}).
Having the ability to write to all of the pages in a superpage
simultaneously, the SSD can fully exploit the internal parallelism
offered by multiple planes/chips, which in turn maximizes
write throughput.

\subsection{Memory Channel}
\label{sec:ssdarch:channel}

Each flash memory channel has its own data and control
connection to the SSD controller, much like a main
memory channel has to the DRAM 
controller\chI{~\cite{kim.ch14, muralidhara.micro11, mutlu.isca08, mutlu.micro07, moscibroda.podc08, kim.hpca10, kim.cal16, subramanian.iccd14, subramanian.tpds16, 
hassan.hpca17, kim.micro10, hassan.hpca16, subramanian.hpca13, ipek.isca08, ghose.isca13}}. 
The connection for each channel
is typically an 8- or 16-bit wide bus between the controller
and one of the flash memory chips~\cite{agrawal.atc08}. Both data and flash
commands can be sent over the bus.

Each channel also contains its own control signal pins
to indicate the type of data or command that is on the bus.
The \emph{address latch enable} (ALE) pin signals that the controller
is sending an address, while the \emph{command latch enable}
(CLE) pin signals that the controller is sending a flash
command. Every rising edge of the \emph{write enable} (WE) signal
indicates that the flash memory should write the piece of
data currently being sent on the bus by the SSD controller.
Similarly, every rising edge of the \emph{read enable} (RE) signal
indicates that the flash memory should send the next piece
of data from the flash memory to the SSD controller.

Each flash memory die connected to a memory channel
has its own \emph{chip enable} (CE) signal, which selects the die
that the controller currently wants to communicate with.
On a channel, the bus broadcasts address, data, and flash
commands to all dies within the channel, but only the die
whose CE signal is active reads the information from the bus
and executes the corresponding operation.

\subsection{SSD Controller}
\label{sec:ssdarch:ctrl}

The SSD controller, shown in Figure~\ref{fig:F1}b, is responsible for
\chI{\chI{(1)}~handling I/O requests received from the host, 
\chI{(2)~ensuring data integrity and efficient storage, and (3)~}managing the 
underlying NAND flash memory.}
To perform these
tasks, the controller runs firmware, which is often referred
to as the \emph{flash translation layer} (FTL). FTL tasks are executed
on one or more embedded processors that exist inside
the controller. The controller has access to DRAM, which
can be used to store various controller metadata (e.g., how
host memory addresses map to physical SSD addresses) and
to cache relevant (e.g., frequently accessed) SSD pages~\cite{meza.sigmetrics15, rollins.micron11}. 

When the controller handles I/O requests, it performs
a number of operations on \chI{both the requests and the data. For requests,
the controller \emph{schedules} them in a manner that ensures correctness and provides
high/reasonable performance. For data, the controller \emph{scrambles}
the data to improve raw bit error rates, performs \emph{ECC
encoding/decoding}, and in some cases \chI{\emph{compresses/decompresses}
and/or \emph{encrypts/decrypts}} the
data and employs \emph{superpage-level data parity}.
To manage the NAND flash memory, the controller runs \emph{firmware} that
maps host data to physical NAND flash pages, performs \emph{garbage
collection} on flash pages that have been invalidated, applies
\emph{wear leveling} to evenly distribute the impact of writes on NAND flash
reliability across all pages, and manages bad NAND flash blocks.} We briefly
examine the various tasks of the SSD controller.

\subsubsection{\chI{Scheduling Requests}}
\label{sec:ssdarch:ctrl:scheduling}

\chI{The controller receives I/O requests over a \emph{host controller interface}
\chI{(shown as \emph{Host Interface} in Figure~\ref{fig:F1}b)},
which consists of a system I/O bus and the protocol used to communicate along 
the bus.  When an application running on the host \chI{system} needs to access the
SSD, it generates an I/O request, which is \chI{sent by the host} over the host controller
interface.  The SSD controller receives the I/O request, and inserts the 
request into a queue.  The controller uses a \emph{scheduling policy} to 
determine the order in which the controller processes the requests that 
\chI{are} in the queue.  \chI{The controller then sends the request selected for
scheduling to}
the FTL \chI{(part of the \emph{Firmware} shown in Figure~\ref{fig:F1}b)}.}

\chI{The host controller interface determines how requests are sent to the SSD
and how the requests are queued for scheduling.
Two of the most common host controller interfaces used by \chI{modern} SSDs are
the Advanced Host Controller Interface (AHCI)~\cite{ahci.1.3.1.spec} and 
NVM Express (NVMe)\chV{~\cite{nvme.1.3.spec, tavakkol.fast18, tavakkol.isca18}}.
AHCI builds upon the Serial Advanced Technology Attachment (SATA) system bus
protocol~\cite{sata.3.3.spec}, which was originally designed to connect the host
system to magnetic hard disk drives.  AHCI allows the host to use advanced
features with SATA, such as \emph{native command queuing} (NCQ).  When an
application executing on the host generates an I/O request, the application 
sends the request to the operating system (OS).  The OS sends the request over
the SATA bus to the SSD controller, and the controller adds the request to a
single \emph{command queue}.  NCQ allows the controller to schedule the queued
I/O requests in a different order than the order in which requests were 
received (i.e., requests are scheduled \emph{out of order}).  As a result, 
the controller can choose requests from the queue \chI{in a manner that
maximizes} the overall SSD
performance (e.g., \chI{a \chI{younger} request can be scheduler earlier than
an
older request that requires access to a \chI{plane} that is \chI
{occupied with} serving another
request}).
A major drawback of AHCI and SATA is the limited throughput they enable 
for SSDs~\cite{xu.systor15}, as the protocols were originally designed to match the much
lower throughput of magnetic hard disk drives.
\chI{For example, a modern magnetic hard drive has a sustained read throughput of \SI{300}{\mega\byte\per\second}~\cite{seagate.cheetah15k.spec},
whereas a modern SSD has a read throughput of \SI{3500}{\mega\byte\per\second}~\cite{samsung.960pro.spec}.}
However, AHCI and SATA are widely deployed in modern \chI{computing systems},
and \chI{they currently} remain
a common choice for the SSD host controller interface.}

\chI{To alleviate the throughput bottleneck of AHCI and SATA, many manufacturers
\chI{have started adopting} host controller interfaces that use the PCI
Express (PCIe) system
bus~\cite{pcie.3.1a.spec}.  A popular standard interface for the PCIe bus is the NVM
Express (NVMe) interface~\cite{nvme.1.3.spec}.  Unlike AHCI, which requires an
application to send I/O requests through the OS, NVMe directly exposes multiple
SSD I/O queues to the applications executing on the host.  By directly exposing 
the queues to the applications, NVMe simplifies the software I/O stack, 
eliminating most OS involvement\chI{~\cite{xu.systor15}}, which in turn reduces communication overheads.
An SSD using the NVMe interface maintains a separate set of queues for
\emph{each} application (as opposed to the single
queue \chI{used for all applications with} AHCI) \chI{within the host interface}.  With more
queues, the controller \chI{(1)~has a larger number of requests to select from
during scheduling, increasing its ability to utilize idle resources (i.e., 
\emph{channels}, \emph{dies}, \emph{planes}; see \chI{Section~\ref{sec:ssdarch:flash}}); and
(2)~can more easily manage and control} the amount of interference that
an application experiences from
other concurrently-executing applications. Currently, NVMe is used by modern
SSDs that are designed \chI{mainly} for high-performance systems (e.g.,
enterprise servers, data centers\chI{~\cite{xu.sigmetrics15,xu.systor15}}).}
\chV{Recent work describes the state-of-the-art request scheduling algorithms
in more detail~\cite{tavakkol.fast18, tavakkol.isca18}.}

\subsubsection{Flash Translation Layer}
\label{sec:ssdarch:ctrl:ftl}

The main duty of the FTL \chI{(which is part of the \emph{Firmware} shown in Figure~\ref{fig:F1})} is to
manage the mapping of \emph{logical addresses} (i.e., the address
space utilized by the host) to \emph{physical addresses} in the
underlying flash memory (i.e., the address space for actual
locations where the data is stored, visible only to the SSD
controller) for each page of data~\cite{gupta.asplos09, chung.jsa09}. By providing this
indirection between address spaces, the FTL can \emph{remap} the
logical address to a different physical address (i.e., move
the data to a different physical address) \emph{without} notifying
the host. Whenever a page of data is written to by the host
or moved for underlying SSD maintenance operations (e.g.,
\chI{garbage collection~\cite{chang.tecs04, yang.smartcomp14}; see 
\chI{Section~\ref{sec:ssdarch:ctrl:gc}}}), the old data (i.e., the
physical location where the overwritten data resides) is simply
marked as invalid in the physical block's \emph{metadata}, and
the new data is written to a page in the flash block that is
currently open for writes (see Section~\ref{sec:flash:pgmerase} for more detail
on how writes are performed).

The FTL is also responsible for \emph{wear leveling}, to ensure that
all of the blocks within the SSD are evenly worn out~\cite{chang.tecs04, yang.smartcomp14}.
By evenly distributing the \emph{wear} (i.e., the number of P/E cycles
that take place) \emph{across} different blocks, the SSD controller
reduces the heterogeneity of the amount of wearout across
these blocks, \chI{thereby} extending the lifetime of the device. \chI{The
wear-leveling
algorithm is} invoked when the current block that
is being written to is full (i.e., no more pages in the block are
available to write to), and \chI{it enables} the controller \chI{to select}
a new block \chI{from the \chI{\emph{free list}} to direct the future writes to}. The wear-leveling
algorithm dictates
which of the blocks from the free list is selected. One
simple approach is to select the block in the free list with the
lowest number of P/E cycles to minimize the variance of the
wearout amount across blocks, though many algorithms have
been developed for wear leveling~\cite{gal.cs05, chang.sac07}.

\subsubsection{\chI{Garbage Collection}}
\label{sec:ssdarch:ctrl:gc}

\chI{When the host issues a write request to a logical address stored in the
SSD, the SSD controller performs the write \emph{out of place} (i.e., \chI
{the} updated
version of the \chI{page data} is written to a different physical page in the
NAND flash
memory), because in-place updates cannot be performed 
(see Section~\ref{sec:flash:pgmerase}).  The old physical page is marked as
\emph{invalid} when the out-of-place write completes.
\chI{\emph{Fragmentation}
refers to the waste of space within a block \chI{due to the presence of
invalid pages}.
In a \emph{fragmented} block, \chI{a fraction of} the pages are
invalid, but these pages are unable to store new data until the
page is erased.  Due to circuit-level limitations, the controller can perform
erase operations \chI{\chI{only at} the granularity of} an \emph{entire block} (see 
Section~\ref{sec:flash:pgmerase} for details).  As a result, until a fragmented
block is erased, the block wastes physical space within the SSD.  Over time,
if fragmented blocks are not erased, the SSD will run out of pages that it
can write new data to.}} \chI{The problem becomes especially severe if the
blocks are highly fragmented (i.e., a large fraction of the pages within a
block are invalid).}

\chI{To reduce the \chI{negative} impact of fragmentation \chI{on usable
SSD storage space}, the FTL periodically performs \chI{a \chI{process
called}}
\emph{garbage collection}.  
\chI{Garbage collection finds highly-fragmented flash blocks in the SSD
and recovers the wasted space due to invalid pages.}
\chI{The basic garbage collection algorithm}~\cite{chang.tecs04, yang.smartcomp14}
(1)~identifies the \chI{highly-fragmented blocks} (which we call the
\emph{selected blocks}),
(2)~migrates any valid pages in a selected block (i.e., each valid page is 
written to a new block, \chI{its virtual-to-physical address mapping is updated, 
and the page in the selected block is marked as invalid}),
(3)~erases each selected block (see Section~\ref{sec:flash:pgmerase}),
and (4)~adds a pointer to each selected block into \chI{the} \emph{free list}
within the FTL.
The \chI{garbage collection} algorithm typically selects blocks with the
\chI{highest number of invalid pages}.
When the controller needs a new block to write pages to, it selects one of the
blocks currently in the free list.}

\chI{\chI{We briefly discuss five optimizations that prior works propose to
improve the performance and/or efficiency of garbage collection~\cite{yang.smartcomp14, qin.dac11, han.uic06, he.eurosys17, agrawal.atc08, choudhuri.codes08,
wu.fast12, gupta.asplos09, luo.msst15}}.
First, the \chI{garbage collection} algorithm can be optimized to determine the most efficient frequency to invoke
garbage collection~\cite{yang.smartcomp14, qin.dac11}, as performing garbage collection too
frequently can delay I/O requests from the host, while not performing garbage
collection frequently enough can cause the controller to stall when there are
no blocks available in the free list.
Second, the algorithm can be optimized to select blocks \chI{in a way that
reduces} the number
of page copy and erase operations required each time the garbage collection 
algorithm is invoked~\cite{han.uic06, qin.dac11}.
Third, some works reduce the latency of garbage collection by using multiple channels
to \chI{perform garbage collection on} multiple blocks in parallel~\cite{he.eurosys17, agrawal.atc08}.
Fourth, the FTL can minimize the latency of I/O requests from the host by
pausing erase and copy operations \chI{that are} being performed for garbage
collection, in
order to service the host requests immediately~\cite{choudhuri.codes08, wu.fast12}.
Fifth, \chI{pages can be grouped together such that all of the pages within a
block become invalid around the same time~\cite{gupta.asplos09, he.eurosys17, luo.msst15}.
For example, the controller can group pages with (1)~a similar degree of 
\emph{write-hotness} (i.e., the frequency at which a page is updated; see 
Section~\ref{sec:mitigation:hotcold}) or (2)~a similar} \emph{death time} (i.e., the time at 
which a page is \chI{overwritten}).}
\chI{Garbage collection remains an active area of research.}

\subsubsection{Flash Reliability Management}
\label{sec:ssdarch:ctrl:reliability}

The SSD controller
performs many background optimizations that improve
flash reliability. These flash reliability management techniques,
as we will discuss in more detail in Section~\ref{sec:mitigation},
can effectively improve flash lifetime at a very low cost,
since the optimizations are usually performed during idle
times, when the interference with the running workload
is minimized. These management techniques sometimes
require small metadata storage in memory (e.g., for storing
\chI{the near-optimal} read reference voltages~\cite{cai.hpca15, cai.dsn15, luo.jsac16}), or
require a timer (e.g., for triggering refreshes in 
time~\cite{cai.iccd12, cai.itj13}).

\subsubsection{Compression}
\label{sec:ssdarch:ctrl:compression}

Compression can reduce the size of the
data written to minimize the number of flash cells worn out
by the original data. Some controllers provide compression,
as well as decompression, which reconstructs the original
data from the compressed data stored in the flash 
memory~\cite{zuck.inflow14, li.fast15}. The controller may contain a \emph{compression engine},
which, for example, performs the LZ77 or LZ78 algorithms.
Compression is optional, as some types of data being stored
by the host (e.g., JPEG images, videos, encrypted files, files
that are already compressed) may not be compressible.

\subsubsection{Data Scrambling and Encryption}
\label{sec:ssdarch:ctrl:scrambling}
\label{sec:ssdarch:ctrl:encryption}

The occurrence of
errors in flash memory is highly dependent on the data values
stored into the memory cells~\cite{cai.date12, cai.iccd13, cai.sigmetrics14}. To reduce
the dependence of the error rate on data values, an SSD
controller first scrambles the data before writing it into the
flash chips~\cite{cha.etrij13, kim.jssc12}. The key idea of scrambling is to probabilistically
ensure that the actual value written to the SSD
contains an equal number of randomly distributed zeroes
and ones, thereby minimizing any data-dependent behavior.
Scrambling is performed using a reversible process, and
the controller \emph{descrambles} the data stored in the SSD during
a read request. The controller employs a \emph{linear feedback shift
register} (LFSR) to perform scrambling and descrambling.
An $n$-bit LFSR generates $2^{n-1}$ bits worth of pseudo-random
numbers without repetition. For each page of data to be written,
the LFSR can be seeded with the \emph{logical} address of that
page, so that the page can be correctly descrambled even if
maintenance operations (e.g., garbage collection) migrate
the page to another physical location, as the logical address
is unchanged. (This also reduces the latency of maintenance
operations, as they do not need to descramble and rescramble
the data when a page is migrated.) The LFSR then generates
a pseudo-random number based on the seed, which is
then XORed with the data to produce the scrambled version
of the data. As the XOR operation is reversible, the same
process can be used to descramble the data.

In addition to the data scrambling employed to minimize
data value dependence, several SSD controllers
include data encryption hardware~\cite{haswell.fms16, codandaramane.fms16, willett.fms15}. An
SSD that contains data encryption hardware within its
controller is known as a \emph{self-encrypting drive} (SED). In the
controller, data encryption hardware typically employs
AES encryption~\cite{codandaramane.fms16, nist.fips01, willett.fms15, daemen.book02}, which performs multiple
rounds of substitutions and permutations to the unencrypted
data in order to encrypt it. AES employs a separate
key for each round~\cite{nist.fips01, daemen.book02}. In an SED, the controller
contains hardware that generates the AES keys for each
round, and performs the substitutions and permutations
to encrypt or decrypt the data using dedicated 
hardware~\cite{haswell.fms16, codandaramane.fms16, willett.fms15}.

\subsubsection{Error-Correcting Codes}
\label{sec:ssdarch:ctrl:ecc}

ECC is used to detect and correct
the raw bit errors that occur within flash memory. A
host writes a page of data, which the SSD controller splits
into one or more chunks. For each chunk, the controller
generates a \emph{codeword}, consisting of the chunk and a correction
code. The strength of protection offered by ECC
is determined by the \emph{coding rate}, which is the chunk size
divided by the codeword size. A higher coding rate provides
weaker protection, but consumes less storage, representing
a key reliability tradeoff in SSDs.

The ECC algorithm employed (typically BCH~\cite{shu.book04, lee.isscc12, hocquenghem.chiffres59, bose.ic60} 
or LDPC\chI{~\cite{shu.book04, zhao.fast13, gallager.ire62, mackay.letters96,mackay.letters97, gallager.tit62}}; see Section~\ref{sec:correction}),
as well as the length of the codeword and the coding rate,
determine the total \emph{error correction capability}, i.e., the
maximum number of raw bit errors that can be corrected
by ECC. ECC engines in contemporary SSDs are able to
correct data with a relatively high raw bit error rate (e.g.,
between $10^{-3}$ and $10^{-2}$~\cite{jesd218.jedec10}) and return data to the host at
an error rate that meets traditional data storage reliability
requirements (e.g., a post-correction error rate of $10^{-15}$ in
the JEDEC standard~\cite{jep122h.jedec16}). The \emph{error correction failure rate}
($P_{ECFR}$) of an ECC implementation, with a codeword length
of $l$ where the codeword has an error correction capability
of $t$ bits, can be modeled as:
\begin{equation}
P_{ECFR} = \sum_{k=t+1}^{l}  \binom{l}{k} (1 - \text{BER})^{(l-k)} \text{BER}^{k}
\label{eq:E1}
\end{equation}
where BER is the bit error rate of the NAND flash memory.
We assume in this equation that errors are independent and
identically distributed.

In addition to the ECC information, a codeword contains
cyclic redundancy checksum (CRC) parity information~\cite{rollins.micron11}.
When data is being read from the NAND flash memory,
there may be times when the ECC algorithm incorrectly
indicates that it has successfully corrected all errors in the
data, when uncorrected errors remain. To ensure that incorrect
data is not returned to the user, the controller performs
a CRC check in hardware to verify that the data is error 
free~\cite{rollins.micron11, peterson.ire61}.

\subsubsection{Data Path Protection}
\label{sec:ssdarch:ctrl:datapathprotection}

\begin{figure}[h]
  \centering
  \includegraphics[width=0.65\columnwidth]{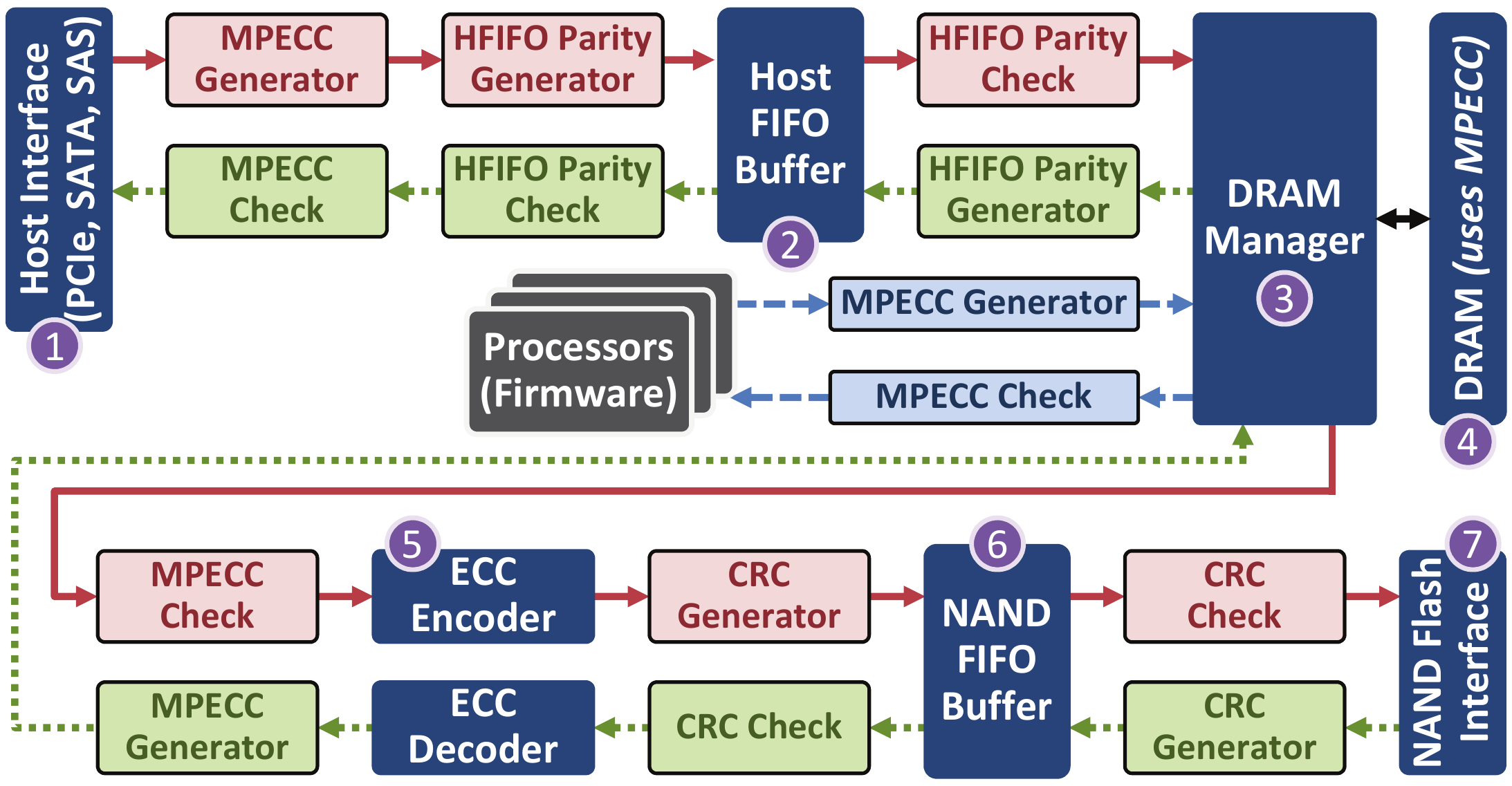}%
  \caption[Data path protection employed within the controller.]
  {Data path protection employed within the controller. \chI{Reproduced from~\cite{cai.arxiv17}.}}%
  \label{fig:F3}%
\end{figure}

In addition to protecting the
data from raw bit errors within the NAND flash memory,
newer SSDs incorporate error detection and correction
mechanisms throughout the SSD controller, in order to
further improve reliability and data integrity~\cite{rollins.micron11}. These
mechanisms are collectively known as \emph{data path protection},
and protect against errors that can be introduced by the various
SRAM and DRAM structures that exist within the SSD.\footnote{See 
Section~\ref{sec:othermem} for a discussion on the possible types of errors that
can be present in DRAM.}
Figure~\ref{fig:F3} illustrates the various structures within the controller
that employ data path protection mechanisms. There are
three data paths that require protection: (1)~the path for data
written by the host to the flash memory, shown as a red solid
line in Figure~\ref{fig:F3}; (2)~the path for data read from the flash memory
by the host, shown as a green dotted line; and (3)~the path
for metadata transferred between the firmware (i.e., FTL)
processors and the DRAM, shown as a blue dashed line.

\FloatBarrier

In the write data path of the controller (the red solid
line shown in Figure~\ref{fig:F3}), data received from the host interface
(\incircle{1} in the figure) is first sent to a host FIFO buffer (\incircle{2}).
Before the data is written into the host FIFO buffer, the data
is appended with \emph{memory protection ECC} (MPECC) and
\emph{host FIFO buffer} (HFIFO) parity~\cite{rollins.micron11}. The MPECC parity is
designed to protect against errors that are introduced when
the data is stored within DRAM (which takes place later
along the data path), while the HFIFO parity is designed
to protect against SRAM errors that are introduced when
the data resides within the host FIFO buffer. When the
data reaches the head of the host FIFO buffer, the controller
fetches the data from the buffer, uses the HFIFO parity
to correct any errors, discards the HFIFO parity, and sends
the data to the DRAM manager (\incircle{3}). The DRAM manager
buffers the data (which still contains the MPECC information)
within DRAM (\incircle{4}), and keeps track of the location of
the buffered data inside the DRAM. When the controller
is ready to write the data to the NAND flash memory, the
DRAM manager reads the data from DRAM. Then, the controller
uses the MPECC information to correct any errors,
and discards the MPECC information. The controller then
encodes the data into an ECC codeword (\incircle{5}), generates CRC
parity for the codeword, and then writes both the codeword
and the CRC parity to a NAND flash FIFO buffer (\incircle{6})~\cite{rollins.micron11}.
When the codeword reaches the head of this buffer, the controller
uses CRC parity to \chI{detect} any errors in the codeword,
and then dispatches the data to the flash interface (\incircle{7}),
which writes the data to the NAND flash memory. \chV{Until the controller
successfully write the data to the NAND flash memory, the data write is not
considered durable. This is because the data stored in the DRAM or FIFO
buffers can be lost if there is a power failure event~\cite{ahmadian.date18}.}
The read
data path of the controller (the green dotted line shown in
Figure~\ref{fig:F3}) performs the same procedure as the write data path,
but in reverse order~\cite{rollins.micron11}.

Aside from buffering data along the write and read paths,
the controller uses the DRAM to store essential metadata,
such as the table that maps each host data address to a physical
block address within the NAND flash memory~\cite{meza.sigmetrics15, rollins.micron11}.
In the metadata path of the controller (the blue dashed
line shown in Figure~\ref{fig:F3}), the metadata is often read from or written
to DRAM by the firmware processors. In order to ensure
correct operation of the SSD, the metadata must not contain
any errors. As a result, the controller uses memory protection
ECC (MPECC) for the metadata stored within DRAM~\cite{luo.dsn14, rollins.micron11}, 
just as it did to buffer data along the write and
read data paths. Due to the lower rate of errors in DRAM
compared to NAND flash memory (see Section~\ref{sec:othermem}), the
employed memory protection ECC algorithms are not as
strong as BCH or LDPC. We describe common ECC algorithms
employed for DRAM error correction in Section~\ref{sec:othermem}.

\subsubsection{Bad Block Management}
\label{sec:ssdarch:ctrl:badblocks}

Due to process variation or
uneven wearout, a small number of flash blocks may have
a much higher raw bit error rate (RBER) than an average
flash block. Mitigating or tolerating the RBER on these flash
blocks often requires a much higher cost than the benefit of
using them. Thus, it is more efficient to identify and record
these blocks as \emph{bad blocks}, and avoid using them to store
useful data. There are two types of bad blocks: \emph{original bad
blocks} (OBBs), which are defective due to manufacturing
issues (e.g., process variation), and \emph{growth bad blocks}
(GBBs), which fail during runtime~\cite{techman.whitepaper16}.

The flash vendor performs extensive testing, known
as \emph{bad block scanning}, to identify OBBs when a flash chip
is manufactured~\cite{micron.tn11}. Initially, all blocks are kept in
the erased state, and contain the value 0xFF in each byte
(see Section~\ref{sec:flash:data}). Inside each OBB, the bad block scanning
procedure writes a specific data value (e.g., 0x00) to
a specific byte location within the block that indicates the
block status. A good block (i.e., a block without defects) is
not modified, and thus its block status byte remains at the
value 0xFF. When the SSD is powered up for the first time,
the SSD controller iterates through all blocks and checks
the value stored in the block status byte of each block. Any
block that does not contain the value 0xFF is marked as bad,
and is recorded in a \emph{bad block table} stored in the controller.
A small number of blocks in each plane are set aside as
\emph{reserved blocks} (i.e., blocks that are not used during normal
operation), and the bad block table automatically remaps
any operation originally destined to an OBB to one of the
reserved blocks. The bad block table remaps an OBB to a
reserved block in the same plane, to ensure that the SSD
maintains the same degree of parallelism when writing to a
superpage, thus avoiding performance loss. Less than 2\% of
all blocks in the SSD are expected to be OBBs~\cite{openmoki.wiki12}.

The SSD identifies growth bad blocks during runtime by
monitoring the status of each block. Each superblock contains
a bit vector indicating which of its blocks are GBBs.
After each program or erase operation to a block, the SSD
reads the \emph{status reporting registers} to check the operation
status. If the operation has failed, the controller marks the
block as a GBB in the superblock bit vector. At this point,
the controller uses superpage-level parity to recover the data
that was stored in the GBB (see \chI{Section~\ref{sec:ssdarch:ctrl:parity}}), and \emph{all data
in the superblock} is copied to a different superblock. The
superblock containing the GBB is then erased. When the
superblock is subsequently opened, blocks marked as GBBs
are \emph{not} used, but the remaining blocks can store new data.

\subsubsection{Superpage-Level Parity}
\label{sec:ssdarch:ctrl:parity}
\label{sec:ssdarch:ctrl:superpageparity}

\begin{figure}[h]
  \centering
  \includegraphics[width=0.65\columnwidth]{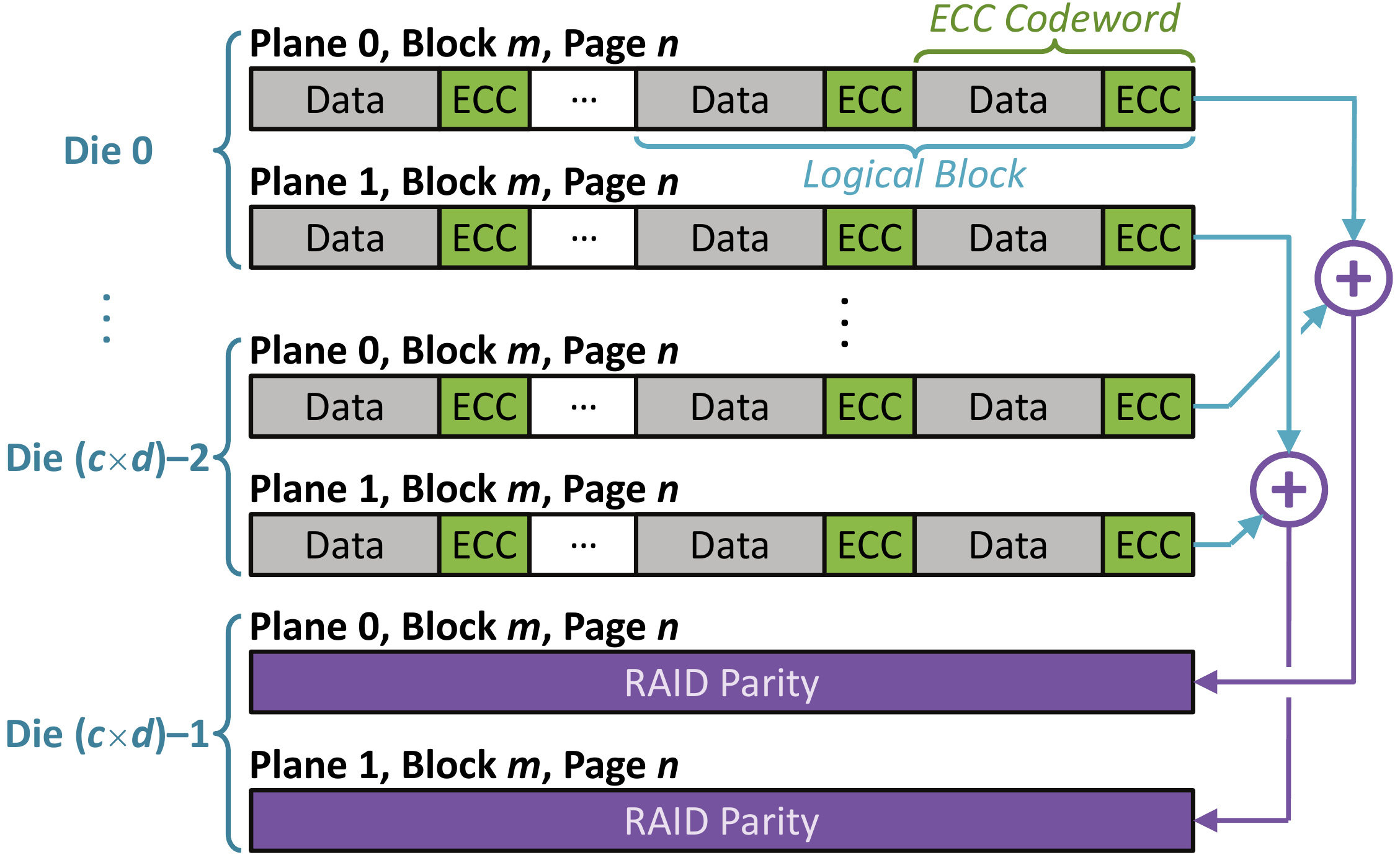}%
  \caption[Example layout of ECC codewords, logical blocks, and
superpage-level parity for superpage \emph{n} in superblock \emph{m}. In this
example, we assume that a logical block contains two codewords.]
  {Example layout of ECC codewords, logical blocks, and
superpage-level parity for superpage \emph{n} in superblock \emph{m}. In this
example, we assume that a logical block contains two codewords. \chI{Reproduced from~\cite{cai.arxiv17}.}}%
  \label{fig:F4}%
\end{figure}

In addition to ECC to protect
against bit-level errors, many SSDs employ RAID-like parity~\cite{dirik.isca09, kang.emsoft06, micron.tn05, patterson.sigmod88}. 
The key idea is to store parity information within
each superpage to protect data from ECC failures that occur
within a single chip or plane. Figure~\ref{fig:F4} shows an example of
how the ECC and parity information are organized within
a superpage. For a superpage that spans across multiple
chips, dies, and planes, the pages stored within one die or
one plane (depending on the implementation) are used to
store parity information for the remaining pages. Without
loss of generality, we assume for the rest of this section that
a superpage that spans $c$ chips and $d$ dies per chip stores parity
information in the pages of a single die (which we call
the \emph{parity die}), and that it stores user data in the pages of
the remaining $(c \times d) - 1$ dies. When all of the user data is
written to the superpage, the SSD controller XORs the data
together one plane at a time (e.g., in Figure~\ref{fig:F4}, all of the pages
in Plane~0 are XORed with each other), which produces the
parity data for that plane. This parity data is written to the
corresponding plane in the parity die, e.g., Plane~0 page in
Die $(c \times d) - 1$ in the figure.

\FloatBarrier

The SSD controller invokes superpage-level parity when
an ECC failure occurs during a host software (e.g., OS, file
system) access to the SSD. The host software accesses data
at the granularity of a \emph{logical block} (LB), which is indexed
by a \emph{logical block address} (LBA). Typically, an LB is \SI{4}{\kilo\byte}
in size, and consists of several ECC codewords (which are
usually \SI{512}{\byte}B to \SI{2}{\kilo\byte} in size) stored consecutively within
a flash memory page, as shown in Figure~\ref{fig:F4}. During the LB
access, a read failure can occur for one of two reasons.
First, it is possible that the LB data is stored within a \emph{hidden}
GBB (i.e., a GBB that has not yet been detected and
excluded by the bad block manager). The probability of
storing data in a hidden GBB is quantified as $P_{HGBB}$. Note
that because bad block management successfully identifies
and excludes most GBBs, $P_{HGBB}$ is much lower than the
total fraction of GBBs within an SSD. Second, it is possible
that at least one ECC codeword within the LB has \emph{failed}
(i.e., the codeword contains an error that cannot be corrected
by ECC). The probability that a codeword fails is
$P_{ECFR}$ (see \chI{Section~\ref{sec:ssdarch:ctrl:ecc}}). For an LB that contains $K$ ECC
codewords, we can model $P_{LBFail}$, the overall probability
that an LB access fails (i.e., the rate at which superpage-level
parity needs to be invoked), as:
\begin{equation}
P_{LBFail} = P_{HGBB} + [1 - P_{HGBB}] \times [1 - (1 - P_{ECFR})^K]
\label{eq:E2}
\end{equation}
In Equation~\ref{eq:E2}, $P_{LBFail}$ consists of (1)~the probability that an LB is
inside a hidden GBB (left side of the addition); and (2)~for
an LB that is not in a hidden GBB, the probability of any
codeword failing (right side of the addition).

When a read failure occurs for an LB in plane $p$, the SSD
controller reconstructs the data using the other LBs in the
same superpage. To do this, the controller reads the LBs
stored in plane $p$ in the other $(c \times d) - 1$ dies of the superpage,
including the LBs in the parity die. The controller
then XORs all of these LBs together, which retrieves the
data that was originally stored in the LB whose access failed.
In order to correctly recover the failed data, all of the LBs
from the $(c \times d) - 1$ dies must be correctly read. The overall
superpage-level parity failure probability $P_{parity}$ (i.e., the
probability that more than one LB contains a failure) for an
SSD with $c$ chips of flash memory, with $d$ dies per chip, can
be modeled as~\cite{patterson.sigmod88}:
\begin{equation}
P_{parity} = P_{LBFail} \times \lbrack 1 - (1 - P_{LBFail})^{(c \times d) - 1} \rbrack
\label{eq:E3}
\end{equation}
Thus, by designating one of the dies to contain parity information
(in a fashion similar to RAID 4~\cite{patterson.sigmod88}), the SSD can
tolerate the \emph{complete failure} of the superpage data in one die
without experiencing data loss during an LB access.

\subsection{Design Tradeoffs for Reliability}
\label{sec:ssdarch:reliability}

Several design decisions impact the SSD \emph{lifetime} (i.e.,
the duration of time that the SSD can be used within a
bounded probability of error without exceeding a given
performance overhead). To capture the tradeoff between
these decisions and lifetime, SSD manufacturers use the
following model:
\begin{equation}
\text{Lifetime (Years)} = \frac{\text{PEC} \times (1 + \text{OP})}{365 \times \text{DWPD} \times \text{WA} \times R_{compress}}
\label{eq:E4}
\end{equation}

In Equation~\ref{eq:E4}, the numerator is the total number of full drive writes
the SSD can endure (i.e., for a drive with an $X$-byte capacity,
the number of times $X$ bytes of data can be written). The number
of full drive writes is calculated as the product of PEC, the
total P/E cycle \emph{endurance} of each flash block (i.e., the number
of P/E cycles the block can sustain before its raw error rate
exceeds the ECC correction capability), and $1+\text{OP}$, where OP
is the \emph{overprovisioning factor} selected by the manufacturer.
Manufacturers overprovision the flash drive by providing
more physical block addresses, or PBAs, to the SSD controller
than the \emph{advertised capacity} of the drive, i.e., the number of
logical block addresses (LBAs) available to the operating system.
Overprovisioning improves performance and endurance,
by providing additional free space in the SSD so that maintenance
operations can take place without stalling host requests.
OP is calculated as:
\begin{equation}
\text{OP} = \frac{\text{PBA count} - \text{LBA count}}{\text{LBA count}}
\label{eq:E5}
\end{equation}

The denominator in Equation~\ref{eq:E4} is the number of full drive writes
per year, which is calculated as the product of days per year
(i.e., 365), DWPD, and the ratio between the total size of
the data written to flash media and the size of the data sent
by the host (i.e., $\text{WA} \times R_{compress}$). DWPD is the number of
full disk writes per day (i.e., the number of times per day the
OS writes the advertised capacity's worth of data). DWPD
is typically less than 1 for read-intensive applications, and
could be greater than 5 for write-intensive applications~\cite{cai.iccd12}.
WA (\emph{write amplification}) is the ratio between the amount
of data written into NAND flash memory by the controller
over the amount of data written by the host machine. Write
amplification occurs because various procedures (e.g.,
garbage collection~\cite{chang.tecs04, yang.smartcomp14}; and remapping-based refresh,
Section~\ref{sec:mitigation:refresh}) in the SSD perform additional writes in the
background. For example, when garbage collection selects a
block to erase, the pages that are remapped to a new block
require background writes. $R_{compress}$, or the compression
ratio, is the ratio between the size of the compressed data
and the size of the uncompressed data, and is a function of
the entropy of the stored data and the efficiency of the compression
algorithms employed in the SSD controller. In Equation~\ref{eq:E4},
DWPD and $R_{compress}$ are largely determined by the workload
and data compressibility, and cannot be changed to optimize
flash lifetime. For controllers that do not implement
compression, we set R compress to 1. However, the SSD controller
can trade off other parameters between one another
to optimize flash lifetime. We discuss the most salient tradeoffs
next.

\begin{figure}[h]
  \centering
  \includegraphics[width=0.65\columnwidth]{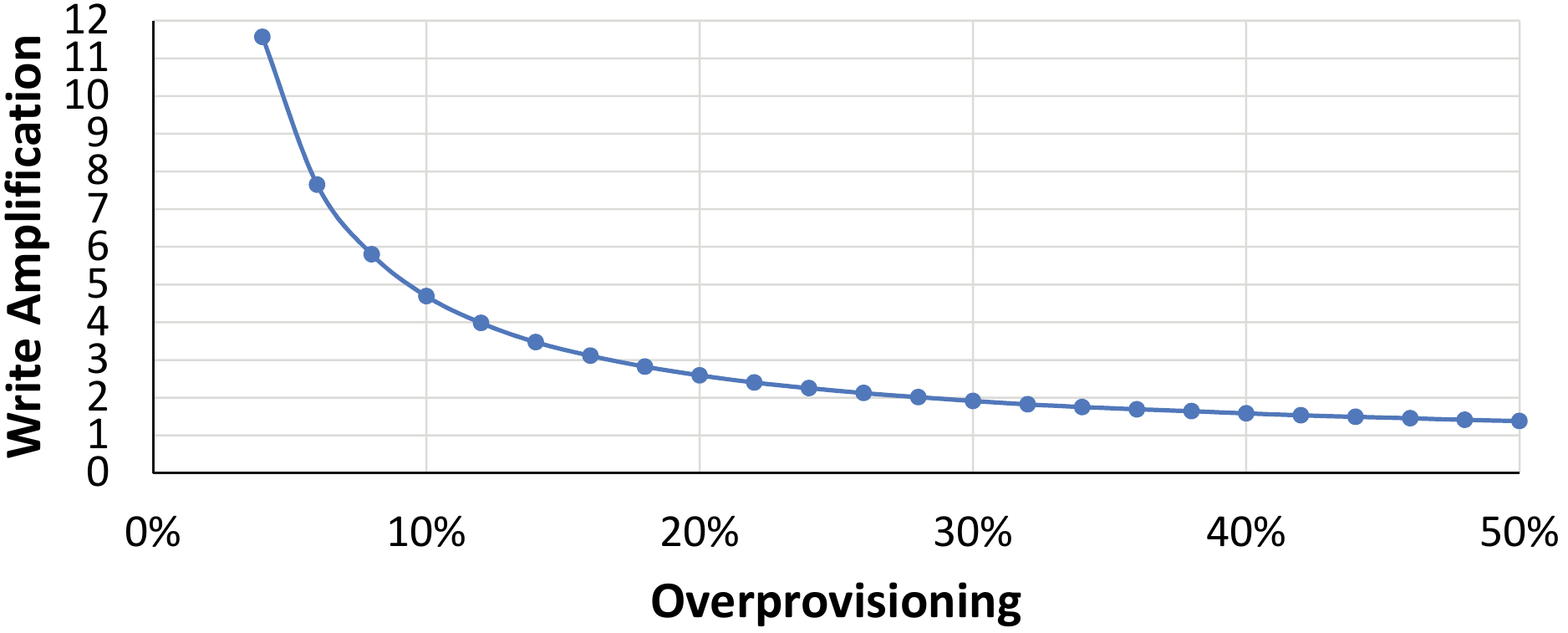}%
  \caption[Relationship between write amplification (WA) and the
overprovisioning factor (OP).]
  {Relationship between write amplification (WA) and the
overprovisioning factor (OP). \chI{Reproduced from~\cite{cai.arxiv17}.}}%
  \label{fig:F5}%
\end{figure}

\paratitle{Tradeoff Between Write Amplification and Overprovisioning}
As mentioned in \chI{Section~\ref{sec:ssdarch:ctrl:gc}}, due to the
granularity mismatch between flash erase and program
operations, garbage collection occasionally remaps remaining
valid pages from a selected block to a new flash block,
in order to avoid block-internal fragmentation. This remapping
causes additional flash memory writes, leading to
\emph{write amplification}. In an SSD with more overprovisioned
capacity, the amount of write amplification decreases,
as the blocks selected for garbage collection are older
and tend to have fewer valid pages. For a greedy garbage collection
algorithm and a random-access workload, the correlation
between WA and OP can be calculated~\cite{desnoyers.systor12, hu.systor09}, as
shown in Figure~\ref{fig:F5}. In an ideal SSD, both WA and OP should
be minimal, i.e., WA = 1 and OP = 0\%, but in reality there
is a tradeoff between these parameters: when one increases,
the other decreases. As Figure~\ref{fig:F5} shows, WA can be reduced by
increasing OP, and with an infinite amount of OP, WA converges
to 1. However, the reduction of WA is smaller when
OP is large, resulting in diminishing returns.

\FloatBarrier

In reality, the relationship between WA and OP is also a
function of the storage space utilization of the SSD. When the
storage space is \emph{not} fully utilized, many more pages are available,
reducing the need to invoke garbage collection, and thus
WA can approach 1 without the need for a large amount of OP.

\paratitle{Tradeoff Between P/E Cycle Endurance and Overprovisioning}
PEC and OP can be traded against each
other by adjusting the amount of redundancy used for error
correction, such as ECC and superpage-level parity (as discussed
in \chI{Section~\ref{sec:ssdarch:ctrl:parity}}). As the error correction capability
increases, PEC increases because the SSD can tolerate the
higher raw bit error rate that occurs at a higher P/E cycle
count. However, this comes at a cost of reducing the amount
of space available for OP, since a stronger error correction
capability requires higher redundancy (i.e., more space).
Table~\ref{tbl:T1} shows the corresponding OP for four different error
correction configurations for an example SSD with \SI{2.0}{\tera\byte}
of advertised capacity and \SI{2.4}{\tera\byte} (20\% extra) of physical
space. In this table, the top two configurations use ECC-1
with a coding rate of 0.93, and the bottom two configurations
use ECC-2 with a coding rate of 0.90, which has higher
redundancy than ECC-1. Thus, the ECC-2 configurations
have a lower OP than the top two. ECC-2, with its higher
redundancy, can correct a greater number of raw bit errors,
which in turn increases the P/E cycle endurance of the SSD.
Similarly, the two configurations with superpage-level parity
have a lower OP than configurations without superpage-level
parity, as parity uses a portion of the overprovisioned
space to store the parity bits.

\begin{table}[h]
\centering
\small
\setlength{\tabcolsep}{0.5em}
\caption{Tradeoff between strength of error correction configuration
and amount of SSD space left for overprovisioning.}
\label{tbl:T1}
\begin{tabular}{|c|c|}
\hline
\textbf{Error Correction Configuration} & \textbf{Overprovisioning Factor} \\ \hhline{|=|=|}
ECC-1 (0.93), no superpage-level parity & 11.6\% \\ \hline
ECC-1 (0.93), with superpage-level parity & 8.1\% \\ \hline
ECC-2 (0.90), no superpage-level parity & 8.0\% \\ \hline
ECC-2 (0.90), with superpage-level parity & 4.6\% \\ \hline
\end{tabular}
\end{table}
\FloatBarrier

When the ECC correction strength is increased, the
amount of overprovisioning in the SSD decreases, which
in turn increases the amount of write amplification that
takes place. Manufacturers must find and use the correct
tradeoff between ECC correction strength and the overprovisioning
factor, based on which of the two is expected
to provide greater reliability for the target applications of
the SSD.


\section{NAND Flash Memory Basics}
\label{sec:flash}

A number of underlying properties of the NAND flash
memory used within the SSD affect SSD management,
performance, and reliability~\cite{mielke.irps08, brewer.book08, bez.procieee03}. In this section,
we present a primer on NAND flash memory and its
operation, to prepare the reader for understanding our
further discussion on error sources (Section~\ref{sec:errors}) and mitigation
mechanisms (Section~\ref{sec:mitigation}). Recall from Section~\ref{sec:ssdarch:flash}
that within each plane, flash cells are organized as multiple
2D arrays known as flash blocks, each of which
contains multiple pages of data, where a page is the granularity
at which the host reads and writes data. We first
discuss how data is stored in NAND flash memory. We
then introduce the three basic operations supported by
NAND flash memory: read, program, and erase.

\subsection{Storing Data in a Flash Cell}
\label{sec:flash:data}

\begin{figure}[h]
  \centering
  \includegraphics[width=0.4\columnwidth]{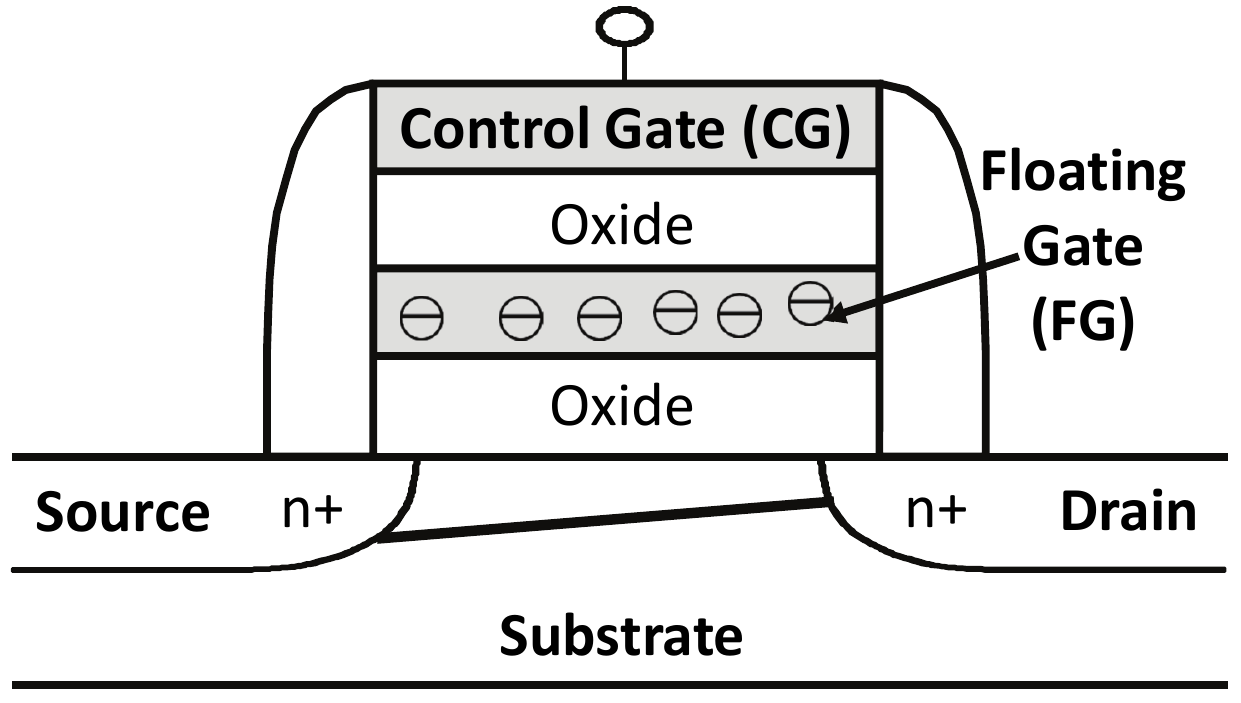}%
  \caption[Flash cell (i.e., floating gate transistor) cross section.]
  {Flash cell (i.e., floating gate transistor) cross section. \chI{Reproduced from~\cite{cai.arxiv17}.}}%
  \label{fig:F6}%
\end{figure}

NAND flash memory stores data as the \emph{threshold voltage}
of each flash cell, which is made up of a \emph{floating gate
transistor}. Figure~\ref{fig:F6} shows a cross section of a floating gate
transistor. On top of a flash cell is the \emph{control gate} (CG) and
below is the floating gate (FG). The floating gate is insulated
on both sides, on top by an inter-poly oxide layer and at the
bottom by a tunnel oxide layer. As a result, the electrons
programmed on the floating gate do not discharge even
when flash memory is powered off.

\FloatBarrier

For \emph{single-level cell} (SLC) NAND flash, each flash cell
stores a 1-bit value, and can be programmed to one of two
threshold voltage states, which we call the ER and P1 states.
\emph{Multi-level cell} (MLC) NAND flash stores a 2-bit value in each
cell, with four possible states (ER, P1, P2, and P3), and \emph{triple-level
cell} (TLC) NAND flash stores a 3-bit value in each cell
with eight possible states (ER, P1--P7). Each state represents
a different value, and is assigned a \emph{voltage window} within
the range of all possible threshold voltages. Due to variation
across \emph{program} operations, the threshold voltage of flash cells
programmed to the same state is initially distributed across
this voltage window.

\begin{figure}[h]
  \centering
  \includegraphics[width=0.75\columnwidth]{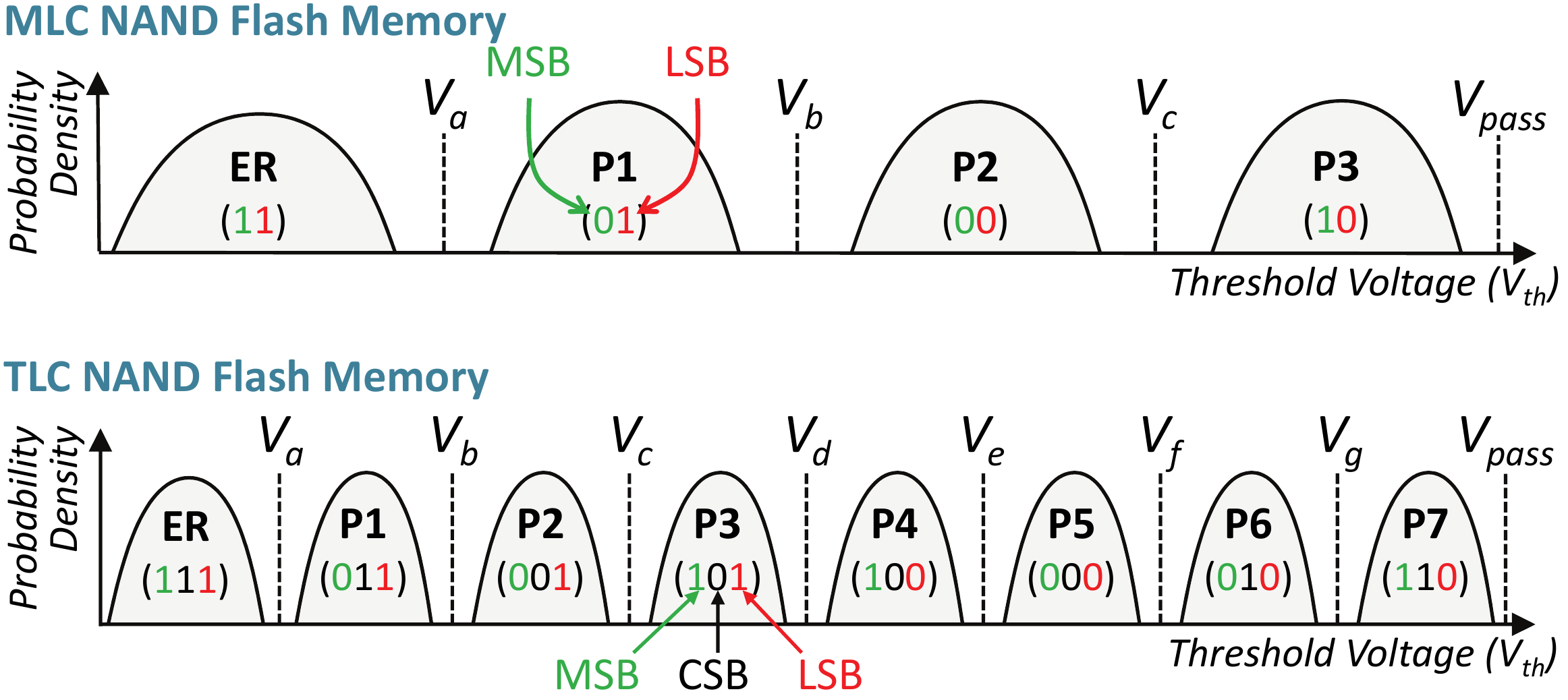}%
  \caption[Threshold voltage distribution of MLC (top) and TLC (bottom)
NAND flash memory.]
  {Threshold voltage distribution of MLC (top) and TLC (bottom)
NAND flash memory. \chI{Reproduced from~\cite{cai.arxiv17}.}}%
  \label{fig:F7}%
\end{figure}

Figure~\ref{fig:F7} illustrates the threshold voltage distribution of
MLC (top) and TLC (bottom) NAND flash memories. The
x-axis shows the threshold voltage ($V_{th}$), which spans a certain
voltage range. The y-axis shows the probability density
of each voltage level across all flash memory cells. The
threshold voltage distribution of each threshold voltage
state can be represented as a probability density curve that
spans over the state's voltage window.

\FloatBarrier

We label the distribution curve for each state with the
name of the state and a corresponding bit value. Note that
some manufacturers may choose to use a different mapping
of values to different states. The bit values of adjacent
states are separated by a Hamming distance of 1. We break
down the bit values for MLC into the most significant bit
(MSB) and least significant bit (LSB), while TLC is broken
down into the MSB, the center significant bit (CSB), and
the LSB. The boundaries between neighboring threshold
voltage windows, which are labeled as $V_a$, $V_b$, and $V_c$ for the
MLC distribution in Figure~\ref{fig:F7}, are referred to as \emph{read reference
voltages}. These voltages are used by the SSD controller to
identify the voltage window (i.e., state) of each cell upon
reading the cell.

\subsection{Flash Block Design}
\label{sec:flash:block}

\begin{figure}[h]
  \centering
  \includegraphics[width=0.5\columnwidth]{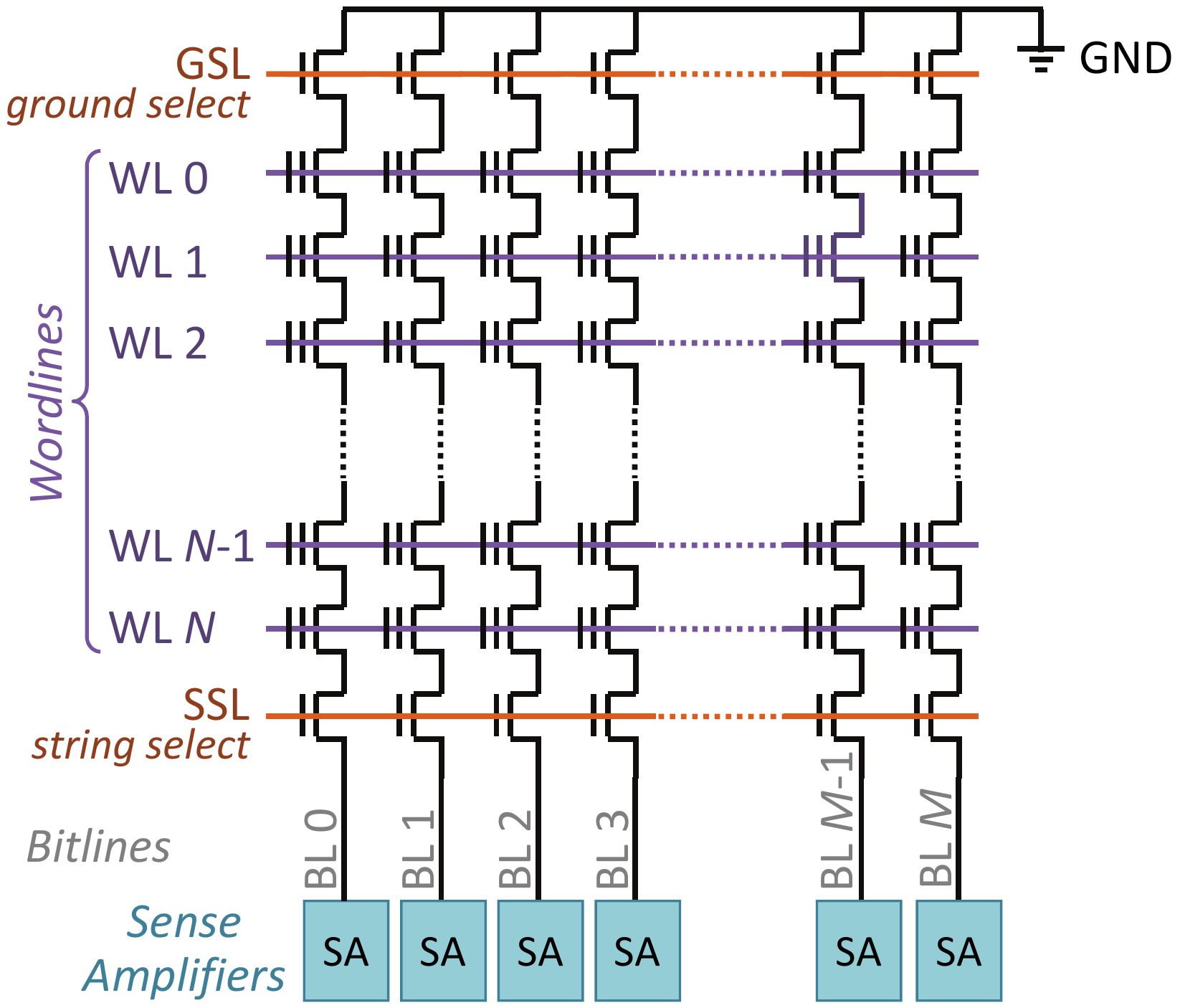}%
  \caption[Internal organization of a flash block.]
  {Internal organization of a flash block. \chI{Reproduced from~\cite{cai.arxiv17}.}}%
  \label{fig:F8}%
\end{figure}

Figure~\ref{fig:F8} shows the high-level internal organization of a
NAND flash memory block. Each block contains multiple
rows of cells (typically 128--512 rows). Each row of cells is
connected together by a common \emph{wordline} (WL, shown horizontally
in Figure~\ref{fig:F8}), typically spanning 32K--64K cells. All of
the cells along the wordline are logically combined to form
a page in an SLC NAND flash memory. For an MLC NAND
flash memory, the MSBs of all cells on the same wordline are
combined to form an \emph{MSB page}, and the LSBs of all cells on
the wordline are combined to form an \emph{LSB page}. Similarly,
a TLC NAND flash memory logically combines the MSBs
on each wordline to form an MSB page, the CSBs on each
wordline to form a \emph{CSB page}, and the LSBs on each wordline
to form an LSB page. In MLC NAND flash memory, each
flash block contains 256--1024 flash pages, each of which
are typically \SIrange{8}{16}{\kilo\byte} in size.

\FloatBarrier

Within a block, all cells in the same column are connected
in series to form a \emph{bitline} (BL, shown vertically in
Figure~\ref{fig:F8}) or \emph{string}. All cells in a bitline share a common ground
(GND) on one end, and a common \emph{sense amplifier} (SA) on
the other for reading the threshold voltage of one of the cells
when decoding data. Bitline operations are controlled by
turning the \emph{ground select line} (GSL) and \emph{string select line}
(SSL) transistor of each bitline on or off. The SSL transistor
is used to enable operations on a bitline, and the GSL
transistor is used to connect the bitline to ground during a
read operation~\cite{mohan.thesis10}. The use of a common bitline across
multiple rows reduces the amount of circuit area required
for read and write operations to a block, improving storage
density.

\subsection{Read Operation}
\label{sec:flash:read}

Data can be read from NAND flash memory by applying
read reference voltages onto the control gate of each cell, to
sense the cell's threshold voltage. To read the value stored
in a single-level cell, we need to distinguish only the state
with a bit value of 1 from the state with a bit value of 0.
This requires us to use only a single read reference voltage.
Likewise, to read the LSB of a multi-level cell, we need to
distinguish only the states where the LSB value is 1 (ER and
P1) from the states where the LSB value is 0 (P2 and P3),
which we can do with a single read reference voltage ($V_b$ in
the top half of Figure~\ref{fig:F7}). To read the MSB page, we need to distinguish
the states with an MSB value of 1 (ER and P3) from
those with an MSB value of 0 (P1 and P2). Therefore, we
need to determine whether the threshold voltage of the cell
falls between $V_a$ and $V_c$, requiring us to apply each of these
two read reference voltages (which can require up to two
consecutive read operations) to determine the MSB.

Reading data from a triple-level cell is similar to the data
read procedure for a multi-level cell. Reading the LSB for TLC
again requires applying only a single read reference voltage
($V_d$ in the bottom half of Figure~\ref{fig:F7}). Reading the CSB requires two
read reference voltages to be applied, and reading the MSB
requires four read reference voltages to be applied.

\begin{figure}[h]
  \centering
  \includegraphics[width=0.65\columnwidth]{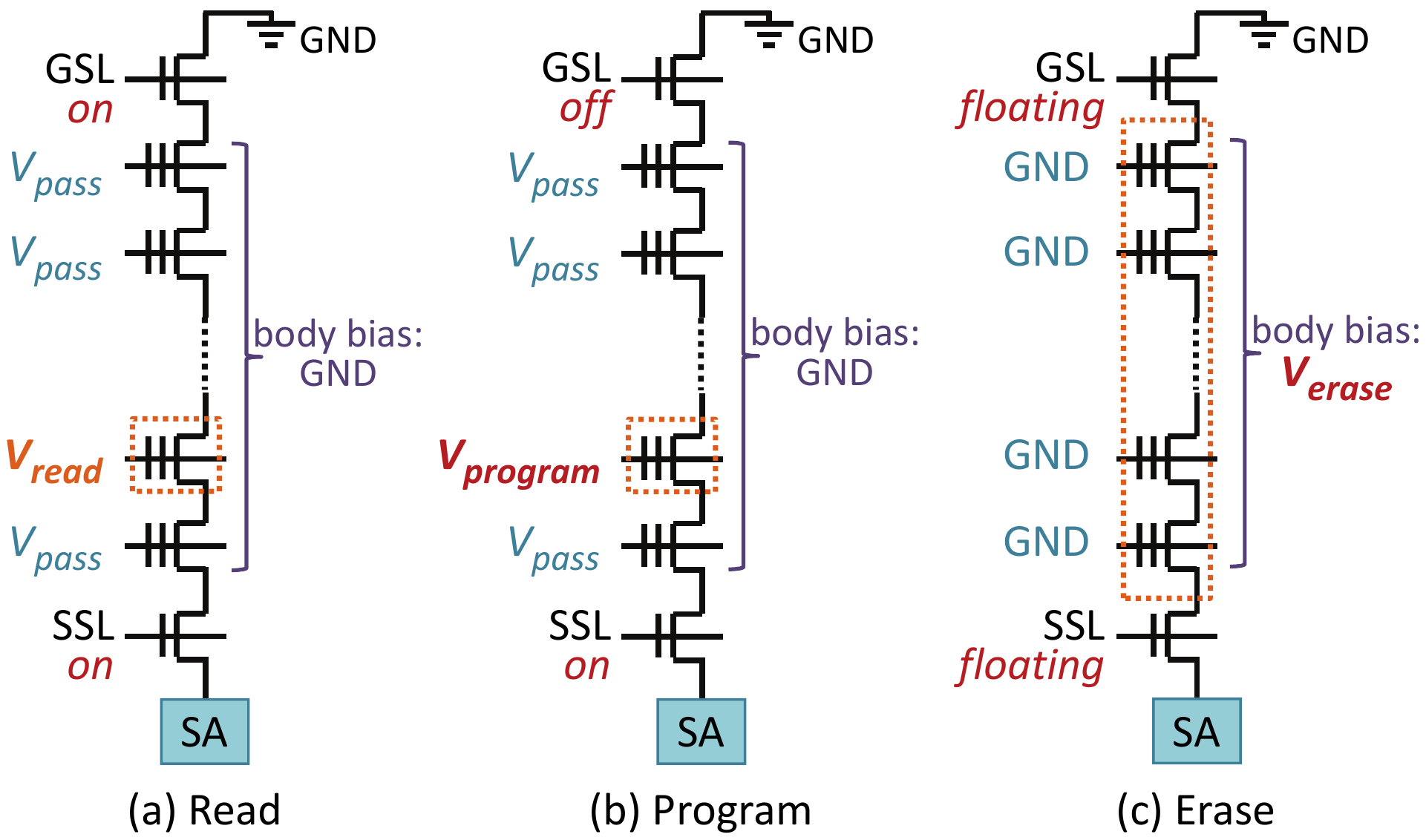}%
  \caption[Voltages applied to flash cell transistors on a bitline to
perform (a)~read, (b)~program, and (c)~erase operations.]
  {Voltages applied to flash cell transistors on a bitline to
perform (a)~read, (b)~program, and (c)~erase operations. \chI{Reproduced from~\cite{cai.arxiv17}.}}%
  \label{fig:F9}%
\end{figure}

As Figure~\ref{fig:F8} shows, cells from multiple wordlines (WL in the
figure) are connected in series on a \emph{shared} bitline (BL) to the
sense amplifier, which drives the value that is being read from
the block onto the memory channel for the plane. In order to
read from a single cell on the bitline, \emph{all of the other cells} (i.e.,
\emph{unread} cells) on the same bitline must be switched on to allow
the value that is being read to propagate through to the sense
amplifier. The NAND flash memory achieves this by applying
the \emph{pass-through voltage} onto the wordlines of the unread cells,
as shown in Figure~\ref{fig:F9}a. When the pass-through voltage (i.e., the
maximum possible threshold voltage $V_{pass}$) is applied to a flash
cell, the source and the drain of the cell transistor are connected,
regardless of the voltage of the floating gate. Modern
flash memories guarantee that all \emph{unread} cells are \emph{passed through}
to minimize errors during the read operation~\cite{cai.dsn15}.

\FloatBarrier

\subsection{Program and Erase Operations}
\label{sec:flash:pgmerase}

The threshold voltage of a floating gate transistor is controlled
through the injection and ejection of electrons through
the tunnel oxide of the transistor, which is enabled by the
Fowler–Nordheim (FN) tunneling effect~\cite{fowler.royalsociety28, bez.procieee03, pavan.procieee97}. The
tunneling current ($J_{FN}$)~\cite{brewer.book08, pavan.procieee97} can be modeled as:
\begin{equation}
J_{FN} = \alpha_{FN} E_{ox}^2 e^{-\beta_{FN} / E_{ox}}
\label{eq:E6}
\end{equation}
In Equation~\ref{eq:E6}, $\alpha_{FN}$ and $\beta_{FN}$ are constants, and $E_{ox}$ is the electric field
strength in the tunnel oxide. As Equation~\ref{eq:E6} shows, $J_{FN}$ is exponentially
correlated with $E_{ox}$.

During a program operation, electrons are injected into
the floating gate of the flash cell from the substrate when
applying a high positive voltage to the control gate (see Figure~\ref{fig:F6}
for a diagram of the flash cell). The pass-through voltage is
applied to all of the other cells on the same bitline as the
cell that is being programmed as shown in Figure~\ref{fig:F9}b. When
data is programmed, charge is transferred into the floating
gate through FN tunneling by repeatedly pulsing the programming
voltage, in a procedure known as \emph{incremental
step-pulse programming} (ISPP)~\cite{mielke.irps08, suh.jssc95, bez.procieee03, wang.ics14}. During
ISPP, a high programming voltage ($V_{program}$) is applied for
a very short period, which we refer to as a \emph{step-pulse}. ISPP
then verifies the current voltage of the cell using the voltage
$V_{verify}$. ISPP repeats the process of applying a step-pulse and
verifying the voltage until the cell reaches the desired target
voltage. In the modern all-bitline NAND flash memory,
all flash cells in a single wordline are programmed concurrently.
During programming, when a cell along the wordline
reaches its target voltage but other cells have yet to reach
their target voltage, ISPP \emph{inhibits} programming pulses to
the cell by turning off the SSL transistor of the cell's bitline.

In SLC NAND flash and older MLC NAND flash, \emph{one-shot
programming} is used, where all of the ISPP step-pulses
required to program a cell are applied back to back until all
cells in the wordline are fully programmed. One-shot programming
does \emph{not} interleave the program operations to
a wordline with the program operations to another wordline.
In newer MLC NAND flash, the lack of interleaving
between program operations can introduce a significant
amount of cell-to-cell program interference on the cells of
immediately-adjacent wordlines (see Section~\ref{sec:errors:celltocell}).

\begin{figure}[h]
  \centering
  \includegraphics[width=0.55\columnwidth]{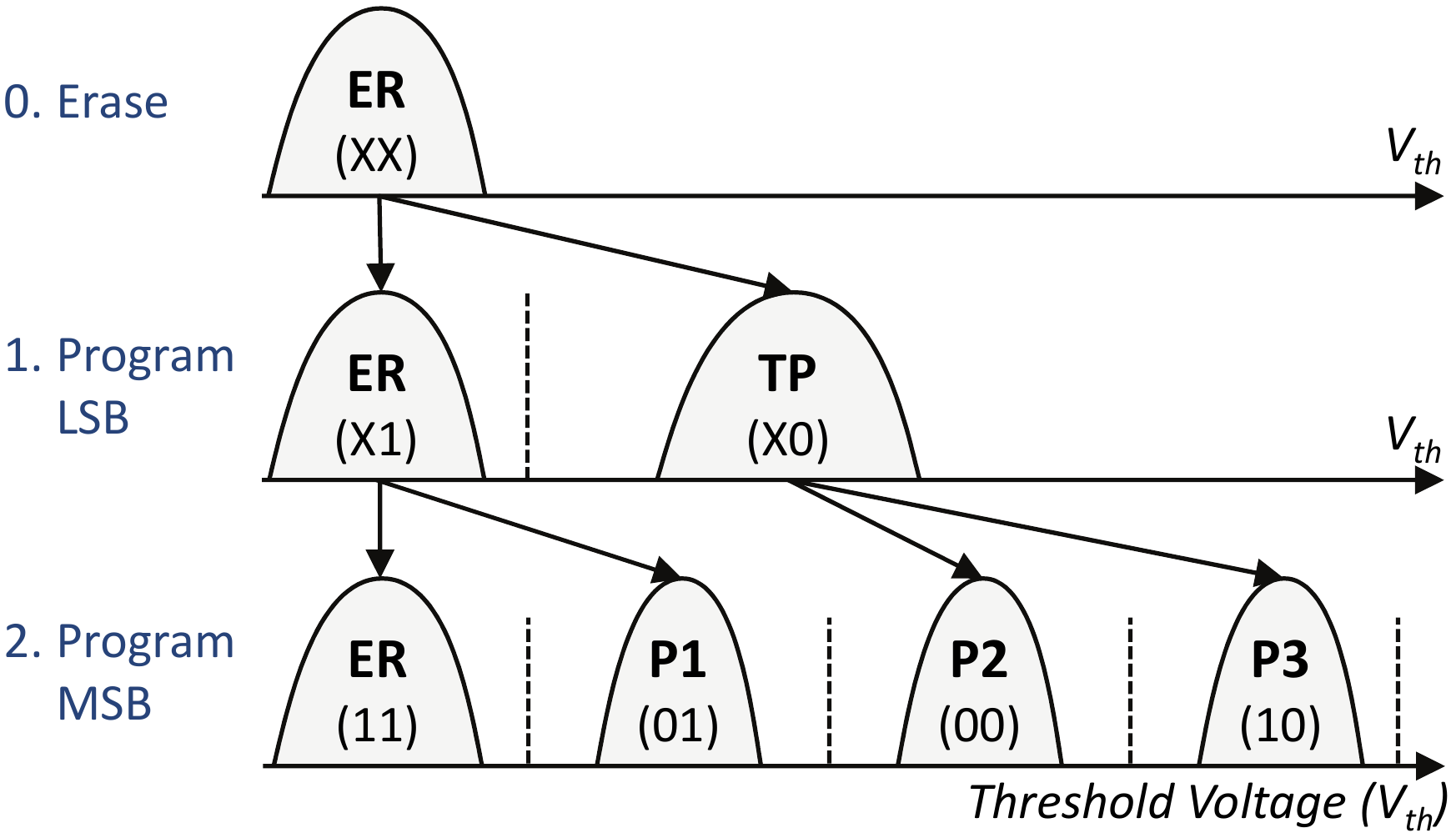}%
  \caption[Two-step programming algorithm for MLC flash.]
  {Two-step programming algorithm for MLC flash. \chI{Reproduced from~\cite{cai.arxiv17}.}}%
  \label{fig:F10}%
\end{figure}

To reduce the impact of program interference, the controller
employs \emph{two-step programming} for sub-\SI{40}{\nano\meter} MLC
NAND flash~\cite{park.jssc08, cai.iccd13}: it first programs the LSBs into the
erased cells of an unprogrammed wordline, and then programs
the MSBs of the cells using a separate program operation~\cite{park.jssc08, park.dac16, cai.date13, cai.hpca17}. 
Between the programming of the
LSBs and the MSBs, the controller programs the LSBs of
the cells in the wordline immediately above~\cite{park.jssc08, park.dac16, cai.date13, cai.hpca17}. 
Figure~\ref{fig:F10} illustrates the two-step programming algorithm.
In the first step, a flash cell is \emph{partially programmed}
based on its LSB value, either staying in the ER state if the
LSB value is 1, or moving to a temporary state (TP) if the LSB
value is 0. The TP state has a mean voltage that falls between
states P1 and P2. In the second step, the LSB data is first
read back into an internal buffer register within the flash
chip to determine the cell's current threshold voltage state,
and then further programming pulses are applied based on
the MSB data to increase the cell's threshold voltage to fall
within the voltage window of its final state. Programming
in MLC NAND flash is discussed in detail in~\cite{cai.hpca17} and~\cite{cai.date13}.

\FloatBarrier

\begin{figure}[h]
  \centering
  \includegraphics[width=0.7\columnwidth]{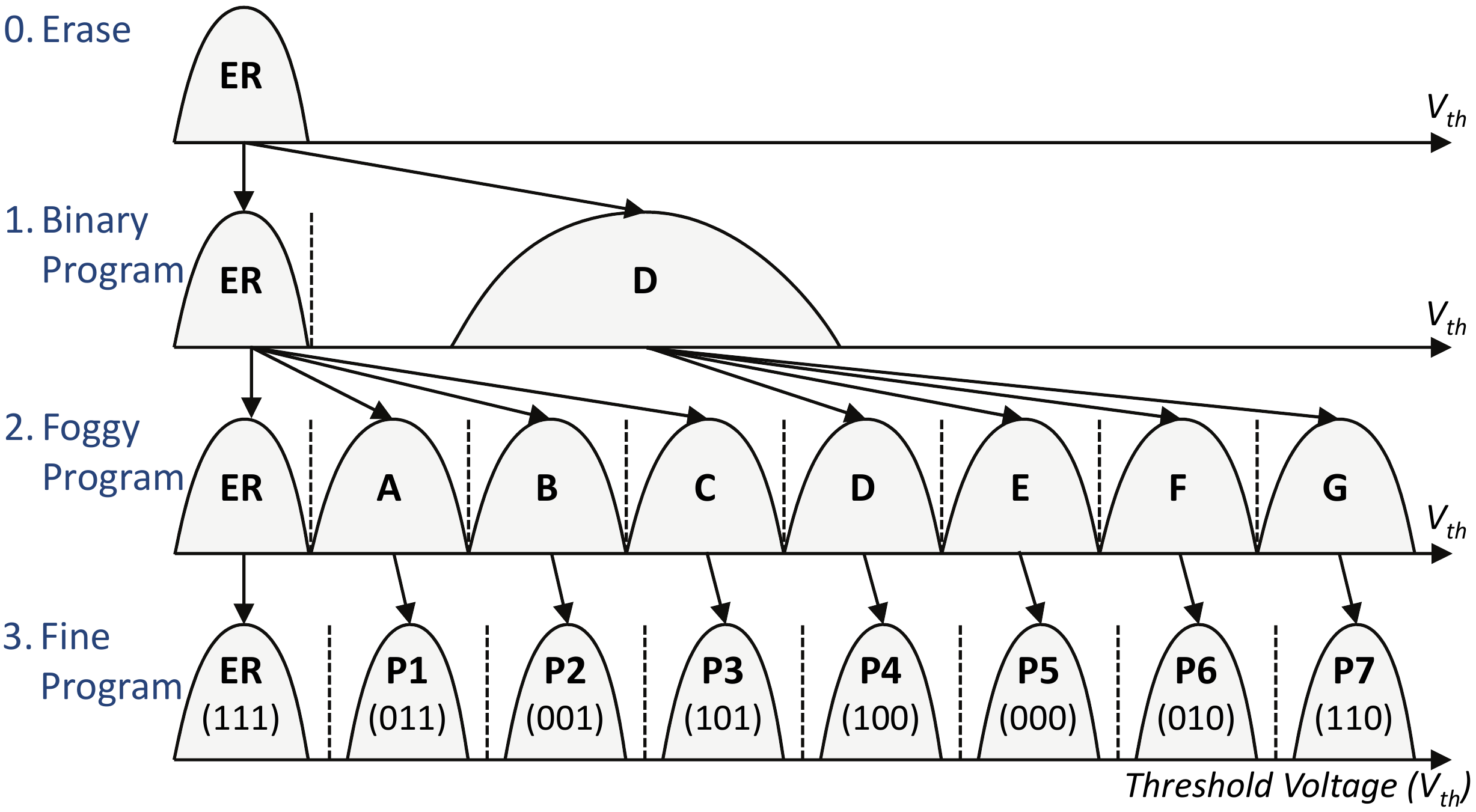}%
  \caption[Foggy-fine programming algorithm for TLC flash.]
  {Foggy-fine programming algorithm for TLC flash. \chI{Reproduced from~\cite{cai.arxiv17}.}}%
  \label{fig:F11}%
\end{figure}

TLC NAND flash takes a similar approach to the two-step
programming of MLC, with a mechanism known as
\emph{foggy-fine programming}~\cite{li.patent14}, which is illustrated in Figure~\ref{fig:F11}.
The flash cell is first partially programmed based on its LSB
value, using a \emph{binary} programming step in which very large
ISPP step-pulses are used to significantly increase the voltage
level. Then, the flash cell is partially programmed again based
on its CSB and MSB values to a new set of temporary states
(these steps are referred to as \emph{foggy} programming, which uses
smaller ISPP step-pulses than binary programming). Due to
the higher potential for errors during TLC programming as a
result of the narrower voltage windows, all of the programmed
bit values are buffered after the binary and foggy programming
steps into SLC buffers that are reserved in each chip/plane. 
Finally, \emph{fine} programming takes place, where these bit
values are read from the SLC buffers, and the smallest ISPP
step-pulses are applied to set each cell to its final threshold
voltage state. The purpose of this last fine programming step
is to fine tune the threshold voltage such that the threshold
voltage distributions are tightened (bottom of Figure~\ref{fig:F11}).

\FloatBarrier

Though programming sets a flash cell to a specific
threshold voltage using programming pulses, the voltage
of the cell can drift over time after programming. When no
external voltage is applied to any of the electrodes (i.e., CG,
source, and drain) of a flash cell, an electric field still exists
between the FG and the substrate, generated by the charge
present in the FG. This is called the \emph{intrinsic electric field}~\cite{brewer.book08}, 
and it generates \emph{stress-induced leakage current} (SILC)~\cite{bez.procieee03, naruke.iedm88, degraeve.ted04}, 
a weak tunneling current that leaks charge
away from the FG. As a result, the voltage that a cell is programmed
to may not be the same as the voltage read for that
cell at a subsequent time.

In NAND flash, a cell can be reprogrammed with new
data \emph{only after} the existing data in the cell is erased. This is
because ISPP can only \emph{increase} the voltage of the cell. The
erase operation resets the threshold voltage state of \emph{all cells
in the flash block} to the ER state. During an erase operation,
electrons are ejected from the FG of the flash cell into
the substrate by inducing a high negative voltage on the cell
transistor. The negative voltage is induced by setting the CG
of the transistor to GND, and biasing the transistor body
(i.e., the substrate) to a high voltage ($V_{erase}$), as shown in
Figure~\ref{fig:F9}c. Because all cells in a flash block share a common
transistor substrate (i.e., the bodies of all transistors in the
block are connected together), a flash block must be erased
in its entirety~\cite{mohan.thesis10}.

\chapter[Flash Memory Reliability: Background and Related Work]{Flash Memory Reliability:\\
Background and Related Work}
\label{sec:related}

In this chapter, we provide the background and related work on reliability
issues in NAND flash memory. First, we provide a thorough introduction on NAND
flash memory error characteristics concluded from prior work
(Section~\ref{sec:errors}). Second, we survey state-of-the-art techniques in
modern SSDs that mitigate NAND flash memory errors
(Section~\ref{sec:mitigation}). Third, we introduce state-of-the-art error
correction and recovery techniques that tolerate NAND flash memory errors 
(Section~\ref{sec:correction}). Fourth, we introduce the state-of-the-art 3D
NAND flash memory technology and the emerging reliability issues for 3D NAND
devices using this technology (Section~\ref{sec:background:3d}). Fifth, we
discuss similar errors in other memory technologies and how we tolerate them
in modern computing systems (Section~\ref{sec:othermem}).


\section{NAND Flash Memory Error Characteristics}
\label{sec:errors}

\begin{figure}[h]
  \centering
  \includegraphics[width=0.8\columnwidth]{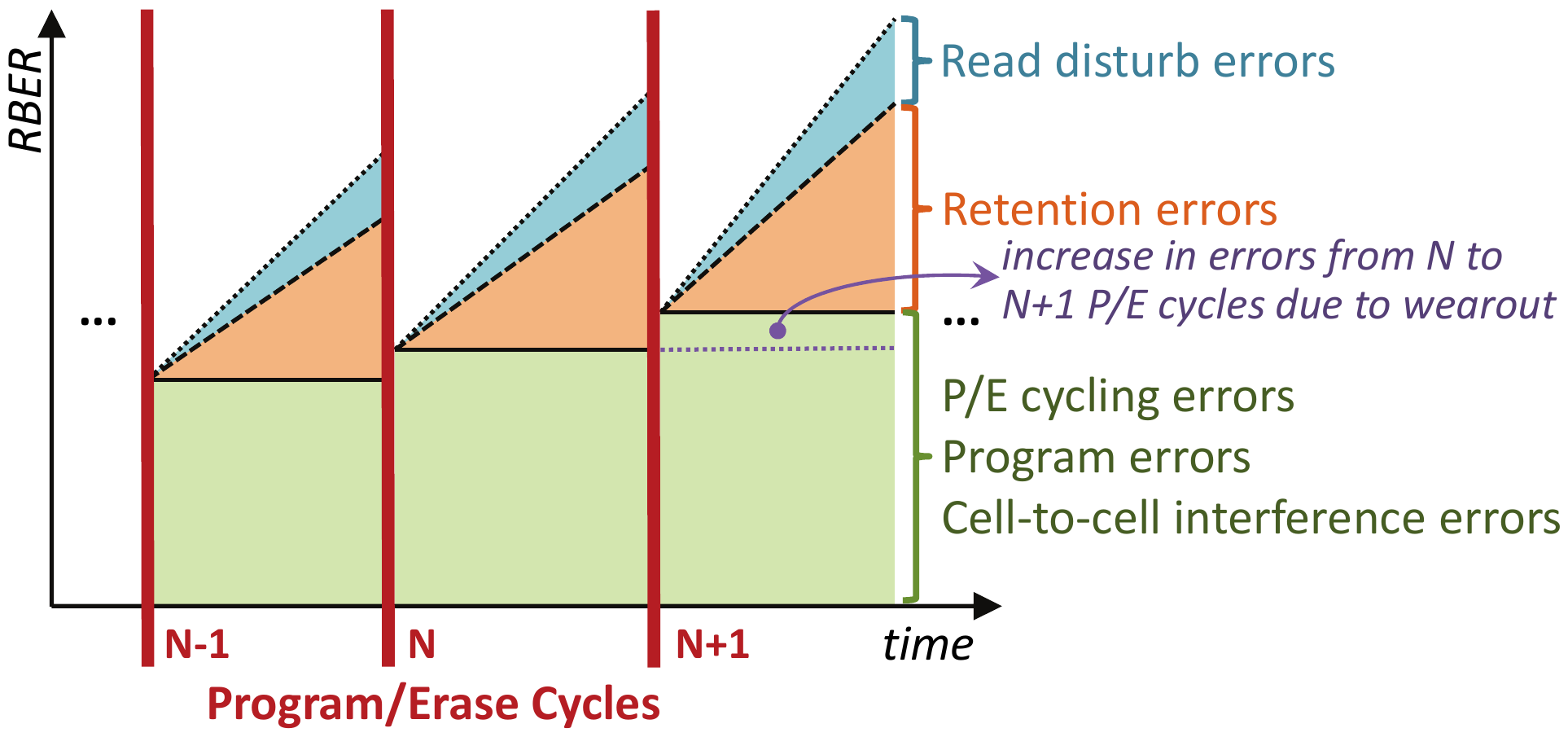}%
  \caption[Pictorial depiction of errors accumulating within a NAND
flash block as P/E cycle count increases.]
  {Pictorial depiction of errors accumulating within a NAND
flash block as P/E cycle count increases. \chI{Reproduced from~\cite{cai.arxiv17}.}}%
  \label{fig:F12}%
\end{figure}

Each block in NAND flash memory is used in a cyclic fashion,
as is illustrated by the observed raw bit error rates seen
over the lifetime of a flash memory block in Figure~\ref{fig:F12}. At the
beginning of a \emph{cycle}, known as a \emph{program/erase (P/E) cycle},
an erased block is \emph{opened} (i.e., selected for programming).
Data is then programmed into the open block one page at
a time. After all of the pages are programmed, the block is
closed, and none of the pages can be reprogrammed until
the whole block is erased. At any point before erasing, read
operations can be performed on a \emph{valid} programmed page
(i.e., a page containing data that has not been modified
by the host). A page is marked as invalid when the data
stored at that page's logical address by the host is modified.
As ISPP can only inject more charge into the floating gate
but cannot remove charge from the gate, it is not possible
to modify data to a new arbitrary value \emph{in place} within
existing NAND flash memories. Once the block is erased,
the P/E cycling behavior repeats until the block is \emph{worn out}
(i.e., the block can no longer avoid data loss over the course
of the minimum data retention period guaranteed by the
manufacturer). Although the 5x-nm (i.e., \SIrange{50}{59}{\nano\meter})
generation of MLC NAND flash could endure $\sim$10,000 P/E
cycles per block before being worn out, modern 1x-nm
(i.e., \SIrange{15}{19}{\nano\meter}) MLC and TLC NAND flash can endure
only $\sim$3,000 and $\sim$1,000 P/E cycles per block, 
respectively~\cite{parnell.globecom14, maislos.fms11, yoon.fms12, koh.imw09}.

\FloatBarrier

\begin{figure}[h]
  \centering
  \includegraphics[width=0.9\columnwidth]{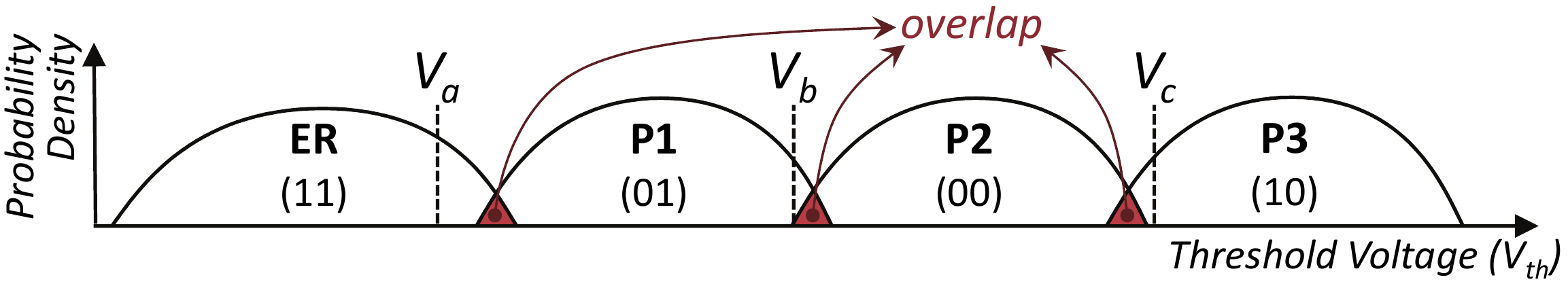}%
  \caption[Threshold voltage distribution shifts and widening can
cause the distributions of two neighboring states to overlap with
each other (compare to Figure~\ref{fig:F7}), leading to read errors.]
{Threshold voltage distribution shifts and widening can
cause the distributions of two neighboring states to overlap with
each other (compare to Figure~\ref{fig:F7}), leading to read errors. \chI{Reproduced from~\cite{cai.arxiv17}.}}%
  \label{fig:F13}%
\end{figure}

As shown in Figure~\ref{fig:F12}, several different types of errors can
be introduced at any point during the P/E cycling process:
\emph{P/E cycling errors}, \emph{program errors}, errors due to \emph{cell-to-cell program
interference}, \emph{data retention errors}, and errors due to \emph{read
disturb}. As discussed in Section~\ref{sec:flash:data}, the threshold voltage
of flash cells programmed to the same state is distributed
across a voltage window due to variation across program
operations and across different flash cells. Several types of
errors introduced during the P/E cycling process, such as
data retention and read disturb, cause the threshold voltage
distribution of each state to shift and widen. Due to the shift
and widening, the tails of the distributions of each state can
enter the margin that originally existed between each of the
two neighboring states' distributions. Thus, the threshold
voltage distributions of different states can start overlapping,
as shown in Figure~\ref{fig:F13}. When the distributions overlap
with each other, the read reference voltages can no longer
correctly identify the state of some flash cells in the overlapping
region, leading to \emph{raw bit errors} during a read operation.

\FloatBarrier

In this section, we discuss the causes of each type of error
in detail.
We later discuss mitigation techniques for these flash
memory errors in Section~\ref{sec:mitigation}, and provide procedures to
recover in the event of data loss in Section~\ref{sec:correction}.

\subsection{P/E Cycling Errors}
\label{sec:errors:pe}

A P/E cycling error occurs when either (1)~an erase operation
fails to reset a cell to the ER state; or (2)~when a program
operation fails to set the cell to the desired target state.
P/E cycling errors occur because electrons become trapped
in the tunnel oxide after stress from repeated P/E cycles.
Errors due to such electron trapping (which we refer to as
\emph{P/E cycling noise}) continue to accumulate over the lifetime
of a NAND flash block. This behavior is called \emph{wearout},
and it refers to the phenomenon where, as more writes are
performed to a block, there are a greater number of raw bit
errors that must be corrected, exhausting more of the fixed
error correction capability of the ECC (see Section~\ref{sec:ssdarch:ctrl}).

More findings on the nature of wearout
and the impact of wearout on NAND flash memory errors and
lifetime can be found in prior \chIII{works from our research group}~\cite{cai.date12, cai.date13,
luo.jsac16, cai.thesis12}.
\chV{Recent work studies P/E cycling errors and proposes to change the ER
state distribution to securely hide data in flash~\cite{zuck.fast18}.}

\subsection{Program Errors}
\label{sec:errors:pgm}

\begin{figure}[h]
  \centering
  \includegraphics[width=0.7\columnwidth]{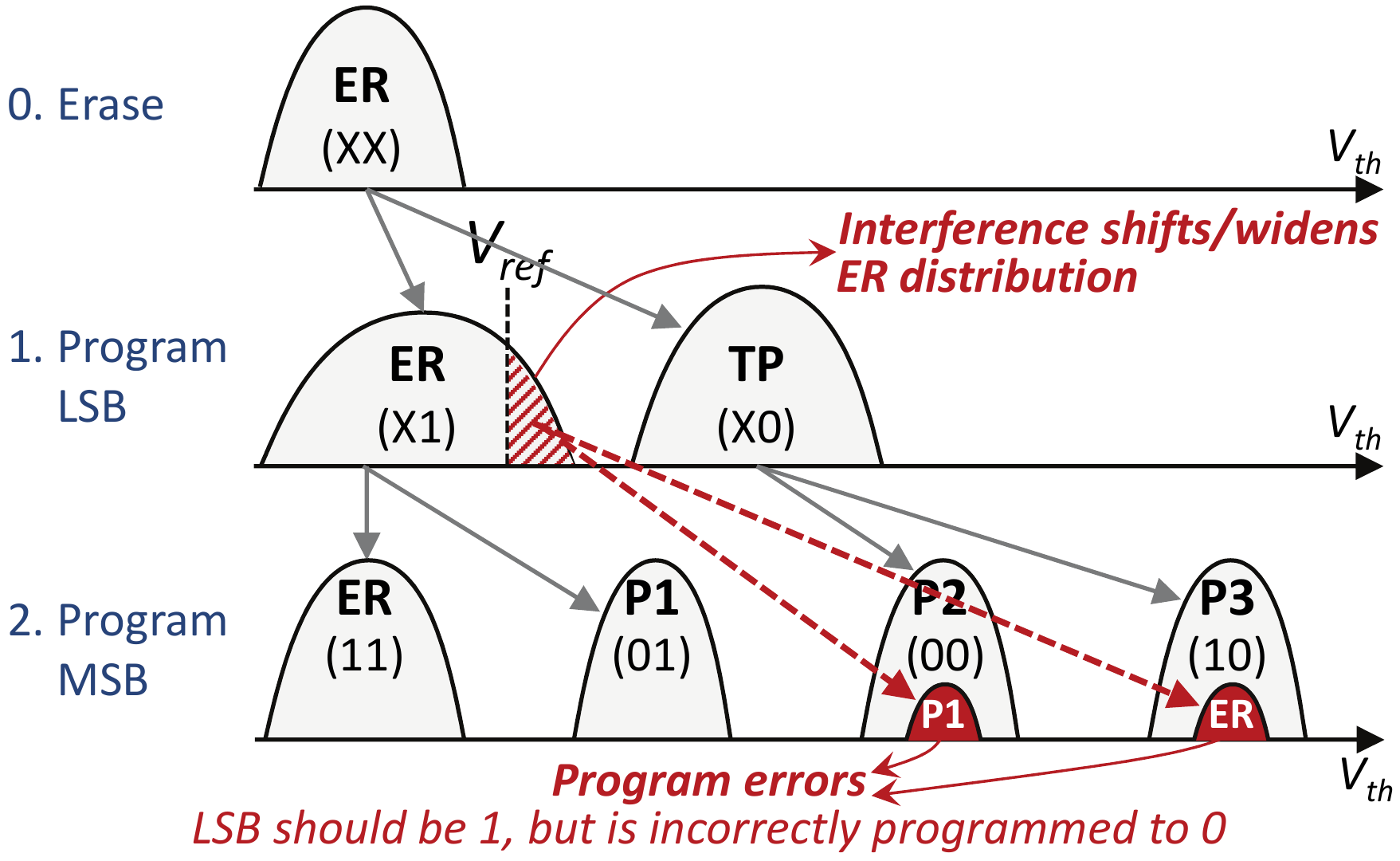}%
  \caption[Impact of program errors during two-step programming on
cell threshold voltage distribution.]
{Impact of program errors during two-step programming on
cell threshold voltage distribution. \chI{Reproduced from~\cite{cai.arxiv17}.}}%
  \label{fig:F15}%
\end{figure}

Program errors occur when data read directly from the
NAND flash array contains errors, and the erroneous values
are used to program the new data. Program errors occur in two
major cases: (1)~partial programming during two-step or foggy-fine
programming, and (2)~\emph{copyback} (i.e., when data is copied
inside the NAND flash memory during a maintenance operation)~\cite{hu.ics11}. 
During two-step programming for MLC NAND
flash memory (see Figure~\ref{fig:F10}), in between the LSB and MSB programming
steps of a cell, threshold voltage shifts can occur
on the partially-programmed cell. These shifts occur because
several other read and program operations to cells in \emph{other}
pages within the same block may take place, causing interference
to the partially-programmed cell. Figure~\ref{fig:F15} illustrates
how the threshold distribution of the ER state widens and
shifts to the right after the LSB value is programmed (step~1
in the figure). The widening and shifting of the distribution
causes some cells that were originally partially programmed
to the ER state (with an LSB value of 1) to be misread as being
in the TP state (with an LSB value of 0) during the \emph{second}
programming step (step~2 in the figure). As shown in Figure~\ref{fig:F15},
the misread LSB value leads to a program error when the
final cell threshold voltage is programmed~\cite{cai.hpca17, luo.jsac16, parnell.globecom14}.
Some cells that should have been programmed to the P1 state
(representing the value 01) are instead programmed to the
P2 state (with the value 00), and some cells that should have
been programmed to the ER state (representing the value 11)
are instead programmed to the P3 state (with the value 10).

\FloatBarrier

More findings on the nature of program errors and the
impact of program errors on NAND flash memory lifetime
can be found in prior \chIII{works from our research group}~\cite{cai.hpca17, luo.jsac16}.

\subsection{Cell-to-Cell Program Interference Errors}
\label{sec:errors:celltocell}

Program interference refers to the phenomenon where the
programming of a flash cell induces errors on adjacent flash
cells within a flash block~\cite{cai.iccd13, cai.sigmetrics14, lee.iedl02, cooke.fms07, grupp.micro09}. The interference
occurs due to \emph{parasitic capacitance coupling} between
these cells. As a result, when the threshold voltage of an adjacent
flash cell increases, the threshold voltage of the \emph{victim
cell} increases as well. The unintended threshold voltage shifts
can eventually move a cell into a different state than the one it
was originally programmed to, leading to a bit error.

\begin{figure}[h]
  \centering
  \includegraphics[width=0.55\columnwidth]{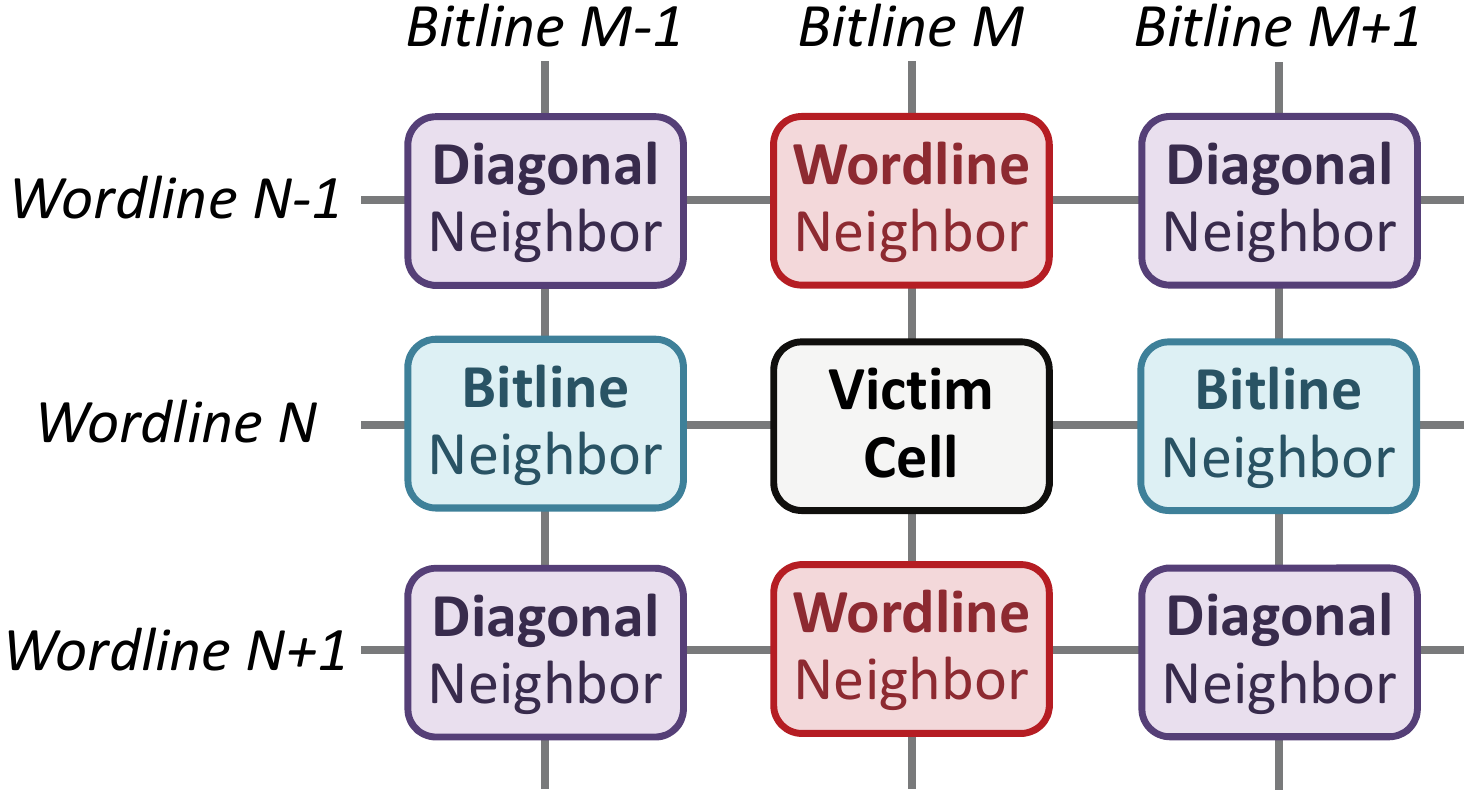}%
  \caption[Immediately-adjacent cells that can induce program
interference on a victim cell that is on wordline~\emph{N} and bitline~\emph{M}.]
{Immediately-adjacent cells that can induce program
interference on a victim cell that is on wordline~\emph{N} and bitline~\emph{M}. 
\chI{Reproduced from~\cite{cai.arxiv17}.}}%
  \label{fig:F16}%
\end{figure}

\chIII{Prior works} have shown, based on \chIII{the} experimental analysis of
modern MLC NAND flash memory chips, that the threshold
voltage change of the victim cell can be accurately modeled
as a linear combination of the threshold voltage changes of
the adjacent cells when they are programmed, using linear
regression with least-square-error estimation~\cite{cai.iccd13, cai.sigmetrics14}.
The cells that are physically located immediately next to the
victim cell (called the \emph{immediately-adjacent cells}) are the
major contributors to the cell-to-cell interference of a victim
cell~\cite{cai.iccd13}. Figure~\ref{fig:F16} shows the eight immediately-adjacent cells
for a victim cell in 2D planar NAND flash memory.

\FloatBarrier

The amount of interference that program operations to
the immediately-adjacent cells can induce on the victim cell
is expressed as:
\begin{equation}
\Delta V_{victim} = \sum_{X} K_X \Delta V_X
\label{eq:E7}
\end{equation}
where $\Delta V_{victim}$ is the change in voltage of the victim cell
due to cell-to-cell program interference, $K_X$ is the \emph{coupling
coefficient} between cell $X$ and the victim cell, and $\Delta V_{X}$ is
the threshold voltage change of cell $X$ during programming.
The coupling coefficient is directly related
to the effective capacitance $C$ between cell $X$ and the victim
cell, which can be calculated as:
\begin{equation}
C = \varepsilon S / d
\label{eq:E8}
\end{equation}
where $\varepsilon$ is the permittivity, $S$ is the effective cell area of cell
$X$ that faces the victim cell, and $d$ is the distance between the
cells. Of the immediately-adjacent cells, the wordline neighbor
cells have the greatest coupling capacitance with the victim
cell, as they likely have a large effective facing area to,
and a small distance from, the victim cell compared to other
surrounding cells.

The coupling
coefficient grows as the feature size decreases~\cite{cai.iccd13, cai.sigmetrics14}.
As NAND flash memory process technology scales down
to smaller feature sizes, cells become smaller and get closer
to each other, which increases the effective capacitance
between them. As a result, at smaller feature sizes, it is easier
for an immediately-adjacent cell to induce program interference
on a victim cell. We conclude that (1)~the program interference
an immediately-adjacent cell induces on a victim cell
is primarily determined by the distance between the cells and
the immediately-adjacent cell's effective area facing the victim
cell; and (2)~the wordline neighbor cell causes the highest
such interference, based on empirical measurements.

More findings on the nature of cell-to-cell program
interference and the impact of cell-to-cell program interference
on NAND flash memory errors and lifetime can be
found in prior \chIII{works from our research group}~\cite{cai.iccd13, cai.sigmetrics14, cai.hpca17, cai.thesis12}.

\subsection{Data Retention Errors}
\label{sec:errors:retention}

Retention errors are caused by charge leakage over time
after a flash cell is programmed, and are the dominant source
of flash memory errors, as demonstrated previously~\cite{mielke.irps08, cai.date12, cai.iccd12, cai.hpca15, cai.itj13, tanakamaru.isscc11}. 
As flash memory process technology
scales to smaller feature sizes, the capacitance of a flash
cell, and the number of electrons stored on it, decreases.
State-of-the-art (i.e., 1x-nm) MLC flash memory cells can
store only $\sim$100 electrons~\cite{yoon.fms12}. Gaining or losing several electrons
on a cell can significantly change the cell's voltage level
and eventually alter its state. Charge leakage is caused by the
unavoidable trapping of charge in the tunnel oxide~\cite{cai.hpca15, lee.irps03}.
The amount of trapped charge increases with the electrical
stress induced by repeated program and erase operations,
which degrade the insulating property of the oxide.

Two failure mechanisms of the tunnel oxide lead to retention
loss. \emph{Trap-assisted tunneling} (TAT) occurs because the
trapped charge forms an electrical tunnel, which exacerbates
the weak tunneling current, SILC (see Section~\ref{sec:flash:pgmerase}).
As a result of this TAT effect, the electrons present in the
floating gate (FG) leak away much faster through the intrinsic
electric field. Hence, the threshold voltage of the flash
cell decreases over time. As the flash cell wears out with
increasing P/E cycles, the amount of trapped charge also
increases~\cite{cai.hpca15, lee.irps03}, and so does the TAT effect. At high P/E
cycles, the amount of trapped charge is large enough to form
percolation paths that significantly hamper the insulating
properties of the gate dielectric~\cite{degraeve.ted04, cai.hpca15}, resulting in retention
failure. \emph{Charge detrapping}, where charge previously
trapped in the tunnel oxide is freed spontaneously, can also
occur over time~\cite{degraeve.ted04, cai.hpca15, lee.irps03, yamada.irps00}. The charge polarity can
be either negative (i.e., electrons) or positive (i.e., holes).
Hence, charge detrapping can either decrease or increase the
threshold voltage of a flash cell, depending on the polarity of
the detrapped charge.

More findings on the nature of data retention and the
impact of data retention behavior on NAND flash memory
errors and lifetime can be found in prior \chIII{works from our research group}~\cite{cai.date12, cai.iccd12, cai.hpca15, cai.itj13, cai.thesis12}.

\subsection{Read Disturb Errors}
\label{sec:errors:readdisturb}

Read disturb is a phenomenon in NAND flash memory
where reading data from a flash cell can cause the threshold
voltages of other (unread) cells in the same block to shift to
a higher value~\cite{mielke.irps08, cai.date12, cai.dsn15, papandreou.glsvlsi14, cooke.fms07, grupp.micro09, takeuchi.jssc99}. While a
single threshold voltage shift is small, such shifts can accumulate
over time, eventually becoming large enough to alter the
state of some cells and hence generate \emph{read disturb errors}.

The failure mechanism of a read disturb error is similar
to the mechanism of a normal program operation. A program
operation applies a high programming voltage (e.g.,
+\SI{15}{\volt}) to the cell to change the cell's threshold voltage to
the desired range. Similarly, a read operation applies a \emph{high
pass-through voltage} (e.g., +\SI{6}{\volt}) to \emph{all other cells} that share
the same bitline with the cell that is being read. Although
the pass-through voltage is not as high as the programming
voltage, it still generates a \emph{weak programming effect} on the
cells it is applied to~\cite{cai.dsn15}, which can unintentionally change
these cells' threshold voltages.

More findings on the nature of read disturb and the
impact of read disturb on NAND flash memory errors and
lifetime can be found in prior \chIII{works from our research group}~\cite{cai.dsn15}.


\subsection{Self-Recovery Effect}
\label{sec:background:recovery}


NAND flash memory has a limited lifetime because of transistor \emph{wearout}
as a flash cell is repeatedly programmed and erased.
After each additional P/E cycle, a greater number of
electrons get \emph{\chI{inadvertently} trapped} within the flash
cell, which changes the threshold voltage of the
transistor\chI{~\cite{cai.procieee17, cai.arxiv17}}.  This threshold voltage
change introduces \chI{errors} and, thus, reduces the flash lifetime.
Some of these \chI{inadvertently-trapped} electrons
gradually \emph{escape} \chI{during the}
idle time between consecutive P/E cycles, \chI{i.e., \emph{the dwell time}}.  The escape (i.e., \emph{detrapping}) of the
\chI{inadvertently-trapped} electrons is known as the \emph{self-recovery 
effect}~\cite{mielke.irps06}, as it \chI{partially} undoes (i.e., \emph{repairs}) the
wearout of the cell.


The self-recovery effect repairs the damage caused by flash wearout during the
time between two P/E cycles, by \emph{detrapping} some of
the \chI{inadvertently-trapped} charge~\cite{mielke.irps06, wu.imw11,
wu.hotstorage11, chen.codes13, mohan.hotstorage10, lee.fast12}. In this \chV{dissertation},
we refer to the delay between consecutive program operations as the \emph{dwell time}.
The amount of repair done by self-recovery is affected by two factors: 
(1)~dwell time and (2)~operating temperature.


During the dwell time of a flash cell, a fraction of the charge that was \chI{inadvertently}
trapped in the tunnel oxide is slowly
detrapped~\cite{mielke.irps06}. The reduction of \chI{inadvertently-trapped} charge
in a cell reduces the number of retention and program variation errors, and \chI{thus} can
extend the NAND flash memory lifetime.
%
For a fixed retention time, \chI{a larger} dwell time reduces the number 
of retention errors~\cite{mielke.irps06}.
\chI{A \emph{recovery cycle} refers to a P/E cycle where the program operation 
is followed by an extended dwell time.}
Since 3D NAND flash memory errors are dominated by
retention errors~\cite{choi.vlsit16, yoon.fms15, mizoguchi.imw17}, reducing the
retention error rate 
\chI{by performing recovery cycles} can
increase flash lifetime significantly.


A higher operating temperature for NAND flash memory increases electron mobility\chI{~\cite{mielke.irps06, cai.hpca15}}.
As a result,
a short retention time at high temperature has the \chI{\emph{same}} retention loss effect
as a longer \chI{retention time at room \chI{temperature~\cite{cai.hpca15}}, which we call the
\emph{effective retention time}}.
Similarly, a short dwell time at high
temperature has the \chI{\emph{same}} self-recovery effect as a \chI{longer dwell
time} at room temperature~\cite{mielke.irps06}, \chI{called the \chI{\emph{effective dwell time}}}.
The equivalence \chI{between time} elapsed at \chI{a certain} temperature and the
\chI{corresponding} effective
time at room temperature can be modeled using
Arrhenius' Law~\cite{arrhenius.zpc1889, jesd91a.jedec03, cai.hpca15,
mielke.irps06}:
\begin{align}
    AF(T_1, T_2) = \frac{t_1}{t_2} = 
    \exp \left( \frac{E_a}{k_B} \cdot \left( \frac{1}{T_1} - \frac{1}
    {T_2} \right) \right)
\label{eqn:arrhenius}
\end{align}

In Equation~\ref{eqn:arrhenius}, $AF$ is the \chI{\emph{acceleration factor}} between $t_1$
and $t_2$, where $t_1$ is the retention or dwell time under temperature $T_1$,
and $t_2$ is the \chI{retention or dwell} time under temperature $T_2$. $k_B$
is the Boltzmann constant,
which is \SI[per-mode=symbol]{8.62e-5}{\electronvolt\per\kelvin}. $E_a$ is the activation energy, which is a
manufacturing-process-dependent constant. For a planar NAND flash memory device, $E_a = $~\SI{1.1}{\electronvolt}~\cite{
jesd218.jedec10}. To our knowledge, there is no public literature
that reports the value of $E_a$ for 3D NAND flash memory.

Prior work has proposed idealized circuit-level models for the self-recovery (or self-healing)
effect~\cite{mohan.hotstorage10, wu.hotstorage11}, demonstrating significant
opportunities for using the self-recovery effect to improve flash reliability and
lifetime. Based on the assumptions about how self-recovery effect works, prior
work has also proposed techniques to exploit this effect to improve flash
lifetime such as heal-leveling~\cite{chang.dac14}, write
throttling~\cite{lee.fast12}, and heat-accelerated
self-recovery~\cite{wu.hotstorage11}. However, these previous results are not yet
convincing enough to show that self-recovery effect can successfully improve
flash lifetime on real devices, because they lack real experimental data and
evidence supporting the self-recovery effect on modern flash devices. Our
characterization in \chIII{Chapter}~\ref{sec:heatwatch} is the first to demonstrate
and comprehensively evaluate the benefit of self-recovery effect using
experimental data from real 3D NAND flash memory chips.


\subsection{Large-Scale Studies on SSD Errors}
\label{sec:errors:largescale}

The error characterization studies we have discussed so
far examine the susceptibility of real NAND flash memory
devices to specific error sources, by conducting controlled
experiments on individual flash devices in controlled environments.
To examine the \chI{\emph{aggregate}} effect of these error
sources on flash devices that operate in the field, several
recent studies have analyzed the reliability of SSDs deployed
at a large scale (\chI{e.g., hundreds} of thousands of SSDs)
in production data centers~\cite{meza.sigmetrics15, schroeder.fast16, narayanan.systor16}. Unlike the controlled
low-level error characterization studies discussed in
Sections~\ref{sec:errors:pe} through~\ref{sec:errors:readdisturb}, these large-scale studies analyze
the observed errors and error rates in an \emph{uncontrolled}
manner, i.e., based on real data center workloads operating
at field conditions \chI{(as opposed to carefully controlling access patterns
and operating conditions)}. As such, these large-scale studies
can study flash memory behavior and reliability using only
a black-box approach, where they are able to access only the
registers used by the SSD to record select statistics. 
\chI{Because of this, their conclusions are usually correlational in nature,
as opposed to identifying the underlying causes behind the observations.}
On the
other hand, these studies incorporate the effects of a real
system, including the system software stack and real workloads~\cite{meza.sigmetrics15}
\chI{and real operational conditions in data centers}, 
on the flash memory devices, which is not present
in the controlled small-scale studies.

These \chI{recent} large-scale studies have made a number of observations
across large sets of SSDs \chI{employed in the data centers of large
internet companies: Facebook~\cite{meza.sigmetrics15}, Google~\cite{schroeder.fast16}, and Microsoft~\cite{narayanan.systor16}}.
We highlight \chI{six} key observations from these studies \chI{about the
\emph{SSD failure rate}, which is the fraction of SSDs that have experienced at least one
uncorrectable error.}

\begin{figure}[h]
  \centering
  \includegraphics[width=0.41\columnwidth]{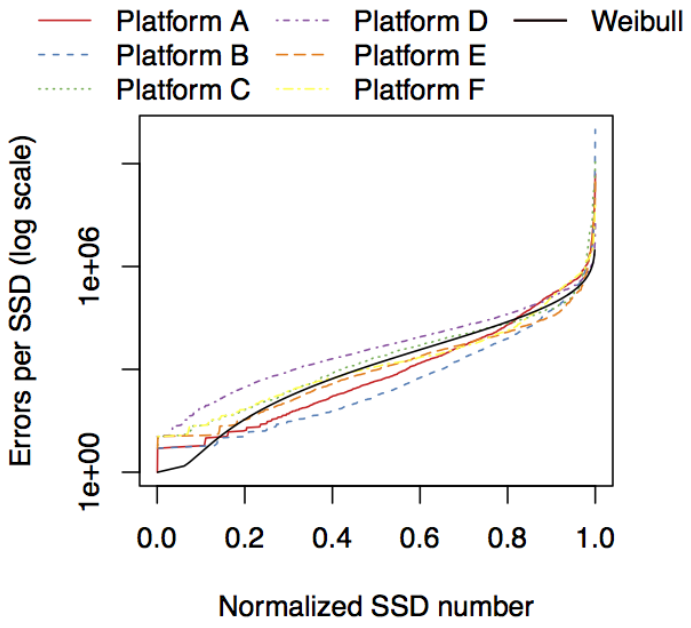}%
  \caption[Distribution of uncorrectable errors across SSDs used in 
  Facebook's data centers.]
  {\chI{Distribution of uncorrectable errors across SSDs used in 
  Facebook's data centers.  Reproduced from~\cite{meza.sigmetrics15}.}}%
  \label{fig:large-scale-distribution}%
\end{figure}

\chI{First, the number of uncorrectable errors observed varies significantly 
for each SSD.  Figure~\ref{fig:large-scale-distribution} shows the
distribution of uncorrectable errors per SSD across a large set of SSDs
used by Facebook.  The
distributions are grouped into six different \emph{platforms}
that are deployed in Facebook's data center.\chI{\footnote{\chI{Each platform has a different combination of SSDs, host controller
interfaces, and workloads. The six platforms are described in detail in~\cite{meza.sigmetrics15}.}}}
For every platform, \chIII{prior work observes} that the top 10\% of SSDs, when sorted by
their uncorrectable error count, account for over 80\% of the total uncorrectable
errors observed across all SSDs for that platform.  \chIII{Prior work finds} that the
distribution of uncorrectable errors across all SSDs belonging to a platform
follows a Weibull distribution, which \chIII{is shown} using a solid black line in
Figure~\ref{fig:large-scale-distribution}.}

\FloatBarrier

\begin{figure}[h]
  \centering
  \includegraphics[width=0.35\columnwidth]{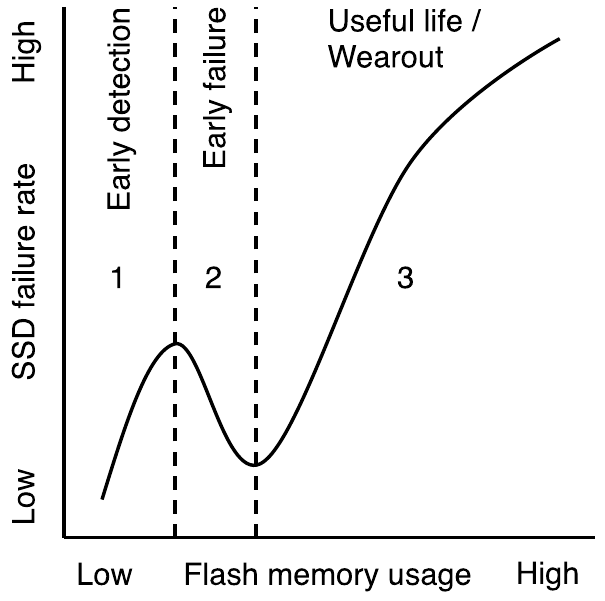}%
  \caption[Pictorial and abstract depiction of the pattern of SSD failure rates observed in real SSDs operating
  in a modern data center.  An SSD fails at different rates
  during distinct periods throughout the SSD lifetime.]
  {\chI{\chI{\chI{Pictorial and abstract \chI{depiction} of the pattern} of SSD failure rates observed in real SSDs operating
  in a modern data center.}  An SSD fails at different rates
  during distinct periods throughout the SSD lifetime. Reproduced from~\cite{meza.sigmetrics15}.}}%
  \label{fig:lifecycle}%
\end{figure}

\begin{figure}[h]
  \centering
  \includegraphics[width=0.32\columnwidth]{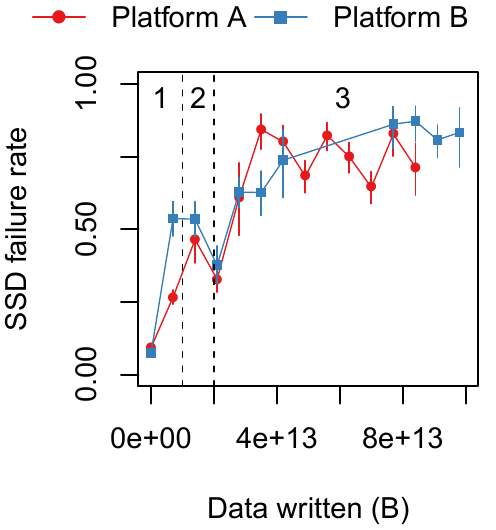}%
  \hfill
  \includegraphics[width=0.32\columnwidth]{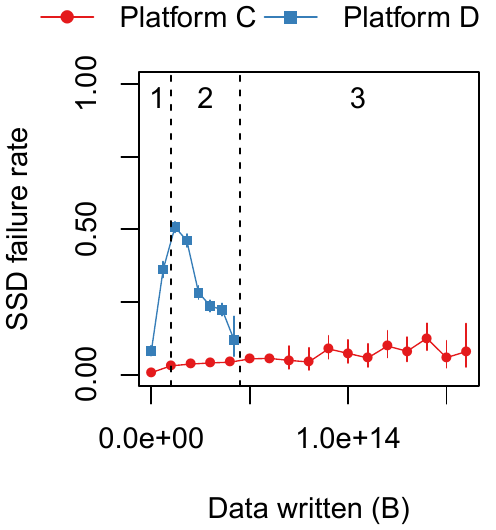}%
  \hfill
  \includegraphics[width=0.32\columnwidth]{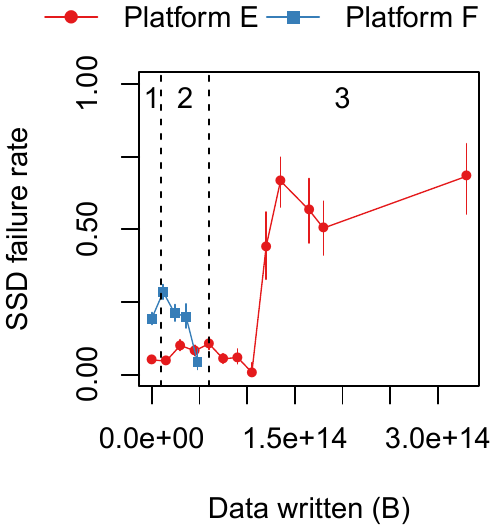}%
  \caption[SSD failure rate vs.\ the amount of data written to the SSD.
  The three \chI{periods of failure rates, shown \chI{pictorially and abstractly}
  in Figure~\ref{fig:lifecycle},} are annotated on each graph:
  (1)~early detection, (2)~early failure, and (3)~useful life/wearout.]
  {\chI{SSD failure rate vs.\ the amount of data written to the SSD.
  The three \chI{periods of failure rates, shown \chI{pictorially and abstractly}
  in Figure~\ref{fig:lifecycle},} are annotated on each graph:
  (1)~early detection, (2)~early failure, and (3)~useful life/wearout. 
  Reproduced from~\cite{meza.sigmetrics15}.}}%
  \label{fig:lifecycle-write}%
\end{figure}

\chI{\chI{Second, the SSD failure rate does} \emph{not}
increase monotonically with the P/E cycle count.  Instead, \chIII{prior work observes}
several \emph{distinct} periods of reliability, as illustrated \chI{pictorially and abstractly} in 
Figure~\ref{fig:lifecycle}, \chI{which is based on data obtained from analyzing
errors in SSDs used in Facebook's data centers~\cite{meza.sigmetrics15}}.  The failure rate increases when the SSDs are
relatively new (shown as the \emph{early detection} period in
Figure~\ref{fig:lifecycle}), as the SSD controller identifies unreliable
NAND flash cells during the initial read and write operations to the devices
\chI{and removes them from the address space (see \chI{Section~\ref{sec:ssdarch:ctrl:badblocks}})}.  
As the SSDs are used \chI{more}, they enter the
\emph{early failure} period, where failures are less likely to occur.
When the SSDs approach the end of their lifetime (\emph{useful life/wearout}
in the figure), the failure rate increases again, as more cells become 
unreliable due to wearout.  Figure~\ref{fig:lifecycle-write} shows how the 
measured failure rate changes as more writes are performed to the SSDs
\chI{(i.e., how real data collected from Facebook's SSDs \chI{corresponds to} the
pictorial depiction in Figure~\ref{fig:lifecycle}) for the
same six platforms shown in Figure~\ref{fig:large-scale-distribution}}. 
\chI{\chIII{Prior work observes} that the failure rates in each platform exhibit the
distinct periods that are illustrated in Figure~\ref{fig:lifecycle}.
For example, \chI{let us consider} the SSDs in Platforms A and B\chI{, which} have more data written to
their cells than SSDs in other platforms.  \chIII{Prior work observes} from 
Figure~\ref{fig:lifecycle-write} that for SSDs in Platform~A,
there is an 81.7\% \chI{increase from} the failure rate during the
early detection period \chI{to} the failure rate during the wearout period~\cite{meza.sigmetrics15}.}}

\FloatBarrier

\chI{Third}, the raw bit error rate grows with the age of the
device even if the P/E cycle count is held constant, indicating
that mechanisms such as silicon aging \chI{likely contribute}
to the error rate~\cite{narayanan.systor16}. 

\chI{Fourth}, the observed failure rate of SSDs
has been noted to be significantly higher than the failure rates
specified by the manufacturers~\cite{schroeder.fast16}. 

\begin{figure}[h]
  \centering
  \includegraphics[width=0.32\columnwidth]{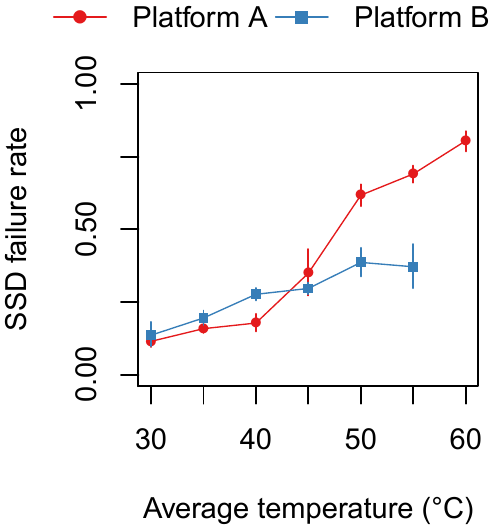}%
  \hfill
  \includegraphics[width=0.32\columnwidth]{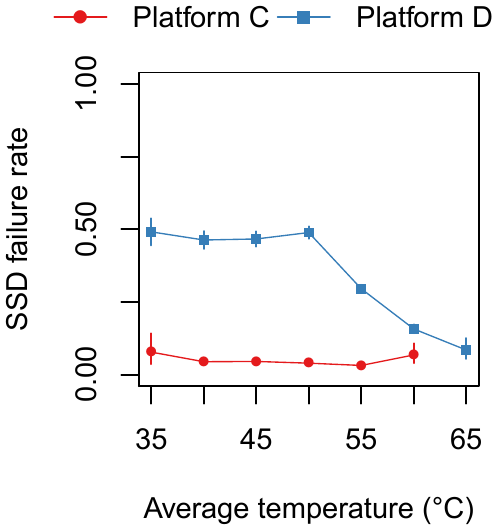}%
  \hfill
  \includegraphics[width=0.32\columnwidth]{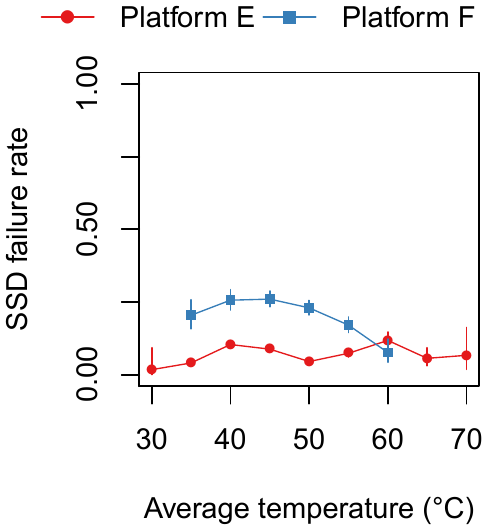}%
  \caption[SSD failure rate vs.\ operating temperature.]{\chI{SSD failure rate vs.\ operating temperature.
  Reproduced from~\cite{meza.sigmetrics15}.}}%
  \label{fig:datacenter-temp}%
\end{figure}

\chI{Fifth}, higher operating
temperatures can lead to higher failure rates, but modern SSDs
employ throttling techniques that reduce the access rates to
the underlying flash chips, which can greatly reduce the negative
reliability impact of higher temperatures~\cite{meza.sigmetrics15}. 
\chI{For example, Figure~\ref{fig:datacenter-temp} shows the
SSD failure rate as the SSD operating temperature varies, for SSDs from the
same six platforms shown in Figure~\ref{fig:large-scale-distribution}~\cite{meza.sigmetrics15}.
\chIII{Prior work observes} that at an operating temperature range of \SIrange{30}{40}{\celsius},
SSDs \chI{either (1)~have} similar failure rates \chI{across the different temperatures, 
or (2)~experience} slight increases in the failure rate as
the temperature increases.  As the temperature increases beyond \SI{40}{\celsius},
the SSDs fall into three categories:
(1)~temperature-sensitive with increasing failure rate (Platforms~A and B),
(2)~less temperature-sensitive (Platforms~C and E), and
(3)~temperature-sensitive with decreasing failure rate (Platforms~D and F).
There are two factors that affect the temperature sensitivity of each platform:
(1)~some, but not all, of the platforms employ techniques to throttle SSD
activity at high operating temperatures to reduce the failure rate \chI{(e.g., Platform~D)}; and
(2)~the platform configuration (e.g., the number of SSDs in each machine,
system airflow) can shorten or prolong the effects of higher operating 
temperatures.}

\FloatBarrier

\chI{Sixth}, while
SSD failure rates are higher than specified \chI{by the manufacturers}, the overall occurrence
of \emph{uncorrectable} errors is lower than expected\chI{~\cite{meza.sigmetrics15}} because
(1)~effective bad block management policies (see \chI{Section~\ref{sec:ssdarch:ctrl:badblocks}})
are implemented in SSD controllers; and (2)~certain types of
error sources, such as read disturb~\cite{meza.sigmetrics15, narayanan.systor16} and incomplete
erase operations~\cite{narayanan.systor16}, have yet to become a major source of
uncorrectable errors at the system level.


\section{Error Mitigation}
\label{sec:mitigation}

Several different types of errors can occur in NAND flash
memory, as we described in Section~\ref{sec:errors}. As NAND flash memory
continues to scale to smaller technology nodes, the magnitude
of these errors has been increasing~\cite{parnell.globecom14, maislos.fms11, yoon.fms12}.
This, in turn, uses up the limited error correction capability
of ECC more rapidly than in past flash memory generations
and shortens the lifetime of modern SSDs. To overcome the
decrease in lifetime, a number of error mitigation techniques \chI{have been designed.
These techniques} exploit intrinsic properties of the different types of
errors to reduce the rate at which they lead to raw bit \chI{errors.}
In this section, we discuss how the flash
controller mitigates each of the error types via \chI{various} proposed error
mitigation mechanisms. Table~\ref{tbl:T3} shows the techniques we
overview and which errors (from Section~\ref{sec:errors}) they mitigate.

\begin{table}[h]
\centering
\small
\setlength{\tabcolsep}{0.26em}
\setlength\arrayrulewidth{0.75pt}
\caption{List of different types of errors mitigated by \chI{various} NAND flash
error mitigation mechanisms.}
\label{tbl:T3}
\begin{tabular}{|c||c|c|c|c|c|}
\hline
& \multicolumn{5}{c|}{\cellcolor{gray!20}\textbf{\emph{Error Type}}} \\
& \cellcolor{gray!20} & \cellcolor{gray!20} & \cellcolor{gray!20} & \cellcolor{gray!20} & \cellcolor{gray!20} \\
& \cellcolor{gray!20} & \cellcolor{gray!20} & \cellcolor{gray!20} & \cellcolor{gray!20} & \cellcolor{gray!20} \\
& \cellcolor{gray!20} & \cellcolor{gray!20} & \cellcolor{gray!20} & \cellcolor{gray!20} & \cellcolor{gray!20} \\
& \cellcolor{gray!20} & \cellcolor{gray!20} & \cellcolor{gray!20} & \cellcolor{gray!20} & \cellcolor{gray!20} \\
& \cellcolor{gray!20} & \cellcolor{gray!20} & \cellcolor{gray!20} & \cellcolor{gray!20} & \cellcolor{gray!20} \\
& \cellcolor{gray!20} & \cellcolor{gray!20} & \cellcolor{gray!20} & \cellcolor{gray!20} & \cellcolor{gray!20} \\
& \cellcolor{gray!20} & \cellcolor{gray!20} & \cellcolor{gray!20} & \cellcolor{gray!20} & \cellcolor{gray!20} \\
\textbf{Mitigation} & \cellcolor{gray!20} & \cellcolor{gray!20} & \cellcolor{gray!20} & \cellcolor{gray!20} & \cellcolor{gray!20} \\
\textbf{Mechanism}
& \multirow{-9}{*}{\cellcolor{gray!20}\rotatebox{90}{\parbox[b]{11em}{\makecell[l]{\textbf{\emph{P/E Cycling}}\\\cite{cai.date12, cai.date13, luo.jsac16} (\S\ref{sec:errors:pe})}}}}
& \multirow{-9}{*}{\cellcolor{gray!20}\rotatebox{90}{\parbox[b]{11em}{\makecell[l]{\textbf{\emph{Program}}\\\cite{cai.hpca17, luo.jsac16, parnell.globecom14} (\S\ref{sec:errors:pgm})}}}}
& \multirow{-9}{*}{\cellcolor{gray!20}\rotatebox{90}{\parbox[b]{11em}{\makecell[l]{\textbf{\emph{Cell-to-Cell Interference}}\\\cite{cai.date12, cai.iccd13, cai.sigmetrics14, lee.iedl02} (\S\ref{sec:errors:celltocell})}}}}
& \multirow{-9}{*}{\cellcolor{gray!20}\rotatebox{90}{\parbox[b]{11em}{\makecell[l]{\textbf{\emph{Data Retention}}\\\cite{mielke.irps08, cai.date12, cai.iccd12, cai.hpca15, cai.itj13} (\S\ref{sec:errors:retention})}}}}
& \multirow{-9}{*}{\cellcolor{gray!20}\rotatebox{90}{\parbox[b]{11em}{\makecell[l]{\textbf{\emph{Read Disturb}}\\\cite{mielke.irps08, cai.date12, cai.dsn15, grupp.micro09} (\S\ref{sec:errors:readdisturb})}}}} \\ \hhline{|=#=|=|=|=|=|}
\textbf{Shadow Program Sequencing} & & & \multirow{2}{*}{X} & & \\
\cite{cai.iccd13, cai.hpca17} (Section~\ref{sec:mitigation:shadow}) & & & & & \\ \hline
\textbf{Neighbor-Cell Assisted Error} & & & \multirow{2}{*}{X} & & \\
\textbf{Correction}~\cite{cai.sigmetrics14} (Section~\ref{sec:mitigation:nac}) & & & & & \\ \hline
\textbf{Refresh} & & & & \multirow{2}{*}{X} & \multirow{2}{*}{X} \\
\cite{cai.iccd12, cai.itj13, pan.hpca12, mohan.tr12} (Section~\ref{sec:mitigation:refresh}) & & & & & \\ \hline
\textbf{Read-Retry} & \multirow{2}{*}{X} & & & \multirow{2}{*}{X} & \multirow{2}{*}{X} \\
\cite{cai.date13, yang.fms11, fukami.di17} (Section~\ref{sec:mitigation:retry}) & & & & & \\ \hline
\textbf{Voltage Optimization} & \multirow{2}{*}{X} & & & \multirow{2}{*}{X} & \multirow{2}{*}{X} \\
\cite{cai.hpca15, cai.dsn15, jeong.fast14} (Section~\ref{sec:mitigation:voltage}) & & & & & \\ \hline
\textbf{Hot Data Management} & \multirow{2}{*}{X} & \multirow{2}{*}{X} & \multirow{2}{*}{X} & \multirow{2}{*}{X} & \multirow{2}{*}{X} \\
\cite{luo.msst15, ha.apsys13, ha.tcad16} (Section~\ref{sec:mitigation:hotcold}) & & & & & \\ \hline
\textbf{Adaptive Error Mitigation} & \multirow{2}{*}{X} & \multirow{2}{*}{X} & \multirow{2}{*}{X} & \multirow{2}{*}{X} & \multirow{2}{*}{X} \\
\cite{cai.patent16.9419655, haratsch.fms16, wu.mascots10, wilson.mascots14, chen.vts09} (Section~\ref{sec:mitigation:adaptive}) & & & & & \\ \hline
\end{tabular}
\end{table}
\FloatBarrier

\subsection{Shadow Program Sequencing}
\label{sec:mitigation:shadow}

As discussed in Section~\ref{sec:errors:celltocell}, cell-to-cell program interference
is a function of the distance between the cells of the
wordline that is being programmed and the cells of the victim
wordline. The impact of program interference is greatest on
a victim wordline when either of the victim's immediately-adjacent
wordlines is programmed (e.g., if we program WL1
in Figure~\ref{fig:F8}, WL0 and WL2 experience the greatest amount of
interference). Early MLC flash memories used one-shot programming,
where both the LSB and MSB pages of a wordline
are programmed at the same time. As flash memory scaled to
smaller process technologies, one-shot programming resulted
in much larger amounts of cell-to-cell program interference.
As a result, manufacturers introduced two-step programming
for MLC NAND flash (see Section~\ref{sec:flash:pgmerase}), where the SSD controller
writes values of the two pages within a wordline in two
independent steps.

\begin{figure}[h]
  \centering
  \includegraphics[width=0.7\columnwidth]{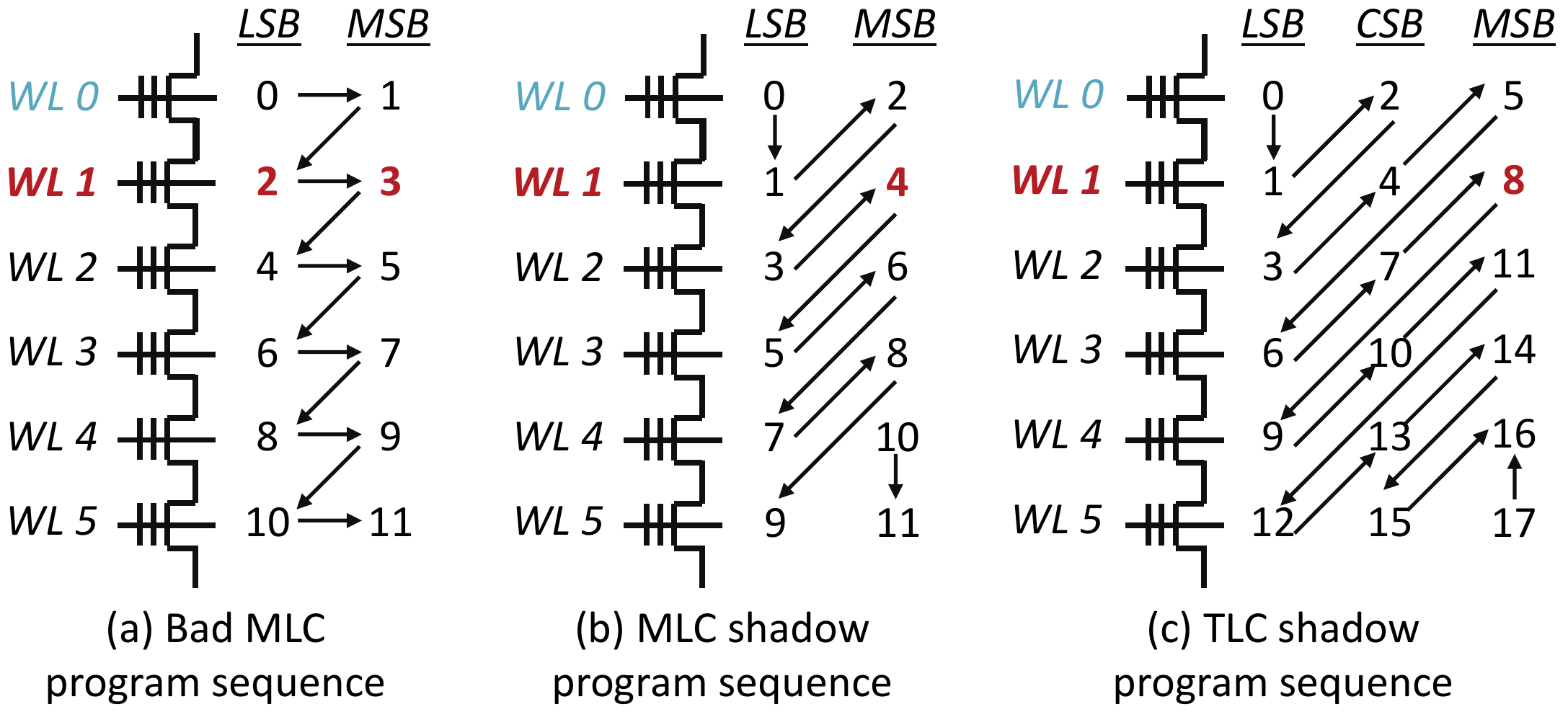}%
  \caption[Order in which the pages of each wordline (WL) are
programmed using (a)~a bad programming sequence, and using
shadow sequencing for (b)~MLC and (c)~TLC NAND flash. The bold
page programming operations for WL1 induce cell-to-cell program
interference when WL0 is fully programmed.]
{Order in which the pages of each wordline (WL) are
programmed using (a)~a bad programming sequence, and using
shadow sequencing for (b)~MLC and (c)~TLC NAND flash. The bold
page programming operations for WL1 induce cell-to-cell program
interference when WL0 is fully programmed. \chI{Reproduced from~\cite{cai.arxiv17}.}}%
  \label{fig:F20}%
\end{figure}

The SSD controller minimizes the interference that
occurs during two-step programming by using \emph{shadow program
sequencing}~\cite{park.dac16, cai.iccd13, cai.hpca17} to determine the order that
data is written to different pages in a block. If we program
the LSB and MSB pages of the same wordline back to back,
as shown in Figure~\ref{fig:F20}a, both programming steps induce
interference on a \emph{fully-programmed wordline} (i.e., a wordline
where both the LSB and MSB pages are already written).
For example, if the controller programs both pages of WL1
back to back, shown as bold page programming operations
in Figure~\ref{fig:F20}a, the program operations induce a high amount
of interference on WL0, which is fully programmed. The key
idea of shadow program sequencing is to ensure that a fully-programmed
wordline experiences interference minimally,
i.e., \emph{only} during MSB page programming (and \emph{not} during
LSB page programming). In shadow program sequencing, \chIII{prior work
assigns} a unique page number to each page within a block, as
shown in Figure~\ref{fig:F20}b. The LSB page of wordline~$i$ is numbered
page $2i - 1$, and the MSB page is numbered page $2i + 2$. The
only exceptions to the numbering are the LSB page of wordline~0 
(page~0) and the MSB page of the last wordline~$n$ (page
$2n + 1$). Two-step programming writes to pages in \emph{increasing}
order of page number inside a block~\cite{park.dac16, cai.iccd13, cai.hpca17}, such that
a \emph{fully-programmed} wordline experiences interference only
from the MSB page programming of the wordline directly
above it, shown as the bold page programming operation in
Figure~\ref{fig:F20}b. With this programming order/sequence, the LSB
page of the wordline above, and both pages of the wordline
below, do \emph{not} cause interference to fully-programmed data~\cite{park.dac16, cai.iccd13, cai.hpca17}, 
as these two pages are programmed \emph{before}
programming the MSB page of the given wordline. Foggy-fine
programming in TLC NAND flash (see Section~\ref{sec:flash:pgmerase})
uses a similar ordering to reduce cell-to-cell program interference,
as shown in Figure~\ref{fig:F20}c.

\FloatBarrier

Shadow program sequencing is an effective solution
to minimize cell-to-cell program interference on fully-programmed
wordlines during two-step programming, and
is employed in commercial SSDs today.

\subsection{Neighbor-Cell Assisted Error Correction}
\label{sec:mitigation:nac}

The threshold voltage shift that occurs due to program
interference is highly correlated with the values stored in
the cells of the \emph{immediately-adjacent wordlines}, as we discussed
in Section~\ref{sec:errors:celltocell}. Due to this correlation, knowing
the value programmed in the immediately-adjacent cell
(i.e., a \emph{neighbor cell}) makes it easier to correctly determine
the value stored in the flash cell that is being read~\cite{cai.sigmetrics14}. We
describe a \chIII{recently-proposed} error correction method that
takes advantage of this observation, called \emph{neighbor-cell-assisted
error correction} (NAC). The key idea of NAC is to
use the data values stored in the cells of the immediately-adjacent
wordline to determine a better set of read reference
voltages for the wordline that is being read. Doing so leads
to a more accurate identification of the logical data value
that is being read, as the data in the immediately-adjacent
wordline was \emph{partially responsible} for shifting the threshold
voltage of the cells in the wordline that is being read when
the immediately-adjacent wordline was programmed.

\begin{figure}[h]
  \centering
  \includegraphics[width=0.65\columnwidth]{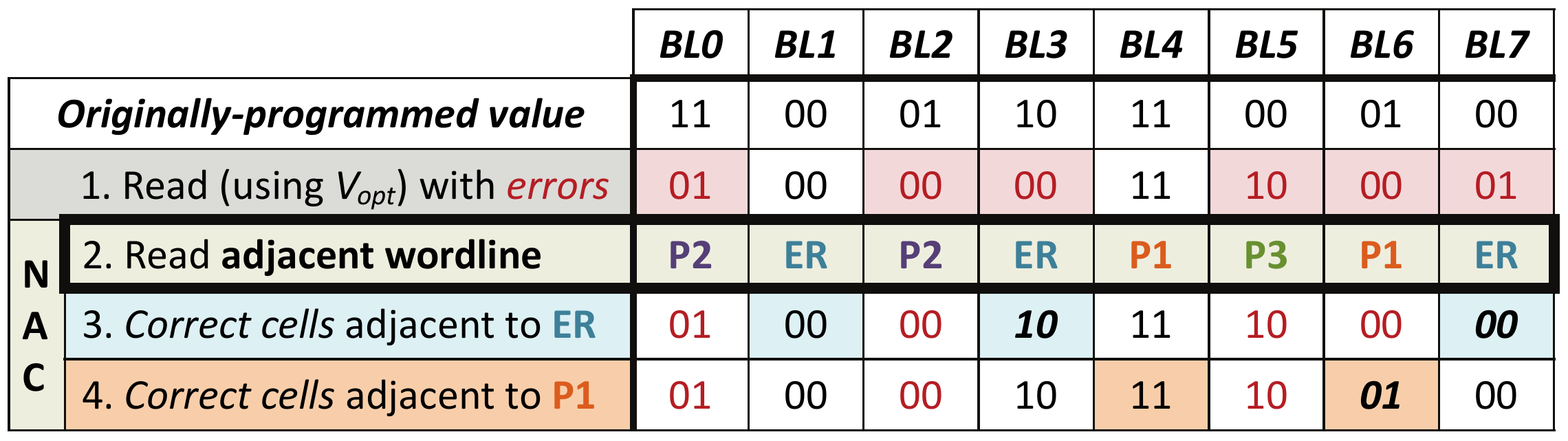}%
  \caption[Overview of neighbor-cell-assisted error correction (NAC).]
  {Overview of neighbor-cell-assisted error correction (NAC). \chI{Reproduced from~\cite{cai.arxiv17}.}}%
  \label{fig:F21}%
\end{figure}

Figure~\ref{fig:F21} shows an operational example of NAC that
is applied to eight bitlines (BL) of an MLC flash wordline.
The SSD controller first reads a flash page from a
wordline using the standard read reference voltages (step~1 in
Figure~\ref{fig:F21}). The bit values read from the wordline are then buffered
in the controller. If there are no errors uncorrectable by ECC,
the read was successful, and nothing else is done. However,
if there are errors that are \emph{uncorrectable} by ECC, we assume
that the threshold voltage distribution of the page shifted due to
cell-to-cell program interference, triggering further correction.
In this case, NAC reads the LSB and MSB pages of the wordline
\emph{immediately above} the requested page (i.e., the \emph{adjacent} wordline
that was programmed \emph{after} the requested page) to classify
the cells of the requested page (step~2). NAC then identifies the
cells adjacent to (i.e., connected to the same bitline as) the ER
cells (i.e., cells in the immediately above wordline that are in
the ER state), such as the cells on BL1, BL3, and BL7 in Figure~\ref{fig:F21}.
NAC rereads these cells using read reference voltages that \emph{compensate
for} the threshold voltage shift caused by programming
the adjacent cell to the ER state (step~3). If ECC can correct the
remaining errors, the controller returns the corrected page to
the host. If ECC fails again, the process is repeated using a different
set of read reference voltages for cells that are adjacent
to the P1 cells (step~4). If ECC continues to fail, the process is
repeated for cells that are adjacent to P2 and P3 cells (steps~5
and 6, respectively, which are not shown in the figure) until
either ECC is able to correct the page or all possible adjacent
values are exhausted.

\FloatBarrier

NAC extends the lifetime of an SSD by reducing the
number of errors that need to be corrected using the limited
correction capability of ECC. With the use of experimental
data collected from real MLC NAND flash memory
chips, \chIII{prior work shows} that NAC extends the NAND flash memory
lifetime by 33\%~\cite{cai.sigmetrics14}. \chIII{Previous work from our research group}~\cite{cai.sigmetrics14} provides a
detailed description of NAC, including a theoretical treatment
of why it works and a practical implementation that
minimizes the number of reads performed, even in the case
when the neighboring wordline itself has errors.

\subsection{Refresh Mechanisms}
\label{sec:mitigation:refresh}

As we see in Figure~\ref{fig:F12}, during the time period after a flash
page is programmed, retention (Section~\ref{sec:errors:retention}) and read
disturb (Section~\ref{sec:errors:readdisturb}) can cause an increasing number of
raw bit errors to accumulate over time. This is particularly
problematic for a page that is not updated frequently. Due
to the limited error correction capability, the accumulation
of these errors can potentially lead to data loss for a
page with a \emph{high retention age} (i.e., a page that has not been
programmed for a long time). To avoid data loss, \emph{refresh
mechanisms} have been proposed, where the stored data is
periodically read, corrected, and reprogrammed, in order
to eliminate the retention and read disturb errors that
have accumulated prior to this periodic 
read/correction/reprogramming (i.e., refresh). The concept of refresh in
flash memory is thus conceptually similar to the refresh
mechanisms found in DRAM~\cite{liu.isca12, chang.hpca14, liu.isca13, jesd79.jedec13}. By
performing refresh and limiting the number of retention
and read disturb errors that can accumulate, the lifetime of
the SSD increases significantly. In this section, we describe
three types of refresh mechanisms used in modern SSDs:
remapping-based refresh, in-place refresh, and read reclaim.

\paratitle{Remapping-Based Refresh}
Flash cells must first be
erased before they can be reprogrammed, due to the fact
the programming a cell via ISPP can only increase the
charge level of the cell but not reduce it (Section~\ref{sec:flash:pgmerase}).
The key idea of \emph{remapping-based refresh} is to periodically
read data from each valid flash block, correct any data
errors, and \emph{remap the data to a different physical location},
in order to prevent the data from accumulating too many
retention errors~\cite{cai.iccd12, cai.itj13, pan.hpca12, mohan.tr12, cai.thesis12}. During each refresh
interval, a block with valid data that needs to be refreshed
is selected. The valid data in the selected block is read out
page by page and moved to the SSD controller. The ECC
engine in the SSD controller corrects the errors in the read
data, including retention errors that have accumulated
since the last refresh. A new block is then selected from
the free list (see \chI{Section~\ref{sec:ssdarch:ctrl:ftl}}), the error-free data is programmed
to a page within the new block, and the logical
address is remapped to point to the newly-programmed
physical page. By reducing the accumulation of retention
and read disturb errors, remapping-based refresh increases
SSD lifetime by an average of 9x for a variety of disk workloads~\cite{cai.iccd12, cai.itj13}.

Prior work proposes extensions to the basic remapping-based
refresh approach. One work, \emph{refresh SSDs}, proposes a
refresh scheduling algorithm based on an earliest deadline
first policy to guarantee that all data is refreshed in time~\cite{mohan.tr12}. 
The \emph{quasi-nonvolatile SSD} proposes to use remapping-based
refresh to choose between improving flash endurance
and reducing the flash programming latency (by using
larger ISPP step-pulses)~\cite{pan.hpca12}. In the quasi-nonvolatile SSD,
refresh requests are deprioritized, scheduled at idle times,
and can be interrupted after refreshing any page within a
block, to minimize the delays that refresh can cause for
the response time of pending workload requests to the
SSD. A refresh operation can also be triggered proactively
based on the data read latency observed for a page, which
is indicative of how many errors the page has experienced~\cite{cai.patent16.9424179}. 
Triggering refresh \emph{proactively} based on the observed
read latency (as opposed to doing so \emph{periodically}) improves
SSD latency and throughput~\cite{cai.patent16.9424179}. Whenever the read
latency for a page within a block exceeds a fixed threshold,
the valid data in the block is refreshed, i.e., remapped to a
new block~\cite{cai.patent16.9424179}.

\paratitle{In-Place Refresh}
A major drawback of remapping-based
refresh is that it performs \emph{additional writes} to the NAND
flash memory, accelerating wearout. To reduce the wearout
overhead of refresh, \chIII{prior work proposes} \emph{in-place refresh}~\cite{cai.iccd12, cai.itj13, cai.thesis12}. As
data sits unmodified in the SSD, data retention errors dominate~\cite{cai.date12, cai.itj13, tanakamaru.isscc11}, 
leading to charge loss and causing the
threshold voltage distribution to shift to the left, as we showed
in Section~\ref{sec:errors:retention}. The key idea of in-place refresh is to incrementally
replenish the lost charge of each page \emph{at its current
location}, i.e., in place, without the need for remapping.

\begin{figure}[h]
  \centering
  \includegraphics[width=0.7\columnwidth]{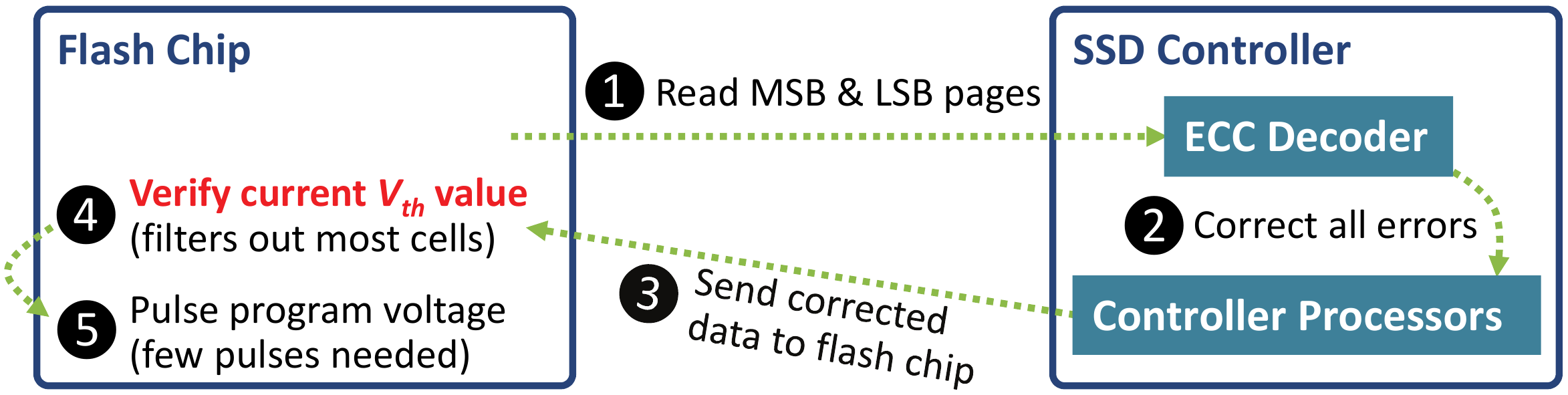}%
  \caption[Overview of in-place refresh mechanism for MLC NAND
flash memory.]
  {Overview of in-place refresh mechanism for MLC NAND
flash memory. \chI{Reproduced from~\cite{cai.arxiv17}.}}%
  \label{fig:F22}%
\end{figure}

Figure~\ref{fig:F22} shows a high-level overview of in-place refresh for
a wordline. The SSD controller first reads all of the pages
in the wordline (\incircle{1} in Figure~\ref{fig:F22}). The controller invokes the
ECC decoder to correct the errors within each page (\incircle{2}), and
sends the corrected data back to the flash chips (\incircle{3}). In-place
refresh then invokes a modified version of the ISPP mechanism
(see Section~\ref{sec:flash:pgmerase}), which we call \emph{Verify-ISPP} (V-ISPP),
to compensate for retention errors by restoring the charge
that was lost. In V-ISPP, we first verify the voltage currently
programmed in a flash cell (\incircle{4}). If the current voltage of the
cell is \emph{lower} than the target threshold voltage of the state that
the cell should be in, V-ISPP pulses the programming voltage
in steps, gradually injecting charge into the cell until the
cell returns to the target threshold voltage (\incircle{5}). If the current
voltage of the cell is \emph{higher} than the target threshold voltage,
V-ISPP inhibits the programming pulses to the cell.

\FloatBarrier

When the controller invokes in-place refresh, it is unable
to use shadow program sequencing (Section~\ref{sec:mitigation:shadow}), as all of the
pages within the wordline have already been programmed.
However, unlike traditional ISPP, V-ISPP does not introduce
a high amount of cell-to-cell program interference (Section~\ref{sec:errors:celltocell}) 
for two reasons. First, V-ISPP programs \emph{only} those cells
that have retention errors, which typically account for less
than 1\% of the total number of cells in a wordline selected
for refresh~\cite{cai.iccd12}. Second, for the small number of cells that
are selected to be refreshed, their threshold voltage is usually
only slightly lower than the target threshold voltage,
which means that only a few programming pulses need to
be applied. As cell-to-cell interference is linearly correlated
with the threshold voltage change to immediately-adjacent
cells~\cite{cai.iccd13, cai.sigmetrics14}, the small voltage change on these in-place
refreshed cells leads to only a small interference effect.

One issue with in-place refresh is that it is unable to
correct retention errors for cells in lower-voltage states.
Retention errors cause the threshold voltage of a cell in a
lower-voltage state to \emph{increase}, but V-ISPP \emph{cannot} decrease the threshold
voltage of a cell. To achieve a balance between the wearout
overhead due to remapping-based refresh and errors that
increase the threshold voltage due to in-place refresh, \chIII{prior work proposes}
\emph{hybrid in-place refresh}~\cite{cai.iccd12, cai.itj13, cai.thesis12}. The key idea is to use
in-place refresh when the number of program errors (caused
due to reprogramming) is within the correction capability of
ECC, but to use remapping-based refresh if the number of
program errors is too large to tolerate. To accomplish this, the
controller tracks the number of \emph{right-shift errors} (i.e., errors
that move a cell to a higher-voltage state)~\cite{cai.iccd12, cai.itj13}. If the
number of right-shift errors remains under a certain threshold,
the controller performs in-place refresh; otherwise, it
performs remapping-based refresh. Such a hybrid in-place
refresh mechanism increases SSD lifetime by an average of
31x for a variety of disk workloads~\cite{cai.iccd12, cai.itj13}.

\paratitle{Read Reclaim to Reduce Read Disturb Errors}
We can
also mitigate read disturb errors using an idea similar to
remapping-based refresh, known as \emph{read reclaim}. The key
idea of read reclaim is to remap the data in a block to a new
flash block, if the block has experienced a high number of
reads~\cite{ha.apsys13, ha.tcad16, kim.patent12}. To bound the number of read disturb
errors, some flash vendors specify a maximum number of
tolerable reads for a flash block, at which point read reclaim
rewrites the data to a new block (just as is done for remapping-
based refresh).

\paratitle{Adaptive Refresh and Read Reclaim Mechanisms}
For
the refresh and read reclaim mechanisms discussed above,
the SSD controller can (1)~invoke the mechanisms at fixed
regular intervals; or (2)~\emph{adapt} the rate at which it invokes the
mechanisms, based on various conditions that impact the
rate at which data retention and read disturb errors occur.
By adapting the mechanisms based on the current conditions
of the SSD, the controller can reduce the overhead
of performing refresh or read reclaim. The controller can
adaptively adjust the rate that the mechanisms are invoked
based on (1)~the wearout (i.e., the current P/E cycle count) of
the NAND flash memory~\cite{cai.iccd12, cai.itj13}; or (2)~the temperature
of the SSD~\cite{cai.date12, cai.hpca15}.

As we discuss in Section~\ref{sec:errors:retention}, for data with a given
retention age, the number of retention errors grows as the
P/E cycle count increases. Exploiting this P/E cycle dependent
behavior of retention time, the SSD controller can perform
refresh less frequently (e.g., once every year) when
the P/E cycle count is low, and more frequently (e.g., once
every week) when the P/E cycle count is high, as proposed
and described in prior \chIII{works from our research group}~\cite{cai.iccd12, cai.itj13}. Similarly, for
data with a given read disturb count, as the P/E cycle count
increases, the number of read disturb errors increases as
well~\cite{cai.dsn15}. As a result, the SSD controller can perform read
reclaim less frequently (i.e., it increases the maximum number
of tolerable reads per block before read reclaim is triggered)
when the P/E cycle count is low, and more frequently
when the P/E cycle count is high.

Prior works demonstrate that for a given retention time,
the number of data retention errors increases as the NAND
flash memory's operating temperature increases~\cite{cai.date12, cai.hpca15}.
To compensate for the increased number of retention errors
at high temperature, a state-of-the-art SSD controller adapts
the rate at which it triggers refresh. The SSD contains sensors
that monitor the current environmental temperature
every few milliseconds~\cite{meza.sigmetrics15, werner.fms10}. The controller then
uses the Arrhenius equation~\cite{mohan.tr12, arrhenius.zpc1889, xu.apl03} to estimate
the rate at which retention errors accumulate at the current
temperature of the SSD. Based on the error rate estimate,
the controller decides if it needs to increase the rate
at which it triggers refresh to ensure that the data is not lost.

By employing adaptive refresh and/or read reclaim mechanisms,
the SSD controller can successfully reduce the mechanism
overheads while effectively mitigating the larger number
of data retention errors that occur under various conditions.

\subsection{Read-Retry}
\label{sec:mitigation:retry}

In earlier generations of NAND flash memory, the read
reference voltage values were fixed at design time~\cite{mielke.irps08, cai.date13}.
However, several types of errors cause the threshold voltage
distribution to shift, as shown in Figure~\ref{fig:F13}. To compensate for
threshold voltage distribution shifts, a mechanism called \emph{read-retry}
has been implemented in modern flash memories (typically
those below \SI{30}{\nano\meter} for planar flash~\cite{cai.date13, shim.vlsit11, yang.fms11, fukami.di17}).

The read-retry mechanism allows the read reference
voltages to dynamically adjust to changes in distributions.
During read-retry, the SSD controller first reads the data out
of NAND flash memory with the default read reference voltage.
It then sends the data for error correction. If ECC successfully
corrects the errors in the data, the read operation
succeeds. Otherwise, the SSD controller reads the memory
again with a \emph{different} read reference voltage. The controller
repeats these steps until it either successfully reads the data
using a certain set of read reference voltages or is unable to
correctly read the data using all of the read reference voltages
that are available to the mechanism.

While read-retry is widely implemented today, it can
significantly increase the overall read operation latency due
to the multiple read attempts it causes~\cite{cai.hpca15}. Mechanisms
have been proposed to reduce the number of read-retry
attempts while taking advantage of the effective capability
of read-retry for reducing read errors, and read-retry has
also been used to enable mitigation mechanisms for various
other types of errors, as we describe in Section~\ref{sec:mitigation:voltage}. As a
result, read-retry is an essential mechanism in modern SSDs
to mitigate read errors (i.e., errors that manifest themselves
during a read operation).

\subsection{Voltage Optimization}
\label{sec:mitigation:voltage}

Many raw bit errors in NAND flash memory are affected
by the various voltages used within the memory to enable
reading of values. We give two examples. First, a suboptimal
\emph{read reference voltage} can lead to a large number of read
errors (Section~\ref{sec:errors}), especially after the threshold voltage distribution
shifts. Second, as we saw in Section~\ref{sec:errors:readdisturb}, the \emph{pass-through
voltage} can have a significant effect on the number
of read disturb errors that occur. As a result, optimizing these
voltages such that they minimize the total number of errors
that are induced can greatly mitigate error counts. In this section,
we discuss mechanisms that can discover and employ
the optimal\footnote{Or, more precisely, near-optimal, if the read-retry steps are too
coarse grained to find the optimal voltage.} read reference and pass-through voltages.

\begin{figure}[h]
  \centering
  \includegraphics[width=0.7\columnwidth]{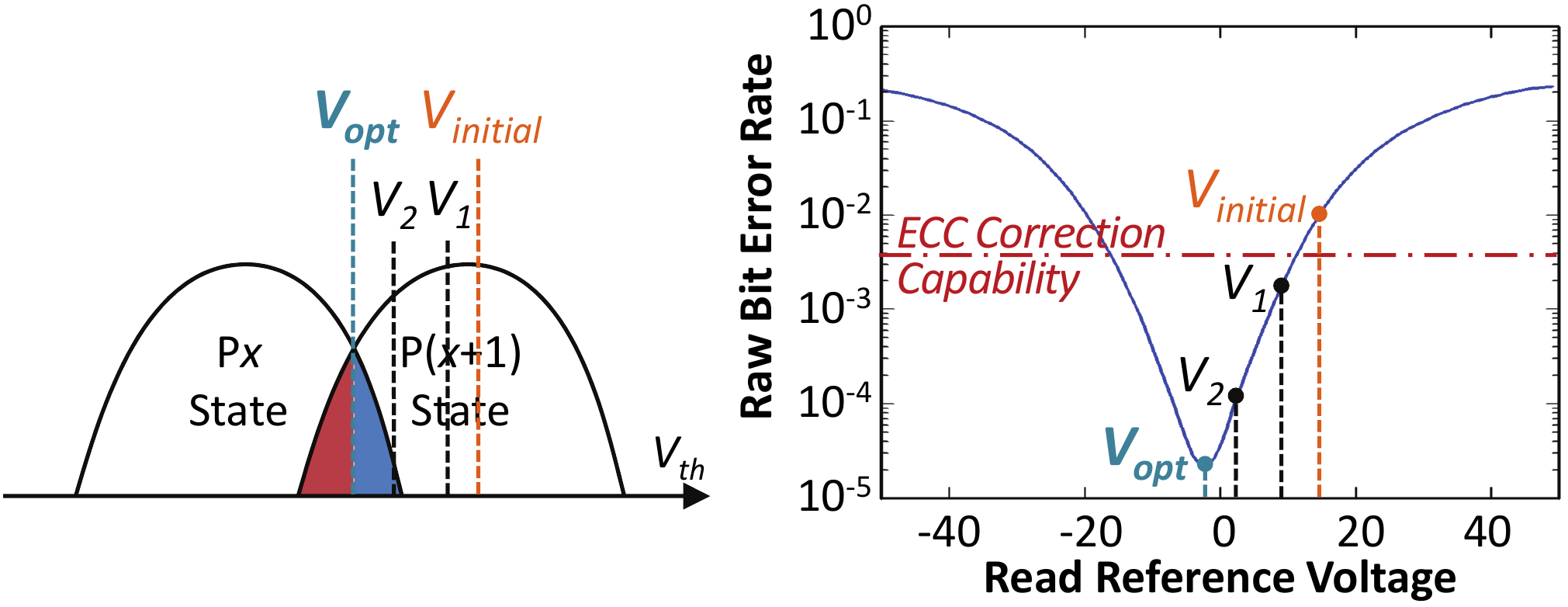}%
  \caption[Finding the optimal read reference voltage after the
threshold voltage distributions overlap (left), and raw bit error rate
as a function of the selected read reference voltage (right).]
  {Finding the optimal read reference voltage after the
threshold voltage distributions overlap (left), and raw bit error rate
as a function of the selected read reference voltage (right). \chI{Reproduced from~\cite{cai.arxiv17}.}}%
  \label{fig:F23}%
\end{figure}

\sloppypar
\paratitle{Optimizing Read Reference Voltages Using Disparity-Based Approximation and Sampling}
As we discussed in
Section~\ref{sec:mitigation:retry}, when the threshold voltage distribution shifts,
it is important to move the read reference voltage to the
point where the number of read errors is minimized. After
the shift occurs and the threshold voltage distribution of
each state widens, the distributions of different states may
overlap with each other, causing many of the cells within
the overlapping regions to be misread. The number of errors
due to misread cells can be minimized by setting the read
reference voltage to be exactly at the point where the distributions
of two neighboring states intersect, which we call
the \emph{optimal read reference voltage} ($V_{opt}$)~\cite{cai.iccd13, cai.sigmetrics14, cai.hpca15, luo.jsac16, papandreou.glsvlsi14}, 
illustrated in Figure~\ref{fig:F23}. Once the optimal read reference
voltage is applied, the raw bit error rate is minimized,
improving the reliability of the device.

\FloatBarrier

One approach to finding $V_{opt}$ is to adaptively learn and
apply the optimal read reference voltage for each flash block
through sampling~\cite{cai.hpca15, chen.patent16, cohen.patent14, wu.patent15}. The key idea is
to periodically (1)~use \emph{disparity} information (i.e., the ratio
of 1s to 0s in the data) to attempt to find a read reference
voltage for which the error rate is lower than the ECC correction
capability; and to (2)~use \emph{sampling} to efficiently tune
the read reference voltage to its optimal value to reduce the
read operation latency. Prior characterization of real NAND
flash memory~\cite{cai.hpca15, papandreou.glsvlsi14} found that the value of $V_{opt}$ does
\emph{not} shift greatly over a short period of time (e.g., a day), and
that all pages within a block experience \emph{similar} amounts of
threshold voltage shifts, as they have the same amount of
wearout and are programmed around the same time~\cite{cai.hpca15, papandreou.glsvlsi14}. 
Therefore, we can invoke \chIII{the} $V_{opt}$ learning mechanism
periodically (e.g., daily) to efficiently tune the \emph{initial
read reference voltage} (i.e., the first read reference voltage
used when the controller invokes the read-retry mechanism,
described in Section~\ref{sec:mitigation:retry}) for each flash block, ensuring that
the initial voltage used by read-retry stays close to $V_{opt}$ even
as the threshold voltage distribution shifts.

The SSD controller searches for $V_{opt}$ by counting the
number of errors that need to be corrected by ECC during
a read. However, there may be times where the initial
read reference voltage ($V_{initial}$) is set to a value at which the
number of errors during a read exceeds the ECC correction
capability, such as the raw bit error rate for $V_{initial}$ in Figure~\ref{fig:F23}
(right). When the ECC correction capability is exceeded, the
SSD controller is unable to count how many errors exist in
the raw data. The SSD controller uses \emph{disparity-based read
reference voltage approximation}~\cite{chen.patent16, cohen.patent14, wu.patent15} for each
flash block to try to bring $V_{initial}$ to a region where the number
of errors does not exceed the ECC correction capability.
Disparity-based read reference voltage approximation takes
advantage of data scrambling. Recall from \chI{Section~\ref{sec:ssdarch:ctrl:scrambling}} that
to minimize data value dependencies for the error rate, the
SSD controller scrambles the data written to the SSD to
probabilistically ensure that an equal number of 0s and 1s
exist in the flash memory cells. The key idea of disparity-based
read reference voltage approximation is to find the
read reference voltages that result in approximately 50\%
of the cells reading out bit value 0, and the other 50\% of
the cells reading out bit value 1. To achieve this, the SSD
controller employs a binary search algorithm, which tracks
the ratio of 0s to 1s for each read reference voltage it tries.
The binary search tests various read reference voltage values,
using the ratios of previously tested voltages to narrow
down the range where the read reference voltage can have
an equal ratio of 0s to 1s. The binary search algorithm continues
narrowing down the range until it finds a read reference
voltage that satisfies the ratio.

\begin{figure}[h]
  \centering
  \includegraphics[width=0.65\columnwidth]{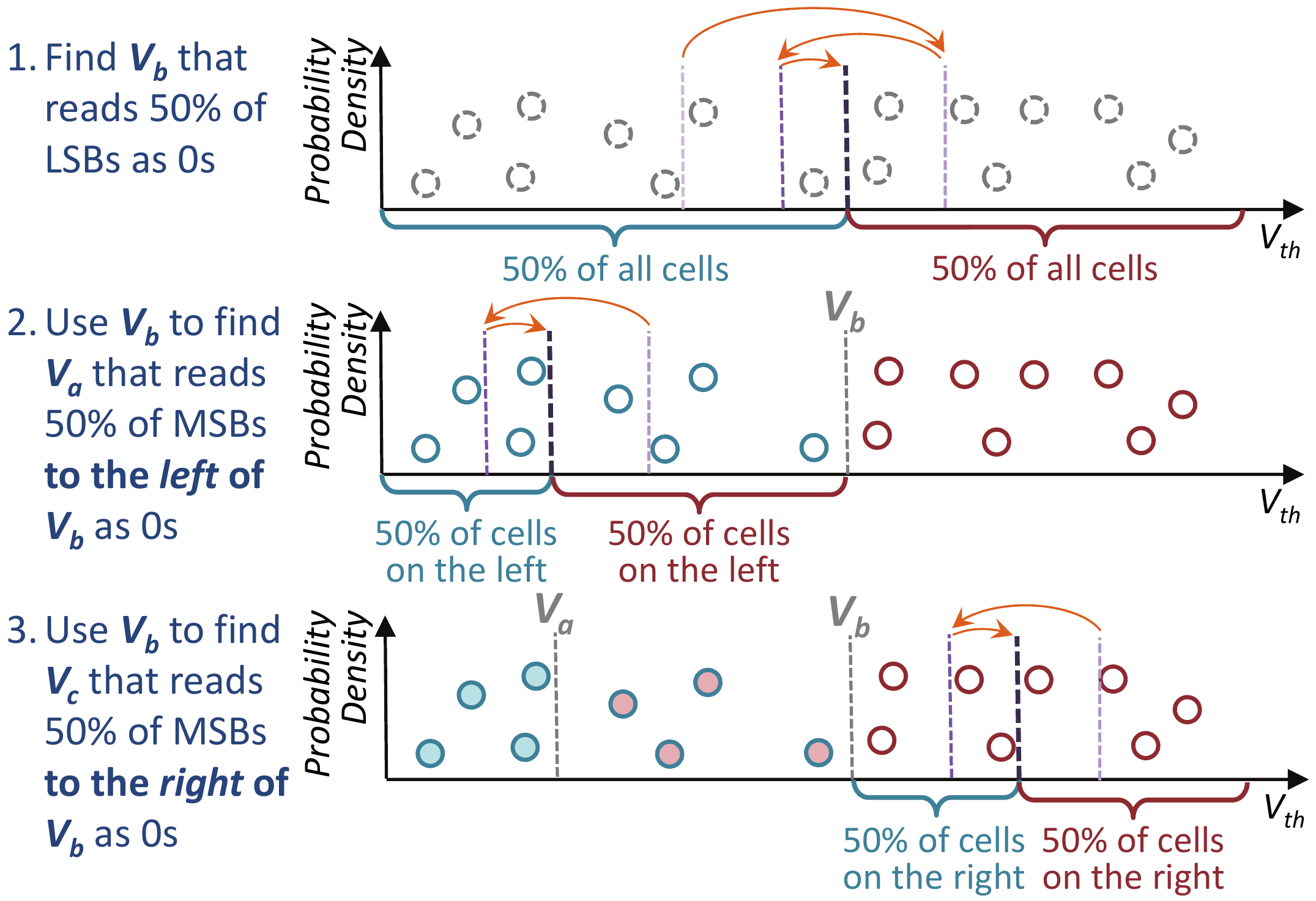}%
  \caption[Disparity-based read reference voltage approximation to
find $V_{initial}$ for MLC NAND flash memory. Each circle represents a
cell, where a dashed border indicates that the LSB is undetermined,
a solid border indicates that the LSB is known, a hollow circle
indicates that the MSB is unknown, and a filled circle indicates that
the MSB is known.]
  {Disparity-based read reference voltage approximation to
find $V_{initial}$ for MLC NAND flash memory. Each circle represents a
cell, where a dashed border indicates that the LSB is undetermined,
a solid border indicates that the LSB is known, a hollow circle
indicates that the MSB is unknown, and a filled circle indicates that
the MSB is known. \chI{Reproduced from~\cite{cai.arxiv17}.}}%
  \label{fig:F24}%
\end{figure}

The usage of the binary search algorithm depends on the
type of NAND flash memory used within the SSD. For SLC
NAND flash, the controller searches for only a single read
reference voltage. For MLC NAND flash, there are three read
reference voltages: the LSB is determined using $V_b$, and the
MSB is determined using both $V_a$ and $V_c$ (see Section~\ref{sec:flash:read}).
Figure~\ref{fig:F24} illustrates the search procedure for MLC NAND flash.
First, the controller uses binary search to find $V_b$, choosing a
voltage that reads the LSB of 50\% of the cells as data value 0
(step~1 in Figure~\ref{fig:F24}). For the MSB, the controller uses the discovered
$V_b$ value to help search for $V_a$ and $V_c$. Due to scrambling,
cells should be equally distributed across each of the
four voltage states. The controller uses binary search to set
$V_a$ such that 25\% of the cells are in the ER state, by ensuring
that half of the cells \emph{to the left of $V_b$} are read with an MSB of
0 (step~2). Likewise, the controller uses binary search to set
$V_c$ such that 25\% of the cells are in the P3 state, by ensuring
that half of the cells \emph{to the right of $V_b$} are read with an
MSB of 0 (step~3). This procedure is extended in a similar
way to approximate the voltages for TLC NAND flash.

\FloatBarrier

If disparity-based approximation finds a value for $V_{initial}$
where the number of errors during a read can be counted by
the SSD controller, the controller invokes \emph{sampling-based
adaptive $V_{opt}$ discovery}~\cite{cai.hpca15} to minimize the error count, and
thus reduce the read latency. Sampling-based adaptive $V_{opt}$
discovery learns and records $V_{opt}$ for the \emph{last-programmed
page} in each block. \chIII{Prior work samples} only the last-programmed
page because it is the page with the lowest data retention
age in the flash block. As retention errors cause the higher-voltage
states to shift to the left (i.e., to lower voltages), the
last-programmed page usually provides an \emph{upper bound} of
$V_{opt}$ for the entire block.

During sampling-based adaptive $V_{opt}$ discovery, the SSD
controller first reads the last-programmed page using $V_{initial}$,
and attempts to correct the errors in the raw data read from
the page. Next, it records the number of raw bit errors as
the current lowest error count $N_{ERR}$, and sets the applied
read reference voltage ($V_{ref}$) as $V_{initial}$. Since $V_{opt}$ typically
decreases over retention age, the controller first attempts
to lower the read reference voltage for the last-programmed
page, decreasing the voltage to $V_{ref} - \Delta V$ and reading the
page. If the number of corrected errors in the new read is
less than or equal to the old $N_{ERR}$, the controller updates
$N_{ERR}$ and $V_{ref}$ with the new values. The controller continues
to lower the read reference voltage until the number
of corrected errors in the data is greater than the old $N_{ERR}$
or the lowest possible read reference voltage is reached.
Since the optimal threshold voltage might increase in rare
cases, the controller also tests increasing the read reference
voltage. It increases the voltage to $V_{ref} + \Delta V$ and reads
the last-programmed page to see if $N_{ERR}$ decreases. Again, it
repeats increasing $V_{ref}$ until the number of corrected errors
in the data is greater than the old $N_{ERR}$ or the highest possible
read reference voltage is reached. The controller sets the
initial read reference voltage of the block as the value of $V_{ref}$
at the end of this process so that the next time an uncorrectable
error occurs, read-retry starts at a $V_{initial}$ that is hopefully
closer to the optimal read reference voltage ($V_{opt}$).

During the course of the day, as more retention errors
(the dominant source of errors on already-programmed
blocks) accumulate, the threshold voltage distribution shifts
to the left (i.e., voltages decrease), and \chIII{the} initial read reference
voltage (i.e., $V_{initial}$) is now an upper bound for the read-retry
voltages. Therefore, whenever read-retry is invoked,
the controller now needs to only decrease the read reference
voltages (as opposed to traditional read-retry, which
tries \emph{both} lower and higher voltages~\cite{cai.hpca15}). Sampling-based
adaptive $V_{opt}$ discovery improves the \emph{endurance} (i.e., the
number of P/E cycles before the ECC correction capability is
exceeded) of the NAND flash memory by 64\% and reduces
error correction latency by 10\%~\cite{cai.hpca15}, and is employed in
some modern SSDs today.

\paratitle{Other Approaches to Optimizing Read Reference Voltages}
One drawback of the sampling-based adaptive technique is
that it requires time and storage overhead to find and record
the per-block initial voltages. To avoid this, the SSD controller
can employ an accurate \emph{online threshold voltage distribution
model}~\cite{cai.date13, cai.thesis12}, which can efficiently track and
predict the shift in the distribution over time. The model
represents the threshold voltage distribution of each state as
a probability density function (PDF), and the controller can
use the model to calculate the intersection of the different
PDFs. The controller uses the PDF in place of the threshold
voltage sampling, determining $V_{opt}$ by calculating the intersection
of the distribution of each state in the model.
\chIII{Chapter~\ref{sec:vthmodel} demonstrates an example of this approach.}

Other prior work examines adapting read reference voltages
based on P/E cycle count, retention age, or read disturb.
In one such work, the controller periodically learns
read reference voltages by testing three read reference voltages
on six pages per block, which the work demonstrates
to be sufficiently accurate~\cite{papandreou.glsvlsi14}. Similarly, error correction
using LDPC soft decoding (see Section~\ref{sec:correction:ldpc}) requires reading
the same page using multiple sets of read reference
voltages to provide fine-grained information on the probability
of each cell representing a bit value 0 or a bit value 1.
Another prior work optimizes the read reference voltages to
increase the ECC correction capability without increasing
the coding rate~\cite{wang.jsac14}.

\paratitle{Optimizing Pass-Through Voltage to Reduce Read Disturb Errors}
As we discussed in Section~\ref{sec:errors:readdisturb}, the vulnerability of a
cell to read disturb is directly correlated with the voltage difference
($V_{pass} - V_{th}$) through the cell oxide~\cite{cai.dsn15}. Traditionally,
a single $V_{pass}$ value is used \emph{globally} for the entire flash memory,
and the value of $V_{pass}$ must be higher than \emph{all} potential
threshold voltages within the chip to ensure that unread
cells along a bitline are turned on during a read operation
(see Section~\ref{sec:flash:read}). To reduce the impact of read disturb,
we can tune $V_{pass}$ to reduce the size of the voltage difference
($V_{pass} - V_{th}$). However, it is difficult to reduce $V_{pass}$ \emph{globally},
as any cell with a value of $V_{th} > V_{pass}$ introduces an
error during a read operation (which we call a \emph{pass-through
error}).

\begin{figure}[h]
  \centering
  \includegraphics[width=0.65\columnwidth]{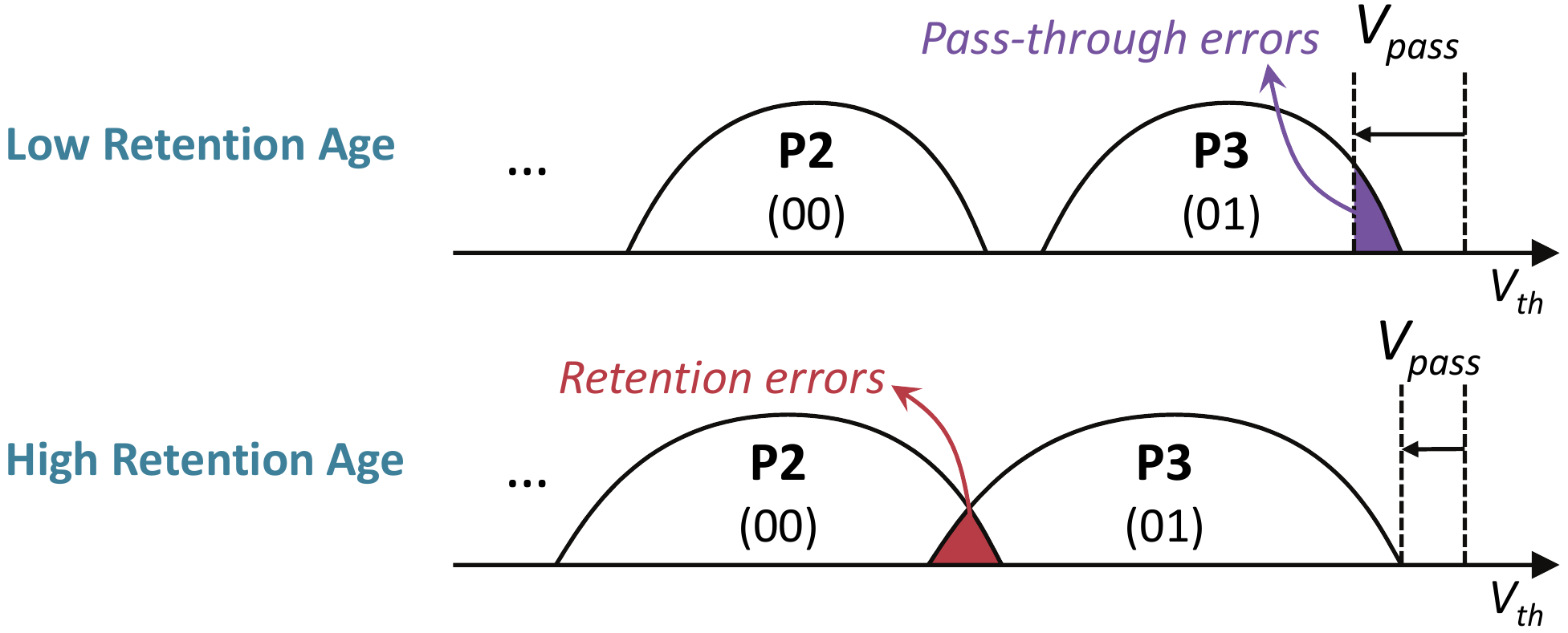}%
  \caption[Dynamic pass-through voltage tuning at different
retention ages.]
  {Dynamic pass-through voltage tuning at different
retention ages. \chI{Reproduced from~\cite{cai.arxiv17}.}}%
  \label{fig:F25}%
\end{figure}

\chIII{In prior work, we} propose a mechanism that can dynamically lower
$V_{pass}$ while ensuring that it can correct any new pass-through
errors introduced. The key idea of the mechanism is to lower
$V_{pass}$ only for those blocks where ECC has enough leftover
error correction capability (see \chI{Section~\ref{sec:ssdarch:ctrl:ecc}}) to correct the
newly introduced pass-through errors. When the retention
age of the data within a block is low, \chIII{prior work finds} that the raw
bit error rate of the block is much lower than the rate for
the block when the retention age is high, as the number of
data retention and read disturb errors remains low at low
retention age~\cite{cai.dsn15, ha.tcad16}. As a result, a block with a low retention
age has significant \emph{unused} ECC correction capability,
which we can use to correct the pass-through errors we
introduce when we lower $V_{pass}$, as shown in Figure~\ref{fig:F25}. Thus,
when a block has a low retention age, the controller lowers
$V_{pass}$ aggressively, making it much less likely for read disturbs
to induce an uncorrectable error. When a block has
a high retention age, the controller also lowers $V_{pass}$, but
does not reduce the voltage aggressively, since the limited
ECC correction capability now needs to correct retention
errors, and might not have enough unused correction capability
to correct many new pass-through errors. By reducing
$V_{pass}$ aggressively when a block has a low retention age, we
can extend the time before the ECC correction capability is
exhausted, improving the flash lifetime.

\FloatBarrier

\chIII{The previously-proposed} read disturb mitigation mechanism~\cite{cai.dsn15} learns the
minimum pass-through voltage for each block, such that
all data within the block can be read correctly with ECC.
\chIII{The previously-proposed} learning mechanism works online and is triggered
periodically (e.g., daily). The mechanism is implemented in
the controller, and has two components. It first finds the
size of the ECC margin $M$ (i.e., the unused correction capability)
that can be exploited to tolerate additional read errors
for each block. Once it knows the available margin $M$, \chIII{the previously-proposed}
mechanism calibrates $V_{pass}$ on a per-block basis to find the
lowest value of $V_{pass}$ that introduces no more than $M$ additional
raw errors (i.e., there are no more than M cells where
$V_{th} > V_{pass}$). \chIII{The} findings on MLC NAND flash memory
show that the mechanism can improve flash endurance by
an average of 21\% for a variety of disk workloads~\cite{cai.dsn15}.

\paratitle{Programming and Erase Voltages}
Prior work also examines
tuning the programming and erase voltages to extend
flash endurance~\cite{jeong.fast14}. By decreasing the two voltages when
the P/E cycle count is low, the accumulated wearout for
each program or erase operation is reduced, which, in turn,
increases the overall flash endurance. Decreasing the programming
voltage, however, comes at the cost of increasing
the time required to perform ISPP, which, in turn, increases
the overall SSD write latency~\cite{jeong.fast14}.

\subsection{Hot Data Management}
\label{sec:mitigation:hotcold}

The data stored in \chI{different locations of an} SSD can be accessed by the host at
different rates.
These pages exhibit high temporal write
locality, and are called \emph{write-hot} pages.
Likewise, pages with a high
amount of temporal read locality \chI{(i.e., pages that are accessed
by a large fraction of the read operations)} are called \emph{read-hot} pages.
A number of issues can arise when an SSD does not distinguish
between write-hot pages and \emph{write-cold} pages (i.e.,
pages with low temporal write locality), or between read-hot
pages and \emph{read-cold} pages (i.e., pages with low temporal
read locality). For example, if write-hot pages and write-cold
pages are \chI{stored} within the same block, refresh
mechanisms \chI{(which operate at the block level; see Section~\ref{sec:mitigation:refresh}) \emph{cannot}} avoid refreshes to pages that were overwritten
recently. \chI{This increases} not only \chI{the} energy consumption
but also \chI{the} write amplification due to remapping-based refresh~\cite{luo.msst15}.
Likewise, if read-hot and read-cold pages are \chI{stored}
within the same block, read-cold pages are unnecessarily
exposed to a high number of read disturb errors~\cite{ha.apsys13, ha.tcad16}.
\emph{Hot data management} refers to a set of mechanisms that can
identify \chI{and exploit} write-hot or read-hot pages in the SSD. 
The key idea \chI{common to such mechanisms}
is to apply special SSD management policies by placing hot
pages and cold pages into \emph{separate} flash blocks.
\chIII{Chapter~\ref{sec:warm} demonstrates an example technique using this idea.}

\chIII{Prior} work~\cite{wang.date12} proposes to reuse the correctly functioning
flash pages within bad blocks (see \chI{Section~\ref{sec:ssdarch:ctrl:badblocks}}) to store
write-cold data. This technique increases the total number
of usable blocks available for overprovisioning, and extends
flash lifetime by delaying the point at which each flash chip
reaches the upper limit of bad blocks it can tolerate.

RedFTL identifies and replicates read-hot pages across
multiple flash blocks, allowing the controller to evenly
distribute read requests to these pages across the replicas~\cite{ha.apsys13}. 
Other works reduce the number of read reclaims
(see Section~\ref{sec:mitigation:refresh}) that need to be performed by mapping
read-hot data to particular flash blocks and lowering the
maximum possible threshold voltage for such blocks~\cite{cai.patent15.9218885, ha.tcad16}.
By lowering the maximum possible threshold voltage
for these blocks, the SSD controller can use a lower $V_{pass}$
value (see Section~\ref{sec:mitigation:voltage}) on the blocks without introducing
any additional errors during a read operation. To lower the
maximum threshold voltage in these blocks, the width of
the voltage window for each voltage state is decreased, and
each voltage window shifts to the left~\cite{cai.patent15.9218885, ha.tcad16}. Another
work applies stronger ECC encodings to \emph{only} read-hot
blocks based on the total read count of the block, in order
to increase SSD endurance without significantly reducing
the amount of overprovisioning~\cite{cai.patent15.192110} (see Section~\ref{sec:ssdarch:reliability} for
a discussion on the tradeoff between ECC strength and
overprovisioning).

\subsection{Adaptive Error Mitigation Mechanisms}
\label{sec:mitigation:adaptive}

Due to the many different factors that contribute to raw
bit errors, error rates in NAND flash memory can be highly
variable. Adaptive error mitigation mechanisms are capable of
adapting error tolerance capability to the error rate. They provide
stronger error tolerance capability when the error rate is
higher, improving flash lifetime significantly. When the error
rate is low, adaptive error mitigation techniques reduce error
tolerance capability to lower the cost of the error mitigation
techniques. In this section, we examine two types of adaptive
techniques: (1)~multi-rate ECC and (2)~dynamic cell levels.

\begin{figure}[h]
  \centering
  \includegraphics[width=0.65\columnwidth]{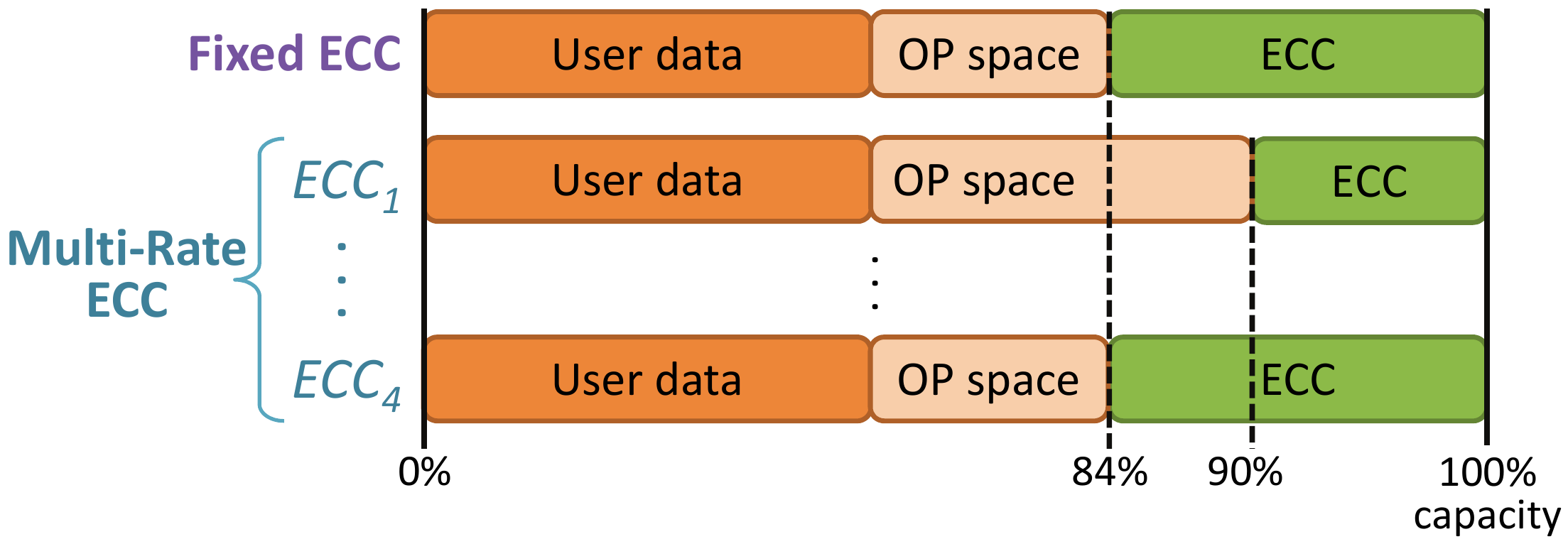}%
  \caption[Comparison of space used for user data, overprovisioning,
and ECC between a fixed ECC and a multi-rate ECC mechanism.]
  {Comparison of space used for user data, overprovisioning,
and ECC between a fixed ECC and a multi-rate ECC mechanism. \chI{Reproduced from~\cite{cai.arxiv17}.}}%
  \label{fig:F26}%
\end{figure}

\paratitle{Multi-Rate ECC}
Some works propose to employ
multiple ECC algorithms in the SSD controller~\cite{cai.patent16.9419655, haratsch.fms16, huang.atc14, wu.mascots10, chen.vts09}. 
Recall from Section~\ref{sec:ssdarch:reliability} that there is a
tradeoff between ECC strength (i.e., the coding rate; see
\chI{Section~\ref{sec:ssdarch:ctrl:ecc}}) and overprovisioning, as a codeword (which
contains a data chunk \emph{and} its corresponding ECC information)
uses more bits when stronger ECC is employed. The
key idea of multi-rate ECC is to employ a weaker codeword
(i.e., one that uses fewer bits for ECC) when the SSD is relatively
new and has a smaller number of raw bit errors, and
to use the saved SSD space to provide additional overprovisioning,
as shown in Figure~\ref{fig:F26}.

\begin{figure}[h]
  \centering
  \includegraphics[width=0.4\columnwidth]{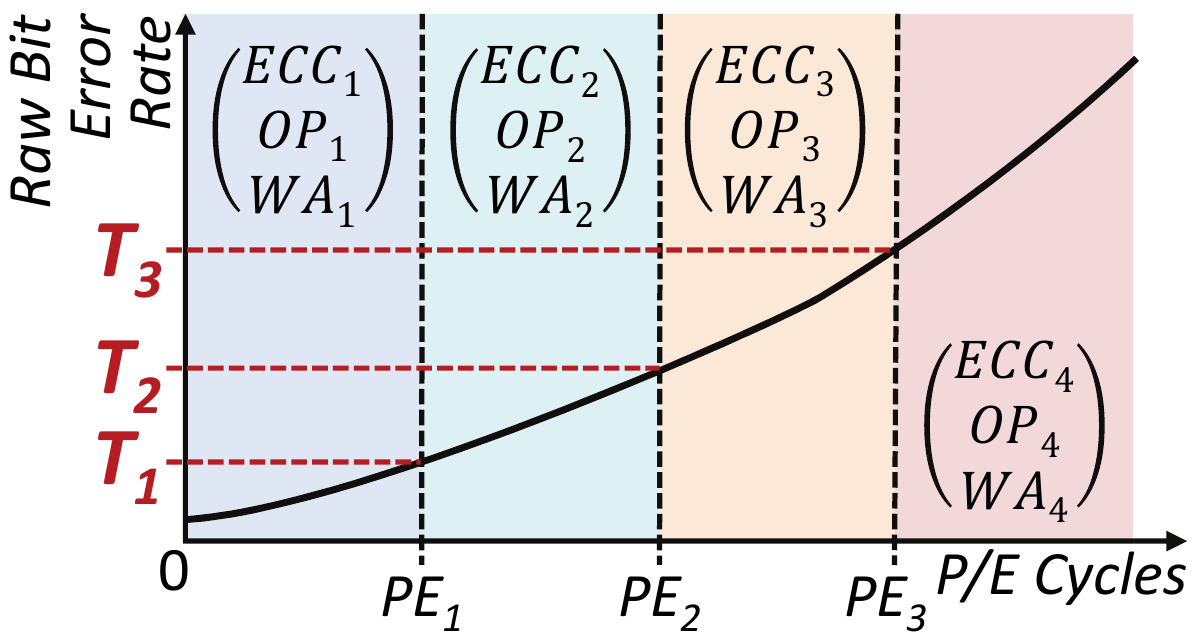}%
  \caption[Illustration of how multi-rate ECC switches to different
ECC codewords (i.e., ECC$_i$) as the RBER grows. OP$_i$ is the
overprovisioning factor used for engine ECC$_i$, and WA$_i$ is the
resulting write amplification value.]
  {Illustration of how multi-rate ECC switches to different
ECC codewords (i.e., ECC$_i$) as the RBER grows. OP$_i$ is the
overprovisioning factor used for engine ECC$_i$, and WA$_i$ is the
resulting write amplification value. \chI{Reproduced from~\cite{cai.arxiv17}.}}%
  \label{fig:F27}%
\end{figure}
\FloatBarrier

Let us assume that the controller contains a configurable
ECC engine that can support $n$ different types of ECC
codewords, which we call $\text{ECC}_i$. Figure~\ref{fig:F26} shows an example
of multi-rate ECC that uses four ECC engines, where $\text{ECC}_1$
provides the weakest protection but has the smallest codeword,
while $\text{ECC}_4$ provides the strongest protection with
the largest codeword. We need to ensure that the NAND
flash memory has enough space to fit the largest codewords,
e.g., those for $\text{ECC}_4$ in Figure~\ref{fig:F26}. Initially, when the raw bit
error rate (RBER) is low, the controller employs $\text{ECC}_1$,
as shown in Figure~\ref{fig:F27}. The smaller codeword size for $\text{ECC}_1$
provides additional space for overprovisioning, as shown
in Figure~\ref{fig:F26}, and thus reduces the effects of write amplification.
Multi-rate ECC works on an interval-by-interval
basis. Every interval (in this case, a predefined number
of P/E cycles), the controller measures the RBER. When
the RBER exceeds the threshold set for transitioning from
a weaker ECC to a stronger ECC, the controller switches
to the stronger ECC. For example, when the SSD exceeds
the first RBER threshold for switching ($T_1$ in Figure~\ref{fig:F27}),
the controller starts switching from $\text{ECC}_1$ to $\text{ECC}_2$. When
switching between ECC engines, the controller uses the
$\text{ECC}_1$ engine to decode data the next time the data is read
out, and stores a new codeword using the $\text{ECC}_2$ engine.
This process is repeated during the lifetime of flash memory
for each stronger engine $\text{ECC}_i$, where each engine has
a corresponding threshold that triggers switching~\cite{cai.patent16.9419655, haratsch.fms16, chen.vts09}, 
as shown in Figure~\ref{fig:F27}.

Multi-rate ECC allows the same maximum P/E cycle
count for each block as if $\text{ECC}_n$ was used throughout the
lifetime of the SSD, but reduces write amplification and
improves performance during the periods where the lower
strength engines are employed, by providing additional
overprovisioning (see Section~\ref{sec:ssdarch:reliability}) during those times.
As the lower-strength engines use smaller codewords
(e.g., $\text{ECC}_1$ versus $\text{ECC}_4$ in Figure~\ref{fig:F26}), the resulting free space
can instead be employed to further increase the amount of
overprovisioning within the NAND flash memory, which in
turn increases the total lifetime of the SSD. We compute
the lifetime improvement by modifying Equation~\ref{eq:E4} (Section~\ref{sec:ssdarch:reliability})
to account for each engine, as follows:
\begin{equation}
\text{Lifetime} = \sum_{i = 1}^{n} \frac{\text{PEC}_i \times (1 + \text{OP}_i)}{365 \times \text{DWPD} \times \text{WA}_i \times R_{compress}}
\label{eq:E9}
\end{equation}
In Equation~\ref{eq:E9}, $\text{WA}_i$ and $\text{OP}_i$ are the write amplification and overprovisioning
factor for $\text{ECC}_i$, and $\text{PEC}_i$ is the number of P/E
cycles that $\text{ECC}_i$ is used for. Manufacturers can set parameters
to maximize SSD lifetime in Equation~\ref{eq:E9}, by optimizing the values
of $\text{WA}_i$ and $\text{OP}_i$.

\begin{figure}[h]
  \centering
  \includegraphics[width=0.65\columnwidth]{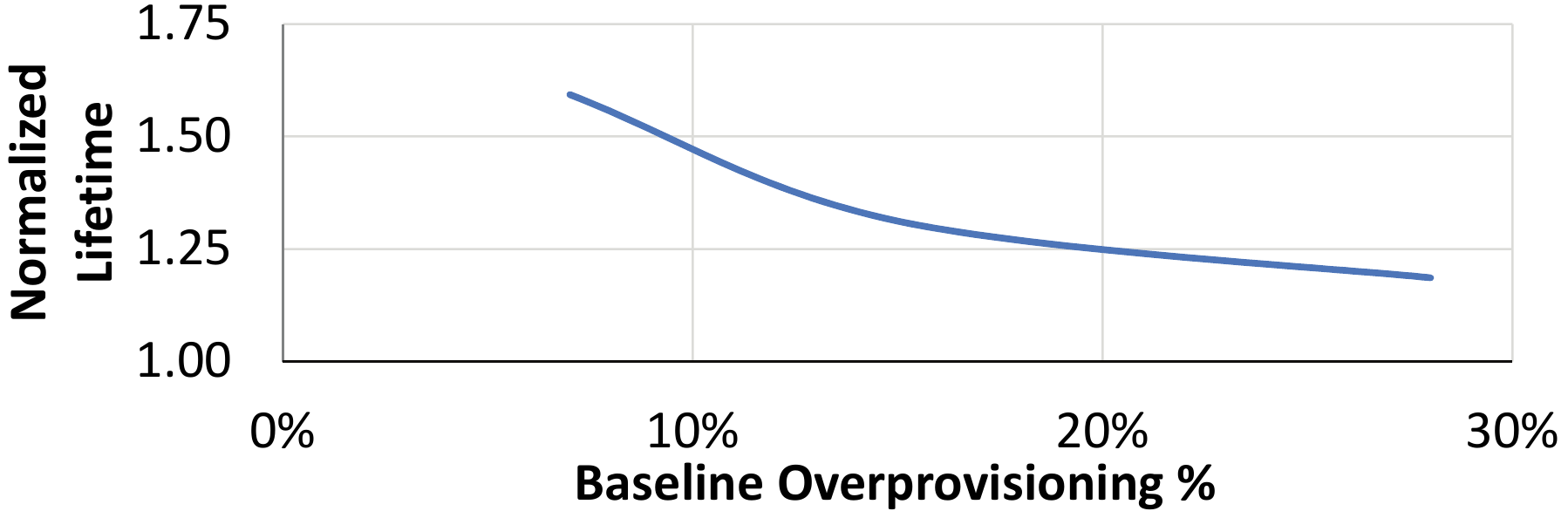}%
  \caption[Lifetime improvements of using multi-rate ECC over using
a fixed ECC coding rate.]
{Lifetime improvements of using multi-rate ECC over using
a fixed ECC coding rate. \chI{Reproduced from~\cite{cai.arxiv17}.}}%
  \label{fig:F28}%
\end{figure}

Figure~\ref{fig:F28} shows the lifetime improvements for a four-engine
multi-rate ECC, with the coding rates for the four
ECC engines ($\text{ECC}_1$--$\text{ECC}_4$) set to 0.90, 0.88, 0.86, and 0.84
(recall that a \emph{lower} coding rate provides stronger protection;
see Section~\ref{sec:ssdarch:reliability}), over a fixed ECC engine that employs a
coding rate of 0.84. We see that the lifetime improvements
of using multi-rate ECC are: (1)~significant, with a 31.2\%
increase if the baseline NAND flash memory has 15\% overprovisioning;
and (2)~greater when the SSD initially has a
smaller amount of overprovisioning.

\FloatBarrier

\paratitle{Dynamic Cell Levels}
A major reason that errors occur
in NAND flash memory is because the threshold voltage distribution
of each state overlaps more with those of neighboring
states as the distributions widen over time. Distribution
overlaps are a greater problem when more states are encoded
within the same voltage range. Hence, TLC flash has a much
lower endurance than MLC, and MLC has a much lower
endurance than SLC (assuming the same process technology
node). If we can increase the margins between the
states' threshold voltage distributions, the amount of overlap
can be reduced significantly, which in turn reduces the
number of errors.

\begin{figure}[h]
  \centering
  \includegraphics[width=0.75\columnwidth]{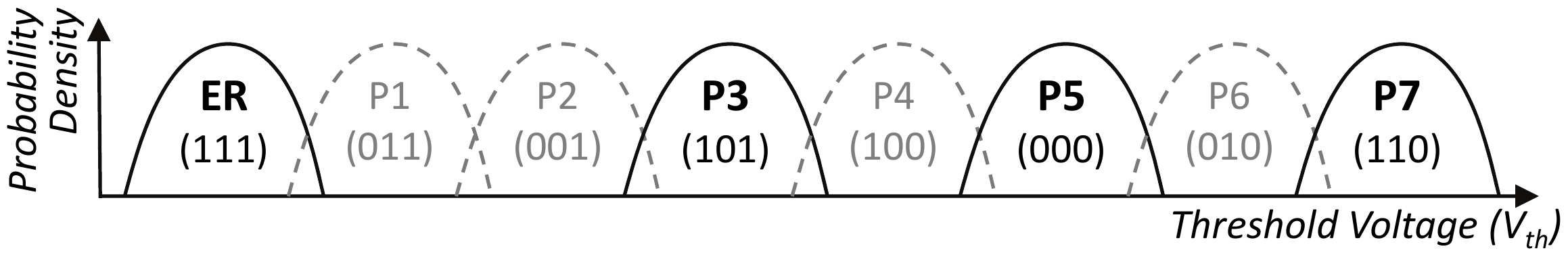}%
  \caption[States used when a TLC cell (with 8 states) is downgraded
to an MLC cell (with 4 states).]
  {States used when a TLC cell (with 8 states) is downgraded
to an MLC cell (with 4 states). \chI{Reproduced from~\cite{cai.arxiv17}.}}%
  \label{fig:F29}%
\end{figure}

Prior work proposes to increase margins by \emph{dynamically}
reducing the number of bits stored within a cell, e.g.,
by going from three bits that encode eight states (TLC)
to two bits that encode four states (equivalent to MLC),
or to one bit that encodes two states (equivalent to SLC)~\cite{cai.patent15.9218885, wilson.mascots14}. 
Recall that TLC uses the ER state and states
P1--P7, which are spaced out approximately equally.
When we \emph{downgrade} a flash block (i.e., reduce the number
of states its cells can represent) from eight states to
four, the cells in the block now employ only the ER state
and states P3, P5, and P7. As we can see from Figure~\ref{fig:F29}, this
provides large margins between states P3, P5, and P7, and
provides an even larger margin between ER and P3. The
SSD controller maintains a list of all of the blocks that
have been downgraded. For each read operation, the SSD
controller checks if the target block is in the downgraded
block list, and uses this information to interpret the data
that it reads out from the wordline of the block.

\FloatBarrier

A cell can be downgraded to reduce various types of
errors (e.g., wearout, read disturb). To reduce wearout, a
cell is downgraded when it has high wearout. To reduce read
disturb, a cell can be downgraded if it stores \emph{read-hot} data
(i.e., the most frequently read data in the SSD). By using
fewer states for a block that holds read-hot data, we can
reduce the impact of read disturb because it becomes harder
for the read disturb mechanism to affect the distributions
enough for them to overlap. As an optimization, the SSD
controller can employ various hot-cold data partitioning
mechanisms (e.g.,~\cite{luo.msst15, cai.patent15.9218885, ha.apsys13, cai.patent15.192110}) to keep read-hot
data in specially designated blocks~\cite{cai.patent15.9218885, ha.apsys13, ha.tcad16, cai.patent15.192110},
allowing the controller to reduce the size of the downgraded
block list and isolate the impact of read disturb from \emph{read-cold}
(i.e., infrequently read) data.

Another approach to dynamically increasing the distribution
margins is to perform program and erase operations
more slowly when the SSD write request throughput is low~\cite{cai.patent15.9218885, jeong.fast14}. 
Slower program/erase operations allow the final
voltage of a cell to be programmed more precisely, and
reduce the amount of oxide degradation that occurs during
programming. As a result, the distribution of each state is
initially much narrower, and subsequent widening of the
distributions results in much lower overlap for a given P/E
cycle count. This technique improves the SSD lifetime by
an average of 61.2\% for a variety of disk workloads~\cite{jeong.fast14}.
Unfortunately, the slower program/erase operations come
at the cost of higher SSD latency, and are thus not applied
during periods of high write traffic. One way to mitigate
the impact of the higher write latency is to perform slower
program/erase operations only during garbage collection,
which ensures that the higher latency occurs only when the
SSD is idle~\cite{cai.patent15.9218885}. As a result, read and write requests from
the host do not experience any additional delays.


\section{Error Correction and Data Recovery Techniques}
\label{sec:correction}

Now that we have described a variety of error mitigation
mechanisms that can target various types of error sources,
we turn our attention to the error correction flow that is
employed in modern SSDs as well as \emph{data recovery techniques}
that can be employed when the error correction flow
fails to produce correct data.
\chI{In this section, we briefly overview the major error correction steps
an SSD performs when reading data. 
We first discuss two ECC encodings that are typically used by modern SSDs:
Bose--Chaudhuri--Hocquenghem (BCH) codes~\cite{shu.book04, lee.isscc12, hocquenghem.chiffres59, bose.ic60} and 
low-density parity-check (LDPC) codes\chI{~\cite{shu.book04, gallager.ire62, mackay.letters96,mackay.letters97, gallager.tit62}}
(Section~\ref{sec:correction:ecc}).
Next, we go through example error correction flows for an SSD that uses
either BCH codes or LDPC codes (Section~\ref{sec:correction:flow}).
Then, we compare the error correction strength (i.e., the
number of errors that ECC can correct) when we employ
BCH codes or LDPC codes in an SSD (Section~\ref{sec:correction:strength}). Finally,}
we discuss techniques that can rescue data from an SSD
when the BCH/LDPC decoding fails to correct all errors
(Section~\ref{sec:correction:recovery}).

\subsection{\chI{Error-Correcting Codes Used in SSDs}}
\label{sec:correction:ecc}

Modern SSDs typically employ one of two types of
ECC. Bose--Chaudhuri--Hocquenghem (BCH) codes
allow for the correction of multiple bit errors~\cite{shu.book04, lee.isscc12, hocquenghem.chiffres59, bose.ic60}, 
and are used to correct the errors observed during
a \emph{single} read from the NAND flash memory~\cite{lee.isscc12}. Low-density
parity-check (LDPC) codes employ information
accumulated over \emph{multiple} read operations to determine
the likelihood of each cell containing a bit value 1 or a bit
value 0\chI{~\cite{shu.book04, gallager.ire62, mackay.letters96,mackay.letters97, gallager.tit62}}, providing stronger protection at
the cost of greater decoding latency and storage overhead~\cite{zhao.fast13, wang.jsac14}.
\chI{Next, we describe the basics of BCH and LDPC codes.}

\subsubsection{\chI{Bose--Chaudhuri--Hocquenghem (BCH) Codes}}
\label{sec:correction:ecc:bch}



\chI{BCH codes\chI{~\cite{shu.book04, lee.isscc12, hocquenghem.chiffres59, bose.ic60}} \chI{have been} widely used in modern SSDs \chI{during} the past decade due to \chI{their}
ability to detect and correct multi-bit errors while keeping \chI{the latency
and hardware cost of encoding and decoding low}~\cite{lee.isscc12, micheloni.isscc06, chen.tit81,
marelli.chapter16}. 
\chI{For SSDs, BCH codes are designed to be \emph{systematic}, which means that the 
\chI{original data message is embedded \chI{\emph{verbatim}} within the codeword}.}
\chI{Within an $n$-bit codeword} \chI{(see 
\chI{Section~\ref{sec:ssdarch:ctrl:ecc}}), error-correcting codes use} the first $k$ bits
\chI{of the codeword, \chI{called \emph{data bits},} to hold the \chI{data} message bits,}
and the remaining $(n-k)$ bits, \chI{called \emph{check bits},} \chI{to hold}
error correction information \chI{\chI{that} protects} the data bits. 
BCH codes are designed \chI{to \emph{guarantee} that they}
correct up to a certain number of raw bit errors (e.g., \chI{$t$} error bits)
within each codeword, \chI{which depends on the values chosen for $n$ and $k$}.
A stronger error correction strength (i.e., a larger \chI{$t$})
requires more redundant \chI{check} bits (i.e., $(n-k)$) or a longer codeword length (i.e.,
$n$).}

\chI{A BCH code\chI{~\cite{shu.book04, lee.isscc12, hocquenghem.chiffres59, bose.ic60}} is a linear block code that consists of check bits generated by
an algorithm.  The codeword generation algorithm ensures that the check bits are selected such that
the check bits can be used during a parity check to detect and correct up to
\chI{$t$} bit errors in the codeword.  A BCH code is defined by (1)~a generator matrix 
$G$, which informs the generation algorithm of how to generate each check bit
using the data bits; and (2)~a parity check matrix $H$, which can be applied to
the codeword to detect if any errors exist.  In order for a BCH code to guarantee
that it can correct \chI{$t$} errors within each codeword, the minimum separation \chI{$d$}
(i.e., the Hamming distance) between valid codewords must be at least
\chI{$d = 2t+1$}~\cite{shu.book04}.}


\paratitle{\chI{BCH Encoding}}
\chI{\chI{The codeword generation algorithm encodes a $k$-bit \chI{data} message $m$ into an
$n$-bit BCH codeword $c$, by computing the dot product of \chI{$m$} and the 
generator matrix $G$ (i.e., $c = m \cdot G$).  $G$ is defined
within a finite Galois field \chI{$GF(2^d) = \{0, \alpha^0, \alpha^1, \ldots,
\alpha^{2^d-1}\}$, where $\alpha$ is a \emph{primitive element} of the field
and $d$ is a positive integer}~\cite{dolecek.fms14}.
An SSD manufacturer constructs $G$} from a set of
polynomials \chI{$g_1(x), g_2(x), \ldots g_{2t}(x)$}, where $g_i(\alpha^i) = 0$. 
\chI{Each polynomial generates a \emph{parity bit}, which is 
used during decoding to determine if any errors were introduced.}
The
$i$-th row of $G$ encodes the $i$-th polynomial $g_i(x)$. When decoding, the
codeword $c$ can be viewed as a polynomial $c(x)$. Since $c(x)$ is generated by
$g_i(x)$ which has a root $\alpha^i$, $\alpha^i$ should also be a root of
$c(x)$. The parity check matrix $H$ is constructed such that $cH^t$ calculates
$c(\alpha_i)$. Thus, the element \chI{in the $i$-th row and $j$-th column of $H$} is
$H_{ij} = \alpha^{(j-1) (i+1)}$. \chI{This allows the decoder to use $H$ to
quickly determine if any of the parity bits do not match, which indicates that
there are errors within the codeword.}  \chI{BCH codes in SSDs are
typically designed to be \emph{systematic}, which \chI{guarantees that a
verbatim copy of the \chI{data} message is embedded within}
the codeword. To form a systematic BCH code,} the generator
matrix and the parity check matrix are transformed such that \chI{they
contain the identity matrix.}

\paratitle{\chI{BCH Decoding}}
\chI{When the SSD controller is servicing a read request, it must extract
the data bits (i.e., the $k$-bit \chI{data} message $m$) from the BCH codeword that is 
stored in the NAND flash memory chips.
Once the controller retrieves the codeword, which we call $r$, from NAND flash 
memory, it sends $r$ to a BCH decoder.  The decoder performs five steps, as 
illustrated in Figure~\ref{fig:bch-decoding-flow}, which correct the retrieved 
codeword $r$ to obtain the originally-written codeword $c$, and then extract 
the \chI{data} message $m$ from $c$.
In Step~1, the decoder uses \emph{syndrome calculation} to detect if any errors exist
within the retrieved codeword $r$.  If no errors are detected, the decoder uses
the retrieved codeword as the original codeword, $c$, and skips
to Step~5.  Otherwise, the decoder continues on to correct the errors and
recover $c$.
In Step~2, the decoder uses the syndromes from Step~1 to construct an
\emph{error location polynomial}, which encodes the locations of each
detected bit error within $r$.
In Step~3, the decoder extracts the specific location of each detected bit error from the
error location polynomial.
In Step~4, the decoder corrects each detected bit error in the retrieved codeword $r$
to recover the original codeword $c$.
In Step~5, the decoder extracts the \chI{data} message from the original codeword $c$.
We describe the algorithms \chI{most commonly used by BCH decoders in 
SSDs~\cite{lee.isscc12, choi.tvlsi09, liu.sips06}} for each step in detail below.}

\begin{figure}[h]
  \centering
  \includegraphics[width=0.85\linewidth]{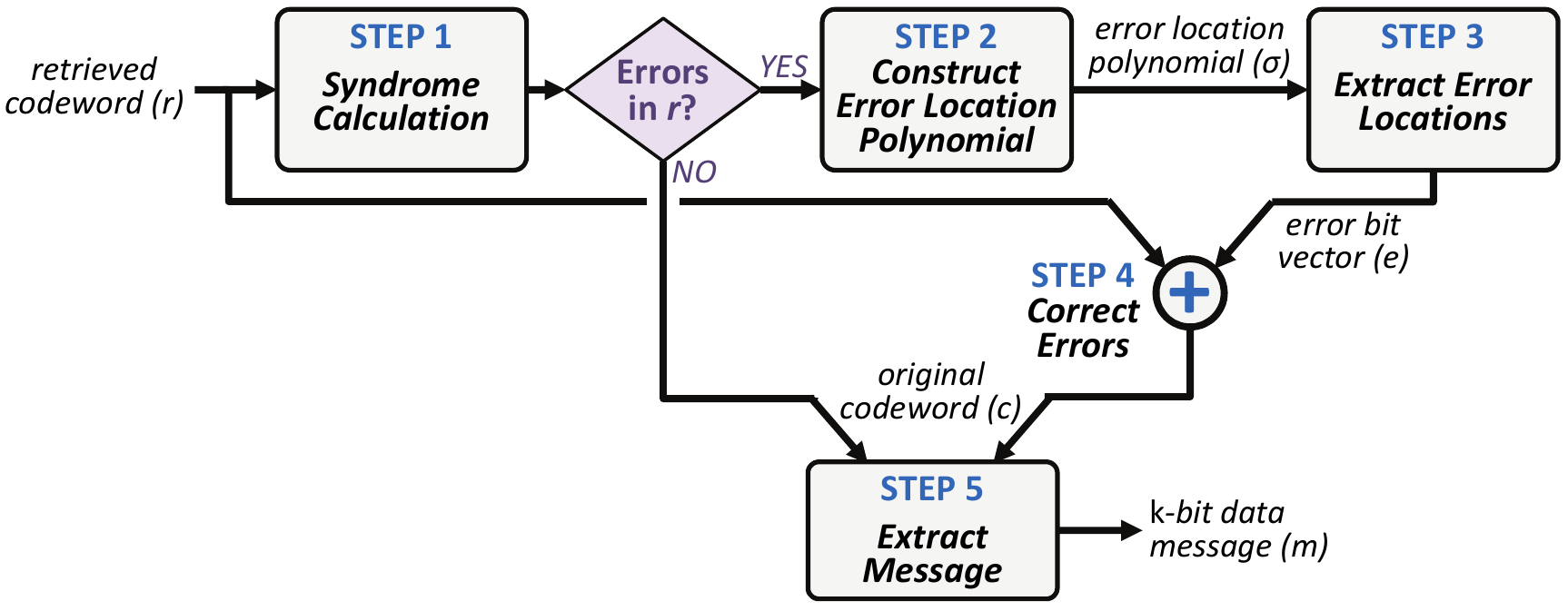}%
  \caption{\chI{BCH decoding steps.}}%
  \label{fig:bch-decoding-flow}%
\end{figure}
\FloatBarrier

\emph{Step~1---Syndrome Calculation:}  
\chI{To determine whether the retrieved codeword $r$ contains any errors,
the decoder computes the \emph{syndrome vector}, $S$, which indicates
how many of the parity check polynomials no longer match with the parity  
bits originally computed during encoding.} 
The $i$-th syndrome, $S_i$, is set to one if \chI{parity bit $i$ does
\chI{\emph{not}} match its corresponding polynomial, and to zero otherwise.}
To calculate $S$, \chI{the decoder calculates the dot product of $r$ and
the parity check matrix $H$ (i.e., $S = r \cdot H$).  If every syndrome in
S is set to 0, the decoder does not detect any errors within the codeword,
and skips to Step~5.  Otherwise, the decoder proceeds to Step~2.}

\vspace{3pt}%
\chI{\emph{Step~2---Constructing the Error Location Polynomial:}}
\chI{A state-of-the-art BCH decoder uses the} Berlekamp--Massey
algorithm\chI{~\cite{berlekamp.isit67, massey.tit69, chen.tit81, ryan.cup09}} to construct an error location
polynomial, $\sigma(x)$, whose roots encode the error locations of the 
\chI{codeword}:
\begin{equation}
  \sigma(x) =
    1 + \sigma_1 \cdot x + \sigma_2 \cdot x^2 + \ldots + \sigma_b \cdot x^b
  \label{eqn:error-location-polynomial}
\end{equation}
In Equation~\ref{eqn:error-location-polynomial}, \chI{$b$} is the number of raw bit
errors in the codeword.

The polynomial is constructed \chI{using} an iterative
process.  Since $b$ is not known \chI{initially}, the algorithm 
\chI{initially assumes that} $b = 0$ (i.e., $\sigma(x) = 1$). Then, it
updates $\sigma(x)$ by adding \chI{a \emph{correction term} to the
equation in each iteration,} until $\sigma(x)$ successfully
encodes all \chI{of the errors that were detected during syndrome calculation}. 
In each iteration, a \chI{new} correction
term is calculated \chI{using both the syndromes from Step~1 and the $\sigma(x)$
equations from prior iterations of the algorithm, as long as these prior values 
of $\sigma(x)$ satisfy certain conditions.}
\chI{This algorithm successfully finds $\sigma(x)$ after $n = (t+b) / 2$ iterations,
where \chI{$t$} is the maximum number of bit errors correctable by the BCH 
code~\cite{dolecek.fms14}.}

Note that (1)~the highest order of the polynomial, $b$, is \chI{directly correlated with} the
number of errors in the codeword; (2)~the number of iterations, $n$, is also
proportional to the number of errors; (3)~each iteration is compute-intensive, as
it involves several multiply and add operations; and (4)~this algorithm cannot
be parallelized across iterations, as the computation in each iteration is
dependent on the previous ones.

\vspace{3pt}%
\emph{Step~3---Extracting Bit Error Locations from the Error Polynomial:}
\chI{A state-of-the-art decoder applies the Chien 
search~\cite{chien.tit64, shu.book04} on} the error
location polynomial to find \chI{the location of \chI{\emph{all}} raw bit errors that have
been detected \chI{during Step~1} in the retrieved codeword $r$}. Each
\chI{bit error} location is encoded with a known function $f$~\cite{ryan.cup09}.
\chI{The error polynomial from Step~2 is constructed such that if the $i$-th 
bit of the codeword has an error, the error location polynomial
$\sigma(f(i)) = 0$; otherwise, if the $i$-th bit does \chI{\emph{not}} have an error,
$\sigma(f(i)) \neq 0$.}
\chI{The} Chien search simply uses trial-and-error
(i.e., tests if $\sigma(f(i))$ is zero), \chI{testing each bit in the codeword
starting at bit~0.
As the decoder needs to correct only the first $k$ bits of the codeword that
contain the data message $m$, the Chien search
needs to evaluate only $k$ different values of $\sigma(f(i))$.
The algorithm builds a bit vector $e$, which is the same length as the 
retrieved codeword $r$, where the $i$-th bit of $e$ is set to one if bit~$i$
of $r$ contains a bit error, and is set to zero if bit~$i$ of $r$ does \emph{not}
contain an error, or if $i \geq k$ (since \chI{there is no} need to correct the
parity bits).}

Note that (1)~\chI{the calculation of $\sigma(f(i))$ is compute-intensive}, but
can be parallelized \chI{because the calculation of each bit $i$ is independent
of the other bits}, and
(2)~the complexity of Step~3 is \chI{linearly} correlated with the number of 
\chI{detected errors in the codeword}.

\vspace{3pt}%
\chI{\emph{Step~4---Correcting the Bit Errors:}
The decoder corrects each detected bit error location
by flipping the bit at that location in the
retrieved codeword $r$.  
This simply involves XORing $r$ with the error vector $e$
created in Step~3.
After the errors are corrected, the decoder now has 
the estimated value of the originally-written codeword $c$
(i.e., $c = r \oplus e$).
\chI{The decoded version of $c$ is only an \emph{estimate}
of the original codeword, since if $r$ contains more bit errors
than the maximum number of errors (\chI{$t$}) that the BCH can correct,
there may be some uncorrectable errors that were \chI{\emph{not}} detected 
during syndrome calculation (Step~1).}  In such cases, the decoder cannot
\emph{guarantee} that it has determined the actual original codeword.
In a modern SSD, the bit error rate of a codeword \emph{after} BCH correction 
is \chI{expected to be} less than $10^{-15}$~\cite{jep122h.jedec16}.}

\vspace{3pt}%
\chI{\emph{Step~5---Extracting the Message from the Codeword:}
As we discuss above, during BCH codeword encoding, the
generator matrix $G$ contains the identity matrix, to ensure that 
\chI{\chI{the $k$-bit message} $m$ is embedded verbatim into the codeword $c$.}
Therefore, the decoder recovers \chI{$m$} by simply
truncating the last $(n-k)$ bits from the $n$-bit codeword $c$.}

\paratitle{\chI{BCH Decoder Latency Analysis}}
\chI{We can model the latency of the state-of-the-art BCH decoder 
($T_{BCH}^{dec}$) that we described above as:}
\begin{equation}
  T_{BCH}^{dec} = T_{Syndrome} + N \cdot T_{Berlekamp}
    + \frac{k}{p} \cdot T_{Chien}
  \label{eqn:bch-decoding-latency}
\end{equation}
In Equation~\ref{eqn:bch-decoding-latency}, $T_{Syndrome}$ is the latency for
calculating the syndrome, which is determined by the size of the parity check
matrix $H$; $T_{Berlekamp}$ is the latency \chI{of} one iteration of the
Berlekamp--Massey algorithm; $N$ is the total number of iterations that the
Berlekamp--Massey algorithm \chI{performs}; $T_{Chien}$ is the latency for deciding
whether or not \chI{a single bit location contains an error, using the
Chien search; $k$ is the length of the \chI{data} message $m$; and 
$p$} is the number of bits that are processed in parallel
in Step~3.
In this equation, $T_{Syndrome}$, $T_{Berlekamp}$, $k$, and $p$ are constants
\chI{for a BCH decoder implementation}, while $N$ and $T_{Chien}$ are proportional to the
raw bit error count of the \chI{codeword}.
\chI{Note that Steps~4 and 5 can typically be implemented such that they take less than one
\chI{clock} cycle in modern hardware, and thus their latencies are \chI{\emph{not}} included in
Equation~\ref{eqn:bch-decoding-latency}.}

\subsubsection{\chI{Low-Density Parity-Check (LDPC) Codes}}
\label{sec:correction:ecc:ldpc}

\chI{LDPC codes\chI{~\cite{shu.book04, gallager.ire62, mackay.letters96,mackay.letters97, gallager.tit62}} are now used widely in modern SSDs, as LDPC codes provide 
a stronger error correction capability than BCH codes, albeit at a greater
storage cost~\cite{zhao.fast13, wang.jsac14}.
LDPC codes are one type of \emph{capacity-approaching codes}, \chI{which
are error-correcting codes that come close to the \emph{Shannon limit}, \chI{i.e.,} the 
maximum number of data message bits ($k_{max}$) that can be delivered without errors 
for a certain codeword size ($n$) under a given error 
rate~\cite{shannon.bell48a, shannon.bell48b}.}
Unlike BCH codes, LDPC codes cannot \emph{guarantee} that they will correct a
minimum number of raw bit errors.
Instead, \chI{a good LDPC code guarantees} that the \emph{failure rate} (i.e., the fraction 
of all reads where \chI{the LDPC code} cannot successfully correct the data) is less than a
target rate for a given number of bit errors.}
\chI{Like BCH codes, LDPC codes for SSDs are designed to be \emph{systematic}, 
\chI{i.e., to} contain \chI{the data message verbatim}
within the codeword.}

\chI{An LDPC code\chI{~\cite{shu.book04, gallager.ire62, mackay.letters96,mackay.letters97, gallager.tit62}} is a linear code that, like a BCH code, consists of check bits generated by
an algorithm.  For an LDPC code, these check bits are used to form a bipartite graph,
where one side of the graph contains nodes that represent each bit in the codeword, and the other 
side of the graph contains nodes that represent the parity check equations used 
to generate each parity bit.  When a codeword containing errors is retrieved from memory, 
an LDPC decoder applies \emph{belief propagation}~\cite{pearl.aaai82} to iteratively identify the
bits within the codeword that are \emph{most likely} to contain a bit error.}

\chI{An LDPC code is defined using a binary \chI{parity} check matrix $H$, where $H$ is very
sparse (i.e., there are few ones in the matrix).  Figure~\ref{fig:ldpc}a shows an
example $H$ matrix for a \chI{seven}-bit codeword $c$ (see \chI{Section~\ref{sec:ssdarch:ctrl:ecc}}). 
\chI{For an $n$-bit codeword that encodes a $k$-bit \chI{data} message, $H$ is sized to
be an $(n-k) \times n$ matrix.}
Within the matrix, each \emph{row} 
represents a \emph{parity check equation}, while each \emph{column} represents one of the
\chI{seven} bits in the codeword.  As our example matrix has three rows, this means that
our error correction uses three parity check equations (denoted as $f$).  A bit value 1 in row $i$,
column $j$ indicates that parity check equation $f_i$ contains bit $c_j$.
Each parity check equation XORs all of the codeword bits in the equation to see
whether the output is zero.
For example, parity check equation \chI{$f_1$} from the $H$ matrix in
Figure~\ref{fig:ldpc}a is:}
\begin{equation}
f_1 = c_1 \oplus c_2 \oplus c_4 \oplus c_5 = 0
\end{equation}
\chI{This means that $c$ is a valid codeword only if $H \cdot c^T = 0$,
where $c^T$ is the transpose matrix of \chI{the codeword} $c$.}

\begin{figure}[h]
  \centering
  \includegraphics[width=.75\columnwidth]{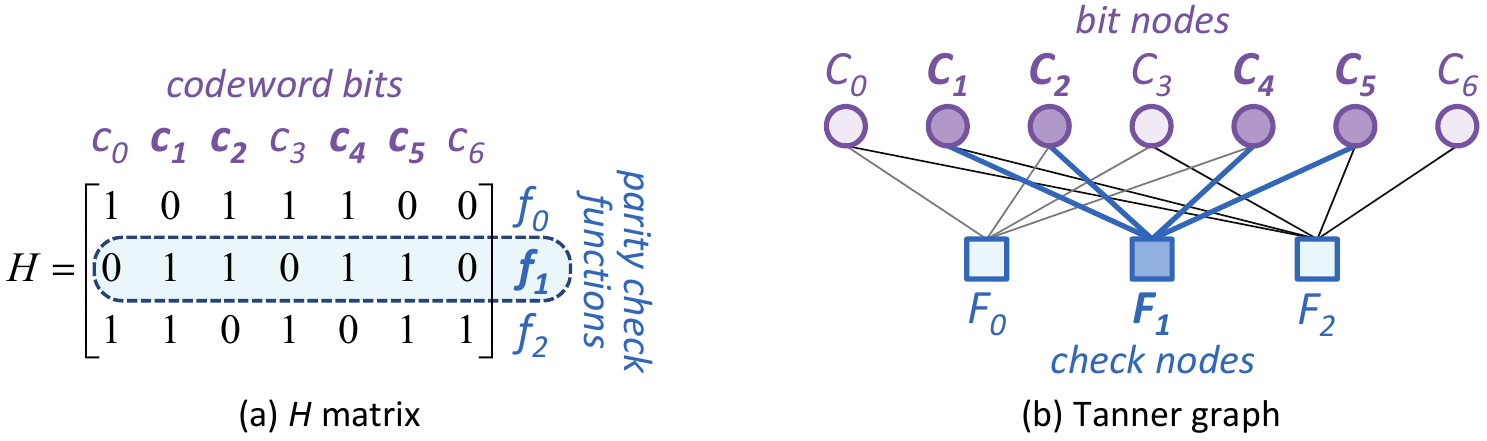}%
  \caption{\chI{Example LDPC code for a \chI{seven}-bit codeword with
  \chI{a four-bit \chI{data} message \chI{(stored in bits $c_0$, $c_1$, $c_2$, and $c_3$)} 
  and} three parity check equations
  \chI{(i.e., $n=7$, $k=4$)},
  represented as (a)~an \emph{H} matrix and (b)~a Tanner graph.}}%
  \label{fig:ldpc}%
\end{figure}
\FloatBarrier

\chI{\chI{In order to perform belief propagation, $H$} can be represented 
using a \emph{Tanner graph}~\cite{tanner.tit81}.  \chI{A Tanner graph is a} bipartite graph
that contains \emph{check nodes}, which represent the parity check equations,
and \emph{bit nodes}, which represent the bits in the codeword.
An edge connects a check node $F_i$ to a bit node $C_j$ only if parity check 
equation $f_i$ contains bit $c_j$.  Figure~\ref{fig:ldpc}b shows the Tanner 
graph that corresponds to the $H$ matrix in Figure~\ref{fig:ldpc}a.
\chI{For example, since parity check equation $f_1$ uses codeword bits
$c_1$, $c_2$, $c_4$, and $c_5$, the $F_1$ check node in Figure~\ref{fig:ldpc}b
is connected to bit nodes $C_1$, $C_2$, $C_4$, and $C_5$.}}

\paratitle{\chI{LDPC Encoding}}
\chI{As was the case with BCH, the LDPC codeword generation algorithm encodes
a $k$-bit \chI{data} message $m$ into an $n$-bit LDPC codeword $c$ by computing the dot
product of $m$ and a generator matrix $G$ (i.e., $c = m \cdot G$).
For an LDPC code, the generator matrix is designed to (1)~preserve 
\chI{$m$ verbatim within the codeword,}
and (2)~generate the parity bits for each parity check equation in $H$.
Thus, $G$ is defined using the parity check matrix $H$.
With linear algebra \chI{based} transformations, $H$ can be expressed in the form
$H=[A, I_{(n-k)}]$, where $H$ is composed of $A$, an $(n-k) \times k$ binary matrix,
and $I_{(n-k)}$, an $(n-k) \times (n-k)$ identity matrix~\cite{johnson.ldpc}.
The generator matrix $G$ can then be created using the composition
$G = [I_k, A^T]$, where $A^T$ is the transpose matrix of $A$.}

\paratitle{\chI{LDPC Decoding}}
\chI{When the SSD controller is servicing a read request, it must extract 
the $k$-bit \chI{data} message from the LDPC codeword $r$ that is stored in NAND flash memory.
\chI{In an SSD, an LDPC decoder performs multiple \emph{levels} of 
decoding~\cite{zhao.fast13, dolecek.fms14, varnica.fms13},}
which correct the retrieved codeword $r$ to obtain the originally-written 
codeword $c$ and extract the \chI{data} message $m$ from $c$.
\chI{Initially, the decoder performs a single level of \emph{hard decoding}, 
where it uses the information from a single read operation on the codeword to 
attempt to correct the codeword bit errors.  If the decoder cannot correct all 
errors using hard decoding, it then initiates the first level of \emph{soft decoding}, where a 
\emph{second} read operation is performed on the \emph{same} codeword using a 
\emph{different} set of read reference voltages.  The \chI{second} read 
provides \chI{\emph{additional}} information on the \emph{probability} 
that each bit in the codeword is a zero or a one.  An LDPC \chI{decoder} typically uses 
multiple levels of soft decoding, where each new level \chI{performs} an additional 
read operation to \chI{calculate a more accurate probability for each bit 
value}.  We discuss multi-level soft decoding in detail in 
Section~\ref{sec:correction:ldpcflow}.}


\chI{For each level, the} decoder performs \chI{five} steps, as illustrated in
Figure~\ref{fig:ldpc-decoding-flow}.
\chI{\chI{\chI{At each level}, the
decoder uses \chI{two pieces of information to determine which bits are 
\emph{most likely} to contain errors: (1)~the \emph{probability} that each bit 
in $r$ is a zero or a one, and
(2)~the parity check equations.}}
In \chI{Step~1} \chI{(Figure~\ref{fig:ldpc-decoding-flow})}, the decoder 
computes an initial \emph{log likelihood ratio} (LLR) for each bit of
the stored codeword.  We refer to the initial codeword LLR values as $L$,
where $L_j$ is the LLR value for bit~$j$ of the codeword.
$L_j$ expresses the likelihood (i.e., \emph{confidence})
that bit~$j$ \emph{should be} a zero 
or a one,
based on the current 
threshold voltage of the NAND flash cell where bit~$j$ is stored.
The decoder uses $L$ as the initial \emph{LLR message} generated using the bit nodes.
\chI{An LLR message consists of the LLR values for each bit, which are updated by and 
communicated between the check nodes and bit nodes during each step of belief
propagation.\footnote{\chI{Note that an LLR message is \chI{\emph{not}} the same as the $k$-bit
data message.  The \emph{data message} refers to the actual data stored
within the SSD, which, \chI{when read,} is modeled in information theory as a message 
\chI{that is transmitted} across a noisy communication channel.
\chI{In contrast, an} \emph{LLR message} refers to the updated LLR values for each bit of
the codeword that are exchanged between the check nodes and the bit nodes 
during belief propagation. \chI{Thus, there is no relationship between a data 
message and an LLR message.}}}}
In \chI{Steps~2 through 4}, the belief propagation algorithm~\cite{pearl.aaai82} 
iteratively updates the LLR message, using the Tanner graph to identify those bits that
are most likely to be incorrect (i.e., the codeword bits whose \chI{(1)~}\chI{bit nodes
are connected to the largest number of check nodes that currently contain a parity error,}
and \chI{(2)~}LLR values
indicate low confidence).
Several decoding algorithms exist to perform belief propagation
for LDPC codes.  The most commonly-used belief propagation
algorithm is the \emph{min-sum algorithm}~\cite{fossorier.tcomm99, chen.tcomm02},
a simplified version of the original sum-product algorithm 
for LDPC\chI{~\cite{gallager.ire62, gallager.tit62}} with near-equivalent error correction capability~\cite{anastasopoulos.globecom01}.
During each iteration of the min-sum algorithm, the decoder
\chI{identifies a set of codeword bits that likely contain errors and \chI{thus} need to be
flipped.  \chI{The decoder accomplishes this} by}
(1)~\chI{having each check node use its parity check information to determine
how much the LLR value of each bit should be updated by,} 
using the most recent LLR messages from the bit nodes;
(2)~\chI{having each bit node gather the LLR updates from each bit to
generate a new LLR value for the bit,} 
using the most recent LLR messages from the check nodes; and
(3)~\chI{using} the parity check equations to see if the values predicted by the new
LLR message for each node are correct.
\chI{The min-sum algorithm terminates \chI{under} one of two conditions:
(1)~the predicted bit values after the most recent iteration are all correct,
which means that the decoder now has an estimate of the original codeword $c$,
and can advance to Step~5; or
(2)~the algorithm exceeds a predetermined number of iterations, at which point
the decoder moves onto the next decoding level, \chI{or returns a decoding failure
if the maximum number of decoding levels have been performed}.
In Step~5, once the errors are corrected, and the decoder has the original codeword $c$, 
the decoder extracts the $k$-bit data message $m$ from the codeword.}
\chI{We describe the steps used by a state-of-the-art decoder in detail below, which uses
an optimized version of the min-sum algorithm that 
can be implemented efficiently in hardware~\cite{gunnam.icc07, gunnam.fms14}.}

\begin{figure}[h]
  \centering
  \includegraphics[width=0.92\linewidth]{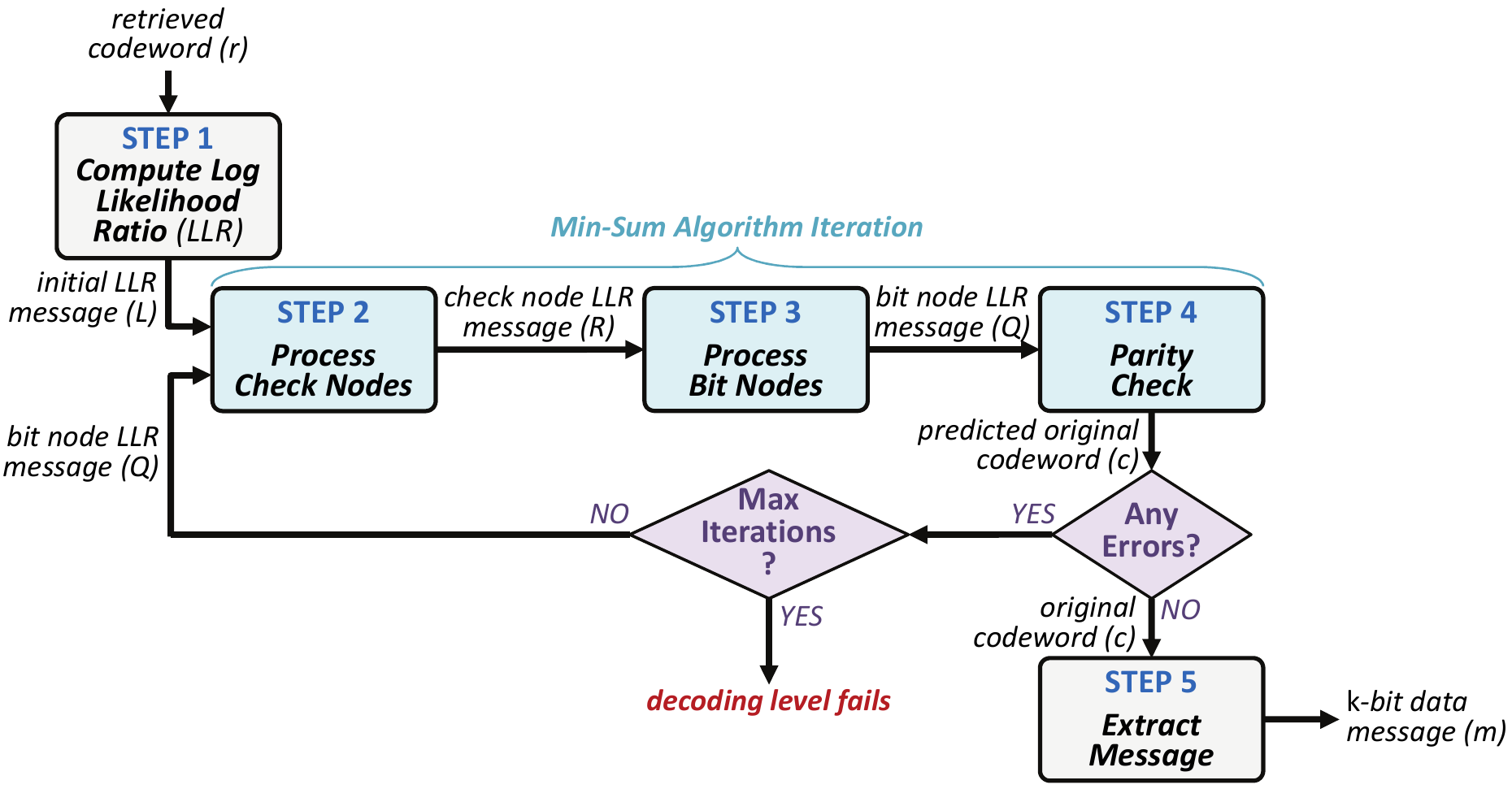}%
  \caption{\chI{LDPC decoding steps \chI{for a single level of hard or soft decoding}.}}%
  \label{fig:ldpc-decoding-flow}%
\end{figure}
\FloatBarrier

\vspace{3pt}%
\chI{\emph{\chI{Step~1}---Computing the Log Likelihood Ratio (LLR):}}
%
\chI{The LDPC decoder uses the \emph{probability} (i.e.,
\emph{likelihood}) that a bit is a zero or a one to identify errors, instead of
using the bit values directly.}
\chI{The \chI{\emph{log likelihood ratio} (LLR)} is the probability \chI{that a certain
bit is} zero, i.e., $P(x = 0| V_{th})$, over the probability \chI{that
the bit is} one, i.e., $P(x = 1| V_{th})$, given a certain threshold
voltage range ($V_{th}$) bounded by two threshold voltage values
(i.e., the maximum and the minimum voltage of the threshold
voltage range)~\cite{zhao.fast13, wang.jsac14}:}
\begin{equation}
\text{LLR} = \log \frac{P(x = 0| V_{th})}{P(x = 1| V_{th})}
\label{eq:E10}
\end{equation}
\chI{The sign of the LLR value indicates whether the bit is likely to be a 
zero (when the LLR value is positive) or a one (when the LLR value is negative).
A \emph{larger} magnitude (i.e., absolute value) of the LLR value indicates 
a \emph{greater} confidence
that a bit should be zero or one, while an LLR value closer to zero indicates
low confidence.  The bits whose LLR values have the smallest magnitudes are the
ones \chI{that are} most likely to contain errors.}

%
\chI{There are several alternatives
for how to compute the LLR values. A common approach for
LLR computation is to treat a flash cell as a communication
channel, where the channel takes an input program signal
(i.e., the target threshold voltage for the cell) and outputs
an observed signal (i.e., the current threshold voltage of the
cell)~\cite{cai.date13}. The observed signal differs from the input signal
due to the various types of NAND flash memory errors. The
communication channel model allows us to break down
the threshold voltage of a cell into two components: (1)~the
expected signal; and (2)~the additive signal noise due to
errors. By enabling the modeling of these two components
separately, the communication channel model allows us to
estimate the current threshold voltage distribution of each
state~\cite{cai.date13}. The threshold voltage distributions can be used to
predict how likely a cell within a certain voltage region is to
belong to a particular voltage state.}

\chI{One popular variant of the communication channel
model assumes that the threshold voltage distribution of
each state can be modeled as a Gaussian distribution~\cite{cai.date13}.
If we use the mean observed threshold voltage of each state
(denoted as $\mu$) to represent the signal, we find that the P/E
cycling noise (i.e., the shift in the distribution of threshold
voltages due to the accumulation of charge from repeated
programming operations; see Section~\ref{sec:errors:pe}) can be modeled
as \emph{additive white Gaussian noise} (AWGN)~\cite{cai.date13}, which
is represented by the standard deviation of the distribution
(denoted as $\sigma$). The closed-form AWGN-based model can be
used to determine the LLR value for a cell with threshold
voltage $y$, as follows:}
\begin{equation}
\text{LLR}(y) = \frac{\mu_1^2 - \mu_0^2}{2 \sigma^2} + \frac{y(\mu_0 - \mu_1)}{\sigma^2}
\label{eq:E11}
\end{equation}
\chI{where $\mu_{0}$ and $\mu_{1}$ are the mean threshold voltages for the distributions
of the threshold voltage states for bit value 0 and
bit value 1, respectively, and $\sigma$ is the standard deviation of
both distributions (assuming that the standard deviation
of each threshold voltage state distribution is equal). Since
\chI{the SSD controller} uses threshold voltage ranges to categorize
a flash cell, we can substitute $\mu_{R_j}$, the mean threshold
voltage of the threshold voltage range $R_j$, in place of $y$ in Equation~\ref{eq:E11}.}

\chI{The AWGN-based LLR model in Equation~\ref{eq:E11} provides only an
estimate of the LLR, because (1)~the actual threshold voltage
distributions observed in NAND flash memory are \emph{not} perfectly
Gaussian in nature~\cite{cai.date13, luo.jsac16}; (2)~the controller uses
the mean voltage of the threshold voltage range to \emph{approximate}
the actual threshold voltage of a cell; and (3)~the standard
deviations of each threshold voltage state distribution
are \emph{not} perfectly equal.
A number of methods have been proposed to improve
upon the AWGN-based LLR estimate by: (1)~using nonlinear
transformations to convert the AWGN-based LLR into a
more accurate LLR value~\cite{wu.patent17.9201729}; (2)~scaling and rounding the
AWGN-based LLR to compensate for the estimation error~\cite{wu.patent17.9582361}; 
(3)~initially using the AWGN-based LLR to read the
data, and, if the read fails, using the ECC information from
the failed read attempt to optimize the LLR and to perform
the read again with the optimized LLR~\cite{cohen.patent15}; and (4)~using
online and offline training to empirically determine the
LLR values under a wide range of conditions (e.g., P/E cycle
count, retention time, read disturb count)~\cite{wu.patent15.9213599}. The SSD
controller can either compute the LLR values at runtime, or
statically store precomputed LLR values in a table.}

\chI{Once the decoder calculates the LLR values for each bit of the codeword,
which we call the initial LLR message $L$, the decoder starts the first 
iteration of the min-sum algorithm (\chI{Steps 2--4} below).}

\vspace{3pt}%
\chI{\emph{\chI{Step~2}---Check Node Processing:}
In every iteration of the min-sum algorithm, each} \chI{check node $i$ \chI{(see Figure~\ref{fig:ldpc})}
generates a revised check node LLR message $R_{ij}$ to send to each bit node $j$ \chI{(see Figure~\ref{fig:ldpc})} that is
connected to check node $i$.  \chI{The decoder computes} $R_{ij}$ as:}
\begin{equation}
R_{ij} = \delta_{ij} \kappa_{ij}
\end{equation}
\chI{where $\delta_{ij}$ is the sign of the \chI{LLR} message, and $\kappa_{ij}$ is the
magnitude of the \chI{LLR} message.}
\chI{The decoder determines the values of both $\delta_{ij}$ and $\kappa_{ij}$
using the bit node LLR message $Q'_{ji}$.
\chI{At a high level, each check node collects LLR values sent from each bit
node ($Q'_{ji}$), and then determines how much each bit's LLR value should be
adjusted using the parity information available at the check node.  These LLR 
value updates are then bundled together into the LLR message $R_{ij}$.}
During the first iteration of the min-sum algorithm,
the decoder sets $Q'_{ji} = L_j$,
the initial LLR value from \chI{Step~1}.  In subsequent iterations, the decoder uses the 
value of $Q'_{ji}$
that was generated in \chI{Step~3} of the \emph{previous} iteration.
The decoder calculates $\delta_{ij}$, the sign of the check node LLR
message, as:}
\begin{equation}
\delta_{ij} = \displaystyle\prod_{J} \text{sgn}(Q'_{Ji})
\end{equation}
\chI{where $J$ represents all bit nodes connected to check node $i$ \emph{except}
for bit node $j$.  
\chI{The sign of a bit node indicates whether the value of a bit is predicted to be a zero
(if the sign is positive) or a one (if the sign is negative).}
\chI{The decoder calculates $\kappa_{ij}$, the 
magnitude of the check node LLR message, as:}
\begin{equation}
\kappa_{ij} = \displaystyle\min_J |Q'_{Ji}|
\end{equation}
\chI{In essence, the smaller the magnitude of $Q'_{ji}$ is, the more
uncertain we are about whether the bit should be a \chI{zero or a one}.  \chI{At} each check node,
\chI{the decoder updates the LLR value of each bit node $j$, adjusting the LLR by} the smallest value of
$Q'$ for \chI{any of the other bits connected to the check node} 
(i.e., the LLR value of the most uncertain bit \chI{aside from bit~$j$}).}

\vspace{3pt}%
\chI{\emph{\chI{Step~3}---Bit Node Processing:}}
\chI{Once each check node generates the
LLR messages for each bit node, we combine the LLR messages received by each 
bit node to update the LLR value of the bit.  \chI{The decoder first generates} the LLR messages
to be used by the check nodes in the next iteration \chI{of the min-sum algorithm}.  \chI{The decoder calculates} the bit node
LLR message $Q_{ji}$ to send from bit node $j$ to check node $i$ as follows:}
\begin{equation}
Q_{ji} = L_j + \displaystyle\sum_{I} R_{Ij}
\end{equation}
\chI{where $I$ represents all check nodes connected to bit node $j$ \emph{except}
for check node $i$, \chI{and $L_j$ is the original LLR value for bit~$j$ generated 
in \chI{Step~1}}.  In essence, for each check node, the bit node LLR message combines the LLR 
messages from the \emph{other check nodes} to ensure that all \chI{of the LLR value updates}
are propagated globally \chI{across all of the check nodes}.}

\vspace{3pt}%
\chI{\emph{\chI{Step~4}---Parity Check:}}
\chI{After the bit node processing is
complete, the decoder uses the revised \chI{LLR} information to predict the value of
each bit.  For bit node $j$, the predicted bit value $P_j$ is calculated as:}
\begin{equation}
P_j = L_j + \displaystyle\sum_{i} R_{ij}
\end{equation}
\chI{where $i$ represents \emph{all} check nodes connected to bit node $j$,
\chI{\emph{including} check node $i$, and $L_j$ is the original LLR value for bit~$j$ generated 
in \chI{Step~1}}.}
\chI{If $P_j$ is positive, bit~$j$ \chI{of the original codeword $c$} is predicted to be a \chI{zero}; otherwise, bit~$j$ is 
predicted to be a \chI{one}.
Once the predicted values have been computed for all bits
\chI{of $c$}, the $H$ matrix is
used to check the parity, by \chI{computing $H \cdot c^T$.  If $H \cdot c^T = 0$},}
then the predicted bit values are correct, \chI{the min-sum algorithm 
terminates, and the decoder goes to \chI{Step~5}}.  Otherwise,
at least one bit is still incorrect, and the decoder goes back to \chI{Step~2} to 
perform the next iteration \chI{of the min-sum algorithm}.  
\chI{In the next iteration, the min-sum algorithm uses the updated LLR values
from the current iteration to identify the next set of bits that are most likely 
incorrect and need to be flipped.}

\chI{The current decoding level fails to correct the data when the decoder 
cannot determine the correct codeword \chI{bit values} after a predetermined number of 
min-sum algorithm iterations.  If the decoder has more soft decoding levels 
left to perform, it
advances to the next soft decoding level.  For the new level, the SSD controller 
performs an additional read operation using a different set of read reference 
voltages than \chI{the ones} it used for the prior decoding levels.  The decoder then goes 
back to Step~1 to generate the new LLR information, using the output of 
\emph{all} of the read operations performed for each decoding level so far.
We discuss how the number of decoding levels and the read reference voltages
are determined, as well as what happens if \emph{all} soft decoding levels fail, 
in Section~\ref{sec:correction:ldpcflow}.}

\vspace{3pt}%
\chI{\emph{\chI{Step~5}---Extracting the Message from the Codeword:}
As we discuss above, during LDPC codeword encoding, the
generator matrix $G$ contains the identity matrix, to ensure that the
codeword $c$ includes \chI{a verbatim} version of $m$.
Therefore, the decoder recovers the $k$-bit \chI{data} message $m$ by simply
truncating the last $(n-k)$ bits from the $n$-bit codeword $c$.}

\subsection{\chI{Error Correction Flow}}
\label{sec:correction:flow}

\chI{For both BCH and LDPC codes, the SSD controller performs several stages
of error correction to retrieve the data, known as the \emph{error correction
flow}.  The error correction flow is invoked when the SSD performs a read operation.
The SSD starts the read operation by using the initial read reference
voltages ($V_{initial}$; see Section~\ref{sec:mitigation:voltage}) to read the raw
data stored within a page of NAND flash memory into the
controller.  Once the raw data is read, the controller starts
error correction.}

\begin{figure}[h]   
\begin{algorithm}[H]    
\caption{Example BCH/LDPC Error Correction Procedure}   
\label{alg:A1}    
\centering    
\includegraphics[trim=0 240 0 0, clip, width=0.49\linewidth]{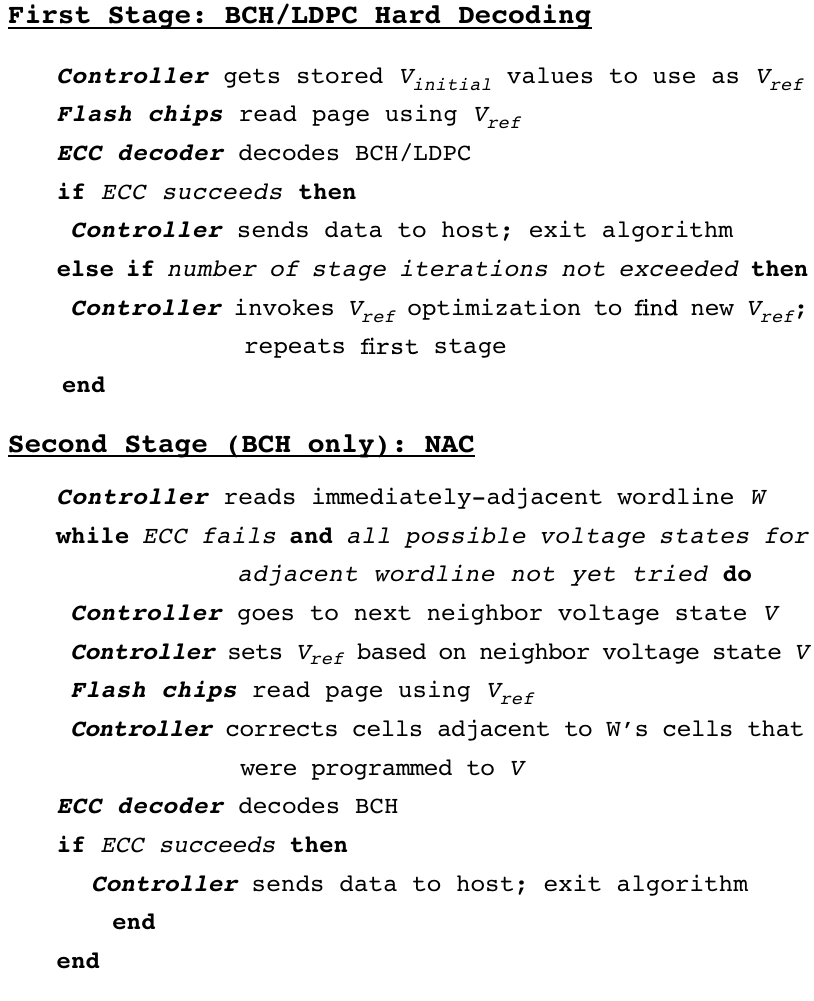}   
~   
\includegraphics[trim=0 -50 0 290, clip, width=0.49\linewidth]{figs/Alg1.pdf}   
\end{algorithm}   
\end{figure}
\FloatBarrier


\chI{Algorithm~\ref{alg:A1} lists the three stages of an example error correction 
flow, which can be used to decode either BCH codes or LDPC codes.
In the first stage, the ECC engine performs \emph{hard decoding} on the raw
data.  In hard decoding, the ECC engine uses only the
\emph{hard} bit value information (i.e., either a 1 or a 0) read for a
cell using a \emph{single} set of read reference voltages.
If the first stage succeeds (i.e., the controller detects that the error rate
of the data after correction is lower than a predetermined threshold),
the flow finishes.  If the first stage fails, then the flow moves on to the
second stage of error correction.  The second stage differs significantly for
BCH and for LDPC, which we discuss below.  If the second stage succeeds,
the flow terminates; otherwise, the flow moves to the third stage of error
correction.  In the third stage, the controller tries to
correct the errors using the more expensive superpage-level
parity recovery (\chI{see Section~\ref{sec:ssdarch:ctrl:parity}}). The steps for superpage-level
parity recovery are shown in the third stage of Algorithm~\ref{alg:A1}.
If the data can be extracted successfully from the other pages
in the superpage, the data from the target page can be recovered.
Whenever data is successfully decoded or recovered,
the data is sent to the host (and it is also reprogrammed into
a new physical page to ensure that the \emph{corrected} data values
are stored for the logical page). Otherwise, the SSD controller
reports an uncorrectable error to the host.}

\chI{Figure~\ref{fig:F30} compares the error correction flow with BCH
codes to the flow with LDPC codes.
Next, we discuss \chI{the flows used with} both BCH codes
(Section~\ref{sec:correction:bchflow}) and LDPC codes
(Section~\ref{sec:correction:ldpcflow}).}

\begin{figure}[h]
  \centering
  \includegraphics[width=0.8\columnwidth]{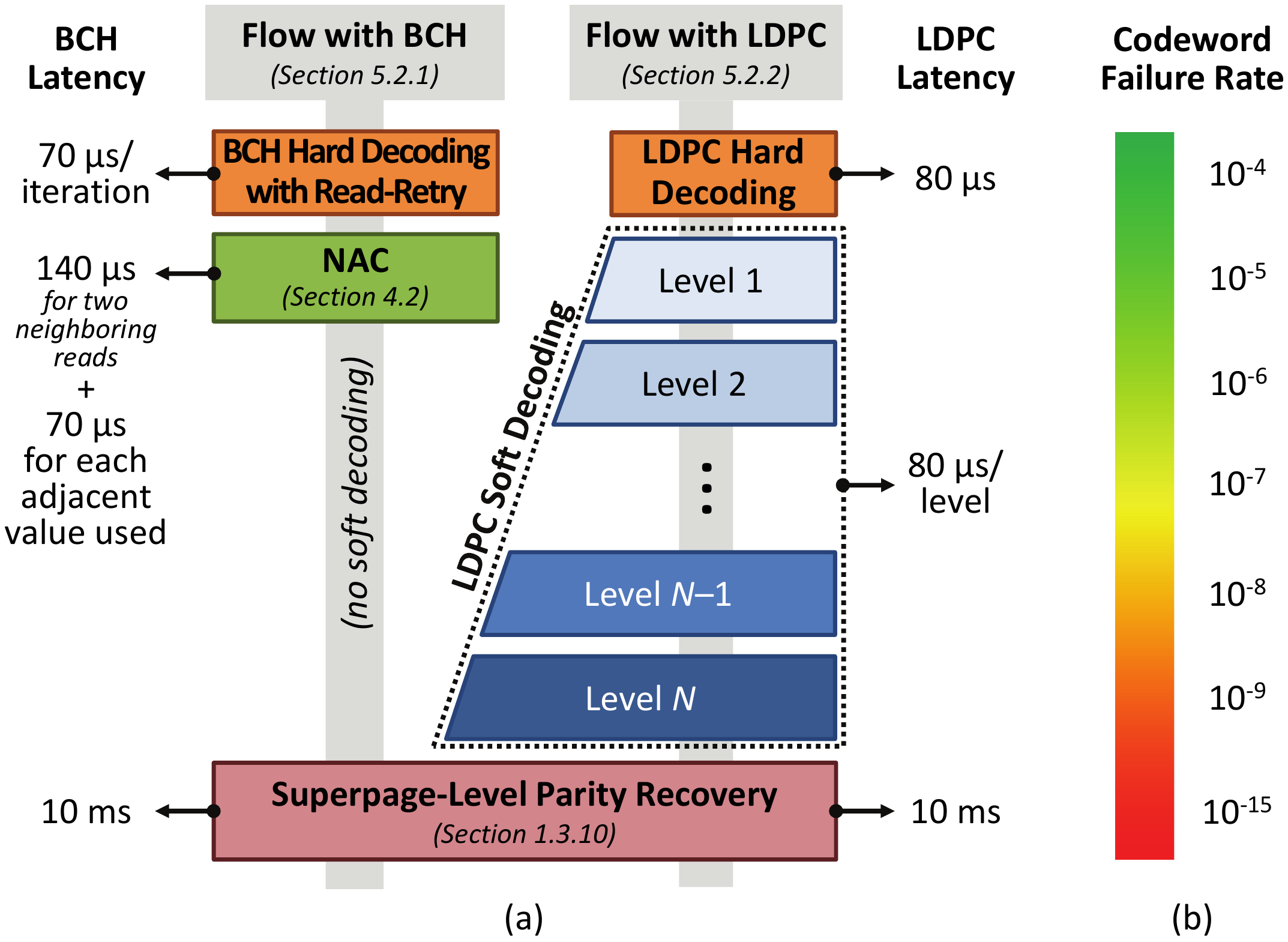}%
  \caption[(a)~Example error correction flow using BCH codes and
LDPC codes, \chI{with average latency of each BCH/LDPC stage.} 
(b)~The corresponding codeword failure rate for each LDPC stage.]
{(a)~Example error correction flow using BCH codes and
LDPC codes, \chI{with average latency of each BCH/LDPC stage.} 
(b)~The corresponding codeword failure rate for each LDPC stage. 
\chI{\chI{Adapted} from~\cite{cai.arxiv17}.}}%
  \label{fig:F30}%
\end{figure}
\FloatBarrier

\subsubsection{\chI{Flow Stages for BCH Codes}}
\label{sec:correction:bchflow}
\label{sec:correction:bch}

\chI{An example flow of
the stages for BCH decoding is
shown on the left-hand side of Figure~\ref{fig:F30}a. In the first stage,
the ECC engine performs BCH hard decoding on the raw data,
which reports the total number of bit errors in the data.
If the data cannot be corrected by the implemented BCH
codes, many controllers invoke read-retry (Section~\ref{sec:mitigation:retry}) or
read reference voltage optimization (Section~\ref{sec:mitigation:voltage}) to find a
new set of read reference voltages ($V_{ref}$) that lower the raw
bit error rate of the data from the error rate when using
$V_{initial}$. The controller uses the new $V_{ref}$ values to read the
data again, and then repeats the BCH decoding.}
\chI{We discuss the algorithm used to perform decoding for
BCH codes in Section~\ref{sec:correction:ecc:bch}.}

If the controller exhausts the maximum number of read
attempts (specified as a parameter in the controller), it
employs correction techniques such as neighbor-cell-assisted
correction (NAC; see Section~\ref{sec:mitigation:nac}) to further reduce the error
rate, as shown in the second BCH stage of Algorithm~\ref{alg:A1}. If NAC
cannot successfully read the data, the controller then tries to
correct the errors using the more expensive superpage-level
parity recovery \chI{(see \chI{Section~\ref{sec:ssdarch:ctrl:parity}})}.

\subsubsection{\chI{Flow Stages for LDPC Codes}}
\label{sec:correction:ldpcflow}
\label{sec:correction:ldpc}

\chI{An example flow of
the stages for LDPC decoding is
shown on the right-hand side of Figure~\ref{fig:F30}a.}
LDPC decoding consists of three major steps. First,
the SSD controller performs LDPC hard decoding, where
the controller reads the data using the optimal read reference
voltages. The process for LDPC hard decoding is similar
to that of BCH hard decoding (as shown in the first stage
of Algorithm~\ref{alg:A1}), but does not typically invoke read-retry if
the first read attempt fails. Second, if LDPC hard decoding
cannot correct all of the errors, the controller uses LDPC
\emph{soft decoding} to decode the data (which we describe in detail
below). Third, if LDPC soft decoding also cannot correct all
of the errors, the controller invokes superpage-level parity.
\chI{We discuss the algorithm used to perform hard and soft decoding for
LDPC codes in Section~\ref{sec:correction:ecc:ldpc}.}

\paratitle{Soft Decoding}
Unlike BCH codes, which require
the invocation of expensive superpage-level parity recovery
immediately if the hard decoding attempts (\chI{i.e.,} BCH hard
decoding with read-retry or NAC) fail to return correct data,
LDPC decoding fails more gracefully: it can perform multiple
levels of \emph{soft decoding} (\chI{shown in the second stage of} Algorithm~\ref{alg:A1})
after hard decoding fails before invoking superpage-level
parity recovery~\cite{zhao.fast13, wang.jsac14}. The key idea of soft decoding is \chI{to}
use \emph{soft} information for each cell (i.e., the \emph{probability} that
the cell contains a 1 or a 0) obtained from \emph{multiple} reads of
the cell via the use of different sets of read reference voltages\chI{~\cite{shu.book04, zhao.fast13, gallager.ire62, mackay.letters96,mackay.letters97, gallager.tit62, dolecek.fms14}}. 
\chI{Soft information is typically represented by
the \emph{log likelihood ratio} (LLR; see Section~\ref{sec:correction:ecc:ldpc}).}

Every additional level of soft decoding (i.e., the use of
a new set of read reference voltages, which we call $V_{ref}^X$ for
level $X$) increases the strength of the error correction, as the
level \emph{adds} new information about the cell (as opposed to
hard decoding, where a new decoding step simply \emph{replaces}
prior information about the cell). The new read reference
voltages, unlike the ones used for hard decoding, are
optimized such that the amount of useful information (or
\emph{mutual information}) provided to the LDPC decoder is maximized~\cite{wang.jsac14}. 
Thus, the use of soft decoding reduces the frequency
at which superpage-level parity needs to be invoked.

Figure~\ref{fig:F31} illustrates the read reference voltages used during
\chI{LDPC hard decoding and during}
the first \chI{two} levels of LDPC soft decoding. At each level, a
new read reference voltage is applied, which divides an existing
threshold voltage range into two ranges. Based on the bit
values read using the various read reference voltages, the SSD
controller bins each cell into a certain $V_{th}$ range, and sends
the bin categorization of all the cells to the LDPC decoder.
For each cell, the decoder applies an LLR value, precomputed
by the SSD manufacturer, which corresponds to the cell's bin
and decodes the data. For example, as shown in the bottom
of Figure~\ref{fig:F31}, the three read reference voltages in \chI{Level~2} soft
decoding form four threshold voltage ranges (i.e., R0--R3).
Each of these ranges corresponds to a different LLR value
(i.e., \chI{$\text{LLR}_2^{R0}$ to  $\text{LLR}_2^{R3}$},
where $\text{LLR}_i^{Rj}$ is the LLR value for range
$R_j$ in \chI{soft decoding} level $i$). Compared with \chI{hard} decoding (shown
at the top of Figure~\ref{fig:F31}), which \chI{has only} two LLR values, 
\chI{Level~2} soft decoding provides more accurate information to the
decoder, and thus has stronger error correction capability.

\begin{figure}[h]
  \centering
  \includegraphics[width=0.72\columnwidth]{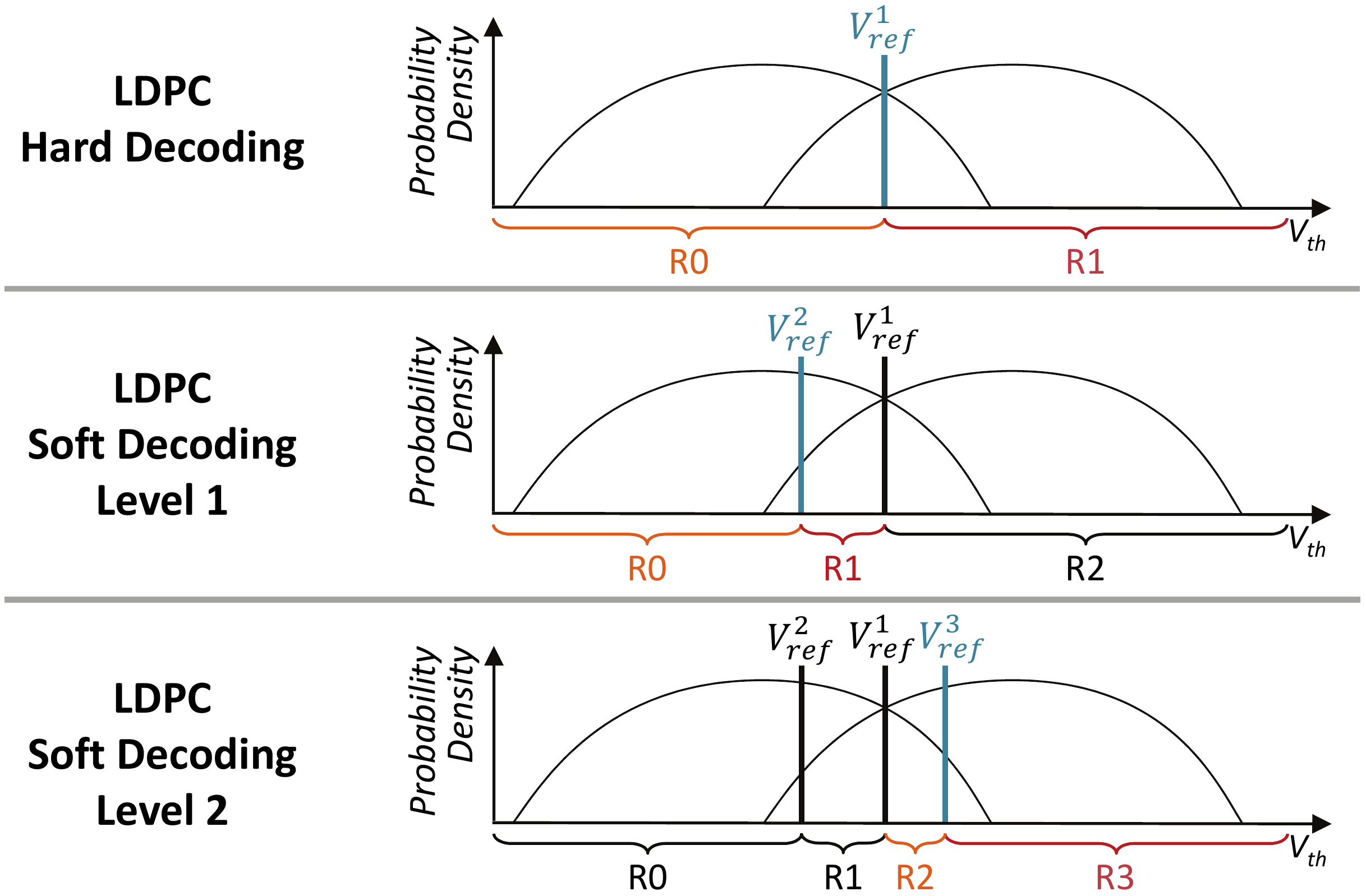}%
  \caption[LDPC hard decoding and the first two levels of LDPC soft decoding, showing the $V_{ref}$
value added at each level, and the resulting threshold voltage
ranges (R0--R3) used for flash cell categorization.]
  {\chI{LDPC hard decoding and the first two} levels of LDPC soft decoding, showing the $V_{ref}$
value added at each level, and the resulting threshold voltage
ranges (R0--R3) used for flash cell categorization. \chI{\chI{Adapted} from~\cite{cai.arxiv17}.}}%
  \label{fig:F31}%
\end{figure}
\FloatBarrier

\paratitle{Determining the Number of Soft Decoding Levels}
If the
final level of soft decoding, i.e., level $N$ in Figure~\ref{fig:F30}a, fails,
the controller attempts to read the data using superpage-level
parity (\chI{see Section~\ref{sec:ssdarch:ctrl:parity}}). The number of levels used for
soft decoding depends on the improved reliability that each
additional level provides, taking into account the latency of
performing additional decoding. Figure~\ref{fig:F30}b shows a rough
estimation of the average latency and the codeword failure
rate for each stage. There is a tradeoff between the number
of levels employed for soft decoding and the expected
read latency. For a smaller number of levels, the additional
reliability can be worth the latency penalty. For example,
while a five-level soft decoding step requires up to \SI{480}{\micro\second}, it
effectively reduces the codeword failure rate by five orders
of magnitude. This not only improves overall reliability,
but also reduces the frequency of triggering expensive
superpage-level parity recovery, which can take around
\SI{10}{\milli\second}~\cite{haratsch.fms16}. However, manufacturers limit the number of
levels, as the benefit of employing an additional soft decoding
level (which requires more read operations) becomes
smaller due to diminishing returns in the number of additional
errors corrected.

\subsection{BCH and LDPC Error Correction Strength}
\label{sec:correction:strength}

BCH and LDPC codes provide different strengths of
error correction. While LDPC codes can offer a stronger
error correction capability, soft LDPC decoding can lead
to a greater latency for error correction. Figure~\ref{fig:F32} compares
the error correction strength of BCH codes, hard LDPC
codes, and soft LDPC codes~\cite{haratsch.fms15}. The x-axis shows the raw
bit error rate (RBER) of the data being corrected, and the
y-axis shows the \emph{uncorrectable bit error rate} (UBER), or the
error rate after correction, once the error correction code
has been applied. The UBER is defined as the ECC codeword
(see \chI{Section~\ref{sec:ssdarch:ctrl:ecc}}) failure rate divided by the codeword
length~\cite{jesd218.jedec10}. To ensure a fair comparison, we choose a similar
codeword length for both BCH and LDPC codes, and use
a similar coding rate (0.935 for BCH, and 0.936 for LDPC)~\cite{haratsch.fms15}. 
We make two observations from Figure~\ref{fig:F32}.

\begin{figure}[h]
  \centering
  \includegraphics[width=0.65\columnwidth]{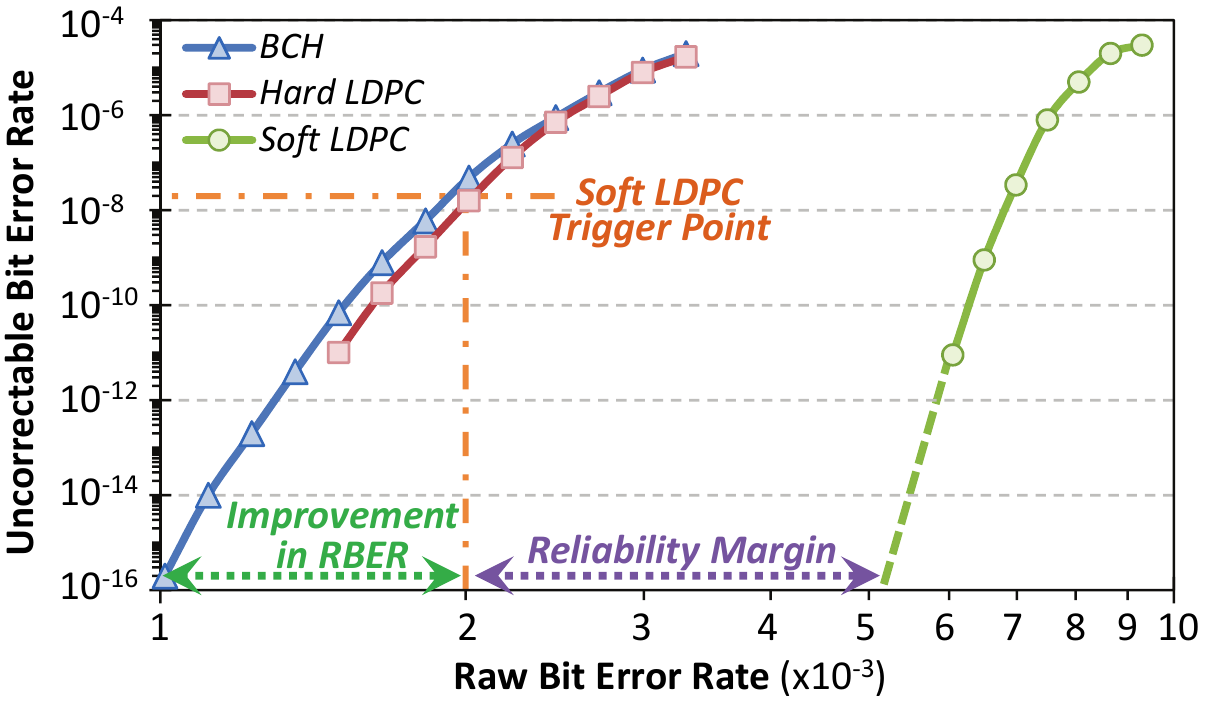}%
  \caption[Raw bit error rate versus uncorrectable bit error rate for
BCH codes, hard LDPC codes, and soft LDPC codes.]
  {Raw bit error rate versus uncorrectable bit error rate for
BCH codes, hard LDPC codes, and soft LDPC codes. \chI{Reproduced from~\cite{cai.arxiv17}.}}%
  \label{fig:F32}%
\end{figure}
\FloatBarrier

First, we observe that the error correction strength of
the hard LDPC code is similar to that of the BCH codes.
Thus, on its own, hard LDPC does not provide a significant
advantage over BCH codes, as it provides an equivalent
degree of error correction with similar latency (i.e.,
one read operation). Second, we observe that soft LDPC
decoding provides a significant advantage in error correction
capability. Contemporary SSD manufacturers target a
UBER of $10^{-16}$~\cite{jesd218.jedec10}. The example BCH code with a coding
rate of 0.935 can successfully correct data with an RBER
of $1.0 \times 10^{-3}$ while remaining within the target UBER. The
example LDPC code with a coding rate of 0.936 is more
successful with soft decoding, and can correct data with an
RBER as high as $5.0 \times 10^{-3}$ while remaining within the target
UBER, based on the error rate extrapolation shown in
Figure~\ref{fig:F32}. While soft LDPC can tolerate up to five times the
raw bit errors as BCH, this comes at a cost of latency (not
shown on the graph), as soft LDPC can require several additional
read operations after hard LDPC decoding fails, while
BCH requires only the original read.

To understand the benefit of LDPC codes over BCH
codes, we need to consider the combined effect of hard
LDPC decoding and soft LDPC decoding. As discussed in
Section~\ref{sec:correction:ldpcflow}, soft LDPC decoding is invoked \emph{only when hard
LDPC decoding fails}. To balance error correction strength
with read performance, SSD manufacturers can require that
the hard LDPC failure rate cannot exceed a certain threshold,
and that the overall read latency (which includes the
error correction time) cannot exceed a certain target~\cite{haratsch.fms16, haratsch.fms15}. 
For example, to limit the impact of error correction
on read performance, a manufacturer can require 99.99\% of
the error correction operations to be completed after a single
read. To meet our example requirement, the hard LDPC
failure rate should not be greater than $10^{-4}$ (i.e., 99.99\%),
which corresponds to an RBER of $2.0 \times 10^{-3}$ and a UBER
of $10^{-8}$ (shown as \emph{Soft LDPC Trigger Point} in Figure~\ref{fig:F32}). For
only the data that contains one or more failed codewords,
soft LDPC is invoked (i.e., soft LDPC is invoked only 0.01\%
of the time). For our example LDPC code with a coding
rate of 0.936, soft LDPC decoding is able to correct these
codewords: for an RBER of $2.0 \times 10^{-3}$, using soft LDPC
results in a UBER well below $10^{-16}$, as shown in Figure~\ref{fig:F32}.

To gauge the combined effectiveness of hard and soft
LDPC codes, we calculate the overhead of using the combined
LDPC decoding over using BCH decoding. If 0.01\%
of the codeword corrections fail, we can assume that in
the worst case, each failed codeword resides in a different
flash page. As the failure of a single codeword in a flash
page causes soft LDPC to be invoked for the entire flash
page, our assumption maximizes the number of flash pages
that require soft LDPC decoding. For an SSD with four
codewords per flash page, our assumption results in up
to 0.04\% of the data reads requiring soft LDPC decoding.
Assuming that the example soft LDPC decoding requires
seven additional reads, this corresponds to 0.28\% more
reads when using combined hard and soft LDPC over BCH
codes. Thus, with a 0.28\% overhead in the number of reads
performed, the combined hard and soft LDPC decoding
provides twice the error correction strength of BCH codes
(shown as \emph{Improvement in RBER} in Figure~\ref{fig:F32}).

In our example, the lifetime of an SSD is limited by
both the UBER and whether more than 0.01\% of the codeword
corrections invoke soft LDPC, to ensure that the
overhead of error correction does not significantly increase
the read latency~\cite{haratsch.fms15}. In this case, when the lifetime
of the SSD ends, we can still read out the data correctly
from the SSD, albeit at an increased read latency. This is
because even though we capped the SSD lifetime to an
RBER of $2.0 \times 10^{-3}$ in our example shown in Figure~\ref{fig:F32}, soft
LDPC is able to correct data with an RBER as high as
$5.0 \times 10^{-3}$ while still maintaining an acceptable UBER
($10^{-16}$) based on the error rate extrapolation shown.
Thus, LDPC codes have a margin, which we call the \emph{reliability
margin} and show in Figure~\ref{fig:F32}. This reliability margin
enables us to trade off lifetime with read latency.

We conclude that with a combination of hard and soft
LDPC decoding, an SSD can offer a significant improvement
in error correction strength over using BCH codes.

\subsection{SSD Data Recovery}
\label{sec:correction:recovery}

When the number of errors in data exceeds the ECC
correction capability and the error correction techniques in
Sections~\ref{sec:correction:bchflow} and~\ref{sec:correction:ldpcflow}
are unable to correct the read data,
then data loss can occur. At this point, the SSD is considered
to have reached the end of its lifetime. In order to avoid such
data loss and \emph{recover} (or, \emph{rescue}) the data from the SSD, we
can harness \chIII{the} understanding of data retention and read
disturb behavior. The SSD controller can employ two conceptually
similar mechanisms, \emph{Retention Failure Recovery}
(RFR)~\cite{cai.hpca15} and \emph{Read Disturb Recovery} (RDR)~\cite{cai.dsn15}, to undo
errors that were introduced into the data as a result of data
retention and read disturb, respectively. The key idea of
both of these mechanisms is to exploit the wide variation of
different flash cells in their susceptibility to data retention
loss and read disturbance effects, respectively, in order to
correct some of the errors \emph{without} the assistance of ECC so
that the remaining error count falls within the ECC error
correction capability.

When a flash page read fails (i.e., uncorrectable errors
exist), RFR and RDR record the current threshold voltages
of each cell in the page using the read-retry mechanism (see
Section~\ref{sec:mitigation:retry}), and identify the cells that are \emph{susceptible} to
generating errors due to retention and read disturb (i.e.,
cells that lie at the tails of the threshold voltage distributions
of each state, where the distributions overlap with
each other), respectively. We observe that some flash cells
are more likely to be affected by retention leakage and read
disturb than others, as a result of process variation~\cite{cai.hpca15, cai.dsn15}. 
We call these cells retention/read disturb \emph{prone}, while
cells that are less likely to be affected are called retention/read 
disturb \emph{resistant}. RFR and RDR classify the susceptible
cells as retention/read disturb prone or resistant by inducing
\emph{even more} retention and read disturb on the failed flash
page, and then recording the new threshold voltages of the
susceptible cells. We classify the susceptible cells by observing
the magnitude of the threshold voltage shift due to the
additional retention/read disturb induction.

\begin{figure}[h]
  \centering
  \includegraphics[width=0.65\columnwidth]{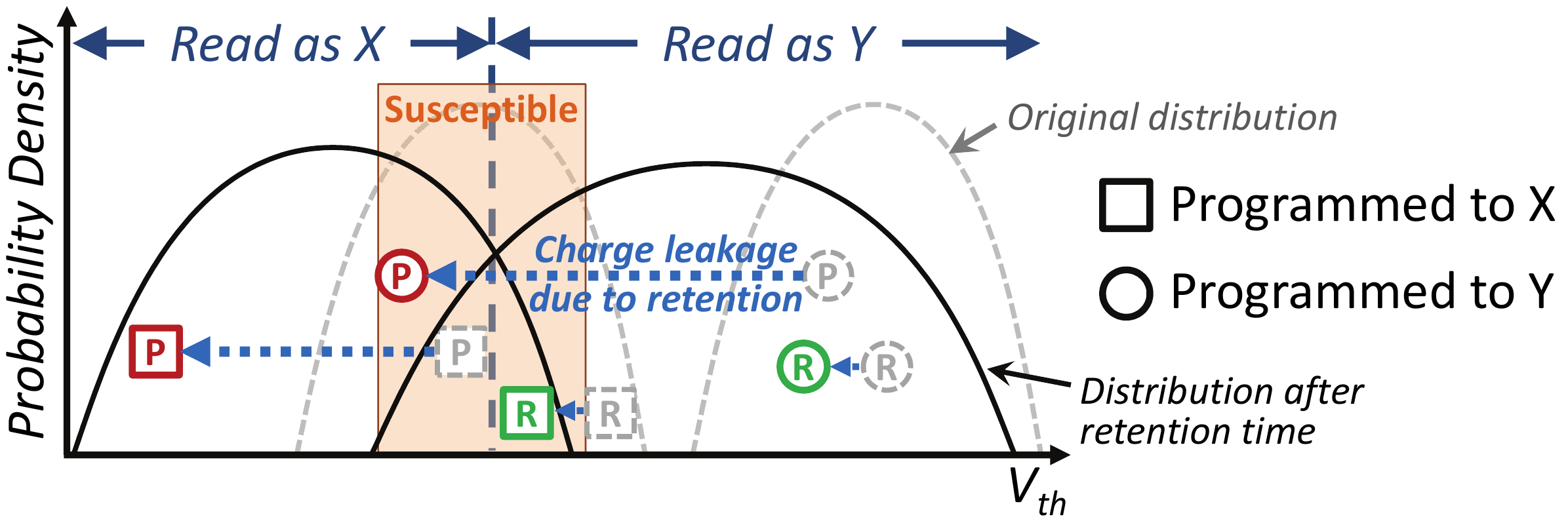}%
  \caption[Some retention-prone (P) and retention-resistant (R) cells
are incorrectly read after charge leakage due to retention time.
RFR identifies and corrects the incorrectly read cells based on their
leakage behavior.]
  {Some retention-prone (P) and retention-resistant (R) cells
are incorrectly read after charge leakage due to retention time.
RFR identifies and corrects the incorrectly read cells based on their
leakage behavior. \chI{Reproduced from~\cite{cai.arxiv17}.}}%
  \label{fig:F33}%
\end{figure}
\FloatBarrier

Figure~\ref{fig:F33} shows how the threshold voltage of a retention-prone
cell (i.e., a \emph{fast-leaking} cell, labeled P in the figure)
decreases over time (i.e., the cell shifts to the left) due to
retention leakage, while the threshold voltage of a retention-
resistant cell (i.e., a \emph{slow-leaking} cell, labeled R in the
figure) does not change significantly over time. Retention
Failure Recovery (RFR) uses this classification of retention-
prone versus retention-resistant cells to correct the
data from the failed page \emph{without} the assistance of ECC.
Without loss of generality, let us assume that we are studying
susceptible cells near the intersection of two threshold
voltage distributions X and Y, where Y contains higher voltages
than X. Figure~\ref{fig:F33} highlights the region of cells considered
susceptible by RFR using a box, labeled \emph{Susceptible}.
A susceptible cell within the box that is retention prone
likely belongs to distribution Y, as a retention-prone cell
shifts rapidly to a lower voltage (see the circled cell labeled
P within the \emph{susceptible} region in the figure). A retention-resistant
cell in the same \emph{susceptible} region likely belongs
to distribution X (see the boxed cell labeled R within the
\emph{susceptible} region in the figure).

Similarly, Read Disturb Recovery (RDR) uses the classification
of read disturb prone versus read disturb resistant
cells to correct data. For RDR, disturb-prone cells shift more
rapidly to higher voltages, and are thus likely to belong to
distribution X, while disturb-resistant cells shift little and
are thus likely to belong to distribution Y. Both RFR and
RDR correct the bit errors for the susceptible cells based on
such \emph{expected} behavior, reducing the number of errors that
ECC needs to correct.

RFR and RDR are highly effective at reducing the error
rate of failed pages, reducing the raw bit error rate by 50\%
and 36\%, respectively, as shown in prior \chIII{works from our research group}~\cite{cai.hpca15, cai.dsn15}, 
where more detailed information and analyses can
be found.


\section{Emerging Reliability Issues for 3D NAND Flash Memory}
\label{sec:background:3d}

\chI{While the demand for NAND flash memory capacity continues to grow,
manufacturers have found it increasingly difficult to rely on manufacturing
process technology scaling to achieve increased capacity~\cite{park.jssc15}.
Due to a combination of limitations in manufacturing process technology
and the increasing reliability issues as manufacturers move to smaller
process technology nodes, planar (i.e., 2D) NAND flash scaling has become
difficult for manufacturers to sustain.  This has led manufacturers to seek
alternative approaches to increase NAND flash memory capacity.}

\chI{Recently, manufacturers have begun to produce SSDs that
contain \emph{three-dimensional} (3D) NAND flash memory\chI{~\cite{yoon.fms15, park.jssc15, kang.isscc16,
im.isscc15, micheloni.procieee17, micheloni.sn16}}.  In 3D NAND flash memory, 
\emph{multiple layers} of flash cells are stacked vertically to increase the density
and to improve the scalability of the memory~\cite{yoon.fms15}. 
In order to achieve this stacking, manufacturers have changed a number of 
underlying properties of the flash memory design.

In this section, we examine
these changes, and discuss how they affect the reliability of the \chI{flash} memory devices.
In Section~\ref{sec:3d:org}, we discuss the flash \chI{memory} cell design commonly used
in contemporary 3D NAND flash memory, and how these cells are organized 
across the multiple layers.
In Section~\ref{sec:3d:errors}, we discuss how the reliability of
3D NAND flash memory compares to the reliability of the planar NAND flash
memory that we have \chIII{discussed} so far in this \chIII{chapter}.
In Section~\ref{sec:3d:mitigation}, we briefly discuss \chI{error} mitigation mechanisms
that cater to emerging reliability issues in 3D NAND flash memory.
\chIII{In Chapters~\ref{sec:3derror} and~\ref{sec:heatwatch}, we perform a
comprehensive error characterization for 3D NAND using real, state-of-the-art
3D NAND devices, and propose new techniques to mitigate raw bit errors in 3D
NAND flash memory.}


\subsection{\chI{3D NAND Flash Design and Operation}}
\label{sec:3d:org}

\chI{As we discuss in Section~\ref{sec:flash:data}, NAND flash memory stores
data as the threshold voltage of each flash cell.  In planar NAND flash memory,
we achieve this using a floating-gate transistor as a flash cell, as shown in
Figure~\ref{fig:F6}.  The floating-gate transistor stores charge in the floating
gate of the cell, which consists of a conductive material.  The floating gate is
surrounded on both sides by an oxide layer.  When high voltage is applied to
the control gate of the transistor, charge can migrate through the oxide layers
into the floating gate due to Fowler-Nordheim (FN) tunneling~\cite{fowler.royalsociety28} (see 
Section~\ref{sec:flash:pgmerase}).}

\chI{Most manufacturers use a \emph{charge trap transistor} as the flash cell
in 3D NAND flash memories, instead of using a floating-gate transistor.
Figure~\ref{fig:3dcell} shows \chI{the} cross section of a charge-trap transistor.
Unlike a floating-gate transistor, which stores data in the form of charge 
within a \emph{conductive} material, a charge trap transistor stores data as
charge within an \emph{insulating} material, known as the \emph{charge trap}.
In a 3D circuit, the charge trap wraps around a cylindrical transistor substrate, 
which contains the source (labeled \emph{S} in Figure~\ref{fig:3dcell}) and 
drain (labeled \emph{D} in the figure), and a control gate wraps around the charge trap.
This arrangement allows the channel between the source and drain to form
\emph{vertically} within the transistor.
As is the case with a floating-gate transistor, a \chI{tunnel oxide layer} exists between
the charge trap and the substrate, and a gate oxide \chI{layer} exists between the charge trap and the control 
gate.}


\begin{figure}[h]
\centering
\includegraphics[width=.35\linewidth]{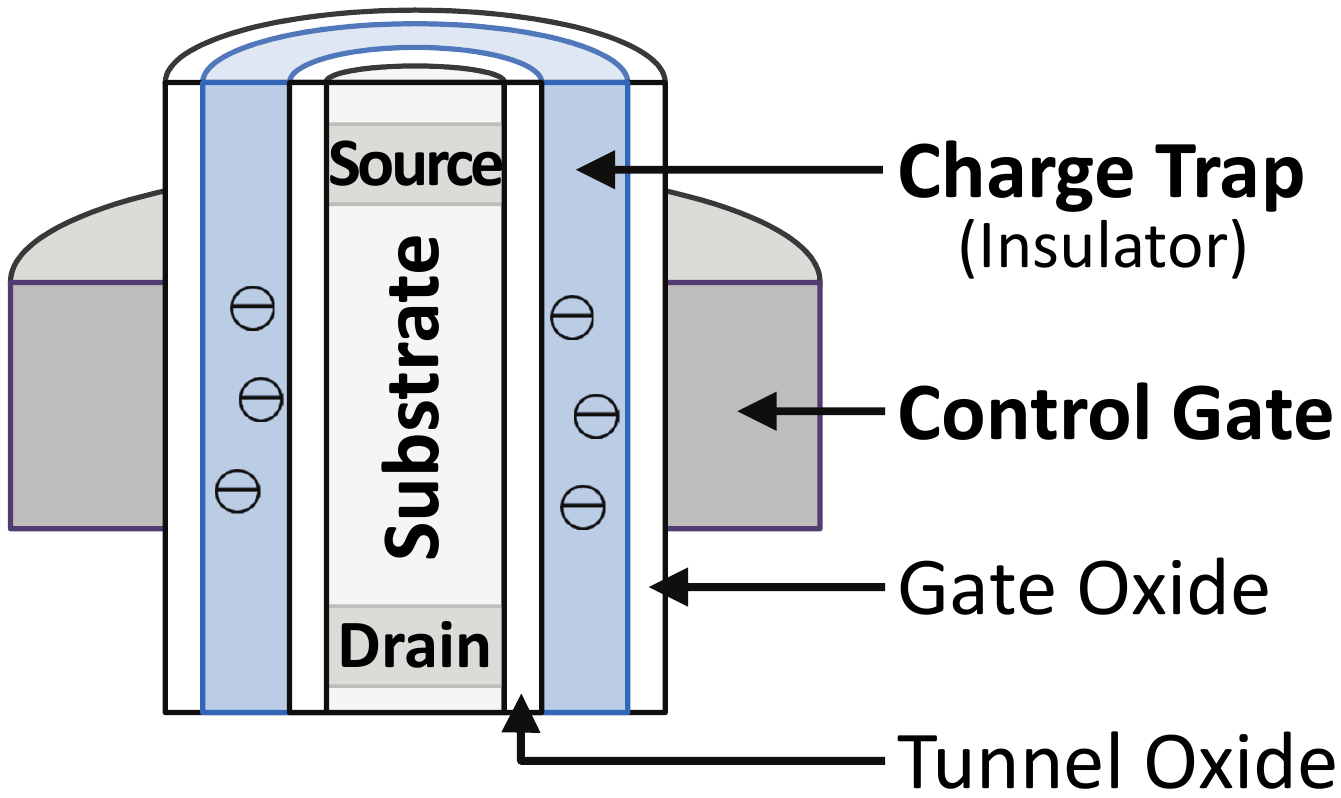}
\caption{\chI{Cross section of a charge trap transistor, used as a flash cell in 3D \chI{charge trap} NAND flash memory.}}
\label{fig:3dcell}
\end{figure}
\FloatBarrier

\chI{Despite the change in cell structure, the mechanism for transferring charge
into and out of the charge trap is similar to the mechanism for transferring
charge into and out of the floating gate.  In 3D NAND flash memory, the charge
trap transistor typically employs FN tunneling to change the threshold voltage 
of the charge trap~\cite{park.jssc15, katsumata.vlsit09}.\footnote{\chI{Note that \chI{\emph{not}} all 
charge trap transistors rely on FN tunneling.  Charge trap transistors used for
NOR flash memory change their threshold voltage using \emph{channel hot
electron injection}, also known as \emph{hot carrier injection}~\cite{luryi.ted84}.}}
When high voltage is applied to the control gate, electrons are injected into
the charge trap from the substrate.  As this behavior is similar to how 
electrons are injected into a floating gate, read, program, and erase operations
remain the same for both planar and 3D NAND flash memory.}


\chI{Figure~\ref{fig:organization} shows how multiple charge trap transistors are
physically organized within 3D NAND flash memory \chI{to form flash blocks, 
wordlines, and bitlines (see Section~\ref{sec:flash:block})}.  \chI{As mentioned above}, the
channel within a charge trap transistor forms vertically, as opposed to the
horizontal channel that forms within a floating-gate transistor.
The vertical orientation of the channel allows us to stack multiple transistors
\emph{on top of each other} (i.e., along the z-axis) within the chip,
using 3D-stacked circuit integration.
The vertically-connected channels form one bitline of a flash block in 3D NAND
flash memory.  Unlike in planar NAND flash memory, where only the substrates
of flash cells on the same bitline are connected together, flash cells along the
same bitline in 3D NAND flash memory share a common substrate and a common
insulator \chI{(i.e., charge trap)}. The FN tunneling induced by the control gate of the transistor
forms a tunnel only in a local region of the insulator, and, thus, electrons are
injected only into that local region.}
Due to the strong insulating properties of the material used for
the insulator, different regions of a single insulator can have different voltages.
\chI{This means that each region of the insulator can store a different
data value, and thus, the data of \emph{multiple} 3D NAND flash memory cells
can be stored reliably in a \emph{single} insulator.}  This is because the FN tunneling
induced by the control gate of the transistor forms a tunnel only in a \chI{\emph{local}} region
of the insulator, and, thus, electrons are injected only into that local region.

\begin{figure}[h]
\centering
\includegraphics[width=.6\linewidth]{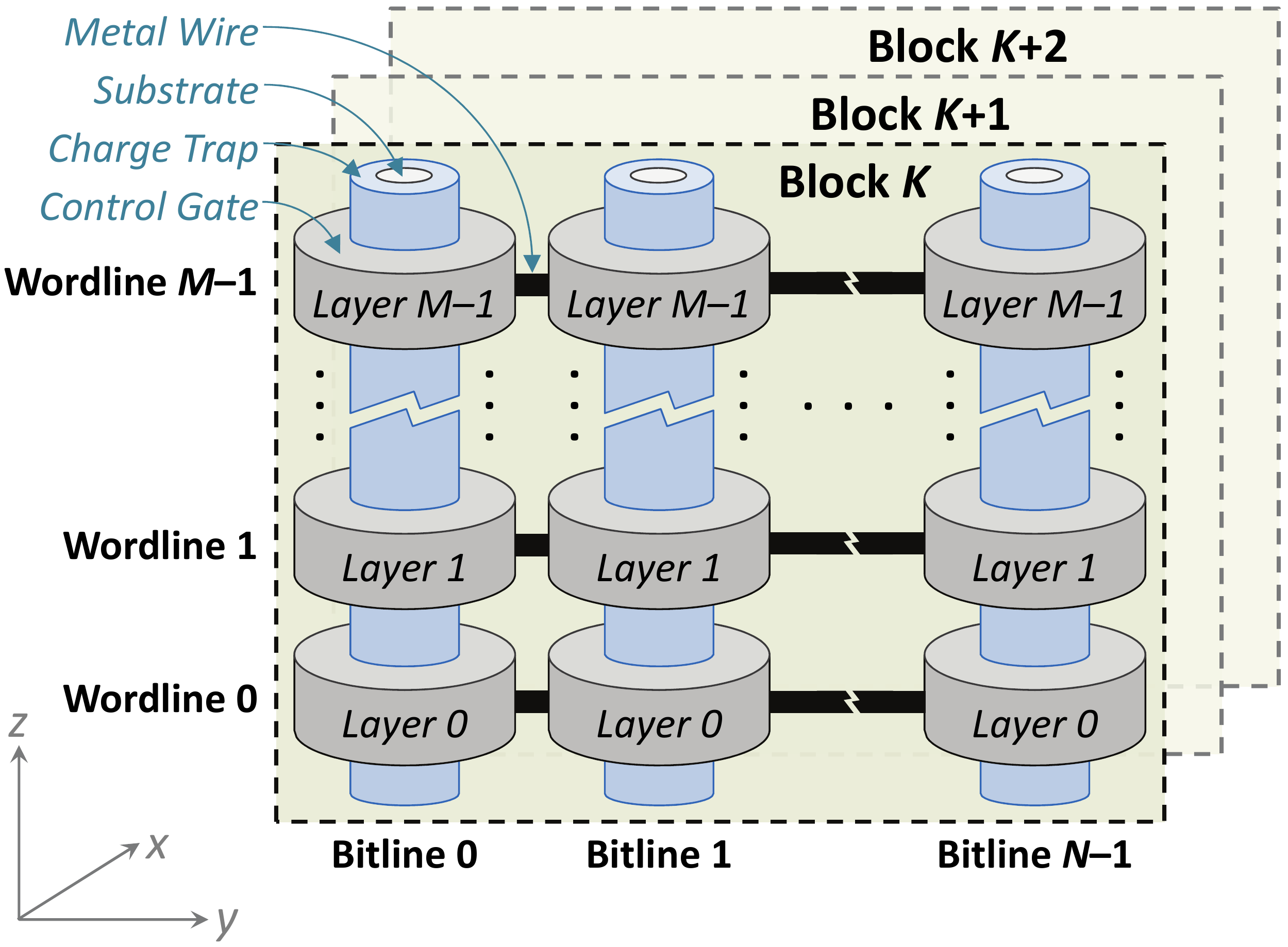}
\caption{\chI{Organization of flash cells in an \emph{M}-layer 3D charge trap NAND flash memory chip,
where each block consists of \emph{M}~wordlines and \emph{N}~bitlines.}}
\label{fig:organization}
\end{figure}
\FloatBarrier

\chI{Each cell along a bitline belongs to a different \emph{layer} of the flash 
memory chip.  Thus, a bitline crosses \chI{\emph{all}} of the layers within the chip.  
Contemporary 3D NAND flash \chI{memory contains} 24--\chI{96 layers~\cite{{yoon.fms15, park.jssc15,
kang.isscc16, kim.isscc17, elliott.fms17, toshiba.3dnand}}}.
Along the y-axis, the control gates of cells within a \emph{single layer} are 
connected together to form one wordline of a flash block.  As we show in
Figure~\ref{fig:organization}, a block in 3D NAND flash memory consists of 
all of the flash cells within the same y-z plane (i.e., all cells that have the
same coordinate along the x-axis).
\chI{Note that, while not depicted in Figure~\ref{fig:organization},
each bitline within a 3D NAND flash block includes a sense 
amplifier and two selection transistors used to select the bitline (i.e., the SSL and GSL
transistors; see Section~\ref{sec:flash:block}).  The sense amplifier and 
selection transistors are connected in series with the charge trap transistors
that belong to the same bitline, in a similar manner to
the connections shown for a planar NAND flash block in Figure~\ref{fig:F8}.}}
\chI{More detail on the circuit-level design of 3D NAND flash memory can be
found in~\cite{katsumata.vlsit09, komori.iedm08, tanaka.vlsit07, jang.vlsit09}.}

\chI{Due to the use of multiple layers of flash cells within a single NAND flash
memory chip, \chI{which greatly increases capacity per unit area,} manufacturers can achieve high cell density \chI{\emph{without}} the need to
use small manufacturing process technologies. \chI{For example, state-of-the-art planar NAND flash
memory uses the \SIrange{15}{19}{\nano\meter} feature size~\cite{luo.jsac16, parnell.globecom14}.
In contrast, contemporary} 3D NAND flash memory uses larger feature sizes (e.g.,
\SIrange{30}{50}{\nano\meter})~\cite{yoon.fms15, samsung.whitepaper14}. 
The larger feature sizes reduce manufacturing costs, as their corresponding
manufacturing process technologies are much more mature and have a higher yield
than the \chI{process} technologies used for small feature sizes.  As we discuss in Section~\ref{sec:3d:errors},
the larger feature size also has an effect on the reliability of 3D NAND flash
memory.}

\subsection{\chI{Errors in 3D NAND Flash \chI{Memory}}}
\label{sec:3d:errors}

\chI{While the high-level behavior of 3D NAND flash memory is similar to the
behavior of 2D planar NAND flash memory, there are a number of differences 
between the reliability of 3D NAND flash and planar NAND flash,
\chI{which we will look into in Chapter~\ref{sec:3derror}}.
There are two reasons for the \chI{differences} in reliability:
(1)~the use of charge trap transistors instead of
floating-gate transistors, and
(2)~moving to a larger manufacturing process technology.
We categorize the changes based on the reason for the change below.}

\paratitle{\chI{Effects of Charge Trap Transistors}}
\chI{Compared to the reliability \chI{issues} discussed in Section~\ref{sec:errors} for 
planar NAND flash memory, the use of charge trap transistors introduces
two key differences:
(1)~\emph{early retention loss}~\cite{choi.vlsit16, yoon.fms15, mizoguchi.imw17}, and
(2)~a \emph{reduction in P/E cycling \chI{errors}}~\cite{park.jssc15, yoon.fms15}.}

\chI{\chI{First,} early retention loss refers to the rapid leaking of electrons from a
flash cell soon after the cell is programmed~\cite{choi.vlsit16, yoon.fms15}.
Early retention loss occurs in 3D NAND flash memory because charge can now
migrate out of the charge trap in \emph{three} dimensions.  In planar NAND flash
memory, charge leakage due to retention occurs across the tunnel oxide,
\chI{which \chI{occupies} two dimensions} (see
Section~\ref{sec:errors:retention}).  In 3D NAND flash memory, charge can 
leak across \emph{both} the tunnel oxide \emph{and} the insulator \chI{that is} used 
for the charge trap, \chI{i.e., across \emph{three} dimensions}.  The additional charge leakage takes place for only a few
seconds after cell programming.  After a few seconds have passed, the impact
of leakage through the charge trap decreases, and the long-term cell retention
behavior is similar to that of flash cells in planar NAND flash memory~\cite{choi.vlsit16, yoon.fms15, mizoguchi.imw17}.}

\chI{\chI{Second, P/E cycling errors} (see Section~\ref{sec:errors:pe}) 
reduce \chI{with 3D NAND flash memory} because the tunneling oxide in charge trap transistors is 
\emph{less} susceptible to breakdown than the oxide in floating-gate transistors
during high-voltage operation~\cite{yoon.fms15, mizoguchi.imw17}.  As a result, the oxide is less likely
to contain trapped electrons once a cell is erased, which in turn makes it less
likely that the cell is subsequently programmed to an incorrect threshold 
voltage.
One benefit of the reduction in P/E cycling errors is that the
endurance (i.e., the maximum P/E cycle count) for a \chI{3D flash memory} cell has increased by
\chI{more than} an order of magnitude\chI{~\cite{parnell.fms16, parnell.fms17}}.}

\paratitle{\chI{Effects of Larger Manufacturing Process Technologies}}
\chI{Due to the use of larger manufacturing process technologies for 3D NAND 
flash memory, many of the \chIII{errors in} 2D planar NAND flash (see
Section~\ref{sec:errors}) are not as prevalent in 3D NAND flash memory.
For example, while read disturb is a prominent source of errors at small feature
sizes (e.g., \SIrange{20}{24}{\nano\meter}), its effects are small at larger feature
sizes~\cite{cai.dsn15}. Likewise, there are much fewer errors due to cell-to-cell 
program interference (see Section~\ref{sec:errors:celltocell}) in 3D NAND flash
memory, as the physical distance between neighboring cells is much larger due to
the increased feature size.
As a result, both cell-to-cell program interference and read disturb are 
\chI{\emph{currently} not} major issues in 3D NAND flash memory 
reliability\chI{~\cite{park.jssc15, yoon.fms15, parnell.fms17}}.}

\chI{One advantage of the lower cell-to-cell program interference is that 
3D NAND flash memory uses the older \emph{one-shot programming} 
algorithm~\cite{parnell.fms16, yoon.fms17, parnell.fms17} (see Section~\ref{sec:flash:pgmerase}).
In planar NAND flash memory, one-shot programming was replaced by two-step
programming (for MLC) and foggy-fine programming (for TLC) in order to reduce 
the impact of cell-to-cell program interference on fully-programmed cells 
\chI{(as we describe in Section~\ref{sec:flash:pgmerase})}.
The lower interference in 3D NAND flash memory makes two-step and foggy-fine
programming unnecessary.  As a result, none of the cells in 3D NAND flash memory
are partially-programmed, significantly reducing the number of program errors (see 
Section~\ref{sec:errors:pgm}) that occur~\cite{parnell.fms17}.}

\chI{Unlike \chI{the effects on reliability} due to the use of a charge trap transistor, \chI{which are likely longer-term},
\chI{the effects on reliability} due to the use of larger manufacturing process technologies
are expected to \chI{be shorter-term}.  As manufacturers seek to further
increase the density of 3D NAND flash memory, they will reach an upper limit \chI{for}
the number of layers that can be integrated within a 3D-stacked \chI{flash memory} 
chip\chI{, which is currently projected to be in the range of 300--512 layers~\cite{lapedus.semieng16, lee.eetimes17}}.  At that
point, manufacturers will once again need to scale down the chip to \chI{\emph{smaller}}
manufacturing process technologies~\cite{yoon.fms15}, \chI{which, in turn,} will reintroduce high amounts
of read disturb and cell-to-cell program interference (just as it happened for
planar NAND flash memory~\cite{park.jssc08, kim.irps10, cai.dsn15, cai.iccd13, cai.sigmetrics14}).}

\subsection{\chI{Changes in Error Mitigation for 3D NAND Flash Memory}}
\label{sec:3d:mitigation}

\chI{Due to the reduction in a number of sources of errors, fewer error mitigation
mechanisms are currently needed for 3D NAND flash memory.  For example,
because the number of errors introduced by cell-to-cell program interference is 
\chI{currently} low, manufacturers have \chI{\emph{reverted}} to using one-shot programming (see
Section~\ref{sec:flash:pgmerase}) for 3D NAND flash\chI{~\cite{parnell.fms16, yoon.fms17, parnell.fms17}}.
As a result of the \chI{currently small} effect of read disturb errors, mitigation and recovery
mechanisms for read disturb (e.g., pass-through voltage optimization in
Section~\ref{sec:mitigation:voltage}, Read Disturb Recovery in
Section~\ref{sec:correction:recovery}) \chI{may} not \chI{be} \chI{needed,}
for the time being.  We expect that once 3D NAND flash memory begins to scale
down to smaller manufacturing process technologies, approaching the current
feature sizes used for planar NAND flash memory, there will be a \chI{significant} need for 3D
NAND flash \chI{memory} to use \chI{many, if not all, of} the error mitigation mechanisms we discuss in
Section~\ref{sec:mitigation}.}

\chI{To our knowledge, no mechanisms have been designed yet to reduce
the impact of early retention loss,
\chI{which is a new error mechanism in 3D NAND flash memory}.  This is in part due to the reduced overall impact
of retention errors in 3D NAND flash memory compared to planar NAND flash
memory~\cite{choi.vlsit16}, \chI{since} a larger cell contains a greater
number of electrons than a smaller cell at the same threshold voltage.
As a result, existing refresh mechanisms (see Section~\ref{sec:mitigation:refresh})
can be used to tolerate errors introduced by early retention loss with little
modification.} \chI{However, as 3D NAND flash memory scales into future smaller
technology nodes, the early retention loss problem may require new mitigation techniques.}

\chI{\chI{While} new error mitigation mechanisms have yet to emerge for
3D NAND flash memory, rigorous studies that examine error characteristics
of and error mitigation techniques for 3D NAND
flash memories are yet to be published.
These studies \chI{(1)}~may expose additional sources of errors that have not yet been
observed, and \chI{that} may be unique to 3D NAND flash memory; \chI{and (2)~can enable
a solid understanding of current error mechanisms in 3D NAND flash memory so
that appropriate specialized mitigation mechanisms can be developed}.
We expect that future works will \chI{experimentally} examine such sources of errors, and will
potentially introduce novel mitigation mechanisms for these errors.}
\chI{Thus, the field (both academia and industry) is currently in much need of
rigorous experimental characterization \chI{and analysis} of 3D NAND flash
memory devices.}
Our characterization in Chapter~\ref{sec:3derror} is the first in open
literature to comprehensively characterization all types of NAND flash memory
errors in 3D NAND using real, state-of-the-art MLC 3D charge trap NAND flash
memory chips.


\section{Similar Errors in Other Memory Technologies}
\label{sec:othermem}

As we discussed in Section~\ref{sec:errors}, there are five major sources
of errors in flash-memory-based SSDs. Many of these error
sources can also be found in other types of memory and
storage technologies. In this section, we take a brief look
at the major reliability issues that exist within DRAM and
in emerging nonvolatile memories. In particular, we focus
on DRAM in our discussion, as modern SSD controllers
have access to dedicated DRAM of considerable capacity
(e.g., \SI{1}{\giga\byte} for every \SI{1}{\tera\byte} of SSD capacity), which exists
within the SSD package (see Section~\ref{sec:ssdarch}). Major sources
of errors in DRAM include data retention, cell-to-cell
interference, and read disturb. There is a wide body of
work on mitigation mechanisms for the \chI{DRAM and
emerging memory technology} errors we describe
in this section, but we explicitly discuss only a select
number of them here, \chI{since a full treatment of
such mechanisms is out of the scope of this current chapter}.

\subsection{Cell-to-Cell Interference Errors in DRAM}
\label{sec:othermem:celltocell}

Another similarity
between the capacitive DRAM cell and the floating gate
cell in NAND flash memory is that they are both vulnerable
to cell-to-cell interference. In DRAM, one important way
in which cell-to-cell interference exhibits itself is the data-dependent
retention behavior, where the retention time of
a DRAM cell is dependent on the values written to \emph{nearby}
DRAM cells\chI{~\cite{liu.isca13, khan.sigmetrics14, khan.dsn16, patel.isca17, khan.cal16, khan.micro17}}. This phenomenon
is called \emph{data pattern dependence} (DPD)~\cite{liu.isca13}. Data pattern
dependence in DRAM is similar to the data-dependent
nature of program interference that exists in NAND flash
memory (see Section~\ref{sec:errors:celltocell}). Within DRAM, data dependence
occurs as a result of parasitic capacitance coupling (between
DRAM cells). Due to this coupling, the amount of charge
stored in one cell's capacitor can inadvertently affect the
amount of charge stored in an adjacent cell's capacitor\chI{~\cite{liu.isca13, khan.sigmetrics14, khan.dsn16, patel.isca17, khan.cal16, khan.micro17}}. 
As DRAM cells become smaller with
technology scaling, cell-to-cell interference worsens because
parasitic capacitance coupling between cells increases~\cite{liu.isca13, khan.sigmetrics14}. 
More findings on cell-to-cell interference and the
data-dependent nature of cell retention times in DRAM,
along with experimental data obtained from modern DRAM
chips, can be found in prior {works from our research group}~\cite{liu.isca13, qureshi.dsn15, khan.sigmetrics14, khan.dsn16, patel.isca17, khan.cal16, chang.thesis17, khan.micro17}.

\subsection{Data Retention Errors in DRAM}
\label{sec:othermem:retention}

DRAM uses the charge
within a capacitor to represent one bit of data. Much like the
floating gate within NAND flash memory, charge leaks from
the DRAM capacitor over time, leading to data retention
issues. Charge leakage in DRAM, if left unmitigated, can lead
to much more rapid data loss than the leakage observed in a
NAND flash cell. While leakage from a NAND flash cell typically
leads to data loss after several days to years of retention
time (see Section~\ref{sec:errors:retention}), leakage from a DRAM cell leads to
data loss after a retention time on the order of \emph{milliseconds} to
\emph{seconds}~\cite{liu.isca13}.

\chI{The retention time of a DRAM cell depends upon several factors,
including (1)~manufacturing process variation and 
(2)~temperature~\cite{liu.isca13}.
Manufacturing process variation affects the amount of current that leaks
from each DRAM cell's capacitor and access transistor~\cite{liu.isca13}.
As a result, the retention time of the cells within a single DRAM chip vary
significantly, resulting in \emph{strong cells} that have high retention
times and \emph{weak cells} that have low retention times within each chip.
The operating temperature affects the rate at which charge leaks from the
capacitor.  As the operating temperature increases, the retention time of a
DRAM cell decreases exponentially~\cite{liu.isca13, hamamoto.ted98}.
Figure~\ref{fig:dram-retention-temperature} shows the change in retention
time as we vary the operating temperature, as measured from real DRAM 
chips~\cite{liu.isca13}.  In Figure~\ref{fig:dram-retention-temperature}, \chIII{prior work
normalizes} the retention time of each cell to its retention time at an
\chI{operating} temperature of \SI{50}{\celsius}.  As the number of cells is
\chI{large}, \chIII{prior work groups} the normalized retention times into bins, and plot the
density of each bin.  \chIII{Prior work draws} two \chI{exponential-fit} curves:
(1)~the \emph{peak} curve, which is drawn through the most populous
bin at each temperature measured; and
(2)~the \emph{tail} curve, which is drawn through the lowest non-zero
bin for each temperature measured.}
\chI{Figure~\ref{fig:dram-retention-temperature} provides us with three major
conclusions about the relationship between DRAM cell retention time and
temperature.
First, both of the exponential-fit curves fit well, which confirms the exponential decrease
in retention time as the operating temperature increases in modern DRAM devices.
Second, the retention times of different DRAM cells are affected very differently
by changes in temperature.
Third, the variation in retention time across cells increases greatly as temperature
increases.
More analysis of factors that affect DRAM retention times can be found in
recent \chIII{works from our research group}\chI{~\cite{liu.isca13,
khan.sigmetrics14, patel.isca17, qureshi.dsn15, khan.dsn16, khan.cal16,
khan.micro17, das.dac18}}.}

\begin{figure}[h]
  \centering
  \includegraphics[width=0.65\columnwidth]{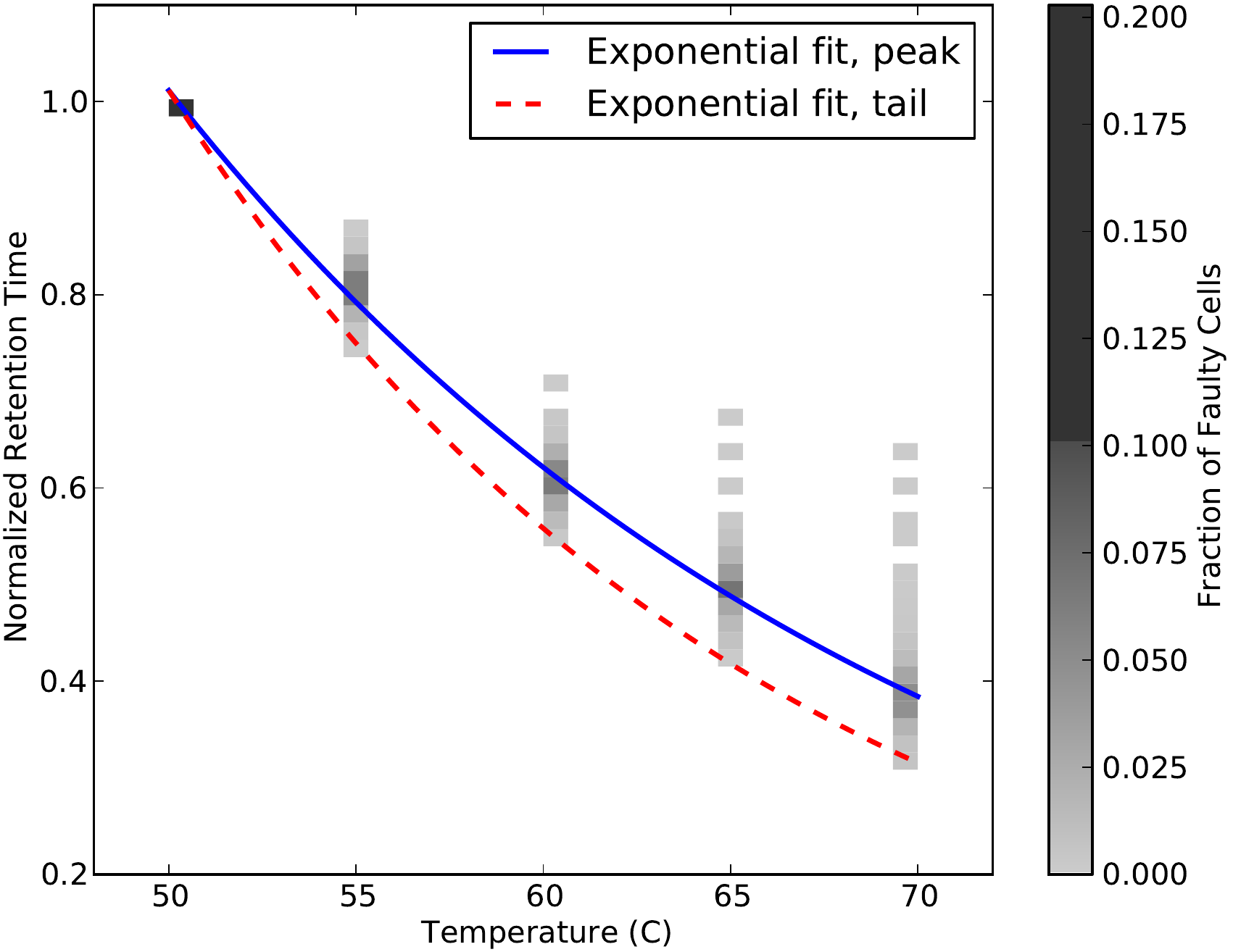}%
  \caption[DRAM retention time vs.\ operating temperature, normalized to 
  the retention time of each DRAM cell at \SI{50}{\celsius}.]
  {\chI{DRAM retention time vs.\ operating temperature, normalized to 
  the retention time of each DRAM cell at \SI{50}{\celsius}.  Reproduced from~\cite{liu.isca13}.}}%
  \label{fig:dram-retention-temperature}%
\end{figure}
\FloatBarrier

Due to the rapid charge leakage from DRAM
cells, a DRAM controller periodically refreshes all DRAM cells
in place~\cite{liu.isca12, chang.hpca14, liu.isca13, jesd79.jedec13, qureshi.dsn15, khan.sigmetrics14, patel.isca17} (similar to
the techniques discussed in Section~\ref{sec:mitigation:refresh}, but at a much smaller
time scale). DRAM standards require a DRAM cell to be
refreshed once every \SI{64}{\milli\second}~\cite{jesd79.jedec13}. As the density of DRAM continues
to increase over successive product generations (e.g., by
128x between 1999 and 2017~\cite{chang.sigmetrics16, chang.thesis17}),
\chI{enabled by the scaling of DRAM to smaller manufacturing process
technology nodes\chI{~\cite{mandelman.ibmjrd02}},} the performance
and energy overheads required to refresh an entire DRAM
module have grown significantly\chI{~\cite{liu.isca12, chang.hpca14}}.
\chI{It is expected that the refresh problem will get worse and \chI{limit}
DRAM density scaling, \chI{as described in a recent work by
Samsung and Intel~\cite{kang.mf14} and by our group~\cite{liu.isca12}}.
\chI{Refresh operations in DRAM cause both
(1)~performance loss and (2)~energy waste, both of which together lead to a
difficult technology scaling challenge.
Refresh operations degrade performance due to three major reasons.
First, refresh operations increase the memory latency, as a request to a DRAM bank that is
refreshing must wait for the refresh latency before it can be serviced.
Second, they reduce the amount of bank-level parallelism available to
requests, as a DRAM bank cannot service requests during refresh.
Third, they decrease the row buffer hit rate, as a refresh operation causes all
open rows in a bank to be closed.}
When a DRAM chip scales to a greater capacity, there are more DRAM rows that
need to be refreshed.  As Figure~\ref{fig:dram-refresh-scaling}a shows,
the amount of time spent on \chI{each refresh operation} scales linearly with the capacity of the
DRAM chip.  The additional time spent on refresh causes the \chI{DRAM data throughput
loss due to refresh} to become more severe in denser DRAM chips, as shown in 
Figure~\ref{fig:dram-refresh-scaling}b.  For a chip with a density of \SI{64}{\giga\bit},
nearly 50\% of the \chI{data} throughput is lost due to the high amount of time spent on
refreshing all of the rows in the chip.  The increased refresh time \chI{also increases the
effect of refresh on
power consumption.}  As \chIII{prior work observes} from Figure~\ref{fig:dram-refresh-scaling}c,
the fraction of DRAM power spent on refresh is expected to be the dominant
component of \chI{the} total DRAM power consumption, \chI{as DRAM chip 
capacity scales to become larger.  For a chip with a density of \SI{64}{\giga\bit},
nearly 50\% of the DRAM chip power is spent on refresh operations.  Thus,} 
refresh poses a clear challenge to DRAM scalability.}

\begin{figure}[h]
  \centering
  \begin{subfigure}[b]{0.32\columnwidth}%
    \includegraphics[width=\textwidth]{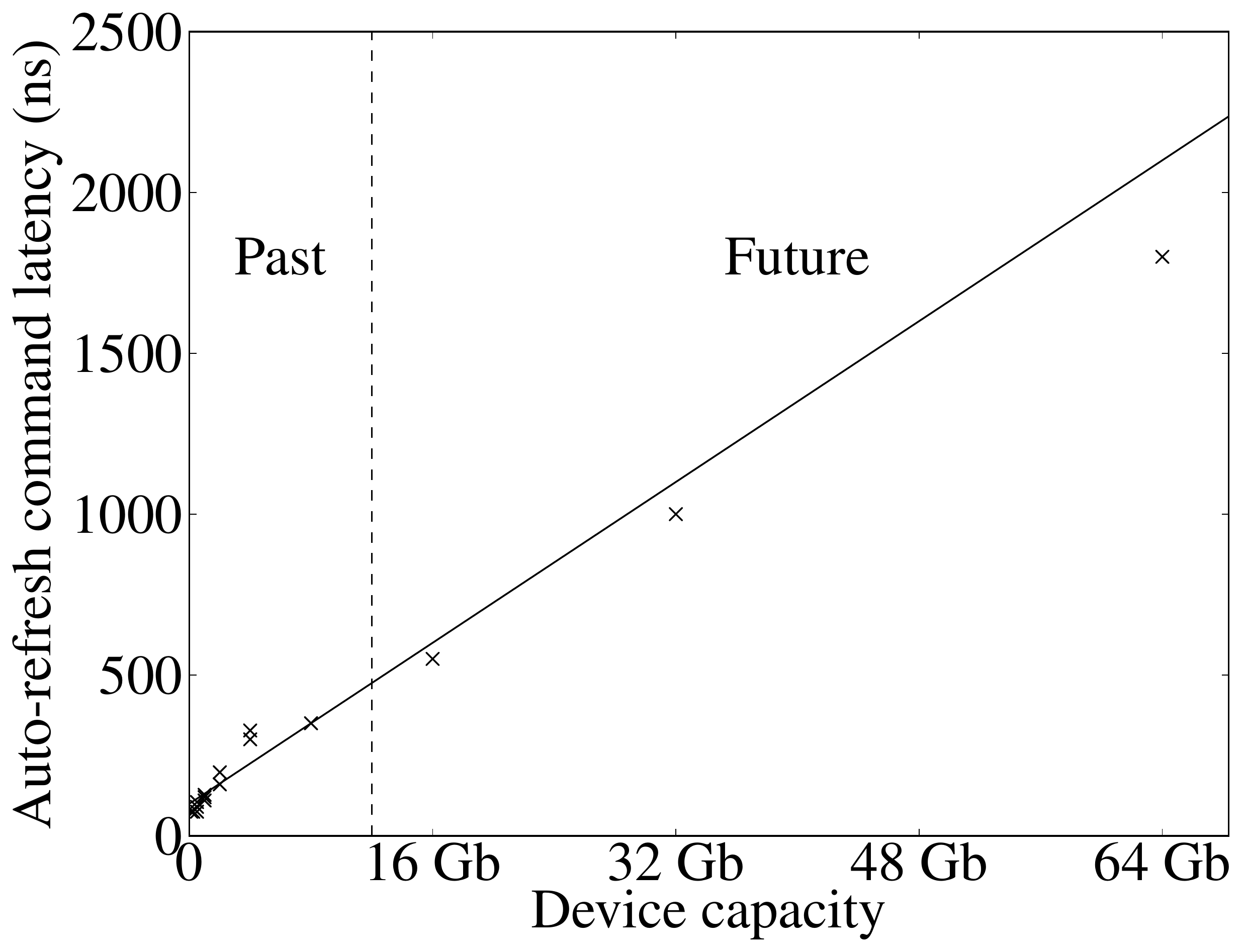}%
    \caption{Refresh latency}%
  \end{subfigure}%
  \hfill%
  \begin{subfigure}[b]{0.32\columnwidth}%
    \includegraphics[width=\textwidth]{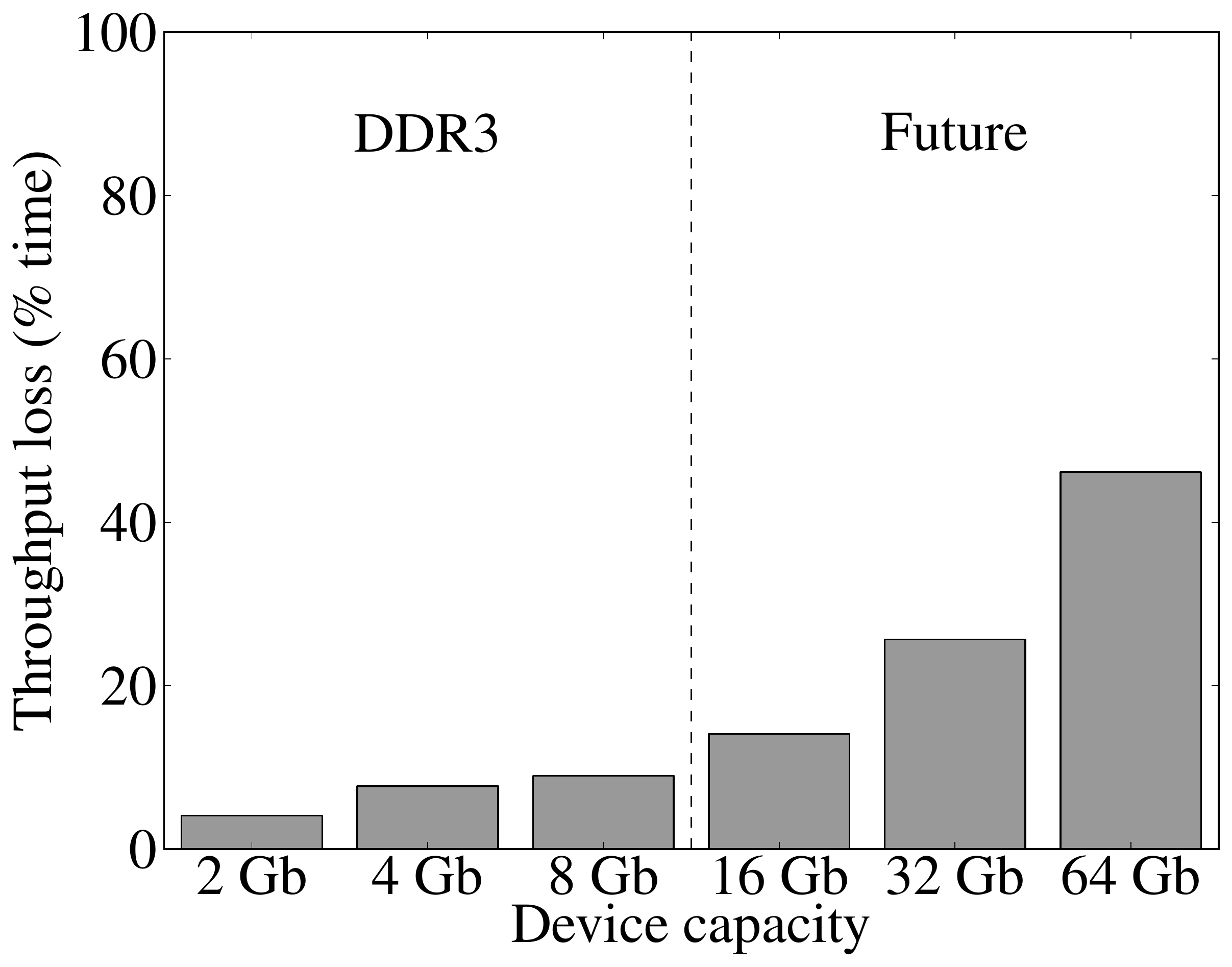}%
    \caption{Throughput loss}%
  \end{subfigure}%
  \hfill%
  \begin{subfigure}[b]{0.32\columnwidth}%
    \includegraphics[width=\textwidth]{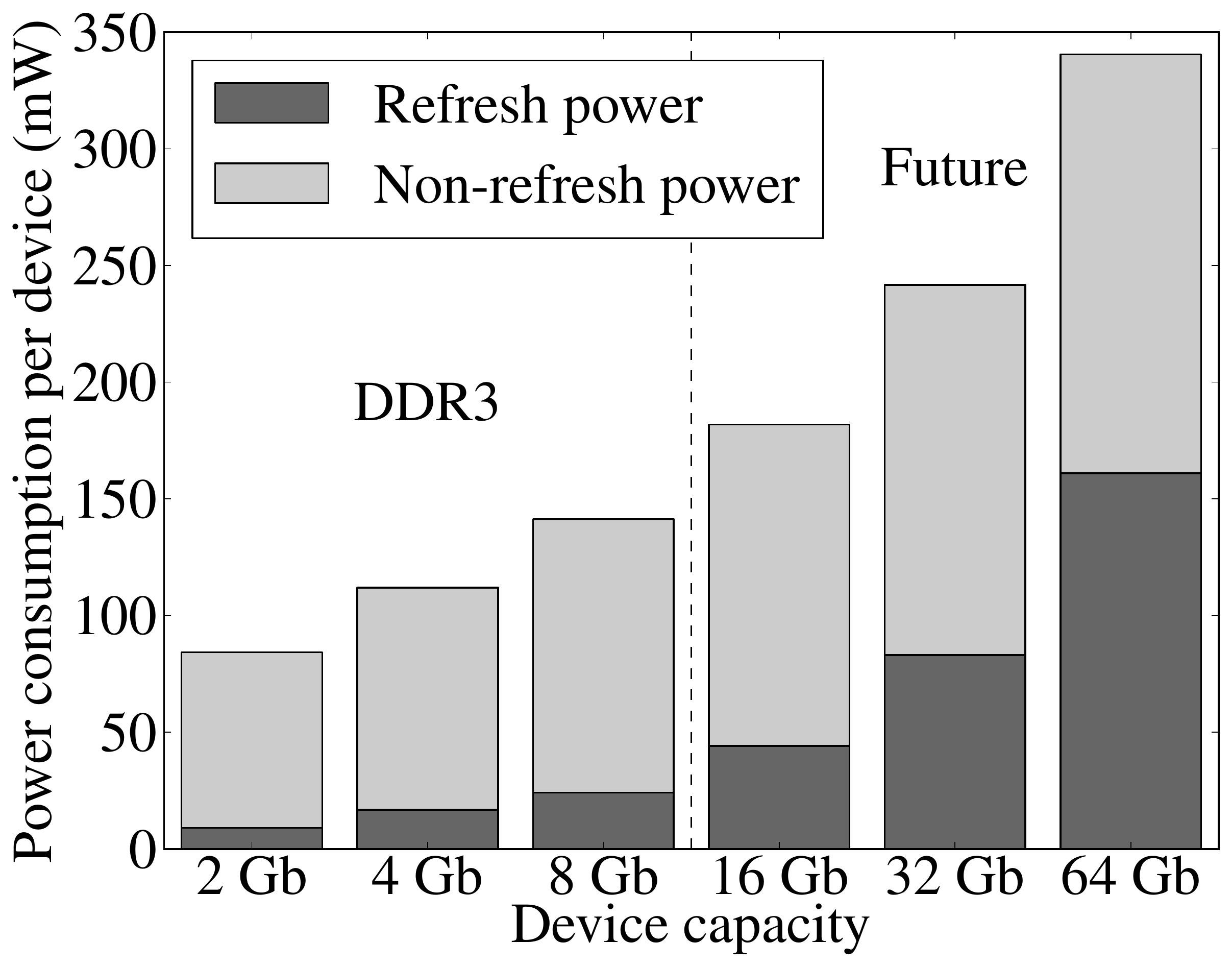}%
    \caption{Power consumption}%
  \end{subfigure}%
  \caption[\chI{Negative performance and power consumption} effects of refresh in contemporary and future DRAM devices.  
  \chI{We expect that as the capacity of each DRAM chip increases, (a)~the refresh latency, 
  (b)~the DRAM throughput lost during refresh operations, and (c)~the power consumed by refresh will all increase.}]
  {\chI{\chI{Negative performance and power consumption} effects of refresh in contemporary and future DRAM devices.  
  \chI{We expect that as the capacity of each DRAM chip increases, (a)~the refresh latency, 
  (b)~the DRAM throughput lost during refresh operations, and (c)~the power consumed by refresh will all increase.}
  Reproduced from~\cite{liu.isca12}.}}%
  \label{fig:dram-refresh-scaling}%
\end{figure}
\FloatBarrier

To combat the growing performance and energy overheads
of refresh, two classes of techniques have been
developed. The first class of techniques reduce the \emph{frequency}
of refresh operations without sacrificing the reliability
of data stored in DRAM (e.g., \chI{\cite{liu.isca12, qureshi.dsn15, khan.sigmetrics14, venkatesan.hpca06, isen.micro09, patel.isca17, khan.cal16, baek.tc14, khan.micro17}}). 
\chI{Various experimental studies of real DRAM chips (e.g., \chI{\cite{liu.isca12, liu.isca13, lee.hpca15, qureshi.dsn15, khan.sigmetrics14, khan.dsn16, patel.isca17, kim.iedl09, hassan.hpca17}})
have studied the \chI{data} retention time \chI{of DRAM cells in modern chips}.
Figure~\ref{fig:dram-retention}
shows the retention time measured from seven different \chI{real} DRAM modules
\chI{(by manufacturers A, B, C, D, and E) at an operating temperature of
\SI{45}{\celsius}}, as a
cumulative distribution (CDF) of the fraction of cells that have a retention time
less than the x-axis value~\cite{liu.isca13}.  \chIII{Prior work observes} from the figure that even
for the DRAM module whose cells have the worst retention time (i.e., the
CDF is the highest), \chI{fewer than only} 0.001\% of the total cells have a retention time
\chI{smaller} than \SI{3}{\second} \chI{at \SI{45}{\celsius}}.  
\chI{As shown in Figure~\ref{fig:dram-retention-temperature}, the retention 
time decreases exponentially as the temperature increases.  We can extrapolate
\chIII{the} observations from Figure~\ref{fig:dram-retention} to the worst-case
operating conditions by using the tail curve from Figure~\ref{fig:dram-retention-temperature}.
DRAM standards specify that the operating temperature of DRAM should not exceed
\SI{85}{\celsius}~\cite{jesd79.jedec13}.  Using the tail curve, \chIII{prior work finds} that a retention
time of \SI{3}{\second} at \SI{45}{\celsius} is equivalent to a retention time
of \SI{246}{\milli\second} at the worst-case temperature of \SI{85}{\celsius}.}
Thus, the vast majority of DRAM cells can retain
data without loss for much longer than the \SI{64}{\milli\second} retention
time specified by DRAM standards.  \chI{The} other experimental studies 
\chI{of DRAM chips have validated this observation} as well\chI{~\cite{liu.isca12, lee.hpca15, qureshi.dsn15, khan.sigmetrics14, khan.dsn16, patel.isca17, kim.iedl09, hassan.hpca17}}.}

\begin{figure}[h]
  \centering
  \includegraphics[width=0.65\columnwidth]{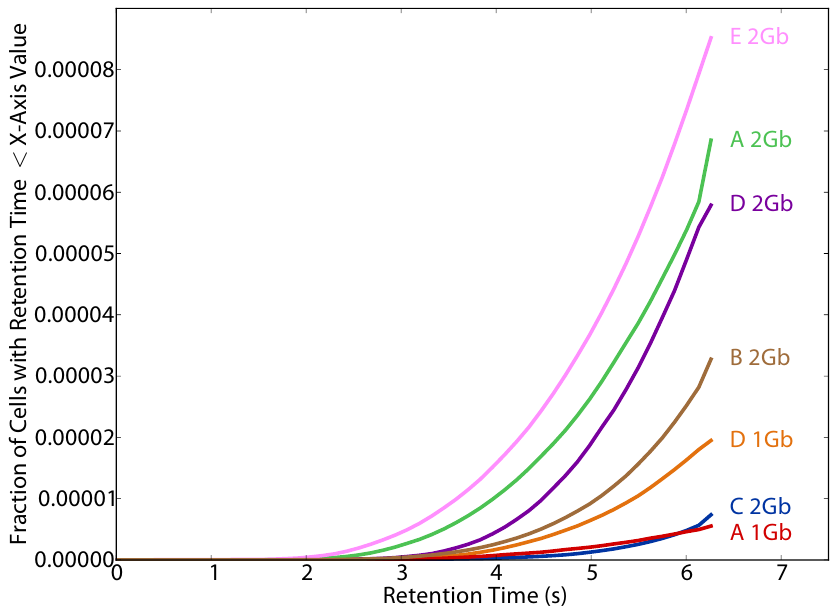}%
  \caption[Cumulative distribution of the number of cells in a DRAM module
  with a retention time less than the value on the x-axis, plotted for seven
  different DRAM modules.]
  {\chI{Cumulative distribution of the number of cells in a DRAM module
  with a retention time less than the value on the x-axis, plotted for seven
  different DRAM modules.  Reproduced from~\cite{liu.isca13}.}}%
  \label{fig:dram-retention}%
\end{figure}
\FloatBarrier

A number of works take advantage of this \chI{variability in data retention time behavior
across DRAM cells,
by introducing heterogeneous refresh rates, i.e., different refresh rates for
different DRAM rows.  Thus, these works can reduce} the frequency at which the vast majority of DRAM \chI{rows} within a
module are refreshed \chI{(e.g., \chI{\cite{liu.isca12, liu.isca13, qureshi.dsn15, khan.sigmetrics14, venkatesan.hpca06, isen.micro09, patel.isca17, khan.cal16, baek.tc14}})}.
\chI{For example, the key idea of RAIDR~\cite{liu.isca12} is to
refresh the \emph{strong} DRAM rows (i.e., those rows that can retain data
for \chI{much} longer than the minimum \SI{64}{\milli\second} retention time in the
DDR4 standard~\cite{jesd79.jedec13}) less frequently, and refresh the \emph{weak}
DRAM rows (i.e., those rows that can retain data only for the minimum retention
time) more frequently.  The major challenge  in such works is how to accurately
identify the retention time of each DRAM row.  To solve this challenge,
many recent works examine (online) DRAM retention time profiling 
techniques~\chI{\cite{patel.isca17, qureshi.dsn15, khan.dsn16, khan.sigmetrics14, liu.isca13, khan.micro17}}.}

The second
class of techniques reduce the interference caused by
refresh requests on demand requests (e.g.,~\cite{chang.hpca14, mukundan.isca13, stuecheli.micro10}). 
These works either change the scheduling order of
refresh requests~\cite{chang.hpca14, mukundan.isca13, stuecheli.micro10} or slightly modify the
DRAM architecture to enable the servicing of refresh and
demand requests in parallel~\cite{chang.hpca14}.

\chI{One \chI{critical challenge} in developing techniques to reduce 
refresh overheads is that it is getting significantly more difficult to
determine the minimum retention time of a DRAM cell,
as \chIII{prior works} have shown experimentally on modern DRAM
chips\chI{~\cite{liu.isca13, khan.sigmetrics14, khan.dsn16, qureshi.dsn15, patel.isca17}}.
Thus, determining the correct rate
at which to refresh DRAM cells has become more difficult, as also
indicated by industry~\cite{kang.mf14}. This is due to two major phenomena, both
of which get worse (i.e., become more prominent) with manufacturing 
process technology scaling. 
The first phenomenon is \emph{variable retention time} (VRT), where the
retention time of some DRAM cells can change drastically over time,
due to a memoryless random process that
results in very fast charge loss via a phenomenon called \emph{trap-assisted
gate-induced drain leakage}\chI{~\cite{liu.isca13, qureshi.dsn15, yaney.iedm87, restle.iedm92}}. 
VRT, as far as we know, is very
difficult to test for, because there seems to be no way of determining
that a cell exhibits VRT until that cell is observed to exhibit VRT,
and the time scale of a cell exhibiting VRT does not seem to
be bounded, based on the current experimental data \chI{on modern
DRAM devices~\cite{liu.isca13, patel.isca17}}. 
The second phenomenon is \emph{data pattern dependence} (DPD), which we discuss
in Section~\ref{sec:othermem:celltocell}.
Both of these phenomena greatly
complicate the accurate determination of minimum data retention
time of DRAM cells.  Therefore, data retention in DRAM continues to be 
a vulnerability that can greatly affect \chI{DRAM technology scaling (and
thus performance and energy consumption) as well as the} reliability and
security of current and future DRAM generations.}

More findings on the
nature of DRAM data retention and associated errors, as
well as relevant experimental data from modern DRAM
chips, can be found in prior \chIII{works from our research group}\chI{~\cite{liu.isca12, chang.hpca14, liu.isca13, lee.hpca15, qureshi.dsn15, khan.sigmetrics14, khan.dsn16, patel.isca17, hassan.hpca17, chang.thesis17, khan.micro17, khan.cal16, mutlu.date17}}.

\subsection{Read Disturb Errors in DRAM}
\label{sec:othermem:readdisturb}

Commodity DRAM
chips that are sold and used in the field today exhibit read
disturb errors~\cite{kim.isca14}, also called \emph{RowHammer}-induced errors~\cite{mutlu.date17}, 
which are \emph{conceptually} similar to the read disturb
errors found in NAND flash memory (see Section~\ref{sec:errors:readdisturb}).
Repeatedly accessing the same row in DRAM can cause
bit flips in data stored in adjacent DRAM rows. In order to
access data within DRAM, the row of cells corresponding
to the requested address must be \emph{activated} (i.e., opened for
read and write operations). This row must be \emph{precharged}
(i.e., closed) when another row in the same DRAM bank
needs to be activated. Through experimental studies on a
large number of real DRAM chips, \chIII{prior work shows} that when a
DRAM row is activated and precharged repeatedly (i.e.,
\emph{hammered}) enough times within a DRAM refresh interval,
one or more bits in physically-adjacent DRAM rows can be
flipped to the wrong value~\cite{kim.isca14}.

\chI{\chIII{Prior work} tested 129~DRAM modules manufactured by three major manufacturers
(A, B, and C) between 2008 and 2014, using an FPGA-based experimental DRAM
testing infrastructure~\cite{hassan.hpca17} (more detail on \chIII{the} experimental
setup, along with a list of all modules and their characteristics, can be found 
in \chIII{the} original RowHammer paper~\cite{kim.isca14}).  
Figure~\ref{fig:rowhammer-date} shows the
rate of RowHammer errors, with the 129~modules that \chIII{prior work} tested
categorized based on their manufacturing date.
\chIII{Prior work finds} that 110 of \chIII{the} tested modules exhibit RowHammer errors, with the
earliest such module dating back to 2010.  In particular, \chIII{prior work finds} that
\emph{all} of the modules manufactured in 2012--2013 that \chIII{prior work} tested are
vulnerable to RowHammer. Like with many NAND flash \chI{memory error 
mechanisms, especially read disturb}, RowHammer
is a recent phenomenon that \chI{especially affects DRAM chips} manufactured with more advanced
manufacturing process technology generations.}

\begin{figure}[h]
  \centering
  \includegraphics[width=0.5\columnwidth]{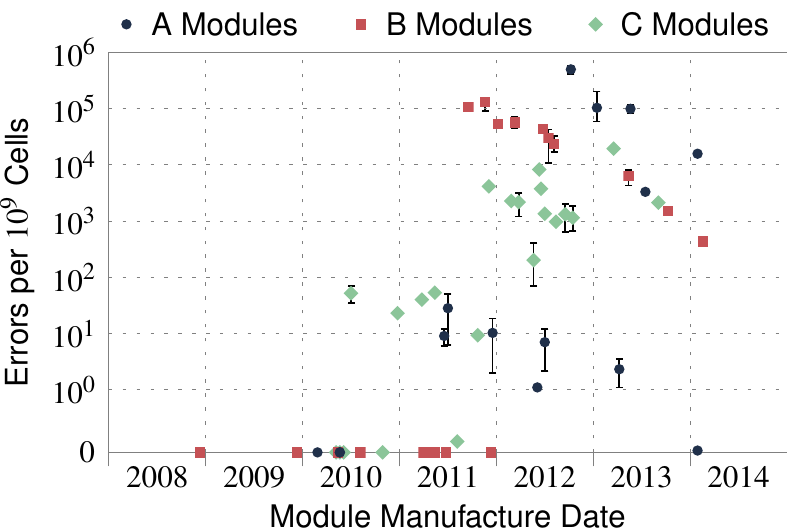}%
  \caption[RowHammer error rate vs.\ manufacturing dates of 129~DRAM
  modules we tested.]
  {\chI{RowHammer error rate vs.\ manufacturing dates of 129~DRAM
  modules we tested.  Reproduced from~\cite{kim.isca14}.}}%
  \label{fig:rowhammer-date}%
\end{figure}
\FloatBarrier

\chI{Figure~\ref{fig:rowhammer-interference} shows the distribution of
the number of rows (plotted in log scale on the y-axis) within a DRAM module that flip the
number of bits along the x-axis, as measured for example DRAM 
modules from three different DRAM manufacturers~\cite{kim.isca14}.
\chIII{Prior work makes} two observations from the figure.  First, the number of bits
flipped when we hammer a row (known as the \emph{aggressor row}) can vary
significantly within a module.  \chI{Second, each module has a different 
distribution of the number of rows.}
Despite these differences, \chIII{prior work finds} that this DRAM failure mode}
affects more than 80\% of the DRAM chips \chIII{prior work} tested~\cite{kim.isca14}.
As indicated above, this read disturb error mechanism in
DRAM is popularly called RowHammer~\cite{mutlu.date17}.

\begin{figure}[h]
  \centering
  \includegraphics[width=0.5\columnwidth]{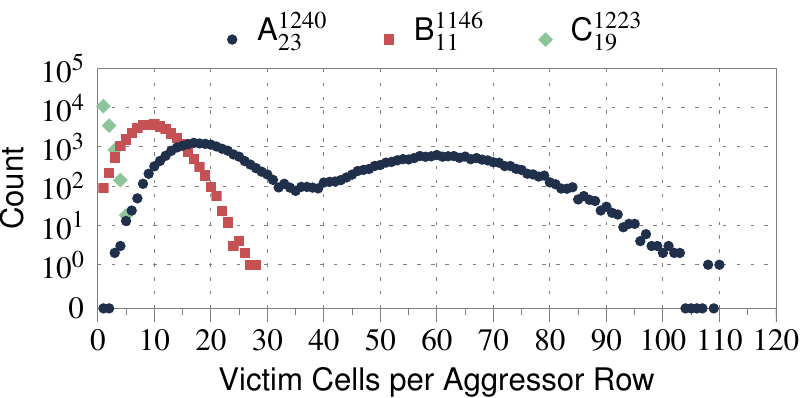}%
  \caption[Number of victim cells (i.e., number of bit errors) when an
  aggressor row is repeatedly activated, for three representative DRAM modules
  from three major manufacturers.  We label the modules in the format $X^{yyww}_n$, 
  where $X$ is the manufacturer (A, B, or C), $yyww$ is the manufacture year ($yy$) and
  week of the year ($ww$), and $n$ is the number of the selected module.]
  {\chI{Number of victim cells (i.e., number of bit errors) when an
  aggressor row is repeatedly activated, for three representative DRAM modules
  from three major manufacturers.  \chI{We label the modules in the format $X^{yyww}_n$, 
  where $X$ is the manufacturer (A, B, or C), $yyww$ is the manufacture year ($yy$) and
  week of the year ($ww$), and $n$ is the number of the selected module.} 
  Reproduced from~\cite{kim.isca14}.}}%
  \label{fig:rowhammer-interference}%
\end{figure}
\FloatBarrier


Various recent works show that RowHammer can be
maliciously exploited by user-level software programs to
(1)~induce errors in existing DRAM modules~\cite{kim.isca14, mutlu.date17}
and (2)~launch attacks to compromise the security of various
systems\chI{~\cite{seaborn.blog15, mutlu.date17, seaborn.blackhat15, gruss.dimva16, bosman.sp16, razavi.security16, veen.ccs16, burleson.dac16, xiao.security16, gruss.arxiv17}}.
For example, by exploiting the RowHammer read disturb
mechanism, a user-level program can gain kernel-level
privileges on real laptop systems~\cite{seaborn.blog15, seaborn.blackhat15}, take over a
server vulnerable to RowHammer~\cite{gruss.dimva16}, take over a victim
virtual machine running on the same system~\cite{bosman.sp16}, and
take over a mobile device~\cite{veen.ccs16}. Thus, the RowHammer
read disturb mechanism is a prime (and perhaps the
first) example of how a circuit-level failure mechanism in
DRAM can cause a practical and widespread system security
vulnerability.  We believe similar (yet \chI{likely} more difficult to
exploit) vulnerabilities exist in MLC NAND flash memory
as well, as described in recent \chIII{work from our research group}~\cite{cai.hpca17}.

\chI{Note that various solutions to RowHammer exist~\cite{kim.isca14, mutlu.date17, kim.thesis15},
but we do not discuss them in detail here.  \chIII{Recent work from our research group}~\cite{mutlu.date17} 
provides a comprehensive overview.  A very promising proposal is to modify
either the memory controller or the DRAM chip such that it probabilistically
refreshes the physically-adjacent rows of a recently-activated row, with very
low probability.  This solution is called \emph{Probabilistic Adjacent Row
Activation} (PARA)\chI{~\cite{kim.isca14}}.  Prior work shows that this low-cost, low-complexity
solution, which does not require any storage overhead, greatly closes the
RowHammer vulnerability~\cite{kim.isca14}.}

The RowHammer effect in DRAM worsens as the manufacturing
process scales down to smaller node sizes~\cite{kim.isca14, mutlu.date17}. 
More findings on RowHammer, along with extensive
experimental data from real DRAM devices, can be found in
prior \chIII{works from our research group}~\cite{kim.isca14, mutlu.date17, kim.thesis15}.

\subsection{Large-Scale DRAM Error Studies}
\label{sec:othermem:largescale}

Like flash memory,
DRAM is employed in a wide range of computing systems,
at scale. Thus, there is a similar need to study the aggregate
behavior of errors observed in a large number of DRAM
chips deployed in the field. Akin to the large-scale flash
memory SSD reliability studies discussed in Section~\ref{sec:errors:largescale}, a
number of experimental studies characterize the reliability
of DRAM at large scale in the field (e.g.,~\cite{meza.dsn15, schroeder.sigmetrics09, sridharan.asplos15, hwang.asplos12, sridharan.sc13}). 
\chI{We highlight three notable results from these studies.}

\chI{First, as \chIII{prior work} saw for large-scale studies of SSDs (see 
Section~\ref{sec:errors:largescale}), the number of errors
observed varies significantly for each DRAM module~\cite{meza.dsn15}.
Figure~\ref{fig:dram-large-scale}a shows the distribution of correctable
errors across the \emph{entire fleet} of servers at Facebook over a
fourteen-month period, omitting the servers that did not exhibit any
correctable DRAM errors.  The x-axis shows the normalized device
number, with devices sorted based on the number of errors they
experienced in a month.  As \chIII{prior work} saw in the case of SSDs,
a small number of servers accounts for the majority of errors.
As \chIII{prior work sees} from Figure~\ref{fig:dram-large-scale}a, the top 1\%
of servers account for 97.8\% of all observed correctable DRAM errors.
The distribution of the number of errors among servers follows a
power law model.  \chIII{Prior work shows} the probability density distribution of
correctable errors in Figure~\ref{fig:dram-large-scale}b, \chI{which indicates}
that the distribution of errors across servers follows a Pareto 
distribution, with a decreasing hazard rate\chI{~\cite{meza.dsn15}}.  This means that a
server that has experienced more errors in the past is likely to 
experience more errors in the future.}

\begin{figure}[h]
  \centering
  \hfill%
  \begin{subfigure}[b]{0.34\columnwidth}%
    \includegraphics[width=\textwidth]{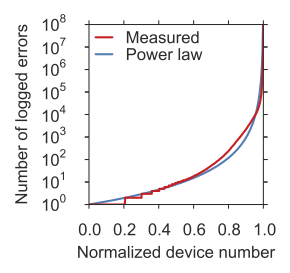}%
    \caption{}%
  \end{subfigure}%
  \hfill%
  \begin{subfigure}[b]{0.34\columnwidth}%
    \includegraphics[width=\textwidth]{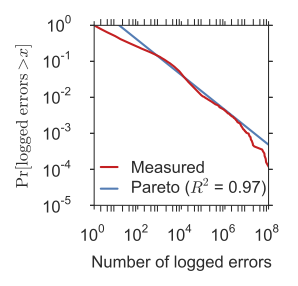}%
    \caption{}%
  \end{subfigure}%
  \hfill%
  \caption[Distribution of memory errors among servers with errors (a),
  which resembles a power law distribution.  Memory errors follow a Pareto
  distribution among servers with errors (b).]
  {\chI{Distribution of memory errors among servers with errors (a),
  which resembles a power law distribution.  Memory errors follow a Pareto
  distribution among servers with errors (b).
  Reproduced from~\cite{meza.dsn15}.}}%
  \label{fig:dram-large-scale}%
\end{figure}
\FloatBarrier

\chI{Second,} unlike SSDs, DRAM does \chI{\emph{not} seem to} show any clearly
discernible trend where higher utilization and age lead to
a greater raw bit error rate~\cite{meza.dsn15}.

\chI{Third,} the increase in the
density of DRAM chips with technology scaling leads to
higher error rates~\cite{meza.dsn15}.
\chI{The latter is illustrated in Figure~\ref{fig:dram-density},
which shows how different DRAM chip densities are related to
device failure rate.  We can see that there is a clear trend of
increasing failure rate with increasing chip density.
\chIII{Prior work finds} that the failure rate increases because despite small
improvements in the reliability of an \emph{individual} cell, the quadratic
increase in the number of cells per chip greatly increases the probability
of observing a single error in the whole chip\chI{~\cite{meza.dsn15}}.}

\begin{figure}[h]
  \centering
  \includegraphics[width=0.37\columnwidth]{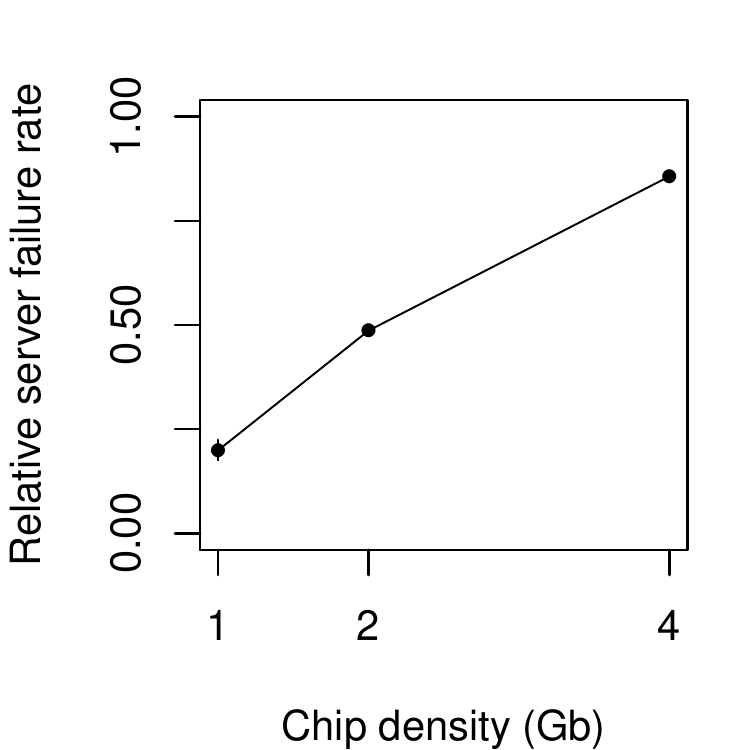}%
  \caption[\chI{Relative failure rate for servers with different chip densities.
  Higher densities (related to newer technology nodes) show a trend of 
  higher failure rates.}]
  {\chI{Relative failure rate for servers with different chip densities.
  Higher densities (related to newer technology nodes) show a trend of 
  higher failure rates.  Reproduced from~\cite{meza.dsn15}.} \chI{See Section~II-E
  of~\cite{meza.dsn15} for the complete definition of the metric plotted on the y-axis,
  i.e., \emph{relative server failure rate}.}}%
  \label{fig:dram-density}%
\end{figure}
\FloatBarrier

\subsection{Latency-Related Errors in DRAM}
\label{sec:othermem:latency}

\chI{Various} experimental
studies examine the tradeoff between DRAM reliability and
latency\chI{~\cite{lee.hpca15, chang.sigmetrics16, lee.sigmetrics17, chandrasekar.date14, hassan.hpca17, chang.thesis17, lee.thesis16, chang.sigmetrics17, kim.hpca18}}.
These works perform extensive experimental studies on
real DRAM chips to identify the effect of (1)~temperature,
(2)~supply voltage, and (3)~manufacturing process variation
that exists in DRAM on the latency and reliability characteristics
of different DRAM cells and chips. The temperature,
supply voltage, and manufacturing process variation
all dictate the amount of time that each cell needs to safely
complete its operations.
\chI{Several of \chIII{the works from our research group}~\cite{lee.hpca15, chang.sigmetrics16, lee.sigmetrics17, chang.sigmetrics17} examine how one can 
\emph{reliably} exploit different effects of variation to improve DRAM 
performance or energy consumption.}

\chI{Adaptive-Latency DRAM (AL-DRAM)~\cite{lee.hpca15} \chI{shows} that significant
variation exists in the access latency of \chI{(1)}~different DRAM modules, as a result of 
\chI{manufacturing process variation; and (2)~the same DRAM module over time,
as a result of varying operating temperature, since at low temperatures DRAM
can be accessed faster}.
\chI{The key idea of AL-DRAM is to adapt the DRAM latency to the operating
temperature and the DRAM module that is being accessed.
Experimental results show that AL-DRAM can reduce DRAM read latency by 32.7\%
and write latency by 55.1\%, averaged across 115~DRAM modules operating at
\SI{55}{\celsius}\chI{~\cite{lee.hpca15}}.}}

\chI{Voltron~\cite{chang.sigmetrics17} identifies the \chI{relationship} between the DRAM supply voltage
and access latency variation.  Voltron uses this \chI{relationship} to identify the
combination of voltage and access latency that minimizes system-level
energy consumption \chI{without exceeding a user-specified threshold for
the maximum acceptable performance loss.
For example, at an average performance loss of only 1.8\%, Voltron reduces
the DRAM energy consumption by 10.5\%, which translates to a
reduction in the overall system energy consumption of 7.3\%,
averaged over seven memory-intensive quad-core workloads\chI{~\cite{chang.sigmetrics17}}.}}

\chI{Flexible-Latency DRAM (FLY-DRAM)~\cite{chang.sigmetrics16} captures access latency
variation across DRAM cells \emph{within} a single DRAM chip due to manufacturing
process variation.
\chI{For example, Figure~\ref{fig:dram-trcd} shows how the bit error rate (BER)
changes if we reduce one of the timing parameters used to control the DRAM 
access latency below the minimum value specified by the manufacturer~\cite{chang.sigmetrics16}.
\chIII{Prior work uses} an FPGA-based experimental DRAM testing infrastructure~\cite{hassan.hpca17} to
measure the BER \chI{of} 30~real DRAM modules, over a total of 7500~rounds of
tests, as we lower the $t_{RCD}$ timing parameter (i.e., how long it takes to
open a DRAM row) below its standard value of
\SI{13.125}{\nano\second}.\chI{\footnote{\chI{More detail on \chIII{the} 
experimental setup, along with a list of all modules and their 
characteristics, can be found in \chIII{the} original FLY-DRAM paper~\cite{chang.sigmetrics16}.}}}
In this figure, \chIII{prior work uses} a box plot to summarize the bit error rate measured during
each round.  For each box, the bottom, middle, and
top lines indicate the 25th, 50th, and 75th percentile of the
population.  The ends of the whiskers indicate the minimum
and maximum BER of all modules for a given $t_{RCD}$ value.
Each round of BER measurement is represented as a single point 
overlaid upon the box.  From the figure, \chIII{prior work makes} three observations.
First, the BER decreases exponentially as we reduce $t_{RCD}$.
Second, there are no errors when $t_{RCD}$ is at \SI{12.5}{\nano\second}
or at \SI{10.0}{\nano\second}, indicating that manufacturers provide a
significant latency \emph{guardband} to provide additional protection against
process variation.
Third, the BER variation \chI{across different models} becomes smaller as $t_{RCD}$ decreases.  \chI{The}
reliability of a module operating at $t_{RCD}=$~\SI{7.5}{\nano\second}
varies significantly based on the DRAM manufacturer and model.
This variation occurs because the number of DRAM cells that experience
an error within a DRAM chip varies significantly from module to module.
\chI{Yet, the BER variation across different modules operating at $t_{RCD}=$~\SI{2.5}{\nano\second}
is much smaller, as most modules fail when the latency is reduced so significantly.}}}

\chI{From other experiments that \chIII{prior work describes} in \chIII{the} FLY-DRAM paper~\cite{chang.sigmetrics16}, 
\chIII{prior work finds} that there is spatial locality in the slower cells, resulting in \emph{fast regions} (i.e.,
regions where all DRAM cells can operate at significantly-reduced access 
latency without \chI{experiencing} errors) and \emph{slow regions} (i.e., regions where \emph{some} of the DRAM
cells \emph{cannot} operate at significantly-reduced access latency without \chI{experiencing} errors)
within each chip.
\chI{To take advantage of this heterogeneity in the reliable access latency of
DRAM cells within a chip,}
FLY-DRAM (1)~categorizes the \chI{cells} into fast and slow regions; and
(2)~lowers the overall DRAM latency by accessing fast regions with a lower 
latency.
FLY-DRAM lowers the timing parameters used for the fast region by as much as
42.8\%\chI{~\cite{chang.sigmetrics16}}.}
\chI{FLY-DRAM improves system performance for a wide variety of real workloads,
with the average improvement for an eight-core system ranging between 
13.3\% and 19.5\%, depending on the amount of variation that exists in each 
module~\cite{chang.sigmetrics16}.}

\begin{figure}[h]
  \centering
  \includegraphics[width=0.55\columnwidth]{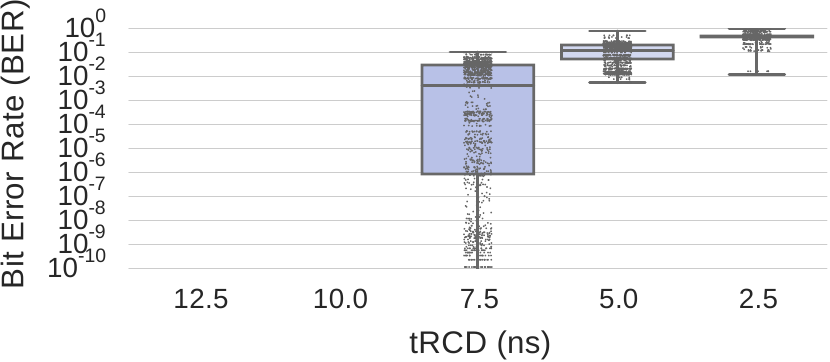}%
  \vspace{-5pt}%
  \caption[Bit error rates of tested DRAM modules as we reduce the
  DRAM access latency (i.e., the $t_{RCD}$ timing parameter).]
  {\chI{Bit error rates of tested DRAM modules as we reduce the
  DRAM access latency (i.e., the $t_{RCD}$ timing parameter).  Reproduced from~\cite{chang.sigmetrics16}.}}%
  \label{fig:dram-trcd}%
\end{figure}
\FloatBarrier

\chI{Design-Induced Variation-Aware DRAM (DIVA-DRAM)~\cite{lee.sigmetrics17} identifies the
latency variation within a single DRAM chip that occurs due to the architectural
design of the chip.  For example, a cell that is further away from the
row decoder requires a longer access time than a cell that is close to the \chI{row
decoder.  Similarly, a cell that is farther away from the wordline driver
requires a larger access time than a cell that is close to the wordline driver.}
DIVA-DRAM uses design-induced variation to reduce the access latency
to different parts of the chip. One can further reduce latency by
sacrificing some amount of reliability and performing error
correction to fix the resulting errors~\cite{lee.sigmetrics17}.
\chI{Experimental results show that DIVA-DRAM can reduce DRAM read latency by 40.0\%
and write latency by 60.5\%\chI{~\cite{lee.sigmetrics17}}.}}
\chI{In an eight-core system running a wide variety of real workloads, 
DIVA-DRAM improves system performance by an average of 13.8\%~\cite{lee.sigmetrics17}.}

More information
about the errors caused by reduced latency \chI{and reduced voltage} operation
in DRAM chips and the tradeoff between reliability and
latency \chI{and voltage} can be found in prior \chIII{works from our research group}~\cite{lee.hpca15, chang.sigmetrics16, lee.sigmetrics17, luo.dsn14, hassan.hpca17, chang.thesis17, lee.thesis16, chang.sigmetrics17}.

\subsection{Error Correction in DRAM}
\label{sec:othermem:ecc}

In order to protect the data
stored within DRAM from various types of errors, some
(but not all) DRAM modules employ ECC~\cite{luo.dsn14}. The ECC
employed within DRAM is much weaker than the ECC
employed in SSDs (see Section~\ref{sec:correction}) for various reasons. First,
DRAM has a much lower access latency, and error correction
mechanisms should be designed to ensure that DRAM
access latency does not increase significantly. Second, the
error rate of a DRAM chip tends to be lower than that of a
flash memory chip. Third, the granularity of access is much
smaller in a DRAM chip than in a flash memory chip, and
hence sophisticated error correction can come at a high
cost. The most common ECC algorithm used in commodity
DRAM modules is \emph{SECDED} (single error correction, double
error detection)~\cite{luo.dsn14}. Another ECC algorithm available for
some commodity DRAM modules is \emph{Chipkill}, which can tolerate
the failure of an \chI{\emph{entire}} DRAM chip within a module~\cite{dell.ibm97} 
\chI{at the expense of higher storage overhead and higher latency}. 
For both SECDED and Chipkill, the ECC information
is stored on one or more extra chips within the DRAM module,
and, on a read request, this information is sent alongside
the data to the memory controller, which performs the
error detection and correction.

As DRAM scales to smaller technology nodes, its error rate
continues to increase\chI{~\cite{mutlu.imw13, mutlu.superfri15, kim.isca14, meza.dsn15, mutlu.date17, kang.mf14, mandelman.ibmjrd02}}.  \chI{Effects}
like read disturb~\cite{kim.isca14}, cell-to-cell interference~\cite{liu.isca13, khan.sigmetrics14, khan.dsn16, patel.isca17, khan.cal16, khan.micro17}, 
and variable retention time~\cite{liu.isca13, qureshi.dsn15, khan.sigmetrics14, patel.isca17} 
become more severe\chI{~\cite{mutlu.imw13, mutlu.superfri15, kim.isca14, mutlu.date17, kang.mf14}}.
As a result, there is an increasing need for (1)~employing ECC
algorithms in \emph{all} DRAM chips/modules; (2)~developing more
sophisticated and efficient ECC algorithms for DRAM chips/modules; 
and (3)~developing error-specific mechanisms for error
correction. To this end, recent work follows various directions.
First, in-DRAM ECC, where correction is performed within the
DRAM module itself (as opposed to in the controller), is proposed~\cite{kang.mf14}.
One work shows how exposing this in-DRAM ECC
information to the memory controller can provide Chipkill-like
error protection at much lower overhead than the traditional
Chipkill mechanism~\cite{nair.isca16}. Second, various works explore and
develop stronger ECC algorithms for DRAM (e.g.,~\cite{kim.sc15, kim.hpca15, wilkerson.isca10}), 
and explore how to make ECC more efficient based
on the current DRAM error rate (e.g., \chI{\cite{dell.ibm97, udipi.isca12, chou.dsn15, alameldeen.isca11, lee.sigmetrics17}}). 
Third, recent work shows how the cost of ECC protection
can be reduced by (1)~exploiting \emph{heterogeneous reliability memory}~\cite{luo.dsn14}, 
where different portions of DRAM use different strengths
of error protection based on the error tolerance of different applications
and different types of data~\cite{luo.dsn14, liu.asplos11}, and (2)~using the
additional DRAM capacity that is otherwise used for ECC to
improve system performance when reliability is not as important
for the given application and/or data~\cite{luo.arxiv17}.

Many of these works that propose error mitigation
mechanisms \chI{for DRAM do \emph{not}} distinguish between the characteristics
of different types of errors. We believe that, in addition to
providing sophisticated and efficient ECC mechanisms in
DRAM, there is also significant value in and opportunity
for exploring \emph{specialized} error mitigation mechanisms that
are \emph{customized for different error types}, just as it is
done for flash memory (as we discussed in Section~\ref{sec:mitigation}). One
such example of a specialized error mitigation mechanism
is targeted to fix the RowHammer read disturb mechanism,
and is called \emph{Probabilistic Adjacent Row Activation} 
(PARA)~\cite{kim.isca14, mutlu.date17}, \chI{as we discussed earlier.  Recall that
the} key idea of PARA is to refresh the rows that
are physically adjacent to an activated row, with a very low
probability. PARA is shown to be very effective in fixing the
RowHammer problem at no storage cost and at very low
performance overhead~\cite{kim.isca14}.
\chI{PARA is a specialized yet very effective solution for fixing a specific 
error mechanism that is important \chI{and \chI{prevalent} in modern DRAM devices}.}

\subsection{Errors in Emerging Nonvolatile Memory Technologies}
\label{sec:othermem:emerging}

DRAM operations are several orders of magnitude faster than
SSD operations, but DRAM has two major disadvantages. First,
DRAM offers orders of magnitude less storage density than
NAND-flash-memory-based SSDs. Second, DRAM is volatile
(i.e., the stored data is lost on a power outage). Emerging
nonvolatile memories, such as \emph{phase-change memory} (PCM)~\cite{lee.isca09, qureshi.isca09, wong.procieee10, lee.micro10, zhou.isca09, lee.comm10, yoon.taco14}, 
\emph{spin-transfer torque
magnetic RAM} (STT-RAM or STT-MRAM)~\cite{naeimi.itj13, kultursay.ispass13}, \emph{metal-oxide
resistive RAM} (RRAM)~\cite{wong.procieee12}, and \emph{memristors}~\cite{chua.tct71, strukov.nature08},
are expected to bridge the gap between DRAM and SSDs, providing
DRAM-like access latency and energy, and at the same
time SSD-like large capacity and nonvolatility (and hence SSD-like
data persistence). 
\chI{These technologies are also expected to be used as part of
\emph{hybrid memory systems} \chI{(also called \emph{heterogeneous
memory systems})}, where one part of the memory consists of
DRAM modules and another part consists of modules of emerging
technologies\chI{~\cite{qureshi.isca09, yoon.taco14, ramos.ics11, yoon.iccd12, qureshi.micro12, 
zhang.pact09, meza.cal12, li.cluster17, meza.weed13, yu.micro17, chou.isca15, 
chou.micro14,jiang.hpca10, phadke.date11, chatterjee.micro12}}.}
PCM-based devices are expected to have
a limited lifetime, as PCM can only endure a certain number
of writes~\cite{lee.isca09, qureshi.isca09, wong.procieee10}, similar to the P/E cycling errors in
NAND-flash-memory-based SSDs (though PCM's write endurance
is higher than that of SSDs). PCM suffers from \chI{(1)}~\chI{\emph{resistance
drift}~\cite{wong.procieee10, pirovano.ted04, ielmini.ted07}}, where the resistance used to represent the value
\chI{becomes} higher over time (and eventually \chI{can introduce} a bit error),
similar to how charge leakage in NAND flash memory and
DRAM lead to retention errors over time; \chI{and
(2)~\chI{\emph{write disturb}~\cite{jiang.dsn14}}, where the heat generated during the programming of one
PCM cell dissipates into neighboring cells and \chI{can change} the value that is
stored within the neighboring cells}. 
\chI{STT-RAM suffers} from \chI{(1)}~\chI{\emph{retention failures}}, where the value
stored for a single bit \chI{(as the magnetic orientation of the layer that 
stores the bit)} can flip over time; and \chI{(2)}~\chI{\emph{read disturb}} (\chI{a
conceptually different phenomenon}
from the read disturb in DRAM and flash memory), where
reading a bit in STT-RAM can inadvertently induce a write to
that same bit~\cite{naeimi.itj13}. Due to the nascent nature of emerging
nonvolatile memory technologies and the lack of availability of
large-capacity devices built with them, extensive and dependable
experimental studies have yet to be conducted on the reliability
of real PCM, STT-RAM, RRAM, and memristor chips.
However, we believe that \chI{error mechanisms conceptually \chI{or
abstractly} similar} to those we
discussed in this chapter for flash memory and DRAM are likely
to be prevalent in emerging technologies as well
\chI{(as supported by some recent studies~\cite{naeimi.itj13, jiang.dsn14, 
zhang.iccd12, khwa.isscc16, athmanathan.jetcas16, sills.vlsic15, sills.vlsit14})}, 
albeit with different underlying mechanisms and error rates.

\chapter{WARM---Write-hotness Aware Retention Management}
\label{sec:warm}

As we have introduced in Section~\ref{sec:errors}, retention \chIII{errors are} one of
the most dominant error sources in NAND flash memory. Retention errors are caused by charge
leakage from the flash cells after the data has been programmed into the
cells. Our prior work has shown that \emph{retention errors not only degrade data
reliability, but also lead to performance \chIII{degradation}} due to increased read-retry
attempts~\cite{cai.hpca15}. In Section~\ref{sec:mitigation}, we have surveyed
many techniques that can mitigate retention errors proposed by prior work,
including refresh, read-retry, voltage optimization, etc. Among these
techniques, refresh is considered to be the most effective in improving flash
lifetime and has recently been implemented in real
SSDs~\cite{knight.techspot15,cai.iccd12,cai.itj13}. However, we observe that, while refresh improves
flash lifetime significantly by relaxing the retention time constraint, \emph{the
refresh operation itself can consume the majority of the improved flash
lifetime}, wasting opportunity for further flash lifetime improvements.

In this chapter, we introduce \emph{\underline{W}rite-hotness \underline{A}ware
\underline{R}etention \underline{M}anagement} (WARM), which exploits workload
\emph{write-hotness} and device \emph{data retention} characteristics to
improve flash lifetime. The goal of WARM is to eliminate redundant refreshes for
write-hot pages with minimal storage and performance overhead.
This work proposes a write hotness-aware flash memory
retention management policy, WARM\@. The \emph{first} key idea of WARM is to effectively
partition pages stored in flash \chIII{memory} into two groups based on the write frequency of
the pages. The \emph{second} key idea of WARM is to apply different management policies
to the two different groups of flash pages/blocks \chIII{to improve the
lifetime of the flash device}.

First, we look into flash memory data retention characteristics and SSD
workload characteristics to show that redundant flash \chIII{refresh operations} are
expensive
(Section~\ref{sec:warm:motivation}). Second, we discuss a novel, lightweight
approach to dynamically identifying and partitioning write-hot versus write-
cold pages (Section~\ref{sec:warm:mechanism:identify}).  Third, we describe how
WARM optimizes flash management policies, such as garbage collection and
wear-leveling, in a partitioned flash memory, and show how WARM integrates with
a refresh mechanism to provide further flash lifetime improvements
(Section~\ref{sec:warm:mechanism:policies}). Fourth, we evaluate the flash
lifetime improvement delivered by WARM, and the hardware and performance
overhead to implement WARM (Section~\ref{sec:warm:evaluation}). Finally, we
conclude \chIII{with} the contributions of WARM
(Section~\ref{sec:warm:conclusion}).

\section{Motivation}
\label{sec:warm:motivation}

As flash memory density has continued to increase, the endurance of flash
devices has been rapidly decreasing.  Relaxing the \emph{internal} retention
time of flash devices can significantly improve upon this endurance
(Section~\ref{sec:warm:motivation:relaxation}), but this cannot simply be externally
exposed, as the relaxation would impact the data integrity guarantee.  Periodically performing data refresh allows the 
flash device to relax the internal retention time while maintaining the data
integrity guarantee~\cite{cai.iccd12, pan.hpca12, mohan.tr12,
liu.fast12, liu.dac13, cai.date12, cai.itj13}.  Unfortunately, for real-world workloads, these refresh
operations consume a large portion of the extra endurance gained from internal
retention time relaxation, as we describe in
Section~\ref{sec:warm:motivation:overhead}.  In order to buy back the endurance,
\emph{we aim to eliminate redundant refresh operations on write-hot data}, as the
write-hot flash pages incur the vast majority of writes
(Section~\ref{sec:warm:motivation:hotness}).  We use the insights from this section
to design a write-hotness aware retention management policy for flash memory,
which we describe in Section~\ref{sec:warm:mechanism}.

\subsection{Retention Time Relaxation}
\label{sec:warm:motivation:relaxation}

Traditionally, data stored within a block is retained for some
amount of time.  This retention time is dependent on a number of factors (e.g., the
number of P/E cycles already performed on the block, process variation).  Flash
devices guarantee a minimum data integrity time.
The endurance of flash memory is a factor of how many P/E cycles can take place
before the internal retention time falls below this minimum guarantee. 

Prior work has shown that P/E cycle endurance of flash memory can be
significantly improved by relaxing the internal retention
time~\cite{cai.iccd12, pan.hpca12, mohan.tr12,
liu.fast12, liu.dac13, cai.date12, cai.itj13}. 
We extrapolate the endurance numbers under different
internal retention times and plot them in
Figure~\ref{fig:retention-endurance}.  The horizontal axis shows the flash
endurance, expressed as the number of P/E cycles before the device experiences
retention failures.  Each bar shows the
number of P/E cycles a flash block can tolerate for a given
internal retention time. With a three-year internal retention
time, which is the typical retention period used in today's flash drives, each
flash cell can endure only 3,000 P/E cycles. However, as we relax the internal
retention time to three days, flash endurance can improve by up to
50$\times$ (i.e., 150,000 P/E cycles).  Hence, relaxing the amount of time that
flash memory is required to internally retain data can potentially
lead to great improvements in endurance.

\begin{figure}[h]
\centering
\includegraphics[trim=50 290 685 5,clip,width=0.6\textwidth]
{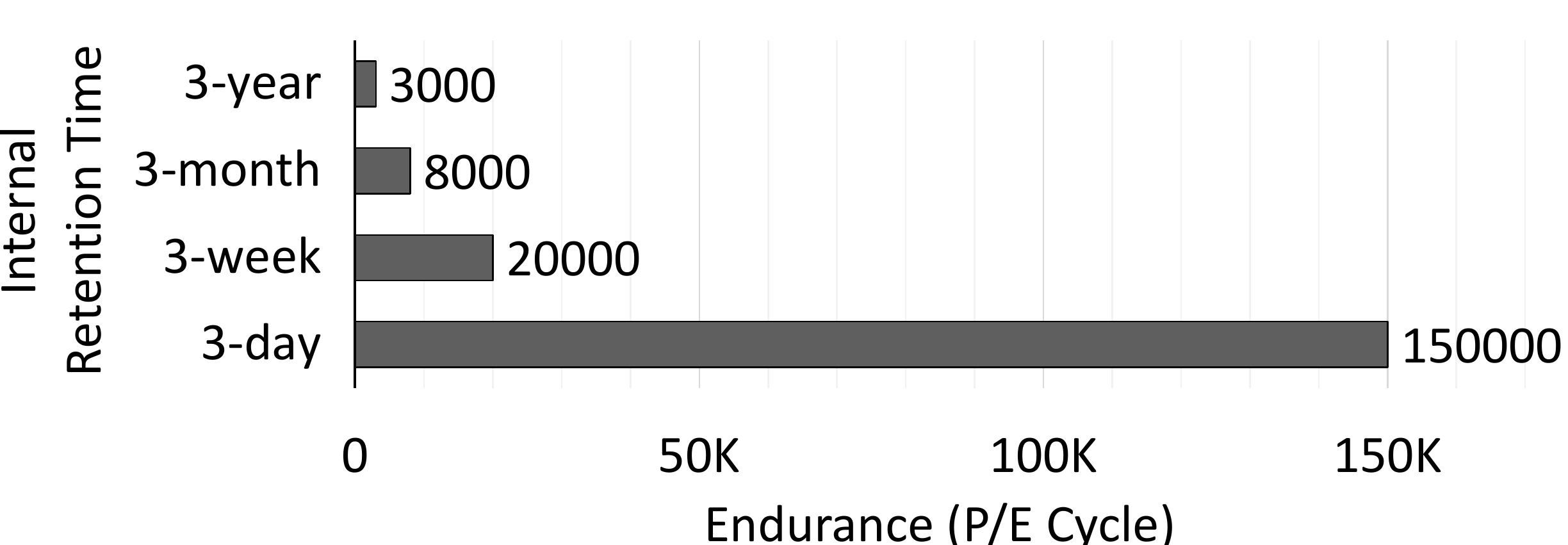}
\caption[P/E cycle endurance from different amounts of internal retention
time without refresh.]
{P/E cycle endurance from different amounts of internal retention
time without refresh.
(Data extrapolated from prior work~\cite{cai.iccd12, cai.itj13}.)}
\label{fig:retention-endurance}
\end{figure}

\subsection{Refresh Overhead Mitigation}
\label{sec:warm:motivation:overhead}

In order to compensate for the reduced internal retention time,
refresh operations are introduced to maintain the data integrity guarantee
provided to the user~\cite{cai.iccd12, cai.itj13}. When the internal
retention time of a flash block expires, the data stored in the block
can be simply remapped to another block (i.e., all
valid pages within a block are read, corrected, and then reprogrammed to a
different block) to extend the duration of data integrity. Several
variants of refresh have been proposed for flash memory~\cite{cai.iccd12,
pan.hpca12, mohan.tr12, liu.fast12, liu.dac13,
cai.itj13}.  Remapping-based flash correct-and-refresh (FCR) involves the
lowest implementation overhead, by triggering refreshes at a fixed refresh
frequency to guarantee the retention time never falls below a predetermined
threshold~\cite{cai.iccd12, cai.itj13}.

Although relaxing internal retention time increases flash endurance,
each refresh operation consumes a portion of this extra endurance, leading to
significantly reduced lifetime improvements.  The curves in
Figure~\ref{fig:consumedPE} plot the relation between the fraction of the extra
endurance cycles consumed by refresh operations for a 256 GB flash drive and the
write intensity of the workload (expressed as the average number of writes the
workload issues to the drive per day) for a refresh mechanism with various
refresh intervals (ranging from three days to three years).  
When the write intensity is as low as
$10^5$ writes/day, refresh operations can consume up to 99\% of the
total endurance when the data is refreshed every three
days (regardless of how recently the data was written). The data points
in Figure~\ref{fig:consumedPE} show the actual fraction of writes that are due to refresh for
each workload that we evaluate in Section~\ref{sec:warm:evaluation}.  Fourteen of the
sixteen workloads are disk traces from real applications, and they all have a
write frequency less than or equal to $10^7$ writes/day. The remaining two
workloads are I/O benchmarks ({\tt iozone} and {\tt postmark}) with higher write
frequencies, which do not represent the typical usage of flash devices.
Unfortunately, refresh operations consume a significant fraction
of the extra endurance for all fourteen real-world workloads.  In
this chapter, we aim to reduce the fraction of endurance consumed as overhead, in order to better utilize the extra
endurance gained from retention time relaxation and thus improve flash lifetime.

\begin{figure}[h]
\centering
\includegraphics[trim=5 5 10 5,clip,width=0.6\textwidth]
{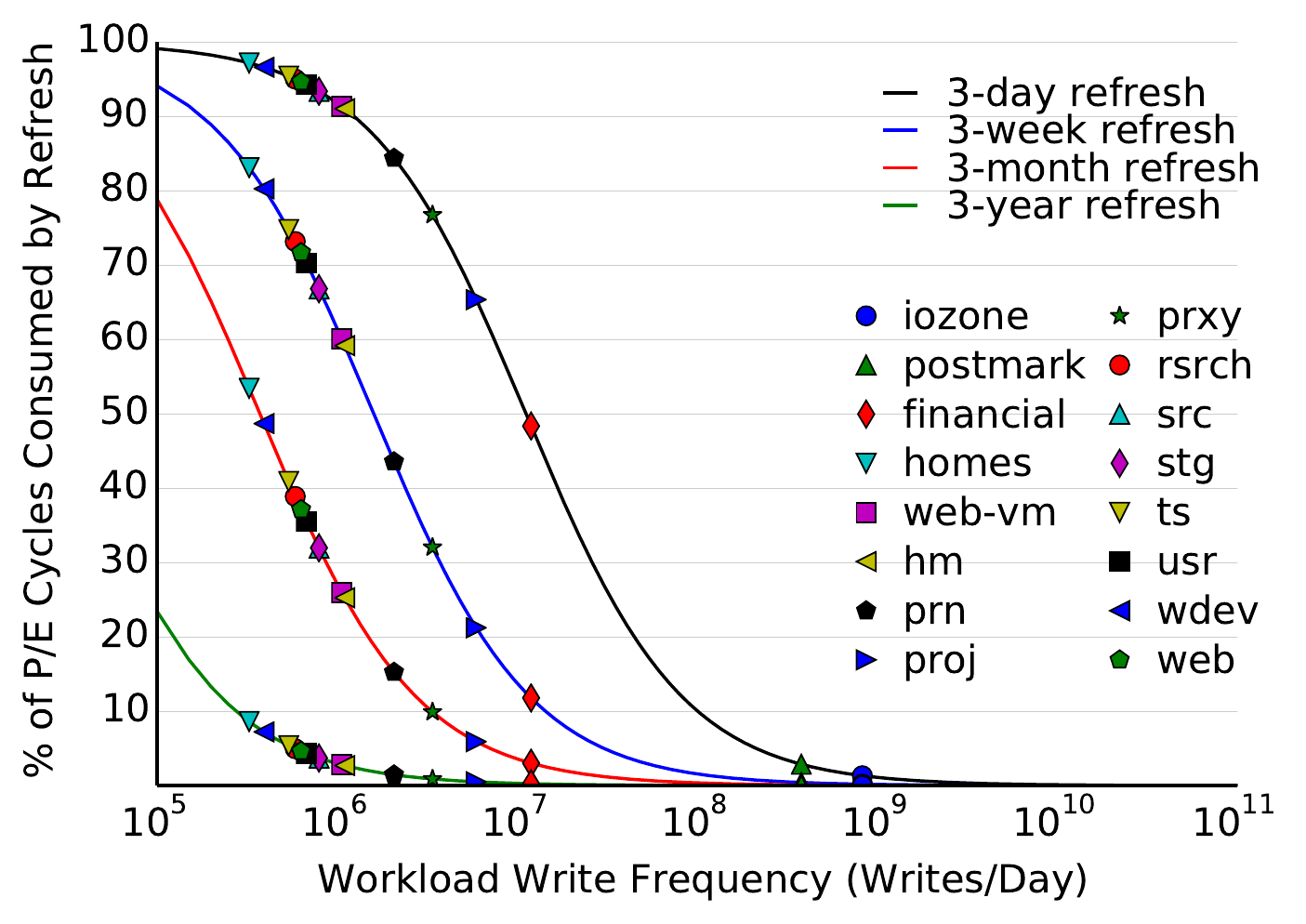}
\caption{Fraction of P/E cycles consumed by refresh operations.}
\label{fig:consumedPE}
\end{figure}

\subsection{Opportunities to Exploit \emph{Write-Hotness}}
\label{sec:warm:motivation:hotness}

Many of the management policies for flash memory are focused on writes,
as write operations reduce the lifetime of the device.  While these
algorithms were designed to evenly distribute device wear-out across blocks,
they crucially ignore the fine-grained behavior of writes across pages within an
application.  We observe that write frequency can be quite heterogeneous across
different pages.  While some pages are often written to (\emph{write-hot
pages}), other pages are infrequently updated (\emph{write-cold pages}).
Figure~\ref{fig:writedist} shows the write distribution for all sixteen of our
applications (described in Table~\ref{tbl:traces}).  We observe that for all but
one of our workloads ({\tt postmark}), only a very small fraction
(i.e., less than 1\%) of the total application data receives the vast majority
of the write requests.  In fact, from Figure~\ref{fig:writedist}, we
observe that for ten of our applications, a very small fraction of all data
pages (i.e., less than 1\%) are the destination of \emph{nearly 100\%} of the
write requests.  Note that our workloads use a total memory footprint of
217.6{\tt GB} each, and that 1\% of the total application data represents
2.176{\tt GB}.  We conclude from this figure that only a small portion of the
pages in each application are \emph{write-hot}, and that the discrepancy between
the write rate to write-hot pages and the write rate to write-cold pages is
highly skewed.  
    
\begin{figure}[h]
\centering
\includegraphics[trim=0 5 0 5,clip,width=\linewidth]
{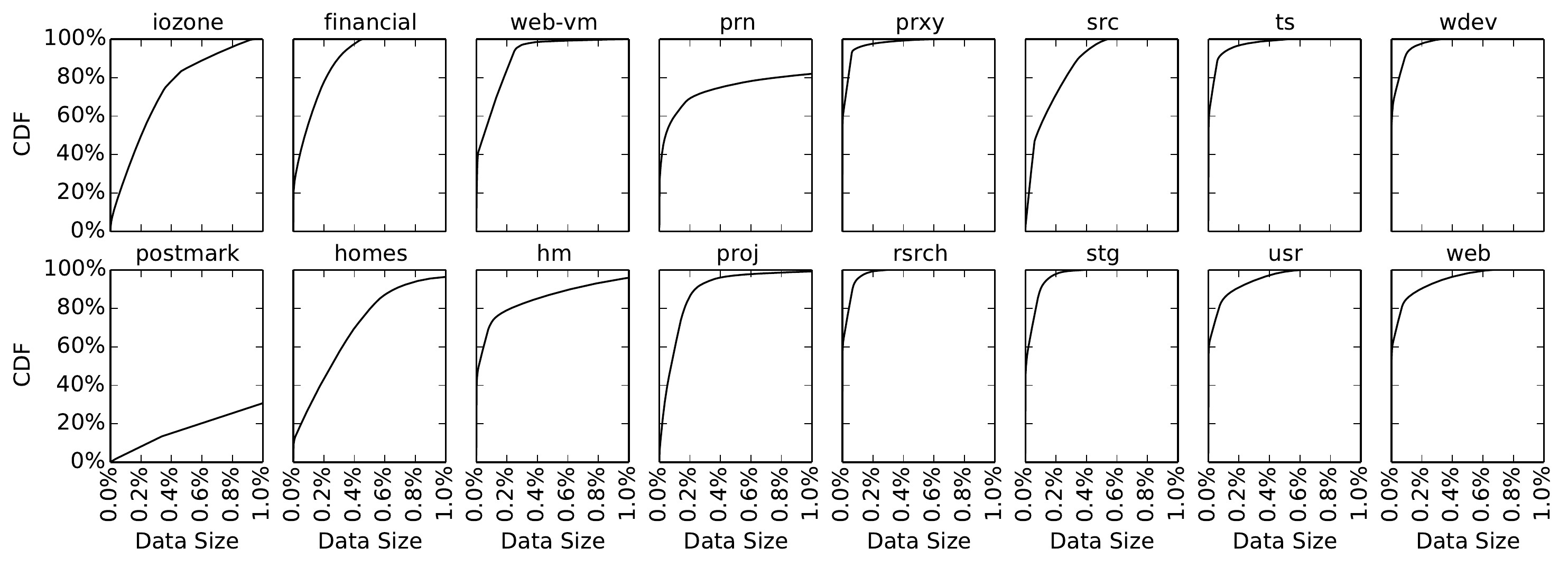}
\caption{Cumulative distribution function of writes to pages for
16 evaluated workload traces.  Total data footprints for our workloads are
217.6{\tt GB}, i.e., 1.0\% on the x-axis represents 2.176{\tt GB} of data.}
\label{fig:writedist}
\end{figure}

In a typical flash device, page allocation policies are oblivious to the
frequency of writes, resulting in blocks that contain a random distribution of
interspersed write-hot and write-cold pages.  We find that such
obliviousness to the program write patterns forces several of our ``general''
flash management algorithms to be inefficient.  One such example is refresh,
where the increased number of program/erase operations greatly limits the
potential endurance gains that can be achieved.  Refreshes are only necessary to
maintain the integrity of data that has not yet been overwritten, as a new write
operation \emph{naturally} refreshes the data by placing the page
into a new block.  In other words, if a page is written to often enough (i.e.,
at a higher frequency than the refresh rate), any refresh operation
to that page will be \emph{redundant}.  

Our goal is to enable such a mechanism that can eliminate refresh operations to
write-hot pages. Unfortunately,
remapping-based refresh operations are performed at a \emph{block
granularity}, which is much coarser than page granularity, as flash devices
only provide a block-granularity erase mechanism.  Therefore, if at least one
page within the block is write-cold, the \emph{whole block} must be refreshed,
foregoing any potential P/E cycle savings from skipping the refresh operation.
As the conventional page allocation algorithm is oblivious to 
write-hotness, there is a very low probability that a block contains \emph{only} write-hot
pages.

Unless we change the page allocation policy, it is impractical to simply modify
the refresh mechanism to skip refreshes for blocks containing only write-hot
pages.  If flash management policies were made aware of the write-hotness of a
page, we could group write-hot pages together in a block such that the entire
block does not
require a refresh operation.  This would allow us to efficiently skip refreshes
to a much larger number of flash blocks that are formed as such.  In addition, since write-cold
pages would be grouped together into blocks, we could also reduce wear-out on
these blocks through more efficient garbage collection.  As write-cold pages
are rarely overwritten, all of the pages within a write-cold block are
more likely to remain valid, requiring much less frequent compaction.  Our goal
for WARM, our proposed management policy, is to \emph{physically separate
write-hot pages from write-cold pages into disjoint sets of flash blocks}, so
that we can significantly improve flash lifetime.


\section{Mechanism}
\label{sec:warm:mechanism}

In this section, we introduce WARM, our proposed write-hotness-aware flash
memory retention management policy.  The first key idea of WARM is to
effectively partition pages stored in flash into two groups based on the
write frequency of the pages. The second key idea of WARM is to
apply different management policies to the two different groups of pages/blocks.
We first discuss a novel, lightweight approach to dynamically identifying and
partitioning write-hot versus write-cold pages (Section~\ref{sec:warm:mechanism:identify}).
We then describe how WARM optimizes flash management policies, such
as garbage collection and wear-leveling, in a partitioned flash memory, and
show how WARM integrates with a refresh mechanism to provide further
flash lifetime improvements (Section~\ref{sec:warm:mechanism:policies}).  We discuss the
hardware overhead required to implement WARM (Section~\ref{sec:warm:mechanism:overhead}).  We show
in Section~\ref{sec:warm:evaluation} that WARM is effective at delivering significant
flash lifetime improvements (by an average of 3.24$\times$ over a conventional
management policy without refresh), and can do so with a minimal performance
overhead (averaging 1.3\%).

\subsection{Partitioning Data Using Write-Hotness}
\label{sec:warm:mechanism:identify}

\subsubsection{Identifying Write-Hot and Write-Cold Data}

\begin{figure}[h]
\centering
\includegraphics[trim=10 365 567 9,clip,width=.9\textwidth]{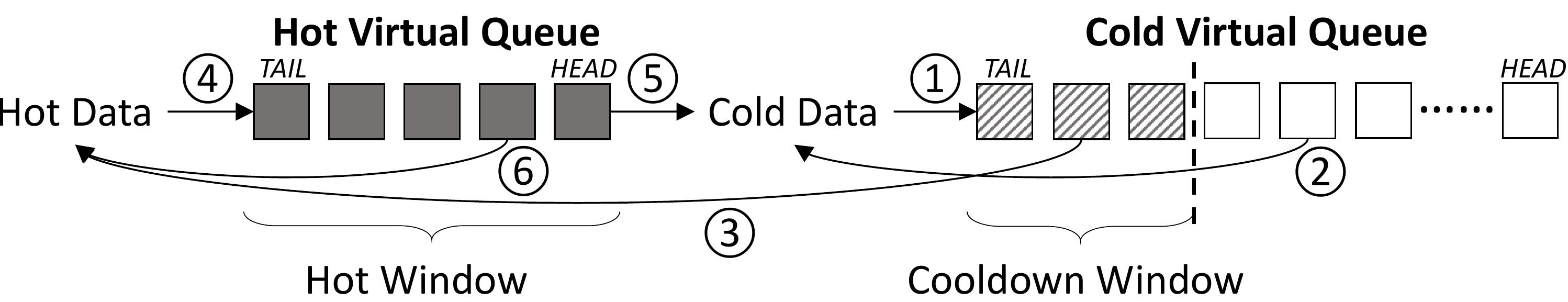}
\caption{Write-hot data identification algorithm using two virtual queues and monitoring windows.}
\label{fig:identify}
\end{figure}

Figure~\ref{fig:identify} illustrates the high-level concept of our
write-hot data identification mechanism.  We maintain two \emph{virtual queues},
one for write-hot data and another for write-cold data, which order all of the
hot and cold data, respectively, by the time of the last write. The
purpose of the virtual queues is to partition write-hot and write-cold data in a
space-efficient way. The partitioning mechanism provides methods of promoting
data from the cold virtual queue to the hot virtual queue, and for demoting data
from the hot virtual queue to the cold virtual queue. The promotion
and demotion decisions are made such that write-hot pages are quickly
identified (after two writes in quick succession to the page), and write-cold
pages are seldom misidentified as write-hot pages (and are quickly demoted if
they are). Note that the cold virtual queue is divided into two parts,
with the part closer to the tail known as the \emph{cooldown window}. The purpose of the cooldown window is to
identify those pages that are most recently written to.  The pages in the
cooldown window are the only ones that can be immediately promoted to the hot virtual queue
(as soon as they receive a write request). We walk through examples for both of these
migration decisions.

Initially, all data is stored in the cold virtual queue.
Any data stored in the cold virtual queue is defined to be \emph{cold}. When
data (which we call Page C) is first identified as cold, a corresponding queue
entry is pushed into the tail of the cold virtual queue
(\circled{1}). This entry progresses forward in the queue as other
cold data is written.  If Page C is written to again after it leaves the cooldown window (\circled{2}), then its queue
entry will be removed from the cold virtual queue and reinserted at the queue
tail (\circled{1}).  This allows the queue to maintain ordering based on the
time of the most recent write to each page.

If a cold page starts to become hot (i.e., it starts being written to
frequently), a \emph{cooldown window} at the tail end of the cold virtual
queue provides these pages with a chance to be promoted into the hot virtual
queue.
The cooldown window monitors the most recently
inserted (i.e., most recently written) cold data.
Let us assume that Page C has just been inserted into the tail of the cold
virtual queue (\circled{1}).  If Page C is written to again while
it is still within the cooldown window, it will be immediately promoted to the
hot virtual queue (\circled{3}). 
If, on the other hand, Page C is not written to
again, then Page C will eventually be pushed out of the cooldown window portion
of the cold virtual queue, at which point Page~C is determined to be
\emph{cold}. Requiring a two-step promotion process from cold to hot
(with the use of a cooldown window) allows us to avoid incorrectly
promoting cold pages due to infrequent writes.  This is important
for two reasons: (1)~hot storage capacity is limited, and (2)~promoted pages
will not be refreshed, which for cold pages could result in data loss.
With our two-step approach, if Page C is cold and is written to
only once, it will remain in the cold queue, though it
will be moved into the cooldown window (\circled{2}) to be monitored for
subsequent write activity.

Any data stored in the hot virtual queue is identified as \emph{hot}.
Newly-identified hot data, which we call Page H, is inserted
into the tail of the hot virtual queue (\circled{4}). The hot virtual queue
length is maximally bounded by a \emph{hot window} size to ensure that the most
recent writes to all hot data pages were performed within a given time period.
(We discuss how this window is sized in Section~\ref{sec:warm:mechanism:identify}.)  The
assumption here is that infrequently-written pages in the hot
virtual queue will eventually progress to the head of the queue (\circled{5}).
If the entry for Page H in the hot virtual queue reaches the
head of the queue and must now be evicted, we demote Page H into the
cooldown window of the cold virtual queue (\circled{1}), and move the
page out of the hot virtual queue.
In contrast, a write to a page in the hot
virtual queue simply moves that page to the tail of the hot virtual queue
(\circled{6}).

\subsubsection{Partitioning the Flash Device}

Figure~\ref{fig:wharr} shows how we apply the identification mechanism from Section~\ref{sec:warm:mechanism:identify} to
perform physical page partitioning inside flash, with labels that correspond to the
actions from Figure~\ref{fig:identify}.  We first separate all of the
flash blocks into two \emph{allocation pools}, one for hot data and another for
cold data.  The \emph{hot pool} contains enough blocks to store every page in the
hot virtual queue (whose sizing is described in Section~\ref{sec:warm:mechanism:identify}), as
well as some extra blocks to tolerate management overhead (e.g., erasing on
garbage collection).  The \emph{cold pool} contains all of the remaining flash
blocks.  Note that blocks can be moved between the two pools when the
queues are resized.

\begin{figure}[h]
\centering
\includegraphics[trim=10 360 890 10,clip,width=0.7\textwidth]{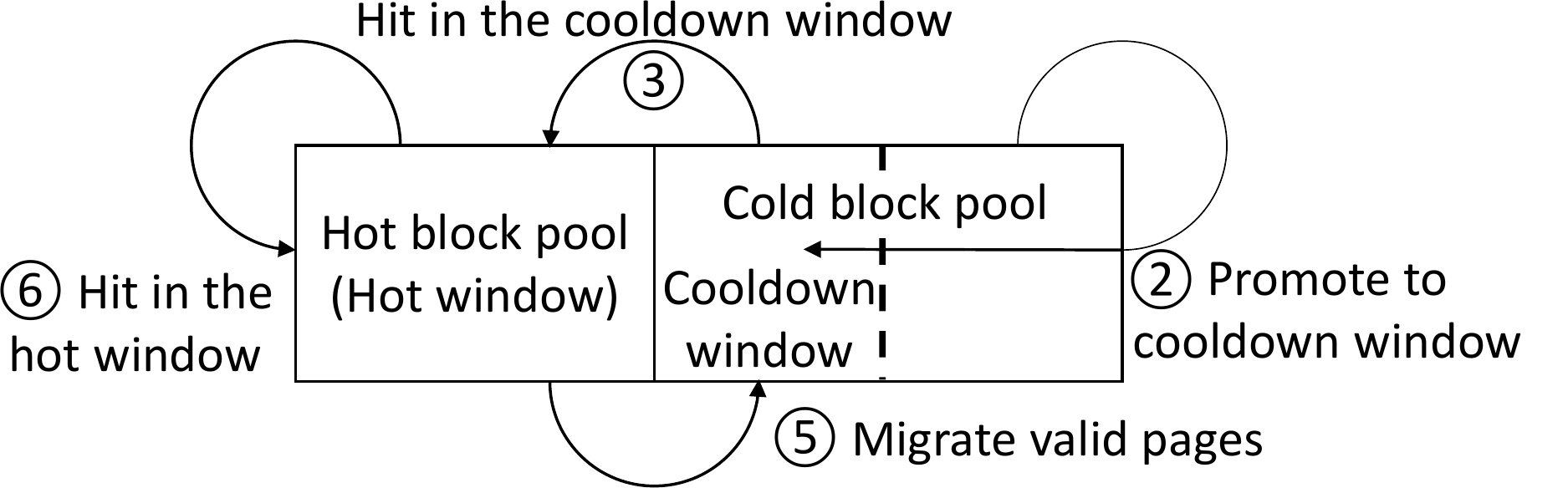}
\caption{Write-hotness aware retention management policy overview.}
\label{fig:wharr}
\end{figure}

To simplify the hardware required to implement the virtual queues, we
exploit the fact that pages are written sequentially into the hot pool
blocks. Consecutive writes to hot pages will be placed in the same
block, which means that a single block in the hot virtual queue will hold \emph{all} of the 
oldest pages.  As a result, we can track the hot virtual queue at a
block granularity instead of a page granularity, which allows us to
significantly reduce the size of the hot virtual queue.

\subsubsection{Tuning the Partition Boundary}

Since the division between hot and cold data can be dependent on both
application and phase characteristics, we need to provide a method for
dynamically adjusting the size of our hot and cold pools periodically.  
Every block is allocated to one of the two pools, so any increase in the
hot pool size will always be paired with a corresponding decrease in the cold
pool size, and vice versa.  Our dynamic sizing mechanism must ensure that: (1)
the hot pool size is such that every page in the hot pool will be written to
more frequently than the hot pool retention time (which is relaxed as the hot pool
does not employ refresh), and (2) the lifetime
of the blocks in the cold pool is maximized.  To this end, we describe an
algorithm that tunes the partitioning of blocks between the hot and cold
pools.

The partitioning algorithm
starts by setting an upper bound for the hot window, to ensure that every page in
the window will be written to at a greater rate than the fixed hot pool retention
time.  Recall that the hot pool retention time is relaxed to 
provide greater endurance (Section~\ref{sec:warm:motivation:relaxation}).  We
estimate this size by collecting the number of writes to the hot pool, to find
the average write frequency and estimate the time it takes to fill the hot
window.  We compare the time to fill the window to the hot pool retention
time, and if the fill time exceeds the retention time, we shrink the hot pool
size to reduce the required fill time. This hot pool size determines
the initial partition boundary between the hot pool and the cold pool.

We then tune this partition boundary to maximize the lifetime of the cold pool, since we do
not relax retention time for the blocks in the cold pool.  Assuming that wear-leveling
evenly distributes the page writes within the cold pool, we can use the
\emph{endurance capacity} metric (i.e., the total number of writes the
cold pool can service),
which is the product of the remaining endurance of a block\footnote{Due 
to wear-leveling, the remaining endurance (i.e., the number of P/E operations
that can still be performed on the block) is the same across all of the blocks.}
and the cold pool size,
to estimate the lifetime of blocks in the cold pool:
\begin{align}
  Endurance\, Capacity = Remaining\, Endurance \times Cold\, Pool\, Size \\
  Lifetime = \frac{Endurance\, Capacity}
                  {Cold\, Write\, Frequency}
          \propto \frac{Cold\, Pool\, Size}
                       {Cold\, Write\, Frequency}
\end{align}

We divide the \emph{endurance capacity} by the cold write frequency (writes per day) to determine the
number of days remaining before the cold pool is worn out.
We use hill climbing to find the partition boundary at which the cold
pool size maximizes the flash lifetime.
The cold write frequency is dependent on cold pool size, because as the cold
pool size increases, the hot pool size correspondingly shrinks, shifting writes of higher
frequency into the cold pool.

Finally, once the partition boundary converges to obtain the maximum lifetime, we must
adjust what portion of the cold pool belongs in the cooldown window.  We
size this window to minimize the ping-ponging of requests between the hot and
cold pools.  For this, we want to maximize the number of hot virtual queue hits
(\circled{6} in Figure~\ref{fig:identify}), while minimizing the number of
requests evicted from the hot window (\circled{5} in Figure~\ref{fig:identify}).
We maintain a counter of each of these events, and then use hill climbing on the
cooldown window size to maximize the utility function $Utility =
(\circled{6} - \circled{5})$.

In our work, we limit the hot pool size to the number of
over-provisioned blocks within the flash device (i.e., the extra blocks
beyond the visible capacity of the device). While the hot pages
are expected to represent only a small portion of the total flash capacity
(see Section~\ref{sec:warm:motivation:hotness}), there may be rare cases where the size
limit prevents the hot pool from holding all of the hot data (i.e., the hot 
pool is significantly undersized).  In such a case, some less-hot pages are
forced to reside in the cold pool, and lose the benefits of WARM (i.e.,
endurance improvements from relaxed retention times).  WARM will not, however,
incur any further write overhead from keeping the less-hot pages in the cold
pool.  For example, the dynamic sizing of the cooldown window prevents the
less-hot pages from going back and forth between the hot and cold pools.

\subsection{Flash Management Policies}
\label{sec:warm:mechanism:policies}

WARM partitions all of the blocks in a flash device into two pools, storing
write-hot data in the blocks belonging to the \emph{hot pool}, and storing
write-cold data in the blocks belonging to the \emph{cold pool}.
Because of the different degrees of write-hotness of the data in each pool,
WARM also applies different management policies (i.e., refresh, garbage
collection, and wear-leveling) to each pool, to best extend their lifetime. We next
describe these management policies for each pool, both when WARM is applied alone
and when WARM is applied along with refresh.

\subsubsection{WARM-Only Management}

WARM relaxes the internal retention time of only the blocks in the hot pool,
without requiring a refresh mechanism for the hot pool.
Within the cold pool, WARM applies conventional garbage
collection (i.e., finding the block with the fewest valid pages to minimize
unnecessary data movement) and wear-leveling policies.  Since the flash blocks
in the cold pool contain data with much lower write frequencies, they
(1)~consume a smaller number of P/E cycles, and (2)~experience much lower
fragmentation (which only occurs when a page is updated), thus reducing garbage
collection activities.  As such, the lifetime of blocks in the cold pool
increases even when conventional management policies are applied.

Within the hot pool, WARM applies simple, in-order garbage collection (i.e.,
finding the oldest block) and no wear-leveling policies.  WARM performs writes
to hot pool blocks in \emph{block order} (i.e., it starts on the block with the
lowest ID number, and then advances to the block with the next lowest ID
number) to maintain a sequential ordering by write time.  Writing pages in block order enables garbage
collection in the hot pool to also be performed in block order.  Due to the
higher write frequency in the hot pool, all data in the hot pool is valid for a shorter amount
of time.  Most of the pages in the oldest block are already invalid when
the block is garbage collected, increasing garbage collection efficiency.
Since both writing and garbage collection are performed in block order, each of
the blocks will be \emph{naturally} wear-leveled, as they will all incur the
same number of P/E cycles.  Thus, we do not need to apply any additional
wear-leveling policy.

\subsubsection{Combining WARM with Refresh}
\label{sec:warm:mechanism:policies:refresh}

WARM can also be used in conjunction with a refresh mechanism to reap
additional endurance benefits.  WARM, on its own, can significantly extend the
lifetime of a flash device by enabling retention time relaxation on only the
write-hot pages.  However, these benefits are limited, as the cold pool blocks
will eventually exhaust their endurance at the original internal retention time.
(Recall from Figure~\ref{fig:retention-endurance} that endurance decreases
significantly as the selected internal retention time increases.)
While WARM cannot enable retention time relaxation on the cold pool blocks due to infrequent writes to such blocks, a
refresh mechanism can enable the relaxation, greatly extending the endurance of
the cold pool blocks.  WARM still provides benefits over a refresh mechanism
for the hot pool blocks, since it avoids unnecessary write operations that
refresh operations would incur.

When WARM and refresh are combined, we split the lifetime of the flash device
into two phases.  The flash device starts in the \emph{pre-refresh phase},
during which the same management policies as WARM-only are applied. Note that
during this phase, internal retention time is only relaxed for the hot pool
blocks.  Once the endurance \emph{at the original retention time} is exhausted,
we enter the \emph{refresh phase}, during which the same management policies as
WARM-only are applied \emph{and} a refresh policy (such as FCR~\cite{cai.iccd12}) is
applied to the cold pool to avoid data loss. During this phase, the retention
time is relaxed for all blocks.  Note that during both phases, the internal
retention time for hot pool blocks is \emph{always} relaxed \emph{without}
the need for a refresh policy.

During the refresh phase, WARM also performs global wear-leveling to prevent the
hot pool from being prematurely worn out. The global wear-leveling policy
rotates the \emph{entire} hot pool to a new set of physical flash blocks
(which were previously part of the cold pool) every 1\texttt{K} hot
block P/E cycles. Over time, this rotation will use all of the flash blocks in the device 
for the hot pool for one 1\texttt{K} P/E cycle interval.  Thus, WARM
wears out all of the flash blocks equally despite the
heterogeneity in write-frequency between the two pools.

\subsection{Implementation and Overheads}
\label{sec:warm:mechanism:overhead}

The logic overhead of the proposed mechanism is minimal. Thanks to the
simplicity of the write-hot data identification algorithm, WARM can be
integrated within an existing FTL, allowing it to be implemented in
the flash controller
that already exists in modern flash drives. 

For our dynamic window tuning mechanism, four 32-bit counters are
required.  Two counters track the number of writes to the hot and cold pools.  
A third counter tracks the number of hot virtual queue write
hits (\circled{6} in Figure~\ref{fig:identify}).  The fourth counter tracks the
number of pages moved from the hot virtual queue into the cooldown window
(\circled{5} in Figure~\ref{fig:identify}).

The memory and storage overheads for the proposed mechanism are small.
Recall that the cooldown window can remain relatively small.  We
need to store data that tracks which blocks belong to the cooldown window, and
which blocks belong to the hot pool. From our evaluation, we find that a
128-block maximum cooldown window size is sufficient.  This
requires us to store the block ID of 128~blocks, for a storage overhead of
128$\times$8{\tt B}=1{\tt KB}.  As the blocks belonging to the hot pool are
written to in order of block ID, we require even lower overhead for them.  We allocate
a contiguous series of block IDs to the hot pool, reducing the tracking overhead
to four registers totaling 32{\tt B}: the starting ID of the series, the current
size of the pool, a pointer to the most recently written block (i.e., the block
at the tail of the hot virtual queue), and a pointer to the oldest block yet to
be erased (i.e., the block at the head of the hot virtual queue).  All this
information can be buffered inside the memory of the flash controller in order
to accelerate write operations.

While WARM saves a significant amount of unnecessary refreshes in the hot data pool, the
proposed mechanism has the potential to indirectly generate extra write operations that
consume some endurance cycles. First, WARM generates extra write
operations when demoting a hot page to the cold data pool (\circled{5}). Second,
partitioning flash blocks into two allocation pools can sometimes
increase garbage collection activities. This is because one of the
pools may have a smaller number of blocks available in the free list, requiring
more frequent invocation of garbage collection. All of these
overheads are accounted for in our evaluation in Section~\ref{sec:warm:evaluation}, and our
results factor in all additional writes.  As we show in
Section~\ref{sec:warm:evaluation}, WARM is designed to minimize these overheads such
that lifetime improvements are not overshadowed, and the resulting impact on
response time is minimal.


\section{Methodology}
\label{sec:warm:methodology}

We use DiskSim-4.0~\cite{bucy.pdl08} with SSD
extensions~\cite{agrawal.atc08} to evaluate WARM\@.
Table~\ref{tbl:sim-config} lists the parameters of our simulated NAND
flash-based SSD\@. The latencies (the first four rows of the table) are from real NAND flash
chip measurements~\cite{heidecker.jpl10}. The sizes (rows 5--8) represent a
modern commercial NAND flash specification~\cite{abraham.fms12}. Flash
endurance and refresh period are measured from real NAND flash
devices~\cite{cai.iccd12, cai.itj13}.

\begin{table}[h]
\small
\centering
\caption{Parameters of the simulated flash-based SSD.}
\label{tbl:sim-config}
\begin{tabular}{lc}
  \toprule
  \textbf{Parameter} & \textbf{Value} \\
  \midrule
  Page read to register latency & 25$\mu$s \\
  Page write from register latency & 200$\mu$s \\
  Block erase latency & 1.5ms \\
  Data bus latency & 50$\mu$s \\
  \midrule
  Page/block size & 8{\tt KB}/1{\tt MB} \\
  Die/package size & 8{\tt GB}/64{\tt GB} \\
  Total storage capacity (incl\. over-provisioning) & 256{\tt GB} \\
  Over-provisioning & 15\% \\
  \midrule
  Endurance for 3-year retention time (P/E cycles) & 3,000 \\
  Endurance for 3-day retention time (P/E cycles) & 150,000 \\
  \bottomrule
\end{tabular}
\end{table}

We run each simulation with I/O traces collected from a wide range of real
workloads with different use cases~\cite{umass.storagetraces, koller.tos10, narayanan.tos08}.
We also select two popular synthetic file system benchmarks to stress our
mechanism with higher write rate applications~\cite{norcott.iozone03, katcher.tr97}.
Table~\ref{tbl:traces} lists the name,
source, length, and description of each trace. To compute the lifetime of each
configuration, we assume the trace is repeated until the flash drive fails. We
fill all the usable space of the flash drive with data, to mimic worst-case usage
conditions and to trigger garbage collection activities within the trace duration.
Similar to the approach employed in prior work~\cite{cai.date12, cai.iccd12,
cai.itj13}, the overall flash lifetime is derived using the average write frequency of one run,
which consists of writes generated by the trace and by garbage collection, as well as by
refresh operations during the refresh phase.
We use this methodology since it is
impossible to simulate multi-year-long traces that drain the flash lifetime.

\begin{table}[h]
\centering
\small
\caption{Source and description of simulated traces.}
\label{tbl:traces}
\begin{tabular}{lccc}
  \toprule
  \textbf{Trace} & \textbf{Source} & \textbf{Length} & \textbf{Workload Description} \\
  \midrule
  \multicolumn{4}{c}{\emph{Synthetic Workloads}} \\
  iozone & IOzone~\cite{norcott.iozone03} & 16 min & File system benchmark \\
  postmark & Postmark~\cite{katcher.tr97} & 8.3 min & File system benchmark \\
  \midrule
  \multicolumn{4}{c}{\emph{Real-World Workloads}} \\
  financial & UMass~\cite{umass.storagetraces} & 1 day & Online transaction processing \\
   homes & FIU~\cite{koller.tos10} & 21 days & Research group activities \\
   web-vm & FIU~\cite{koller.tos10} & 21 days & Web mail proxy server \\
   hm    & MSR~\cite{narayanan.tos08} & 7 days & Hardware monitoring \\
   prn   & MSR~\cite{narayanan.tos08} & 7 days & Print server \\
   proj  & MSR~\cite{narayanan.tos08} & 7 days & Project directories \\
   prxy  & MSR~\cite{narayanan.tos08} & 7 days & Firewall/web proxy \\
   rsrch & MSR~\cite{narayanan.tos08} & 7 days & Research projects \\
   src   & MSR~\cite{narayanan.tos08} & 7 days & Source control \\
   stg   & MSR~\cite{narayanan.tos08} & 7 days & Web staging \\
   ts    & MSR~\cite{narayanan.tos08} & 7 days & Terminal server \\
   usr   & MSR~\cite{narayanan.tos08} & 7 days & User home directories \\
   wdev  & MSR~\cite{narayanan.tos08} & 7 days & Test web server \\
   web   & MSR~\cite{narayanan.tos08} & 7 days & Web/SQL server \\
   \bottomrule
 \end{tabular}
\end{table}

\section{Evaluations}
\label{sec:warm:evaluation}

In this section, we evaluate and compare \emph{six} configurations:

\begin{itemize}

\item {\tt Baseline}
does not include WARM or refresh, and uses conventional 
garbage collection and wear-leveling
policies, as described in Section~\ref{sec:warm:motivation}.

\item {\tt WARM} uses the proposed write-hotness aware
retention management policy that we described in Section~\ref{sec:warm:mechanism}. 

\item {\tt FCR}
adds a remapping-based refresh mechanism to {\tt Baseline}.  Our refresh 
mechanism is similar to the remapping-based FCR
described in prior work~\cite{cai.iccd12, cai.itj13}, but refresh is \emph{not} performed in the
pre-refresh phase (see Section~\ref{sec:warm:mechanism:policies:refresh}) to reduce unnecessary overhead.  During the refresh phase (see Section~\ref{sec:warm:mechanism:policies:refresh}), {\tt FCR} refreshes
all valid blocks every three days, which yields the best endurance
improvement.

\item {\tt WARM+FCR} uses
write-hotness aware retention management alongside 3-day refresh
(Section~\ref{sec:warm:mechanism:policies}) to achieve maximum lifetime. 

\item {\tt ARFCR} adds the
ability to progressively increase refresh frequency on top of the remapping-based
refresh mechanism (similar to adaptive-rate FCR~\cite{cai.iccd12, cai.itj13}).  The 
refresh frequency increases as the retention capabilities of the flash memory decrease, in
order to minimize the overhead of write-hotness-oblivious refresh.

\item {\tt WARM+ARFCR} adds WARM
alongside the adaptive-rate refresh mechanism. 
\end{itemize}

To provide insights into our results, we first show the hot pool sizes and the
cooldown window sizes as determined by WARM for each of the configurations
(Section~\ref{sec:warm:evaluation:sizes}).
We then use four metrics to show the benefits and costs associated with our mechanism:

\begin{itemize}

\item We evaluate all configurations in terms of \emph{overall lifetime}
(Section~\ref{sec:warm:evaluation:lifetime}).

\item We evaluate the gain in {\em endurance capacity}, the aggregate number
of write requests that the flash device can endure \emph{across all pages}, for
{\tt WARM} with respect to {\tt Baseline} (Section~\ref{sec:warm:evaluation:endurance}).
We use this metric as an indicator of how many additional writes we can sustain
to the flash device with our mechanism.

\item We evaluate and break down the \emph{total number of writes} consumed by
{\tt FCR} and {\tt WARM+FCR} during the refresh phase, to demonstrate how our
mechanism reduces the write overhead of retention time relaxation
(Section~\ref{sec:warm:evaluation:refresh}).

\item We evaluate the \emph{average response time}, the mean latency for the
flash device to service a host request, for both {\tt Baseline} and {\tt WARM}
to demonstrate the performance overhead of using WARM
(Section~\ref{sec:warm:evaluation:performance}).

\end{itemize}

Finally, we show sensitivity studies on flash memory over-provisioning and the
refresh rate (Section~\ref{sec:warm:evaluation:sensitivity}).

\subsection{Hot Pool and Cooldown Window Sizes}
\label{sec:warm:evaluation:sizes}

Table~\ref{tbl:hot-sizes} lists the hot pool and the cooldown window sizes
learned by WARM for each of our WARM-based configurations.
To allow WARM
to quickly adapt to different workload behaviors, we set the smallest step size
by which the hot pool size can change to 2\% of the total flash drive capacity, and we restrict the
cooldown window sizes to power-of-two block counts. For {\tt WARM+ARFCR},
as the refresh frequency of the flash cells increases
(going to the right in
Table~\ref{tbl:hot-sizes}), the hot pool size generally reduces.
This is because WARM automatically selects a smaller hot pool size to
ensure that the data in the hot pool has a high enough write intensity to skip
refreshes.  Naturally, as the internal retention time of a cell decreases, previously 
write-hot pages with a write rate slower than the new retention time no longer
qualify as hot, thereby reducing the number of pages that need to be maintained
in the hot pool.  
WARM adaptively selects different 
hot pool sizes based on the fraction of write-hot data in each particular
workload.  Similarly, WARM intelligently selects the best cooldown window size for each 
workload, such that it minimizes the number of cold pages that are
misidentified as hot and considered for promotion to the hot pool.
As such, our analysis indicates that WARM can
intelligently and adaptively adjust the hot pool size and the cooldown window size to achieve maximum
lifetime.

\begin{table}[h]
\centering
\small
\caption{Hot pool and cooldown window sizes as set dynamically by WARM\@. H\%: Hot pool size as a percentage of total flash drive capacity. CW\@: Cooldown window size in number of blocks.}
\label{tbl:hot-sizes}
\begin{tabular}{m{1.6cm} >{\centering}m{0.6cm}<{\centering}>{\centering}m{0.6cm}<{\centering} >{\centering}m{0.6cm}<{\centering}>{\centering}m{0.6cm}<{\centering} >{\centering}m{0.6cm}<{\centering}>{\centering}m{0.6cm}<{\centering} >{\centering}m{0.6cm}<{\centering}>{\centering}m{0.6cm}<{\centering} >{\centering}m{0.6cm}<{\centering}>{\centering}m{0.6cm}<{\centering}}
  \toprule
  \multirow{3}{*}{\textbf{Trace}} & \multicolumn{2}{c}{\multirow{2}{*}{\textbf{WARM}}} & \multicolumn{2}{c}{\textbf{WARM}} & \multicolumn{6}{c}{\textbf{WARM+ARFCR}} \tabularnewline%
  & & & \multicolumn{2}{c}{\textbf{+FCR}} & \multicolumn{2}{c}{\textbf{3-month}} & \multicolumn{2}{c}{\textbf{3-week}} & \multicolumn{2}{c}{\textbf{3-day}} \tabularnewline%
  & \textbf{H\%} & \textbf{CW} & \textbf{H\%} & \textbf{CW} & \textbf{H\%} & \textbf{CW} & \textbf{H\%} & \textbf{CW} & \textbf{H\%} & \textbf{CW} \tabularnewline%
  \midrule
    iozone & 10 & 8 & 10 & 8 & 10 & 8 & 10 & 8 & 10 & 8 \tabularnewline%
    postmark & 2 & 128 & 4 & 128 & 4 & 128 & 4 & 128 & 4 & 128 \tabularnewline%
    financial & 10 & 4 & 10 & 4 & 10 & 4 & 10 & 4 & 10 & 4 \tabularnewline%
    homes & 10 & 128 & 4 & 32 & 10 & 4 & 10 & 4 & 4 & 32 \tabularnewline%
    web-vm & 10 & 128 & 10 & 32 & 10 & 4 & 10 & 4 & 10 & 32 \tabularnewline%
    hm & 10 & 128 & 10 & 128 & 10 & 32 & 10 & 32 & 10 & 128 \tabularnewline%
    prn & 10 & 128 & 10 & 128 & 8 & 4 & 10 & 128 & 10 & 128 \tabularnewline%
    proj & 10 & 4 & 10 & 4 & 10 & 4 & 10 & 4 & 10 & 4 \tabularnewline%
    prxy & 10 & 4 & 10 & 4 & 10 & 4 & 10 & 4 & 10 & 4 \tabularnewline%
    rsrch & 6 & 128 & 6 & 128 & 10 & 4 & 10 & 4 & 6 & 128 \tabularnewline%
    src & 10 & 128 & 8 & 128 & 10 & 32 & 10 & 32 & 8 & 128 \tabularnewline%
    stg & 10 & 128 & 8 & 128 & 10 & 4 & 10 & 4 & 8 & 128 \tabularnewline%
    ts & 10 & 128 & 6 & 128 & 10 & 4 & 10 & 4 & 6 & 128 \tabularnewline%
    usr & 6 & 128 & 6 & 128 & 10 & 4 & 10 & 4 & 6 & 128 \tabularnewline%
    wdev & 6 & 128 & 4 & 128 & 10 & 128 & 10 & 128 & 4 & 128 \tabularnewline%
    web & 6 & 128 & 6 & 128 & 10 & 4 & 10 & 4 & 6 & 128 \tabularnewline%
  \bottomrule
\end{tabular}
\end{table}

\subsection{Lifetime Improvement}
\label{sec:warm:evaluation:lifetime}

Figure~\ref{fig:lifetime} shows the lifetime in days (using a logarithmic scale)
for all six of our evaluated configurations. 
Figure~\ref{fig:lifetime_improvement:warm} shows the lifetime 
improvement of {\tt WARM} when normalized to the lifetime of {\tt Baseline}.
The mean lifetime improvement for {\tt
WARM} across all of our workloads is 3.24$\times$ over {\tt Baseline}. 
In addition, {\tt WARM+FCR}
improves the mean lifetime over {\tt FCR} alone by 1.30$\times$ 
(Figure~\ref{fig:lifetime_improvement:warm_fcr}), leading to a mean
improvement for combined WARM and FCR over {\tt Baseline} of 10.4$\times$ (as opposed to 8.0$\times$ with
{\tt FCR} alone). {\tt WARM+ARFCR} improves the mean lifetime over {\tt ARFCR} by
1.21$\times$ (Figure~\ref{fig:lifetime_improvement:warm_arfcr}),
leading to a mean improvement for combined WARM and ARFCR over {\tt Baseline} of 12.9$\times$
(as opposed to 10.7$\times$ with {\tt ARFCR} alone). Even for our worst performing
workload, {\tt postmark}, in which the amount of hot data and the fraction of
writes due to refresh are very low (as discussed in Sections~\ref{sec:warm:evaluation:endurance}
and~\ref{sec:warm:evaluation:refresh}), the overall lifetime improves by 8\% when {\tt WARM}
is applied without refresh, and remains unaffected with respect to {\tt FCR}
when {\tt WARM+FCR} is applied.  We conclude that WARM can adjust to workload
behavior and effectively improve overall flash lifetime, either when used on its own or
when used together with a refresh mechanism, without adverse impacts.

\begin{figure}[h]
\centering
\includegraphics[trim=15 160 0 0,clip,width=\linewidth]{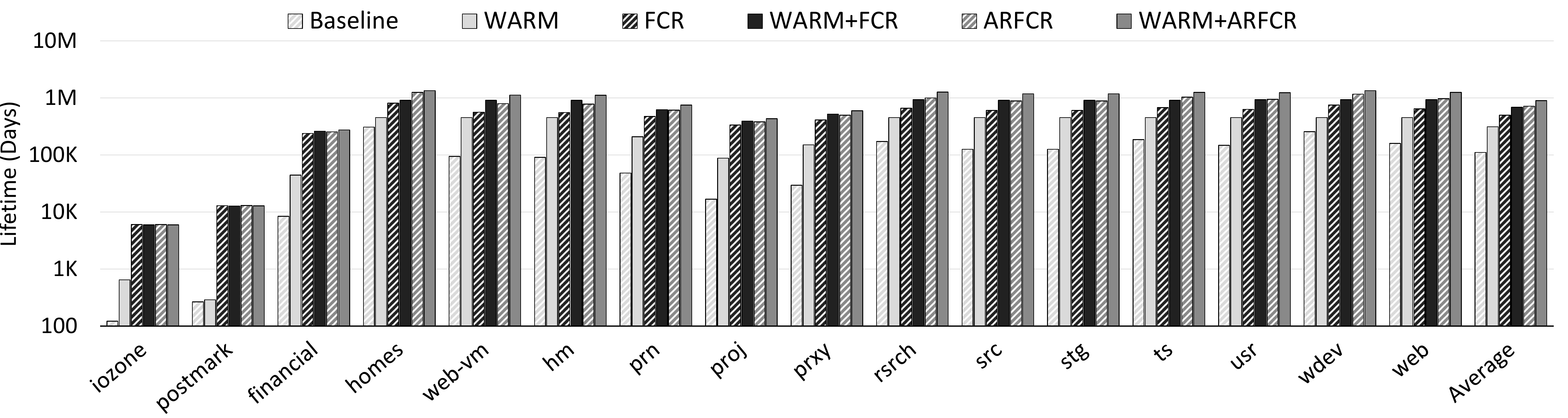}
\caption{Absolute flash memory lifetime for {\tt Baseline}, {\tt WARM}, {\tt
FCR}, {\tt WARM+FCR}, {\tt ARFCR}, and {\tt WARM+ARFCR} configurations.  Note that the y-axis uses a log scale.}
\label{fig:lifetime}
\end{figure}
\FloatBarrier

\begin{figure}[h]
\centering
\begin{subfigure}[h]{0.48\linewidth}
  \includegraphics[trim=10 200 800 0,clip,width=\linewidth]{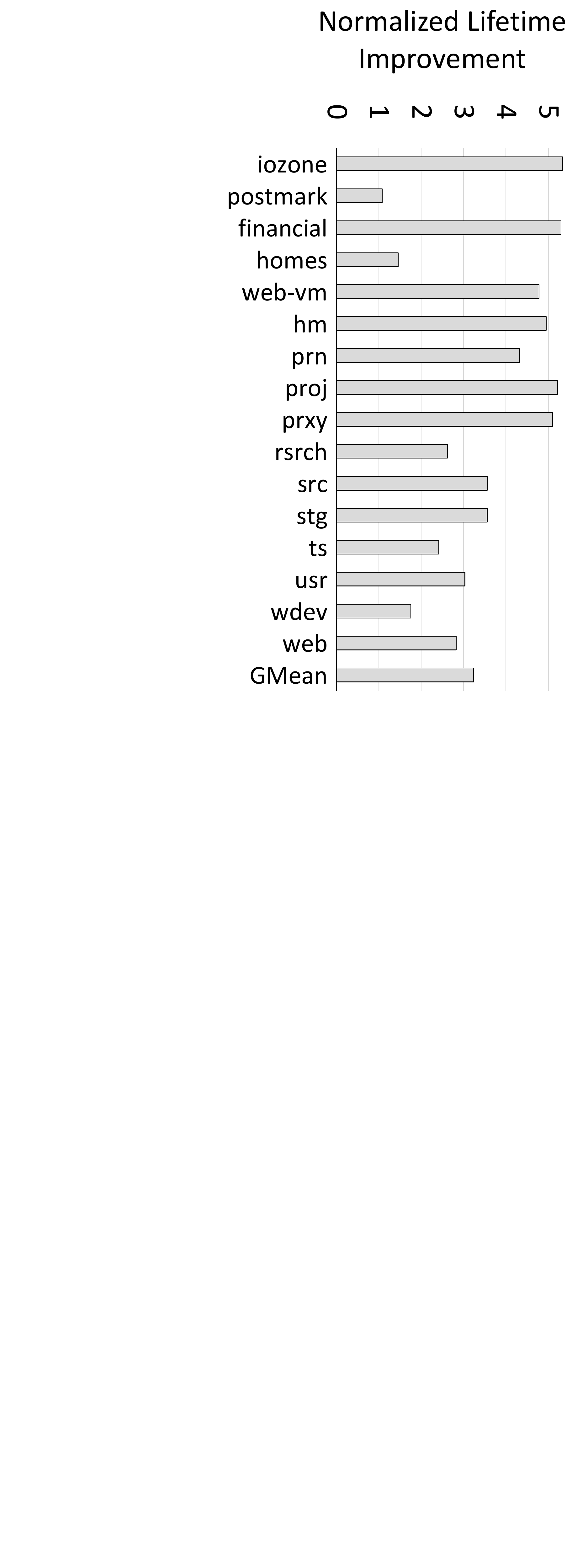}
  \caption{{\tt WARM} over {\tt Baseline}}
\label{fig:lifetime_improvement:warm}
\end{subfigure}%
\begin{subfigure}[h]{0.48\linewidth}
  \includegraphics[trim=10 200 800 0,clip,width=\linewidth]{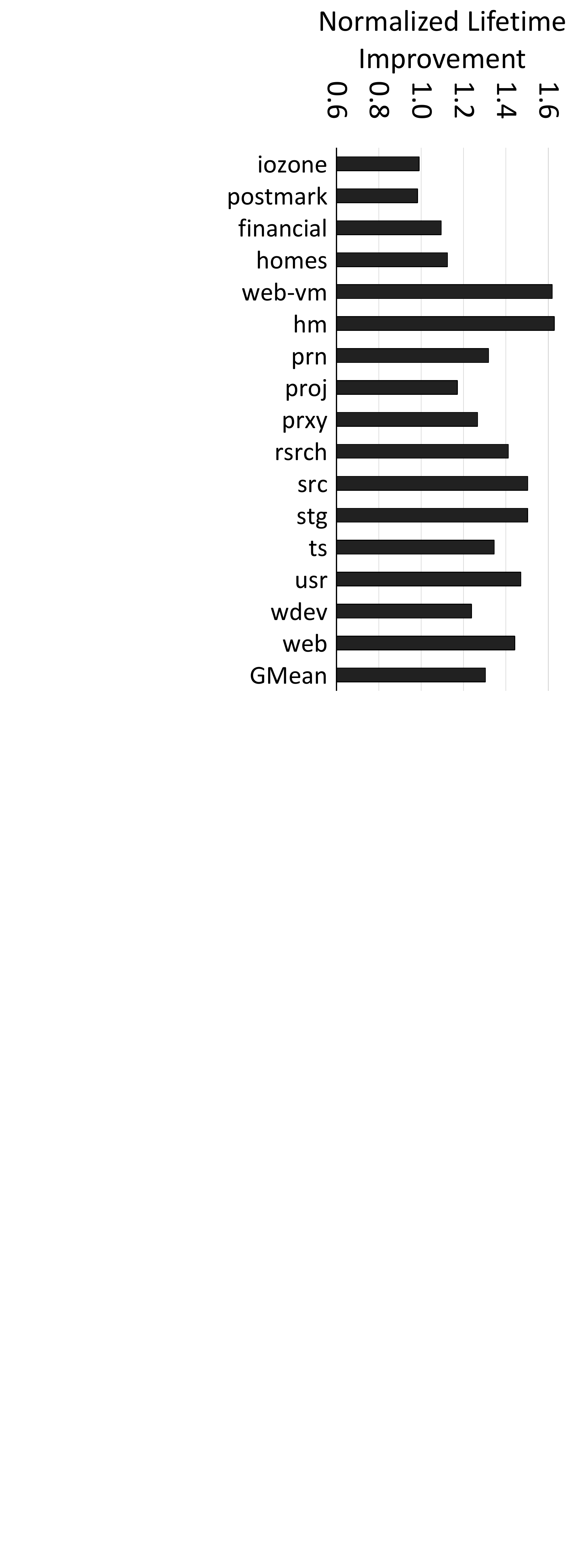}
  \caption{{\tt WARM+FCR} over {\tt FCR}}
\label{fig:lifetime_improvement:warm_fcr}
\end{subfigure}%

\begin{subfigure}[h]{0.48\linewidth}
  \includegraphics[trim=10 200 800 0,clip,width=\linewidth]{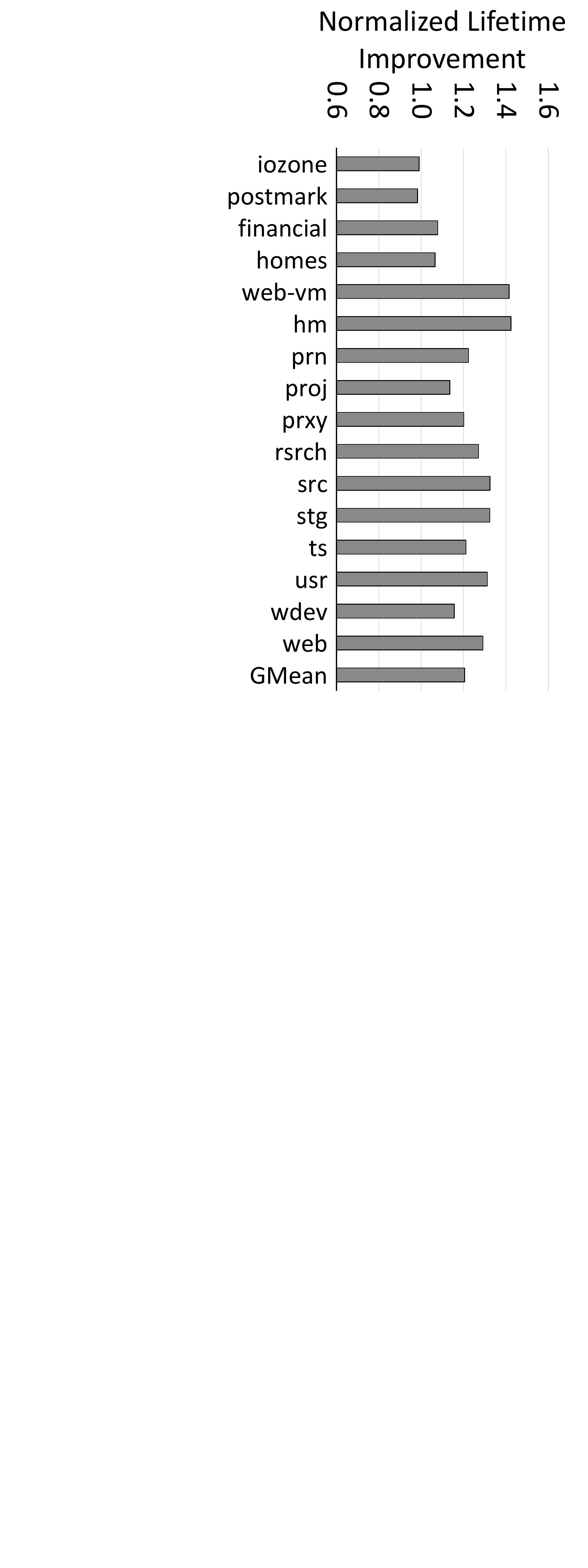}
  \caption{{\tt WARM+ARFCR} over {\tt ARFCR}}
\label{fig:lifetime_improvement:warm_arfcr}
\end{subfigure}%
\caption{Normalized flash memory lifetime improvement when WARM is applied on top of {\tt Baseline},
  {\tt FCR}, and {\tt ARFCR} configurations.}
\label{fig:lifetime_improvement}
\end{figure}

\subsection{Improvement in Endurance Capacity}
\label{sec:warm:evaluation:endurance}

Figure~\ref{fig:endurance} plots the normalized \emph{endurance capacity} of {\tt
WARM} for each workload split up by the endurance for both the hot and cold data pools.
The endurance capacity is defined as the total number of write operations the entire flash device
can sustain before wear-out.
On average, {\tt WARM} improves
the total endurance capacity by 3.6$\times$ over {\tt
Baseline}. Note that the endurance capacity varies across different workloads, in
relation to the number of hot writes that can be identified by the mechanism. For
example, {\tt postmark} contains only a limited amount of write-hot data (as is
shown in Figure~\ref{fig:writedist}), which results in only minor endurance
capacity improvement (8\%).  Unlike the other workloads, the majority of the
endurance capacity for {\tt postmark} remains within the cold pool, as the workload
exhibits very low write locality.

\begin{figure}[h]
\centering
\includegraphics[trim=10 220 680 5,clip,width=.7\linewidth]{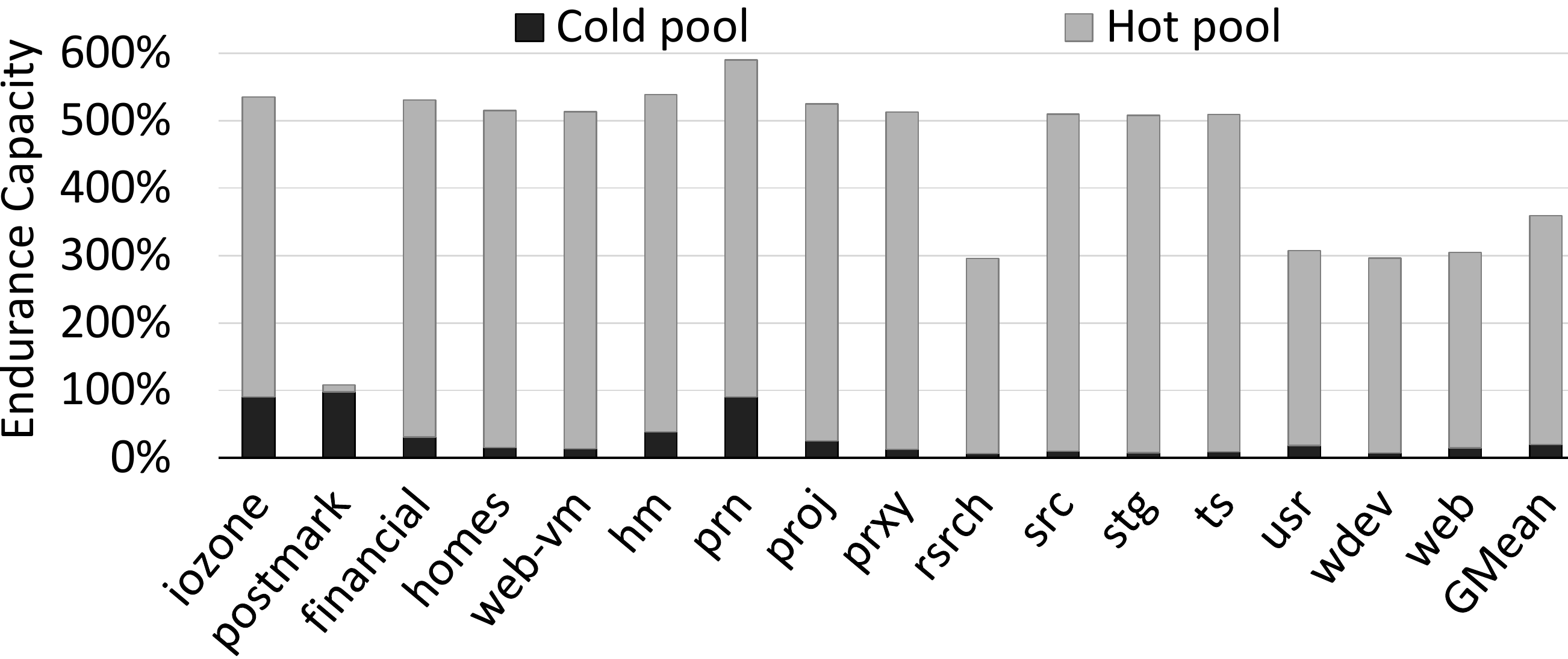}
\caption{{\tt WARM} endurance capacity, normalized to {\tt Baseline}.}
\label{fig:endurance}
\end{figure}
\FloatBarrier

In contrast, the endurance capacity for all of our other workloads mainly comes from the hot
pool, despite the size of the hot pool being significantly smaller than that of the cold pool.  WARM in
essence ``converts'' blocks from normal internal retention time (those in the cold pool)
into relaxed internal retention time (hot pool) for the write-hot portion of data.  Blocks
with a relaxed retention time can tolerate a much larger number of writes (as shown
in Figure~\ref{fig:retention-endurance}).
As Figure~\ref{fig:writedist} shows, the vast majority of 
overall writes are to a small fraction of pages that are write-hot. This allows {\tt WARM} to improve
the overall flash endurance capacity by using a small number of
blocks with a relaxed retention time to house the write-hot pages.  We conclude that {\tt WARM} can effectively
improve endurance capacity even when applied on its own.

\subsection{Reduction of Refresh Operations}
\label{sec:warm:evaluation:refresh}

Figure~\ref{fig:refreshes} breaks down the percentage of endurance (P/E cycles) used for the
host's write requests, for management operations, and for refresh requests during the refresh phase.  Two
bars are shown side by side for each application. The first bar shows the
number of total writes for {\tt FCR}, normalized to 100\%.  The second
bar shows a similar breakdown for {\tt WARM+FCR}, \emph{normalized to the number
of writes for {\tt FCR}}.  Although the two synthetic workloads ({\tt iozone} and {\tt postmark}) do not show much reduction in total write frequency (because host
writes dominate their flash endurance usage, as shown in
Figure~\ref{fig:consumedPE}), the number of writes across all sixteen 
of our workloads is reduced by an average of 5.3\%.

\begin{figure}[h]
\centering
\includegraphics[trim=10 5 10 10,clip,width=.7\linewidth]{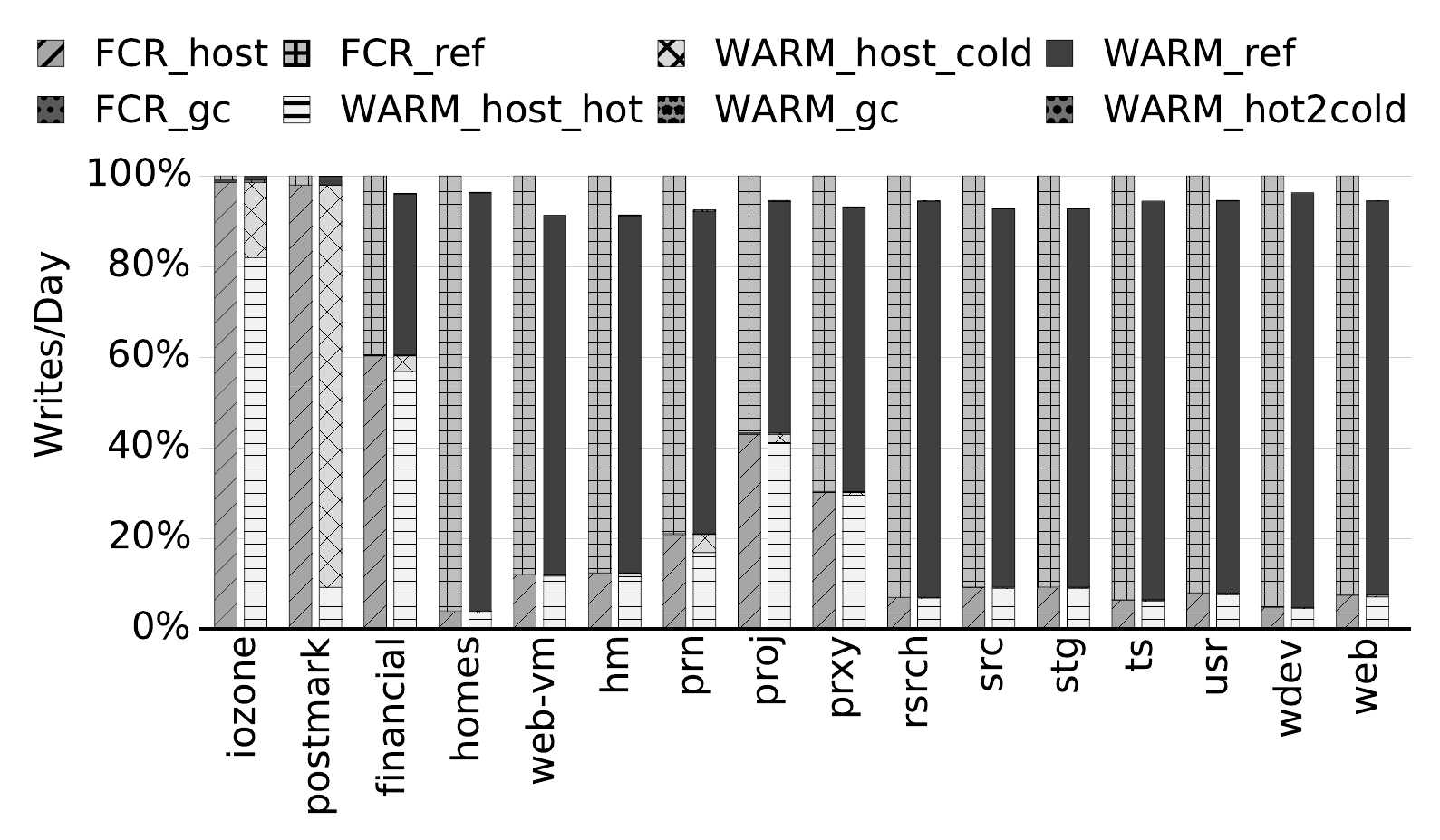}
\caption{Flash writes for {\tt FCR} (left bar) and {\tt WARM+FCR} (right bar),
broken down into host writes to the hot/cold pool ({\tt 
host{\char`_}hot/host{\char`_}cold}), garbage collection writes ({\tt gc}), refresh writes
({\tt ref}), and writes generated by WARM for migrations from the hot pool to 
the cold pool ({\tt hot2cold}).}
\label{fig:refreshes} 
\end{figure}
\FloatBarrier

From the breakdown of the write requests, we can see that the reduction in write count
mainly comes from the decreased number of refresh requests after applying WARM\@. 
In contrast, the additional overhead in WARM due to migrating data from the hot pool
to the cold pool is minimal (shown as {\tt WARM\_hot2cold}).  This suggests that the write locality 
behavior within many of the hot pool pages
lasts throughout the lifetime of the application, and thus these pages do not 
need to be evicted from the hot pool.\footnote{Migrations from the cold pool to the
hot pool are not broken down separately, as such migrations are performed during the 
host write request itself and do not incur additional writes}, as explained in Section~\ref{sec:warm:mechanism:identify}.  We
conclude that {\tt WARM+FCR}, by providing refresh-free retention time
relaxation for hot data, can reduce a significant fraction of unnecessary
refresh writes, and that {\tt WARM+FCR} can utilize the flash endurance more
effectively during the refresh phase.

\subsection{Impact on Performance}
\label{sec:warm:evaluation:performance}

As we discussed in Section~\ref{sec:warm:mechanism:overhead}, {\tt WARM} has the potential
to generate additional write operations.  First, when a page is demoted from the
hot pool to the cold pool (which happens when another page is being promoted into
the hot pool), an extra write will be required to move the page
into a block in the cold pool.\footnote{In contrast, promoting a page from the cold pool to
the hot pool does not incur additional writes, as promotion only occurs when
that page is being written.  Since a write was needed regardless, the promotion is free.}  Second, as one of the pools may have fewer
blocks available in its free list (which is dependent on how our partitioning
algorithm splits up the flash blocks into the hot and cold pools), garbage
collection may need to be invoked more frequently when a new page is required.
To understand the impact of these additional writes, we evaluate how WARM
affects the average response time of the FTL\@.

Figure~\ref{fig:resptime} shows the average response time for {\tt WARM},
normalized to the {\tt Baseline} response time. Across all of our workloads, the
average performance reduces by only 1.3\%. Even in the worst case ({\tt
homes}), {\tt WARM} only has a performance penalty of 5.8\% over {\tt
Baseline}.  The relatively significant overhead for {\tt homes} is due to the write-hot portion of its
data changing frequently over time within the trace.  This is likely because the user operates on
different files, which effectively shifts the write locality to an entirely different set of pages
whenever a new file is operated on. The shifting of the write-hot page
set evicts \emph{all} of the write-hot pages from the hot pool, which as we stated
above incurs several additional writes.

\begin{figure}[h]
\centering
\includegraphics[trim=5 275 680 0,clip,width=.7\linewidth]{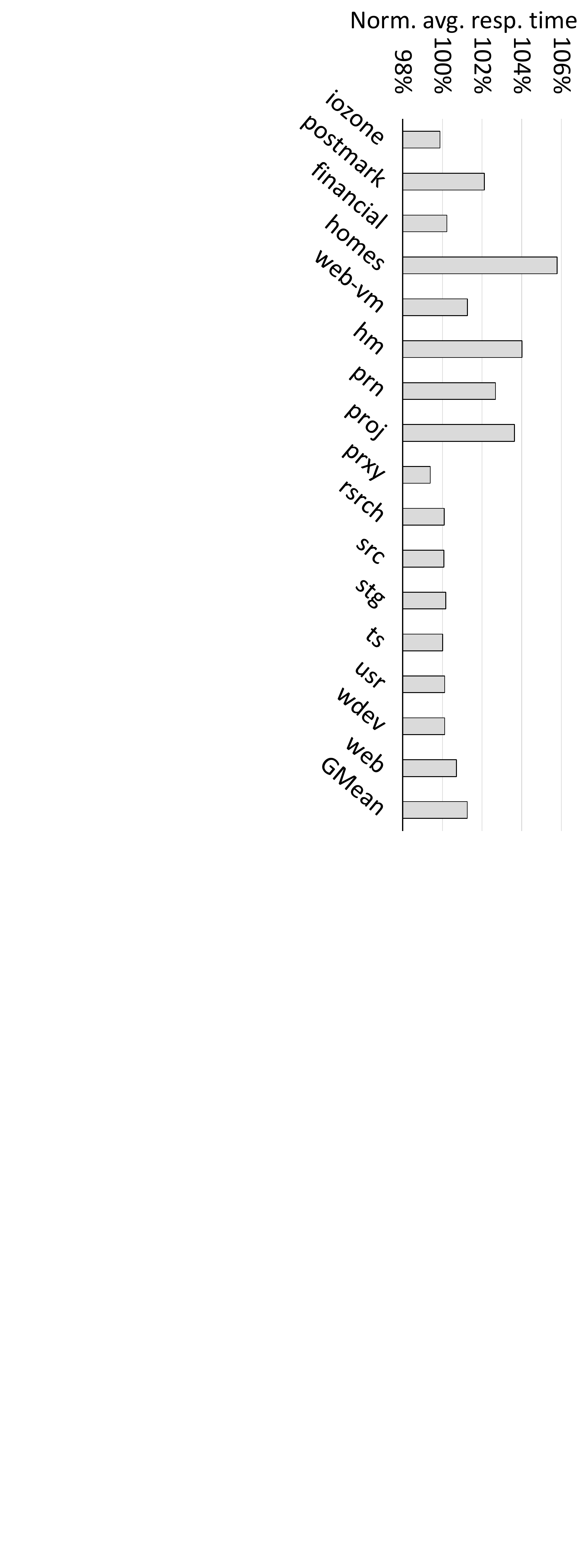}
\caption{{\tt WARM} average response time, normalized to {\tt Baseline}.}
\label{fig:resptime}
\end{figure}
\FloatBarrier

For most of the workloads, 
any performance degradation is negligible ($<$2\%), and is
a result of the increased garbage collection that occurs in the hot pool
due to its small free list size.  For some other workloads, such as {\tt prxy},
we find that the performance actually \emph{improves} slightly with {\tt WARM}, because of the
reduction in data movement induced by garbage collection.  This savings is thanks to grouping
write-cold data together, which greatly lessens the degree of fragmentation within the majority
of the flash blocks (those within the cold pool).
Overall, we conclude that across all of our workloads, the performance 
penalty of using {\tt WARM} is minimal.

\subsection{Sensitivity Studies}
\label{sec:warm:evaluation:sensitivity}

Figure~\ref{fig:sweep-op} compares the flash memory lifetime under different
capacity over-provisioning assumptions. In high-end server-class flash drives, the amount
of capacity over-provisioning is higher than that in consumer-class flash drives to
provide an overall longer lifetime and higher reliability. In this figure, we
evaluate the lifetime improvement of the same six configurations using 30\% of the flash blocks for
over-provisioning to represent a server-class flash drive (all other parameters
from Table~\ref{tbl:sim-config} remain the same).
We also show the lifetime of the six configurations on
a consumer-class flash drive with 15\% over-provisioning (which we assumed in our evaluations until now). We show
that the lifetime improvement of WARM become more significant as over-provisioning
increases.  The lifetime improvement delivered by {\tt WARM} over {\tt
Baseline}, for example, increases to 4.1$\times$, while the
improvement of {\tt WARM+ARFCR} over {\tt Baseline} increases to 14.4$\times$. We conclude that WARM
delivers higher lifetime improvements as over-provisioning increases.

\begin{figure}[h]
\centering
\includegraphics[trim=15 120 700 0,clip,width=0.7\linewidth]{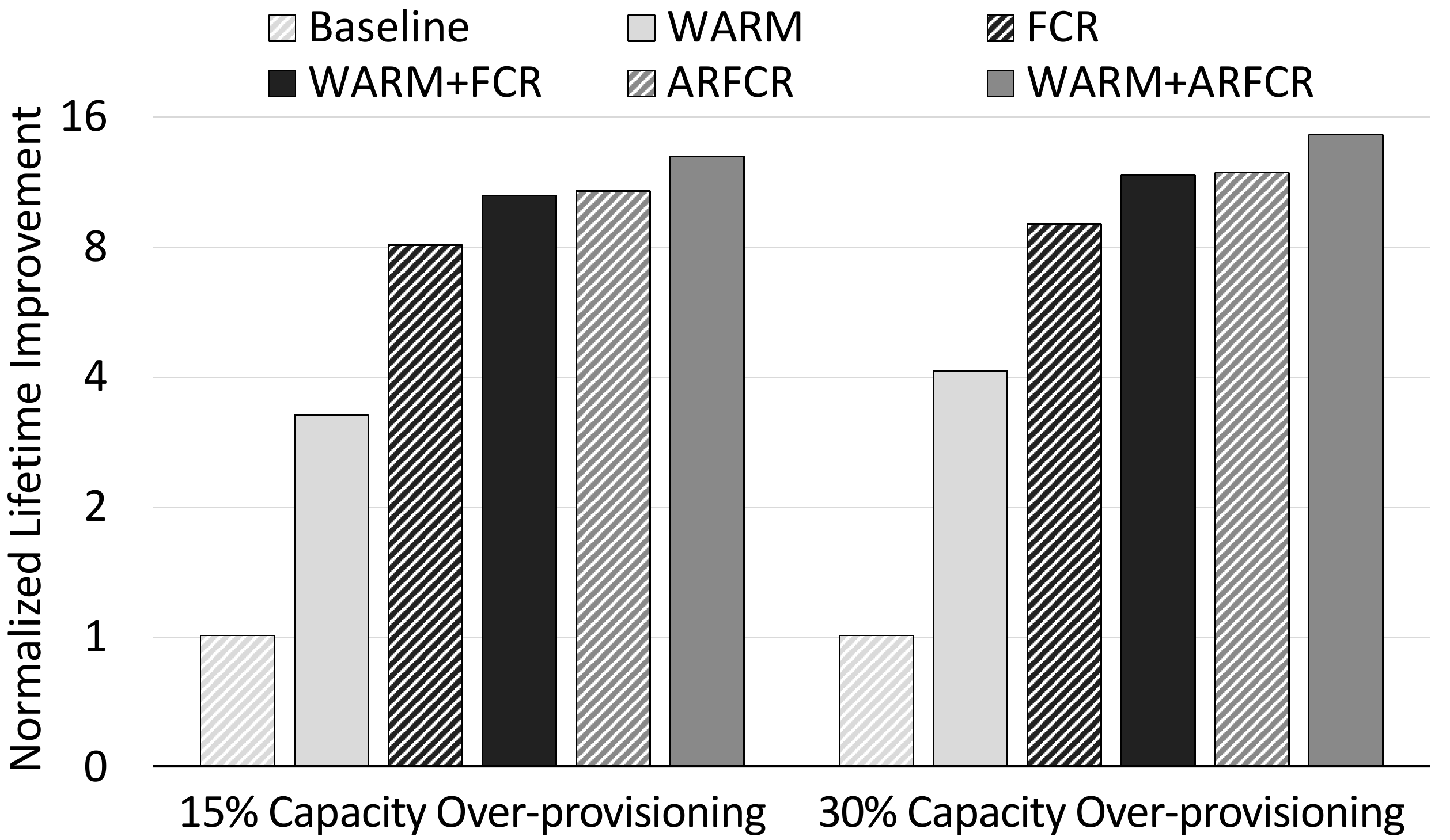}
\caption{Flash memory lifetime improvement for {\tt WARM}, {\tt
FCR}, {\tt WARM+FCR}, {\tt ARFCR}, and {\tt WARM+ARFCR} configurations under
different amounts of over-provisioning, normalized to the {\tt Baseline}
lifetime \emph{for each over-provisioning amount}.  Note that the y-axis uses a log scale.}
\label{fig:sweep-op}
\end{figure}
\FloatBarrier

Figure~\ref{fig:sweep-refresh} compares the flash memory lifetime improvement
for {\tt WARM+FCR} over {\tt FCR} under different refresh rate assumptions.
Our evaluation has so far assumed a three-day refresh period for {\tt FCR}. In
this figure, we change this assumption to three-month and three-week refresh periods, and compare
the corresponding lifetime improvement. As we see from this figure, the lifetime
improvement delivered by {\tt WARM+FCR} drops significantly as the refresh period
becomes longer. This is because a smaller fraction of the endurance is consumed by
refresh operations as the rate of refresh decreases (as shown in
Figure~\ref{fig:consumedPE}), which is where our major savings come from.

\begin{figure}[h]
\centering
\includegraphics[trim=5 230 880 0,clip,width=.5\linewidth]{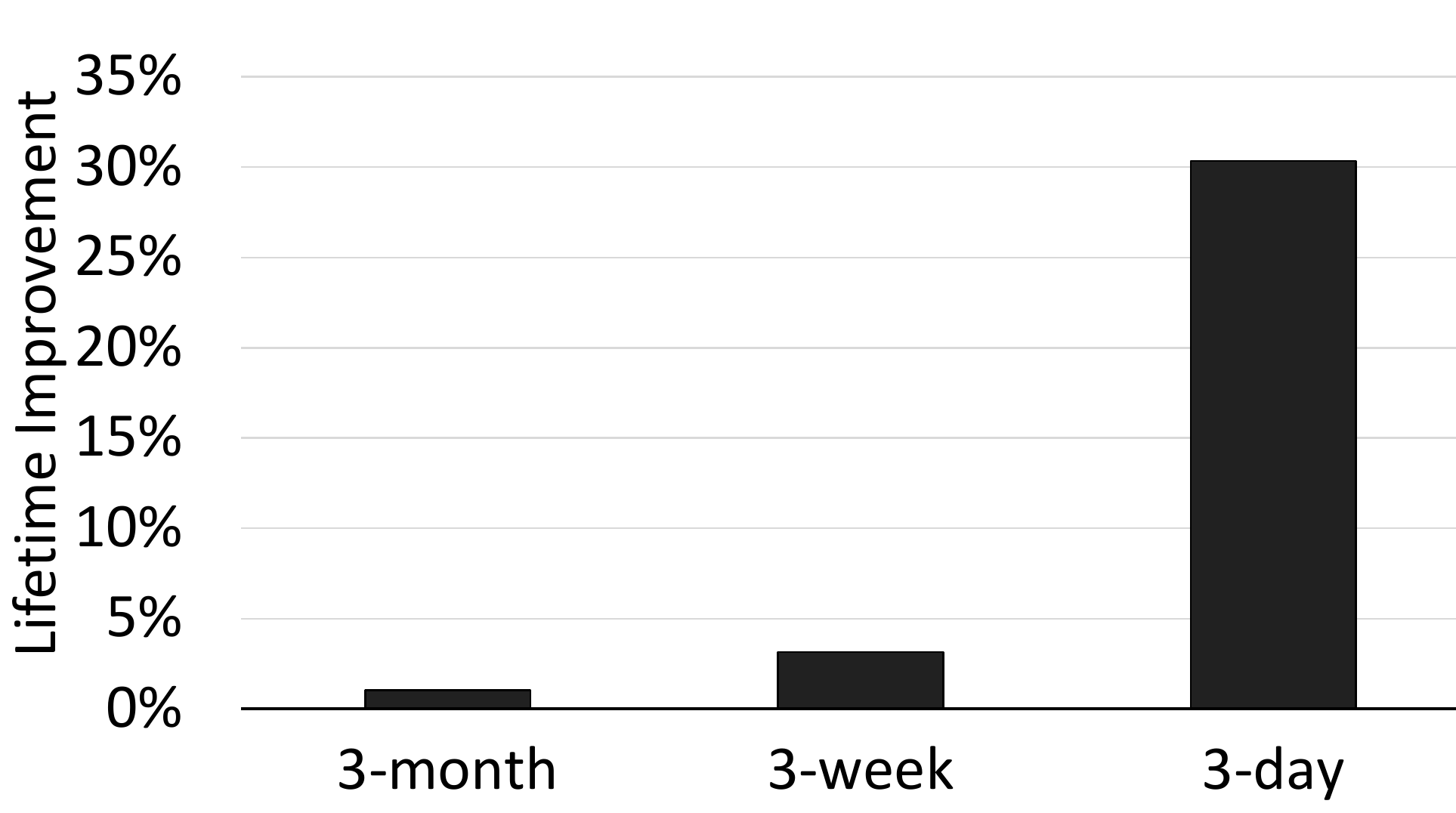}
\caption{Flash memory lifetime improvements for {\tt WARM+FCR} over {\tt FCR}
under different refresh rate assumptions.}
\label{fig:sweep-refresh}
\end{figure}
\FloatBarrier

Figure~\ref{fig:consumedPE-warm-sweep} illustrates how {\tt WARM+FCR}
reduces the fraction of P/E cycles consumed by refresh operations, over {\tt FCR}
only, as we sweep over longer refresh periods.
Note that the x-axis
in the figure uses a log scale.  The solid lines in the figure illustrate the
fraction of P/E cycles consumed by refresh for {\tt FCR} only, as was shown in
Figure~\ref{fig:consumedPE}.
The figure shows that for as the refresh interval increases,
{\tt WARM+FCR}  is effective at reducing the number of writes that are consumed
by refresh, but that these make up a smaller portion of the total P/E cycles, hence
the smaller improvements over {\tt FCR} alone.
As flash memory becomes denser and less reliable, we expect it to
require \emph{more frequent} refreshes in order to maintain a useful
lifetime, at which point WARM can deliver greater improvements. We conclude
that {\tt WARM+FCR} delivers higher lifetime improvements as the refresh rate
increases.

\begin{figure}[h]
\centering
\begin{subfigure}[h]{0.5\linewidth}
\includegraphics[trim=5 5 10 5,clip,width=\linewidth]{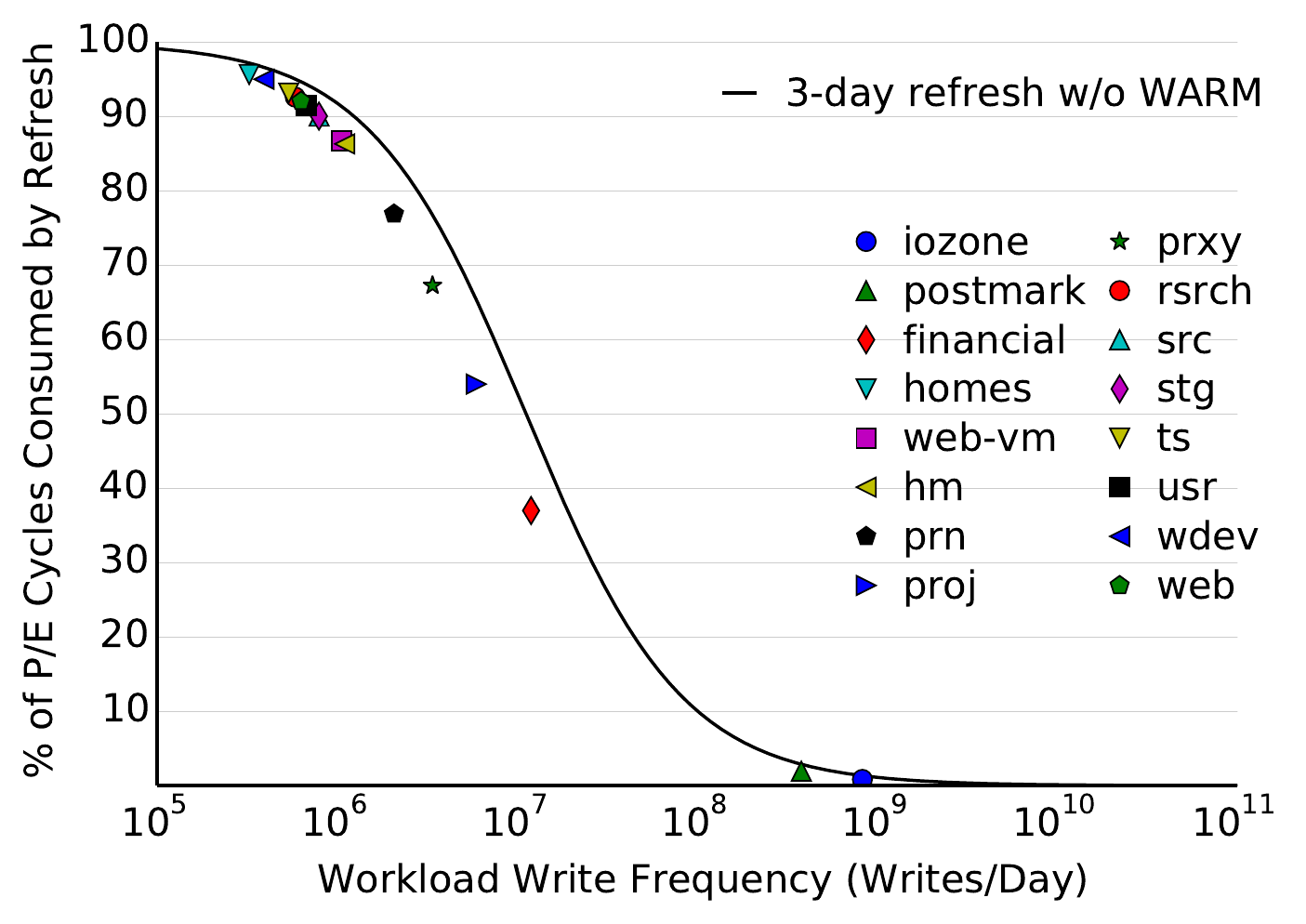}
  \caption{3-day refresh.}
\label{fig:consumedPE-warm-sweep:3-day}
\end{subfigure}%
~
\begin{subfigure}[h]{0.5\linewidth}
\includegraphics[trim=5 5 10 5,clip,width=\linewidth]{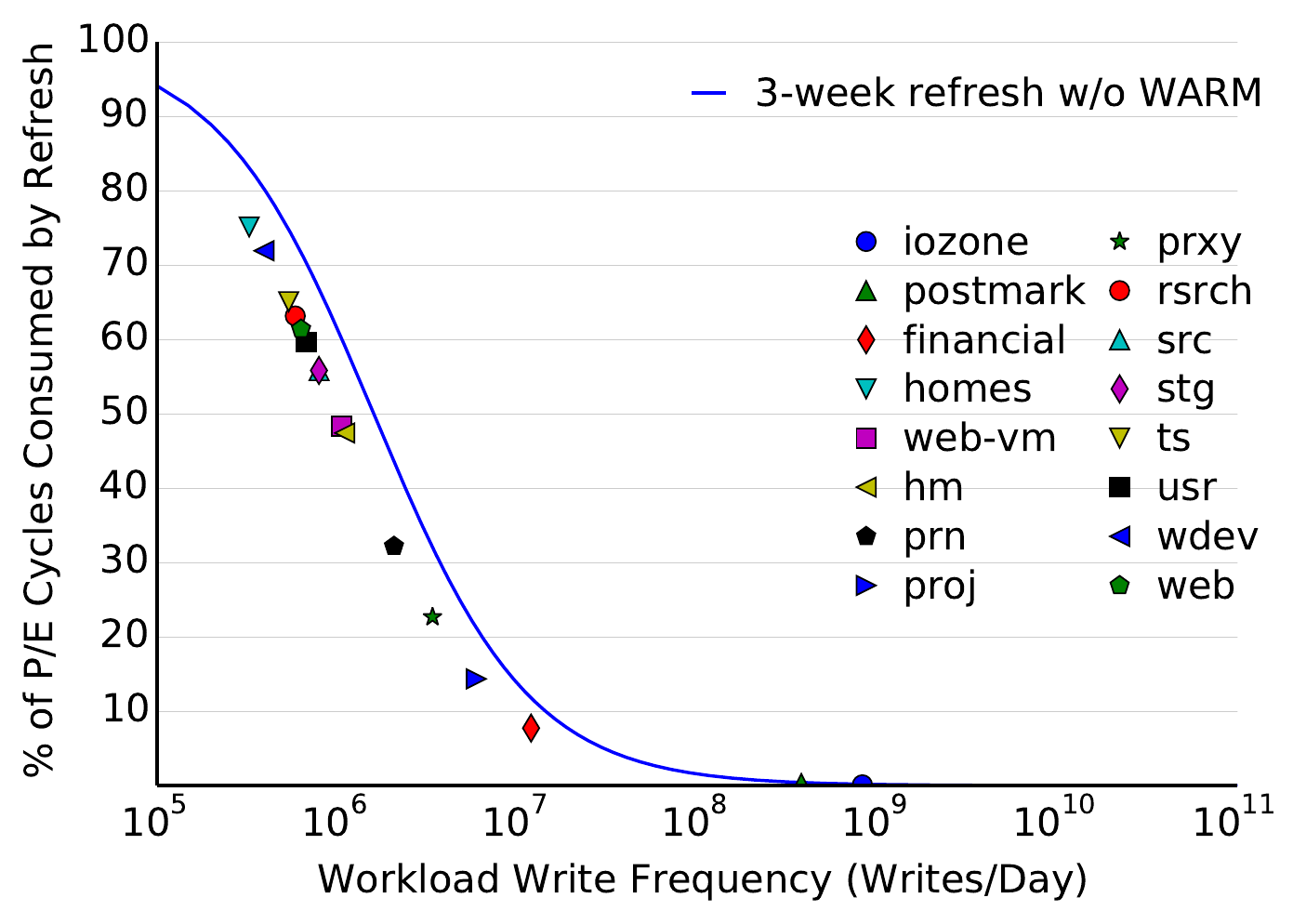}
  \caption{3-week refresh.}
\label{fig:consumedPE-warm-sweep:3-week}
\end{subfigure}%

\begin{subfigure}[h]{0.5\linewidth}
\includegraphics[trim=5 5 10 5,clip,width=\linewidth]{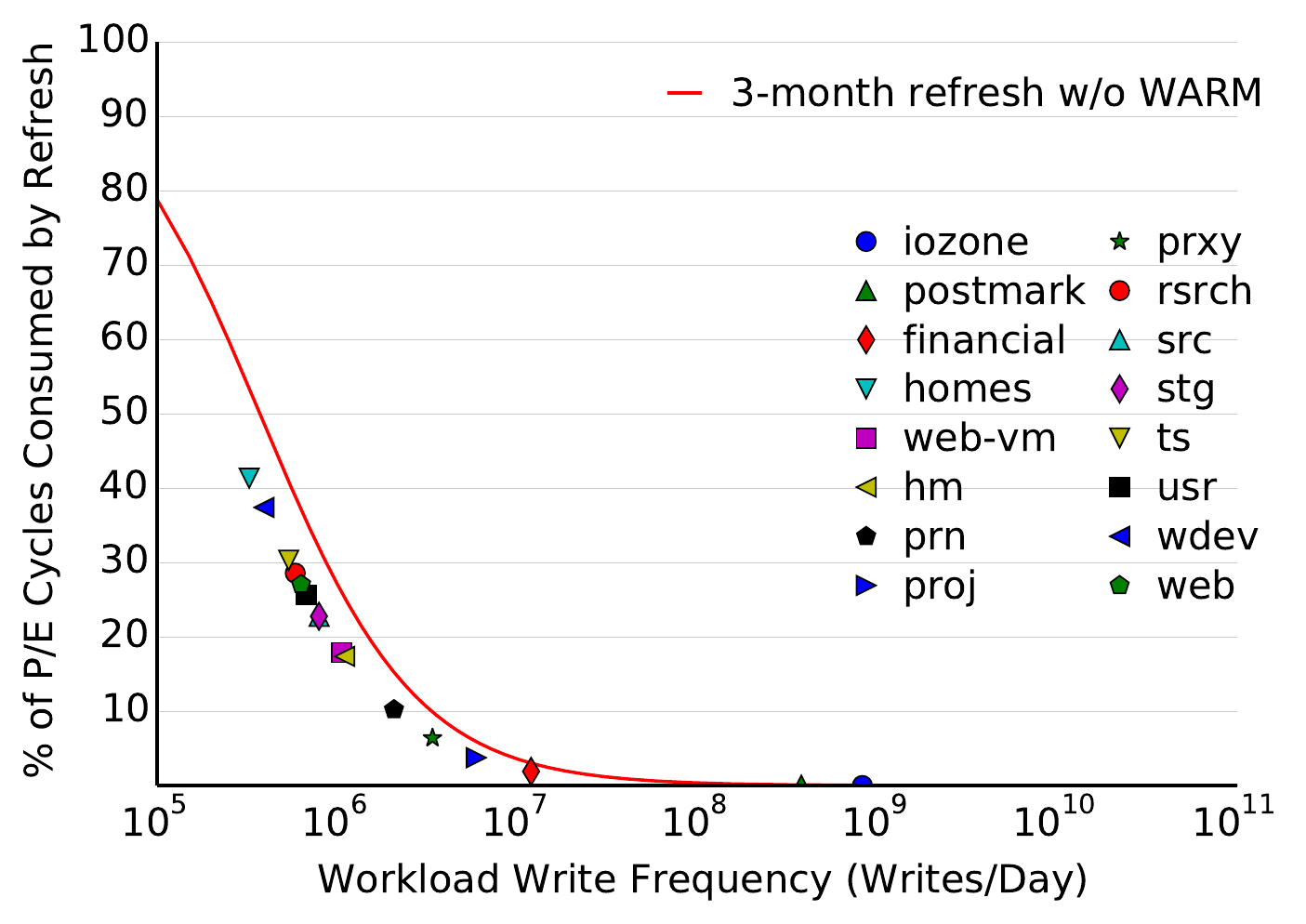}
  \caption{3-month refresh.}
\label{fig:consumedPE-warm-sweep:3-month}
\end{subfigure}%
\caption{Fraction of P/E cycles consumed by refresh operations \emph{after} applying {\tt WARM+FCR} for a (a)~3-day, (b)~3-week, or (c)~3-month refresh period.  Solid trend lines show the fraction consumed by {\tt FCR} only, from Figure~\ref{fig:consumedPE}, for comparison.  Note that the x-axis uses a log scale.}
\label{fig:consumedPE-warm-sweep}
\end{figure}
\FloatBarrier

\section{\chIV{Limitations}}

\chIV{
WARM relies on the existence of write-hot data in many existing write-intensive
workloads. If the workload is not write-intensive, or if all data in the workload
is equally write-hot, WARM cannot identify any write-hot data and skip any
refreshes. In these unlikely cases, WARM does not provide any benefit or
penalty in flash lifetime.
}


\section{Conclusion}
\label{sec:warm:conclusion}

In this chapter, we introduce WARM, a write-hotness aware retention management policy for NAND
flash memory that is designed to extend its lifetime. We find that pages with
different degrees of \emph{write-hotness} have widely ranging retention time
requirements. WARM allows us to relax the flash retention time for
write-hot data without the need for refresh, by exploiting the high write frequency
of this data. On its own, WARM
improves the lifetime of flash by an average of 3.24$\times$ over a
conventionally-managed flash device, across a range of real I/O workload
traces.  When combined with refresh mechanisms, WARM
eliminates redundant refresh operations to recently written data.  In conjunction with
an adaptive refresh mechanism, WARM extends the average flash lifetime by 12.9$\times$,
which represents a 1.21$\times$ increase over using the adaptive refresh mechanism alone. We
conclude that WARM is an effective policy for improving overall flash
lifetime with or without refresh mechanisms.


\chapter{Online Flash Channel Modeling and Its Applications}
\label{sec:vthmodel}

As we have introduced in Section~\ref{sec:flash}, the threshold voltage value
of a flash cell is used to represent the data that is stored within the cell.
\chIII{In} Section~\ref{sec:errors}, we have shown that the threshold voltage of
the cell changes (i.e., shifts) as a result of many different types of
circuit-level noise. Some cells' threshold voltages might shift enough to
cross over to neighboring voltage windows. The value of such a cell would be
misread on a flash memory channel read, causing an error. Thus, \emph{knowing how
the threshold voltage distribution shifts within the flash controller can help
\chIII{to mitigate} these errors}.

In this chapter, we introduce a new framework that exploits the unused
computing resources in the flash controller to enable greater device awareness
by exploiting \emph{an online flash channel model}. Our goal is to build
an \emph{accurate} and \emph{easy-to-compute}
model of the threshold voltage distribution of modern MLC NAND flash memory.
This model must be practical to implement, as we intend to use it \emph{online}
to design flash controllers that can adapt to the changing NAND flash memory
behavior.  Our model can (1)~\emph{statically} determine the threshold voltage
distribution at a given level of wear-out (i.e., a given P/E cycle  count), and
(2)~\emph{dynamically} predict how this threshold voltage distribution shifts
over time as a result of the P/E cycling effect. The key idea is to learn this
online flash channel model with low overhead and use this model in multiple
components within the flash controller to help improve flash reliability.

Section~\ref{sec:vthmodel:motivation} describes the motivation of developing
an online flash channel model. To build our flash channel model, we first use
the methodology described in Section~\ref{sec:vthmodel:overview:methodology}
to characterize the threshold voltage distribution (i.e., the flash channel)
under different P/E cycles using real \SI{1X}{\nano\meter} MLC NAND flash
chips. Second, in Section~\ref{sec:static}, we construct a static threshold
voltage distribution model that can fit the characterized distribution under
any given P/E cycle count. Third, in Section~\ref{sec:dynamic}, we construct a
dynamic P/E cycling model that predicts how each parameter of the static
distribution model changes after some number of further P/E cycles. Finally,
in Section~\ref{sec:application}, we demonstrate several example use cases in
the flash controller that utilize the complete model to enhance the
performance and reliability of the NAND flash memory device. We conclude the
contributions of our online flash channel model in
Section~\ref{sec:vthmodel:conclusion}.


\section{Motivation}
\label{sec:vthmodel:motivation}

Having online information on the current threshold voltage
distribution across all of the flash cells within a flash memory chip
(i.e., the \emph{static} distribution), as well as how this
distribution changes over time (i.e., the \emph{dynamic}
distribution), is important to quantify errors and develop techniques
to improve the reliability of the flash device.  First, the static
distribution can be used to determine the number of errors that would
occur for any read reference voltage that is applied.  This data can
be used by the flash controller to select the read reference voltage
that minimizes the error rate.  Lowering the error rate increases the
lifetime of the flash device, as it delays the time at which the
number of errors becomes too large for the built-in ECC mechanism to
successfully correct.  Second, knowing how the dynamic distribution
changes over time (i.e., as more writes are performed) is important,
as it can guide flash controller mechanisms that adjust various flash
parameters online (e.g., ECC strength~\cite{haratsch.fms15, liu.fast12, wu.mascots10}, read reference
voltages~\cite{cai.hpca15}, pass-through voltage~\cite{cai.dsn15}) to
increase the flash memory lifetime.  Prior proposals to adjust these
parameters (e.g.,~\cite{cai.hpca15, cai.dsn15}) rely on a
trial-and-error approach to select parameters with low error rates,
which can be inaccurate, high-latency, and suboptimal in terms of
lifetime improvement.  In both cases (static and dynamic), the
threshold voltage distribution must be determined at runtime by the
flash controller.  Therefore, it is critical to design a \emph{practical} and
\emph{low-complexity} mechanism to determine the distribution and the
shifts in the distribution.

In order for the model to be useful to help flash controller
algorithms, it should have certain properties. First, the model needs
to be \emph{accurate} for all threshold voltages and at all P/E
cycles.  This is because inaccurate information can lead a flash
controller to make suboptimal decisions, hurting lifetime
improvements.  Second, the model needs to be \emph{easy to compute},
because the flash controller has only limited computational resources.
While we base our model off of a distribution that is easy to compute,
we can further reduce online computation with the dynamic component of
our model, by performing only a few online static characterizations of
the threshold voltage distribution, and then using the simpler dynamic
model to predict shifts in these initial characterizations at very low
cost over P/E cycles.

\section{Characterization Methodology}
\label{sec:vthmodel:overview:methodology}

To build our model, we perform an experimental characterization of the
threshold voltage distribution on real state-of-the-art 1X-nm (i.e.,
15-19nm) MLC NAND flash chips.  This characterization is essential to
verify that the model we develop accurately captures the behavior of a
real, modern device. 

We collect experimental characterization data on the threshold voltage distribution
using an FPGA-based NAND flash testing
platform~\cite{cai.fccm11} with state-of-the-art 1X-nm MLC NAND flash
chips. We use the read-retry technique~\cite{cai.date13, cai.hpca15} (described in Section~\ref{sec:mitigation:retry}) to sweep
\emph{all possible} read reference voltages and determine the threshold voltage value
for each cell. We program and erase these blocks to 11 different wear levels, up
to 20K P/E cycles, using known pseudo-random data. The manufacturer-specified P/E cycle endurance for the
tested flash chips is 3000 P/E cycles. All tests are performed at room temperature
with a 5-second dwell time.\footnote{\emph{Dwell time} is the time duration between an
erase operation and the next program operation to the same flash cell.}

Figure~\ref{fig:methodology} shows the threshold voltage distribution
for each of the cell states. The read-retry capability on the MLC NAND
flash memory chip allows us to fine-tune each read reference voltage
($V_a$, $V_b$, and $V_c$) to one of 101 different steps (a total of
303 read reference voltage steps, labeled as $V_1$ to $V_{303}$ from
left to right). Note that $V_1$  does not extend
all the way to the lowest possible threshold voltage for the ER
state, and $V_{303}$ does not extend to the highest possible threshold voltage for the P3 state.
In this chapter, we normalize the threshold voltage
values such that the distance between most of the adjacent
read reference voltage steps is one, as the exact values are
proprietary information. 
The distances between steps
$V_{101}$ and $V_{102}$ and between steps $V_{202}$ and $V_{203}$ are much
larger than the typical distance between voltage steps,\footnote{Some flash
vendors choose to provide fewer read reference voltages near the
peak of the distribution of each state.  This is because flash cells near the
have threshold voltages far away from the
default read reference voltage, and hence are less likely to
have errors.} as shown in Figure~\ref{fig:methodology}.
As a result, the voltage step $V_{303}$ has a normalized voltage value that is greater than 303. Overall, the
threshold voltage range is divided by these read reference voltages
into 304 bins, labeled as $bin_0$ to $bin_{303}$. Each flash cell can
be classified into one of these bins based on the threshold voltage
value read from the cell. If the read reference voltage is higher than
the threshold voltage of the cell, the value read out from the flash device is 1,
otherwise the value read out is 0. For a cell whose threshold voltage
falls between two neighboring read reference voltages
($V_{k}$ and $V_{k+1}$), the cell is placed into $bin_k$, as
illustrated in Figure~\ref{fig:methodology}.

\begin{figure}[h]
\centering
\includegraphics[trim=0 135 0 32,clip,width=.75\linewidth]{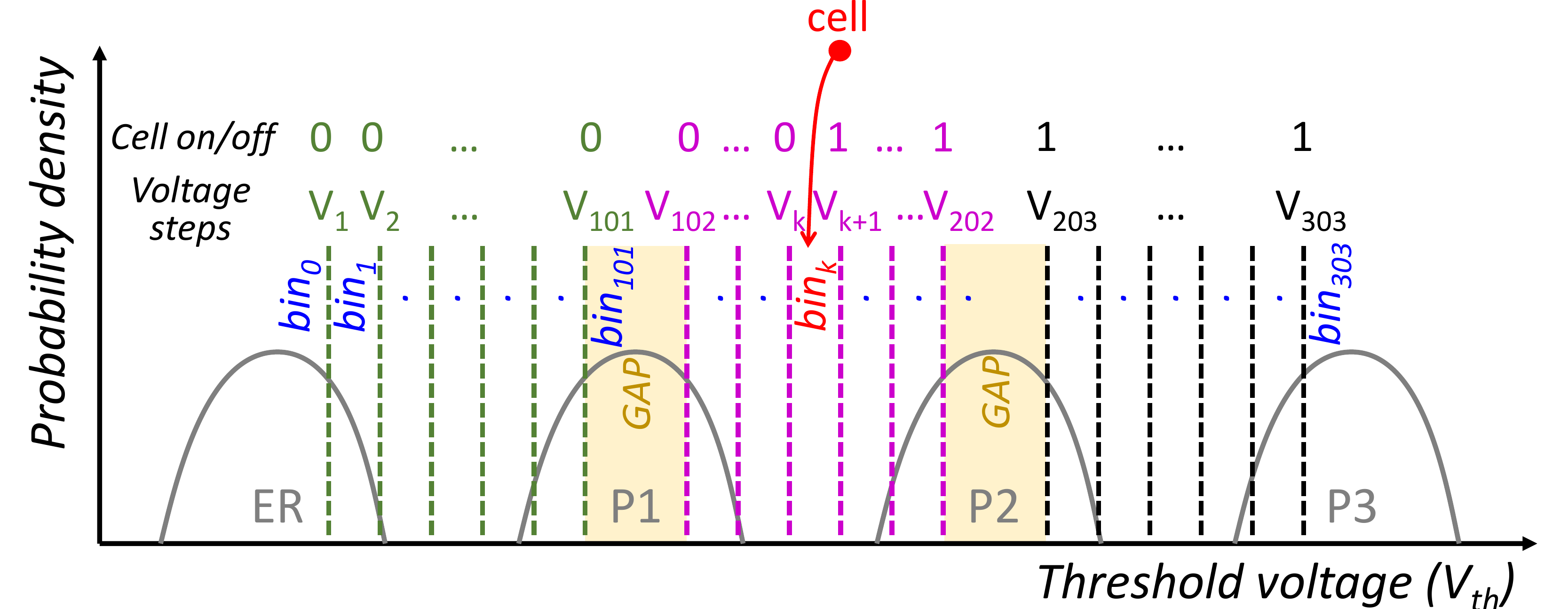}%
\caption{
  Methodology for finding the threshold voltage of an MLC NAND flash memory
  cell.
}
\label{fig:methodology}
\end{figure}

After classifying every cell using the methodology described above, we
count the number of flash cells with state $X \in$ \{ER, P1, P2, P3\}
in bin $k$ as $H_k(X)$. Equation~\ref{eqn:pkx} shows how we then
normalize the bin counts as the probability density of each bin,
$P_k(X)$. Note that in our characterization, we
assign each flash cell to the threshold voltage distribution of the \emph{correct state} that
it was originally programmed to, as we know the data value that we
programmed. The characterized bin density can be viewed as a
discretized version of the measured distribution, which our model is
constructed to fit.

\begin{align}
  P_k(X) = \frac{H_k(X)}{\Sigma_{i=0}^{303}H_i(X)}
\label{eqn:pkx}
\end{align}


\section{Static Distribution Model}
\label{sec:static}

We construct a {\em static} threshold voltage distribution model that
can fit the characterized threshold voltage distribution well under
any P/E cycle count, based on data collected using the methodology \chI{described} in
Section~\ref{sec:vthmodel:overview:methodology}.  Recall that this model needs
to be (1)~accurate for all threshold voltages and at any given P/E
cycle, and (2)~easy to evaluate within the flash
controller. While a more complex model can satisfy the accuracy
requirements, it can be difficult to compute the model on the fly
given the limited computational resources in a flash controller. In
this section, we first describe two state-of-the-art models, each of
which meets only one of our two requirements. The first \chI{previously-proposed model~\cite{cai.date13}}, based
on a \emph{Gaussian distribution}, is simple and
easy to compute, but is not accurate enough for raw bit error rate
estimation (Section~\ref{sec:static:gaussian}). The second \chI{previously-proposed model~\cite{parnell.globecom14}},
based on a \emph{normal-Laplace
  distribution}, is accurate, but requires
significant computational resources, taking 10.7x the computation time
of the Gaussian-based model (Section~\ref{sec:static:laplace}).  We
propose a new model, based on \chI{our modified version of the}
\emph{Student's t-distribution}~\cite{speigel.book92},
which satisfies both of our requirements, maintaining the accuracy of
the normal-Laplace-based model while requiring 4.41x less computation
time (Section~\ref{sec:static:student}). Finally, we validate and
compare the three models (Section~\ref{sec:static:compare}).

\subsection{Gaussian-based Model}
\label{sec:static:gaussian}


The Gaussian-based model assumes that the threshold voltage
distribution of each state follows a Gaussian (i.e., normal)
distribution~\cite{cai.date13}. Equation~\ref{eqn:gkx} shows how the
Gaussian\chI{-based} model estimates the probability density for state $X$ (i.e., ER, P1, P2,
and P3) in each bin $k$, denoted as $G_k(X)$:
\begin{align}
  G_k(X) = GCDF(V_k, \mu_X, \sigma_X) - GCDF(V_{k-1}, \mu_X, \sigma_X) \label{eqn:gkx}
\end{align}

The density $G_k(X)$ is calculated as the difference between the
Gaussian cumulative distribution function ($GCDF$) of the bin's two
boundaries, $V_k$ and $V_{k-1}$. The Gaussian\chI{-based} model has two variables
for each state: $\mu_X$ is the mean of the distribution, and
$\sigma_X$ is the standard deviation of the distribution. In total,
the Gaussian threshold voltage distribution model has eight
parameters.

The intuition behind using a Gaussian distribution is twofold. First,
the threshold voltage distribution is a result of physical noise and
manufacturing process variation, which naturally follow a Gaussian
distribution.  During a program operation, the flash controller uses
ISPP (see Section~\ref{sec:flash:pgmerase}), iteratively
increasing the threshold voltage until the desired threshold voltage
level is achieved. Each programming step increases the threshold
voltage of a cell by a small random amount.  As \chI{programming} subjects the cell
to random physical noise, the threshold voltage distribution of each
state naturally approximates a Gaussian
distribution~\cite{cai.date13}.

Second, \chI{the Gaussian-based model can be computed quickly, and
is easily implementable}
in the flash controller hardware \chI{if we use}
a \emph{z-table}, a lookup table that stores the
\emph{precomputed} cumulative distribution function of the standard
Gaussian distribution.  Equation~\ref{eqn:gcdf} shows how the z-table
simplifies the computation of $GCDF$. First, we calculate the z-scores
$Z = \frac{V-\mu}{\sigma}$ for $V_k$ and $V_{k-1}$. Then, we calculate
$\Phi(Z)$, the precomputed cumulative distribution function of $Z$, by
looking up the z-score in the z-table.  The two z-scores (one each for
$V_k$ and $V_{k-1}$) are then combined to get $G_k(X)$, using
Equation~\ref{eqn:gkx}.
\begin{align}
  GCDF(V, \mu, \sigma) = \Phi(Z) = \textit{z-table}(Z) \label{eqn:gcdf}
\end{align}

The goal of static modeling is to fit the estimated distribution
$G_k(X)$ to the measured distribution $P_k(X)$. We use
Kullback-Leibler divergence~\cite{kullback.math51} to estimate \chI{the
accuracy of the model}
(i.e., the error between the estimated and
measured distributions). The Kullback-Leibler divergence between the measured and the estimated
probability density for each bin ($P_k$ and $G_k$, respectively) can be mathematically defined as:
\begin{align}
  D_{K-L} = \sum_{k=1}^{N_{bin}} P_k \log(\frac{P_k}{G_k})
\end{align}

We use the Nelder-Mead simplex
method~\cite{nelder.computer65} to minimize the error, in order to
learn the model under different P/E cycles.  We use a reasonable
initial guess of the parameters for the Nelder-Mead simplex method,
allowing us to quickly approach the best fit.


Figure~\ref{fig:static-gauss} shows the distribution measured by our
experimental characterization using \chI{markers}, and shows how the
Gaussian-based model (the curves depicted with solid or dashed lines) fits to
this data \chI{at different P/E cycle counts}. The x-axis is the normalized threshold voltage, and the
y-axis is the probability density function at each normalized
threshold voltage in log scale. 
\chI{In this figure, \chI{from left to right, we show the threshold voltage
distribution of the ER state, the P1 state, the P2 state, and the P3 state.}
We show the modeled distributions of the ER and P2 states
using solid lines, and the modeled distributions of the P1 and P3 states using dashed lines.}

\begin{figure}[h]
  \centering
  \begin{subfigure}[h]{.4\linewidth}
    \centering
    \includegraphics[trim=0 7 0 0,clip,width=\linewidth]{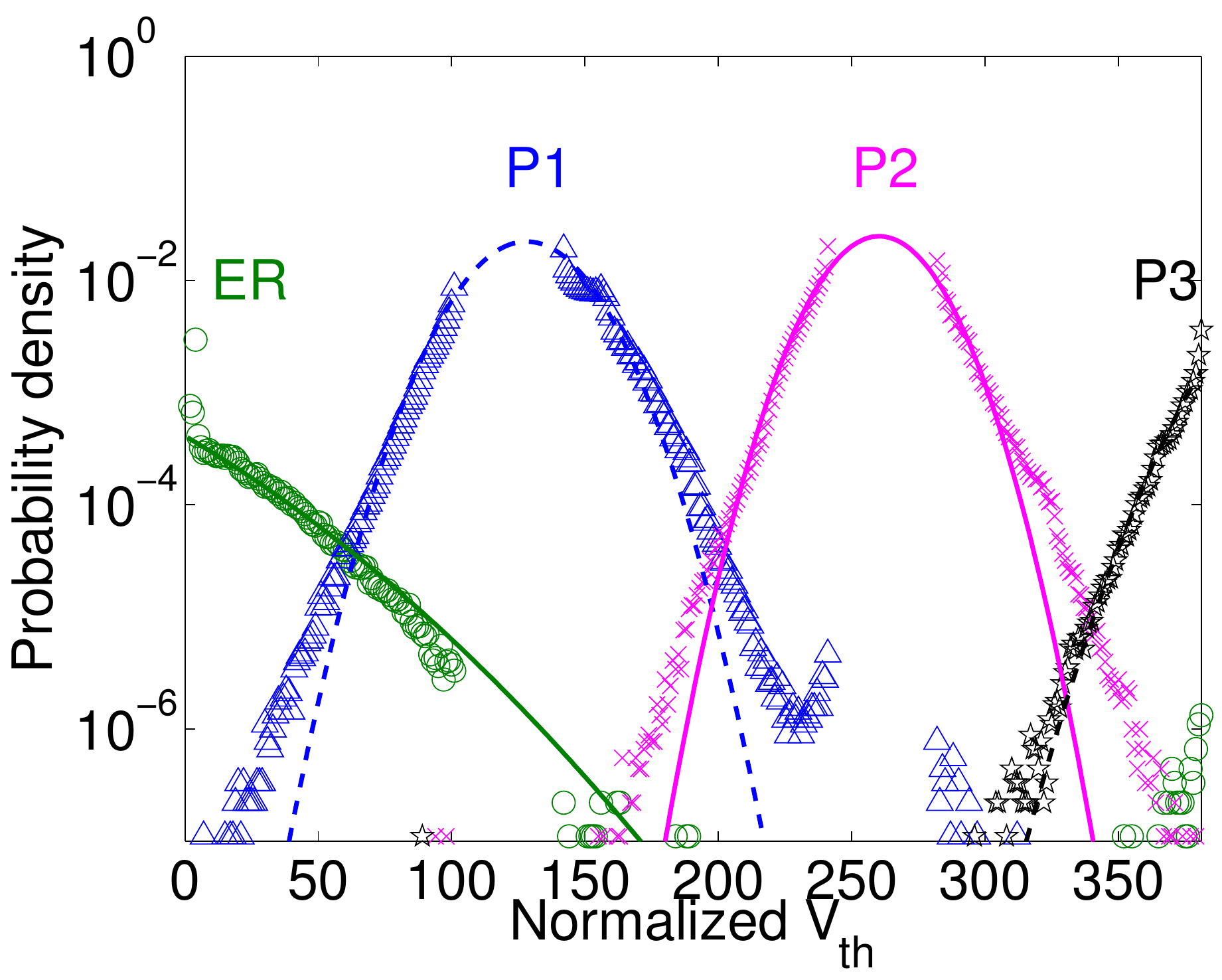}%
    \caption[]%
    {{2.5K P/E cycles}}
    \label{fig:static-gauss-2.5k}
  \end{subfigure}
  \begin{subfigure}[h]{.4\linewidth}
    \centering
    \includegraphics[trim=0 7 0 0,clip,width=\linewidth]{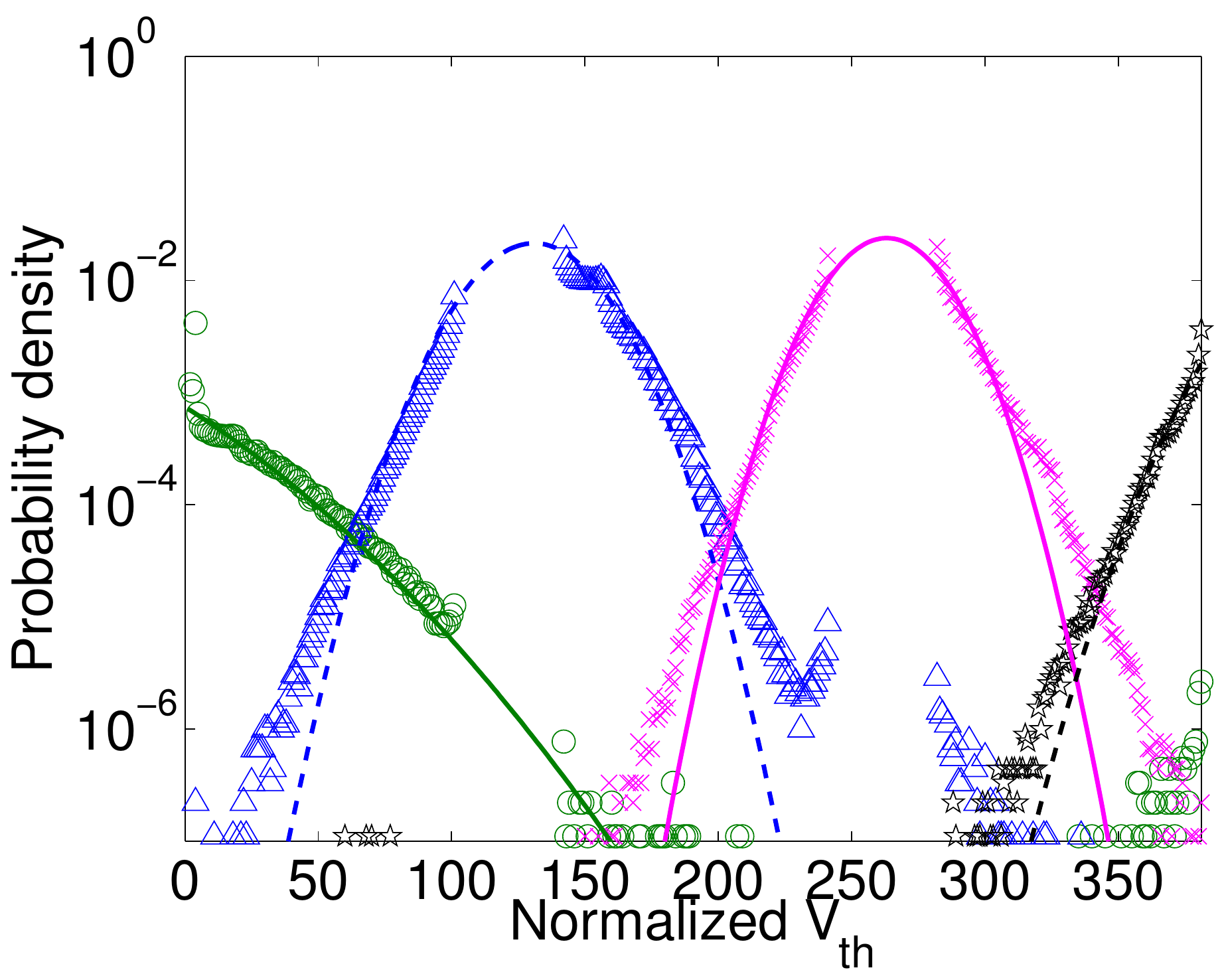}%
    \caption[]%
    {{5K P/E cycles}}
    \label{fig:static-gauss-5k}
  \end{subfigure}
  \begin{subfigure}[h]{.4\linewidth}
    \centering
    \vspace{8pt}
    \includegraphics[trim=0 7 0 0,clip,width=\linewidth]{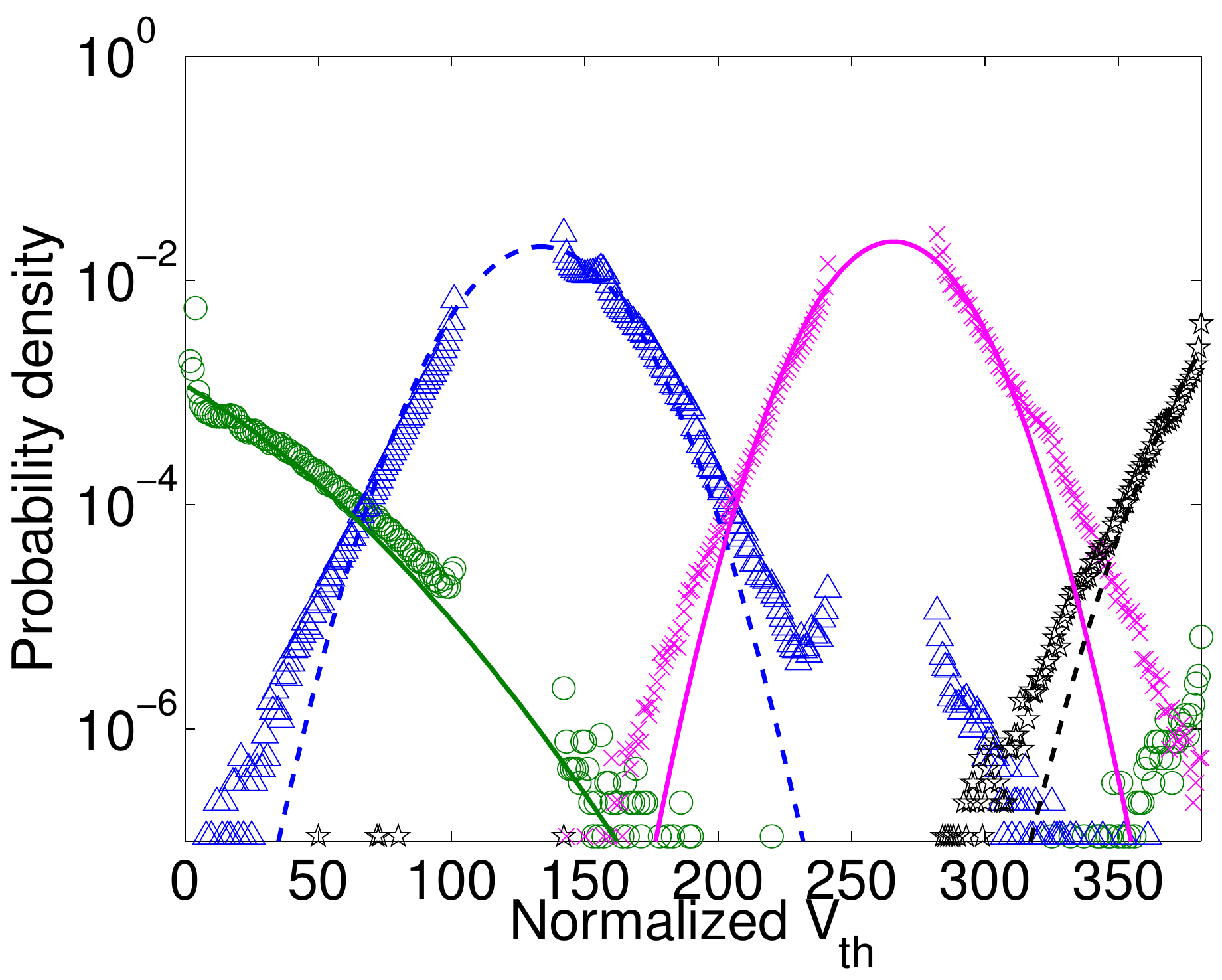}%
    \caption[]%
    {{10K P/E cycles}}
    \label{fig:static-gauss-10k}
  \end{subfigure}
  \begin{subfigure}[h]{.4\linewidth}
    \centering
    \vspace{8pt}
    \includegraphics[trim=0 7 0 0,clip,width=\linewidth]{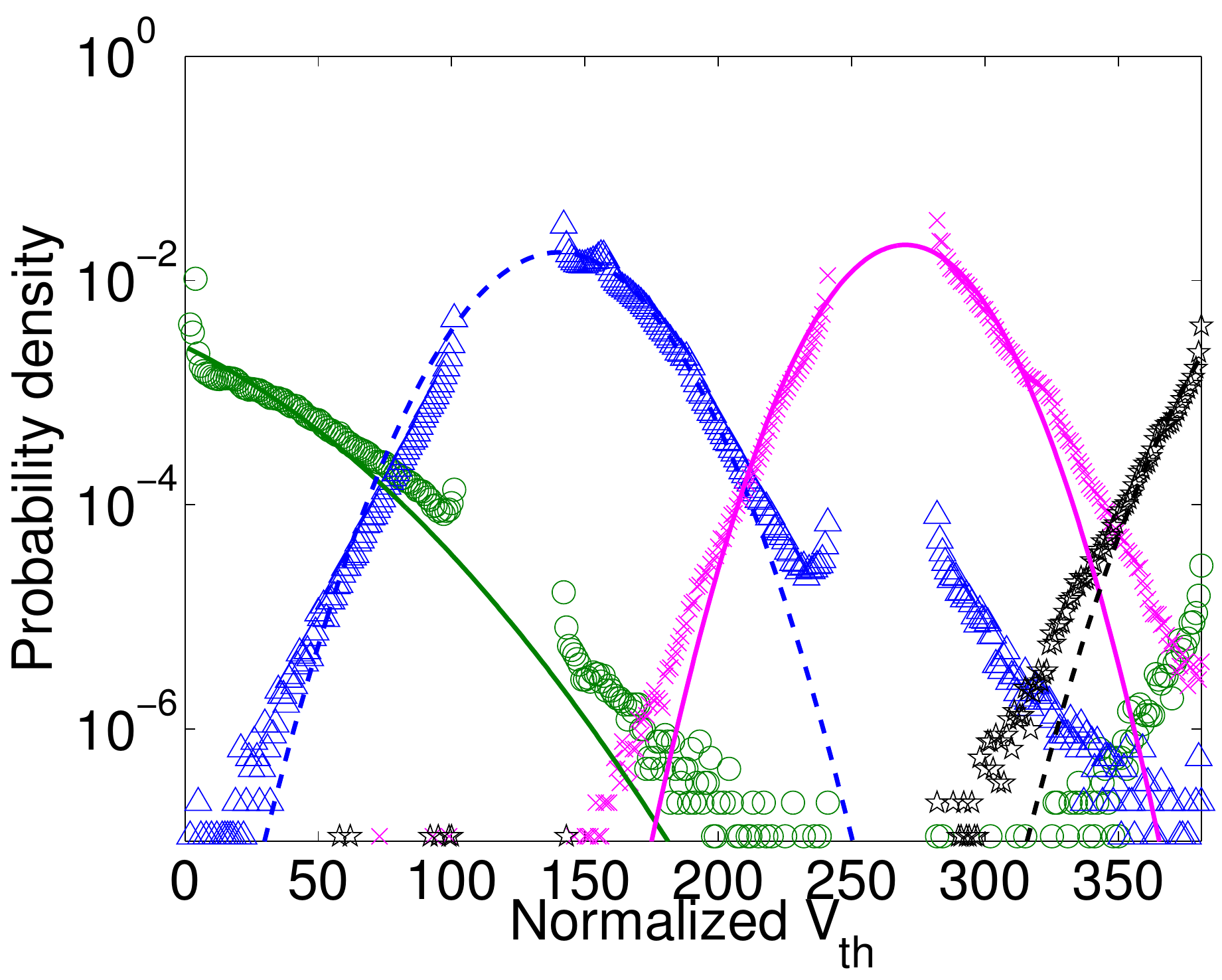}%
    \caption[]%
    {{20K P/E cycles}}
    \label{fig:static-gauss-20k}
  \end{subfigure}
  \caption[]%
  {Gaussian-based model (solid/dashed \chI{lines}) vs.\ data measured from real NAND flash
  chips (markers) under different P/E cycle counts.}
  \label{fig:static-gauss}
\end{figure}


We observe that the Gaussian-based model has two limitations, which
are demonstrated in Figure~\ref{fig:static-gauss}.  First, the
threshold voltage distribution \chI{of each state as} measured from real flash chips has a \emph{fatter tail} than that of
a Gaussian distribution, and the left and right tails of each state
have different sizes.  We observe the fat tail by comparing the
measured distribution \chI{of each state to the modeled distribution} when the probability density is low
(i.e., less than $10^{-4}$), and find that the measured distribution
has a much greater density than the modeled distribution at the
tail. This is because the Gaussian distribution has only two
parameters for each state, which capture only the center ($\mu$) and
the width ($\sigma$) of the distribution. We observe the asymmetric
tail by comparing \chI{the} densities of the left and right tails of the P2
state distribution.  Unfortunately, the Gaussian distribution has no
way to fine tune the ratio between the left and right tails, or the
ratio between the tails and the body of the distribution.

Second, the measured distribution demonstrates large second peaks \chI{in the distributions of the ER and P1 states}, which are not captured by
the Gaussian-based model.  These second peaks are evidence of a significant number of
program errors \chI{(see Section~\ref{sec:flash:pgmerase})}.
Figure~\ref{fig:static-gauss} shows that
the ER state distribution (\chI{the leftmost distribution}) has a \emph{second} peak that
shows up under the P3 state distribution, and that the P1
state distribution (\chI{the second distribution from the left}) has a second peak under the P2 state
distribution. These second peaks occur as a result of the two-step
programming mechanism used in MLC NAND flash memory.  As we discuss in
Section~\ref{sec:flash:pgmerase}, \emph{program errors} can be
introduced \chI{for the ER and P1 states} as a result of intermediate operations that take place
while a cell is partially programmed, which causes the LSB to be
misread.


\chI{
As we observe in Figure~\ref{fig:static-gauss}, both types of inaccuracies
occur \emph{throughout all P/E cycle counts} (from 2.5K to 20K), and
are \chI{not, as prior work \chI{had shown}~\cite{parnell.globecom14},}
exclusive to high wear-out scenarios (e.g., when the P/E cycle count is higher 
than the vendor-specified lifetime).  
\chI{The magnitudes of the fatter tails
and program error peaks increase
as wear-out (i.e., P/E cycle count) increases.}
}
As we can see, even though
the Gaussian-based model captures the general trend for the threshold
voltage distribution and is easy to compute, it is limited in its
accuracy, especially at higher P/E cycles.

Section~\ref{sec:static:compare} quantifies the modeling error and
computational requirements of the Gaussian-based model.

\subsection{Normal-Laplace-based Model}
\label{sec:static:laplace}

To overcome the limitations of the Gaussian-based model, prior
work~\cite{parnell.globecom14} proposes to modify the model to
increase its accuracy. This modified threshold voltage distribution
model {\em assumes} that the distribution of each state follows a
normal-Laplace distribution, and accounts for the peaks that result
from misprogramming some cells \chI{that should be} in the ER and P1 states into the P3 and
P2 states, respectively~\cite{parnell.globecom14}.

The normal-Laplace distribution combines the normal (Gaussian)
distribution with the Laplace distribution, which adds an exponential
component to both tails of the distribution. As we observe in
Figure~\ref{fig:static-gauss}, this is similar to the measured
behavior of the threshold voltage distribution.  Note that the figure
is in log scale, and \chI{as a result, the exponential 
component at the tails of the model appears as a straight line in the figure}.
By combining the two \chI{probability} distributions, we can maintain
the Gaussian distribution at the center of the distribution, and also
model the fat tail more accurately.


However, computing the normal-Laplace distribution becomes much more
complex than the Gaussian distribution, as \chI{the normal-Laplace distribution} is \chI{\emph{not}} a simple superposition of 
the Gaussian and Laplace distributions.  Equation~\ref{eqn:ncdf} shows how we compute the
cumulative distribution function for the normal-Laplace
distribution~\cite{reed.springer06}:
\begin{align}
    NCDF&(V, \mu, \sigma, \alpha, \beta, \lambda) \nonumber \\
  & = \Phi(Z) - \phi(Z)\frac{\beta R(\alpha\sigma - Z) - \alpha R(\beta\sigma + Z)}{\alpha + \beta}
\label{eqn:ncdf}
\end{align}
This distribution adds two new parameters,
$\alpha$ and $\beta$, which can be adjusted to model the right and left tail sizes,
respectively. In Equation~\ref{eqn:ncdf}, $Z = \frac{V - \mu}{\sigma}$ is the
z-score; $\Phi$ and $\phi$ are the cumulative distribution function and
probability density function of the standard Gaussian distribution, respectively; and $R(x) = \frac{1
- \Phi(x)}{\phi(x)}$ is Mills' ratio for the Gaussian distribution. $\Phi(Z)$ can be
obtained by looking up the z-table, as was done for Equation~\ref{eqn:gcdf}. $\phi(Z)$ can be
approximated as $\phi(Z) = \frac{\Phi(Z+\delta) - \Phi(Z-\delta)}{2\delta}$.

The normal-Laplace-based model adds two further parameters,
$\lambda_{ER}$ and $\lambda_{P1}$, to model the probability of program
errors \chI{occurring for cells programmed to the ER and P1 states,}
respectively. This model assumes that the
threshold voltage distribution of the cells with program errors has
the same shape (i.e., the same parameters) as the distribution of the
state the cells were incorrectly programmed into (e.g., the cells that
should be in the ER state but were programmed into the P3 state will
have a distribution with the same shape as the correct cells in the P3
state).  This is because once the cells are incorrectly programmed to
another state, they are treated as if they belong to that other state,
\chI{and} thus it is natural for them to follow the same distribution as the
correct cells in that state. Equation~\ref{eqn:nkx} shows how the
normal-Laplace model estimates the probability density for state $X$ being
misprogrammed to state $Y$ in bin $k$, which is denoted as $N_k(X)$:
\begin{align}
  N_k(X) = \; & (1 - \lambda_X) NCDF(V_k, \mu_X, \sigma_X, \alpha_X, \beta_X) \nonumber \\
           & + \lambda_X NCDF(V_k, \mu_Y, \sigma_Y, \alpha_Y, \beta_Y) \nonumber \\
           & - (1 - \lambda_X) NCDF(V_{k-1}, \mu_X, \sigma_X, \alpha_X, \beta_X) \nonumber \\
           & - \lambda_X NCDF(V_{k-1}, \mu_Y, \sigma_Y, \alpha_Y, \beta_Y)
\label{eqn:nkx}
\end{align}
This density is calculated as the difference of the $NCDF$ at the
bin's two boundaries, $V_k$ and $V_{k-1}$. The normal-Laplace-based
model allows each state to have at most five parameters (20 parameters
over all four states).  $\mu$ and $\sigma$ are the mean and standard
deviation, respectively; $\alpha$ and $\beta$ are the tail sizes; and
$\lambda_X$ is the probability that a cell that should actually be in
state $X$ is incorrectly programmed.

Following prior work~\cite{parnell.globecom14}, we eliminate four unnecessary parameters of the model, which include
$\lambda_{P2}$, $\lambda_{P3}$, $\beta_{ER}$, and
$\alpha_{P3}$. $\lambda_{P2}$ and $\lambda_{P3}$ are \chI{estimated as} zero, as
program errors for cells that should be in the P2 or
P3 states seldom occur. We also assume that the left and right tails
are the same size for the ER and P3 states (i.e., $\beta_{ER} =
\alpha_{ER}$ and $\alpha_{P3} = \beta_{P3}$), \chI{because the read-retry mechanism
prevents us from measuring the left tail of the ER state and the right tail of the P3 state}.
\chI{As we did in Section~\ref{sec:static:gaussian}, and following} prior work~\cite{parnell.globecom14}, we use Kullback-Leibler divergence error~\cite{kullback.math51} as the objective
\chI{function, and we} use the Nelder-Mead simplex method~\cite{nelder.computer65} with a reasonable initial guess to
learn the best parameters under different P/E cycles. 

Figure~\ref{fig:static-gmix} shows the modeled distribution \chI{of each state} as curves
with solid or dashed lines, \chI{and shows} the distribution measured from real chips using
markers.  As we can see, {\em the normal-Laplace-based model fits the
measured distribution much better than the Gaussian-based model}. The
modeled tails for the ER, P2, and P3 states follow the measured
distribution very closely, thanks to the tail size parameters. Also,
the distributions of the ER and P1 states take the program error rate
into account, and allow the model to correctly include two peaks for
the distributions of the ER and P1 states.

\begin{figure}[h]
  \centering
  \begin{subfigure}[h]{.4\linewidth}
    \centering
    \includegraphics[trim=0 7 0 0,clip,width=\linewidth] {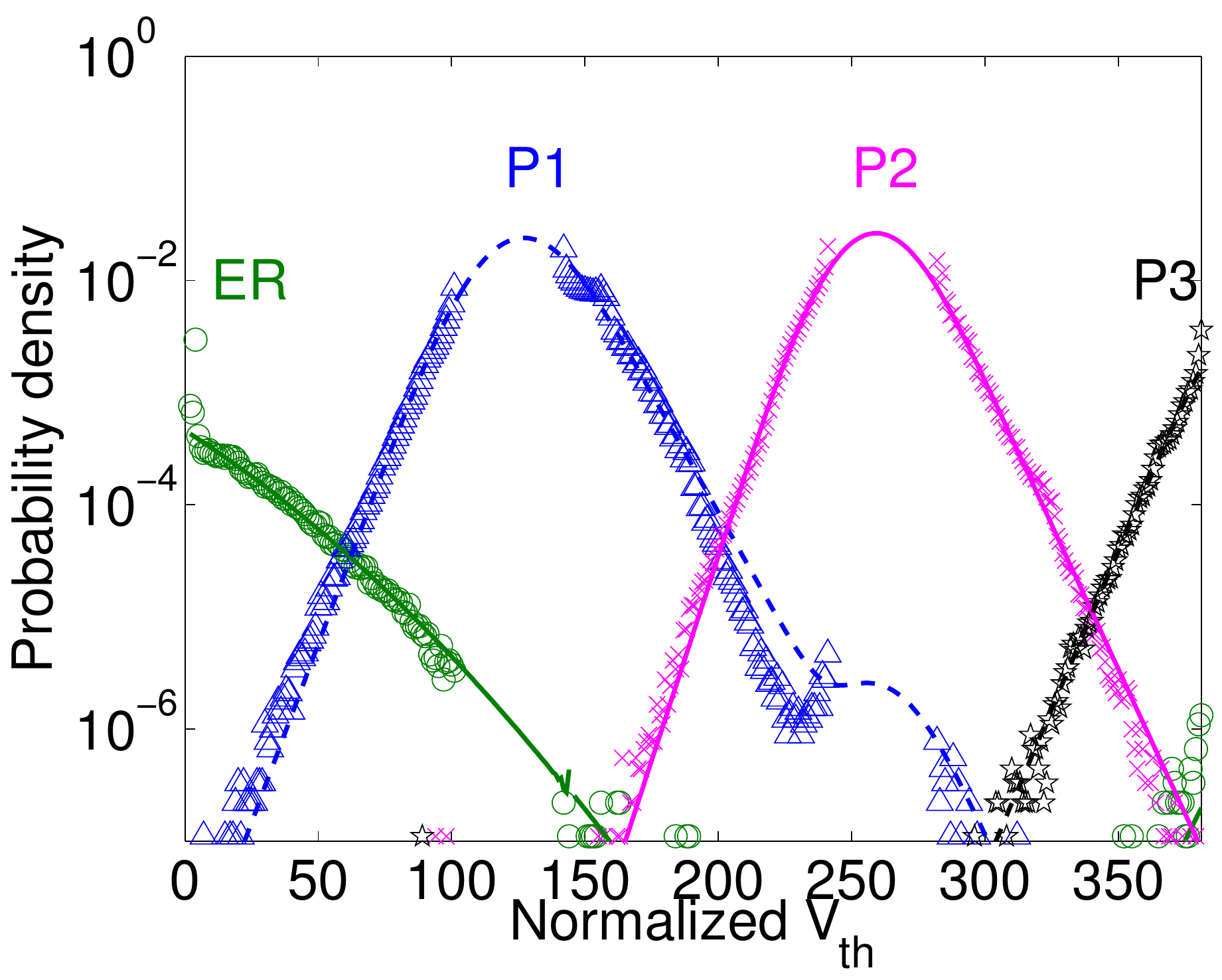}%
    \caption[]%
    {{2.5K P/E cycles}}
    \label{fig:static-gmix-2.5k}
  \end{subfigure}
  \begin{subfigure}[h]{.4\linewidth}
    \centering
    \includegraphics[trim=0 7 0 0,clip,width=\linewidth] {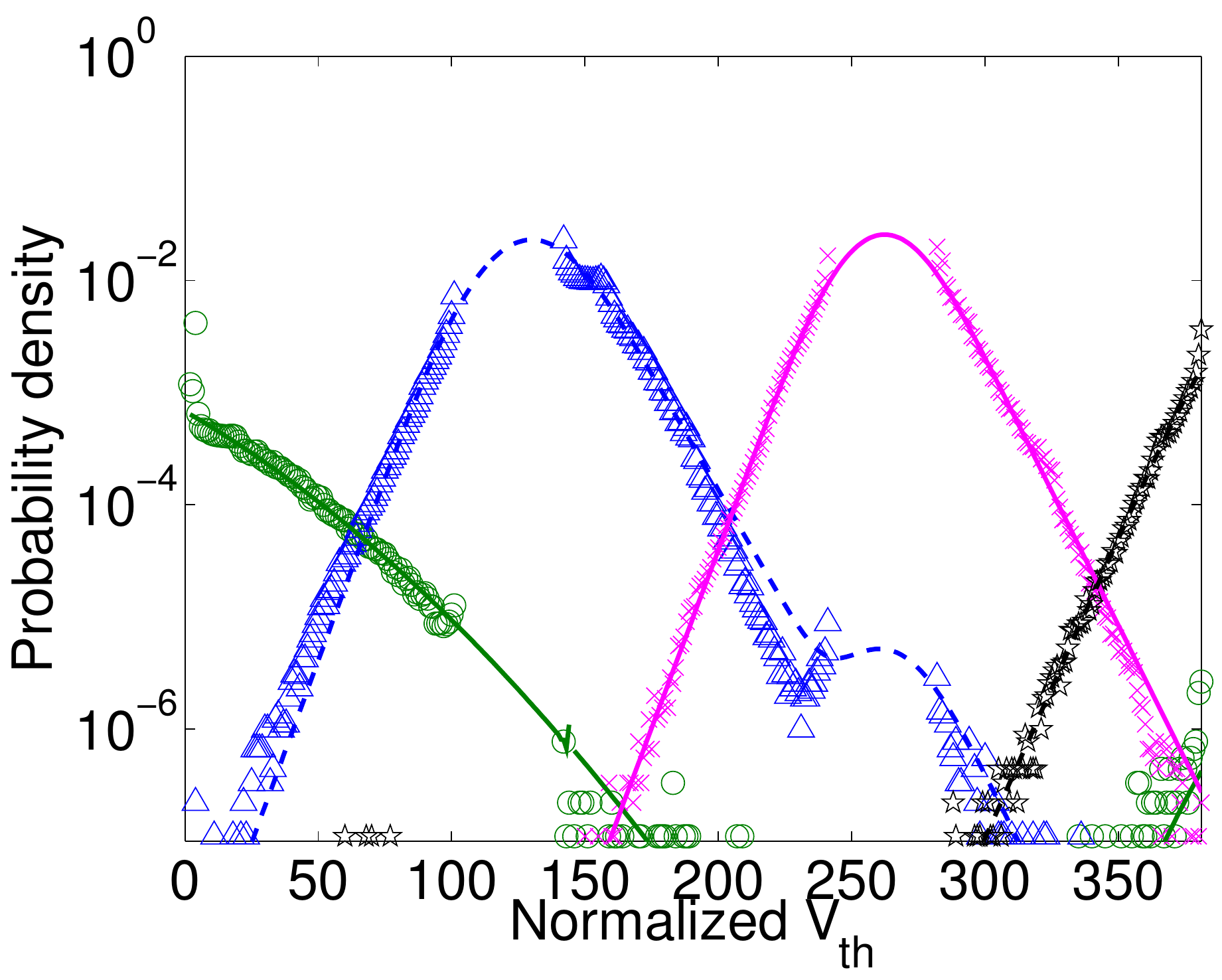}%
    \caption[]%
    {{5K P/E cycles}}
    \label{fig:static-gmix-5k}
  \end{subfigure}
  \begin{subfigure}[h]{.4\linewidth}
    \centering
    \vspace{8pt}
    \includegraphics[trim=0 7 0 0,clip,width=\linewidth]{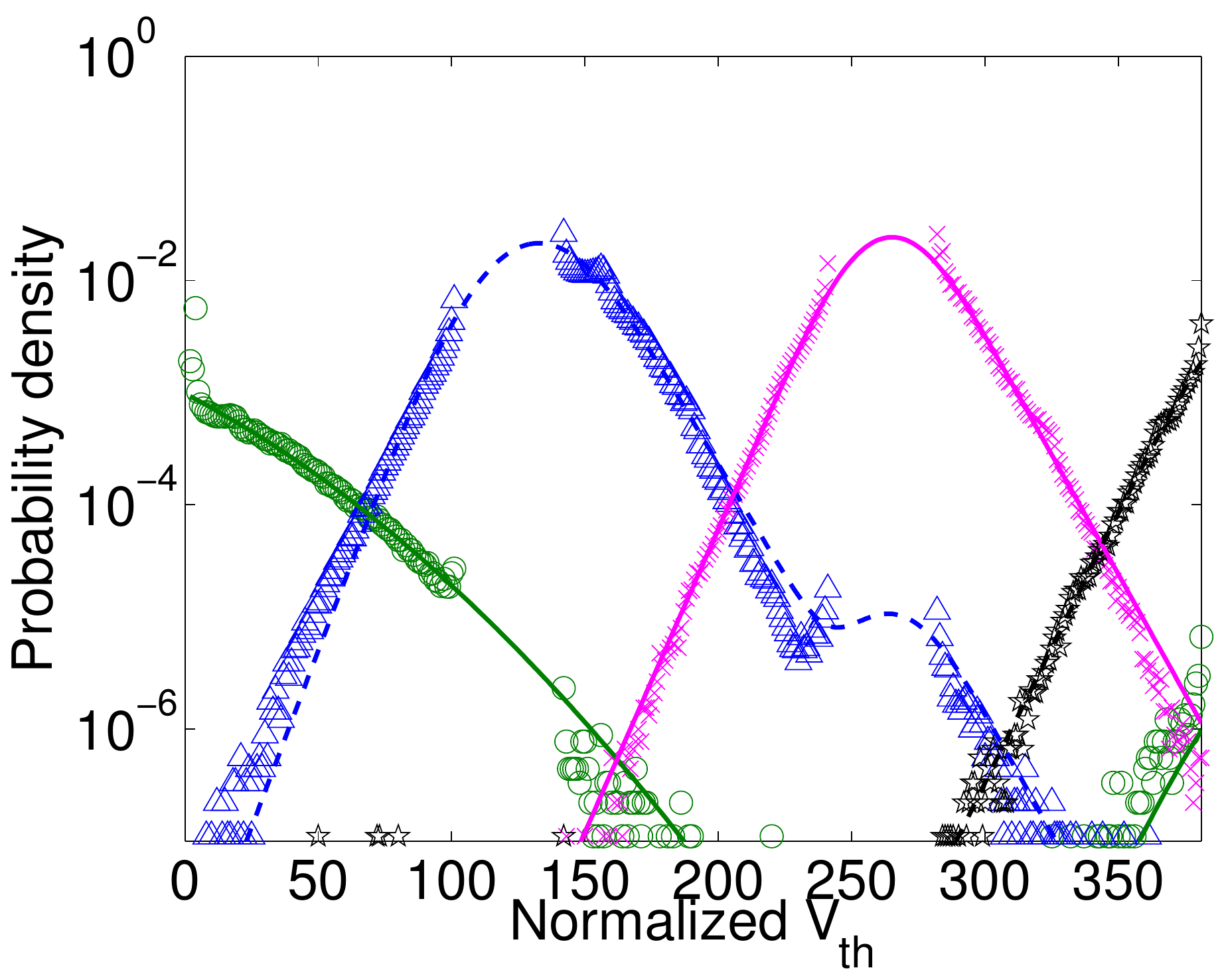}%
    \caption[]%
    {{10K P/E cycles}}
    \label{fig:static-gmix-10k}
  \end{subfigure}
  \begin{subfigure}[h]{.4\linewidth}
    \centering
    \vspace{8pt}
    \includegraphics[trim=0 7 0 0,clip,width=\linewidth]{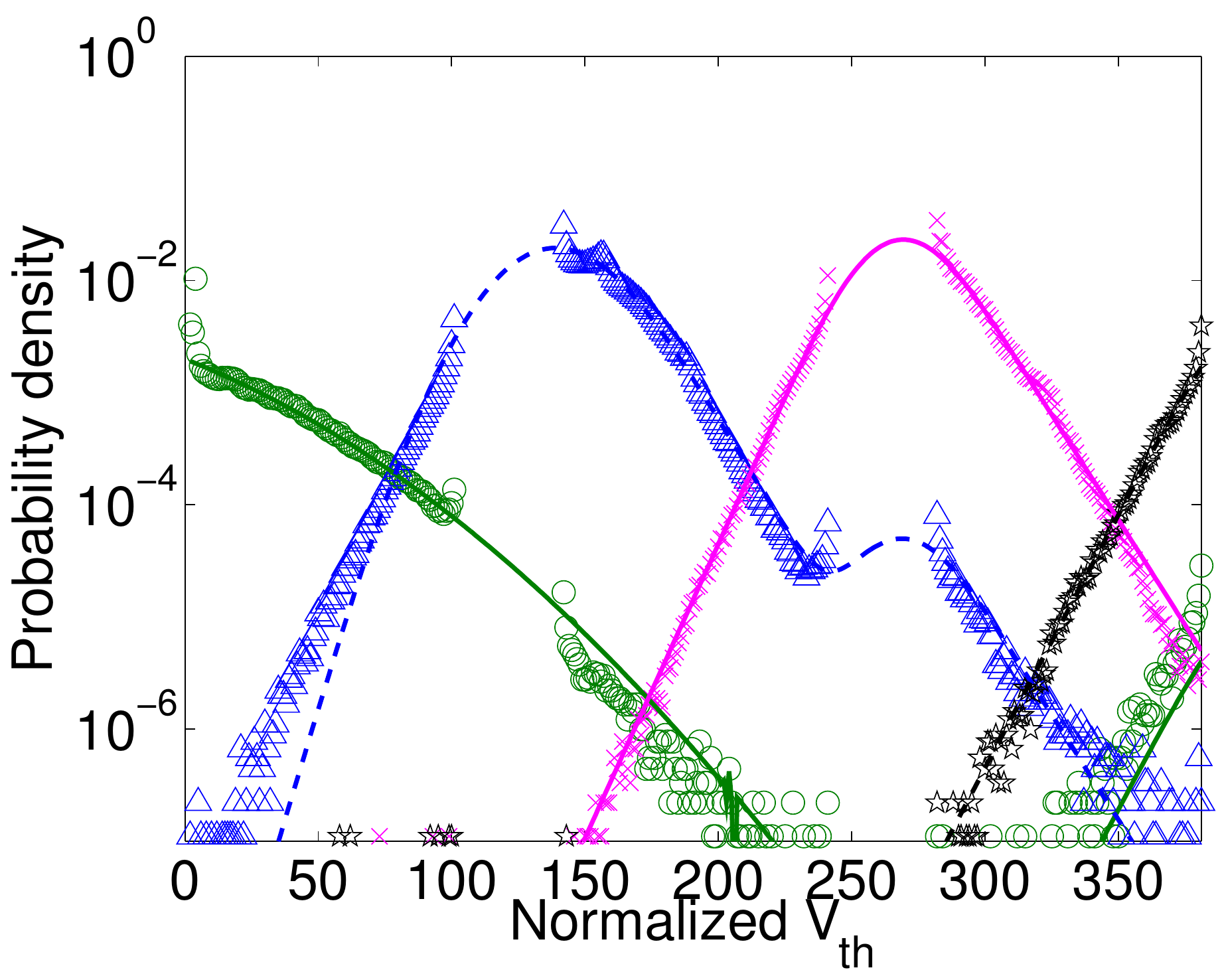}%
    \caption[]%
    {{20K P/E cycles}}
    \label{fig:static-gmix-20k}
  \end{subfigure}
  \caption[]%
  {Normal-Laplace-based model (solid/dashed \chI{lines}) vs.\ data measured from real NAND
  flash chips (markers) under different P/E cycle counts.}
  \label{fig:static-gmix}
\end{figure}



Unfortunately, although the normal-Laplace model is based on the Gaussian
model, the computational requirements of the model are much more
complex. This is not only because the model adds three more parameters for
each state, but also because we now cannot eliminate $\mu$ and
$\sigma$ using the z-score. Thus, directly computing the model requires
many more floating point operations than the Gaussian model (as we
demonstrate in Section~\ref{sec:static:compare}). As such, even though
the normal-Laplace model fits the measured threshold voltage distribution
accurately, it is less practical to implement.

Section~\ref{sec:static:compare} quantifies the modeling error and
computational requirements of the normal-Laplace-based model.

\FloatBarrier

\subsection{Student's t-based Model}
\label{sec:static:student}

Recall that we need a threshold voltage model that is both accurate and easy to
compute.  As we can see, the Gaussian-based model is simple and fast, but does
not meet the accuracy requirement. In contrast, the normal-Laplace-based model
fixes the accuracy problem, but uses significantly more complex calculations.
We \chI{thus aim to develop} a model that meets both of our requirements at the same time.

We propose to \chI{modify the Student's t-distribution~\cite{speigel.book92} so \chI{that} it can be used to} model the
threshold voltage distribution. \yixin{The Student's t-distribution is a
well-known distribution used in statistics that describes samples drawn from a
normally-distributed population. The Student's t-distribution is typically used
to estimate the true mean of a large, normally-distributed population whose 
standard deviation is unknown, using only a small sample from the population.
Compared to the standard normal distribution, the Student's
t-distribution uses an extra parameter, $\nu$, to represent the \emph{degrees of
freedom} (i.e., the ratio of the sample size relative to the population size). As
$\nu$ increases (i.e., the sample size becomes larger), the Student's
t-distribution moves closer to a standard normal distribution.}
However, instead of using the distribution for its original
purpose, we use this distribution for a completely different role. We find
that $\nu$ can be used to tune the size of the distribution tail. When $\nu \to
+\infty$, the Student's t-distribution becomes a standard Gaussian
distribution, which has a smaller tail.  When $\nu \to 0$, the distribution
instead has a fatter tail. We generalize the standard Student's t-distribution
using the z-score $Z = \frac{V-\mu}{\sigma}$, such that the center and the
width of the distribution can be scaled (as was done for the Gaussian
distribution). We also allow the left and right tails of the distribution to
have different values of $\nu$, which we denote as \chI{$\beta$ and $\alpha$ for the
left and right tails}, respectively. Thus, our modified Student's t-distribution
can fit our measured threshold voltage distribution better than the original
Student's t-distribution.


\chI{We use \emph{precomputation} to simplify the calculation of the cumulative
distribution function for our modified Student's t-distribution ($TCDF$). }
\chI{Similar to the precomputed z-tables available for the Gaussian-based model,
we look up values in the precomputed \emph{t-tables} commonly available for the 
Student's t-distribution to determine the $TCDF$ values.}
Each t-table contains $TCDF$ values over a
range of $Z$ values \emph{for a single $\nu$}.\footnote{\chI{The t-table can also
be \chI{thought of} as a two-dimensional array, where each entry corresponds to a unique
pair of ($Z$, $\nu$) values.}}  Equation~\ref{eqn:tcdf} shows
how we calculate $TCDF$ using the precomputed t-table: 
\begin{align}
  TCDF(V, \mu, \sigma, \alpha, \beta) = \begin{cases}
    \textit{t-table}_{\beta}(Z) & V \leq \mu \\
    \textit{t-table}_{\alpha}(Z) & V > \mu
  \end{cases}
  \label{eqn:tcdf}
\end{align}
We first compare $V$ with $\mu$ to observe whether $V$ is on the left
side or the right side of the distribution. Then, depending on the
result of the comparison, we use the corresponding tail parameter
$\alpha$ or $\beta$ as $\nu$ to select the correct t-table. Finally,
we compute the z-score $Z$ to look up $TCDF$ in the selected t-table.

Equation~\ref{eqn:tkx} shows how our Student's t-based model estimates the
density for cells that should be in state $X$ but are incorrectly programmed to
state $Y$ in bin $k$, which is denoted as $T_k(X)$:
\begin{align}
  T_k(X) = \; & (1 - \lambda_X) TCDF(V_k, \mu_X, \sigma_X, \alpha_X, \beta_X) \nonumber \\
           & + \lambda_X TCDF(V_k, \mu_Y, \sigma_Y, \alpha_Y, \beta_Y) \nonumber \\
           & - (1 - \lambda_X) TCDF(V_{k-1}, \mu_X, \sigma_X, \alpha_X, \beta_X) \nonumber \\
           & - \lambda_X TCDF(V_{k-1}, \mu_Y, \sigma_Y, \alpha_Y, \beta_Y)
\label{eqn:tkx}
\end{align}
Similar to the normal-Laplace-based model (Section~\ref{sec:static:laplace}), our Student's t-based model uses
$\lambda$ to estimate such program errors caused by the two-step
programming mechanism (see Section~\ref{sec:flash:pgmerase}).
Again, like the normal-Laplace-based model, our Student's t-based
model assumes that the distribution of these cells has the same
parameters as the cells correctly programmed into state $Y$.

We set $\lambda_{P2}$ and $\lambda_{P3}$ to zero, $\beta_{ER} = \alpha_{ER}$,
and $\alpha_{P3} = \beta_{P3}$ for the same reasons as the
normal-Laplace-based model \chI{(see Section~\ref{sec:static:laplace})}.  Putting everything together, we use
Kullback-Leibler divergence error~\cite{kullback.math51} as our objective function, and use the
Nelder-Mead simplex method~\cite{nelder.computer65} with a reasonable initial guess to learn the best
parameters for the model under different P/E cycles, as described in
Section~\ref{sec:static:gaussian}.

Figure~\ref{fig:static-gtmix} shows \chI{our} modeled Student's t-based
distribution as curves with solid or dashed lines, once again showing the
distribution measured from real chips with markers. The figure shows that \chI{our}
Student's t-based model fits perfectly when the probability density is
greater than $10^{-4}$. The differences between our Student's t-based
model and the measured distribution are within $10^{-6}$.  The
difference becomes non-trivial only for the left tail of the P1 state
and the right tail of P2 state. The accuracy improvements over the
Gaussian model are similar \chI{to} those \chI{of} the normal-Laplace-based
model.  This shows that our Student's t-based model, like the
normal-Laplace-based model, makes good use of its extra parameters
(both models have 16 parameters) to cater to the program errors and
fat tails that the measured distribution has.

\begin{figure}[h]
  \centering
  \begin{subfigure}[h]{.4\linewidth}
    \centering
    \includegraphics[trim=0 7 0 0,clip,width=\linewidth] {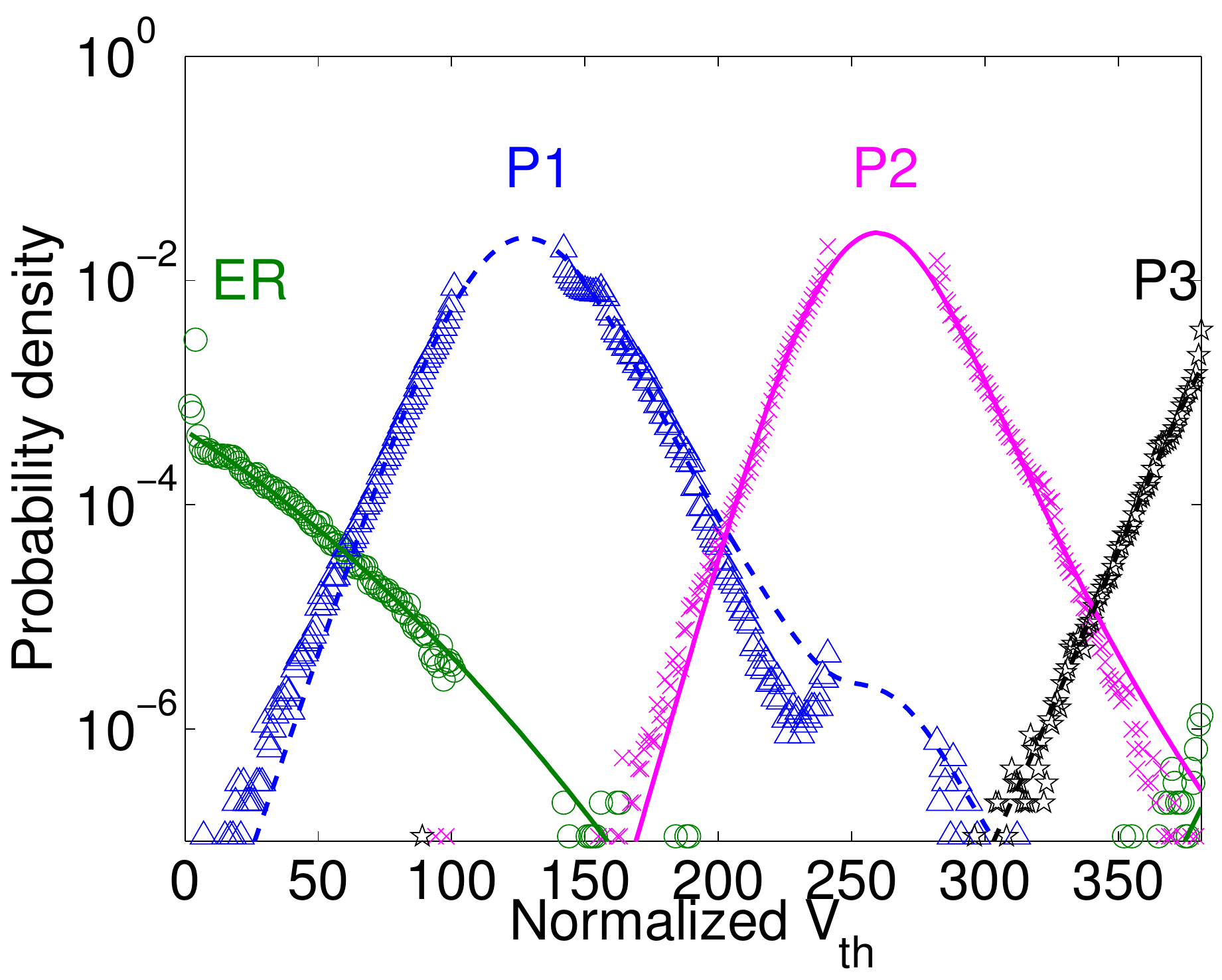}%
    \caption[]%
    {{2.5K P/E cycles}}
    \label{fig:static-gtmix-2.5k}
  \end{subfigure}
  \begin{subfigure}[h]{.4\linewidth}
    \centering
    \includegraphics[trim=0 7 0 0,clip,width=\linewidth] {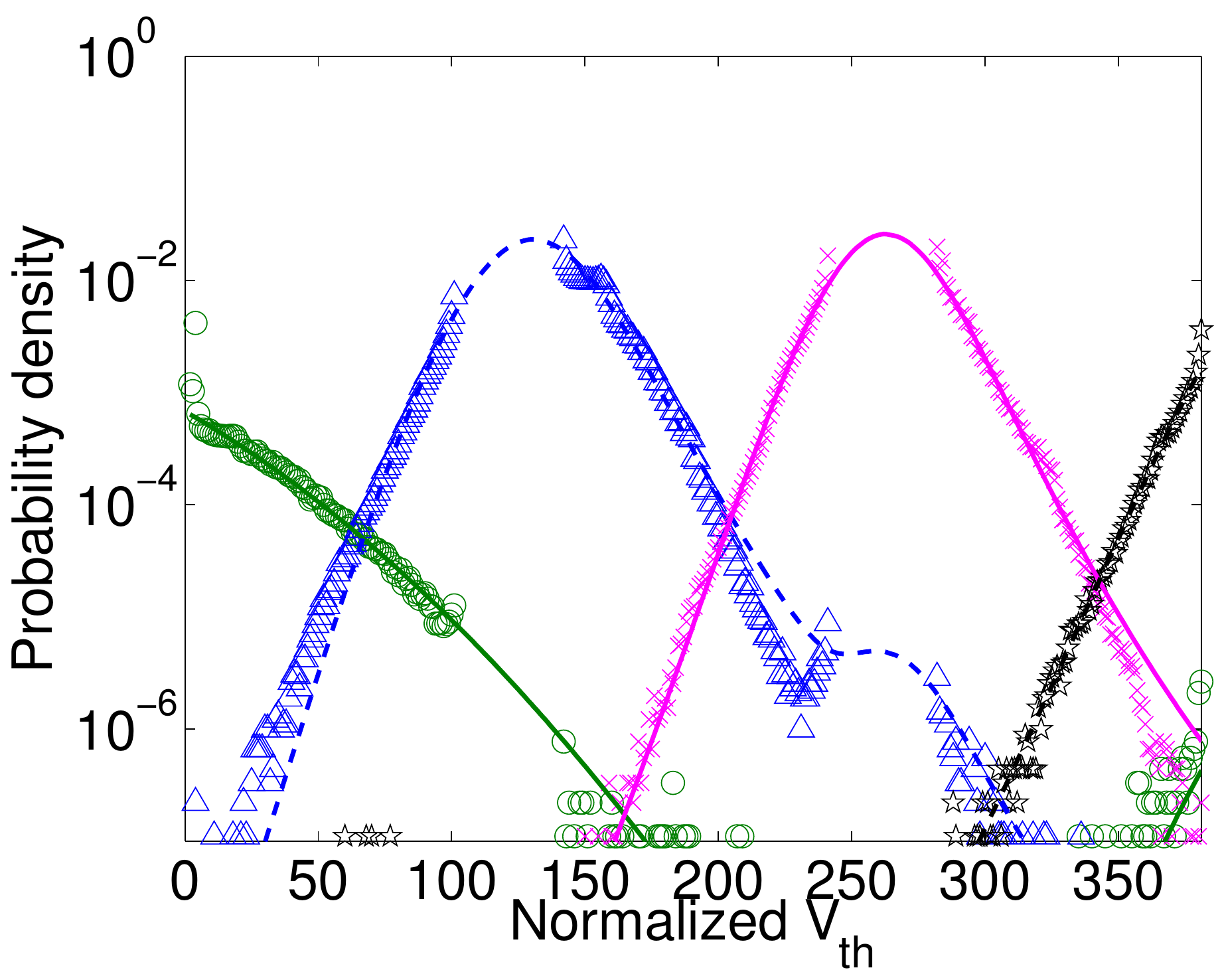}%
    \caption[]%
    {{5K P/E cycles}}
    \label{fig:static-gtmix-5k}
  \end{subfigure}
  \begin{subfigure}[h]{.4\linewidth}
    \centering
    \vspace{8pt}
    \includegraphics[trim=0 7 0 0,clip,width=\linewidth]{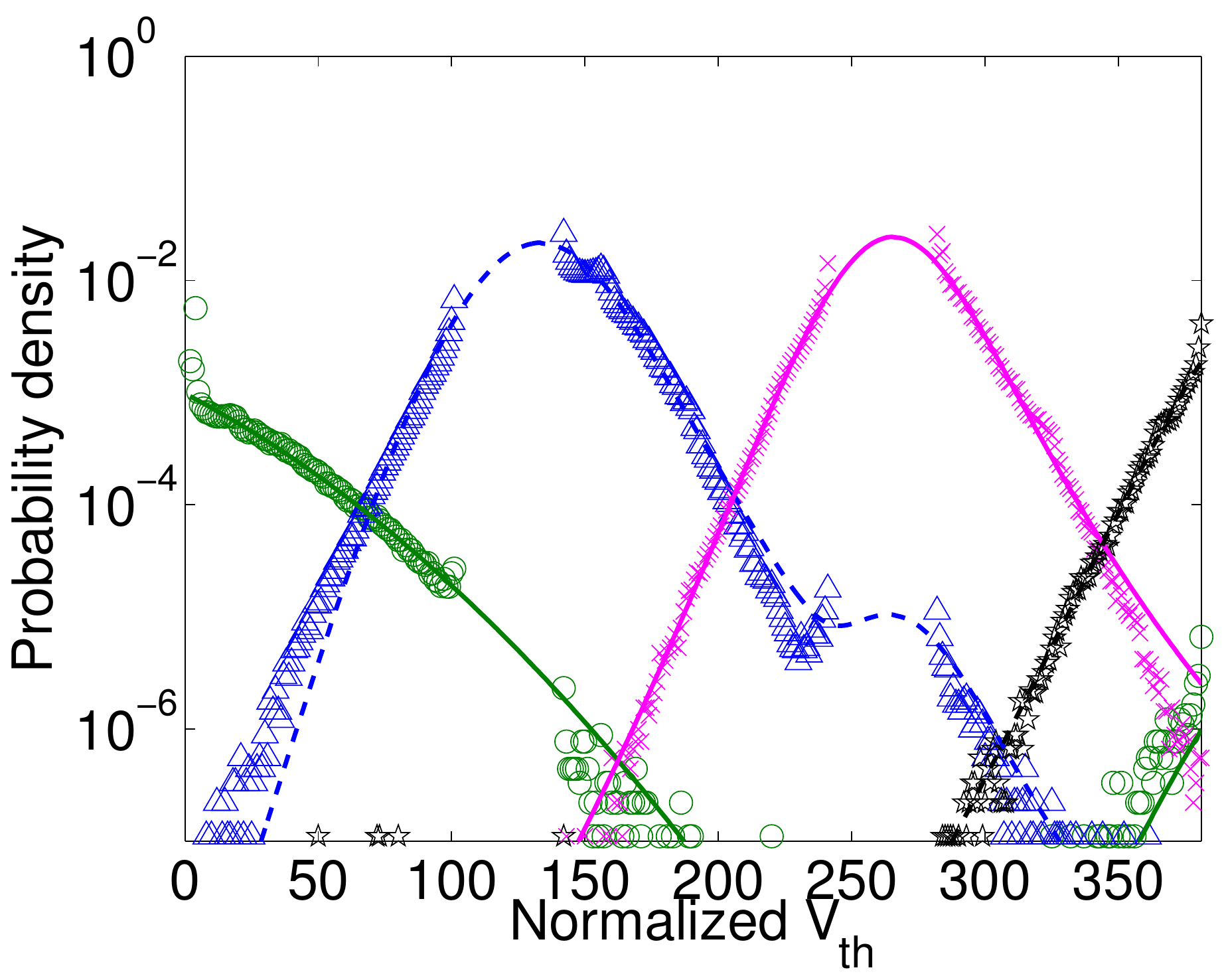}%
    \caption[]%
    {{10K P/E cycles}}
    \label{fig:static-gtmix-10k}
  \end{subfigure}
  \begin{subfigure}[h]{.4\linewidth}
    \centering
    \vspace{8pt}
    \includegraphics[trim=0 7 0 0,clip,width=\linewidth]{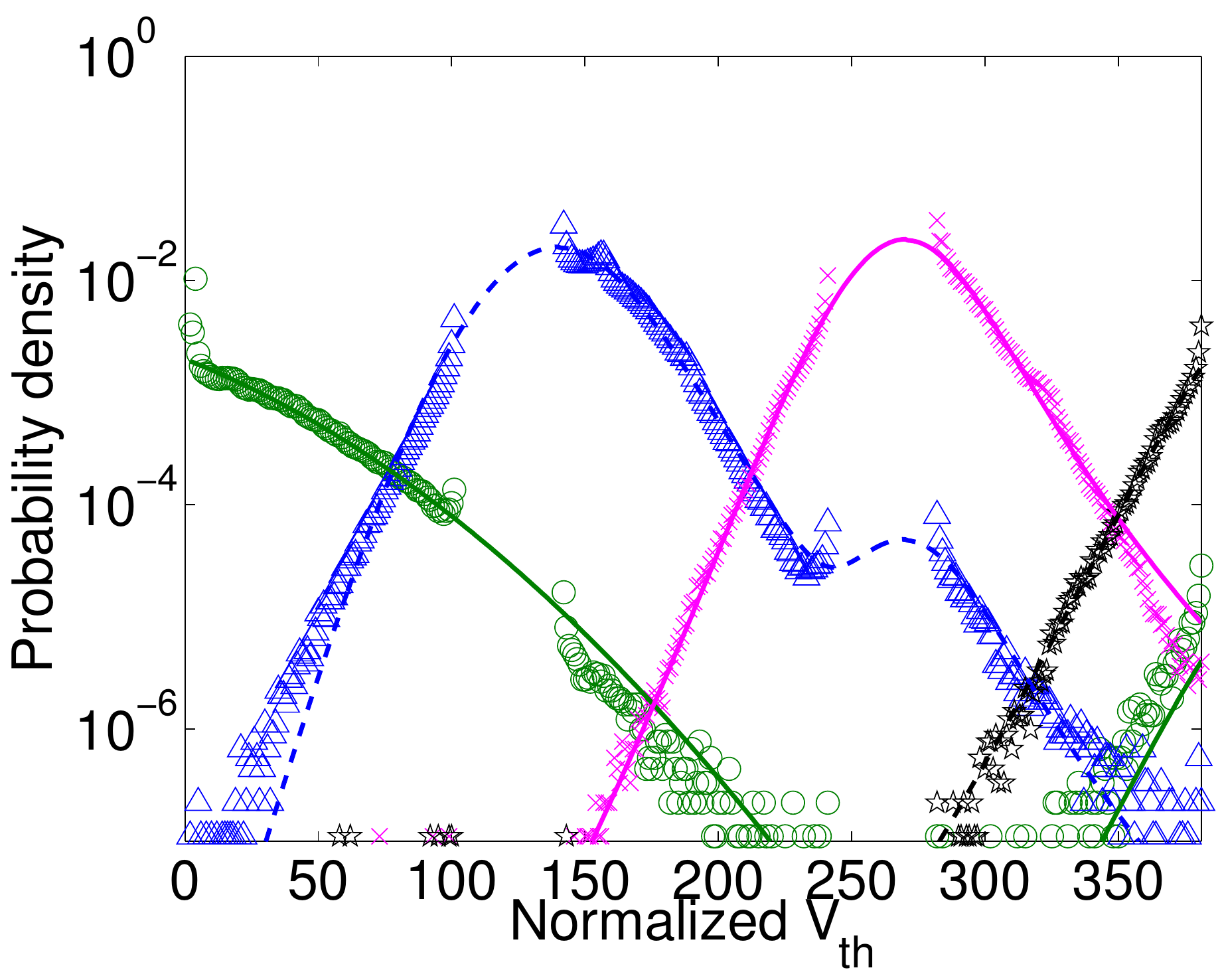}%
    \caption[]%
    {{20K P/E cycles}}
    \label{fig:static-gtmix-20k}
  \end{subfigure}
  \caption[]%
  {Our new Student's t-based model (solid/dashed \chI{lines}) vs.\ data measured from real NAND flash
  chips (markers) under different P/E cycle counts.}
  \label{fig:static-gtmix}
\end{figure}
\FloatBarrier


\subsection{Model Validation and Comparison}
\label{sec:static:compare}

{\bf Accuracy.} To quantitatively compare the accuracy of each model
and validate them, we compute the Kullback-Leibler (K-L)
divergence~\cite{kullback.math51} between the modeled and the measured
\chI{distributions}, as K-L divergence measures the difference between two
distributions (see Section~\ref{sec:static:gaussian}). Table~\ref{tbl:static-error} and Figure~\ref{fig:static-error} show the modeling error
of the three models across a range of P/E cycle counts. \chI{We} observe
two types of behavior shared by all three models. First, as the P/E
cycle count increases, the modeling error increases. Second, the
increase in modeling error is more rapid at \chI{\emph{smaller}} P/E cycle \chI{counts},
and slower at higher P/E cycle \chI{counts}.  As we see in
Section~\ref{sec:dynamic}, this is because the threshold voltage
distribution is affected by the P/E cycling effect more significantly
\chI{at smaller} P/E cycle counts.

\begin{table}[h]
\centering
\small
\caption{
  Modeling error of the evaluated threshold voltage distribution models,
  at various P/E cycle counts.
}
\setlength{\tabcolsep}{.3em}
\begin{tabular}{lccccccccccc}
  \toprule
  \textbf{P/E Cycles}      & 0 & 2.5K & 5K & 7.5K & 10K & 12K & 14K & 16K & 18K & 20K & \textbf{AVG} \\
  \midrule
  \textbf{Gaussian}       & .99\% & 1.8\% & 1.6\% & 1.8\% & 1.9\% & 2.4\% & 3.1\% & 8.7\% & 2.1\% & 2.3\% & \textbf{2.6\%} \\
  \textbf{Normal-Laplace} & .34\% & .46\% & .55\% & .61\% & .63\% & .67\% & .68\% & .70\% & .67\% & .67\% & \textbf{.61\%} \\
  \textbf{Student's t}    & .37\% & .51\% & .61\% & .68\% & .70\% & .76\% & .76\% & .78\% & .76\% & .78\% & \textbf{.68\%} \\
  \bottomrule
\end{tabular}%
\label{tbl:static-error}
\end{table}

\begin{figure}[h]
\centering
\includegraphics[trim=20 0 30 0,clip,width=.7\linewidth]{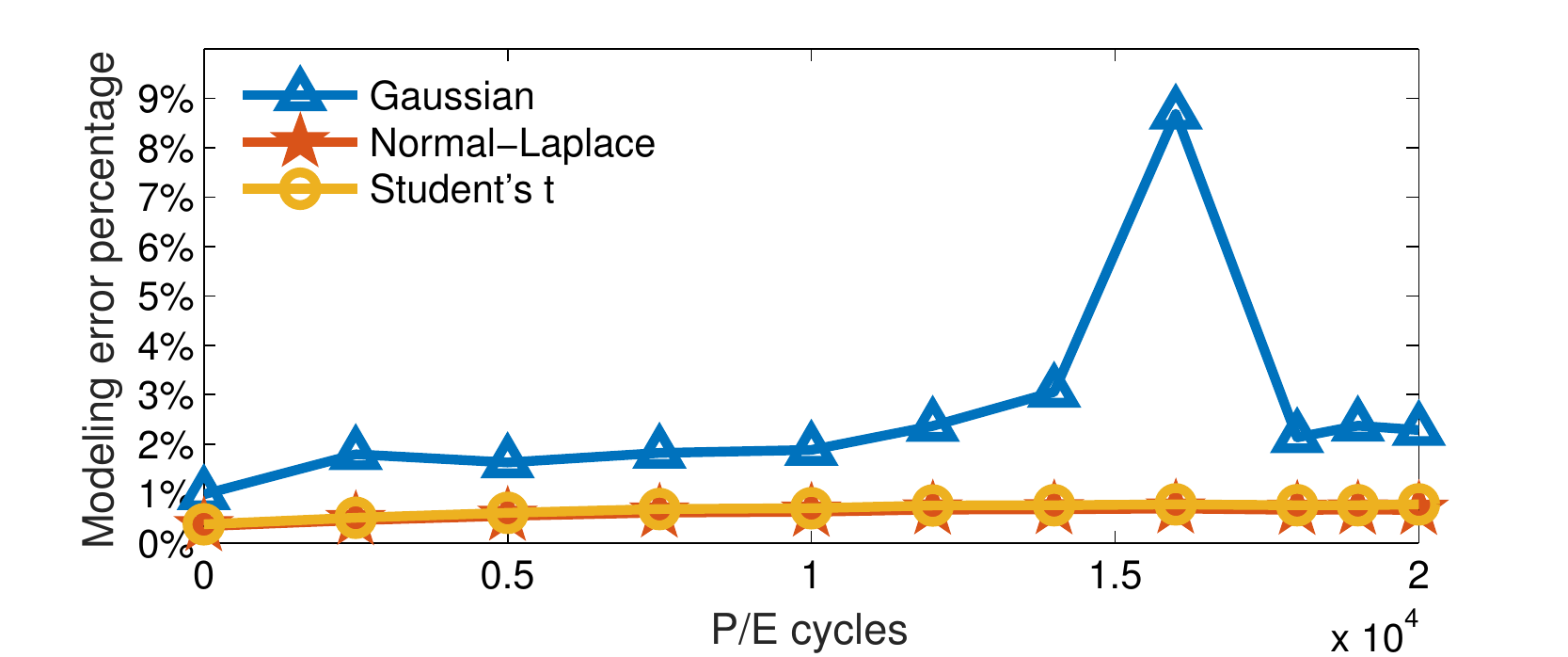}%
\caption{
  Modeling error of the evaluated threshold voltage distribution models,
  at various P/E cycle counts.
}
\label{fig:static-error}
\end{figure}
\FloatBarrier

Comparing the three \chI{models} in Figure~\ref{fig:static-error}, we make
two observations. First, the average \chI{Kullback-Leibler divergence} error for
the Gaussian-based model is 2.64\%, which is 4.32x and 3.88x greater than
\chI{the error of} the normal-Laplace-based and our Student's t-based models,
respectively.  We also see that this error can be as large as 8.7\%, leading to high inaccuracy.
This is mainly due to two limitations of the Gaussian-based model.  As
mentioned in Section~\ref{sec:static:gaussian}, the Gaussian-based
model (1)~cannot adjust its tail size to fit with the fat and
asymmetric \chI{tails} of the observed \chI{distributions of the voltage states}, and (2)~does not
account for misprogrammed cells that form a second peak in the
\chI{distributions of the ER and P1 states}.

Second, the modeling errors of the normal-Laplace-based and our Student's
t-based models are very close (averaged across all tested P/E cycles,
0.61\% for the normal-Laplace-based model, and 0.68\% for our
Student's t-based model). The maximum difference in error between
these two models is 0.11\%, at 20K P/E cycles (already well beyond the
rated lifetime of the flash chip, which is 3K P/E cycles).

{\bf Complexity.} \chI{All three of the models require \chI{online} \emph{iterative
computation} \chI{whenever the flash controller needs to generate a characterization
of the threshold voltage distribution at a new P/E cycle count.  For each iterative computation,} 
hundreds to thousands of iterations of the model computation algorithm must
be executed before the model reaches high accuracy (i.e., the model converges,
or in other words, reaches \emph{convergence}).}
As we alluded to in
Section~\ref{sec:static:laplace}, while the normal-Laplace-based model
is accurate, it requires significant computation \chI{during each iteration}, 
and cannot be
precomputed and stored in a lookup table, making it less practical for
use within a flash controller. To compare the complexity of the three
models, we summarize their computation overhead in terms of the number
of floating-point operations and table lookups performed for each
iteration, as well as their storage overhead in terms of lookup table
size. Table~\ref{tbl:complexity} compares the three models. As we can
see, the normal-Laplace-based model requires 91,200 operations per
iteration (which involves computing $N_k(X)$ for four states \chI{in each of the 304
threshold voltage bins}). 
Assuming that each floating-point operation
takes the same number of cycles, the normal-Laplace-based model is
10.71x slower than the Gaussian-based model.  In contrast, our Student's
t-based model takes 4.41x less computation time than the
normal-Laplace-based model, with near-identical accuracy.  Our
Student's t-based model is only 2.43x slower than the Gaussian-based
model, but has a 74\% smaller modeling error.

\begin{table}[h]
\centering
\small
\caption{
  Computation and storage complexity comparison for the three evaluated threshold distribution models.
}
\begin{tabular}{cccc}
  \toprule
  \textbf{Model} & Gaussian & Normal-Laplace & Student's t \\
  \midrule
  \textbf{Operations} & 8512 & 91200 & 20672 \\
  \textbf{Storage} & 640B & 3.84KB & 25.6KB \\
  \bottomrule
\end{tabular}%
\label{tbl:complexity}
\end{table}
\FloatBarrier

The third row of the table shows the storage overhead for the z-table
or t-tables used by each model \chI{(which are populated in the
\chI{flash controller's} DRAM when the flash device is powered up)}. The Gaussian-based model needs only
640B to store the useful range of the z-table. The
normal-Laplace-based model requires a larger lookup range for the
z-table, increasing the storage overhead to 3.84KB.  Our Student's
t-based model requires storing multiple t-tables \chI{(one table per value
of $\nu$)}, and uses 25.6KB of storage in total.  We find that all three storage
overhead values are negligible, as these tables are easily stored
within the flash controller's DRAM, which is usually \chI{sized to be a fixed fraction} of the flash storage
capacity (e.g., 1GB memory for a 512GB drive).

{\bf Latency.} \chI{
The flash controller builds a threshold voltage distribution model in two 
steps: \emph{characterization} and \emph{model computation}.  First, we identify the threshold voltages of each cell in \chI{a sampled}
flash wordline by performing 303~read operations, one read for each \chI{read reference voltage level}
(using the approach described in Section~\ref{sec:vthmodel:overview:methodology}).
This \emph{characterization} takes 30.3~ms for the wordline, assuming a typical read
latency of 100$\mu$s.  Second, once characterization is complete, the controller 
\emph{computes} the model, using a combination of the precomputed tables stored in 
DRAM and the characterized voltages.
To calculate the overhead, we assume that each of the 
models takes 1000~iterations to converge, and that computation is performed on
a 1GHz embedded processor that completes one instruction per cycle.
}

\chI{
Figure~\ref{fig:latency} shows the overall latency for the three models
we evaluate, broken down into characterization latency (which is the same for
all three models) and model computation latency.
}
The computation
overhead of the normal-Laplace-based model dominates its overall latency, while
the computation overhead of our Student's t-based model is much smaller than
the characterization latency.  As a result, our Student's t-based model has a
58.0\% lower overall latency than the normal-Laplace-based model. Since the
fixed characterization latency dominates \chI{overall latency in} both our Student's t-based model
and the Gaussian-based model, our model is only 31.3\% slower in overall
latency than the Gaussian-based model, while it reduces modeling error by 74\%.


\begin{figure}[h]
\centering
\includegraphics[trim=0 208 0 5, clip, width=.7\linewidth]{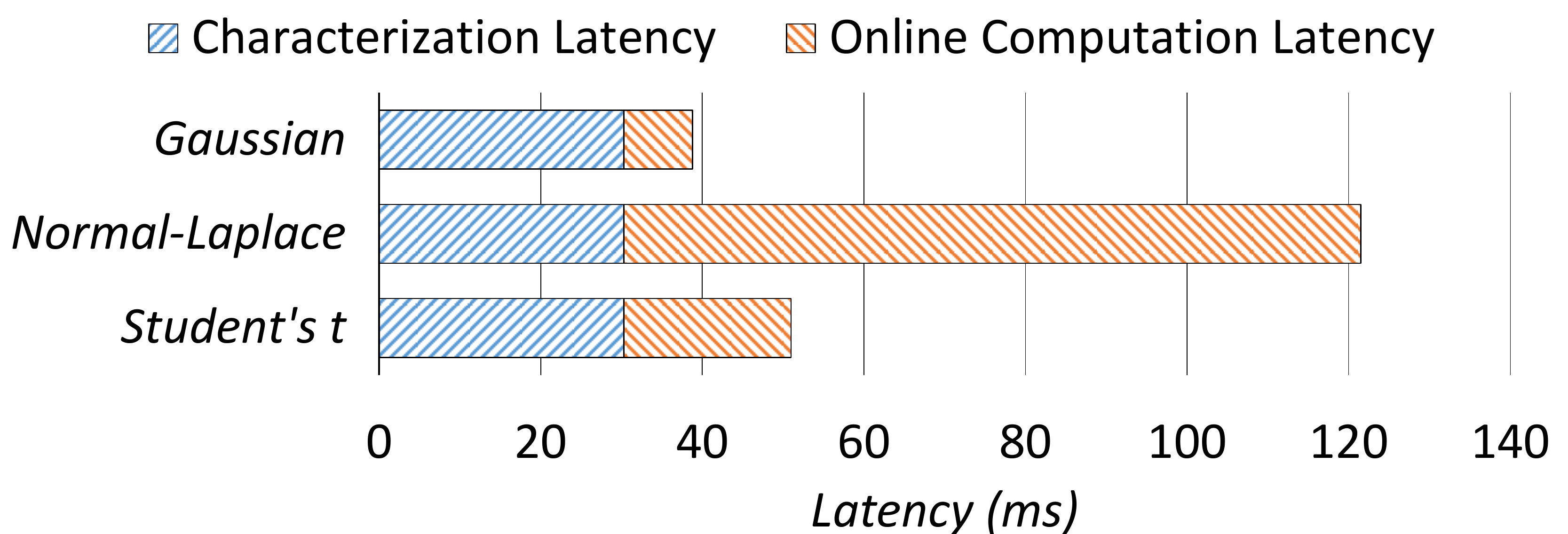}%
\caption{
  Overall latency breakdown of the three evaluated threshold voltage 
  distribution models for static modeling.
}
\label{fig:latency}
\end{figure}
\FloatBarrier

\chI{
The frequency with which the characterization and modeling procedure is triggered depends purely on
the \chI{application making use of} the threshold voltage distribution model.  Note that the choice of
model should not change the frequency at which the procedure is executed (as
each model provides an equivalent end result). As an example, we can determine
the amortized overhead per 4KB read/write operation for one application of our
model, which predicts the optimal read reference voltage (see 
Section~\ref{sec:application:opt}). The prediction mechanism requires us to
repeat the characterization and modeling procedure only once every 1000~P/E cycles. For a flash device with
512 pages per block, if we conservatively assume a read-to-write ratio of 1:1, 
the average overhead amortized over each read/write operation is 49.8ns using our static 
Student's t-based model~\cite{luo.tr16}.
}

\chI{
{\bf Summary.} In summary, the majority of the accuracy improvement over the Gaussian-based model comes
from (1)~accounting for the program errors for the erased and P1 states, and 
(2)~accounting for the fat tails of each state. Our Student's t-based model, as well
as the previously-proposed normal-Laplace-based model, both contain these
improvements, and hence achieve similar accuracy.

\chI{Our} Student's t-distribution based model has much lower complexity than the
normal-Laplace-based model due to its ability to exploit precomputation. We show
in Section 4.3 that the CDF of the Student's t-distribution can be simplified into
a simple table lookup using the z-score, $Z$, and the \chI{degrees} of freedom, $\nu$.
We are unaware of a similar precomputation-based approach that can be applied to
the normal-Laplace model.
}


We conclude that our new Student's t-based model achieves the high
accuracy of the normal-Laplace-based model while requiring significantly
less complexity and latency to compute. As such, we believe that our Student's t-based
model meets the requirements of accuracy and simplicity, and is a
practical model for implementation within the flash controller.


\section{Dynamic Modeling}
\label{sec:dynamic}

We now construct a \emph{dynamic} threshold voltage distribution model, building off
of our Student's t-based static model in
Section~\ref{sec:static:student}, to capture how the threshold voltage distribution \emph{changes} as the
program/erase (P/E) cycle count increases.  Again, we must ensure that this
dynamic model is accurate, and that it is easy to compute, as we aim to
implement the model within the flash controller.  To construct the model, we
first analyze how each of the individual parameters making up our Student's
t-based model change over the P/E cycle count
(Section~\ref{sec:dynamic:trend}).  By analyzing the meaning of each parameter
and observing how it changes, we gain new insights on how the
threshold voltage shifts with increasing P/E cycle count.  We then use these
new insights to construct a model using the power law, which can
successfully predict the \chI{\emph{future}} changes to each of these parameters based on the
\chI{\emph{current}} threshold voltage distribution (Section~\ref{sec:dynamic:fitting}).
Finally, we validate this model (Section~\ref{sec:dynamic:validation}).

\subsection{Static Model Trends Over P/E Cycles}
\label{sec:dynamic:trend}

In order to analyze and observe how the parameters for our Student's t-based
model change as P/E cycle count increases, we first need to understand what
each parameter means.  As we discuss in Section~\ref{sec:static:student}, our
Student's t-based model has 16 parameters. Four of them are the mean values for
each state $X$'s threshold voltage distribution ($\mu_X$).  Another four
parameters are the standard deviation values of \chI{the threshold voltage distribution of} each state $X$ ($\sigma_X$).
Three of them are the left tail sizes of the P1, P2, and P3 state distributions
($\beta_X$), and another three are the right tail sizes of the ER, P1, and P2
state distributions ($\alpha_X$). (Recall from \chI{Section~\ref{sec:static:laplace}}
that the left tail of the ER state and the right tail of the P3 state cannot be
observed experimentally, so we assume that they equal the right tail of the ER
state and the left tail of the P3 state, respectively.)  The remaining two
parameters are the probability of program errors, \chI{occurring for
cells programmed into the ER and P1 states ($\lambda_X$)}.

\textbf{Mean.}
The mean value of each state represents the center of the distribution. In our
Student's t-based model, the majority of the mass of the threshold voltage
\chI{distribution} for each state is
near the center.  Thus, \chI{a} change in the mean reflects how \chI{the} P/E cycle count \chI{generally}
affects \chI{the threshold voltages of \emph{all} cells in each state}. 

Figure~\ref{fig:dynamic-mean} plots the mean values obtained from \chI{sample Student's
t-based models constructed over a range of 20K P/E cycles}, shown as circles.  The x-axis
shows the P/E cycle count, while the y-axis shows the normalized threshold
voltage of the mean. Each graph plots the mean value for a different state,
which is labeled at the top of the graph. We make three observations from this
figure. First, the mean value of each state's distribution increases monotonically
with P/E cycle count.  Second, the mean \chI{value increases} faster
at lower P/E cycle counts, then \chI{slows} down to a constant rate of increase after
5K P/E cycles. Third, the mean value shifts \chI{more quickly for}
lower threshold voltage states (ER, P1).  

\begin{figure}[h]
\centering
\includegraphics[trim=22 10 35 10,clip,width=.7\linewidth]{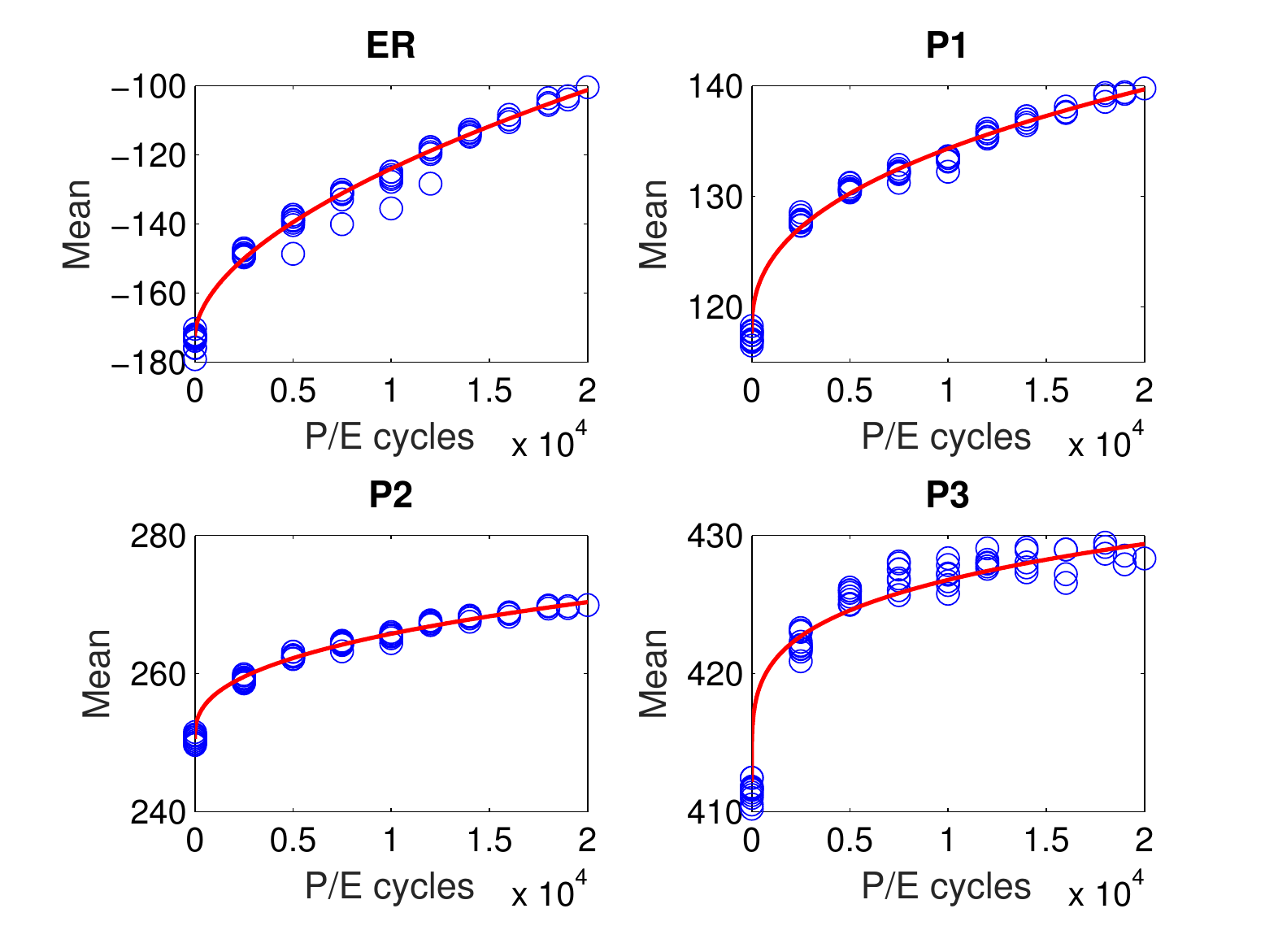}%
\caption{
  Change in mean value of each state's threshold voltage distribution as P/E
  cycle count increases, for the static Student's t-based model (blue circles)
  and the dynamic model (red line).
}
\label{fig:dynamic-mean}
\end{figure}
\FloatBarrier

\textbf{Standard Deviation.}
The standard deviation of each state represents the width of the distribution.
Similar to the Gaussian distribution, the Student's t-distribution contains the
vast majority ($\sim$95\%) of its mass within two standard deviations. Thus,
the change in standard deviation reflects how P/E cycle count affects the
threshold voltage variation among flash cells.

Figure~\ref{fig:dynamic-stdev} plots the standard deviation values obtained
from our Student's t-based model as circles. For this figure, the y-axis shows
the standard deviation in terms of normalized threshold voltage.  We make three
observations from this figure.  First, the standard \chI{deviation of each state's distribution increases}
monotonically with P/E cycle count. Second, the standard
deviations of the P1 and P2 states increase linearly \chI{with P/E cycle count}. 
Third,
like the \chI{mean}, the \chI{standard deviation increases} faster at lower P/E
cycle counts, then \chI{slows} down to a constant rate of increase after 5K P/E
cycles.

\begin{figure}[h]
\centering
\includegraphics[trim=27 10 35 10,clip,width=.7\linewidth]{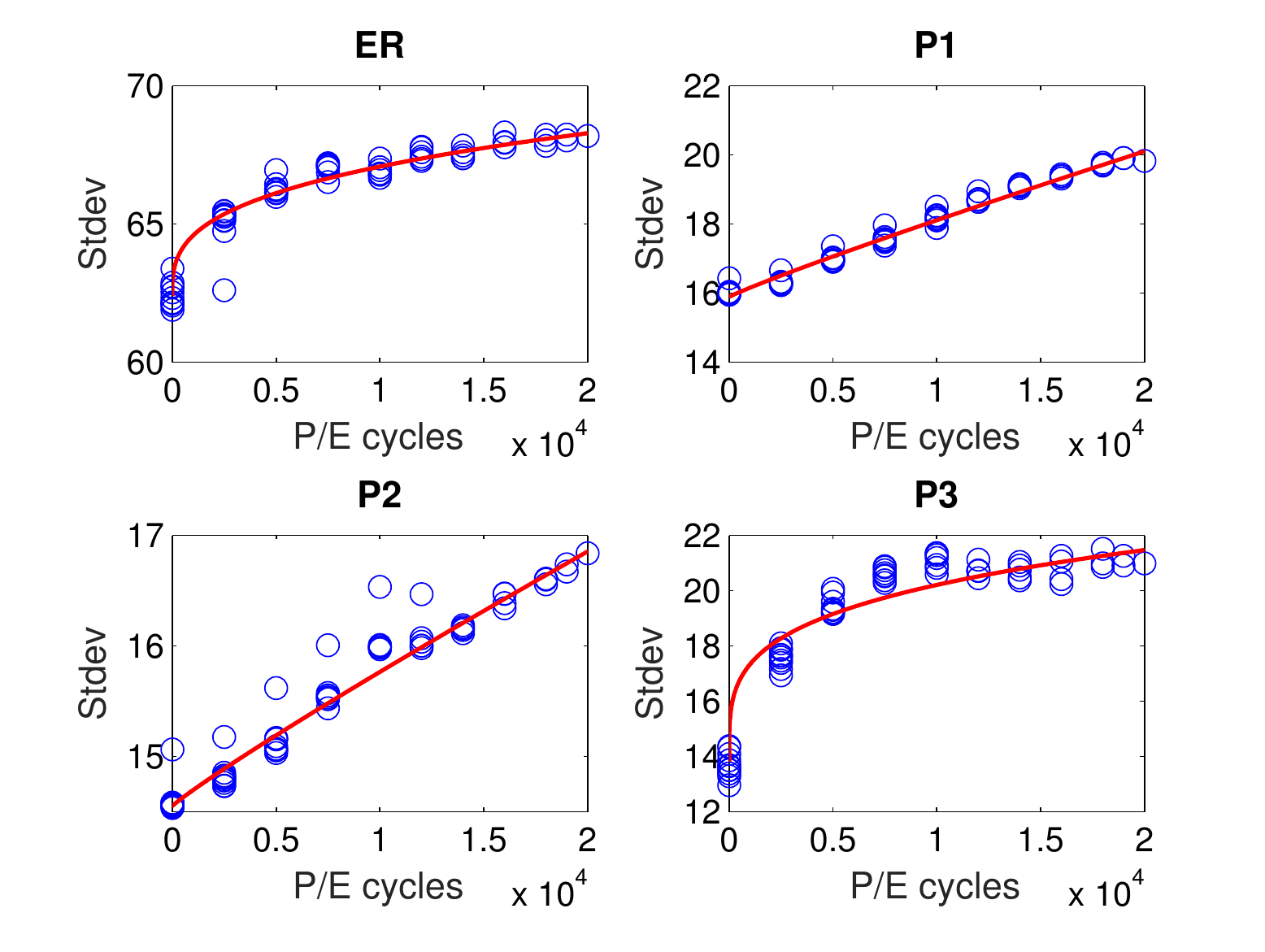}%
\caption{
  Change in standard deviation of each state's threshold voltage distribution as P/E cycle count increases, for the static
  Student's t-based model (blue circles) and the dynamic model (red line).
}
\label{fig:dynamic-stdev}
\end{figure}
\FloatBarrier

\textbf{Tail Values.}
The tail values of each state represent the size \chI{and shape} of the distribution tail.
Recall from Section~\ref{sec:static:student} that \chI{we use} $\nu$, which actually
represents the degrees of freedom, to control how fat the tail of the
model is.  Thus, the tail value reflects how the P/E cycle count affects the
number of outlier cells (i.e., the number of cells that lie at the tail).

Figure~\ref{fig:dynamic-tail} plots the tail values obtained from our Student's
t-based model as circles. In this figure, the y-axis shows the value of $\nu$,
where a lower value of $\nu$ corresponds to a fatter tail.  We make three
observations. \chI{First, the range of values for the tail sizes of the ER and P3
states is much smaller in comparison to the tail sizes of the distributions of
the other states.}
Second,
the sizes of both tails for the P1 state increase with P/E cycle count.
Third, the tail sizes of the P2 state decrease as P/E cycle count increases. 

\begin{figure}[h]
\centering
\includegraphics[trim=30 30 32 15,clip,width=.7\linewidth]{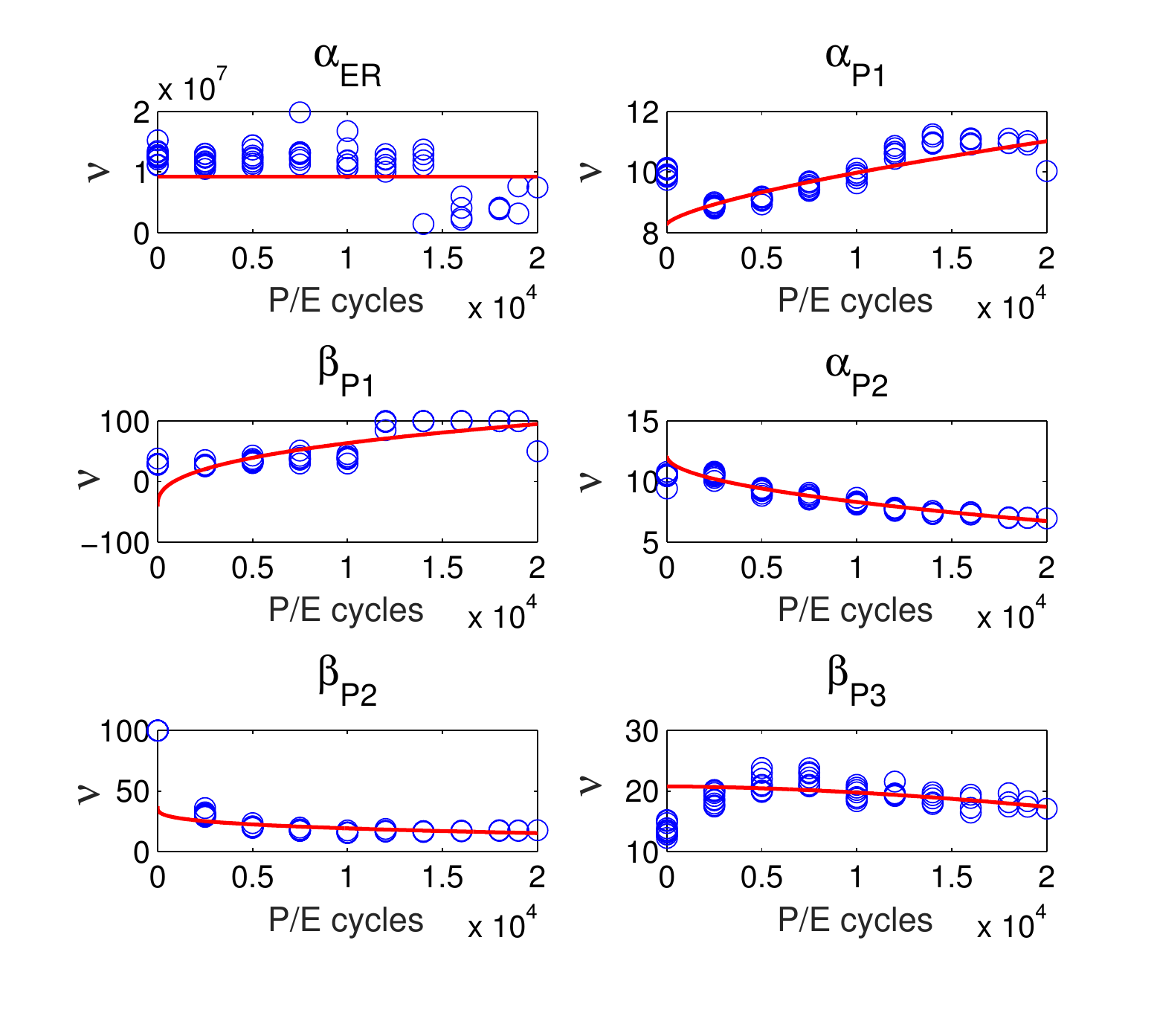}%
\caption{
  Change in tail values ($\nu$) of each state's threshold voltage distribution as P/E cycle count increases, for the static
  Student's t-based model (blue circles) and the dynamic model (red line).}
\label{fig:dynamic-tail}
\end{figure}
\FloatBarrier

\textbf{Probability of Program Errors.}
The program error probability \chI{$\lambda_X$} represents the percentage of cells that should be
programmed into \chI{state} $X$\chI{,} \chI{but} are instead misprogrammed to a different state, as a
result of two-step programming (see Section~\ref{sec:flash:pgmerase}).  In
our model, we assume that only certain types of program errors exist (ER$\to$P3
and P1$\to$P2), as program errors flip the value of \chI{only} the LSB \chI{within a cell} and can only
\emph{increase} the threshold voltage.

Figure~\ref{fig:dynamic-perr} plots the program error probability obtained from
our Student's t-based model as circles. For this graph, the y-axis shows the
\chI{log$_{10}$} value of the program error probability.  We make two observations. First,
the program error rate increases with P/E cycle count. Second, the number
of program errors increases more rapidly at lower P/E cycle counts, and then
slows down to a constant rate of increase at higher P/E cycle counts. 

\begin{figure}[h]
\centering
\includegraphics[trim=20 0 20 0,clip,width=.7\linewidth]{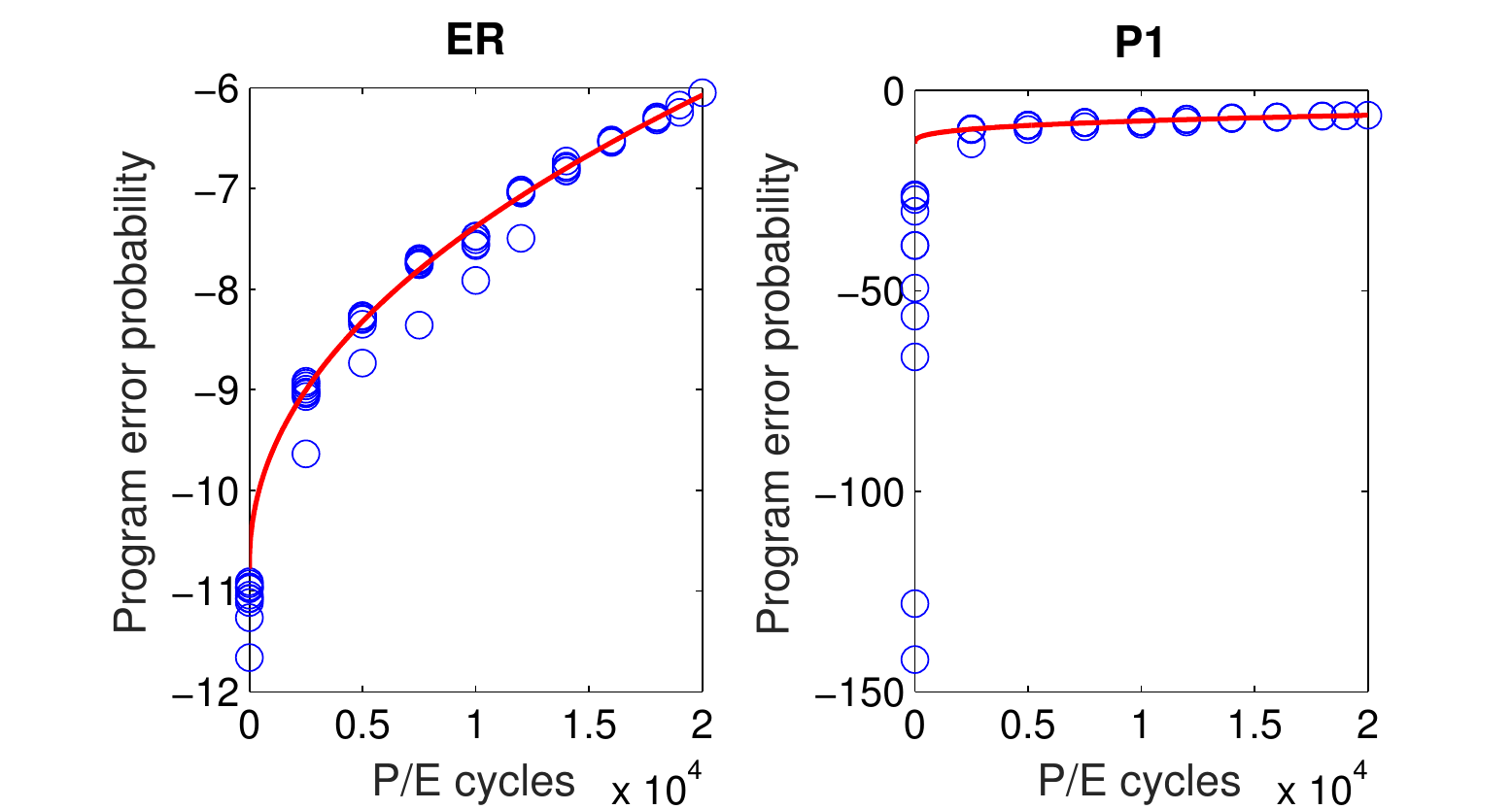}%
\caption{
  Change in log value of the program error probability as P/E cycle count increases, for the static
  Student's t-based model (blue circles) and the dynamic model (red line).
}
\label{fig:dynamic-perr}
\end{figure}
\FloatBarrier

\subsection{Power Law-based Model} \label{sec:dynamic:fitting}


Now that we have characterized how each of the parameters for our Student's
t-based model changes with respect to the P/E cycle count, we use this
characterization to develop a \emph{dynamic} model of the threshold voltage
distribution.  A dynamic model can reduce the \chI{total} computation effort for \chI{the} threshold
voltage distribution significantly, by requiring \chI{as little as a \emph{single} static model
\emph{characterization} for the entire lifetime of the flash device}.  
The dynamic model \chI{takes the \chI{static characterization-based} model(s) \chI{generated in} the past,
and simply adjusts the model} parameters at higher P/E cycle counts 
\chI{based on its prediction of how each parameter would change with P/E cycle count}
 (without requiring any further characterization).  \chI{Without the dynamic model}, 
\chI{\chI{a static model of the characterization must be generated} \emph{every time} a
new threshold voltage distribution 
is requested by the controller (e.g., after a fixed number of P/E cycles have occurred), with each
characterization requiring} a large number of read-retry operations (see
Section~\ref{sec:static:compare}).
These read-retry operations increase the accuracy of the model, but 
interfere with and slow down host commands.  Our goal is to build a dynamic
model that is accurate and easy to compute (such that it requires only a \chI{small number of
characterizations}), so \chI{that} it
can be used within the flash controller.


In Section~\ref{sec:dynamic:trend}, we observe that all of the parameters can
increase, decrease, or remain relatively constant.  We also observe that the rate
at which increases and decreases occur differs between \chI{lower P/E cycle counts
and higher P/E cycle counts}.  Our dynamic model must be able to represent all
of these behaviors.  We find that the \emph{power law} satisfies all of these
characteristics.  Equation~\ref{eqn:power} shows the power law function, which
models each parameter from our Student's t-based model, $Y$, as a function of
the P/E cycle count ($x$):
\begin{align}
  Y = a \times x^b + c
\label{eqn:power}
\end{align}
The power, $b$, can \chI{be set} to a positive value to represent an increasing
trend, or can \chI{be set} to a negative value to represent a decreasing trend. $b$
can also \chI{control} the difference in slope at different P/E cycle counts. For
example, when $b<1$, the modeled parameter $Y$ changes faster at \chI{\emph{lower}} P/E
cycle counts, \chI{and} when $b>1$, $Y$ changes faster at \chI{\emph{higher}} P/E cycle counts.


\chI{To observe how well the power law models changes to the parameters of our Student's t-based model, we fit
the power law to the values of each of the parameters as measured over several
P/E cycle counts (see Section~\ref{sec:dynamic:trend}).}%
\footnote{We exclude 0 P/E cycle results when modeling, as they show a
completely different behavior than results at any other P/E cycle count.} We
use mean squared error (MSE) to estimate the error, where the divergence
between the measured and estimated parameters ($Y_i$ and $\hat{Y_i}$,
respectively) can be mathematically defined as: $MSE = \frac{1}{n} \sum_{i=1}^n
(Y_i - \hat{Y_i})^2$. \chI{We} use the Nelder-Mead
simplex method~\cite{nelder.computer65}, with a reasonable initial guess, to fit
the trend.


Figures~\ref{fig:dynamic-mean},~\ref{fig:dynamic-stdev},~\ref{fig:dynamic-tail},
and~\ref{fig:dynamic-perr} show \chI{the power law-based models
fit to the trends of} each \chI{of our parameters as solid lines}. \chI{We fit the power
law to \chI{the} static model parameter values generated over a range of 20K P/E cycles.}
We observe that \chI{the predictions from the power law} fit very well with the
actual parameters measured from \chI{our} Student's t-based model, which are shown as
blue circles.  We next quantify the accuracy of our \chI{power law-based dynamic} model.

\subsection{Model Validation} \label{sec:dynamic:validation}

We validate our dynamic model by using it to predict the threshold
voltage distribution at 20K P/E cycles.  
\chI{We perform threshold voltage distribution characterizations at 2.5K,
5K, 7.5K, and 10K P/E cycles, and use these parameters to predict the
distribution at 20K P/E cycles.}
Figure~\ref{fig:dynamic-pec20000} shows the comparison between the actual
characterized distribution (markers) and the distribution predicted by our
\chI{dynamic} model (solid or dashed curves) at 20K P/E cycles.
The modeling error for the dynamic model is only 2.72\%, which is close to the
modeling error of directly using a static Gaussian-based model at 20K P/E cycles.  The dynamic model avoids
the need to perform the extensive read-retry characterization that \chI{all static models, including the Gaussian-based
model,} would require.  

\begin{figure}[h]
\centering
\includegraphics[trim=0 0 0 0,clip,width=.5\linewidth]{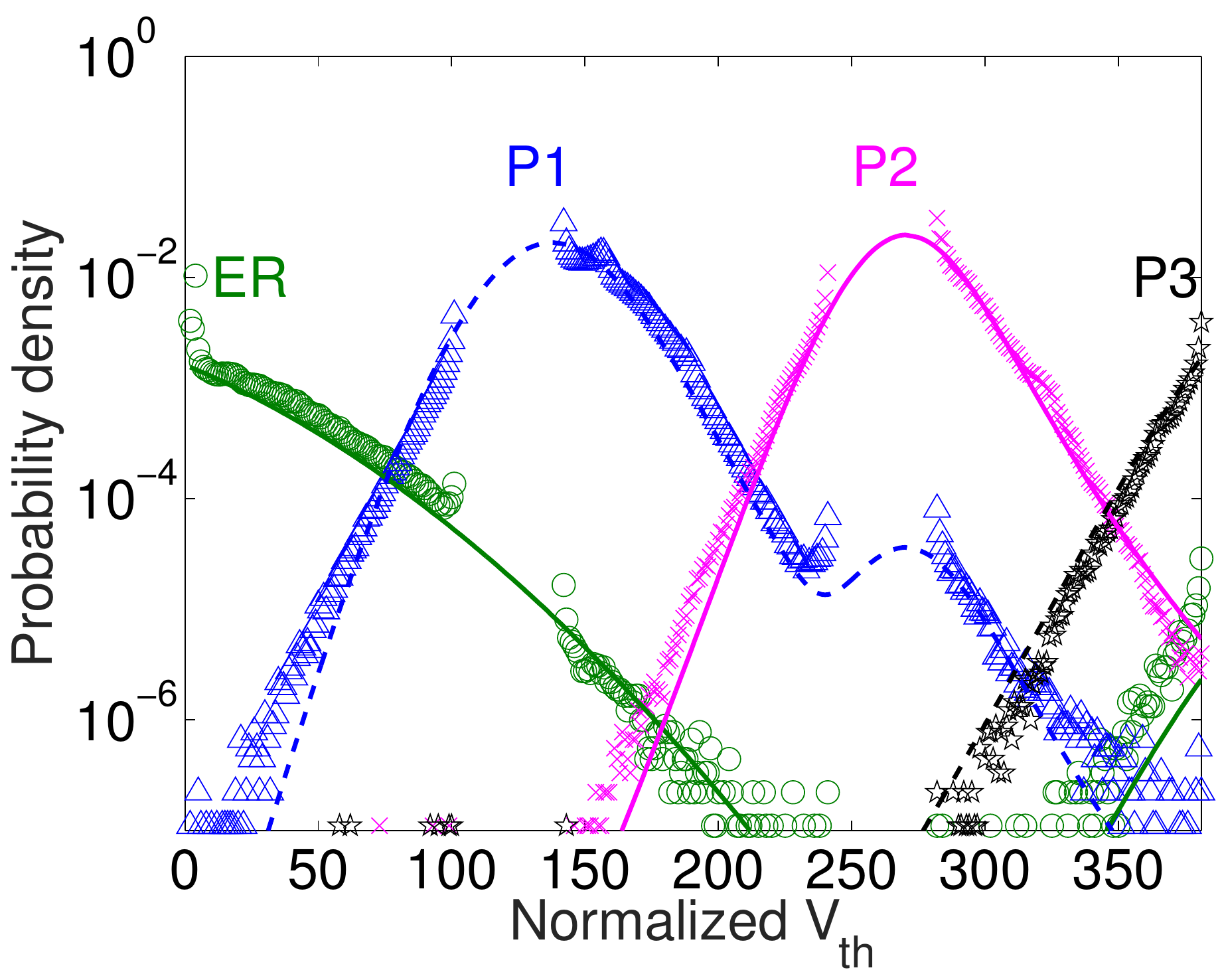}%
\caption{
Threshold voltage distribution as predicted by \chI{our} dynamic model for 20K P/E cycles, using
  characterization data \chI{\chI{from} 2.5K, 5K, 7.5K, and 10K P/E cycles}, shown
  as solid/dashed lines.  Markers represent data
  measured from real NAND flash chips at 20K P/E cycles.
}
\label{fig:dynamic-pec20000}
\end{figure}
\FloatBarrier

\chI{
Figure~\ref{fig:dynamic-error} shows how the modeling error of our
dynamic model decreases for a prediction at 20K P/E cycles as the
number of characterized data points increases.  The number of
characterized data points represents the $N$ earliest static models out of
a range that consists of static models for 2.5K, 5K, 7.5K, 10K, 12K, 14K, 16K, 18K, and 19K P/E cycles.}
\chI{Note that we start
with three characterization data points, \chI{which allows the dynamic model to observe a trend in the change of each parameter.}}
This figure shows that the error rate decreases rapidly as we increase
the number of data points used to train the dynamic model. 

\begin{figure}[h]
\centering
\includegraphics[trim=10 0 30 8,clip,width=.7\linewidth]{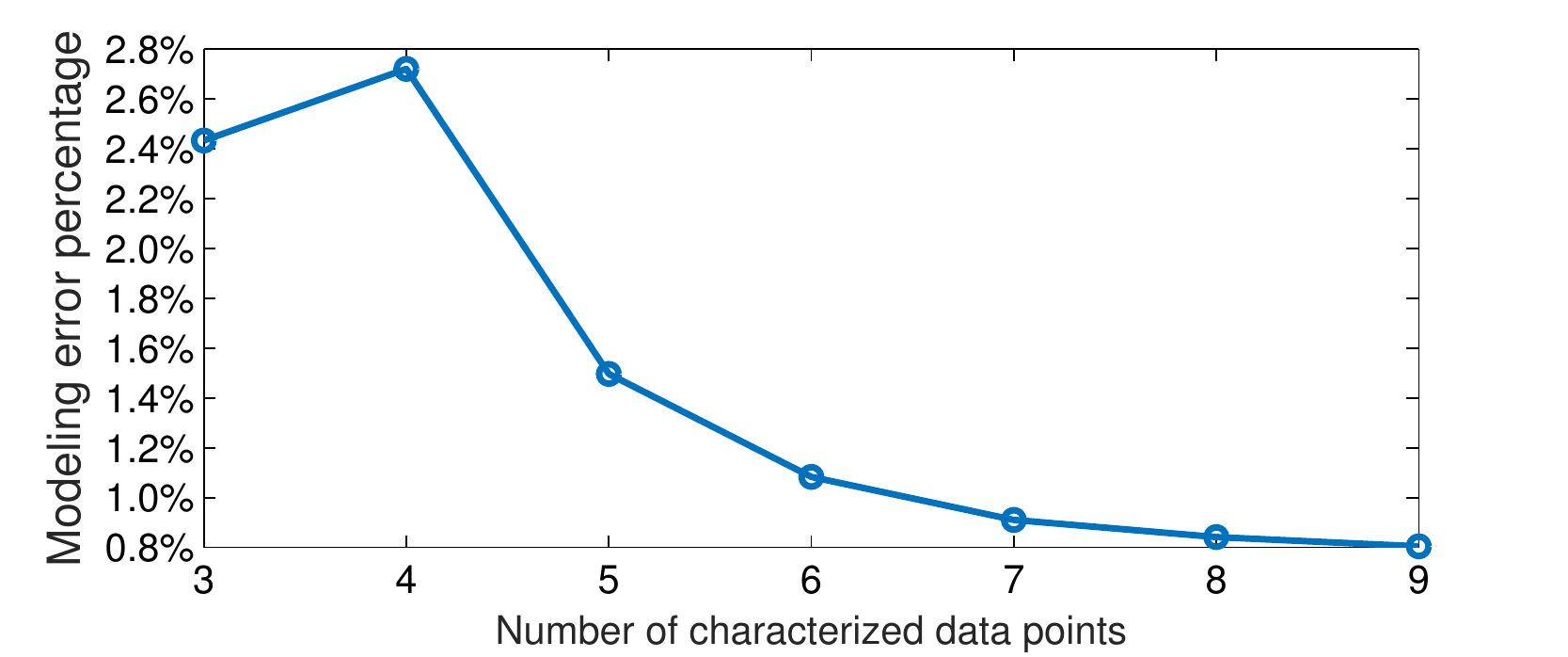}%
\caption{
Modeling error of predicted threshold voltage distribution for \chI{our} dynamic model at 20K P/E
  cycles, using characterization data from \emph{N} different P/E cycles.
}
\label{fig:dynamic-error}
\end{figure}
\FloatBarrier

We conclude that our dynamic model successfully predicts \chI{an accurate
threshold voltage distribution} for \chI{a P/E cycle count} it has not observed, based
on only prior characterization data, and is thus practical for use in the flash
controller.


\section{Example Applications}
\label{sec:application}

Now that we have developed our threshold voltage distribution model, we 
demonstrate \chI{three} example applications within the flash controller that take
advantage of the model to enhance the reliability of the flash device.
\chI{The first
application, described in Section~\ref{sec:application:rber}, uses our model to
accurately estimate the raw bit error rate. The second application, described in
Section~\ref{sec:application:opt}, uses our model to accurately predict the optimal read reference
voltage. The third application,
described in Section~\ref{sec:application:lifetime}, uses our model to estimate the expected
lifetime of the flash memory device, to safely achieve higher P/E cycle endurance than
manufacturer specification.
}

\subsection{Raw Bit Error Rate Estimation}
\label{sec:application:rber}


The raw bit error rate \chI{(i.e., the probability of
reading an incorrect state for a flash cell), or RBER,}
is important not only because it is a measure of the reliability of a flash device, but
because it also can be used to determine the lifetime
and performance of the flash drive~\cite{cai.hpca15}.  The raw bit error rate
can be used to enable several optimizations in the flash controller.  For
example, accurately estimating the current raw bit error rate allows us to
safely utilize the \emph{currently unused} ECC correction capability to
accelerate program operations~\cite{jeong.fast14}, relax the retention
time~\cite{cai.hpca15, liu.fast12}, and reduce the
effects of read disturbance~\cite{cai.dsn15}.  Accurate estimation of the
raw bit error rate enables other optimizations, such as predicting the 
optimal read reference voltage\chI{~\cite{cai.iccd13}} or performing error rate based wear-leveling.


To estimate the raw bit error rate based on the \chI{static} threshold voltage distribution
model, we use the \chI{static} model to calculate the cumulative distribution function \chI{(CDF)} for
each state at each of our read reference voltages ($V_a$, $V_b$, and $V_c$),
and use this data to determine how many cells are misread.  For example, if
there are cells in the \chI{distribution of the ER state} whose threshold voltages are
greater than $V_a$, they will be misread.  By calculating the ER state CDF up
to $V_a$, we know what percentage of cells will be \emph{correctly} read.  We
subtract this \chI{value} from 1 to obtain the percentage of cells that will be \chI{\emph{misread}}
(and will thus contribute to the raw bit error rate).


Figure~\ref{fig:app-rber} shows the actual \chI{measured} raw bit error rate and the modeled raw bit
error rates using the three static models from Section~\ref{sec:static}, for different P/E cycle counts. The
x-axis shows P/E cycle \chI{count}, and the y-axis shows the \chI{measured or model-predicted} raw bit error rate. The three
graphs show the \chI{average error rate for only the LSB pages, only the MSB pages, and for all of the pages}. We
make two observations from this data. First, the normal-Laplace-based and our Student's t-based models
give a much better estimate of the raw bit error rate than the Gaussian-based model. 
\chI{Averaged across \chI{all P/E cycle counts}, our Student's t-based model estimates the 
RBER for all pages within 13.0\% of the actual measured RBER, while the 
normal-Laplace-based model is within 14.9\% and the Gaussian-based model is
only within 44.7\%.}
This is
due to the limitations of the Gaussian-based model, as it cannot adjust the tail
size or take program errors into account. Second, the normal-Laplace-based and our Student's t-based models
tend to overestimate the error rate, which is usually safe \chI{for the
purposes} of many optimizations, because overestimation results in more than
adequate ECC correction capability \chI{to remain available} for these errors.  In contrast, the Gaussian-based
model always \emph{underestimates} the raw bit error rate, which, \chI{if
used for an optimization that relies on an RBER estimation,} can \chI{cause the
number of errors to exceed} the correction capability of \chI{ECC,
resulting} in uncorrectable errors during reads.
\chI{We conclude that our Student's t-based model is effective at providing 
an accurate estimate of the raw bit error rate for use by the flash controller.}

\begin{figure}[h]
\centering
\includegraphics[trim=45 0 60 0,clip,width=.8\linewidth]{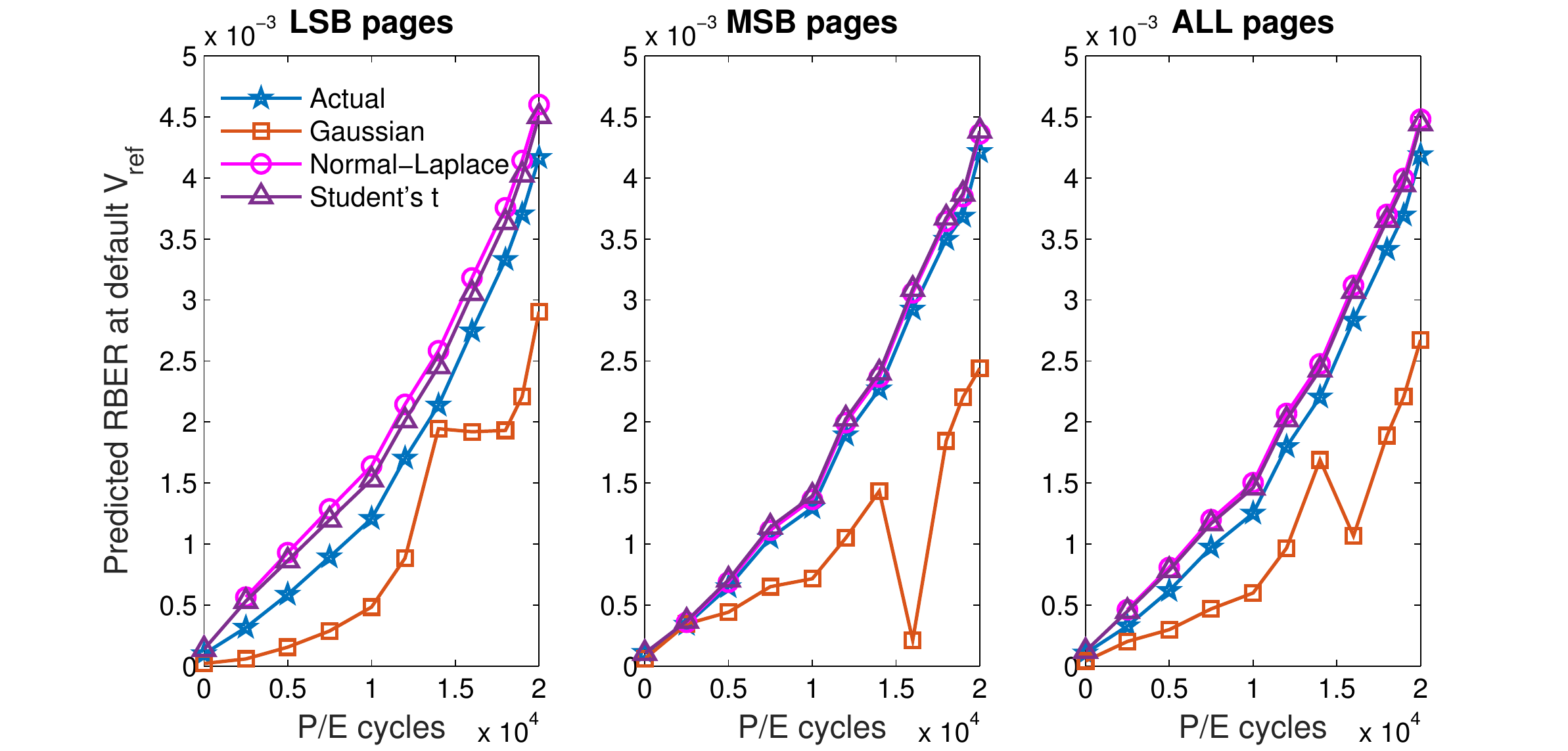}%
\caption{
  Actual and modeled raw bit error rate using the three evaluated threshold
  voltage distribution models when reading with \chI{fixed \emph{default} read
  reference voltages ($V_{ref}$)}, across
  different P/E cycle counts.}
\label{fig:app-rber}
\end{figure}
\FloatBarrier

\subsection{Optimal Read Reference Voltage Prediction}
\label{sec:application:opt}


As we discussed in Section~\ref{sec:errors}, when the threshold
voltage distribution shifts, it is important to move the read reference voltage
to the point where the number of read errors \chI{is} minimized.  After the
\chI{shift occurs, the threshold voltage distributions of each state} may
overlap with each other, \chI{causing many of the cells within the
overlapping regions to be misread}.  The number of errors \chI{due to misread cells}
can be minimized by setting the read reference voltage to be at the point where
the distributions of two neighboring states \emph{intersect}, which we call the \emph{optimal
read reference voltage} \chI{($V_{opt}$)~\cite{cai.iccd13}}.  
Once the optimal read reference voltage
is applied, the raw bit error rate is minimized, improving \chI{the} reliability of the
device. Furthermore, since fewer errors are corrected, and fewer read-retries
are needed, read latency is also significantly reduced~\cite{cai.hpca15}.


Prior work proposes to
learn and record the optimal read reference voltage
periodically~\cite{cai.hpca15, papandreou.glsvlsi14, tabrizi.icc15} by sampling
\chI{the threshold voltages of \emph{some} of the cells in} each flash block, but this sampling requires time and storage
overhead.  
With our new distribution model, we can determine the optimal read
reference voltage from the model and predict how it changes, without
\chI{having to exhaustively} learn it for each block.
From our threshold voltage distribution model, we can  predict the optimal
read reference voltage by finding the point at which the probability density
functions of \chI{the distributions of two neighboring states} are the same (i.e., the
intersection of the two distributions).


Figure~\ref{fig:app-vopt} plots
the actual \chI{measured} and modeled optimal read reference voltage using the three 
static models from Section~\ref{sec:static}, \chI{at} different P/E cycle 
counts.\footnote{Note that the default read reference voltages are ($V_a$,$V_b$,$V_c$) =
(50,190,330). We observe that the actual optimal read reference voltage can be higher than
the default read reference voltage by as much as \chI{27} voltage steps.}
Each graph shows \chI{the voltage chosen for one of the three read
reference voltages ($V_a$, $V_b$, and $V_c$) used to distinguish between the
distributions of} two neighboring states.
The x-axis shows the P/E cycle count, while the y-axis shows the normalized
optimal read reference voltage.  We
make \chI{three} observations from this result. First, the normal-Laplace-based and \chI{our} Student's t-based
models slightly overestimate all three optimal read reference voltages. Second,
the Gaussian-based model underestimates the optimal read reference voltages in most
cases, and has glitches of underestimation as large as 17 voltage steps. We
suspect that this is because the Gaussian-based model cannot capture the asymmetric
tail sizes of the distribution. 
\chI{Third, at 0 P/E cycles, the read reference voltages predicted using
the normal-Laplace-based model deviate significantly from the actual optimal 
read reference voltages.  We find that the normal-Laplace-based model has
\chI{difficulty} converging to a good value at 0 P/E cycles, while our Student's t-based
model does not experience any such \chI{difficulty}.}


\begin{figure}[h]
\centering
\includegraphics[trim=50 0 60 0,clip,width=.8\linewidth]{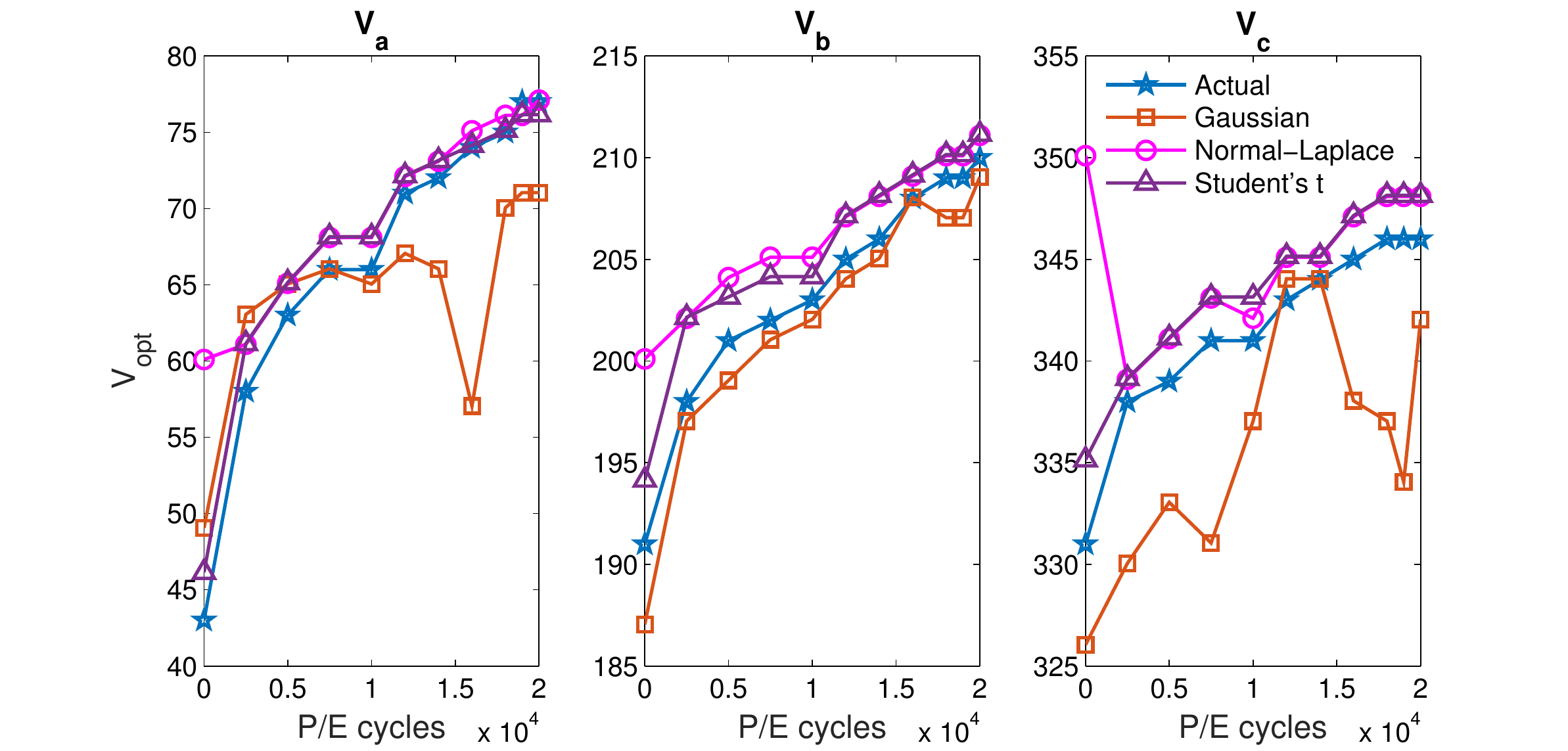}%
\caption{
  Actual and modeled \emph{optimal} read reference voltages \chI{($V_{opt}$)} using the three evaluated threshold voltage distribution models at different P/E cycle counts.
}
\label{fig:app-vopt}
\end{figure}
\FloatBarrier

\yixin{Figure~\ref{fig:app-vopt-rber} shows the RBER when
we use the actual optimal read reference voltage to read data, as well as the RBER when we use the optimal read
reference voltages predicted by each of the three static models
from Section~\ref{sec:static}, at different P/E cycle 
counts. 
\chI{
As we did for Figure~\ref{fig:app-rber}, we show the \chI{average} error rate for only the LSB
page\chI{s,} only the
MSB page\chI{s}, and \chI{for all of the pages}.
}
We observe that the prediction generated from the Gaussian-based model results in
a significantly higher MSB error rate than the actual optimal voltage. The normal-Laplace-based
and our Student's t-based models generate read reference voltage predictions that result in near-optimal RBER
\chI{(within 1.5\% and 1.1\%, respectively, of the optimal RBER)}, despite some
difference between the actual optimal read reference voltage and the model-predicted voltages.}

\begin{figure}[h]
\centering
\includegraphics[trim=50 0 60 0,clip,width=.8\linewidth]{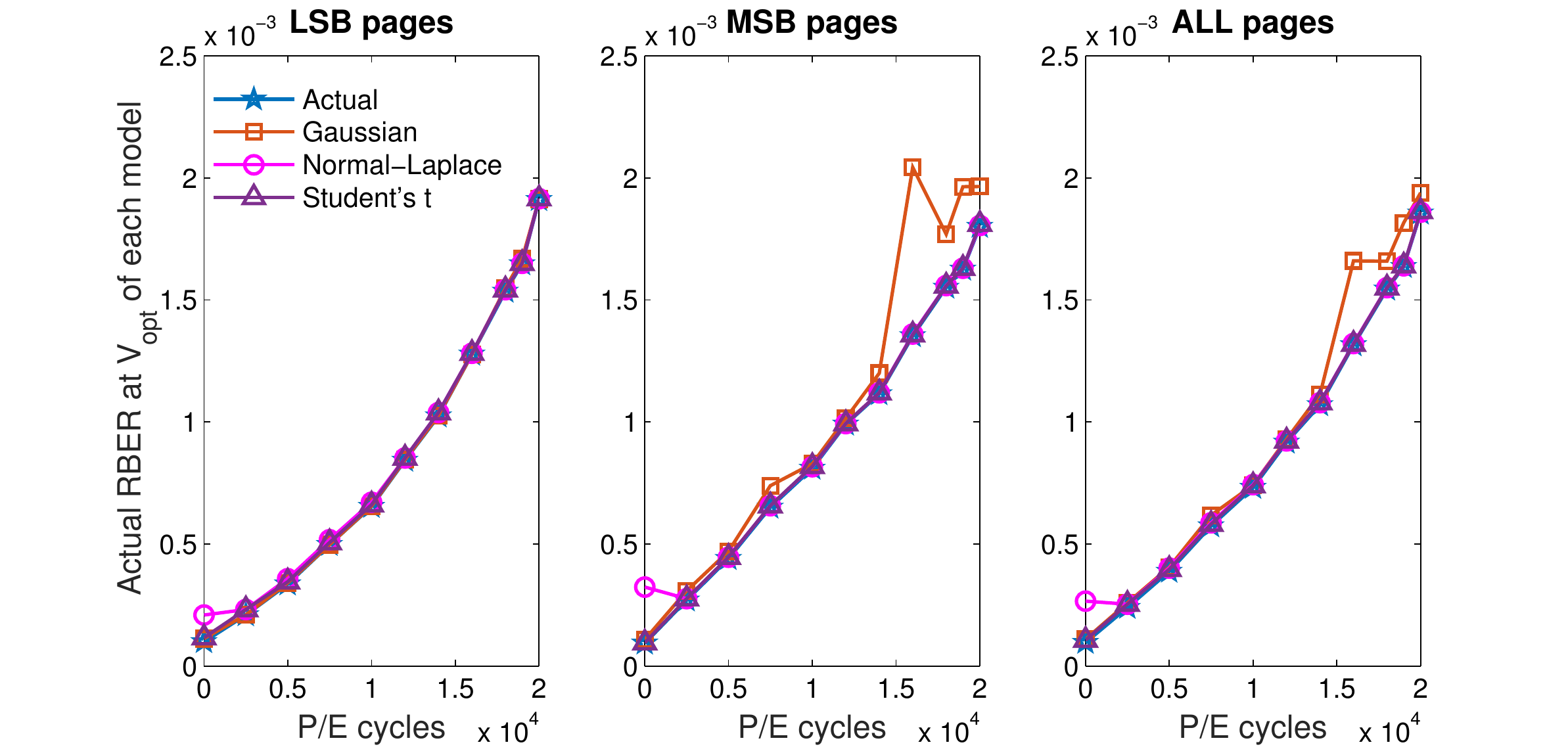}%
\caption{
  RBER achieved by actual and modeled \emph{optimal} read reference voltages \chI{($V_{opt}$)} using the three evaluated threshold voltage distribution models at different P/E cycle counts.
}
\label{fig:app-vopt-rber}
\end{figure}
\FloatBarrier

We evaluate how using the optimal voltages predicted by each model can improve
flash lifetime \chI{compared} to using the default read reference voltages.  We 
assume that we have a state-of-the-art LDPC decoder, which can tolerate a raw 
bit error rate as high as $5\times10^{-3}$~\cite{haratsch.fms14} while still
\chI{keeping} the required uncorrectable error rate \chI{below}
$10^{-15}$~\cite{jep122h.jedec16} during the flash device's lifetime.\footnote{To tolerate
variation in the raw bit error rate, we assume that 10\% of the total ECC
correction capability is reserved, lowering the maximum tolerable raw bit error
rate.} We also assume 
that our flash device refreshes its data every three weeks, limiting the number 
of retention and read disturb errors that occur~\cite{cai.iccd12}.  Using the
actual (i.e., ideal) optimal read reference voltage, flash lifetime 
improves by 50.6\%.  Both our Student's t-based model and the 
normal-Laplace-based model come very close to the ideal improvement, providing
a \emph{48.9\% lifetime improvement} \chI{with our Student's t-based model}.  Due to its lower accuracy, the
Gaussian-based model achieves only a 38.5\% improvement.

\subsection{Expected Lifetime Estimation} \label{sec:application:lifetime}


Due to the \chI{increasing} raw bit error rate \chI{at higher P/E cycle counts}, flash memory can
endure only a limited number of writes.  To make sure that all data stored in the
flash drive is reliable over the course of a predefined device lifetime (typically
several years), enterprise users limit the number of writes to each flash
drive. Due to process variation, different flash chips can have different raw bit
error rates and thus different P/E cycle endurance.  However, flash vendors
conservatively \chI{set} the flash drive's \chI{P/E cycle} endurance to the \emph{worst case}
\chI{(i.e., to the lowest endurance value out of all of the chips that they produce)}, as they do
not know how fast each individual flash chip wears out over time. In fact, prior work has
tested six commercial flash drives, and found that they all surpassed their official
endurance specifications by an average of 81\%~\cite{gasior.tr15}.


If the flash controller can monitor how fast each flash chip wears out due to
\chI{flash} writes, the
users can determine the \emph{actual} endurance limit of the flash drive, and write more
data to it without worrying about prematurely wearing out the drive and losing data.
The \chI{model} proposed in this chapter
\chI{enables} raw bit error rate prediction for \emph{future P/E cycle counts}.
Thus, the controller can predict the endurance limit of each flash chip
by iterating through our dynamic model to predict the point at which the raw bit
error rate exceeds the \chI{ECC correction capability}
(i.e., when the lifetime \chI{\emph{actually}} ends). The flash controller then communicates this prediction
to the \chI{file system} to allow higher write intensity to the flash drive.

We estimate the lifetime improvements using this technique, with the same
assumptions we made in Section~\ref{sec:application:opt} and the data shown in
Section~\ref{sec:application:rber}. With our dynamic model, we safely achieve \emph{69.9\% higher P/E cycle endurance}
than manufacturer specification. This translates to 69.9\% more tolerable writes per
day, \chI{if we assume that the flash device will be used for the same number of years (i.e., lifetime) as before}.




\subsection{Soft Information Estimation for LDPC Codes}
\label{sec:future:ldpc}

To tolerate flash errors more efficiently, today's flash controllers
use LDPC codes to detect and correct multiple raw bit
errors in the data read from the flash memory channel~\cite{wang.jsac14, dong.tocs13, zhao.fast13}. An LDPC code can use \emph{soft information}
about each bit to increase the probability of correcting
the raw bit errors. This soft information is provided by the flash
controller, which estimates the probability of each bit being a 1 or a 0
using the threshold voltage of a cell. A modern flash controller
typically obtains this probability from a Gaussian-based model for the
threshold voltage distribution, \chI{since} the soft information can be \chI{computed}
as a quadratic function of the threshold voltage.  However, as we have
shown in Section~\ref{sec:static}, the Gaussian-based model
underestimates the probability density at the tail of the
distribution, and does not model program errors.  Thus, the model can
provide inaccurate information to the LDPC decoder. This compromises
the error correction capability of the LDPC codes, thus reducing the
reliability and performance of the flash drive.

With the models proposed in this chapter, we now have an accurate
threshold voltage distribution model that adapts to the P/E cycle of
each block, and can be implemented within the flash controller.  Using
our Student's t-based model, we can accurately and efficiently compute
the probability density \chI{for any threshold voltage range} to provide \emph{accurate} soft
information to the flash controller. By increasing the accuracy of
this soft information, we effectively increase the error correction
capability of the LDPC code, which can lead to longer flash lifetime
and better read performance~\cite{dong.tocs13, wang.jsac14, zhao.fast13}.  We leave the precise
implementation of such a mechanism for future work.

\subsection{Improving Flash Performance}
\label{sec:future:performance}

\yixin{
While the applications \chI{of our threshold voltage distribution model
that} we have discussed \chI{in Sections~\ref{sec:application}
and~\ref{sec:future:ldpc}} aim to improve
reliability, they can also \chI{improve flash performance}. For example, by \chI{predicting and} applying the optimal read
reference voltage (Section~\ref{sec:application:opt}), we can greatly lower the probability that read-retries need to be
performed for a read operation, which also reduces the number of \chI{ECC} decoding iterations, both of
which lead to a lower read latency~\cite{cai.hpca15}. Other applications can 
also take advantage of our model to improve flash performance.  \chI{For} example, we can
minimize the ECC decoding latency by \chI{adaptively} applying a weaker ECC code when the raw bit
error rate indicated by our model is low~\cite{yuan.icassp15, haratsch.fms14,
haratsch.fms15, wu.mascots10}. We expect and hope future
work to evaluate the performance benefits of these applications, and to propose
other new applications of our online model that can improve flash performance.
}


\section{Related Work} \label{sec:vthmodel:related}

To our knowledge, our work in this chapter is the first to (1)~propose a threshold
voltage distribution model that is both highly accurate and
computationally efficient, (2)~propose a dynamic threshold voltage
distribution model that predicts how the parameters of this model
change with increasing program/erase cycle count, and (3)~demonstrate
several \chI{new} practical uses of this threshold voltage distribution model
within a flash controller to improve flash memory reliability.

We have already comprehensively compared our Student's t-based static model to
the two most relevant models based on real characterization results, the
Gaussian-based model~\cite{cai.date13, motwani.globecom15} and the
normal-Laplace-based model~\cite{parnell.globecom14}, in
Sections~\ref{sec:static:gaussian},~\ref{sec:static:laplace},
and~\ref{sec:application}. \chI{We show that our Student's t-based model
has an error rate within 0.11\% of the error rate \chI{of} the highly-accurate
normal-Laplace model, while requiring 4.41x less computation \chI{time}.}
Several prior works fit the
threshold voltage distribution \chI{to other models that are either less
accurate or more complex, such as the
beta distribution~\cite{cai.date13}, gamma
distribution~\cite{cai.date13}, log-normal distribution~\cite{cai.date13},
Weibull distribution~\cite{cai.date13}, and beta-binomial probability
distribution~\cite{taranalli.arxiv16}.} Other prior works model the threshold
voltage distribution based on idealized circuit-level models~\cite{dong.tocs13,
motwani.globecom15, pan.fast11}.  These models capture some of the desired
threshold voltage distribution behavior, but are less accurate than those
derived from real characterization.

A few works also propose dynamic models of the threshold voltage distribution
shifts based on the power law~\cite{cai.date13, cai.iccd13, parnell.globecom14}. While these
models are sufficient for offline analysis, they are
unsuitable for deployment in today's flash controllers, as they fail to achieve
high accuracy and low \chI{computational} complexity at the same time. 
Our dynamic model also uses the power law, but is based on our new, accurate, and
low-complexity \chI{Student's t-based} static model. We show that our model has an error rate
of only 2.72\% when estimating the distribution at 20K P/E cycles, even though
it uses characterization data collected \chI{at only four different P/E cycle counts from the past (up to 10K P/E cycles)}.
While other dynamic models based on idealized circuit models exist~\cite{dong.tocs13, pan.fast11}, they
are not validated with real characterization data, and cannot achieve the
same accuracy as our model.



Prior works propose and evaluate techniques for raw bit error
rate estimation~\cite{prodromakis.microproc15, parnell.globecom14},
optimal read reference voltage estimation~\cite{papandreou.todaes15, cai.hpca15,
papandreou.glsvlsi14, tabrizi.icc15}, and LDPC soft
decoding~\cite{wang.jsac14, dong.tocs13, zhao.fast13}. These works utilize a threshold
voltage distribution model \chI{only} offline, or do not utilize a threshold voltage distribution model at
all. We show that, by utilizing our model, we can effectively and practically
guide such flash reliability mechanisms \emph{online} in the flash controller. We also
provide a new mechanism to {\em exploit} process variation for \emph{higher} flash
endurance, by predicting and safely utilizing the remaining lifetime of a flash device online.
Prior works propose to only {\em tolerate} error rate variation and process variation to
improve flash lifetime~\cite{motwani.icnc13, motwani.icnc15, li.tcs14}.

We note that several prior works have already extensively studied the impact of
retention behavior on the threshold voltage distribution using real
hardware~\cite{cai.hpca15, cai.date12, cai.iccd12, cai.itj13}. They show
that commonly-employed refresh mechanisms in flash
devices can successfully mitigate most of the impact of retention on the
threshold voltage \chI{distribution}~\cite{cai.iccd12, cai.itj13, luo.msst15}.  As a
result, we expect that even without capturing the effects of retention, our
proposed threshold voltage distribution model will work well in practice.

\section{\chIV{Limitations}}

\chIV{Our online model captures only P/E cycling and two-step programming effects,
which are two of the most dominant error sources in \SI{1X}{\nano\meter} MLC
planar NAND flash memory. \chV{Currently, our online model} does not
model and mitigate retention and read disturb errors because they are
successfully mitigated by commonly-employed flash refresh
mechanisms~\cite{cai.iccd12, mohan.tr12, pan.hpca12}, which can be combined
with our online model to provide greater reduction in raw bit errors.
Furthermore, our online model can be extended to model the threshold voltage
shift due to retention or read disturb. Online modeling effectively reduces
retention errors when refresh becomes less effective in 3D NAND, as we show in
Chapter~\ref{sec:3derror} and~\ref{sec:heatwatch}.}


\section{Conclusion}
\label{sec:vthmodel:conclusion}

In this chapter, we introduce a new threshold voltage distribution model for
modern NAND flash memory devices. Our model is based on a new
experimental characterization of the threshold voltage distribution
and how it shifts over time using state-of-the-art 1X-nm MLC NAND
flash chips. Our characterization shows that the threshold voltage
distribution can be approximated using our modified version of the Student's t-distribution,
and \chI{that the amount by which the distribution shifts as the P/E cycle count increases is} governed by the
power law. Our new model, which combines these two observations in its
static and dynamic components, is capable of accurately capturing the
current and predicting the future threshold voltage distribution of
flash memory cells. We show that our model achieves low modeling
error, and is computationally \chI{simple enough to implement online} in a flash
controller. We demonstrate various applications of our model in a
flash controller. We show that these applications improve flash lifetime by
48.9\% and/or enable the flash device to safely utilize 69.9\% more P/E cycles than
\chI{manufacturer specification}.
We conclude that our proposed threshold
voltage distribution model for modern MLC NAND flash memory devices is
practical and effective.
We hope that this dissertation inspires future work to improve upon our online flash
channel model, and to develop and evaluate new techniques that take advantage of
such a model to increase flash memory reliability and performance.



\chapter{3D NAND Flash Memory Error Characterization and Mitigation}
\label{sec:3derror}

\chV{In order for} planar NAND flash memory to continually increase the SSD capacity and
decrease the \emph{cost-per-bit} of the SSD, flash vendors have to aggressively scale
NAND flash memory to smaller manufacturing process technologies. This,
however, comes at the cost of the decreasing flash
reliability~\cite{mielke.irps08, cai.date12, cai.procieee17}, as we have shown
in Section~\ref{sec:errors}.  Due to a
combination of manufacturing process limitations and decreasing reliability,
it has become increasingly difficult for manufacturers to continue to scale
the density of planar NAND flash memory~\cite{park.jssc15}.

To overcome this scaling challenge, 3D NAND flash memory
has recently been introduced~\cite{park.jssc15, kang.isscc16,
im.isscc15} (see Section~\ref{sec:background:3d} for a comparison between
planar NAND and 3D NAND technology). Previous publicly-available experimental
studies on NAND flash memory errors using real flash memory chips
(e.g.,~\cite{mielke.irps08, cai.date12, cai.date13, cai.iccd12,
cai.iccd13, cai.sigmetrics14, cai.hpca15, cai.dsn15, cai.hpca17,
parnell.globecom14, luo.jsac16}) have all been on planar NAND devices.  As
3D NAND flash memory is already being deployed at a large scale in new computer
systems, there is a lack of available knowledge on the error
characteristics of real 3D NAND flash chips, which makes it harder to estimate
the reliability characteristics of systems that employ such chips.

In this chapter, our goal is to (1)~identify and
understand the \chI{\emph{new} error characteristics of 3D NAND flash memory
(i.e., those that did not exist previously in planar NAND flash memory),}
\chI{and} (2)~propose new mechanisms to mitigate prevailing 3D
NAND flash errors. We aim to achieve these goals via rigorous experimental
characterization of real, state-of-the-art 3D NAND flash memory chips
\chI{from a major flash vendor}.
Based on our comprehensive characterization and
analysis, we identify \emph{three new error characteristics} that were not
previously observed in planar NAND flash memory, but are fundamental to the new
architecture of 3D NAND flash memory.
(1)~3D NAND flash exhibits \emph{layer-to-layer process
variation}, a new phenomenon specific to the 3D nature of the device, where the
average error rate of each 3D-stacked layer in a
chip is significantly different (Section~\ref{sec:3derror:variation}).
\chI{We are the \emph{first} to provide detailed experimental characterization
results of layer-to-layer process variation in real flash devices in open literature.}
(2)~3D NAND flash memory experiences
\emph{early retention loss}, a new phenomenon where the number of
errors due to charge leakage increases quickly within several hours after
programming, but then increases at a much slower rate
(Section~\ref{sec:3derror:retention}).
\chI{We are the \emph{first} \chV{to perform} an extended duration observation
of early retention loss.  While prior studies examine the impact of early 
retention loss over only the first 5~minutes after data is written,
we examine the impact of early retention loss over 24~days.}
(3)~3D NAND flash \chI{memory} experiences \emph{retention interference}, a new
phenomenon where the rate at which charge leaks from a flash cell is
dependent on the amount of charge stored in neighboring flash cells
(Section~\ref{sec:3derror:retention:interference}).

Our experimental observations indicate that we must revisit the error models
and the error mitigation mechanisms devised for planar NAND flash, as they are
no longer accurate for 3D NAND flash behavior. To this end, we develop
\emph{new analytical model\chI{s}} 
of (1)~the layer-to-layer process variation in 3D NAND flash memory
(Section~\ref{sec:3derror:model:variation}), and 
(2)~retention loss in 3D NAND flash memory (Section~\ref{sec:3derror:model:retention}). Both models are
useful for developing
techniques to mitigate raw bit errors in 3D NAND flash memory. Our models estimate the
raw bit error rate (RBER), threshold voltage distribution, and the 
\emph{optimal read reference voltage} (i.e., the voltage at which the raw bit
errors are minimized when applied during a read operation) for each flash \chI{page}.

We propose \emph{four new techniques} to
mitigate the unique layer-to-layer process variation and early retention loss
errors observed in 3D NAND flash memory. 
Our first technique, LaVAR, reduces process variation by fine-tuning the
read reference voltage independently for each layer (Section~\ref{sec:3derror:mitigation:variation}).
Our second technique, LI-RAID, is a new RAID scheme that eliminates the page with
the worst-case reliability within \chI{each} block by changing how we pair up
pages from different \chI{flash blocks}
(Section~\ref{sec:3derror:mitigation:raid}).
Our third technique, ReMAR, reduces retention errors in 3D
NAND flash memory by tracking the retention age of the data using our retention
model and adapting \chI{the read reference voltage} to the data age
(Section~\ref{sec:3derror:mitigation:retention}). Our fourth technique, ReNAC,
predicts and adapts \chI{the read reference voltage} to the
amount of retention interference during each read operation
(Section~\ref{sec:3derror:mitigation:nac}).
\chI{These four techniques are complementary, and can be combined together
to significantly improve NAND flash reliability.  Compared to a state-of-the-art
baseline, our techniques provide a combined 3D NAND flash memory lifetime
improvement of 85.0\%.  Alternatively, in the case where a NAND flash 
manufacturer wants to keep the lifetime of the 3D NAND flash memory device 
constant, our techniques reduce the storage overhead required to hold error
correction information by 78.9\%.}


\section{Characterization of 3D NAND Flash Memory Errors}
\label{sec:3derror:errors}


\chI{\textbf{Our goal} is to identify and understand new error characteristics
in 3D NAND flash memory, through rigorous experimental characterization of
real, state-of-the-art 3D NAND flash memory chips.}
We use the observations and \chVI{analyses obtained from such} characterization to
(1)~compare how the reliability of a 3D NAND flash memory chip differs from that of
a planar NAND flash memory chip,
(2)~develop a model of how \chI{each new} error source affects the error rate of
3D NAND flash memory,
(3)~understand if and how these reliability characteristics will change with
future generations of 3D NAND flash memory, and
(4)~develop mechanisms that can mitigate \chI{new error sources} in 3D NAND flash memory.



For our characterization, we use the methodology discussed in
Section~\ref{sec:3derror:methodology}. First, we perform a detailed
characterization and analysis of three error characteristics that are
drastically different in 3D NAND flash memory than in planar NAND flash \chVI{memory}:
\chV{layer-to-layer} process variation
(Section~\ref{sec:3derror:variation}), \chV{early retention loss}
(Section~\ref{sec:3derror:retention}), and retention interference
(Section~\ref{sec:3derror:retention:interference}).
\chI{In addition to identifying \emph{new} error sources in 3D NAND flash 
memory, we use our methodology to corroborate and quantify 3D NAND error
characteristics that are a result of error sources that were \chVI{\emph{previously}}
identified in planar NAND flash memory, including
retention loss~\cite{cai.book18, cai.procieee17, cai.iccd12, cai.hpca15, choi.vlsit16, park.jssc15, cai.itj13},
P/E cycling~\cite{cai.book18, cai.procieee17, cai.date13, parnell.globecom14, luo.jsac16, park.jssc15},
program interference~\cite{cai.book18, cai.procieee17, cai.iccd13, cai.sigmetrics14, park.jssc15, cai.hpca17},
read disturb~\cite{cai.book18, cai.procieee17, parnell.globecom14, cai.dsn15}, and
process variation~\cite{prabhu.trust11, cai.date12}.
We summarize our findings for these error types in
Section~\ref{sec:3derror:summary}, and provide detailed results on our characterization
of these previously-identified error sources in Appendix~\ref{sec:3derror:appendix}.}



\subsection{Methodology}
\label{sec:3derror:methodology}

We experimentally characterize several real, state-of-the-art 3D MLC
NAND flash memory chips from a single vendor.\footnote{The trends we observe
from the characterization are expected be similar for 3D charge trap flash
\chVI{memory}
manufactured by different vendors, as their 3D flash \chX{memory organizations} are
similar in design.}$^{,}$\footnote{We normalize the actual \chVI{number of}
stacked layers of the chips
and leave out the exact process technology
to protect the anonymity of the flash vendor and to avoid revealing proprietary
information.} We use a NAND flash characterization platform similar to prior
work~\chVII{\cite{cai.fccm11, cai.date12, cai.date13, cai.iccd12, cai.itj13,
cai.iccd13, cai.sigmetrics14, cai.hpca15, cai.dsn15, cai.hpca17,
parnell.globecom14, luo.jsac16, cai.procieee17, cai.book18, luo.hpca18}} and
in Section~\ref{sec:vthmodel:overview:methodology},
which allows us to issue \emph{read-retry} commands directly to the
flash chip. The read-retry command~\cite{cai.procieee17, cai.date13}
allows us to fine-tune the read
reference voltage used for each read operation. The smallest amount by
which we can change the read reference voltage is called a
\emph{voltage step}. We conduct all experiments at room temperature
(\SI{20}{\celsius}).

We use two metrics to evaluate 3D NAND \chVI{flash memory} reliability.
\chVI{First,} we show the
\emph{raw bit error rate} (RBER), which is the rate at which errors
occur in the data \emph{before error correction}. We show the RBER for
when we read data using the \emph{optimal read reference voltage}
($V_{opt}$), which is the read reference voltage that generates the
fewest errors in the data.\footnote{We show RBER at the optimal read
reference voltage to accurately represent the reliability of NAND
flash memory, as SSD controllers tune the read reference voltage to a
near-optimal point to extend the NAND flash lifetime~\cite{cai.hpca15,
papandreou.glsvlsi14, luo.jsac16, cai.procieee17}.}

Second, we show how the various error sources change the {\em
threshold voltage distribution}. These \chVI{changes} (i.e., shifting
and widening) in threshold voltage distribution directly \chVI{lead} to raw
bit errors in the flash memory. To obtain the distribution, we first
use the read-retry command to sweep over all possible voltage values,
to identify the threshold voltage of each cell.\footnote{\chVI{We}
refer to prior work and Section~\ref{sec:vthmodel:overview:methodology}
for more detail \chVII{on} the methodology
to obtain the threshold voltage distribution~\cite{parnell.globecom14,
luo.jsac16, cai.date13}.}  Then, we use this data to calculate the probability
density of each state at every possible threshold voltage value.
As part of our analysis, we fit the threshold voltage
distribution of each state to a Gaussian distribution.  We
use the \emph{mean} of the Gaussian model to represent how the
distribution shifts as a result of errors, and we
use the \emph{standard deviation} of the model to represent how the
distribution widens. Throughout this chapter, we present normalized voltage
values, as the actual voltage values are proprietary to NAND flash memory
\chX{vendors}. A normalized voltage of $1$ represents a single \chV{fixed} voltage step.

We show two examples in Figure~\ref{fig:distribution-shape} to
visualize how well this simple Gaussian model captures the change in
the measured threshold voltage distribution.
Figure~\ref{fig:distribution-shape} shows the measured and modeled
distributions under two conditions: (1)~\chVII{after} 0~P/E cycles,
0-day retention \chVII{time}~\chVI{\cite{cai.hpca15}}, and 0~read disturbs (i.e., the data contains few
errors); and (2)~\chVII{after} 10K~P/E cycles, 3-day
retention \chVII{time}~\chVI{\cite{cai.hpca15}}, and 900K~read
disturbs (i.e., the data contains a high number of errors). \chVII{Dotted
points} plot
the \chVI{measured threshold voltage \chVII{distributions}} from \chX{the real} 3D NAND
\chVII{memory chips}. \chV{Note that we are
unable to show the ER state distribution when \sg{the P/E cycle count} is low (\chVI{i.e.,} the
black dots), because the erase operation cleanly resets the threshold voltage
to a negative value that is lower than the observable voltage range \chVII
{under a low P/E cycle count}.}
We use a solid
line to show a fitted Gaussian distribution for each state. The
\chVII{Kullback-Leibler divergence \chXI{error values}~\cite{luo.jsac16, parnell.globecom14}
of the fitted \chVII{Gaussian} distributions 
are 0.034 and 0.23.}\footnote{\chVII{A KL-divergence error of $x$ means that
the model loses $x$ natural \chVIII{units} of information (i.e., nats) due to modeling
error.}} We observe, from this figure,
that after the chip \chVIII{is
used}, the threshold voltage distribution \chVI{shifts due to P/E cycling,
retention loss, and read disturb}, reducing the
error margins between neighboring states, \chVI{and} leading to more raw bit errors in the
data. Thus, \chVII{depicting and understanding} how \chVII{threshold voltage}
distributions are affected by various \chVI{factors helps}
us understand how raw bit errors occur and \chVI{thus devise mechanisms to}
mitigate \chVI{various} errors more effectively.

\begin{figure}[h]
\centering
\includegraphics[trim=0 10 0 10,clip,width=\figscale\linewidth]{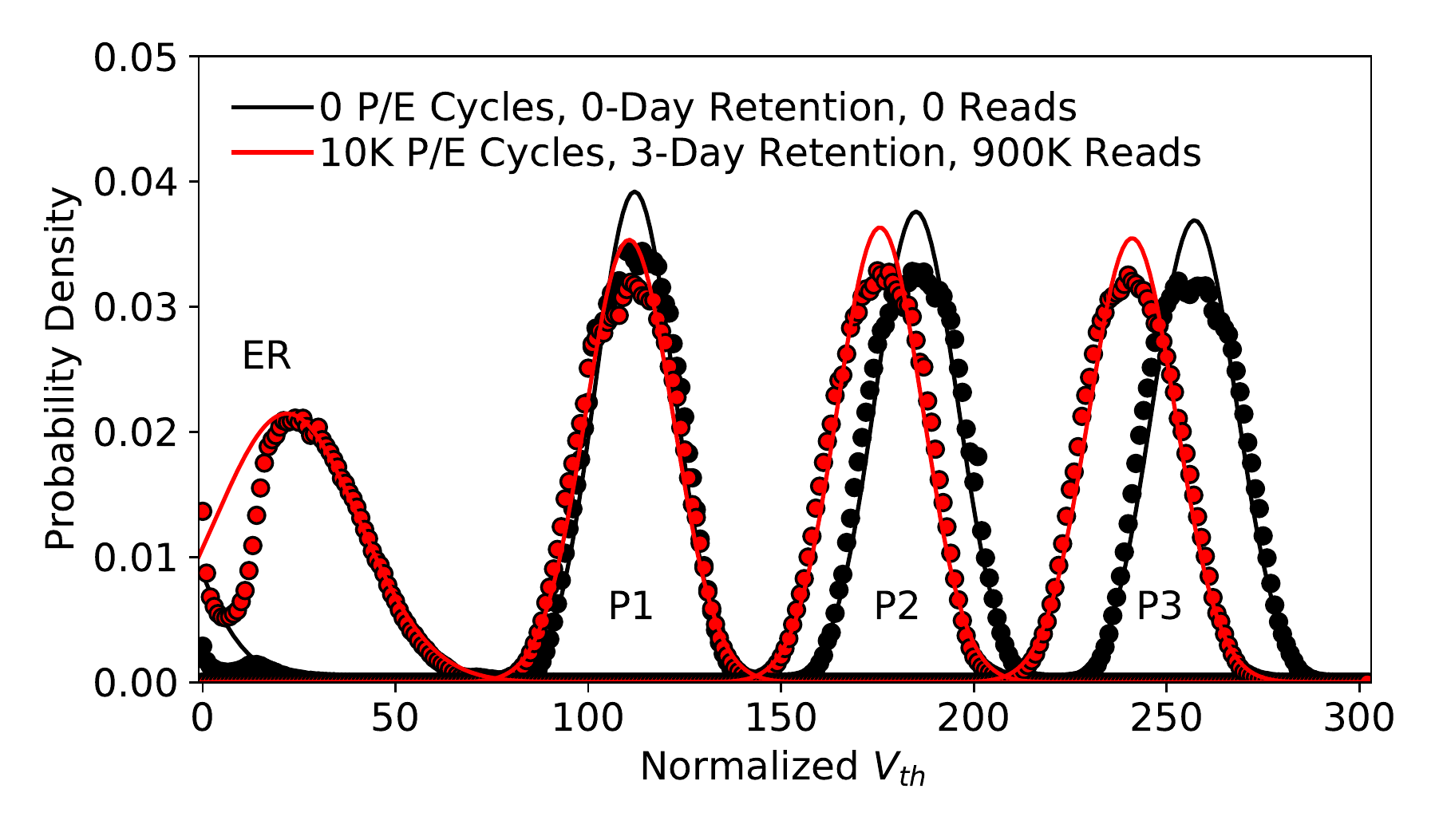}
\caption{3D NAND threshold voltage distribution before (black) and after (red)
the data is subject to a high number of errors \chVI{(due to P/E cycling,
retention loss, and read disturb)}.}
\label{fig:distribution-shape}
\end{figure}

In the following sections, we directly show the mean and the
standard deviation of the \chVII{\emph{fitted}} threshold voltage distributions
instead of the distribution itself, \chVI{to simplify the presentation of our}
results.

\sph{\textbf{Limitations.} In our experiments, we randomly sampled 27 flash
\chVI{blocks throughout} our characterizations. Note that each sampled flash
block consists of tens of millions of flash cells. Thus, we believe that our
observations are representative of the general behavior that takes place in
the model of 3D NAND chips that we tested. While adding
more data samples (i.e., flash blocks to test) can add to the statistical
strength of our results,
we do not believe that this would change the \emph{general qualitative
findings} that we make and the \emph{models} that we develop in this work.
This is because the new error characteristics we observe are caused by the
underlying architecture of 3D NAND flash memory (see
Section~\ref{sec:background:3d}).}

\sph{Note that we do not characterize \emph{chip-to-chip}
process variation, as an accurate study of such variation requires a large-scale
study of a large number (e.g., hundreds) of 3D NAND flash memory chips, which
we do not have access to. Hence, we leave such a large-scale study for future
work.
}









\subsection{Layer-to-Layer Process Variation}
\label{sec:3derror:variation}

Process variation refers to the variation in the attributes of flash cells
when they are fabricated (see Section~\ref{sec:errors}). Due to
process variation, some flash cells can have a higher RBER than others, making
these cells the limiting factor of overall flash memory reliability.
In 3D NAND flash memory, process variation can occur along all three axes of
the memory (see Figure~\ref{fig:organization}). Among
the three axes, we expect the variation along the z-axis (i.e.,
layer-to-layer variation) to be the most significant,
due to the new challenge of stacking multiple flash cells across layers.
\chI{Prior work has shown that current circuit etching technologies are unable
to produce identical 3D NAND cells when punching through multiple stacked
layers, leading to significant variation in the error characteristics of flash
cells that reside in different layers~\cite{wang.tecs17, hung.jssc15}.}

To characterize layer-to-layer process variation errors within a
flash block, we first wear out the block by
programming random data to each page in the block until the block
\chX{endures} 10K P/E cycles. Then, we compare the collective
characteristics of the flash cells in
one layer with those in another layer. We repeat this experiment for
flash blocks on multiple chips to verify all of our findings.




\textbf{Observations.} 
Figure~\ref{fig:variation-wlopterr} shows the RBER
variation along the z-axis (i.e., across layers) for a flash block \chX{that has endured} 10K
P/E cycles. \sph{\chV{The chips we use for characterization have between 30 and 40 layers. We
normalize the number of layers from 0 (the top-most
layer) to 100 (the bottom-most layer) by multiplying the actual layer number
with a constant, \sg{to maintain the anonymity of the chip vendors}.}}
\chVII{Figure~\ref{fig:variation-wlopterr}a}
breaks down the errors
according to the \chX{originally-programmed state and the} current state of each cell;
\chVI{Figure~\ref{fig:variation-wlopterr}b} breaks down the
errors into MSB and LSB page errors.
\chVI{In Figure~\ref{fig:variation-wlopterr}b}, the solid curve and
the dotted curve show the results for two blocks that were randomly selected
from two different flash chips. 
\sph{We make five observations from Figure~\ref{fig:variation-wlopterr}.
First, ER~$\leftrightarrow$~P1 and P1~$\leftrightarrow$~P2 errors vary
significantly across layers, while P2~$\leftrightarrow$~P3 errors remain
similar across layers.
\chVII{The variation in ER~$\leftrightarrow$~P1 errors is mainly caused by the
large variation in mean threshold voltage of \chVIII{the} ER state across layers; the
variation in P1~$\leftrightarrow$~P2 is caused by the variation in the
threshold voltage distribution width of the P1 state across layers
(Section~\ref{sec:3derror:appendix:variation}).}
Second, both the MSB and LSB error rates vary significantly across layers.
\chVI{\chVII{We call this
phenomenon} \emph{layer-to-layer process variation}.}
For example, MSB page on normalized layer~55 in the middle (i.e., \emph{Max
MSB}) has an RBER 21$\times$
\chVIII{that of} normalized layer~0.
Third, MSB error rates are much higher than LSB error rates in a majority of
the layers, \chVI{on average by 2.4$\times$}. \chVI{\chVII{We call this phenomenon} \emph{MSB--LSB RBER variation}. MSB error rates are
usually higher than LSB error rates because reading \chVII{an} MSB page requires two
read reference voltages ($V_a$ and $V_c$), whereas reading \chVII{an} LSB page requires
only one ($V_b$).}
Fourth, the top half of the layers have lower error rates than the bottom half.
\chVII{This is likely caused by the variation in the flash cell size \chVIII{across layers}.}
\chVII{Fifth}, the RBER
variation we observe is consistent across two \chX{randomly-selected} blocks from
two different chips. \chVII{This indicates \chVIII{that layer}-to-layer process
variation and MSB--LSB RBER variation are \chVIII{consistent characteristics}
of 3D NAND flash memory.}}

\begin{figure}[h]
\centering
\includegraphics[trim=0 10 0 10,clip,width=\figscale\linewidth]
{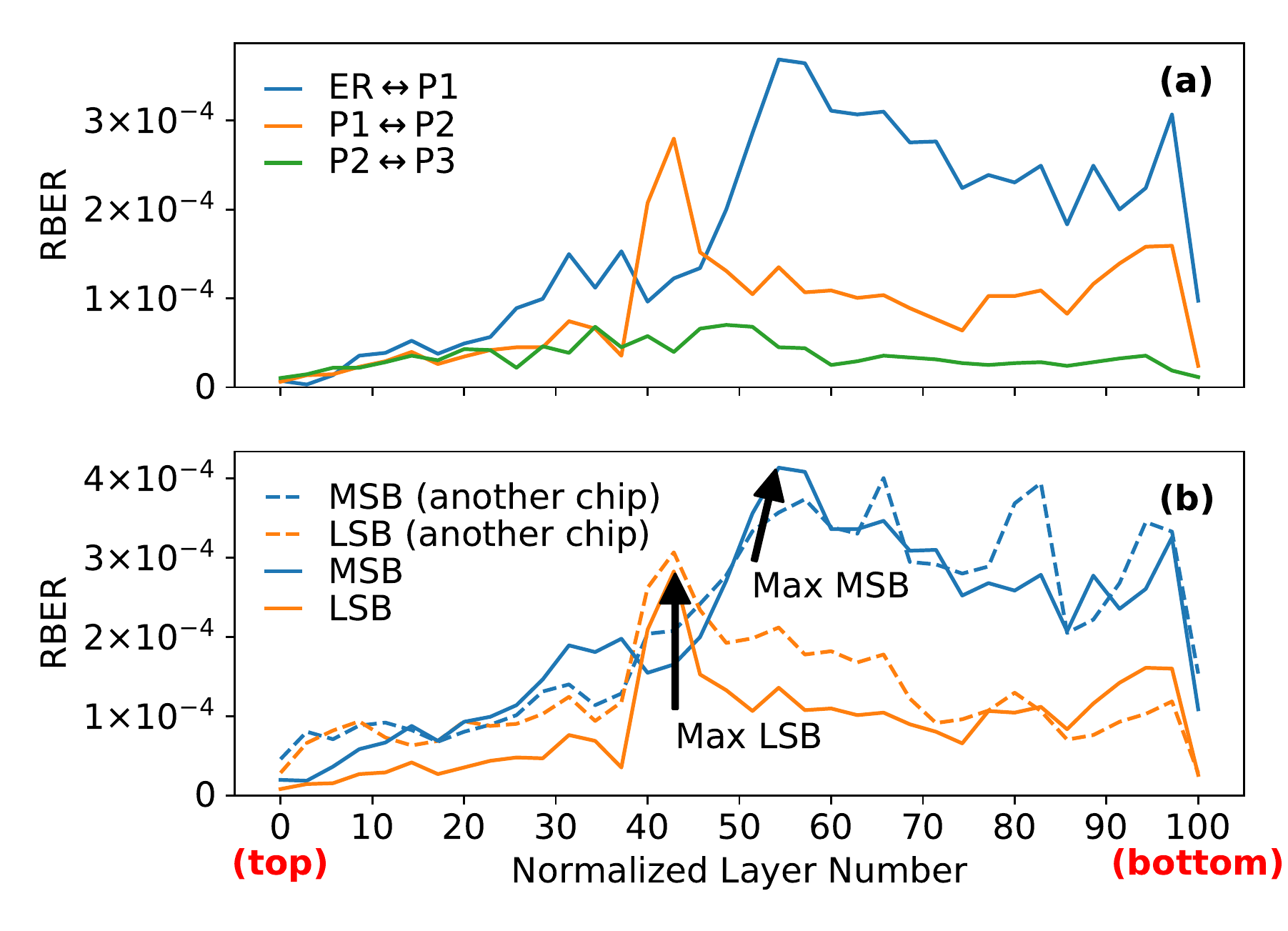}
\caption{\chIX{Variation of RBER across layers.}}
\label{fig:variation-wlopterr}
\end{figure}

Figure~\ref{fig:variation-wloptvrefs} shows how the optimal read reference
voltages vary across layers. Three subfigures show the optimal read reference
voltages for $V_a$, $V_b$, and $V_c$.  We make two observations \chXI{from
Figure~\ref{fig:variation-wloptvrefs}}.  First, the optimal voltages for $V_a$ and $V_b$ vary significantly
across layers, but the optimal \chX{voltage for} $V_c$ does not change by much. \chVI{This is
because process variation mainly affects the threshold voltage distributions of
the ER and P1 states, whereas the threshold voltage distributions of the
P2 and P3 states, \chVII{which are more accurately controlled by ISPP (see
Section~\ref{sec:background}),} are similar across layers. We discuss this
further in Appendix~\ref{sec:3derror:appendix:variation}.}
\sph{Second, \chVI{the optimal read reference voltages for $V_a$ and $V_b$ 
 are lower for cells in the top half of the layers than for cells in the bottom half.}} \chVI{This is because
process variation \chVII{significantly affects} the threshold voltage of the
ER and P1 states (see Appendix~\ref{sec:3derror:appendix:variation}).}

\begin{figure}[h]
\centering
\includegraphics[trim=0 10 0 10,clip,width=\figscale\linewidth]
{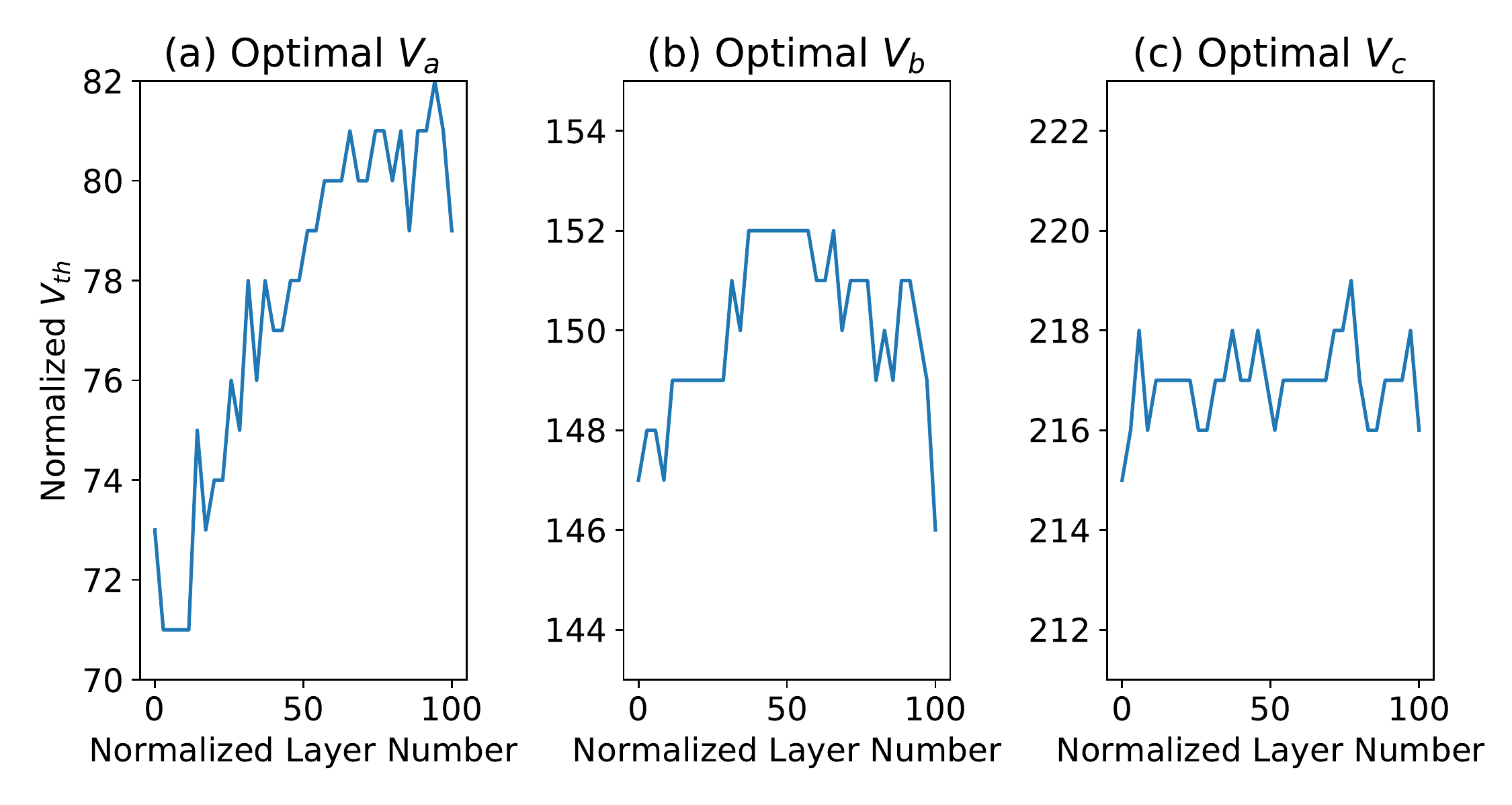}
\caption{\chVII{Variation of optimal} read reference voltage across layers.}
\label{fig:variation-wloptvrefs}
\end{figure}


\textbf{Insights.} We show that the \chVII{\chVIII{phenomena} of}
layer-to-layer process variation \chVIII{and MSB-LSB RBER variation},
which \chVIII{are} unique to 3D
NAND flash memory, \chX{are} significant. We refer to
Appendix~\ref{sec:3derror:appendix:variation} for \chVI{a comparison
between layer-to-layer process variation and}
bitline-to-bitline process variation. In the future, as 3D NAND flash devices
scale along the z-axis, more layers
will be stacked vertically along each bitline. This will \chVI{likely} further
exacerbate
the effect of layer-to-layer process variation, making it even more
important to study and mitigate \chVII{its negative effects}.







\subsection{Early Retention Loss}
\label{sec:3derror:retention}

Retention errors are flash \chVI{memory} errors that accumulate after data has been
programmed to the flash cells~\chVI{\cite{cai.iccd12, cai.hpca15, cai.itj13}}
(see Section~\ref{sec:errors}).
\chI{Because 3D NAND flash memory typically uses a different cell design
(i.e., the charge trap cell described in Section~\ref{sec:background:3d})
than planar NAND flash memory (which uses floating-gate cells), it has
drastically different retention error characteristics.
The charge trap flash cells used in 3D NAND flash memory
suffer from \chVI{\emph{early retention loss}}, i.e., fast charge loss within a few seconds.
This phenomenon \chVI{has been} observed by prior works
using circuit-level characterization~\cite{chen.iedm10, choi.vlsit16}.
However, due to limitations of the \chVI{circuit-level characterization}
methodology used by these prior works, openly-available characterizations of
early retention loss in 3D charge trap NAND flash devices document retention
loss behavior for up to only 5~minutes after the data is written (i.e., for
a maximum \emph{retention time} of 5~minutes).  This limited window is insufficient for
understanding early retention loss under real workloads, which typically have
much longer \chVI{\emph{retention time} \chVII{requirements}}~\cite{luo.msst15}, \chVI{i.e., the length of time 
that has elapsed since programming \chVII{until the data is accessed again}}.}

\chI{Our goal is to experimentally characterize early retention loss in 3D NAND
flash memory for a large range of retention times (e.g., from several minutes
to several weeks).}
\sph{First, we randomly select 11 flash blocks within each chip and write pseudo-random data to each
page within the block to wear the blocks out.  We wear \chX{out} each block to a different
\chX{P/E cycle count}, so that we have error data for every 1K P/E cycles
between 0 and 10K P/E cycles.\footnote{For all
experiments throughout the thesis, we consistently assume a 0.5-second
\emph{dwell time}, which is the length of time between consecutive
program/erase operations~\cite{luo.hpca18}.}
Then, we program pseudo-random data to each flash block, and wait for
up to 24~days under room temperature. To characterize retention loss, we measure
the RBER and the threshold voltage distribution at
nine different retention times, ranging from 7~minutes to 24~days. To
minimize the impact of other errors, and to allow us to include very low retention
times, we characterize only the
first 72~flash pages within each block. We believe that the observations we
make on these flash cells are representative of the entire chip, and we can 
generalize the observations to \chVI{a majority of} 3D NAND \chX{flash memory} cells.}
\chX{We analyze the threshold voltage distribution} in
Appendix~\ref{sec:3derror:appendix:retention}.

\textbf{Observations.} Figure~\ref{fig:retention-3d-vs-2d} shows the comparison
between the retention error rate of 3D NAND and planar NAND flash memory at
10,000 P/E cycles \chVI{\chX{using} both \chX{a} logarithmic time scale \chVII{on the x-axis}
(Figure~\ref{fig:retention-3d-vs-2d}a) and \chX{a} linear time scale \chVII{on the x-axis}
(Figure~\ref{fig:retention-3d-vs-2d}b) for different retention times after
programming}. \sph{To \chVII{make} this comparison,
we perform the same experiment as above for planar NAND flash memory chips.
\chVI{Due to limitations of the available data, we} extend \chVII{our data to
the same retention time range} using \chV{a linear model \sg{that was
proposed by} prior work~\cite{mielke.irps08, luo.hpca18}: $\log(RBER) = A \cdot
\log(t) + B$, where $t$ is the retention time, and $A$ and $B$ are parameters
of the linear model.}} The dotted portions of the lines represent the RBER 
\sg{that is predicted} by the linear model.

\chVI{We make two observations from this figure. First, in
Figure~\ref{fig:retention-3d-vs-2d}a, we} observe that the retention error rate
changes much more slowly for planar NAND \chX{flash memory} than for 3D NAND flash memory.
Although the 3D NAND flash \chX{memory} chip has lower RBER than the planar NAND flash \chX{memory} chip
shortly after programming, the RBER becomes higher on the 3D NAND flash \chX{memory} chip
after \chVII{$7\times 10^3$ seconds ($\sim$2 hours)} of
retention time. This means that
3D NAND flash memory is more susceptible to \chVII{the} retention loss
phenomenon than
planar NAND flash memory. Second, in Figure~\ref{fig:retention-3d-vs-2d}b,
we observe that the RBER of 3D NAND flash memory quickly increases by an order
of magnitude in \chVI{$10^4$ seconds ($\sim$3~hours)}, and by another order of
magnitude in \chVI{$10^6$ seconds ($\sim$11~days)}. However, we do \chVII{\emph{not}}
observe a large difference in retention loss between low and high retention
times \chVI{for \chVII{\emph{planar}} NAND flash memory \chVII{(also
\chVIII{shown} by prior works~\cite{mielke.irps08, cai.hpca15})}.}
This shows that the retention loss
is \emph{steep} when retention time is \emph{low}, but the retention loss
flattens out when the retention time is high. This is a result of the early
retention \chVI{loss phenomenon} in 3D NAND flash memory.

\chX{Early} retention loss
can be caused by two possible reasons. First, the tunnel oxide layer is thinner
in 3D NAND \chVI{flash memory} \chVII{than in planar NAND flash
memory}~\cite{zhang.acsnano14, samsung.whitepaper14}.
Since a 3D charge trap cell uses an insulator to store charge, which is immune
to the short circuiting caused by stress-induced leakage current
(SILC)~\chVI{\cite{naruke.iedm88, degraeve.ted04}}, the tunnel
oxide layer in 3D NAND flash memory is designed to be thinner to improve
\chVI{programming} speed~\cite{park.jssc15}. \chVI{This causes charge to leak
\chVII{very fast soon} after programming.} \chVIII{Second, 
\chX{cells connected on the same bitline
share the same \chX{charge trap} layer. \chX{As a \chXI{result, charge} that is}
programmed to a flash cell quickly leaks to 
\chXII{adjacent cells that are on the same bitline}
due to \chIX{\emph{electron diffusion} through the shared \chX{charge trap}
layer}~\cite{choi.vlsit16}, which we discuss further
in Section~\ref{sec:3derror:retention:interference}.}

\begin{figure}[h]
\centering
\iftwocolumn
  \includegraphics[trim=0 0 0 0,clip,width=\linewidth]{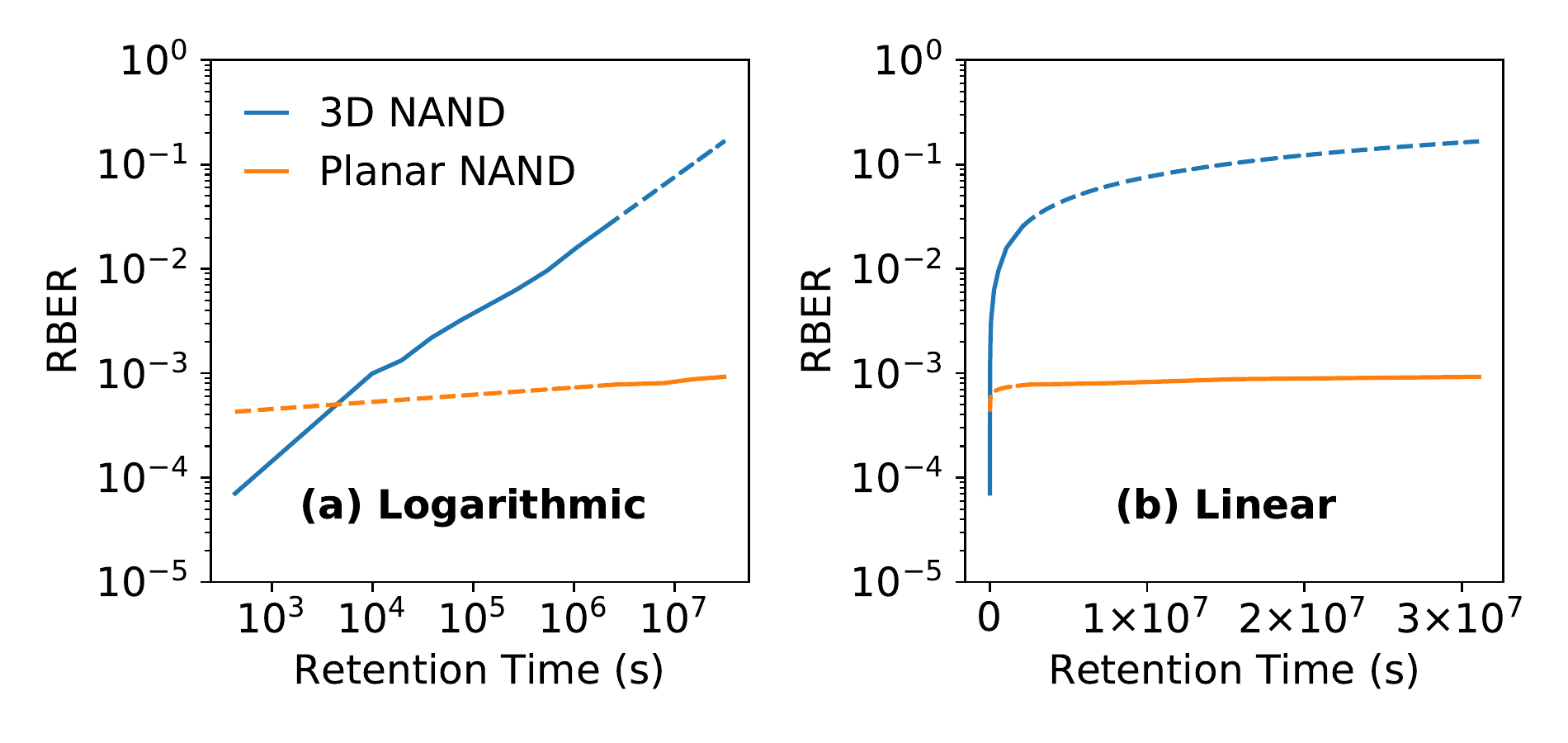}
\else
  \includegraphics[trim=0 0 0 0,clip,width=.7\linewidth]{figs/retention-3d-vs-2d-pec-10000.pdf}
\fi
\caption{Retention error rate comparison between 3D NAND and planar NAND flash
memory \chVI{at 10K P/E cycles}. \chVII{Dotted portions of lines represent
the RBER predicted by the linear model \chVIII{proposed by prior
work~\cite{mielke.irps08, luo.hpca18}}. 
\chX{We show the retention time on the x-axis using both (a)~a \emph{logarithmic}
time scale and (b)~a \emph{linear} time scale.}}}
\label{fig:retention-3d-vs-2d}
\end{figure}

Figure~\ref{fig:retention-optvrefs} plots how the optimal read reference
voltage changes with retention time. The three subfigures show the optimal
voltages for $V_a$, $V_b$, and $V_c$. We make three observations from this
figure.  First, the relation between the optimal read reference voltages
\chVI{of} $V_b$ or $V_c$ and the retention time can be modeled
as~\cite{mielke.irps08, luo.hpca18}: $V = A \cdot \log(t) + B$, similar to the
logarithm of RBER \chVII{(which we \chX{discuss} above)}. Second, the
optimal read reference voltages for $V_b$ and $V_c$ decrease significantly as
retention time increases, whereas $V_a$ remains relatively constant. Third,
\chVI{due to the early retention loss phenomenon,}
the optimal read reference voltages \chVI{for} $V_b$ and $V_c$ change rapidly when the
retention time is low \chVII{(e.g., $V_c$ changes by 5 voltage steps within
the first 3 hours)}, but
they change slowly when the retention time is high \chVII{(e.g., $V_c$ changes
by another 5 voltage steps after 11 days)}.

\begin{figure}[h]
\centering
\includegraphics[trim=0 0 0 0,clip,width=\figscale\linewidth]
{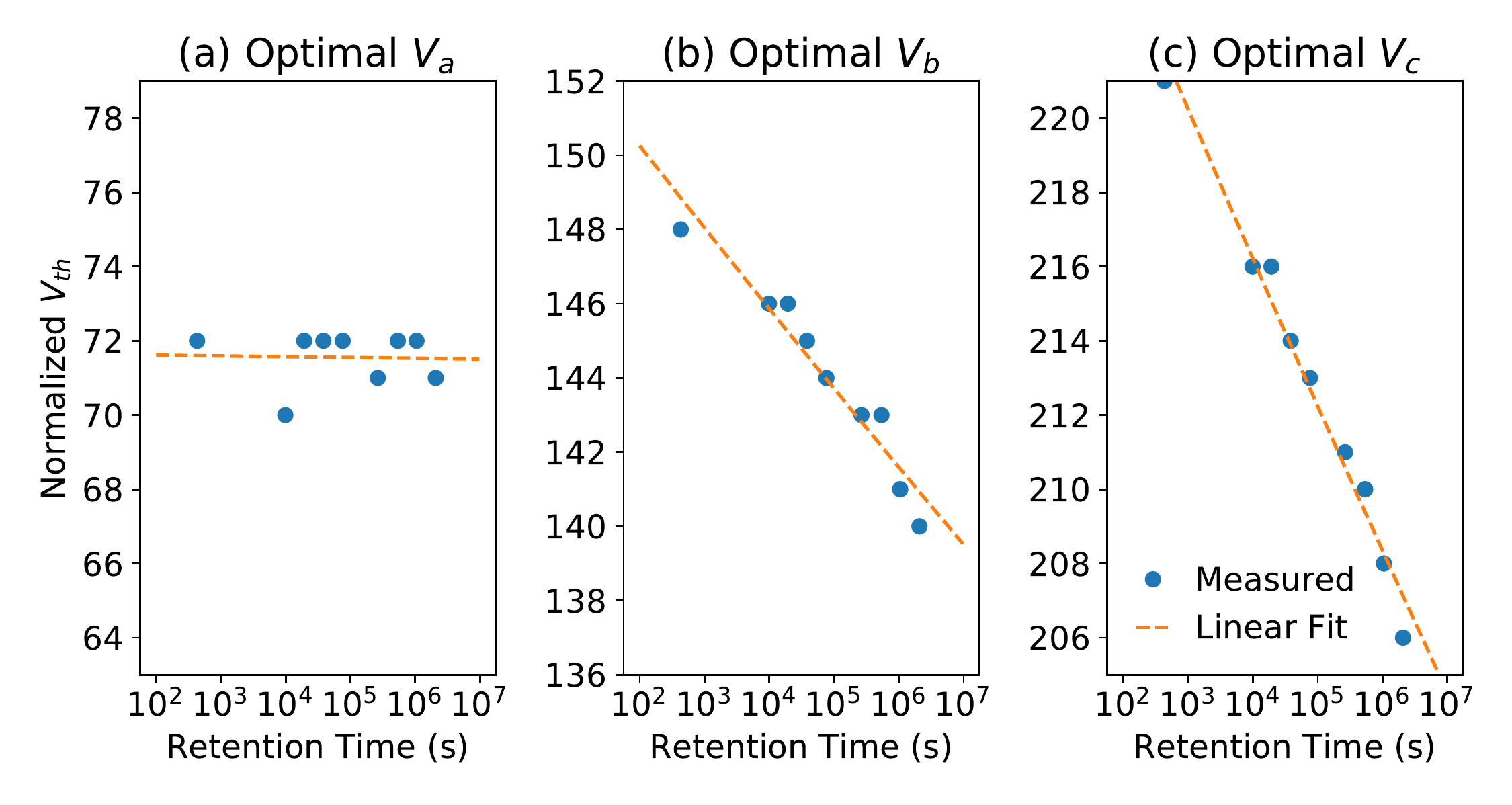}
\caption{Optimal read reference voltages \chVIII{for different} retention
times.
\chVII{Note that \chX{the x-axis uses a} logarithmic time scale.}}
\label{fig:retention-optvrefs}
\end{figure}

\textbf{Insights.} We compare the errors caused by retention loss in 3D NAND \chX{flash memory} to
that in planar NAND \chX{flash memory}, using our results in Figure~\ref{fig:retention-3d-vs-2d}
\chX{and} the results reported in prior work~\cite{mielke.irps08,
cai.hpca15, cai.iccd12}. We find two major differences in 3D NAND \chVI{flash
memory, which we summarize below. More}
results and insights are in Appendix~\ref{sec:3derror:appendix:retention}.
First, \chVI{3D NAND flash memory is more susceptible to retention errors than planar
NAND flash memory, and} its error \chVI{rate} increases much faster when the
retention time is low than when the retention time is high. \chVI{This is a result of
the early retention \chVI{loss} phenomenon in 3D NAND flash memory, which is
due to the use of a different flash cell design and thus is likely \chVII{to}
remain in future generations of 3D NAND flash memory.}
Second, the optimal read reference \chVIII{voltages} for $V_b$ \chVII{and $V_c$} in 3D
NAND flash memory
\chX{change} \chVIII{significantly} with retention time. However, in planar NAND
flash memory, the optimal
voltage \chVI{for} $V_b$ does \chX{\emph{not}} change by much~\cite{cai.hpca15}, \chVIII
{indicating that retention loss is a more pressing phenomenon in 3D NAND
flash memory}. This makes
adjusting the optimal read reference voltages even more important for 3D NAND
flash memory than for planar NAND flash memory. \chVI{We conclude that it is
necessary to develop novel mechanisms to mitigate \chVII{the} early retention
loss \chVII{phenomenon} in 3D NAND flash memory.}

\subsection{Retention Interference}
\label{sec:3derror:retention:interference}

Retention interference is the phenomenon that the speed of retention loss for a
cell depends on the threshold voltage of a
\emph{\chX{vertically}-\chXI{adjacent neighbor} cell} \chVI{whose charge trap layer
is directly connected to the victim cell along the bitline}. Retention
interference is unique to 3D NAND flash memory, as cells along the
\chVI{\emph{same}} bitline in 3D NAND flash memory share the same charge trap
layer. If two neighboring cells \chVI{have} different threshold voltages \chVII{over time},
charge can leak away from the cell with a higher threshold voltage to the cell
with a lower threshold voltage~\cite{choi.vlsit16}.
\chXI{Figure~\ref{fig:retention-interference-cell} shows an example of this
phenomenon, where charge leaks from the top cell (which is in a higher-voltage
state) to the bottom cell (which is in a lower-voltage state) through
the \chXII{shared} charge trap layer.  \chXII{This} charge leakage reduces the threshold voltage of the top cell
while increasing the threshold voltage of the bottom cell.}

\begin{figure}[h]
\centering
\iftwocolumn
  \includegraphics[trim=0 225 635 0,clip,width=.55\linewidth]
  {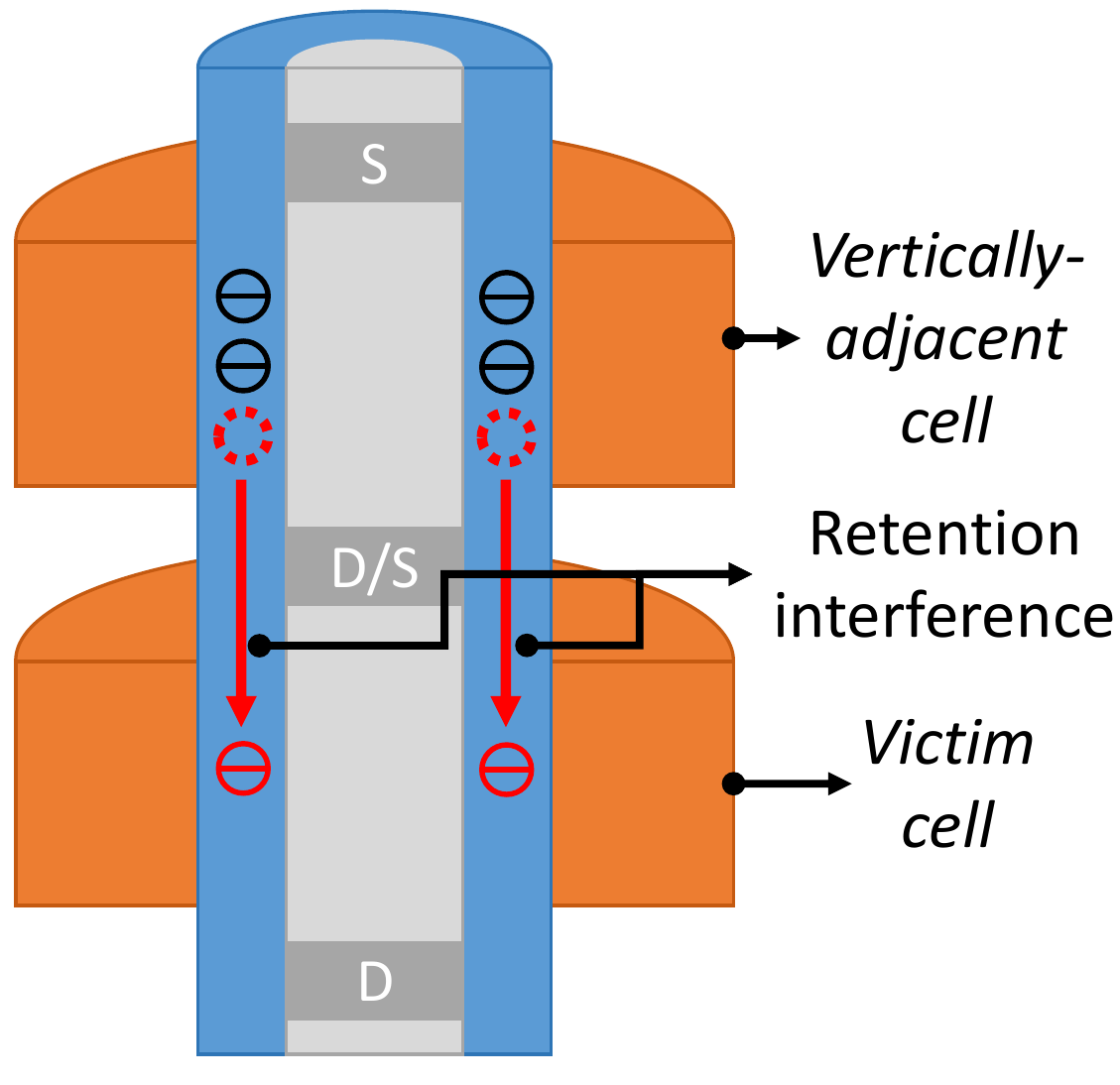}
\else
  \includegraphics[trim=0 225 635 0,clip,width=.35\linewidth]
  {figs/retention-interference-cell.pdf}
\fi
\caption{\chXII{Retention interference phenomenon: a vertically-adjacent cell
leaks charge into a victim cell.}}
\label{fig:retention-interference-cell}
\end{figure}

We use the same data used for retention loss in Section~\ref{sec:3derror:retention} to
observe the \chX{effects} of retention interference. To eliminate any noise due to
program interference, we use \chVI{only} the neighboring cells \chVI{that are
programmed \emph{before} the victim cells} to establish the \chVI{retention}
interference correlation, as these cells \chVI{do \chVII{\emph{not}}
induce program interference on the victim cells}. We also ignore victim cells that
are in the ER state, as they are significantly affected by program interference
\chVI{even though they are programmed after their neighbors~\cite{cai.hpca17}}.
\chVI{Once} program interference \chVI{is eliminated},
the cells should experience a similar threshold voltage shift \chVII{due to
retention loss} \emph{except for}
the effects of retention interference. To find the retention interference, we
first group all \chX{of} the victim cells based on their threshold voltage states and
the states of \chX{their} neighboring cells. Then, we compare the amount by which the
threshold voltages shift over a 24-day retention time, for each group, to
observe how the cells are \chVI{affected} by \chVI{the retention interference
\chVII{caused by}} neighboring cells.

\textbf{Observations.} 
Figure~\ref{fig:retention-interference} shows the average threshold voltage
shift over \chVI{a} 24-day retention time, broken down by the state of the
victim cell (V) and the state of the neighboring cell (N). Each bar represents
a different (V, N) pair. Different shades represent the different states of the
neighboring cell, as labeled in the legend. Every 4 bars are grouped
by the state of the victim cell, as labeled on the \chVI{y-axis}. \chVI{The
length of each bar represents the amount of threshold voltage shift \chVII
{over the 24-day retention time.}}
From Figure~\ref{fig:retention-interference}, we observe that the
threshold voltage shift over retention time is lower when the neighboring
cell is in a higher-voltage state (e.g., the P3~state).


\begin{figure}[h]
\centering
\includegraphics[trim=0 285 0 0,clip,width=\figscale\linewidth]
{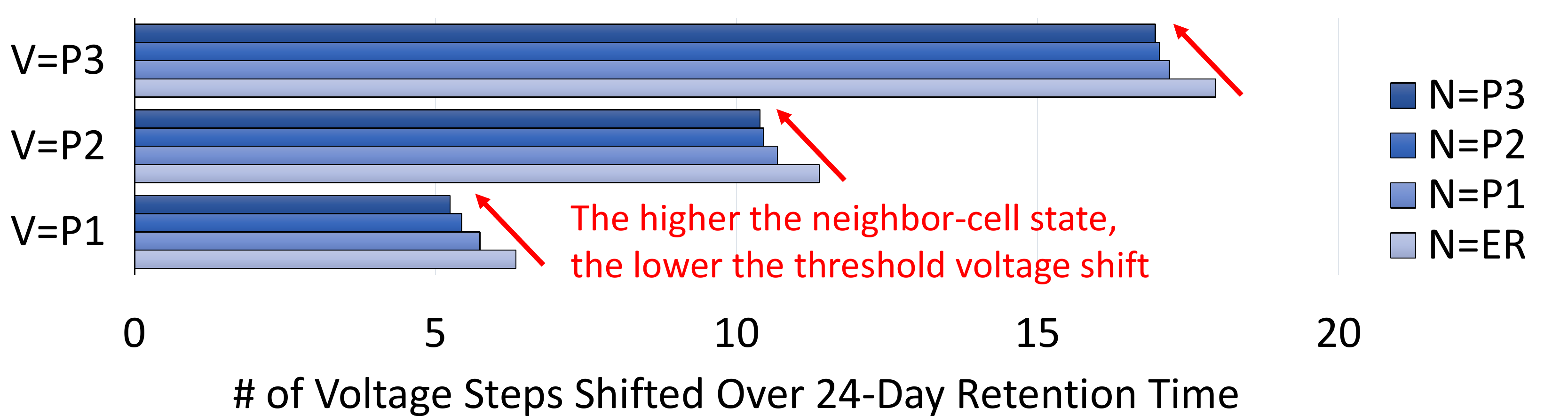}
\caption{Retention interference \chVI{phenomenon observed} at 10K P/E cycles.}
\label{fig:retention-interference}
\end{figure}

\textbf{Insights.} We are the first to \chVI{quantify the retention
interference phenomenon} in 3D NAND flash memory. Our observation from
Figure~\ref{fig:retention-interference} shows that the amount of retention
loss for a flash cell is correlated with its neighboring cell's state. We
expect retention interference to become stronger as we shrink the manufacturing
process technology \chVII{node} in future 3D NAND flash memory devices. This
is because the
distance between neighboring cells will decrease, and fewer electrons will be
stored within each flash cell, increasing the susceptibility of \chVII{a cell} to
interference from neighboring cells.


\subsection{\chVI{Other Error Characteristics}}
\label{sec:3derror:summary}



In addition to the three \chVII{\emph{new}} error sources we find in 3D NAND flash memory, we
also \chI{characterize the behavior} of other \chVII{\emph{known}} error sources in 3D NAND
flash memory \chI{and compare them to their behavior in} planar NAND flash
memory. \chI{We present a high-level summary of our findings for these errors
here, and provide} detailed results and analyses \chVI{for them} in
Appendix~\ref{sec:3derror:appendix}:

\begin{itemize}[topsep=0pt,partopsep=0pt,noitemsep,leftmargin=10pt]

\item Unlike in planar NAND \chVI{flash memory}, we do \emph{not} \chX{find any evidence of}
\emph{program errors}~\chVI{\cite{luo.jsac16, parnell.globecom14,
cai.hpca17}} in 3D NAND \chVI{flash memory}
\chI{(Section~\ref{sec:3derror:programerror})}.

\item \sph{P/E cycling error in 3D NAND flash memory follows a linear trend, which
is similar to that in planar NAND flash memory using an older
manufacturing process technology \chX{node} (e.g.,
\SIrange{20}{24}{\nano\meter})~\cite{cai.date13}. However, in
sub-\SI{20}{\nano\meter} planar NAND flash
memory, P/E cycling error exhibits a \emph{power law}
trend~\cite{parnell.globecom14, luo.jsac16} \chI{(Appendix~\ref{sec:3derror:pecycle})}.}

\item 3D NAND flash memory experiences 40\% \emph{less program interference} than
\SIrange{20}{24}{\nano\meter} planar NAND flash
memory
\chI{(Appendix~\ref{sec:3derror:interference})}.

\item 3D NAND flash memory experiences 96.7\% \emph{weaker read disturb} than
\SIrange{20}{24}{\nano\meter}
planar NAND flash memory.
The impact of read disturb is low enough
in 3D NAND flash memory that it does \emph{not} require significant error
mitigation \chI{(Appendix~\ref{sec:3derror:read:disturb})}.

\end{itemize}
\chI{Note that these differences are
mainly due to the larger manufacturing process technology \chX{nodes} currently used in 3D
NAND flash memory, and thus are not the focus of this thesis.
In comparison, the new error characteristics that we focus on (layer-to-layer
process variation, early retention loss, \chX{and} retention interference) are 
caused by \chVII{the architectural and circuit-level} changes introduced
in 3D NAND flash memory.}



\begin{sidewaystable*}[htbp]
\centering
\iftwocolumn
\else
  \small
\fi
\setlength{\tabcolsep}{5pt}
\begin{tabular}{C{2.4cm}C{5cm}C{4.4cm}C{6cm}}
\toprule
\multirow{2}{\linewidth}{\centering \textbf{Attribute}} & 
\multirow{2}{\linewidth}{\centering \textbf{Observation in 3D NAND}} & 
\textbf{Cause of Difference} & 
\multirow{2}{\linewidth}{\centering \textbf{Future Trend}} \\

& & \textbf{\chVII{in 3D vs. Planar}} & \\ \midrule

\emph{Process Variation} (Section~\ref{sec:3derror:variation}, Appendix~\ref{sec:3derror:appendix:variation},
\ref{sec:3derror:bitline-variation})
  & Layer-to-layer process variation \newline is significant
  & Vertical stacking of flash cells
  & Process variation will increase \newline as we stack more cells vertically \\ \midrule

\multirow{2}{\linewidth}[-4pt]{\centering \emph{Retention Loss} (\chVII{Sections}~\ref{sec:3derror:retention},
\ref{sec:3derror:retention:interference}, Appendix~\ref{sec:3derror:appendix:retention})}
  & Early retention loss
  &\chX{Charge trap} cell
  & Early retention loss will continue \newline if \chX{charge trap} cell is used \\ \cmidrule{2-4}
  & Retention interference
  & Vertical stacking of flash cells
  & Retention interference will increase \newline when smaller process technology \chVI{node} is used \\ \midrule

\emph{P/E Cycling} (Appendix~\ref{sec:3derror:pecycle})
  & Distribution parameters change \chX{with} \newline P/E \chVI{\chX{cycle count} following a} linear trend
  instead of \chVI{a} power-law trend
  & Larger manufacturing process technology \chVI{node}
  & P/E cycle trend will go back to power-law trend when smaller process technology \chVI{node} is used \\ \midrule

  \multirow{2}{\linewidth}[-2pt]{\centering \emph{Program Interference} (Appendix~\ref{sec:3derror:interference})}
  & Wordline-to-wordline interference along \chVI{the} z-axis
  & Vertical stacking of flash cells
  & Will \chVI{continue to} \chVII{exist} in 3D NAND \\ \cmidrule{2-4}
  & 40\% lower program \chVI{interference} 
  & Larger manufacturing process technology \chVI{node}
  & Program \chVI{interference will} increase \newline when smaller process technology \chVI{node} is used \\ \midrule

  \emph{$V_{th}$ Distribution} (\chVII{Section}~\ref{sec:3derror:methodology})
  & ER and P1 states have \newline no programming errors
  & Use of one-shot programming instead of two-step programming
  & Programming errors may \chVI{start occurring} \newline if two-step programming is used \\ \midrule

  \emph{Read Disturb} (Appendix~\ref{sec:3derror:read:disturb})
  & 96.7\% smaller read disturb effect 
  & Larger manufacturing process technology \chVI{node}
  & Read disturb effect will increase \newline when smaller process technology \chVI{node} is used \\

\bottomrule
\end{tabular}
\vspace{5pt}
\caption{Summary \chVI{of error} characteristics of 3D NAND and planar NAND flash memory.}
\label{tbl:summary}
\end{sidewaystable*}

\subsection{Summary}
\label{sec:3derror:summary:summary}

We summarize the key differences between 3D NAND and planar NAND flash memory, in terms of error
characteristics and the expected trends for future 3D NAND \chX{flash memory} devices, in
Table~\ref{tbl:summary}. The first column of this table lists \chX{each} attribute \chX{that} we 
study. The second column shows the key difference in the observation that we
find in 3D NAND flash memory \chVII{versus planar NAND flash memory,} \chX
{for each attribute that we study}. The
third column shows the fundamental cause of each
difference. The last column \chVII{describes} the expected trend of this
difference in
future 3D NAND flash \chX{memory} devices. \chI{We provide the necessary characterizations
and models that help us quantitatively understand these differences in
\chVII{Appendix~\ref{sec:3derror:pecycle}, \ref{sec:3derror:interference},
\ref{sec:3derror:appendix:retention}, 
\ref{sec:3derror:read:variation}, \ref{sec:3derror:read:disturb}, \chX{and} \ref{sec:3derror:appendix:variation}}.}


\section{3D NAND Flash Memory Error Models}
\label{sec:3derror:model}

In \chX{the} previous sections, we have established a basic understanding of the
similarities and differences between 3D NAND and planar NAND flash memory in terms of
\chVII{error characteristics and} reliability. In this section, we quantify these differences by developing
analytical models of the process variation (Section~\ref{sec:3derror:model:variation})
and retention loss (Section~\ref{sec:3derror:model:retention}) \chVIII{phenomena} in 3D NAND flash memory. 
\chVII{These models are useful for at least two major purposes. First,} the
insights \chVIII{obtained from using} these models \chVII{can} motivate \chVII{and enable} us to develop new error mitigation
mechanisms for 3D NAND flash memory. \chVII{Second,} the retention model and the model parameters 
are also useful for comparing the reliability of newer or older
generations of \chX{planar} NAND flash memory with our tested 3D NAND flash \chX{memory} chips.
\sph{\chX{We} focus on developing these models using our existing
characterization data \chVI{from real 3D NAND flash \chVII{memory} chips
\chVII{(some of which was presented in Section~\ref{sec:3derror:errors})}}.
In Section~\ref{sec:3derror:mitigation}, we discuss \chVII{(1)~}how to
efficiently \chVII{\emph{learn}} the \sg{models} for each chip \emph{online}
\chVIII{within the SSD controller}
by performing the characterization and model fitting \chVIII{online},
and \chVII{(2)~}how to use the online \sg{models to
develop mechanisms that improve the lifetime of 3D NAND flash memory}.}

\subsection{\chVI{RBER} Variation Model}
\label{sec:3derror:model:variation}

Since the layer-to-layer variation in 3D NAND \chX{flash memory} causes variation in RBER within a
flash block, 
\chX{it is no longer sufficient to use a single RBER value to represent the
reliability of \emph{all} pages in that block.}
Instead, we model the
variation in per-page RBER within a flash block as a gamma distribution
(i.e., $gamma(x, a, s) = \frac{x^{a-1} e^{-\frac{x}{s}}}
{\Gamma(a) s^a}$). \chV{In this model, $x$ is the RBER;
$a$ is the shape parameter, which \chX{controls} how the RBER distribution is skewed; and
$s$ is the scale parameter, which \chX{controls the width of the RBER distribution}.}

Figure~\ref{fig:rber-distribution} \chV{shows} the probability density for per-page
RBER within a block \chVIII{that has \chX{endured}} 10K P/E cycles. The bars show the measured per-page
RBERs categorized into 50 bins, and the \chX{blue and orange} curves are the fitted gamma
distributions whose parameters are shown on the legend. 
\chX{The blue bars and curve represent the measured and fitted RBER distributions
when the pages are read using the \emph{variation-agnostic $V_{opt}$}.
To find the variation-agnostic $V_{opt}$, we use techniques designed for planar NAND flash
memory to learn a single optimal read reference voltage ($V_{opt}$) for each flash block,
such that the chosen voltage minimizes the overall RBER \emph{across the entire 
block}~\cite{papandreou.glsvlsi14, luo.jsac16}.
The orange bars and curve represent the measured and fitted RBER distributions
when the pages are read using the \emph{variation-aware $V_{opt}$}, \chXI{on a
per-page basis}.
To find the variation-aware $V_{opt}$, we use techniques that are described in
Section~\ref{sec:3derror:mitigation:variation} to efficiently learn an optimal read reference
voltage \emph{for each page in the block}, such that we minimize the \chXI{\emph{per-page}}
RBER.}

\begin{figure}[h]
\centering
\iftwocolumn
  \includegraphics[trim=0 10 0 10,clip,width=\linewidth]
  {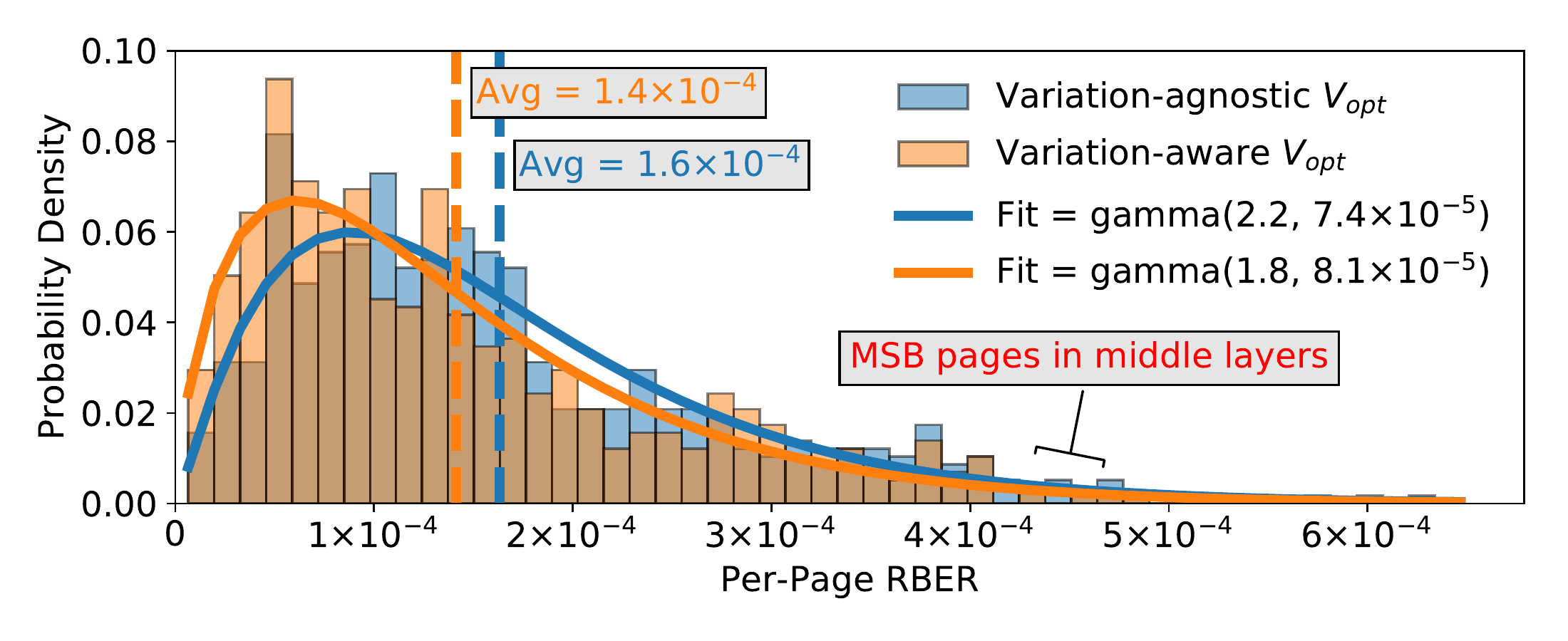}
\else
  \includegraphics[trim=0 10 0 10,clip,width=.85\linewidth]
  {figs/err-distribution-10kPEC.pdf}
\fi
\caption{RBER distribution \chVII{across pages} within a flash block.}
\label{fig:rber-distribution}
\end{figure}

We make \chVI{three} observations from \chX{the}
figure. \chVI{First, the gamma distribution fits well with the \chX{measured} probability
density function of RBER variation across layers: the Kullback-Leibler
divergence \chXI{error value}~\cite{kullback.math51} between the \chX
{measured} and fitted
distributions is \chVII{only} $0.09$.} Second, the average RBER
\chVII{reduces} from $1.6 \times 10^{-4}$ to
$1.4 \times 10^{-4}$ \chVII{when we \chVIII{use} the} variation-aware $V_{opt}$.
Third, \chVII{some} flash pages have a much higher RBER than the average RBER 
(e.g., $>4 \times 10^{-4}$) \chVI{even when \chVIII{we \chX{use} the}
variation-aware $V_{opt}$}.
\chVI{This large gap between the worst-case RBER and the average RBER is caused
by both layer-to-layer process variation and MSB--LSB RBER variation
(see Figure~\ref{fig:variation-wlopterr} in Section~\ref{sec:3derror:variation}).}
\chVI{The pages that have the highest RBER} are MSB pages \chVII{that} reside
in the middle layers.
\chVI{This observation indicates that there is potential to significantly improve
reliability by \chVIII{minimizing} the RBER \chX{variation across flash pages}
(for which we describe a
mechanism in Section~\ref{sec:3derror:mitigation:raid}).}

\subsection{Retention Loss Model}
\label{sec:3derror:model:retention}

\chVI{We} construct a model \chVI{to describe the early retention
loss phenomenon and its impact on RBER ($\log(RBER)$) and threshold voltage 
($V$)} in 3D NAND \chVI{flash memory}, as a function of retention time ($t$) and
\chX{the} P/E cycle \chVI{count} ($PEC$):
$\log(RBER) = A \cdot \log(t) + B$; $V = A \cdot
\log(t) + B$. \chVII{\chX{For both equations,} $A = \alpha \cdot PEC + \beta$ and
$B = \gamma \cdot PEC + \delta$}, \chX{where 
$\alpha$, $\beta$, $\gamma$, and $\delta$ are constants that
change depending on which variable we are solving for.}
\sph{We use ordinary least squares \chXI{method} implemented
in Statsmodel~\cite{seabold.statsmodels10} to fit the model
\chVI{to \chX{our} real characterization data \chX{described} in Section~\ref{sec:3derror:retention}.}
\chVI{Recall that this data is}
collected from 72 flash pages \sg{belonging to} 11 randomly-selected flash blocks.}
Following the \chVI{experimental} observations in Section~\ref{sec:3derror:retention}
and \chVI{in} prior work~\cite{mielke.irps08, luo.hpca18},
we break down our model into two \chX{parts}.
The first \chX{part ($A$)} models the retention loss at a certain P/E cycle \chVI{count} as a
logarithmic function of retention time. The second \chX{part ($B$)} models how
\chVI{the P/E cycle count changes} the parameters of retention loss.

\sph{\chV{Table~\ref{tbl:retention-model} shows all of the parameters we
use to model the RBER and the threshold voltage as a function of the 
\chVI{retention time ($t$) and the P/E cycle count ($PEC$)}. 
In this table, the first column
shows the modeled variable for each row.
\chVII{The second to fifth columns show the parameters (i.e., $\alpha$,
$\beta$, $\gamma$, and $\delta$) fitted to our model. Note that the \chX{model for the}
optimal $V_a$ does not have
$\alpha$ and $\beta$ parameters because \chXI{$V_a$} is insensitive to
retention time.}
The last column shows the \chVII{adjusted coefficient of determination 
(\chXII{\emph{adjusted $R^2$})} of our model.} \chVII{We find that \chXII{our model achieves
high adjusted $R^2$ values for} all variables except for
$\sigma_{ER}$ and $V_a$,
meaning that our model explains $>$89\% of the variation in the
characterized data. The adjusted $R^2$ values are relatively small for
$\sigma_{ER}$ and $V_a$ because these two variables do not change much \chX{with the
retention time or the P/E cycle count}.}
\chVI{\chVII{We conclude that our} model is
accurate and \chX{easy to \chXI{compute}} \chVIII{(as it can be \chX{computed using simple} linear regression)}.
\chX{Thus, our model} is suitable to use 
\chX{online in the SSD controller} (for which we will describe a mechanism in
Section~\ref{sec:3derror:mitigation:retention}).}}}

\begin{table*}[h]
\centering
\setlength{\tabcolsep}{4pt}
\begin{tabular}{ccccccc}
\toprule
\multicolumn{2}{c}{\multirow{3}{*}{\textbf{Variable}}}
	& \multicolumn{4}{c}{\textbf{Model Parameters for:}}
	& \multirow{3}{*}{\textbf{Adjusted $R^2$}} \\
 & & \multicolumn{4}{c}{$Variable = (\alpha \cdot PEC + \beta) \cdot \log(t) +
 \gamma \cdot PEC + \delta$} & \\ \cmidrule{3-6}
 & & $\alpha$ & $\beta$ & $\gamma$ & $\delta$ & \\
\midrule
MSB RBER & $\log(RBER_{MSB})$ &  $5.49\times10^{-6}$ &  0.16 & $1.33\times10^{-4}$ & -13.11 & 97.17\% \\
LSB RBER & $\log(RBER_{LSB})$ &  $7.92\times10^{-6}$ &  0.25 & $3.28\times10^{-5}$ & -12.72 & 90.05\% \\
ER Mean & $\mu_{ER}$ &  $1.01\times10^{-4}$ &  0.74 & $1.52\times10^{-3}$ & -27.27 & 96.86\% \\
P1 Mean & $\mu_{P1}$ & -$1.94\times10^{-5}$ & -0.40 & $3.51\times10^{-4}$ &  114.47 & 95.88\% \\
P2 Mean & $\mu_{P2}$ & -$4.71\times10^{-5}$ & -0.70 & $3.23\times10^{-4}$ &  189.58 & 98.50\% \\
P3 Mean & $\mu_{P3}$ & -$7.37\times10^{-5}$ & -1.20 & $5.75\times10^{-4}$ &  264.85 & 98.29\% \\
ER Stdev & $\sigma_{ER}$ &  $1.20\times10^{-5}$ & -0.10 & $1.63\times10^{-6}$ &  17.01 & 56.33\% \\
P1 Stdev & $\sigma_{P1}$ & -$1.34\times10^{-6}$ &  $9.83\times10^{-3}$ & $7.55\times10^{-5}$ &  10.20 & 93.20\% \\
P2 Stdev & $\sigma_{P2}$ & -$2.12\times10^{-6}$ &  $9.85\times10^{-3}$ & $6.69\times10^{-5}$ &  10.65 & 89.02\% \\
P3 Stdev & $\sigma_{P3}$ &  $2.87\times10^{-6}$ &  $1.40\times10^{-2}$ & $3.30\times10^{-5}$ &  10.83 & 93.00\% \\
Optimal $V_a$ & $V_a$ & --- & --- & $1.20\times10^{-3}$ &  60.52 & 71.20\% \\
Optimal $V_b$ & $V_b$ & -$3.72\times10^{-5}$ & -0.57 & $4.20\times10^{-4}$ &  150.56 & 94.27\% \\
Optimal $V_c$ & $V_c$ & -$6.51\times10^{-5}$ & -1.06 & $4.81\times10^{-4}$ &  227.24 & 97.72\% \\
\bottomrule
\end{tabular}
\vspace{5pt}
\caption{\sph{\chVI{Retention loss model for 3D NAND flash memory and its model parameters}.
\emph{PEC} is P/E cycle lifetime, \emph{t} is retention time.}}
\label{tbl:retention-model}
\end{table*}


\section{3D NAND Error Mitigation Techniques}
\label{sec:3derror:mitigation}

Motivated by our new findings in Section~\ref{sec:3derror:errors},
we aim to \chI{design new techniques that} mitigate
the three unique error effects \chVI{(i.e., layer-to-layer process variation, early
retention loss, and retention interference)} in 3D NAND flash memory. We propose
four error mitigation mechanisms.  To mitigate layer-to-layer process
variation, we propose LaVAR and LI-RAID\@. LaVAR learns \chX{our new} \chXI{RBER} variation
model \chX{(see Section~\ref{sec:3derror:model:variation})} online in the SSD
controller, and uses this model to predict and apply \chVI{an} optimal read reference
voltage that is fine-tuned \chXI{to} each layer (Section~\ref{sec:3derror:mitigation:variation}).
LI-RAID is a new RAID scheme that reduces
the RBER variation induced by layer-to-layer process variation in 3D NAND flash memory
(Section~\ref{sec:3derror:mitigation:raid}). To
mitigate retention loss in 3D NAND flash memory, we propose ReMAR, a new technique that
tracks the retention time information within the SSD controller and uses \chX{our new}
retention \chVI{loss} model \chX{(see Section~\ref{sec:3derror:model:retention})}
to predict and apply the optimal read reference voltage \chX{that is} fine-tuned \chXI{to} the
retention time of the data (Section~\ref{sec:3derror:mitigation:retention}). To mitigate
retention interference, we propose ReNAC, which is adapted from \chVI{neighbor-cell
assisted correction (NAC)~\cite{cai.sigmetrics14},} an existing
technique originally designed to reduce program interference in planar NAND
\chVI{flash memory,} to
also account for retention interference in 3D NAND flash memory
\chI{(Section~\ref{sec:3derror:mitigation:nac})}.

\subsection{LaVAR\@: Layer Variation Aware Reading}
\label{sec:3derror:mitigation:variation}

In planar NAND \sg{flash memory}, existing techniques assume that \chX{the RBER} is
the \chVII{\emph{same}} across all pages within a flash \chVI{memory} block,
\chX{and, thus, a single $V_{opt}$ value can be used for all pages in the block}~\cite{cai.hpca15,
papandreou.glsvlsi14}.
\chX{This approach is called} \chVI{\emph{variation-agnostic $V_{opt}$}}.
However, as our results in Section~\ref{sec:3derror:variation}
show, this assumption no longer \chVII{holds} in 3D NAND \chVI{flash memory} \chVII{due to}
layer-to-layer process variation, \chX{as each page in a block resides in a different layer}. 
\chVI{\chVII{We} aim to improve flash memory
lifetime by mitigating layer-to-layer process variation and reducing \chX{the} RBER. 
The key idea is to identify how much the read reference voltage
must be offset by for \chVII{\emph{each}} layer in a flash chip, to account for the
layer-to-layer process variation, \chVII{instead of using a single read
reference voltage for the \emph{entire} block} \chVIII{irrespective of layers}.
When the SSD controller performs a read request, it \chX{accounts for 
\chXI{(1)~\chXII{per-block variation in RBER,} by predicting a variation-agnostic $V_{opt}$
based on the P/E cycle count of the flash block;
and (2)}~layer-to-layer variation, by adding} the layer-specific 
offset to the variation-agnostic $V_{opt}$ for \chX{the target} block.  \chX{This generates} a
\emph{variation-aware $V_{opt}$} that the controller uses as the
read reference voltage.}

\textbf{Mechanism.} \chVIII{We devise a new
mechanism called \chX{\emph{Layer Variation Aware Reading} (LaVAR)},
which (1)~learns the voltage offsets for each layer and records them in
per-chip tables in the \chX{SSD} controller, and (2)~uses the variation-aware
$V_{opt}$ during a read operation
by reading the appropriate voltage offset for the request from the
\chX{per-chip table that corresponds to the layer} of the request.}
\chVI{LaVAR constructs}
\chV{a model of the optimal read reference voltage ($V_{opt}$)
variation across different layers. Since there are only a limited number of layers,
this model can be represented as a table (i.e., it is a non-parametric model)
of the offset between the $V_{opt}$ \chVII{for} each layer 
(\emph{variation-aware
$V_{opt}$}) and the overall $V_{opt}$ for the entire flash block
(\emph{variation-agnostic $V_{opt}$}). Any previously-proposed model for
$V_{opt}$\chVII{~\cite{luo.jsac16, papandreou.glsvlsi14, cai.hpca15}} can \sg{be used to calculate the} variation-agnostic
$V_{opt}$.}
Since the \chVI{layer-to-layer} process variation is similar across blocks and
is consistent across P/E \chX{cycle counts}, the $V_{opt}$ variation model can be learned 
\emph{offline} for each chip
through an extensive characterization of a single flash block. To do this, the
SSD controller randomly picks a flash block and records the difference between
the variation-aware $V_{opt}$ and the variation-agnostic $V_{opt}$. 
\sph{\chVI{LaVAR uses the existing read-retry functionality in modern NAND flash
\chX{memory} chips (see Section~\ref{sec:3derror:methodology}) to find the 
variation-aware $V_{opt}$ online.}} The
controller then computes and stores the average $V_{opt}$ offset for each
layer in a lookup table \chVIII{stored for each chip}. Note that $V_c$
variation does not need to be modeled,
since $V_c$ is unaffected by layer-to-layer process variation \chVI{(see
Figure~\ref{fig:variation-wloptvrefs} \chVII{in} Section~\ref{sec:3derror:variation})}.

\chVI{When performing a read operation,} the SSD controller
simply looks up the $V_{opt}$ \chVI{offset} 
\chVI{that corresponds to the layer \chVIII{and the chip} that contains the
data being read,}
and \chVI{adds} the offset to the per-block $V_{opt}$ predicted by existing
techniques~\cite{luo.jsac16, papandreou.glsvlsi14, cai.hpca15}. By \chVII{using}
variation-aware $V_{opt}$, LaVAR \chVI{enables the use of} a more accurate
$V_{opt}$ for 3D NAND \chVI{flash memory} than existing techniques, and \chVII
{thus} reduces the RBER (see Figure~\ref{fig:rber-distribution} in
Section~\ref{sec:3derror:model:variation}).

\textbf{Overhead.} \sph{\chV{\chVI{LaVAR} can be implemented fully in the SSD
controller firmware, and, thus, does not require any modification to the
hardware.} Assuming that the 3D NAND \chX{flash memory chip} has $N$~layers and that it takes
\sg{1~Byte to store each} $V_{opt}$ \chVI{offset for each layer}, the memory
overhead of \chVI{storing} the lookup table \chVI{for $V_a$ and $V_b$} in the
SSD controller is $2N$~Bytes. The latency overhead of each read operation is
negligible as LaVAR requires \chVII{only} a table lookup and an addition to obtain
variation-aware $V_{opt}$, which take less than \SI{100}{\nano\second}. Since the lookup table
is shared across \chVIII{\emph{all} blocks in a chip}, it needs to be learned
\chVII{only} once, and \chX{it} can be
\chVI{constructed} gradually in the background. \chVI{Thus,} the performance
overhead \chVI{of LaVAR} is negligible.}

\textbf{Evaluation.} Figure~\ref{fig:vopt-variation} compares the RBER \chVIII
{obtained} by \chVIII{using} LaVAR (\emph{variation-aware
$V_{opt}$})\chVII{~\cite{luo.jsac16, papandreou.glsvlsi14, cai.hpca15}} to
that \chVIII{obtained} by \chVIII{using} an existing read
reference voltage tuning technique (\emph{variation-agnostic $V_{opt}$})
designed for planar NAND \chX{flash memory}.
We evaluate the average RBER \chVIII{obtained} by each
mechanism by simulating \chX{read operations} using our characterization data in
Section~\ref{sec:3derror:variation}. \chXI{Averaged} \chVI{across all P/E cycle counts},
LaVAR reduces \chX{the} RBER by 43.3\%.
The benefit comes from tuning the read reference voltage towards the
\chVI{variation-aware $V_{opt}$} by an offset learned by our model. \chVI{The
RBER} reduction becomes smaller as \chX{the} P/E cycle \chVI{count} increases, because the
overall RBER increases exponentially as the \chX{NAND} flash memory wears out, decreasing
the fraction of process variation errors. \sph{\chV{While the flash lifetime
improvements \chVII{produced} by LaVAR might \chVIII{seem} small (as we show
in Section~\ref{sec:3derror:mitigation:all}), 
(1)~they are achieved with negligible overhead, and
(2)~the RBER reduction enabled by LaVAR throughout
the flash memory lifetime reduces the average flash read latency~\cite{cai.hpca15}.
As the number of layers within
a 3D NAND flash memory \chX{chip} grows (e.g., \chX{vendors}
are already bringing chips with 96~layers to the market~\cite{anandtech.web17}),
we expect that layer-to-layer process variation will increase,
which in turn will increase the magnitude of the lifetime benefits
provided by LaVAR.}}

\begin{figure}[h]
\centering
\iftwocolumn
  \includegraphics[trim=0 0 0 0,clip,width=\linewidth]
  {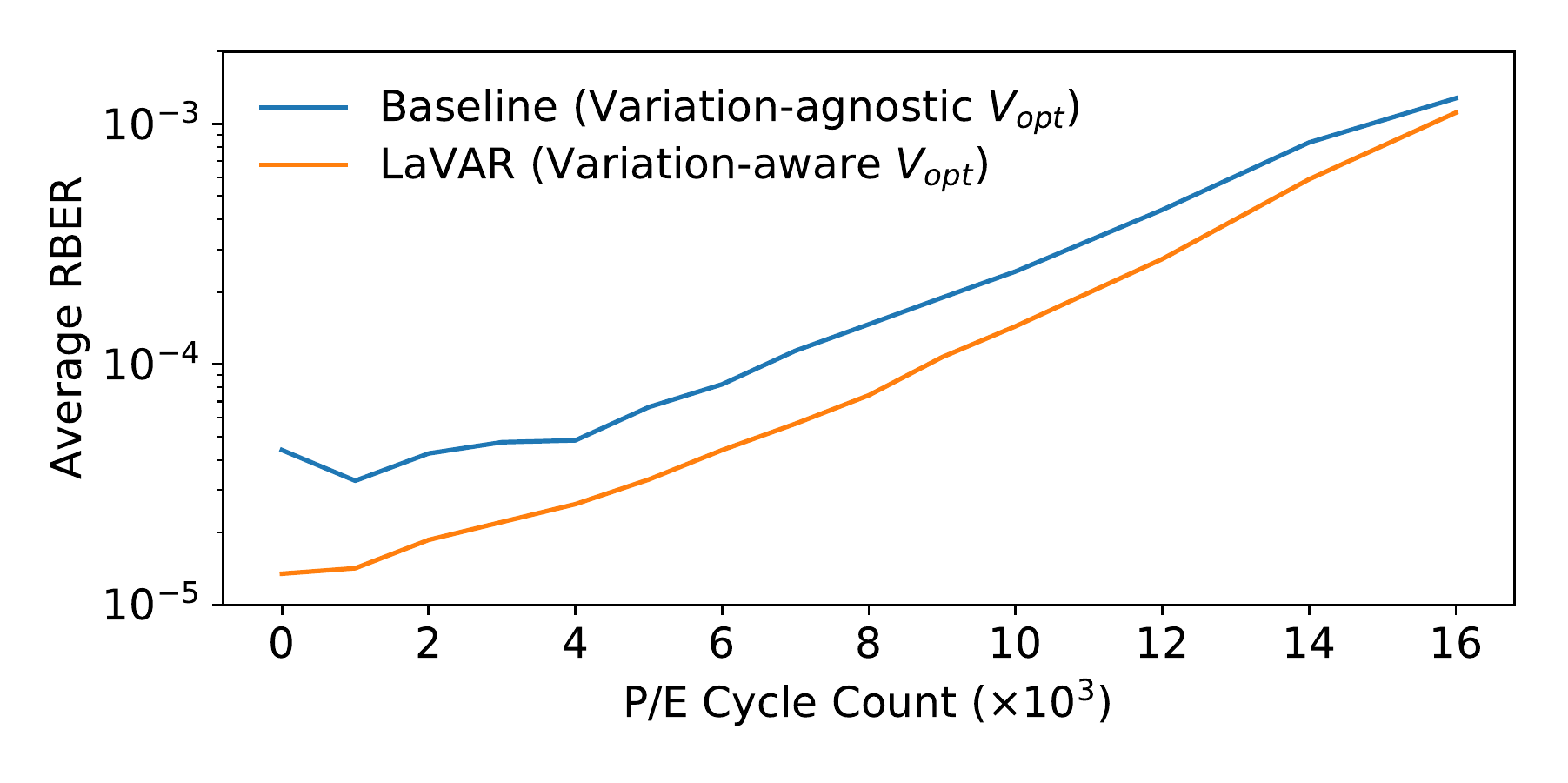}
\else
  \includegraphics[trim=0 0 0 0,clip,width=.65\linewidth]
  {figs/vopt-model-pv.pdf}
\fi
\vspace{-3mm}
\caption{RBER reduction using LaVAR.}
\label{fig:vopt-variation}
\end{figure}

\subsection{LI-RAID\@: Layer-Interleaved RAID}
\label{sec:3derror:mitigation:raid}

As we \chVI{observe} in Section~\ref{sec:3derror:model:variation}, even
after applying the variation-aware $V_{opt}$, \chVI{the per-page RBER is
distributed over a wide range according to a \chVIII{fitted} gamma
distribution}
due to
layer-to-layer process variation and MSB--LSB RBER variation
(see Figure~\ref{fig:variation-wlopterr} in Section~\ref{sec:3derror:variation}). In
enterprise SSDs, \chVI{in addition to ECC, the Redundant Array of Independent Disks
(RAID)~\cite{patterson.sigmod88, balakrishnan.tos10} error recovery
technique} is \chVII{used} across \chVIII{\emph{multiple}} flash chips to tolerate
chip-to-chip process variation \chVIII{in error rates}. \chVI{RAID in modern
SSDs typically combines one flash page from
each flash chip into a logical unit called a \emph{RAID group}, and uses one of the
pages to store the parity information for the entire group.}
However, state-of-the-art RAID schemes do \chVI{\emph{not}} consider
layer-to-layer process variation and MSB--LSB RBER variation.  These schemes
group MSB \chI{or} LSB pages in the \chVI{\emph{same}} layer together \chI{in
a RAID group}. As a
result, the reliability of the SSD is limited by the RBER of the weakest 
(i.e., the least reliable) RAID group that contains the MSB or LSB pages from
the least reliable layer across all chips. We devise a new RAID scheme called
\chX{\emph{Layer-Interleaved RAID} (LI-RAID)}, \chVI{which} eliminates these low-reliability RAID
groups by equalizing the RBER among different RAID groups. \chVI{LI-RAID makes
use of two key ideas: (1)~group flash pages in less reliable layers
with pages in more reliable layers, and (2)~group MSB pages with LSB pages.}

\textbf{Mechanism.}
Instead of
grouping pages in the same layer together \chVI{in the same RAID group}, we
select pages from different chips
and different layers and group them together, such that \chVI{the low-reliability
pages (either due to layer-to-layer process variation or MSB--LSB RBER
variation) are \emph{distributed} to different} RAID groups. Thus,
the new groups formed by LI-RAID have a more evenly-distributed \chVII{RBER}
than the groups formed using
traditional \chVI{layer-unaware} RAID schemes. We assume, without loss of
generality, that there
are $m$ chips in the SSD, and each RAID
group contains $m$ pages, \chI{one from each chip}. We also assume that each block
contains $n$ wordlines, and that the layer numbers of each wordline are in ascending
order  (e.g., the wordline in layer~$i$ has a lower wordline number than its
neighboring wordline in layer~$i+1$). Thus, LI-RAID groups together
the MSB page of wordline~$0$, the LSB page of wordline~$\frac{n}{m}$, the MSB
page of wordline~$2\cdot\frac{n}{m}$, \chVI{the LSB page} \ldots, \chVII
{the MSB page of wordline~$(m-2)\cdot \frac{n}{m}$}, the \chVI{LSB} page
of wordline~$(m-1)\cdot \frac{n}{m}$.
\chI{Figure~\ref{fig:li-raid} shows an example LI-RAID
layout on an SSD with 4 chips and with 4 wordlines within each flash block.
Flash pages in the same RAID group are highlighted in the same color.}
In this way, LI-RAID \chVI{distributes} the less reliable
pages within each chip across different RAID groups, \chVIII{thereby avoiding}
the \chVIII{formation of significantly less reliable} RAID groups that \chVI
{bottleneck} SSD reliability.

\begin{figure*}[h]
\centering

\begin{tabular}{ccccccc}
\toprule
\textbf{Wordline \#} & \textbf{Layer \#} & \textbf{Page} & \textbf{Chip 0} & \textbf{Chip 1} & \textbf{Chip 2} & \textbf{Chip 3} \\
\midrule
0 & 0 & MSB & \cellcolor{olive!25}Group 0 & Blank & \cellcolor{red!25}Group 4 & \cellcolor{green!25}Group 3 \\
0 & 0 & LSB & \cellcolor{blue!25}Group 1 & Blank & \cellcolor{cyan!25}Group 5 & \cellcolor{gray!25}Group 2 \\
1 & 1 & MSB & \cellcolor{gray!25}Group 2 & \cellcolor{blue!25}Group 1 & Blank & \cellcolor{cyan!25}Group 5 \\
1 & 1 & LSB & \cellcolor{green!25}Group 3 & \cellcolor{olive!25}Group 0 & Blank & \cellcolor{red!25}Group 4 \\
2 & 2 & MSB & \cellcolor{red!25}Group 4 & \cellcolor{green!25}Group 3 & \cellcolor{olive!25}Group 0 & Blank \\
2 & 2 & LSB & \cellcolor{cyan!25}Group 5 & \cellcolor{gray!25}Group 2 & \cellcolor{blue!25}Group 1 & Blank \\
3 & 3 & MSB & Blank & \cellcolor{cyan!25}Group 5 & \cellcolor{gray!25}Group 2 & \cellcolor{blue!25}Group 1 \\
3 & 3 & LSB & Blank & \cellcolor{red!25}Group 4 & \cellcolor{green!25}Group 3 & \cellcolor{olive!25}Group 0 \\
\bottomrule
\end{tabular}

\caption{LI-RAID layout example for an SSD with 4 chips and with 4 wordlines
in each flash block.}
\label{fig:li-raid}
\end{figure*}

\sph{\chI{Note that, \chVI{since \chVII{the} order of RAID group number is different
in each flash chip}, the LI-RAID layout may potentially violate the
program sequence recommended by flash vendors,
where wordlines within each flash block
must be programmed \emph{in order} to minimize harmful program
interference~\cite{cai.sigmetrics14, cai.iccd13,
park.dac16, cai.procieee17}.
For example, in Chip~2 in Figure~\ref{fig:li-raid},
\chVIII{Wordline~3 (\chX{Groups 2 and 3}) is programmed \emph{after}
Wordline~2 (\chX{Groups 0 and 1}).
In Chip~2, we leave \chX{Wordline~1} \emph{blank} 
(marked as``Blank'' \chX{in Figure~\ref{fig:li-raid}}).  \chX{Otherwise, 
Wordline~1 would} cause
program interference to \chX{the data in Wordline~2,
which already experiences program interference when Wordline~3
is programmed, significantly increasing the error rate of
Wordline~1~\cite{cai.iccd13, cai.sigmetrics14} (see 
Appendix~\ref{sec:3derror:interference}).}}
\chX{By laying out the data in the proposed manner}, 
LI-RAID provides the same reliability guarantee as
the recommended program sequence, by guaranteeing that any data stored in a
flash page experiences program interference from \chVII{\emph{at most}} one
neighboring wordline.}}

\chI{\textbf{Overhead.} \sph{\chV{The grouping of flash pages by LI-RAID is
implemented entirely in the SSD controller firmware. This requires the
firmware to be aware of the physical-page-to-layer \chVI{mapping.}}
The \chX{flash pages} left blank in LI-RAID incur a small
additional storage overhead compared to a conventional RAID scheme. Only
\emph{one} wordline \chX{(\chX{i.e., two pages in MLC NAND flash memory)} 
within a flash block is left blank, \chX{to mitigate the
impact of program interference on Groups 0 and 1}. 
\chX{Without this blank wordline, the data in Groups 0 and 1 would be the
only data to experience program interference twice: once when Groups 2 and 3
are programmed, and once when the last two groups are programmed.}
In modern NAND flash
memory, each flash block typically contains \chXI{at least 256} flash pages. Thus,
the additional storage overhead \chX{for the blank pages} is less than \chVII{0.8\%}.
LI-RAID does not incur additional computational overhead because it computes
parity in the same way as a conventional RAID \chVI{scheme}, and only
\chVI{\emph{reorganizes}} the RAID
groups differently. Because we do \chVI{\emph{not}} change the data layout across flash
blocks, the flash translation layer (FTL) and the garbage collection (GC)
\chX{algorithms} remain the same as in a conventional RAID scheme.}}

\textbf{Evaluation.} Figure~\ref{fig:99p-rber} plots \chVI{the \emph{worst-case
RBER} (i.e., the highest per-page RBER within a flash block)} when \chVIII{we
use} different error mitigation techniques at 10,000
P/E cycles. \chVI{Recall that the per-page RBER within a flash block follows
a gamma distribution (see Figure~\ref{fig:rber-distribution} in
Section~\ref{sec:3derror:model:variation}).}
Thus, several \chVI{least-reliable} flash pages within a block may become
unusable (i.e., their RBER exceeds the ECC correction capability)
before the \emph{overall} RBER of the flash chip
exceeds the ECC correction capability. We use the worst-case RBER to represent the reliability
of these \chVI{least-reliable} flash pages. In this figure, the baseline
\chVIII{uses} the
per-block variation-agnostic optimal read reference voltage (i.e.,
variation-agnostic $V_{opt}$), achieving a worst-case RBER of $4.8\cdot10^{-4}$. When
we \chVIII{use} the variation-aware $V_{opt}$ proposed in
Section~\ref{sec:3derror:mitigation:variation}, the worst-case RBER is reduced by 9.6\% over the
baseline, to $4.3\cdot10^{-4}$. LI-RAID reduces the
worst-case RBER by 66.9\% over the baseline, to only $1.6\cdot10^{-4}$. Thus, by
grouping flash pages on less reliable layers with pages on more reliable
layers, and by grouping MSB pages with LSB pages, LI-RAID
reduces the probability of unusable pages within a block, \chVI{thereby
\chVIII{reducing}} the
number of retired flash blocks due to ECC failures.

\begin{figure}[h]
\centering
\includegraphics[trim=0 390 295 0,clip,width=\figscale\linewidth]
{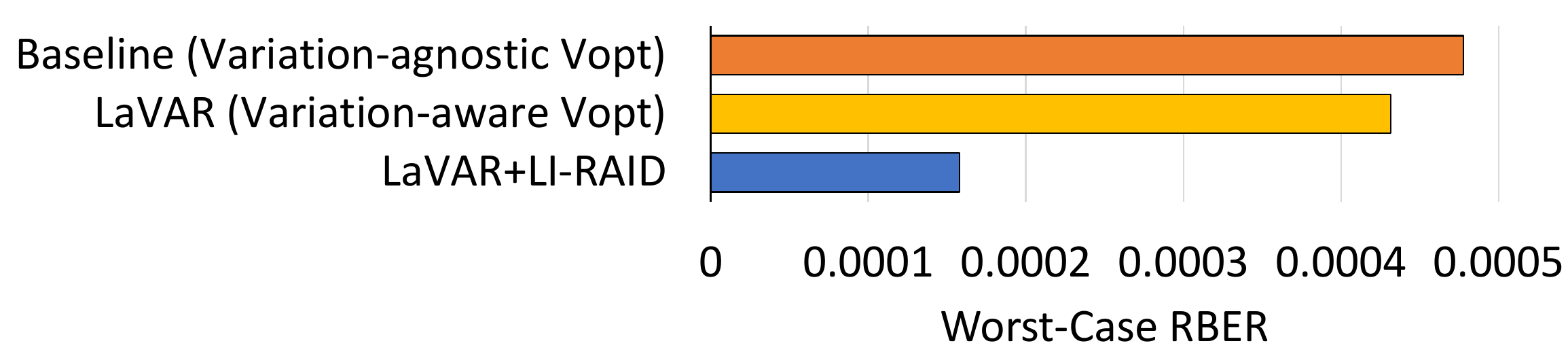}
\caption{\chVII{Effect of LaVAR and LI-RAID on} \chVI{worst-case} RBER at
10,000 P/E cycles.}
\label{fig:99p-rber}
\end{figure}

\sph{\chVI{Note that LaVAR and LI-RAID do
\chVI{\emph{not}} rely on whether the RBER variation is consistent across all chips. LaVAR
learns a different lookup table for each chip.
So, even if there is some chip-to-chip process variation that is present,
our models are effective at dynamically capturing the behavior
of \emph{any} \chVIII{NAND flash memory chips. Conventional RAID tolerates
only chip-to-chip process variation. LI-RAID improves
flash reliability over conventional RAID by eliminating the strong correlation
between RBER and layer number, which we show in
Figure~\ref{fig:variation-wlopterr}. We conclude that both LaVAR and LI-RAID
are effective \chX{at reducing the impact of layer-to-layer variation
on the RBER}.}}}




\subsection{ReMAR\@: Retention Model Aware Reading}
\label{sec:3derror:mitigation:retention}

As we show in Section~\ref{sec:3derror:retention}, due to early
retention loss, retention errors increase much faster \chVI{after programming
a page} in 3D NAND \chVI{flash memory} than \chX{they do} in
planar NAND \chVI{flash memory}. Thus, mitigating retention errors has become
more important in
3D NAND than in planar NAND \chX{flash memory}, as \chX{the errors have a} greater impact on SSD
reliability.
However, as we show in our model in Section~\ref{sec:3derror:model:retention}, \chVIII
{the RBER impact of} early
retention loss is proportional to the logarithm of \chVIII{retention} time.
This means that \chX{a large}
majority of the retention errors and threshold voltage shifts happen \chX{\emph{shortly}}
after programming. As a result, traditional retention \chVI{error} mitigation
techniques
developed for planar NAND \chX{flash memory, which are optimized for
\chXI{much} larger retention times,} may become less effective on 3D NAND
\chVIII{flash
memory}. For example,
\chVI{Flash Correct-and-Refresh (FCR)}~\chVII{\cite{cai.iccd12, cai.itj13}},
a mechanism that remaps all data
periodically, allows planar NAND to tolerate $50\times$ more P/E cycles with
\chVII{a}
3-day refresh period. \sg{However,} \chVI{according to our evaluations, the
P/E cycle lifetime improvement of FCR} reduces to \chVI{\emph{only}}
$2.7\times$ for 3D NAND \chX{flash memory}
due to \chVI{the} early retention loss \chVI{phenomenon}. This motivates us to
explore new ways to mitigate
retention errors in 3D NAND \chVI{flash memory}.

\textbf{Mechanism.} 
\sg{We propose a new mechanism called \chX{\emph{Retention Model Aware Reading} (ReMAR)},
whose key idea} is to accurately track the retention
time of the data and apply the optimal read reference voltage predicted by our
model in Section~\ref{sec:3derror:model:retention}.
First, ReMAR \chVI{constructs} the same linear models proposed in
Section~\ref{sec:3derror:model:retention} \emph{online} to accurately predict the
optimal $V_a$, $V_b$, and $V_c$. Similar to the distribution parameter model
used in Section~\ref{sec:3derror:model:retention}, we model the optimal $V_b$ and $V_c$
as: \chVIII{$V = (\alpha \cdot PEC + \beta) \cdot \log(t) + \gamma \cdot PEC
+ \delta$}. We model the
optimal $V_a$ as: \chVIII{$V_a = \gamma \cdot PEC + \delta$}, since $V_a$ is
\chVI{\emph{not}} affected by
retention time \chX{(as we show empirically in Section~\ref{sec:3derror:retention})}. 
To construct this model \emph{online}, the
controller randomly selects a \chVI{flash block and} records the
optimal read reference voltage of the block \chX{(which the controller learns} by sweeping the read
reference voltages\chVII{,} \chVI{as done} in prior work~\cite{cai.hpca15}), \chVI
{along with the block's P/E cycle count} ($PEC$) and retention time ($t$). Over
time,
these data samples \chVI{would} cover a range of P/E cycle \chVI{counts} and
retention times.\footnote{The SSD controller can also perform \chVI
{additional} characterization if \chVI{a}
certain data range is missing.} Note that as the P/E cycle count of the SSD
increases, the accuracy of the model increases, because more data samples
are collected. Once this online model is constructed, it is used in the
controller to predict the optimal read reference voltage \chVI{to be used for}
each read
operation. To do this, the SSD controller stores the P/E cycle \chX{count} and
the program time of each block as metadata. During each read operation, the
controller computes the retention time for each read by subtracting the program
time from the read time. Using the recorded P/E cycle \chVI{count} and the
computed
retention time of the data, ReMAR applies the online model to predict $V_a$,
$V_b$, and $V_c$. By accurately predicting and applying the optimal read
reference voltages, ReMAR \chVII{increases} the accuracy of read
operations and
thereby \chVI{decreases} the raw bit error rate.

\textbf{Overhead.} \sph{\chV{Like LaVAR, \chVI{ReMAR is}
implemented fully in the SSD controller firmware, and \chVI{does}
\emph{not} require any modifications to the hardware.}
\chVI{Assuming that the flash block size is \SI{5}{\mega\byte}, and \chX{that} \chVII{ReMAR}
stores the program time in the UNIX Epoch time format~\cite{matthew.book08},
which takes up \SI{4}{\byte},
the} memory and storage overhead of \chVII{ReMAR} is 800KB for a 1TB SSD.
The performance overhead of each read operation is small, \chVI{as ReMAR needs \chX{only}
a few dozen CPU cycles (on the order of \SI{100}{\nano\second} in total)} in the
\chX{SSD} controller to compute $V_{opt}$, which is negligible
compared to flash read latency (on the order of \SI{10}{\micro\second}). The performance
overhead of learning the model can be hidden by \chVI{\chX{(1)~}performing \chX{learning
in} the background and \chX{(2)~}deprioritizing the requests issued for
characterization} \chVIII{purposes}.}

\sph{The controller uses the UNIX Epoch time format~\cite{matthew.book08}
for program and read times, such that
the recorded time is valid after reboot. To do this, the controller needs
a real-time clock to keep track of the current time. Without a power source on
the SSD, the controller needs a special command to synchronize the current
time with the host when it boots up. The program time of each block is \chVI{stored}
in the memory of the controller, along with other metadata that already
exists such as the logical address map and the P/E cycle \chX{count} of each block.}

\textbf{Evaluation.} Figure~\ref{fig:vopt-retention} compares the RBER
achieved by ReMAR to \chX{that of} the state-of-the-art
read reference voltage tuning technique~\chVIII{\cite{luo.jsac16}}
\chVI{designed for planar NAND \chX{flash memory}
(\emph{Baseline})}. The results are based on the
characterization data in Section~\ref{sec:3derror:retention}.
We assume that the average retention time of the
data is 24~days. \chVI{The \emph{Baseline} technique is unaware of the 
retention time. Thus, \emph{Baseline}
uses a \emph{retention-agnostic} $V_{opt}$ based on only the P/E cycle
count of the flash page. ReMAR uses a \emph{retention-aware} $V_{opt}$ based
on both the P/E cycle count and the retention time of the flash page.}
On average \chVI{across all P/E cycle counts}, ReMAR reduces \chVI{the} RBER by 51.9\%.
As \chVI{the} P/E cycle \chVI{count} increases, the benefit of ReMAR \chVI{(i.e., the
RBER improvement of ReMAR over \emph{Baseline}) also increases}. \chVII{We
conclude that,} by accurately tracking retention time, and by
using our \chVI{retention loss} model, ReMAR \chVII{accurately adapts the
read reference voltage} to the threshold
voltage \chVII{shifts} \chVIII{that occur} due to retention loss, and \chVI
{hence} \chVII{it effectively reduces the} RBER\@.

\begin{figure}[h]
\centering
\iftwocolumn
  \includegraphics[trim=0 10 0 0,clip,width=\linewidth]
  {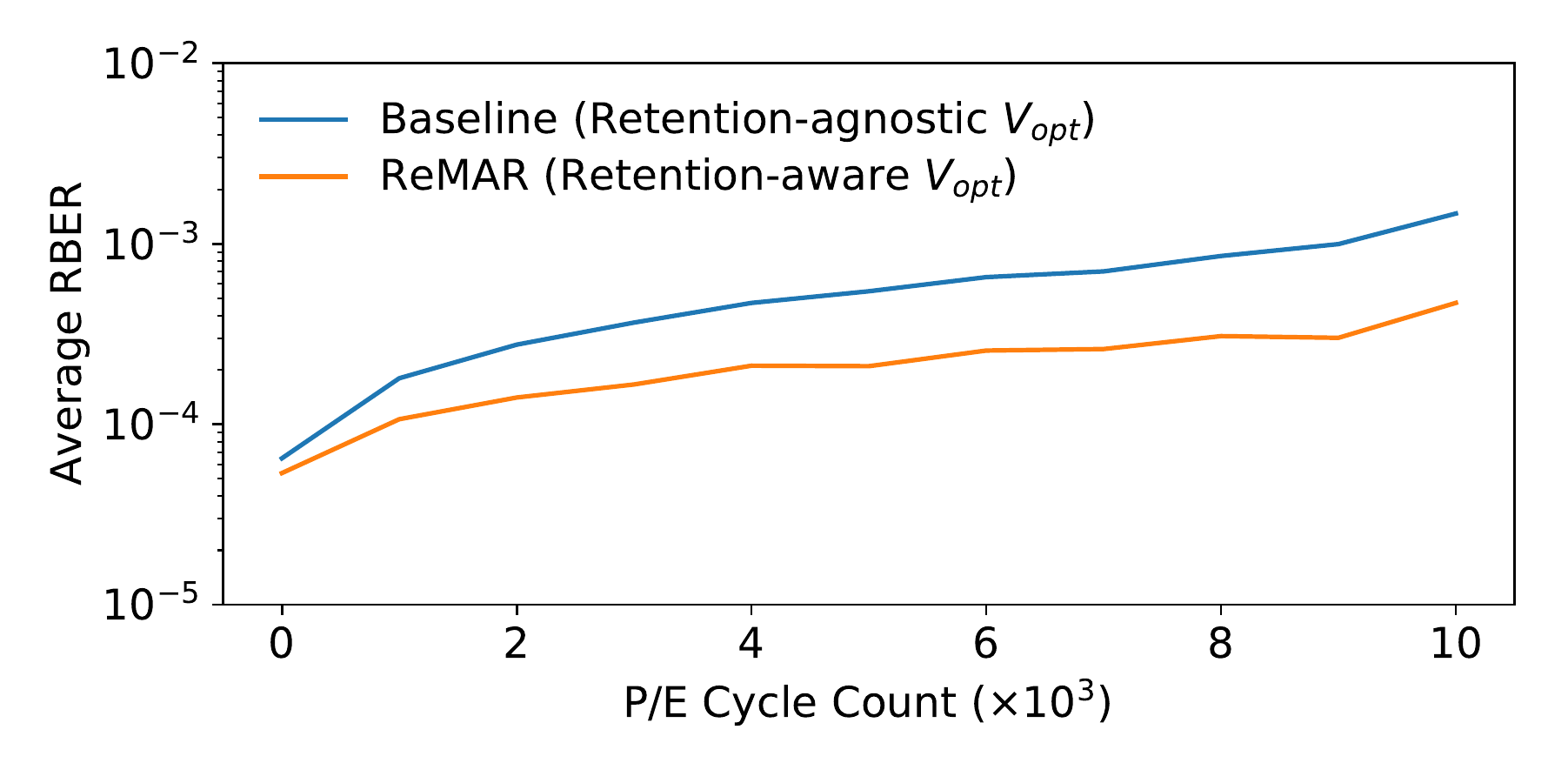}
\else
  \includegraphics[trim=0 10 0 0,clip,width=.65\linewidth]
  {figs/vopt-model-ret.pdf}
\fi
\caption{RBER reduction using ReMAR.}
\label{fig:vopt-retention}
\end{figure}

\subsection{ReNAC\@: Retention Interference Aware Neighbor-Cell Assisted Correction}
\label{sec:3derror:mitigation:nac}

As we observe in Section~\ref{sec:3derror:retention:interference}, due
to retention interference, the amount of threshold voltage shift of a
victim cell during a certain amount of retention time is affected by
the value stored in \chXII{a \chVI{\chX{vertically-\chXI{adjacent neighbor}}
cell}}. This \chVII{phenomenon}
presents a similar
data dependency as that induced by program interference, where the amount
of \chVII{the} threshold voltage shift of a victim cell during programming
operation also depends on the value stored in the \chX{directly-neighboring
cells}~\cite{cai.iccd13, cai.sigmetrics14}.
To mitigate program interference errors, \chV{prior
work proposes neighbor-cell assisted correction
(NAC)~\cite{cai.sigmetrics14}. \chVI{The goal of NAC is to reduce \chVII{the}
raw bit
error rate by reading each cell at the read reference voltage
optimized for the amount of program interference induced by its \chVIII{\chX{directly-}neighboring
cells}.} \sg{\chVI{To achieve this goal,} after error correction fails on a
flash page, NAC \chVI{reads the data stored in the \chVII{\emph{neighboring}} wordline and}
re-reads the \chVI{failed} page using a set of
read reference voltage values that \chVII{are adjusted based on} the data
\chVII{values} stored in \chX{the directly-}neighboring \chVII{cells}~\cite{cai.sigmetrics14}.}} However,
this mechanism does \chVI{\emph{not}}
account for
\chXI{\emph{retention interference}} \chX{induced by the neighboring cells}, which is new in 3D
NAND flash memory. \chVII{We} adapt NAC
for 3D NAND flash memory to account for the new retention interference
\chX{phenomenon, and call this adapted mechanism
\emph{Retention Interference Aware Neighbor-Cell Assisted Correction}
(ReNAC)}.

\textbf{Mechanism.} The key idea of ReNAC is to use the data stored in 
\chX{a vertically-adjacent neighbor cell} to predict the amount of retention interference on
a victim cell. Using similar techniques from
Section~\ref{sec:3derror:model:retention}, ReNAC first develops an online model of
retention interference as a function of the retention time and the neighbor
cell's state. The SSD controller obtains the retention time of each block using a
mechanism \chVII{similar} to ReMAR, and computes and applies the
neighbor-cell-dependent read offset at that retention time from the model.
\sph{\chV{\chVI{For ReNAC, we} are \chVIII{currently} unable to show
any meaningful improvements in flash lifetime for the current generation of 3D
NAND flash memory, because retention interference shifts the threshold
voltage by \chVII{only} less than two \chX{voltage} steps (Figure~\ref{fig:retention-interference}),
which is much smaller than the voltage \chX{changes} due to process variation
(Figure~\ref{fig:variation-wloptvrefs}) and early retention loss 
(Figure~\ref{fig:retention-optvrefs}). However, we expect that retention
interference will increase in future 3D NAND flash memory devices due to \chX{decreasing} cell
sizes and \chX{decreasing} distances between neighboring cells
\chVII{(Table~\ref{tbl:summary}), which, in turn, will likely increase} the
benefit of \chVIII{using} ReNAC.} \chXI{We also expect ReNAC to have a
relatively larger benefit in 3D NAND flash memory chips that use
\emph{triple-level cell} (TLC) or \emph{quadruple-level cell} (QLC)
\chXII{technologies.  A TLC or QLC NAND flash memory chip stores more bits in a cell than 
an MLC NAND flash memory chip}, by splitting up the same voltage range into a greater
number of states (eight for TLC and sixteen for QLC).  \chXII{Doing so} reduces the
voltage margin between neighboring threshold voltage distributions.
\chXII{Therefore}, shifting the read reference voltage by two voltage
steps may affect more cells \chXII{in TLC} and QLC 3D NAND flash memory than in
MLC 3D NAND flash memory, and, thus, \chXII{ReNAC can reduce a greater number of} 
raw bit errors in future TLC or QLC NAND flash memory.}
We leave a quantitative evaluation of ReNAC
\chVIII{on future \chX{3D} NAND flash memory chips to} future work.}

\subsection{\chVI{Putting It All Together: Effect on System Reliability \chXI
{and Performance}}}
\label{sec:3derror:mitigation:all}

\chI{The \chVIII{mechanisms we} propose in this section can be combined together to
achieve significant reductions in average and worst-case RBER\@.}
\chVI{For a consumer-class 3D NAND flash memory device, these reductions
improve \emph{flash memory lifetime}, i.e., the device can tolerate more
P/E cycles before failing. For \chVII{an} enterprise-class device which is expected to
be replaced after a fixed amount of time, these reductions improve the
sustainable workload write intensity \chVIII{or} reduce the ECC storage
overhead.} \chVI{We \chVIII{evaluate these} potential effects of our mechanisms on
storage system \chVIII{reliability and performance}.}

\textbf{Flash Lifetime \chXI{(or Performance)} Improvement.}
\chI{In Figure~\ref{fig:this-work-rber}, we compare and contrast the
reliability \chVII{(i.e., the RBER)} of \chV{five} example SSDs:
(1)~\emph{Baseline}, an SSD that uses a
fixed, default read reference voltage and \chX{employs} a conventional RAID scheme;
(2)~\emph{State-of-the-art}, an SSD that uses the optimal read reference voltage
predicted by existing mechanisms designed for planar
NAND \chVI{flash memory}~\cite{luo.jsac16, parnell.globecom14, cai.hpca15,
papandreou.glsvlsi14}
and \chX{employs} a conventional RAID scheme; \sph{\chV{(3)~\emph{LaVAR}, an SSD that uses
the optimal read reference voltage for each layer predicted by LaVAR in
addition to \emph{State-of-the-art}; (4)~\emph{LaVAR+LI-RAID}, an SSD that
\chX{uses} the LI-RAID scheme in addition to \emph{LaVAR};} and (5)~\emph{This
Work} \chXI{(LaVAR + LI-RAID + ReMAR)}, an SSD
that uses the optimal read reference voltage predicted by LaVAR and ReMAR, and
\chX{also employs} the
LI-RAID scheme.} In this figure, we plot the \chVII{\emph{worst-case RBER}}
\chVI{(i.e., the highest per-page RBER within a flash block)} instead of the
average RBER, because the worst-case RBER \chX{\emph{limits}} \chVIII{the} flash memory
lifetime. Because
RBER increases \sg{with} P/E cycle count, if the worst-case RAID group has a high enough
\chVIII{worst-case} RBER, NAND flash memory can no longer guarantee reliable
operation.}

\begin{figure}[h]
\centering
\includegraphics[trim=0 290 420 0,clip,width=\figscale\linewidth]
{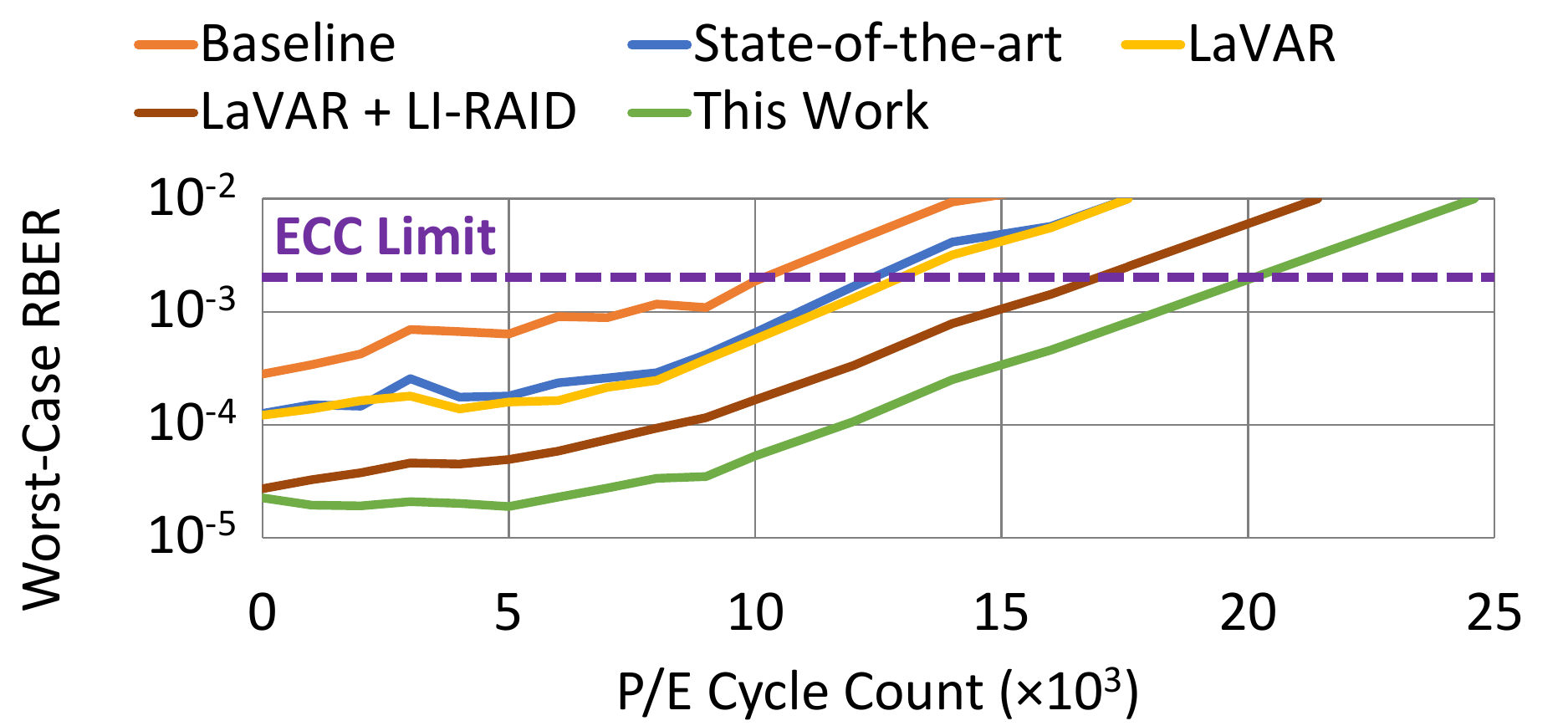}
\caption{\sph{\chVII{Effect of} \chVI{LaVAR, LI-RAID, and ReMAR} \chVII{on
worst-case RBER experienced by any flash block}.}}
\label{fig:this-work-rber}
\end{figure}

\chI{Assuming that the ECC deployed on the SSD can correct \chX{errors up to 
an RBER of $3\cdot10^{-3}$~\cite{cai.hpca15, cai.procieee17}} 
(i.e., the \chXI{\emph{ECC limit}}, \chVI{shown as} \chX{a} purple \chX{dashed} line in
Figure~\ref{fig:this-work-rber}), we can calculate the lifetime of \chVIII
{each SSD we evaluate}.\footnote{\chV{Note that we are \emph{unable} to 
\emph{directly}
measure the flash lifetime improvements on real devices, because manufacturers
do \emph{not} provide us with the ability to modify the SSD firmware
directly, which prevents us from
evaluating our techniques on the real devices themselves.
Unfortunately, we also do \emph{not} have the \chVI{resources} to measure the
lifetime of a large number of real flash chips by emulating the behavior of our mechanisms,
as this would require many additional months to years of effort.}
\chV{Instead, \sg{we follow the precedent of prior work to} evaluate the flash memory
lifetime based on real \chVI{RBER characterization data we obtain from the
testing of real flash memory devices}.}} In our
evaluations, the flash memory lifetime ends when the \chVI{worst-case RBER
exceeds} the ECC limit. We find that \emph{State-of-the-art}, \chV{\emph{LaVAR},
\emph{LaVAR+LI-RAID}, and \emph{This Work} \chVII{improve} flash memory lifetime by
23.8\%, 25.3\%, 57.2\%, and 85.0\%, respectively,} over the
\emph{Baseline}. When the SSD is \chX{used} in a server, which has a fixed device
lifetime, the server has to throttle the write frequency to a certain 
\emph{drive writes per day} (DWPD) to ensure \chVIII{that} the SSD can operate
\chVIII{reliably}
during the fixed lifetime. In this case, \chX{our combined mechanisms (\emph{This Work})
increase the maximum write frequency (i.e., the maximum DWPD) of 
the SSDs in a server by 85.0\%.}
\chVII{Thus, our mechanisms either \chVIII{improve}
lifetime or \chVIII{improve} performance under a fixed lifetime.}

\chVI{\textbf{ECC Storage Overhead Reduction.}}
\chI{In modern SSDs, the storage overhead for error \chVI{correction
increases} in each generation to better tolerate the degraded flash
reliability due to aggressive scaling. For example, to tolerate an RBER of up to
$3\cdot10^{-3}$ for the \emph{Baseline} SSD at the end of its lifetime, a
modern BCH code~\cite{hocquenghem.chiffres59} requires 12.8\% storage overhead for
the redundant ECC bits~\cite{deal.whitepaper09} \chVI{(i.e., \emph{ECC
redundancy})}. \chVI{By deploying all of our
proposed error mitigation techniques in an enterprise-class SSD, the
RBER at the end of the fixed flash memory lifetime is significantly
lower compared to \emph{Baseline}. Thus, we can redesign the ECC deployed in
the SSD to tolerate only up to the reduced RBER, which requires fewer \chVIII
{ECC} bits
\chX{and, thus,} lower ECC \chVI{redundancy} than \chVII{the ECC required for the}
\emph{Baseline}.}
Assuming all five \chX{of the evaluated} SSDs achieve the same lifetime, and
the same reliability (i.e., uncorrectable error rate) at the end of their lifetime,
\chVIII{\emph{State-of-the-art}, \emph{LaVAR}, \emph{LaVAR+LI-RAID}, and 
\emph{This Work} reduce ECC redundancy by 42.2\%, 45.3\%, 68.8\%, and 78.9\%,
respectively, over \emph{Baseline}.}
We leave the evaluation of the performance improvements due to a weaker ECC
requirement~\cite{lee.isscc12, chen.tit81} for future work.}

\sg{We conclude that by combining LaVAR, LI-RAID, and \chVI{ReMAR,}
we can \chX{(1)~}achieve significant improvements in the lifetime of
3D NAND flash memory, \chVIII{\chX{(2)~}enable \chX{higher} write intensity in workloads within
a given lifetime requirement,} \chVII{or \chVIII{(3)}~keep the lifetime constant but
greatly reduce the storage cost of reliability in 3D NAND flash memory}.}



\section{Related Work}
\label{sec:3derror:related}

To our knowledge, this thesis is the first in open literature
to (1)~show the \chVII{differences} between the error characteristics of 3D
NAND \sg{flash memory} and that of planar NAND flash memory through extensive
characterization using real \chVI{3D NAND flash memory} chips, (2)~develop models of layer-to-layer
process variation and early retention loss for 3D NAND flash memory, 
and \sph{(3)}~propose \chVI{and show the benefits of} four new
mechanisms based on the new error characteristics of 3D NAND flash memory.
Due to the importance of NAND flash memory reliability \chVI{in storage systems}, 
there is a large body of related work. We \chVI{treat} this related
work \chVI{in} \chI{five} \chVI{different} categories.

\textbf{3D NAND \chVI{Flash Memory} Error Characterization.} \chVI{Two recent}
works \chVI{compare} the retention loss \chVI{phenomenon} between 3D NAND and
planar NAND \sg{flash memory}~\cite{mizoguchi.imw17, luo.hpca18} through real device
characterization, and \chVI{report} findings \chVI{similar} \sg{to our work in this chapter}
regarding \chVI{the} early retention loss \chVI{phenomenon}.
\chVI{Two other} recent \chVI{works use} \sg{a methodology similar to ours} to characterize
3D NAND devices based on \chVII{\emph{different}} 3D NAND \chVI{flash memory cell}
technologies (\chVI{i.e.,} 3D floating-gate cell
and 3D vertical gate cell)~\cite{xiong.sigmetrics17, xiong.tos18, hung.jssc15}, which are
\chVII{\emph{less common}} than the 3D charge trap NAND \sg{flash memory \chVI{cell
technology} that we test} in this thesis.
\sph{Other recent works~\chVI{\cite{park.jssc15, wang.tecs17, choi.vlsit16,
grossi.bookchapter16, park.iedm12} report several differences
of 3D NAND flash memory from planar NAND flash memory}.
These differences include (1)~smaller
program variation at high P/E cycle \sg{counts}~\cite{park.jssc15}, (2)~smaller program
interference~\cite{park.jssc15}, (3)~layer-to-layer process
variation~\cite{wang.tecs17}, (4)~early retention
loss~\cite{choi.vlsit16, grossi.bookchapter16, park.iedm12}, and (5)~retention
interference~\cite{choi.vlsit16}.
\chV{While prior works have reported 
on the existence of these errors, \emph{none} of them provide a comprehensive characterization
of \emph{all} of the different errors \chVII{using} the \emph{same} \chVII{chips}.
Only one of these prior works~\cite{choi.vlsit16} provides a detailed analysis
based on
\chI{\emph{circuit-level}} \chVI{measurements and characterizations}, and does
so only for early retention loss and
retention interference. Other works provide only a high-level \chVI{\emph{summary}} of
real device characterization~\cite{park.jssc15} or do \chVI{\emph{not}} provide any real device characterization
results at all~~\cite{wang.tecs17, grossi.bookchapter16, park.iedm12}. 
Our work in this chapter performs an extensive detailed analysis of \emph{all} known sources of error in 
3D NAND flash memory chips, which allows us to understand
the relative impact of each error source on the same chip.
We \chVI{report \chVII{the first set of} extensive results on} three error characteristics \chVI{that
are new in 3D NAND flash memory}: layer-to-layer
process variation, early retention loss, and retention interference.}}



\textbf{Planar NAND \chVI{Flash Memory} Error Characterization.}
\chVI{A large body of prior work studies} all types of \chVI{error sources on}
planar NAND \chVI{flash memory},
including P/E cycling errors~\cite{cai.procieee17, cai.date13,
parnell.globecom14, luo.jsac16}, programming errors~\cite{cai.hpca17,
parnell.globecom14, luo.jsac16},
cell-to-cell program interference errors~\cite{cai.iccd13, cai.sigmetrics14},
retention errors~\chVII{\cite{cai.procieee17, cai.iccd12, cai.hpca15, fukami.di17}}, and read disturb
errors~\cite{cai.procieee17, cai.dsn15}. These works characterize how \chX{the} raw bit error
rate and threshold voltage change \chVII{due to} various types of
\chVI{error \chX{sources}}. \chVI{A detailed survey of such prior works on
planar NAND flash memory can be found in \chX{our \chXI{recent} survey articles}~\cite{cai.procieee17, cai.book18}.}
Our \chVII{thesis} \chVI{experimentally studies} all \chX{of} these
\chVI{error mechanisms} in the new 3D NAND \chVI{flash memory context}, and
\chVI{compares} 3D NAND \chVI{flash memory} error characteristics with results in
these \chVI{prior works to \chVII{show}} the
differences between 3D NAND and planar NAND \chVI{flash memory}. Prior work
\chVII{demonstrates} \chVI{the} early retention loss \chVI{phenomenon} in planar NAND \chVI{flash
memory} based on charge trap
transistors~\cite{chen.iedm10}, which is similar to, \chVII{but not as severe as,} \chVI{the} early retention
loss \chVI{phenomenon} in 3D NAND \chVI{flash memory}. \chVII{We} \chVI
{investigate} retention interference and process variation \chVIII{related} \chX{errors, in
addition to these other} error types \chVIII{discovered before in planar NAND flash memory}.

\textbf{Planar NAND Error Modeling and Mitigation.}
Based on characterization results, prior work \chVI{proposes} models for planar
NAND \chVI{flash memory} threshold voltage distribution, and models for \chVI
{estimating the effect of P/E cycling} on the \chVI{threshold voltage}
distribution~\cite{parnell.globecom14, luo.jsac16, cai.date13}. Our work in this chapter uses a
simpler \chVII{threshold voltage} distribution model, \chVI{since more complex models are designed to
handle programming errors in planar NAND flash memory that do not exist in the
3D NAND flash \chVII{memory chips} that we \chX{test}.} \chVII{We develop} a unified model
of retention loss and wearout for \chX{the} RBER, threshold voltage distribution,
\chVI{and $V_{opt}$} in 3D NAND \chVI{flash memory}. There is a large body of prior
work that proposes mechanisms to mitigate planar NAND \chVI{flash memory}
errors~\cite{cai.book18, cai.procieee17, cai.hpca17, cai.hpca15, cai.iccd12,
pan.hpca12, cai.dsn15, cai.iccd13, cai.sigmetrics14, cai.itj13, luo.jsac16,
ha.apsys13, ha.tcad16, jeong.fast14, luo.msst15, wilson.mascots14, li.fast15,
huang.fast17, zhang.fast16, jimenez.fast14, pan.fast11}. \chVI{In
Section~\ref{sec:3derror:mitigation},} we
have \chVI{already} compared \chVI{our mechanisms} to several of \chVI{these
techniques \chVII{that} are state-of-the-art}, and \chVI{have
shown} that \chVII{prior techniques developed for planar NAND flash memory}
are less effective in 3D NAND \chVI{flash memory} than our
techniques due to the new
error characteristics \chVIII{of 3D NAND flash memory}.

\chI{\textbf{3D NAND \chVI{Flash Memory} Error Mitigation.}
Prior work \chVI{proposes} circuit-level and system-level techniques to tolerate
layer-to-layer process variation in 3D NAND \chVI{flash memory}. \sg{Two
recent works} propose to use
different read reference voltages for different layers~\cite{hung.jssc15,
ye.fms17}, which is similar to the LaVAR technique \sg{that we propose} in
Section~\ref{sec:3derror:mitigation:variation}. \sph{\chVI{Unlike} our work in this chapter, these
\chVI{prior} works do not
(1)~design a detailed mechanism like LaVAR to learn and use the $V_{opt}$ in a
lookup table, or (2)~evaluate \chVI{their techniques} using real
characterization data.} Wang et al.\ propose to
apply different read reference \chVII{voltages for} less-reliable pages storing
critical metadata~\cite{wang.tecs17}. As we have shown in
Section~\ref{sec:3derror:mitigation:variation}, while these prior techniques improve
average RBER, they do \chVI{\emph{not}} significantly reduce worst-case RBER,
which limits the flash memory lifetime. In this work, we \chVI{propose} a
series of mitigation techniques that not only significantly reduce the
\chVI{average and }worst-case RBER but also tolerate other new error characteristics
we find in
3D NAND \chX{flash memory}, such as early retention loss and retention interference.}

\textbf{Large-Scale SSD Error Characterization.}
Prior work \chVI{performs} large-scale \chVI{studies of errors found in} flash
memories deployed in data centers~\cite{meza.sigmetrics15, schroeder.fast16,
narayanan.systor16}.
Since \chVI{the operating system is unaware of the raw bit errors in the NAND
flash memory} \chX{devices}, these studies can \chVI{only} \chVII{use} drive-level
statistics
provided by the \chX{SSD} controller, such as overall RBER and uncorrectable error rate, 
average P/E \chX{cycle count}, and \chX{a coarse} estimation of
retention time and read disturb counts. \chVI{In contrast, in our studies, we
have complete} access to the
physical location, P/E cycle \chVI{count}, retention time, \chVI{and} read disturb count of
each read/write
operation, and thus can provide deeper insights \chVI{and controlled
experimental results compared to} large-scale studies, \chVI{which have to
be correlational in nature}.

\chVI{\textbf{DRAM Error Characterization.} Like a flash memory cell, a DRAM cell
stores charge to represent a piece of data. Hence, DRAM has
many error characteristics \chX{that are} similar to NAND flash memory. For example, charge
leaks from \chX{a} DRAM cell over time, at a speed much faster than 
\chX{that for} NAND flash memory 
(i.e., on the order of \emph{milliseconds} to \emph{seconds} \chX{in DRAM}~\chVII{\cite{liu.isca13,
liu.isca12}}),
leading to \emph{data retention errors}. This phenomenon in DRAM is analogous to the
retention loss phenomenon in NAND flash memory (see
Section~\ref{sec:3derror:retention} and Appendix~\ref{sec:3derror:appendix:retention}), and its
effect has been
studied through extensive experimental characterization of DRAM
chips~\chX{\cite{khan.sigmetrics14, khan.dsn16, khan.cal16, khan.micro17, qureshi.dsn15, lee.hpca15, liu.isca13, patel.isca17, kim.hpca18,
jung.dt17, hamamoto.ted98, hassan.hpca17}}. Similar
to the retention interference phenomenon found in 3D NAND flash memory (see
Section~\ref{sec:3derror:retention:interference}), DRAM
exhibits data-dependent retention behavior,
or \emph{data pattern dependence} (DPD)~\cite{liu.isca13}, where the retention time
of a DRAM cell is dependent on the values written to \emph{nearby}
DRAM cells~\cite{liu.isca13, khan.sigmetrics14, khan.dsn16, patel.isca17, khan.cal16, khan.micro17}. Conceptually similar to
the read disturb errors found in NAND flash memory (see
Appendix~\ref{sec:3derror:read:disturb}), commodity DRAM
chips that are sold and used in the field today exhibit read
disturb errors~\cite{kim.isca14}, also called \emph{RowHammer}-induced
errors~\cite{mutlu.date17}. \chVII{These errors are affected by \emph{process
variation}, which we \chVIII{comprehensively} examine} in 3D NAND flash memory
(see Section~\ref{sec:3derror:variation} and Appendix~\ref{sec:3derror:appendix:variation}).
\emph{Process variation} in DRAM \chVII{is shown to also affect access
latency, retention time, and power consumption}\chX{~\cite{khan.sigmetrics14, khan.dsn16, khan.cal16, 
khan.micro17, qureshi.dsn15, ghose.sigmetrics18, lee.hpca15, lee.sigmetrics17, chang.sigmetrics16, chang.sigmetrics17, liu.isca13, patel.isca17, kim.hpca18,
jung.patmos16, mathew.date18, jung.dt17, hassan.hpca17, hamamoto.ted98, chandrasekar.date14,
chang.thesis17, lee.thesis16, liu.isca12}}.}

\section{\chIV{Limitations}}

The observations and analysis results only \chV{apply} to 3D NAND based on a charge
trap cell design. The error characteristics of future 3D NAND chips may also
change when the manufacturing process technology starts shrinking again if
adding more stacked layers become less economically viable. In a smaller
process technology node, two-step programming errors, read disturb errors, and
retention interference errors may become dominant\chV{;} thus our model needs to
be extended to consider all of them. Our techniques
focus on mitigating layer-to-layer process variation and retention errors
that are unique to 3D NAND flash memories. Our techniques do not reduce other
errors caused by read disturb or program interference because prior works
already provide techniques to mitigate them~\cite{cai.iccd13, cai.dsn15}. Our
retention model and ReMAR do not consider self-recovery and temperature
effects, which could affect flash reliability significantly. In
Chapter~\ref{sec:heatwatch}, we extend our retention
model to consider these effects, and provide mechanisms to track the amount of
self-recovery and temperature effects.


\section{Conclusion}
\label{sec:3dnand:conclusion}

\chI{We develop \chVI{a new} understanding of three new error characteristics in
3D NAND flash memory through rigorous experimental characterization of real,
state-of-the-art 3D NAND flash memory chips: layer-to-layer process
variation, early retention loss, and retention interference. We analyze and
show that these new error characteristics are fundamentally caused by changes
introduced in the 3D NAND flash memory architecture \chX{compared to the
planar NAND flash memory architecture}.}
To handle these three new error characteristics in 3D NAND \chVI{flash memory}, we
develop new analytical models for layer-to-layer process variation and early
retention loss in 3D NAND flash memory. \chVI{Our models can accurately
predict/estimate the optimal read reference voltage and the raw bit error
rate based on the retention time and the layer number of \chVII{each flash memory} page. We}
propose four new \chX{error mitigation} techniques that
utilize our \chVI{new} models to improve the reliability of 3D NAND flash
memory. 
\chI{Our evaluations show that our newly-proposed techniques successfully
mitigate the new error patterns that we discover in 3D NAND flash memory.}
We hope that the \chVI{rigorous and comprehensive}
error characterization and \chX{analyses} performed in this work \chX{motivate} future
\chVI{rigorous}
studies on \chX{3D NAND \chVII{flash memory} reliability}, and that \chX{they inspire} new error mitigation
mechanisms that cater to \chVII{the} \chX{new} \chVI{error \chX{characteristics} found in} 3D NAND flash
memory.



\section{Appendix}
\label{sec:3derror:appendix}
\label{sec:3derror:comprehensive}


\subsection{Write-Induced Errors}
\label{sec:3derror:write}

We analyze how each type of write-induced error affects the
RBER and the threshold voltage distribution of 3D NAND flash memory.

\subsubsection{Program Errors}
\label{sec:3derror:programerror}

\chX{Program errors occur when the data is incorrectly written to the
NAND flash memory~\cite{park.jssc08, cai.hpca17, cai.procieee17,
cai.book18}.  Such errors are introduced when multiple programming operations
are required to write data to a single cell.  For example, in many MLC NAND
flash memory devices, \emph{two-step programming}~\cite{park.jssc08,
cai.hpca17} \chXI{is employed. Two-step programming} uses two separate \emph{partial programming} steps to write data
to an MLC NAND flash cell.  In the first step, the flash controller writes
only the LSB to the cell, setting the cell to a temporary voltage state.
In the second step, the controller writes the MSB to the cell, but in order
to perform this write, the controller must first determine the current
voltage state of the cell.  This requires reading the partially-programmed data
from the cell, during which an error may occur.  This error \chXI{causes the
controller to} incorrectly \chXI{set} the final voltage state of the cell during the
second programming step, and, thus, is called a program error.
Prior work~\cite{cai.hpca17} \chXI{shows} that program errors occur in
state-of-the-art planar MLC NAND flash memory.}


\chIX{\chX{Current generations of 3D NAND flash memory use} \emph{one-shot
programming}~\cite{park.jssc08, cai.hpca17, cai.procieee17, cai.book18}, which
programs \chX{\chXI{\emph{both}} the LSB and MSB of a cell at the \chXI{\emph{same time}}.
As a result, current 3D NAND flash memory devices do \chXI{\emph{not} experience} program
errors.
Our measurements in Figure~\ref{fig:distribution-shape} confirm the lack of
program errors \chXI{in 3D NAND flash memory}.  In an MLC NAND flash memory
that has program errors,
the threshold voltage distributions of the ER and P1 states have secondary
peaks near the P2 and P3 states, respectively~\cite{cai.hpca17}.  This is
because program errors affect only the LSB, since only the LSB is being read
during the second programming step.  Since there is no second peak in
Figure~\ref{fig:distribution-shape}, there are no program errors.}}

\chIX{\chX{Program errors may appear in future 3D NAND flash memory devices.
In planar NAND flash memory, two-step programming was introduced when
planar MLC NAND flash memory transitioned to the \SI{40}{\nano\meter} 
manufacturing process technology \chXI{node,} in order to reduce the number of
program interference errors~\cite{park.jssc08}.  A similar transition may
occur in the future to continue scaling the density of 3D NAND flash
memory, especially as it becomes increasingly difficult to add more layers
into a 3D NAND flash memory chip.}
Thus, we conclude that today's 3D NAND flash memories
do \chX{\emph{not}} have program errors, but program errors may appear in future
generations.}

\subsubsection{Program/Erase Cycling Errors}
\label{sec:3derror:pecycle}

A P/E cycling \chVIII{error occurs} \chVI{because of
the natural variation of the threshold voltage of cells in each
state~\cite{mielke.irps08, cai.date13} due to the inaccuracy of each program
and erase operation} (see Section~\ref{sec:errors}). \chVI{Such
inaccuracy during program and erase operations increases as \chX{the} 
P/E cycle count increases.}
To study the impact of P/E cycling errors, we randomly select
a flash block within each 3D NAND chip, and wear out the block by
programming random data to each page in the block until the block reaches 
16K P/E cycles.
Using the methodology described in Section~\ref{sec:3derror:methodology}, 
we obtain the overall RBER and the threshold voltage of each cell at various
P/E cycle counts.\footnote{Due to limitations with our \chVI{experimental testing}
platform, each data
point at a particular P/E cycle count has a retention time of 50 minutes.}

\textbf{Observations.}
Figure~\ref{fig:pec-mean-var} shows how the
mean and standard deviation \chVI{of the threshold voltage distribution
\chX{of each state}} change
\chX{as a function of the P/E cycle count, when we 
\chXI{fit our voltage measurements for each state to a Gaussian model}}.
\chVI{Each subfigure in the top row represents the mean for a
different state; each subfigure in the bottom row represents the standard
deviation for a different state. The blue dots shows the measured data; each
orange line shows a linear trend fitted to the measured data. The x-axis
shows the P/E cycle count; 
\chX{the y-axis shows the mean 
(Figures~\ref{fig:pec-mean-var}a--\ref{fig:pec-mean-var}d) or the
standard deviation 
(Figures~\ref{fig:pec-mean-var}e--\ref{fig:pec-mean-var}h) of the threshold
voltage distribution of each state, in voltage steps.}}
We make four observations from Figure~\ref{fig:pec-mean-var}. 
First, the mean and standard deviation of all states \chX{increase linearly
as the P/E cycle count increases.  We fit a line using linear regression, 
shown as an orange dotted line in each subfigure.}%
\footnote{For the ER state, a linear fit has a
5.9\% higher root mean square error than a power-law
fit. However, we choose the linear fit due to its simplicity.} 
\chVI{Second, 
\chX{the threshold voltage distributions of the ER and P1 states}
shift to higher voltages, while the distributions of the \chX{P2 and
P3 states shift to lower voltages, causing the distributions to
move closer to the middle of the threshold voltage range}.}
Third, the threshold voltage distributions \chX{of} all four states become wider
\chVI{(i.e., the standard deviation increases)} as the P/E cycle count increases.
\chVI{\chX{Since the distributions shift towards the middle of the threshold
voltage range} and become wider as \chX{the} P/E
cycle count increases, the distributions become closer to each other,
\chX{which increases} the raw bit error rate.
Fourth, the magnitude of the \chX{threshold voltage shift and the
widening of the distributions} is much larger for the ER~state than
it is for the other three states (i.e., P1, P2, P3). \chVI{Therefore,
ER$\leftrightarrow$P1 errors \chX{(i.e., an error that shifts a cell
that is originally programmed in the ER state to the P1 state, or vice versa)} 
increase faster than other errors \chX{with the P/E cycle count.}}}


\begin{figure*}[h]
\centering
\includegraphics[trim=10 10 10 10,clip,width=\linewidth]
{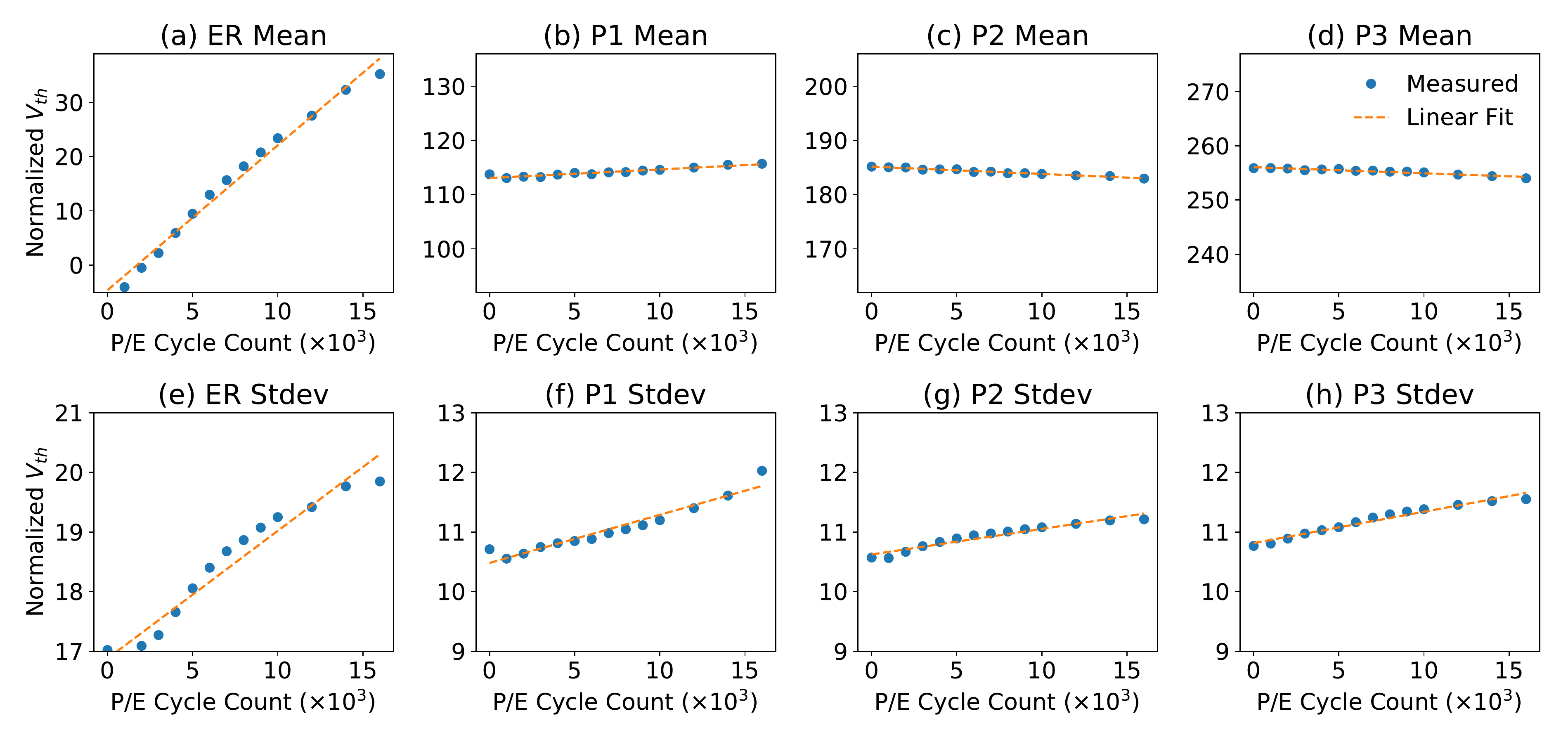}
\caption{\chVI{Mean and standard deviation of \chIX{our Gaussian} threshold
voltage} distribution \chIX{model of each state}, \chVI{versus} P/E cycle count.}
\label{fig:pec-mean-var}
\end{figure*}


Figure~\ref{fig:pec-opterr} shows how the RBER increases as the
P/E cycle count increases.  The top graph breaks down the errors into which \chX{bit}
(i.e., LSB or MSB) they occur in.  The bottom graph
breaks down the errors based on how the error changed the cell state due to a
shift in the cell threshold voltage.  If
the error \chXI{caused} \emph{either} the LSB or MSB (but not both) \chXI{to
be read incorrectly}, we refer to \chX{that error} as a single-bit error
\chX{(ER~$\leftrightarrow$~P1, P1~$\leftrightarrow$~P2, and P2~$\leftrightarrow$~P3
in the graph)}.
If \chX{\emph{both}} the LSB and MSB are \chXI{read incorrectly} as a result
of the \chXII{error}, we refer to that
\chX{error} as a multi-bit error.
We make four observations from Figure~\ref{fig:pec-opterr}.
First, both LSB and MSB errors increase as the P/E cycle count increases,
following an exponential trend. 
Second, ER~$\leftrightarrow$~P1 errors increase at \chX{a} much faster rate as the P/E
cycle count increases,
\chVI{compared} to the other types of cell state changes, and
ER~$\leftrightarrow$~P1 errors become the
dominant MSB error type when the P/E cycle count reaches \chVI{8K P/E cycles (6K
is the cross-over point)}. \chVI{This is because the electrons trapped in the
cell during wearout \chX{prevent the cell from being set to very low threshold
voltages}.
\chX{As a result, the threshold voltage distribution of the ER~state \chXI{shifts
and widens} more than the distributions of the other states, as we see in
Figure~\ref{fig:pec-mean-var}.}}
Third, multi-bit errors are \chXI{less common}, \chVI{but} they occur 
as early as \chXII{at} 1K P/E cycles.
\chVI{\chX{Only} a large difference between the target and actual
threshold voltage can lead to a multi-bit error, which is unlikely to happen.}
Fourth, \chX{MSBs} have a $2.1\times$ higher error rate than \chX{LSBs}, on
average across all P/E cycle counts. 
\chX{This is because the flash controller must use two read reference voltages
to read a cell's MSB, but needs \chXII{\emph{only one}} read reference \chXI{voltage} to read
a cell's LSB.}

\begin{figure}[h]
\centering
\includegraphics[trim=10 10 10 10,clip,width=\figscale\linewidth]
{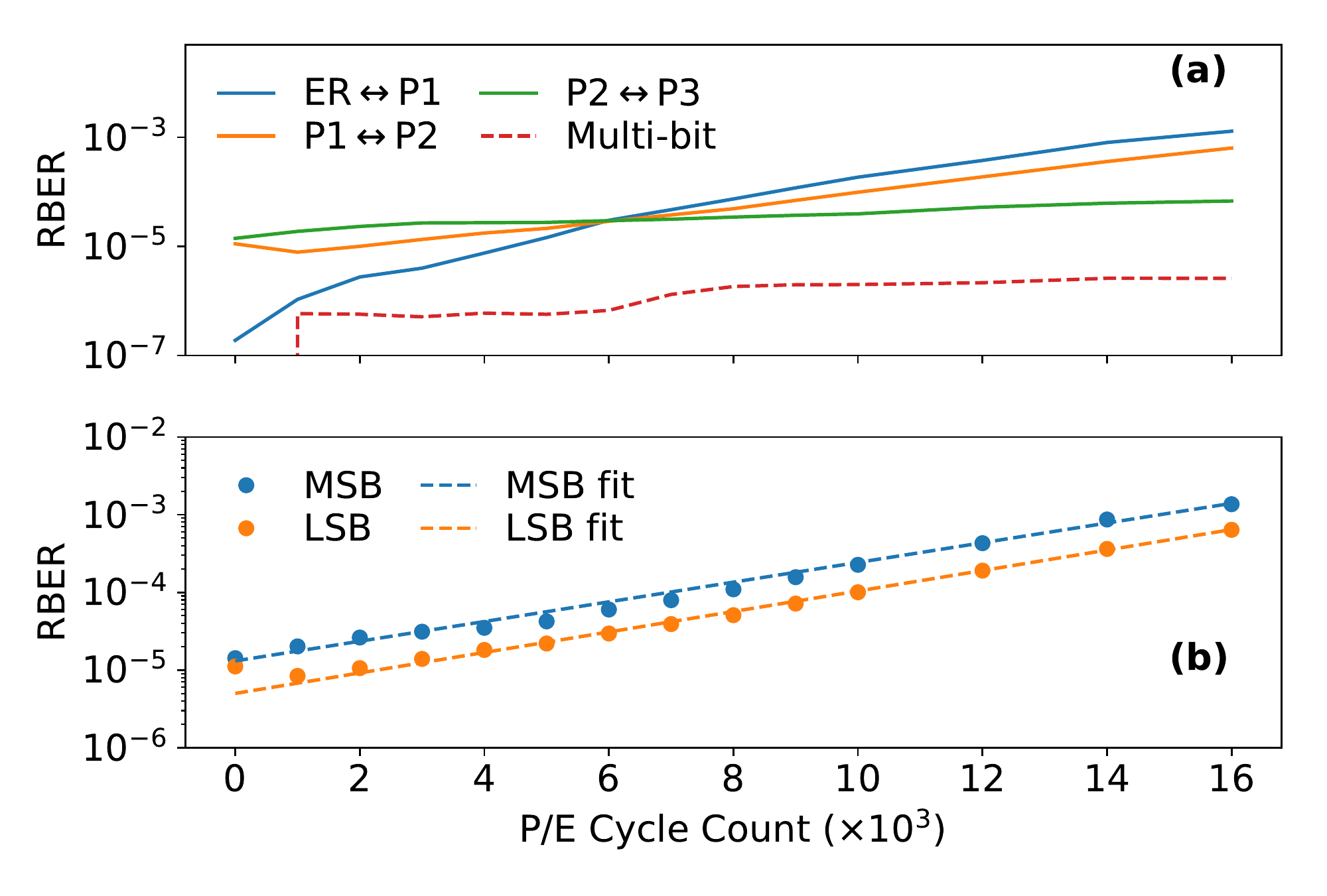}
\caption{\chX{RBER due to} \chVI{P/E cycling} errors vs. P/E \chX{cycle count}.}
\label{fig:pec-opterr}
\end{figure}

Figure~\ref{fig:pec-optvrefs} shows how the optimal read reference voltages
change as the P/E cycle count increases. This figure contains three
subfigures, each of which shows the
optimal voltage for $V_a$, $V_b$, and $V_c$ (see Figure~\ref{fig:F7}). 
We make two observations from this figure.
First, the optimal voltage for $V_a$ increases rapidly as the P/E cycle count
increases: after 16K P/E cycles, the voltage goes up by more than 20 voltage 
steps.
Second, the optimal \chX{voltages} for 
$V_b$ and $V_c$ remain almost constant as the P/E cycle count increases:
neither voltage changes by more than 4~voltage steps after 16K P/E cycles,
\chVI{as expected from the lack of change in P1, P2, \chX{and P3 distribution}
means shown in Figure~\ref{fig:pec-mean-var}.}

\begin{figure}[h]
\centering
\iftwocolumn
  \includegraphics[trim=10 10 10 10,clip,width=\linewidth]
  {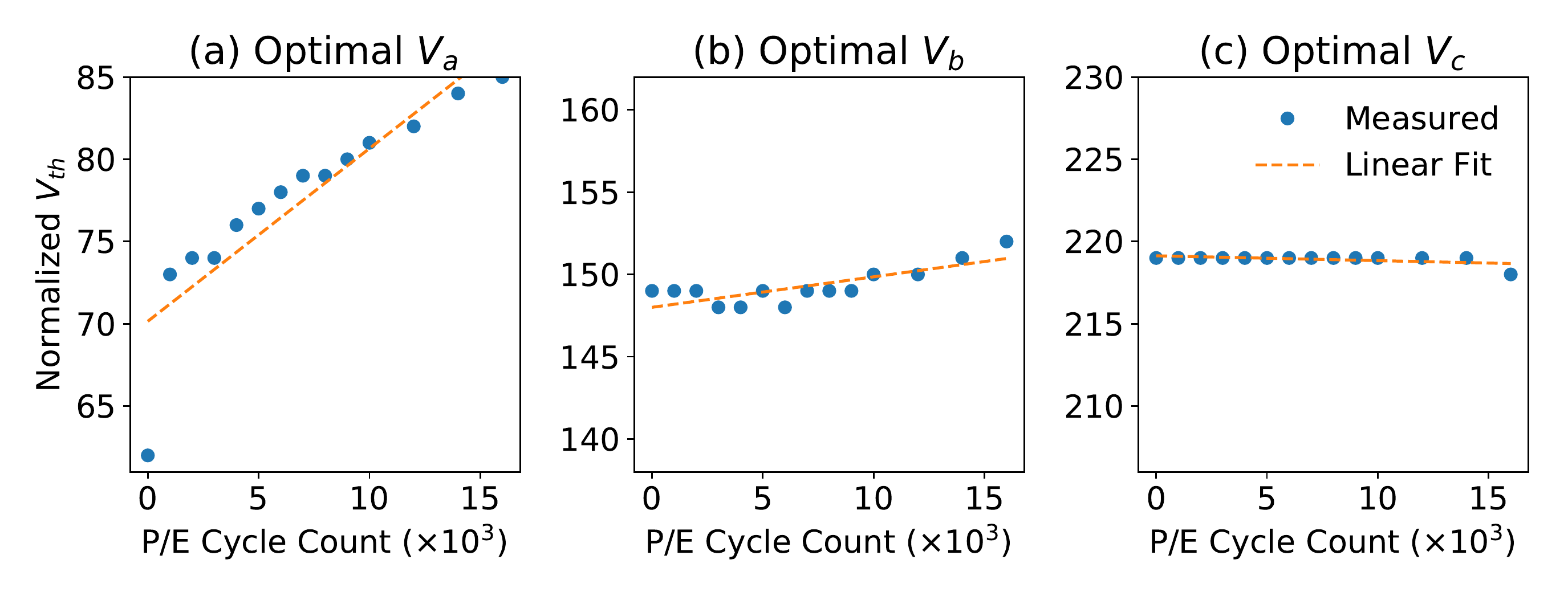}
\else
  \includegraphics[trim=10 10 10 10,clip,width=.8\linewidth]
  {figs/pec-optvrefs.pdf}
\fi
\caption{Optimal read reference voltages vs. P/E cycle \chX{count}.}
\label{fig:pec-optvrefs}
\end{figure}

\textbf{Insights.} To compare the error characteristics of 3D NAND \chX{flash memory} to that of
planar NAND flash memory, we \chX{take the} equivalent observations on planar NAND
\chX{flash memory} reported by prior
works~\cite{cai.date13, parnell.globecom14,
luo.jsac16}, and compare them to our findings for 3D NAND \chVI{flash \chX{memory,}
which we just described}.
We find two key differences.
First, for 3D NAND \chX{flash memory}, the threshold voltage distributions for the P2~state and
the P3~state shift to \chVI{\emph{lower}} voltages as the P/E cycle count increases.  In
contrast, for planar NAND flash memory, the distributions of both states shift
to \emph{higher} voltages~\cite{cai.date13, parnell.globecom14, luo.jsac16}.
\chX{One possible source of this change is the \chXI{increased impact of early retention loss
with P/E cycle count,}}
which lowers the threshold voltage of cells in \chX{higher-voltage} states (i.e.,
P2 and P3)~\cite{choi.vlsit16}. 
Second, for 3D NAND \chX{flash memory}, the \emph{change} in the mean \chVI{threshold voltage} of
each state distribution
exhibits a linear \chX{increase}.  However, in sub-\SI{20}{\nano\meter} planar NAND flash memory,
the change in the mean \chVI{threshold voltage} exhibits a \chX{\emph{power-law}-based
increase with P/E cycle count}~\cite{parnell.globecom14, luo.jsac16}.
In sub-\SI{20}{\nano\meter}
planar NAND flash memory, the mean \chVI{threshold voltage} of each state distribution increases more rapidly
at lower P/E cycle counts than in higher P/E cycle counts, resulting in \chX{the power-law-based}
behavior.  However, we note that planar NAND flash memory using an older manufacturing
process technology (e.g., \SIrange{20}{24}{\nano\meter}) exhibits a linear \chX{increase
with P/E cycle count} for the
distribution mean~\cite{cai.date13}, just as we \chVI{observe} for 3D NAND
\chVI{flash memory}.
Thus, \chVI{there is evidence that} when the manufacturing process technology
scales below
a certain size, the change in the distribution mean transitions from linear
behavior to power-law\chX{-based} behavior \chX{with respect to P/E cycle count}.  
As a result, when future 3D NAND \chX{flash \chXI{memory scales}} down
to a sub-\SI{20}{\nano\meter} manufacturing process technology \chX{node}, we \chVI
{might} expect that it too will exhibit power-law behavior \chXI{for the change in the
distribution mean}.
\chIX{We conclude that the differences we observe between the P/E cycling
effect in 3D NAND flash memory and planar NAND flash memory are mainly caused
by \chXI{the use of a significantly} different manufacturing process technology \chX{node}.}



\subsubsection{Program Interference}
\label{sec:3derror:interference}

When a cell (which we call the \emph{aggressor cell}) is being programmed, 
cell-to-cell program interference can cause
the threshold voltage of nearby flash cells (which we call \emph{victim cells})
to increase unintentionally~\cite{cai.sigmetrics14, cai.iccd13}
(see Section~\ref{sec:errors}).
In 3D NAND \chVI{flash memory}, there are two types of program interference that can occur.
The first, \emph{wordline-to-wordline program interference}, affects
victim cells along the z-axis \chVI{of} the cell \chX{that is} programmed (see Figure~\ref{fig:organization}).
These victim cells are physically next to the cell \chX{that is} programmed, and
belong to the same bitline (and thus the same flash block).
The second, \emph{bitline-to-bitline program interference}, affects
victim cells along the x-axis or y-axis \chVI{of} the cell \chX{that is} programmed.  
Bitline-to-bitline program interference can affect victim cells in the same
wordline (i.e., cells on the y-axis), or it can affect victim cells that
belong to other flash blocks (i.e., cells on the x-axis).

To quantitatively analyze the effect of program interference \chVI{on
cell threshold voltage and raw bit error rate}, we use the
same experimental data that we have for \chVI{P/E cycling} errors
(see Section~\ref{sec:3derror:pecycle}).  
\chX{A correlation exists between the amount by which program interference 
changes the threshold voltage of a victim cell ($\Delta V_{victim}$) and the 
threshold voltage change of the aggressor cell ($\Delta V_{aggressor}$)~\cite{cai.iccd13}.
As a result of this \emph{interference correlation}, the threshold voltage 
of a victim cell is \emph{dependent} on the threshold voltage}
of the aggressor cell. The strength of this correlation can be
quantified as $\frac{\Delta V_{victim}}{\Delta V_{aggressor}}$, which is a
property of the NAND device and is largely dependent on the distance between the 
cells~\cite{lee.iedl02}. \chVI{After programming \chX{randomly-generated} data to
the victim cells and the aggressor cells, we} estimate $\Delta V_{aggressor}$
by calculating the threshold voltage difference between the aggressor cell's
\chVI{threshold voltage in its} final state and \chVI{that in} the ER~state.
We estimate
$\Delta V_{victim}$ by calculating the difference between the
victim cell's threshold voltage with and without program
interference.\footnote{\chXI{The cell threshold voltage \chXII{\emph{without}} program
interference is obtained by reading the cell \chXII{\emph{before}} the next wordline is 
programmed.}}



\textbf{Observations.} 
Figure~\ref{fig:interference-correlation} shows the
interference correlation 
for wordline-to-wordline interference and bitline-to-bitline
interference on a victim cell, for aggressor cells of varying
distance from the victim cell. \chVI{For example, the victim cell
in BL~M, WL~N has \chX{an interference correlation of} 2.7\%
with the \emph{next wordline} aggressor cell in BL~M,
WL~N+1, which means
that, if the threshold voltage of the aggressor cell increases by $\Delta V$,
the threshold voltage of the victim cell \chX{increases by $0.027
\Delta V$} due to wordline-to-wordline program interference.}
We make two observations from this figure. 
\chVI{First, the interference correlation of the \emph{next wordline}
aggressor cell (i.e., 2.7\%) is over an order of magnitude higher than 
\chX{that of} any
other aggressor \chX{cell, of which the maximum
interference correlation is only 0.080\% (the \chXII{\emph{previous wordline}}
aggressor cell in \chXI{BL~M, WL~N-1}).} 
Thus, the program interference to the victim cell,} is dominated by
wordline-to-wordline interference from the \chXI{\emph{next}} wordline.
Second, all of the other types of interference have \chX{much smaller
interference correlation values}.


\begin{figure}[h]
\centering
\iftwocolumn
  \includegraphics[trim=0 180 370 0,clip,width=\linewidth]
  {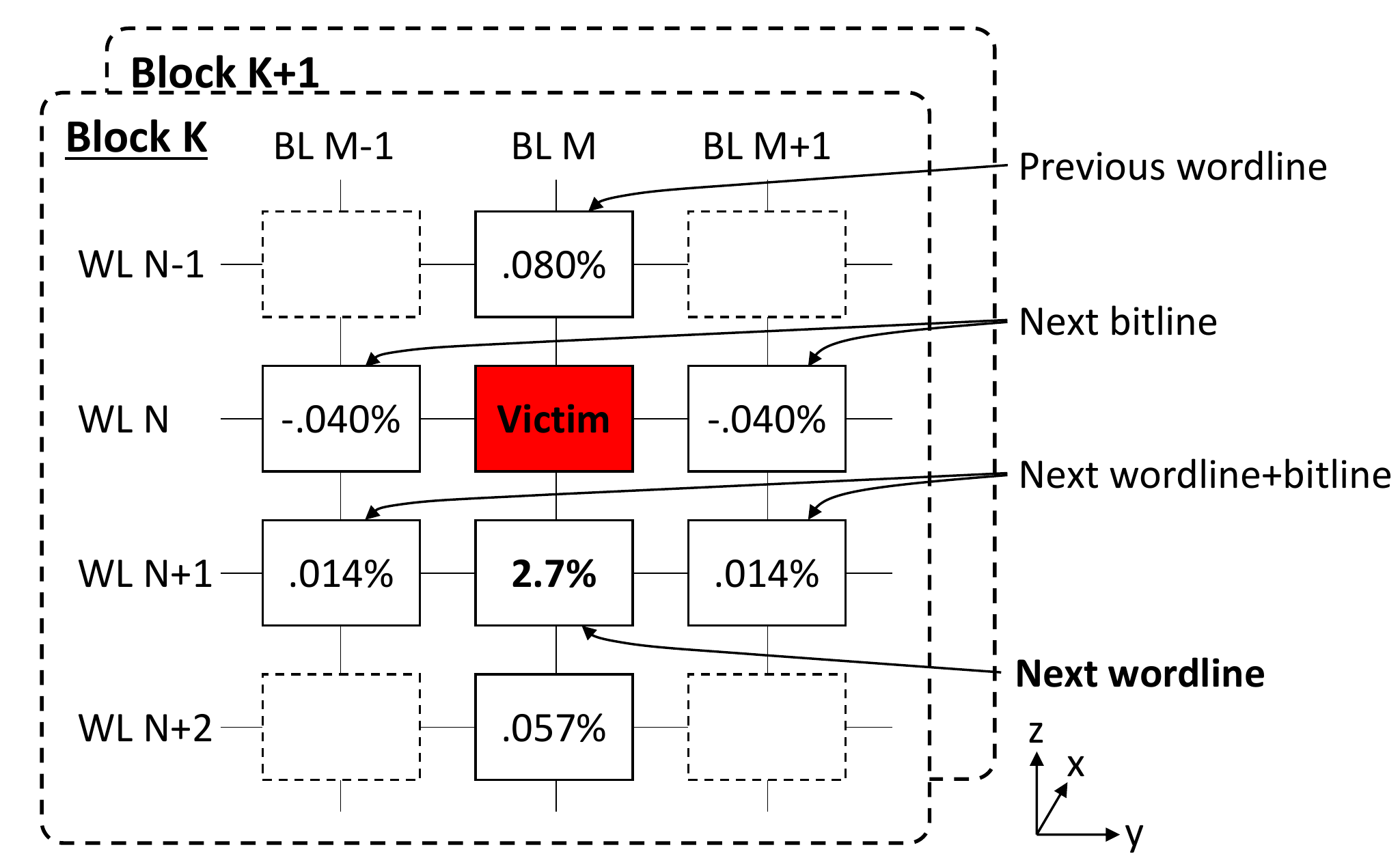}
\else
  \includegraphics[trim=0 180 370 0,clip,width=.8\linewidth]
  {figs/interference-correlation.pdf}
\fi
\caption{Interference correlation for a victim cell, as a result of programming \chX{aggressor cells of varying 
distances from the victim cell}.}
\label{fig:interference-correlation}
\end{figure}

Figure~\ref{fig:interference-trend} \chX{shows how} much the threshold voltage of a victim
cell shifts \chVI{($\Delta V_{victim}$)} when a neighboring aggressor cell is
programmed to the P3~state, which
generates the largest possible program interference.
Each curve represents a certain program interference type \chVI{
(i.e., Next WL or Prev WL)}
and a certain state of the victim cell \chVI{(V)}. \chVI{The curves \chX{that} have
a significant amount of threshold voltage shift (e.g., $>$6 voltage steps) due
to program interference are shown in Figure~\ref{fig:interference-trend}(a);
the curves \chX{that} have \chX{a} small amount of threshold voltage shift are shown in
Figure~\ref{fig:interference-trend}(b).}
We make three observations from Figure~\ref{fig:interference-trend}.
First, the effect of program interference decreases as the 
P/E cycle count increases \chVI{(along the x-axis, from left to right).
\chX{As we discuss in Section~\ref{sec:3derror:pecycle}, electrons trapped in a
flash cell due to wearout prevent the cell from returning to the lowest 
threshold voltage values during an erase operation.  As a result, as the
P/E cycle count increases, the \chXI{mean} threshold voltage \chXI{of 
the ER~state increases.  This causes $\Delta V_{aggressor}$ to decrease
as the P/E cycle count increases, because the starting voltage of the 
aggressor cell increases but its target voltage after programming remains 
the same.  As we discuss above, the interference correlation (i.e., the ratio
between $\Delta V_{aggressor}$ and $\Delta V_{victim}$) is largely a function 
of the distance between flash cells.  Thus, since $\Delta V_{aggressor}$
decreases, $\Delta V_{victim}$ also decreases with the P/E cycle count.}}
Second, the \chX{amount of} program interference induced by an aggressor cell in the next
wordline decreases
when the victim cell is in a higher-voltage state \chVI{(Next WL curves in
Figure~\ref{fig:interference-trend}a, from top to bottom). \chX{This} is likely
because the voltage difference between the aggressor cell and the victim cell
is lower when the victim cell is in a higher-voltage state, reducing the
the threshold voltage shift due to program interference.}
Third, the program interference induced by an aggressor cell in the previous
wordline \chVI{(Prev WL curves in Figure~\ref{fig:interference-trend})}
affects the threshold voltage
distribution of only the ER~state for a victim cell, but it has little effect
on the distributions of the other three states (i.e., P1, P2, P3).
\chX{This is a result of how programming takes place in NAND flash memory.
A program operation can only \emph{increase} the voltage of a cell due to
circuit-level limitations.  When the aggressor cell in the previous wordline is
programmed, the victim cell is already in the ER~state, and the victim cell's
voltage increases due to program interference.  Some time later, the victim cell
is programmed.  If the target state of the victim cell is P1, P2, or P3, the
programming operation needs to further increase the voltage of the cell,
and any effects of program interference from the aggressor cell in the
previous wordline are eliminated.
If, however, the target state of the victim cell is ER, the programming
operation does not change the victim cell's voltage, and the effects of
program interference from the aggressor cell in the previous wordline remain.}

\begin{figure}[h]
\centering
\iftwocolumn
  \includegraphics[trim=0 10 0 10,clip,width=\linewidth]
  {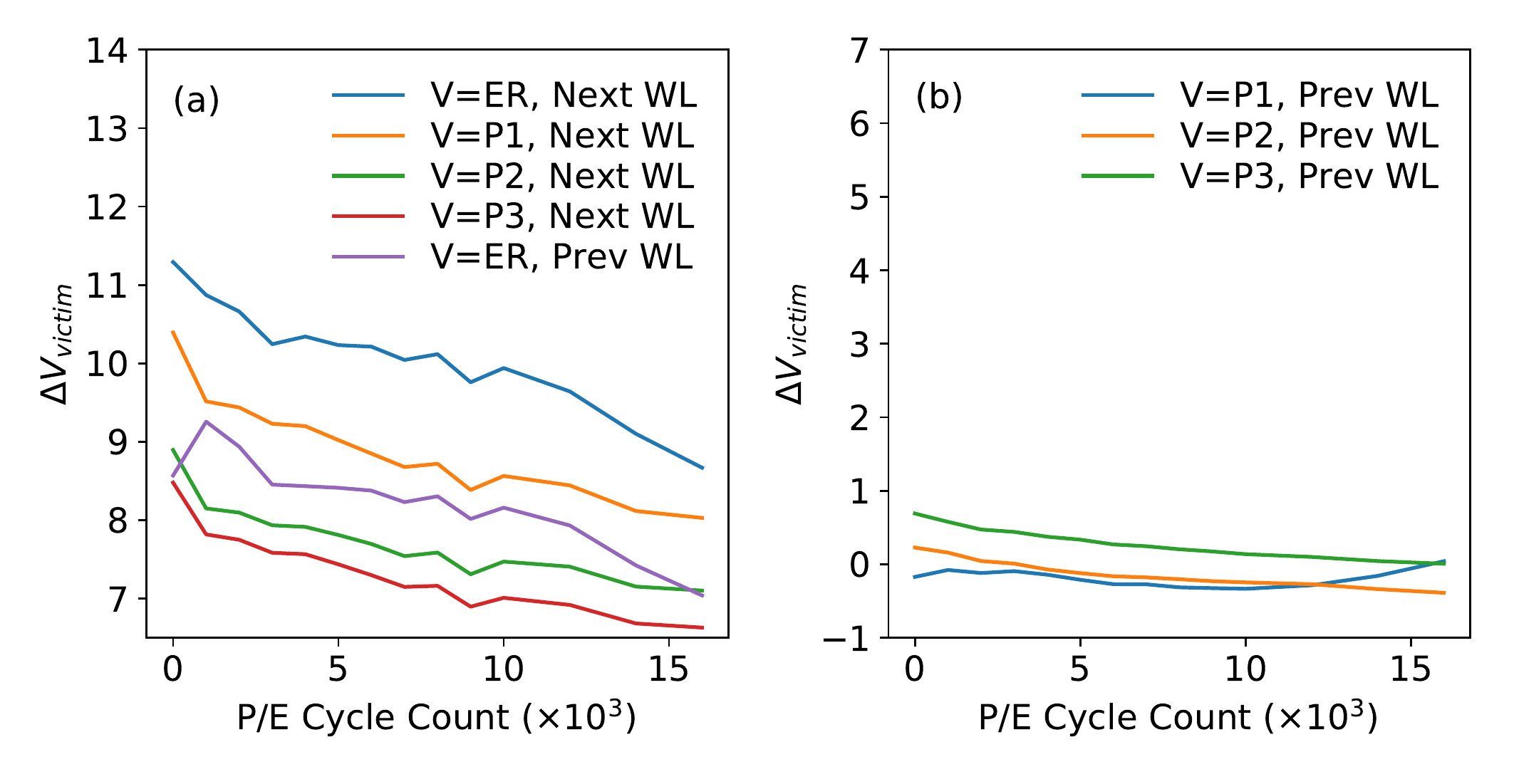}
\else
  \includegraphics[trim=0 10 0 10,clip,width=.7\linewidth]
  {figs/interference-trend.pdf}
\fi
\caption{\chVI{Amount of threshold voltage shift due to program interference}
vs. P/E cycle \chVI{count}.}
\label{fig:interference-trend}
\end{figure}

\textbf{Insights.} We compare the program interference in 3D NAND \chX{flash memory}
to the program interference observed in
planar NAND \chVI{flash memory}, as reported in prior work~\cite{cai.iccd13,
cai.sigmetrics14}. We
find one major \chVI{difference.} 
\chX{The maximum interference correlation of program interference from a 
directly-adjacent cell is 40\% lower in 3D NAND flash memory (2.7\%) than in
state-of-the-art (\SIrange{20}{24}{\nano\meter}) planar NAND flash memory
(4.5\%~\cite{cai.iccd13}).
This is corroborated by findings in prior work~\cite{park.jssc15}, which
shows that 3D NAND flash memory has 84\% lower program interference than
\SIrange{15}{19}{\nano\meter} planar NAND flash memory.
The lower interference correlation in 3D NAND flash memory is due to the
larger manufacturing process technology node
\chXI{(\SIrange{30}{50}{\nano\meter} for the chips we test)} 
that it uses compared to state-of-the-art
planar NAND flash memory.  \chXI{The} amount of interference correlation
between neighboring cells is a function of the distance between the cells~\cite{lee.iedl02}.
In a larger manufacturing process technology node, the flash cells are farther 
away from each other, causing the interference correlation to decrease.
We note that when future 3D NAND flash \chXI{memory chips} use smaller manufacturing
process technology nodes, the impact of programming interference will increase,
similar to what happened in planar NAND flash memory.}

Note that we are the first to compare how the threshold voltage shift caused
by program interference changes with the P/E cycle count. As we discuss in our
first observation for Figure~\ref{fig:interference-trend},  the program
interference \chXI{effect decreases} as the P/E cycle count increases
\chX{because the increasing effects of wearout reduce the value of
$\Delta V_{aggressor}$ during programming}.
\chIX{We conclude that the 40\% reduction in the program interference effect
we observe in 3D NAND flash memory compared to planar NAND flash memory is
mainly caused by the difference in manufacturing process technology.}

\subsection{Early Retention Loss}
\label{sec:3derror:appendix:retention}

In this \chX{section}, we present the results and analysis of retention loss in
3D NAND \chX{flash memory} in addition to the key findings in Section~\ref{sec:3derror:retention}. We use
the same methodology as described in Section~\ref{sec:3derror:retention}.

\textbf{Observations.}
\chX{Figure~\ref{fig:retention-mean-var} shows} how the
mean and the standard deviation of \chVI{the threshold voltage}
distribution \chVI{change} with retention time.
Each subfigure \chVI{in the top row \chX{shows} the mean for a different
state; each subfigure in the bottom row \chX{show} the standard deviation for
a different state. The blue dots \chX{show} the measured data; each orange line
shows a linear \chX{trend line} fitted to the measured data. \chVI{The x-axis shows the
retention time in log scale; the y-axis shows the mean or standard deviation
value in voltage steps.}}
We make five observations from \chVI{this figure}.
\chVI{First}, the threshold voltage distribution shifts \chX{more} when the
retention time is low. \chVI{This is the early retention \chXI{loss} phenomenon, \chX{which
occurs because charge that is trapped near the surface of the charge
trap layer is detrapped soon} after programming.}
\chVI{Second}, \chX{as the retention time increases,} 
the \chX{voltage values of cells in the P1, P2, and P3 states decrease, while
the voltage values of cells in the ER state increase.  This is} because the
cells in \chXI{the} ER state have \emph{negative} threshold \chX{voltages, and hence 
they} \emph{gain} charge over retention time.
\chVI{Third}, the \chX{threshold voltage distributions of the ER and P3 states 
shift faster than the distributions of the P1 and P2 states as the retention time
increases.  This is because the}
ER and P3 states have larger voltage \chX{differences} \chXI{from} the ground than the other
states.}
\chVI{Fourth}, retention loss has little effect on the width of the threshold
voltage distribution (i.e., standard deviation values change by less than 1
voltage step \chX{after 24 days}).  \chX{This is because the effects of retention loss
(i.e., charge leakage) impact cells at a similar rate, causing all of the cells
within the threshold voltage distribution to lose a similar amount of voltage.}
\chVI{Fifth}, the correlation between any distribution parameter ($V$) and the
retention time ($t$) can be modeled as a linear function (\chX{shown by} the dotted lines in
\chXI{Figure~\ref{fig:retention-mean-var}}): $V = A \cdot \log(t) + B$. 
\chX{$A$ and $B$ are constants that change based on which parameter $V$
is modeling (i.e., the threshold voltage distribution mean or standard deviation).}
\chVI{Prior
work shows that planar NAND flash memory has a similar trend
for retention loss, \chX{even though it uses} a different flash cell design. We have
already compared \chX{and evaluated the differences} between 3D NAND and planar NAND flash memory
in retention loss speed in Section~\ref{sec:3derror:retention}, and \chX{provided} more
detail about the linear \chX{function \chXI{that models the}
threshold voltage distribution parameters} in Section~\ref{sec:3derror:model:retention}.}

\begin{figure*}[h]
\centering
\includegraphics[trim=10 10 10 10,clip,width=\linewidth]
{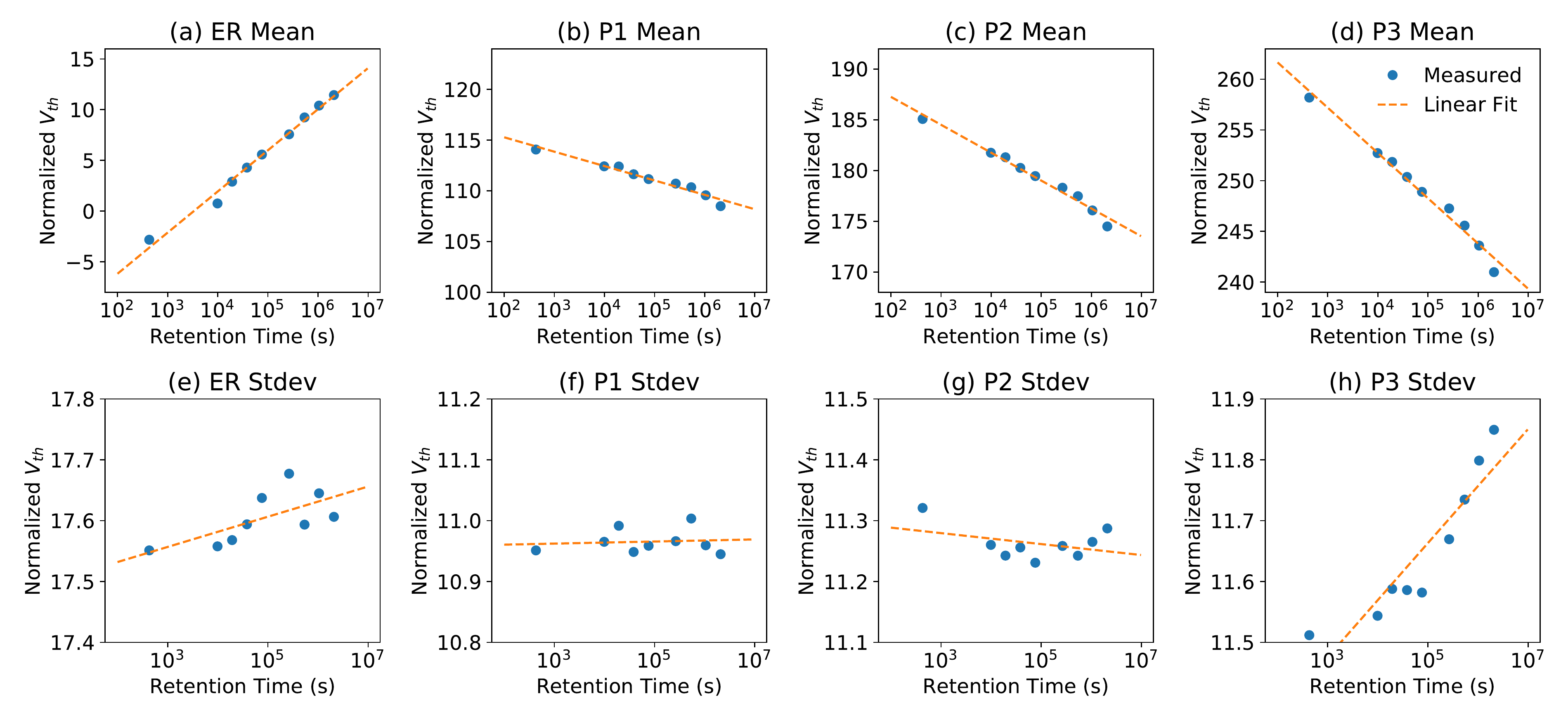}
\caption{Mean \chVI{and standard deviation of \chIX{our Gaussian} threshold voltage} distribution
\chIX{model of each state}, \chVI{versus retention time}.}
\label{fig:retention-mean-var}
\end{figure*}


Figure~\ref{fig:retention-opterr} \chVI{shows how} \chXI{the} RBER
increases with retention time for a block \chX{that has endured} 10K P/E cycles. The top \chVI
{graph} breaks down the errors according to the change in cell state as a
result of the errors; the bottom graph breaks down
the errors into MSB and LSB page errors.
We make \chVI{two} observations from Figure~\ref{fig:retention-opterr}, \chVI
{in addition to our observations in Section~\ref{sec:3derror:retention}}.
\chVI{First}, retention errors are dominated by P2~$\leftrightarrow$~P3
errors, \chVI{because \chX{the threshold voltage distribution of the} P3 state not 
only shifts \chX{more} but also widens \chX{more with}
retention time than \chX{the distributions of} the other states (see Figure~\ref{fig:retention-mean-var}). 
Although \chX{the distribution of the} ER
state also shifts \chX{significantly}, there are fewer ER~$\leftrightarrow$~P1 errors to
begin with.}
\chVI{Second}, the MSB error rate increases
faster than the LSB error rate \chX{as the retention time increases. This is}
\chVI{because \chX{as the distributions of both the ER and P3 states shift more
than those of the P1 and P2 states, cells in the ER and P3 states are more 
likely to have errors.  These errors (ER~$\leftrightarrow$~P1 and
P2~$\leftrightarrow$~P3) affect the MSB of the cell.}

\begin{figure}[h]
\centering
\includegraphics[trim=0 0 0 0,clip,width=\figscale\linewidth]
{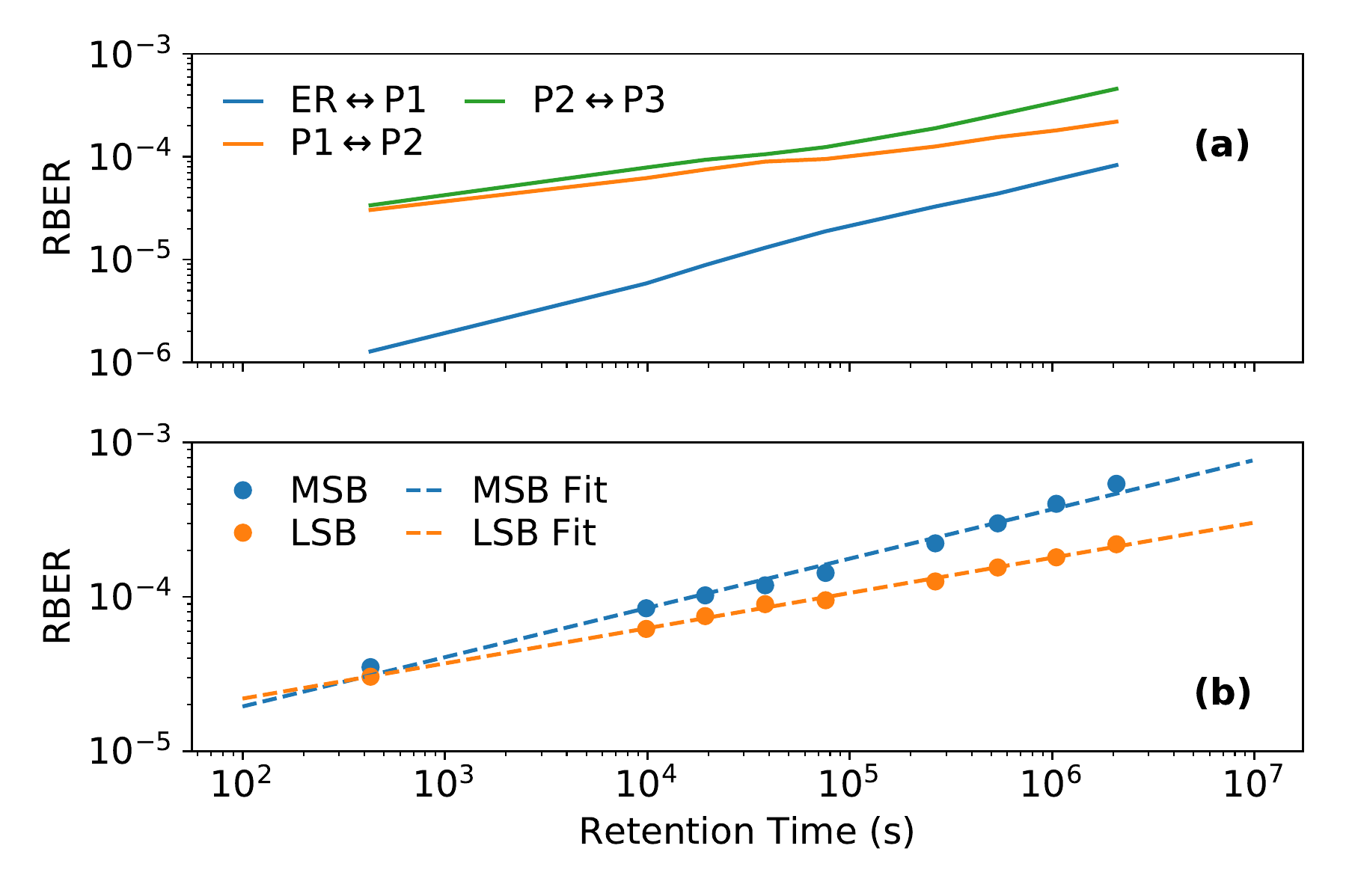}
\caption{RBER \chVI{vs.} retention time, broken down by \chVI{(a)~the state transition
of each flash cell, and (b)~MSB or LSB page}.}
\label{fig:retention-opterr}
\end{figure}






\textbf{Insights.} We compare the errors due to retention loss in 3D NAND \chX{flash memory to those} in planar
NAND flash memory, as reported in prior work~\cite{mielke.irps08, cai.hpca15,
cai.iccd12}. We find another major \chVI{difference} in 3D NAND \chVI{flash
memory} in terms of threshold
voltage \chX{distribution,} in addition to those
discussed in Section~\ref{sec:3derror:retention}. We find that \chX{the}
retention loss \chX{phenomenon we observe} in 3D NAND \chX{flash memory
(1)~}shifts the threshold voltage distributions of the P1, P2 and \chX{P3 states} lower, and 
\chX{(2)~}has little effect on the width of the distribution of each state. In contrast, the
retention loss \chX{phenomenon observed in} planar NAND flash memory 
\chX{(1)~}does \emph{not} shift the P1 and P2 state
distributions by much, and 
\chX{(2)~}increases the width of each state's distribution 
\emph{significantly}~\cite{cai.hpca15}. 
This indicates that a mechanism that
adjusts the optimal read reference voltage to the threshold voltage shift caused
by retention \chX{loss} can be more effective on 3D NAND \chIX{flash memory} than on
planar NAND \chIX{flash memory}, \chXI{because the distributions shift by a 
greater amount (indicating a greater need for voltage adjustment) \chXII{with}
a smaller amount of overlap between two threshold voltage distributions 
(reducing the number of read errors when the optimal read reference voltage is used)}. 
\chIX{We conclude that, due to \chX{the} early
retention loss phenomenon \chX{we observe in 3D NAND flash memory}, 
\chX{the threshold voltage of a flash cell}
changes quickly within several hours after programming, leading to
significant changes in RBER and optimal read reference \chX{voltage values}.}

\subsection{Read-Induced Errors}
\label{sec:3derror:read}

\chX{In this section, we} analyze how each type of read-induced error affects the
RBER and the threshold voltage distribution of 3D NAND flash
memory.

\subsubsection{Read Errors}
\label{sec:3derror:read:variation}

\chX{A read error is a type of read-induced error where two reads to a
flash cell may return different data values if the read reference voltage used
to read the cell is close to the cell's threshold voltage\chXI{~\cite{joe.ted11, 
compagnoni.edl09, ghetti.ted09}} (see Section~\ref{sec:errors}).}
\chXI{A read error adds uncertainty to the outcome of \emph{every} 
read operation performed by the SSD controller.  However, despite the potential
for widespread impact, read} errors are \chX{\emph{not}} well-studied by prior work. 


To quantify read errors, we use the data \chIX{we collected} in
Section~\ref{sec:3derror:retention}. \chX{For each cell, we see if the \emph{actual}} read outcome
\chX{(i.e., the bit value output by the flash controller after a read operation)
matches the \emph{expected}} read outcome \chX{(i.e., the
value that the read should have returned based on the current voltage of the
flash cell).  We determine the expected read outcome} by comparing $V_{ref}$
with $V_{th}$ (i.e., we expect to read 1 if $V_{th} < V_{ref}$,
\chX{because $V_{ref}$ is high enough that it should turn on the cell}).
\chX{We obtain $V_{th}$} by combining the outcomes of multiple reads when sweeping the read
reference voltage, thus \chXI{we expect that the combined output
eliminates the impact of read errors and is thus accurate.}
\chX{We say that a read error occurs if the actual read outcome and the
expected read outcome do \chXI{\emph{not}} match.}



\textbf{Observations.}
Figure~\ref{fig:read-error-offset} shows \chIX{how \chX{the} read
error rate changes \chX{as a function of the} \emph{read offset} (i.e., $V_{ref} - V_{th}$).}
We observe that\chIX{,} as the \chIX{absolute value of} \chX{the} read offset \chIX
{increases, \chX{the} read} error rate decreases exponentially. \chIX{This is likely
because \chX{when} $V_{ref}$ is closer to $V_{th}$ (i.e., \chXI{when} $V_{ref} - V_{th}$ has a
smaller absolute value), the amount of noise (i.e., voltage fluctuations) in
the sense amplifier increases exponentially\chXI{~\cite{compagnoni.edl09, ghetti.ted09}}.  The
\chXI{larger amount of} noise increases the
likelihood that the sense amplifier incorrectly detects whether the cell 
turns on, which leads to a \chXI{larger} probability that a read error occurs.}

\begin{figure}[h]
\centering
\iftwocolumn
  \includegraphics[trim=10 10 10 10,clip,width=\linewidth]
  {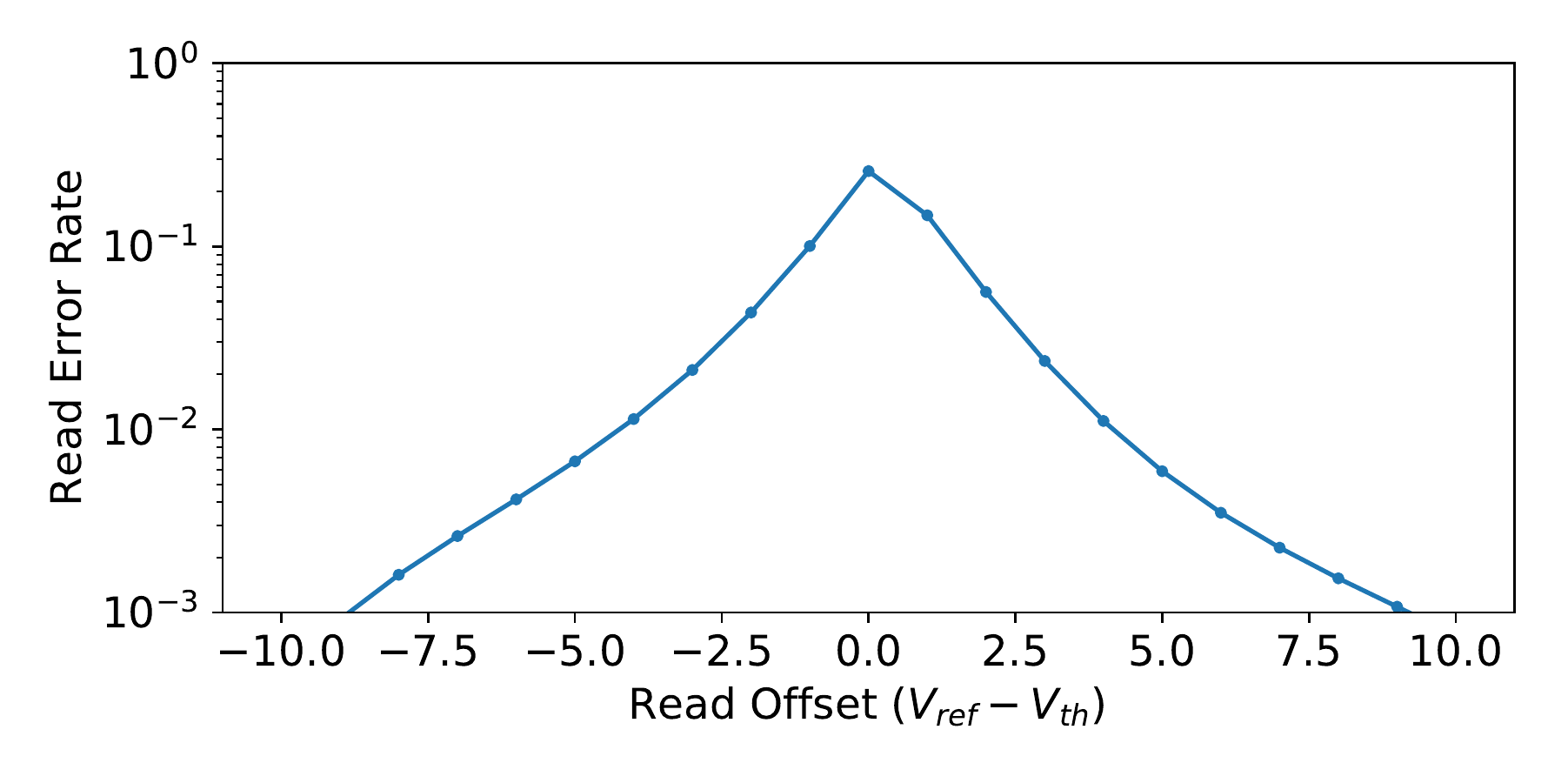}
\else
  \includegraphics[trim=10 10 10 10,clip,width=.6\linewidth]
  {figs/readError-vs-offset-10kPEC.pdf}
\fi
\caption{Read error \chXI{rate} vs.\ read offset \chXI{($V_{ref} - V_{th}$)}.}
\label{fig:read-error-offset}
\end{figure}

Figure~\ref{fig:read-error-ratio} shows the correlation
between the read error rate and the total RBER in a flash page.
We observe that \chXI{the} read error rate is linearly correlated with the overall
RBER\@. This is because, when the RBER is high, the 
threshold voltage distributions of neighboring states overlap with
each other by a greater amount.  This causes a larger number of cells to
be close to the read reference voltage value, increasing the probability
that a read error occurs (see Figure~\ref{fig:read-error-offset}).

\begin{figure}[h]
\centering
\includegraphics[trim=0 10 0 10,clip,width=\figscale\linewidth]
{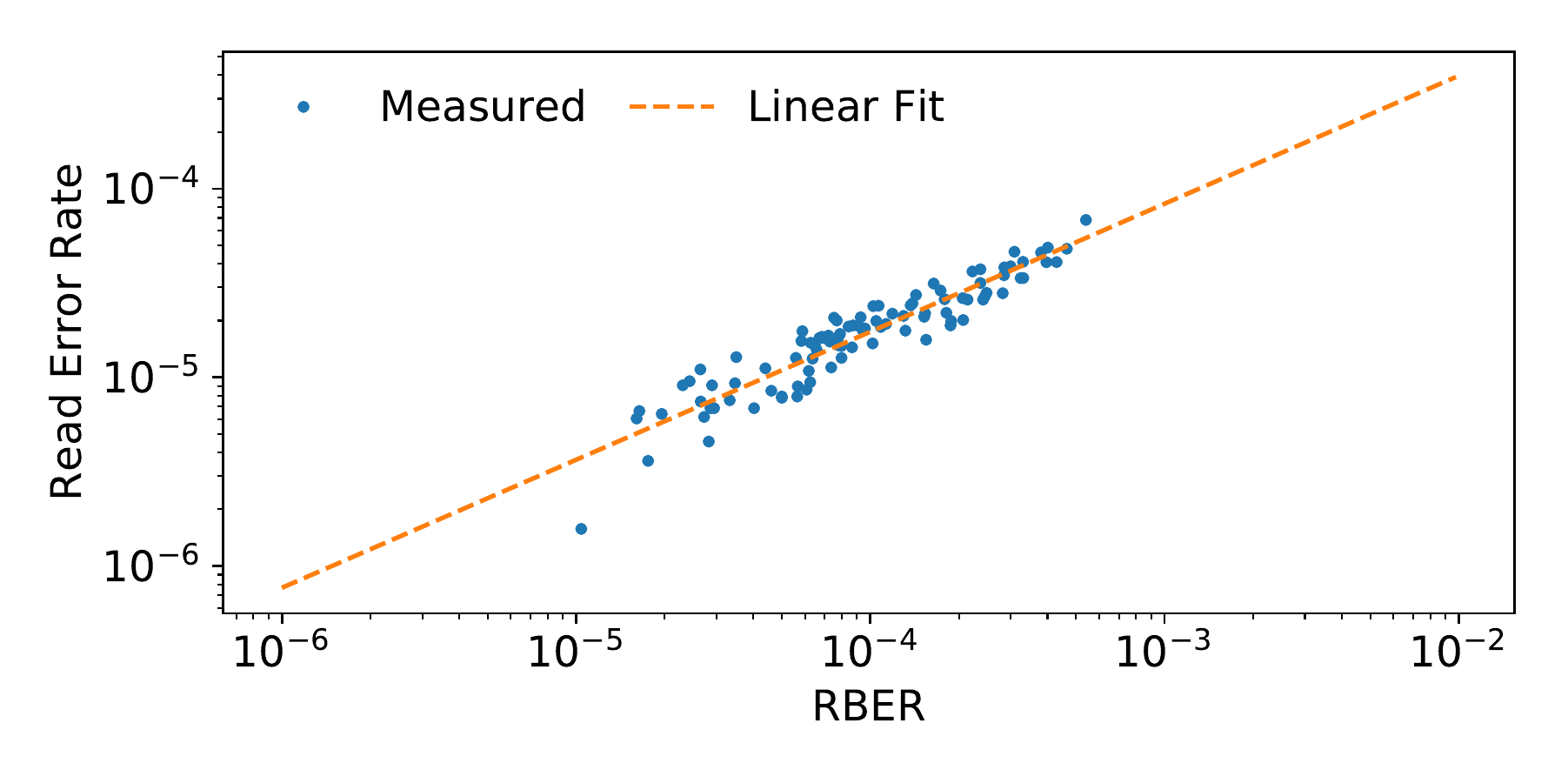}
\caption{\chXI{Relationship between the read error rate and} the RBER.}
\label{fig:read-error-ratio}
\end{figure}

\textbf{Insights.}
We are the first to discover and quantify the extent of read errors,
and to show the
correlation of these errors with the RBER and with the read reference voltage.
\chIX{We conclude that read errors are correlated with the read offset 
\chX{(i.e., $V_{ref} - V_{th}$)} and
the overall RBER of the flash page.}






\subsubsection{Read Disturb Errors}
\label{sec:3derror:read:disturb}

\chXI{Read disturb errors occur when a read operation to one page in a 
flash block may introduce errors in \emph{other, unread} pages in the
same block~\cite{cai.dsn15, papandreou.glsvlsi14}}
(see Section~\ref{sec:errors}). Read disturb
errors are caused by the high pass-through voltage applied \chXI{to cells
in the unread pages}.

To characterize read disturb errors, we first randomly select 11 flash blocks
and wear out each block to 10K P/E cycles \chVI{by repeatedly erasing and
programming pseudorandomly generated data} \chX{into each page of each 
block}. Then, we program
\chX{pseudorandomly-generated} data to \chX{each page of} each flash block. To minimize the
impact of other errors,
especially retention errors due to early retention loss, we wait until the
data has \chVI{a} 2-day retention time before inducing read disturb. This ensures
that, according to our results in Section~\ref{sec:3derror:retention}, \chVI{after
2 days,} retention loss \chVI{has slowed down and} can
only shift the threshold voltage by at most 1 voltage step during the \chVI
{relatively short} characterization process \chVI{($\sim$\SI{9}{\hour})}. To
induce read disturb in the flash block, we
repeatedly read from a wordline within the block for up to 900K times (i.e.,
up to 900K read disturbs). During this process, to characterize \chX{the} read disturb
effect, we obtain the RBER and threshold voltage distribution at ten different read
disturb counts from 0 to 900K.


\textbf{Observations.}
\chX{Figure~\ref{fig:read-mean-var}} shows how the
mean and standard deviation \chVI{of the threshold voltage distribution} change with
read disturb count.
\chVI{Each subfigure in the top row \chX{shows} the mean for a
different state; each subfigure in the bottom row \chX{shows} the standard
deviation for a different state. The blue dots shows the measured data; each
orange line shows a linear \chX{trend line} fitted to the measured data. The x-axis
shows the P/E cycle count; the y-axis shows the distribution parameters in
voltage steps.}
We make three observations from \chVI{this figure}. \chVI{First}, \chX{the} read
disturb effect increases the \chVI{mean threshold voltage of \chX{the} ER state
significantly, by $\sim$8 voltage steps after 900K read disturbs. In contrast,
the mean threshold \chX{voltages} of the programmed states change by \chX{only a small amount}
($<$3 voltage steps).} The increase in the \chVI{mean threshold voltage} is lower for
a higher $V_{th}$ state. 
\chX{This is because the impact of read disturb is correlated with the 
difference between the pass-through voltage (see
Section~\ref{sec:flash}) and the threshold voltage of a
cell.  When the difference is larger (i.e., when the threshold voltage of a
cell is lower), the impact of read disturb increases.}
\chXI{In fact, we observe that the threshold voltage distribution of the P3 state
even shifts to slightly \emph{lower} voltage values during the experiment,
because read disturb has little effect on cells in the P3~state, and the impact
of retention \chXII{loss dominates}.}
\chVI{Second}, the distribution width of each
state (i.e., standard deviation) \emph{decreases} slightly \chVI{as \chX{the} read disturb count
increases, by $<$0.2 voltage steps after 900K read disturbs}. 
\chVI{Third}, the change in \chVI{each} distribution
\chVI{parameter} can be
modeled as a linear function of \chXI{the} read disturb \chVI{count} \chX{(as shown by the orange
dotted lines)}. This shows that read disturb in 3D NAND flash memory follows a
similar linear trend as that observed in planar NAND flash memory by prior
work~\cite{cai.dsn15}.

\begin{figure*}[h]
\centering
\includegraphics[trim=10 10 10 10,clip,width=\linewidth]
{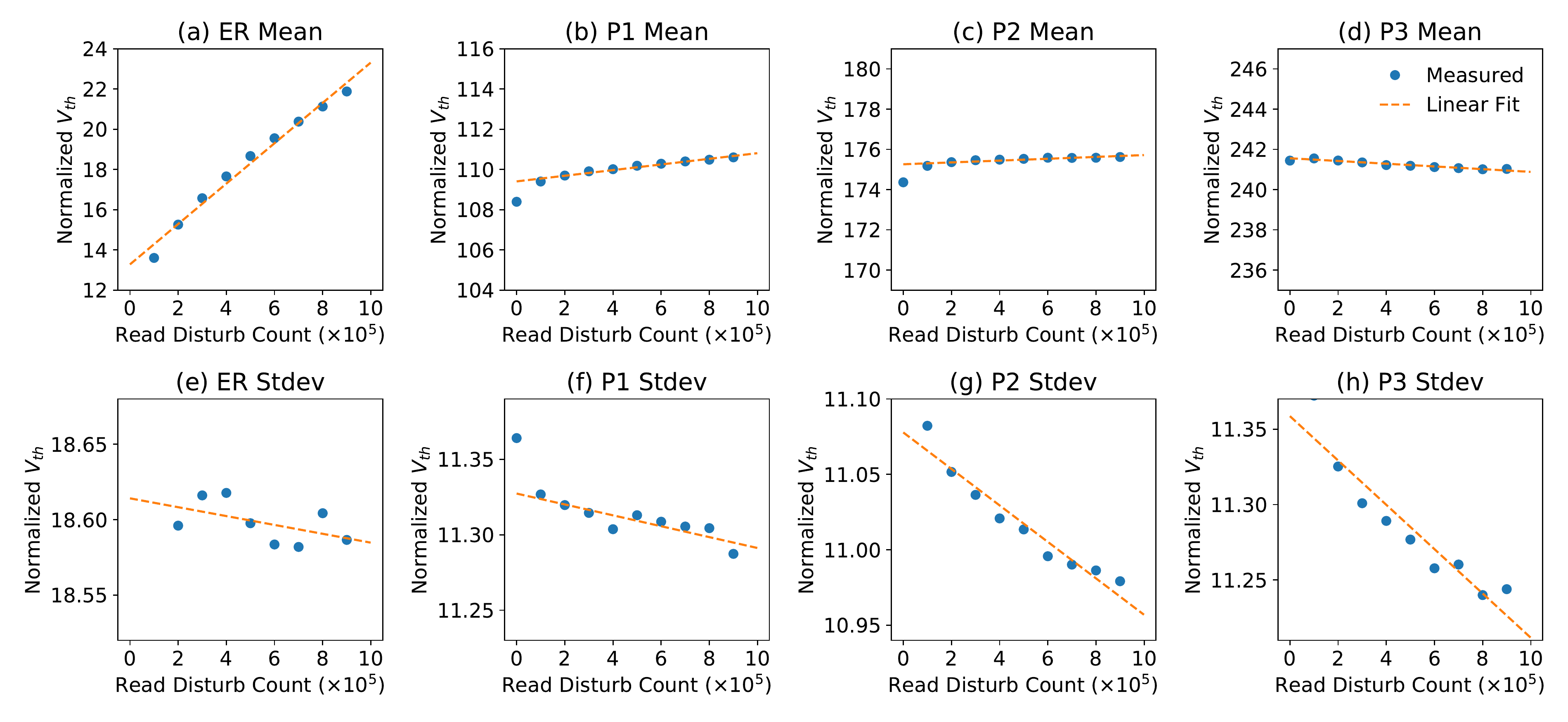}
\caption{\chVI{Mean and standard deviation of threshold voltage distribution
of each state}, \chX{vs.\ }read disturb count.}
\label{fig:read-mean-var}
\end{figure*}


Figure~\ref{fig:read-opterr} plots how RBER increases \chX{with}
read disturb \chVI{count for a flash block \chXI{that has endured} 10K P/E cycles}. The top graph
breaks down the errors according to the change in cell state as a
result of the errors; the bottom graph breaks down
the errors into \chX{MSB and LSB} errors.  We make three
observations from Figure~\ref{fig:read-opterr}.
\chIX{First, ER$\leftrightarrow$P1 errors increase significantly with read
disturb count, whereas P1$\leftrightarrow$P2 and P2$\leftrightarrow$P3 errors
do not. This is because \chX{the} ER state threshold voltage distribution \chX{shifts}
significantly with read disturb count (see Figure~\ref{fig:read-mean-var}),
reducing
the threshold voltage difference between \chXI{the ER and P1 states}. Second, MSB errors
increase much faster than LSB errors with read disturb count because
ER$\leftrightarrow$P1 errors are a type of MSB \chX{error, and they increase}
significantly \chXI{with read disturb count}. Third, the increase in RBER with read disturb count follows a
linear trend (\chX{as shown by} the dotted line in Figure~\ref{fig:read-opterr}b), which
is similar to the observation \chX{made for} planar NAND flash memory by prior
work~\cite{cai.dsn15}.}

\begin{figure}[h]
\centering
\includegraphics[trim=0 10 0 10,clip,width=\figscale\linewidth]
{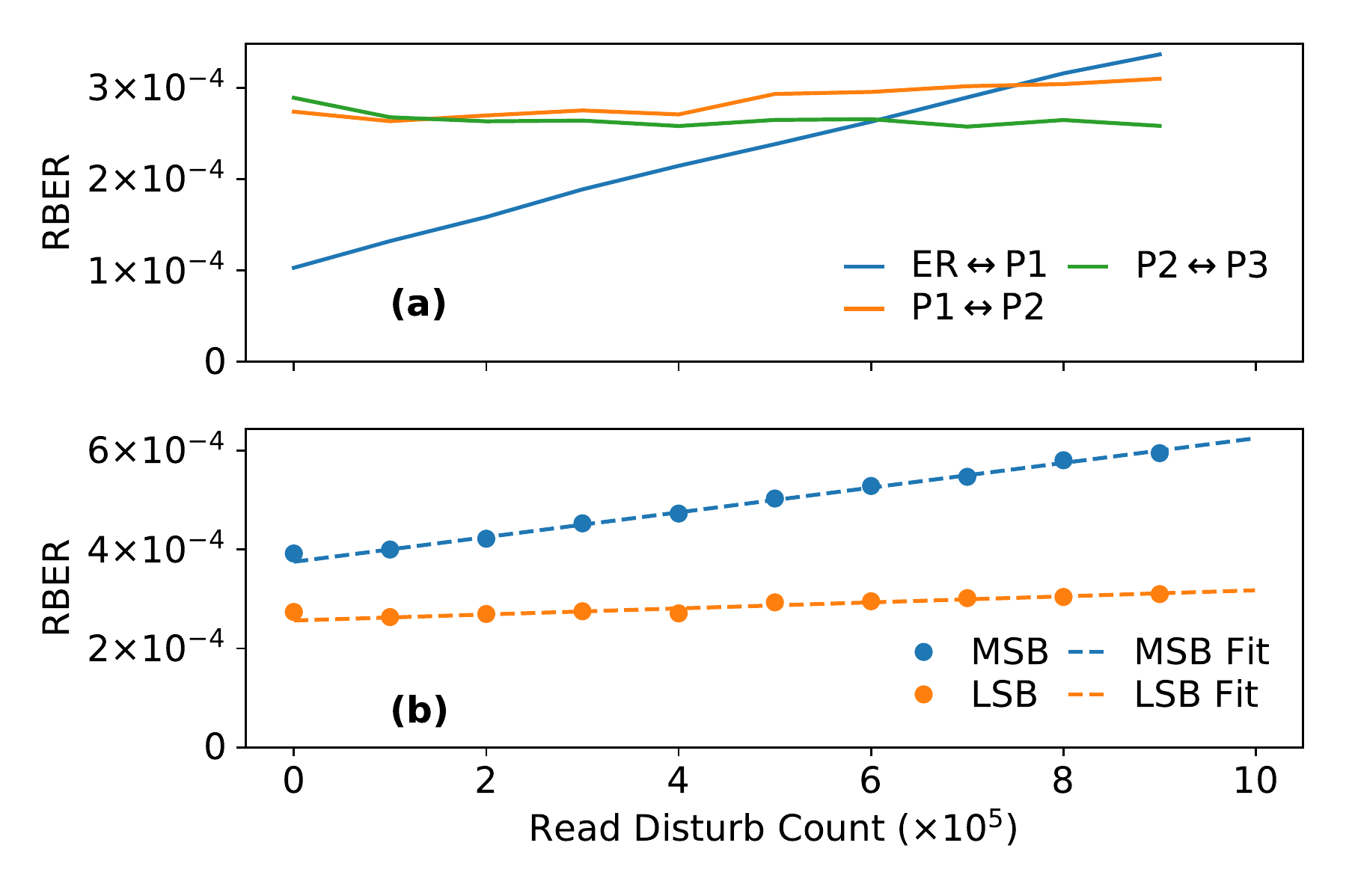}
\caption{RBER vs.\ read disturb count, broken down by (a)~the state
transition of each flash cell, and (b)~MSB or LSB page.}
\label{fig:read-opterr}
\end{figure}

Figure~\ref{fig:read-optvrefs} shows how the optimal read reference voltages
change \chIX{with} read disturb \chIX{count}.
\chX{The three} subfigures show the optimal voltages for $V_a$, $V_b$, and $V_c$. We
make two observations from this figure. \chIX{First, the optimal voltages
for $V_b$ and $V_c$ \chXI{change by relatively little} as \chXI{the} read disturb
count increases ($<$3 voltage steps after 900K read disturbs), whereas the
optimal $V_a$ \chX{changes more with the} read disturb count.
\chXI{This is because read disturb causes the threshold voltage distributions of
lower-voltage states to change by a greater amount, which requires the
read reference voltages separating the lower-voltage states (e.g., $V_a$) to
change more.}
Second, the increase in the optimal $V_a$ follows a linear trend with read
disturb count, because the ER state threshold voltage distribution shifts
linearly (as we see from Figure~\ref{fig:read-mean-var}).}

\begin{figure}[h]
\centering
\iftwocolumn
  \includegraphics[trim=10 10 10 10,clip,width=\linewidth]{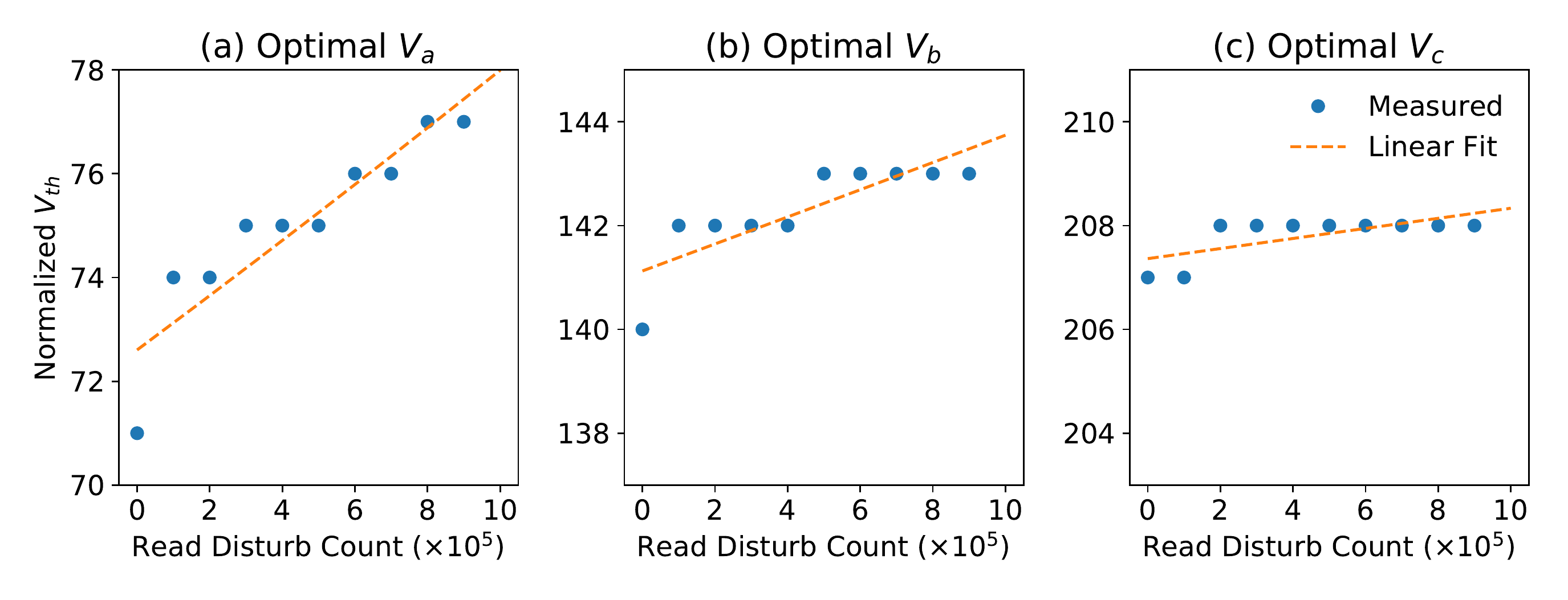}
\else
  \includegraphics[trim=10 10 10 10,clip,width=.8\linewidth]{figs/readDisturb-10kPEC-optvrefs.pdf}
\fi
\caption{Optimal read reference voltages vs.\ read disturb \chXII{count}.}
\label{fig:read-optvrefs}
\end{figure}

\textbf{Insights.} We compare the read disturb effect \chX{that we observe in
3D NAND flash memory to that observed in
planar NAND flash memory} \chXI{by prior} work~\cite{cai.dsn15}. We make the
observation that, although RBER \chX{increases} linearly \chIX{with read disturb count} in
both 3D NAND and planar NAND \chX{flash memory}, the slope of \chXI{the} increase (i.e., the sensitivity \chX{of \chXI{the} RBER} to
read disturb) at 10K P/E cycles is 96.7\% \chXI{\emph{lower}} in 3D NAND \chX{flash memory} than \chX{that} in planar
NAND \chX{flash memory}~\cite{cai.dsn15}. We believe that this difference in \chIX{the} sensitivity to
read disturb \chIX{effect} is due to the use of a larger process technology
\chIX{node} \chX{(\SIrange{30}{40}{\nano\meter})}
in \chX{current} 3D NAND \chIX{flash memory}. 
\chX{The comparable planar NAND flash memory results} from prior work \chX{are
collected on} \SIrange{20}{24}{\nano\meter} planar NAND \chIX{flash memory} \chX{devices~\cite{cai.dsn15}}. 
\chX{We}
expect the \chIX{read disturb effect} in 3D NAND \chX{flash memory} to increase in the future as
\chX{the process technology node size shrinks}.
\chIX{We conclude that the 96.7\% reduction in the read disturb effect
we observe in 3D NAND flash memory compared to planar NAND flash memory is
mainly caused by the difference in manufacturing process technology \chX{nodes}
\chXI{of the two types of NAND flash memories}.}


\subsection{Layer-to-Layer Process Variation}
\label{sec:3derror:appendix:variation}

In this \chX{section}, we present \chX{new results and analyses} of \chXI{the layer-to-layer}
process variation \chXI{phenomenon} in
3D NAND \chX{flash memory,} in addition to the key findings 
\chX{we already presented} in Section~\ref{sec:3derror:variation}. We use
the same methodology as \chIX{we describe} in Section~\ref{sec:3derror:variation}.

\chIX{Figure~\ref{fig:variation-wlmean-var} shows how the}
threshold voltage distribution mean and standard deviation of each state
\chIX{changes with layer number}, for a flash block \chX{that has endured} 10K
P/E cycles. \chIX{Each subfigure in the top row \chX{shows} the mean for a
different state; each subfigure in the bottom row \chX{shows} the standard
deviation for a different state.} We make two observations from \chIX{this
figure}.
First, the ER state \chIX{threshold voltage} \chIX{increases} by as much as
25 voltage steps \chIX{as \chX{the layer number changes}}, while the \chIX{mean
threshold voltages} of the other three states do not \chIX{vary} by much.
\chX{This is because \chXI{the threshold voltage of a cell in ER~state
is} set after
an erase \chXII{operation, and the value it is set to} is a function of
manufacturing process variation and of wearout.
In contrast, \chXI{the threshold voltage of} a cell in one of the other states
(P1, P2, or \chXI{P3) is} set to a \chXII{\emph{fixed}} target voltage value \chXII{\emph{regardless}} of
process variation~\cite{mielke.irps08, bez.procieee03, suh.jssc95, wang.ics14} (see
Section~\ref{sec:flash}).}
\chX{Since only the voltage of the ER~state is affected by layer-to-layer
process variation, only one of the read reference voltages, $V_a$, changes
with the layer number, \chXI{as we already observed} in
Figure~\ref{fig:variation-wloptvrefs}.}
\chIX{Second}, the distribution \chIX{widths of ER and P1 states} \chIX{(i.e.,
\chX{their standard
deviations}) increase} in the \chXI{top layers}, and decrease in the
bottom \chXI{layers}. \chIX{This pattern is similar to the pattern of how \chX{the}
RBER changes
with layer number, which we show in Figure~\ref{fig:variation-wlopterr}
(Section~\ref{sec:3derror:variation}).
\chX{A} wider threshold voltage distribution increases the overlap of
neighboring distributions, leading to more errors \chX{in the top layer}.} However, the
distribution widths of the P2 and P3 states \chXI{mainly} \chIX
{decrease as layer number increases.} \chIX{Unfortunately, we are unable to
\chX{completely} explain why mean threshold voltage and distribution width change differently
with layer number \chXI{for different states} because we do not have \chX
{exact} circuit-level information about
layer-to-layer process variation.}

\begin{figure*}[h]
\centering
\includegraphics[trim=10 10 10 10,clip,width=\linewidth]
{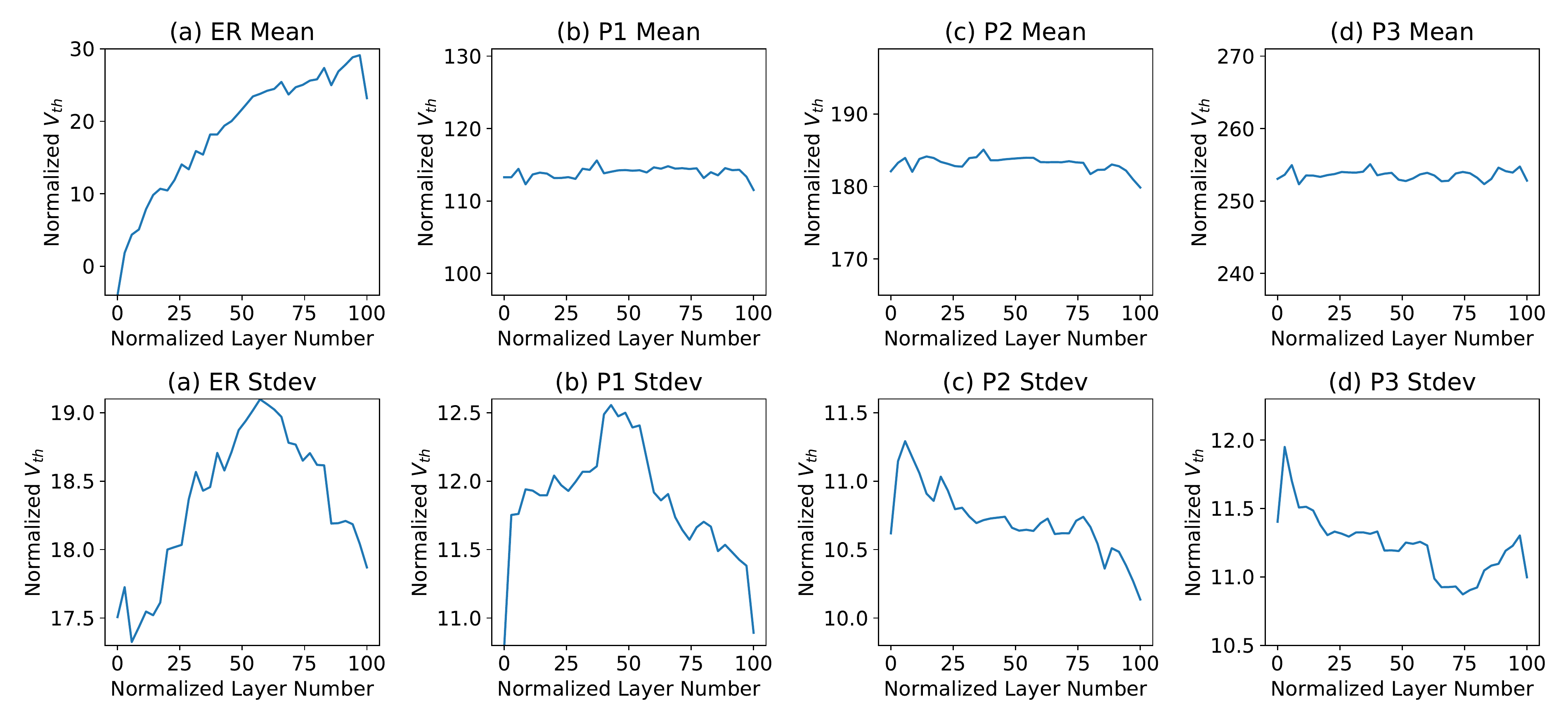}
\caption{\chIX{Mean and standard deviation of our Gaussian threshold voltage
distribution model of each state, versus layer number.}}
\label{fig:variation-wlmean-var}
\end{figure*}


\chIX{We conclude that layer-to-layer process
variation significantly impacts the threshold voltage distribution and leads
to large variations in RBER and optimal read reference voltages across
layers.}

\subsection{Bitline-to-Bitline Process Variation}
\label{sec:3derror:bitline-variation}

We perform \chX{an analysis} \chXI{of} the variation of RBER and threshold
voltage distribution along the y-axis (i.e., across groups of bitlines) for a
flash block \chXI{that has endured} 10K P/E cycles.  \chX{We use a similar
methodology
to our layer-to-layer process variation experiments  (see 
Section~\ref{sec:3derror:variation}).}

\chIX{Figure}~\ref{fig:variation-blmean-var} shows how the
threshold voltage distribution mean and standard deviation of each state
changes with layer number, for a flash block \chX{that has endured} 10K
P/E cycles. Each subfigure in the top row \chX{shows} the mean for a
different state; each subfigure in the bottom row \chX{shows} the standard
deviation for a different state. Note that we normalize the number of bitlines
from 0 to 100, by multiplying the actual bitline number
with a constant, to maintain the anonymity of the chip vendors. We
make two observations from this figure.  First, the variations in mean
threshold voltage and the distribution width (i.e., standard deviation) are
much smaller in this figure compared to that observed in
Figure~\ref{fig:variation-wlmean-var} \chX{for layer-to-layer variation} (Appendix~\ref{sec:3derror:appendix:variation}).
This indicates that bitline-to-bitline process variation is much smaller
compared to layer-to-layer process variation in 3D NAND flash memory. Second,
\chX{we observe that the pattern of the mean threshold voltage repeats
periodically, for every 25 bitlines.  We believe that this indicates a
repetitive architecture in the way that the 3D NAND flash memory chip is
organized (for example, each block may be made up of four arrays of
flash cells that \chXII{are connected together}).  Unfortunately, we cannot
completely explain this behavior without access to circuit-level design
information that is proprietary to NAND flash memory vendors.}


\begin{figure*}[h]
\centering
\includegraphics[trim=10 10 10 10,clip,width=\linewidth]
{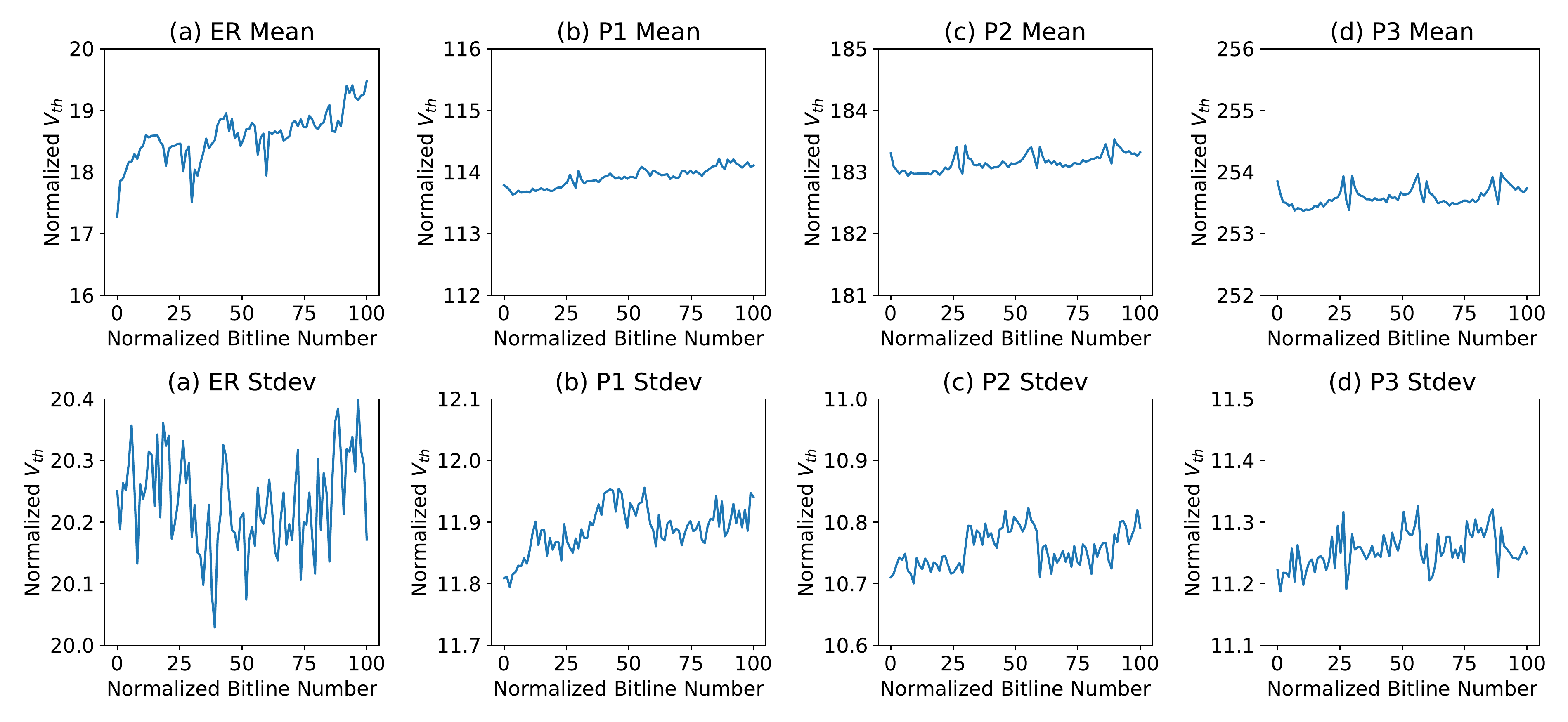}
\caption{\chXI{Mean and standard deviation of our Gaussian threshold voltage distribution model of each state,
versus bitline number.}}
\label{fig:variation-blmean-var}
\end{figure*}

\chIX{\chX{Figures}~\ref{fig:variation-blopterr} and \ref{fig:variation-bloptvrefs} show
how \chX{the RBER and} optimal read reference voltages change with bitline number,
for a flash block \chXI{that has endured} 10K P/E cycles. We observe that
neither RBER nor the
optimal read reference voltages change by much across bitlines. This indicates
that the \chX{changes that} we observe in Figure~\ref{fig:variation-blmean-var} may not be
significant enough to \chX{lead to variation in the reliability of different} bitlines. We conclude that
bitline-to-bitline process variation is much smaller than layer-to-layer
process variation in 3D NAND flash memory.}

\begin{figure}[h]
\centering
\includegraphics[trim=10 10 10 10,clip,width=\figscale\linewidth]
{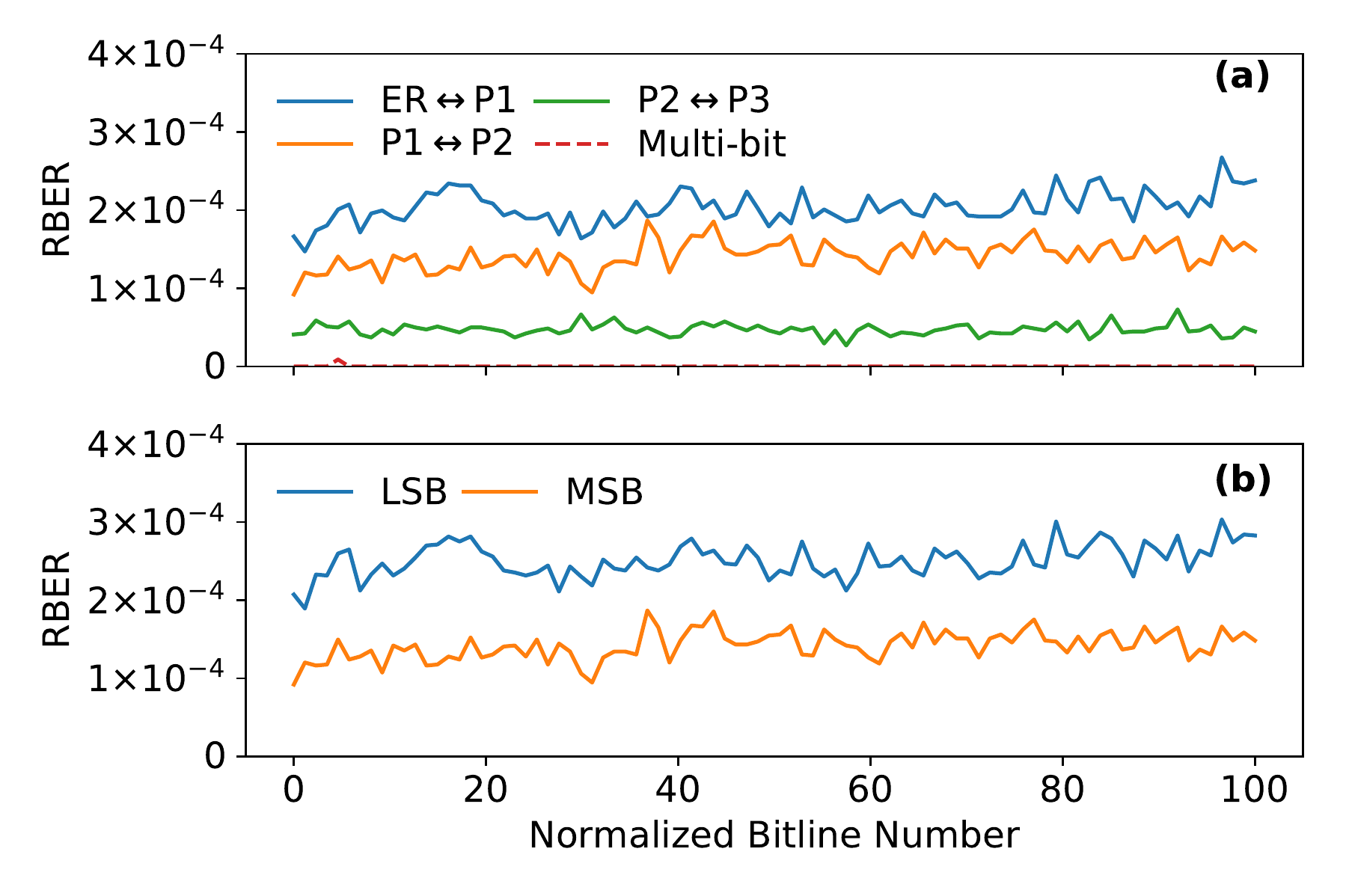}
\caption{\chXI{RBER vs. bitline number, broken down by (a)~the state transition of
each flash cell, and (b)~MSB or LSB page.}}
\label{fig:variation-blopterr}
\end{figure}

\begin{figure}[h]
\centering
\iftwocolumn
  \includegraphics[trim=10 10 10 10,clip,width=\linewidth]
  {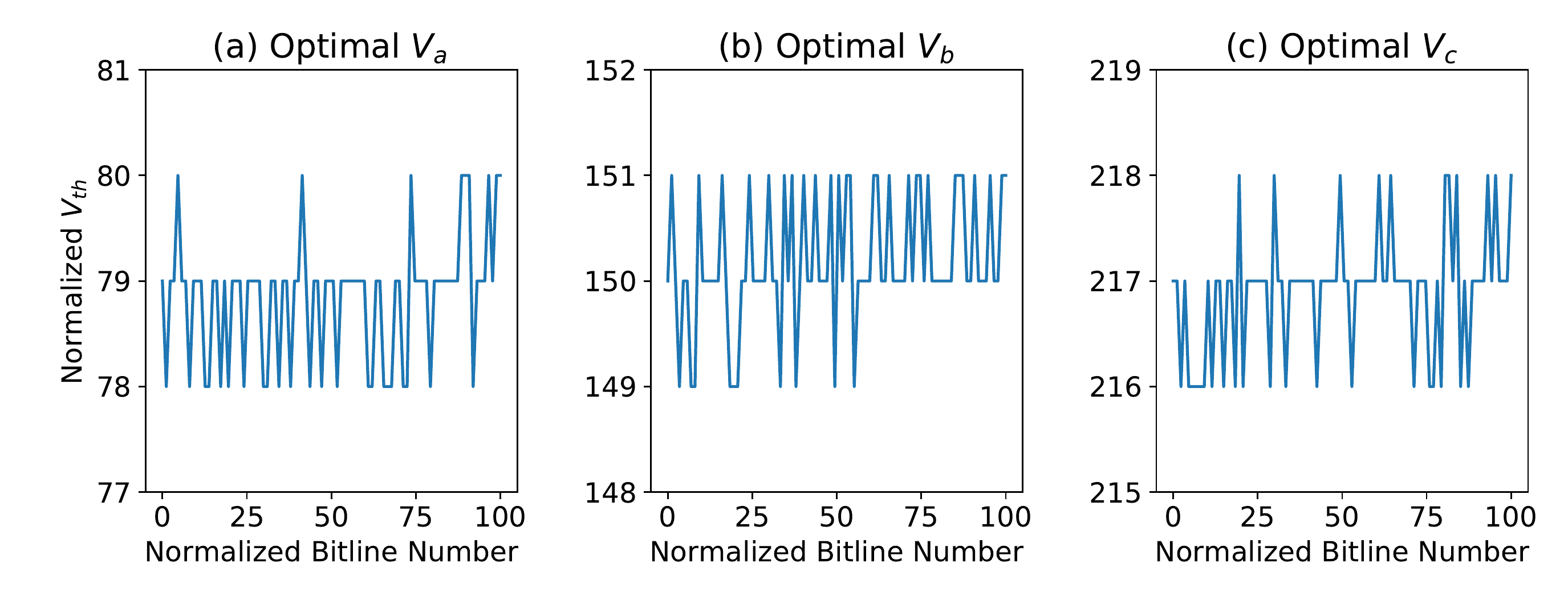}
\else
  \includegraphics[trim=10 10 10 10,clip,width=.8\linewidth]
  {figs/variation-block0-10kPEC-bloptvrefs.pdf}
\fi
\caption{\chXI{Optimal read reference voltages vs. bitline number.}}
\label{fig:variation-bloptvrefs}
\end{figure}





\chapter{HeatWatch: Self-Recovery and Temperature Aware Retention Error Mitigation}
\label{sec:heatwatch}


As we have shown in Chapter~\ref{sec:3derror}, retention errors dominate
3D NAND flash memory errors. Thus, mitigating retention errors in 3D NAND can
yield significant flash reliability improvements. We find promising opportunity
to mitigate flash retention by exploiting \emph{self-recovery}, which we have
introduced in Section~\ref{sec:background:recovery}. \chIII{This} opportunity
has \chIII{largely been} ignored in recent literature, as \chIII{the} self-recovery effect is not
well understood, especially in 3D NAND flash memory.

In this chapter, we \chIII{exploit} the 3D NAND flash
memory's device-level self-recovery behavior to improve flash reliability. The
goal of \chIII{this chapter} is to (1)~perform a detailed experimental characterization of
the self-recovery effect using real, state-of-the-art 3D NAND flash memory
devices, and (2)~exploit our findings by developing \chIII{HeatWatch, a new mechanism} to help
self-recovery in 3D NAND flash \chIII{memory}. The key idea of HeatWatch is to \emph{adapt the read
reference voltage to the dwell time \chIII{(i.e., the idle time between
consecutive program cycles)} and retention time of the workload as well
as the SSD's operating temperature}.

First, we extensively characterize how self-recovery and temperature affects 3D
NAND flash reliability using real, state-of-the-art MLC 3D NAND flash memory
chips (Section~\ref{sec:heatwatch:characterization}). Second, based on the
findings of our characterization, we propose \emph{URT}, a new unified model of
self-recovery, temperature, retention loss, and wearout for 3D NAND flash
memory (Section~\ref{sec:heatwatch:modeling}). Third, \chIII{using} URT, we propose
and
evaluate \emph{HeatWatch}, a mechanism to improve flash reliability and
lifetime by optimizing read operations in 3D NAND flash memory for the amount
of self-recovery allowed by the workload and the operating temperature of the
flash memory (Section~\ref{sec:heatwatch:mechanism}).

\section{Characterizing the Self-Recovery Effect}
\label{sec:heatwatch:characterization}


\label{sec:heatwatch:characterization:goal}

To understand the behavior of the self-recovery effect in 3D NAND flash memory,
we perform an extensive characterization of the effect using \emph{real},
state-of-the-art 3D NAND flash memory chips.
Our goal in this characterization is to answer the following research questions:

\begin{itemize}[topsep=0pt,partopsep=0pt,noitemsep,leftmargin=10pt]

    \item How does the dwell time affect retention and program variation errors?

    \item What is the correlation between dwell time and the magnitude of the
    self-recovery effect?


    \item How does the operating temperature \chI{affect} the number of retention and
    program variation errors?



    \item \chI{How do the benefits of self-recovery change based on the number
    of \chI{performed \emph{recovery cycles}?}}

\end{itemize}
\chI{We make all \chI{of} our characterization data, including results not shown in this
\chV{chapter} for brevity, available in an extended report~\cite{luo.tr18} and 
online~\cite{heatwatch.github}.}


We use the observations and analysis from our characterization to drive the
design of a new \chI{model \chI{of 3D NAND flash memory reliability,}}
\chI{as described} in Section~\ref{sec:heatwatch:modeling}.

\subsection{Characterization Methodology}
\label{sec:heatwatch:characterization:methodology}

To answer \chI{the above} research questions, we design new
experiments to test how flash reliability changes with different dwell
times and temperatures. In these experiments, we use \chI{state-of-the-art
30- to 40-layer 3D charge trap NAND flash \chI{chips from} a major vendor}.\footnote{\label{fnt2}We \chI{do not disclose} the
exact number of stacked layers in the chips, \chI{to protect
the anonymity of the flash vendor, and we cannot disclose the \chI{\emph{exact}} voltage values, 
as this is proprietary information.}} We use a NAND flash
\chI{memory} characterization
platform that fits in the SSD form factor, and contains \chI{a} \chI{special} version of
\chI{the} firmware in the SSD
controller. We use a server machine to issue remote procedure calls
(RPC)~\cite{birrell.tocs84} to the
firmware over a Serial ATA (SATA)~\cite{sata.3.3.spec} connection. These \chI{RPCs} trigger
commands to be sent directly to the flash chips, and can transfer the raw data 
(i.e., data with raw bit errors) directly from the flash chips to the server 
\emph{without} applying \chI{error correction (ECC)}.  This allows us to
observe \chI{the effect} of dwell time
and operating temperature on the \chI{raw bit error rate of each flash chip}.

We use two metrics to evaluate flash reliability. First, we
measure the \emph{raw bit error rate} (RBER), which is the rate at which errors
occur in the data \emph{before error correction}. 
To calculate RBER, we read data from a NAND flash memory chip using the default 
read reference voltage, and
compare the data using a pristine server-side copy of the data that was 
originally written to the chip.
Second, we show the statistical mean of the threshold
voltage distribution of \chI{each} \chI{high-voltage} state (i.e., P1, P2, P3). As we
mention in Section~\ref{sec:errors}, 
retention loss and program variation cause the threshold \chI{voltages of
cells} to shift,
which leads to the raw bit errors.\footnote{We are unable to show
the full threshold voltage distribution for the ER state, because \chI{the} ER state
threshold voltages are negative, and our platform
cannot \chI{measure} negative voltage values. \chI{This is similar to prior
works~\cite{cai.hpca15, cai.dsn15, cai.procieee17, cai.date13}.}} To obtain the
threshold voltage
distribution of a flash page, we perform multiple read operations to sweep the 
range of all positive read reference voltages, using the \emph{read-retry} 
command \chI{on} the NAND flash memory
chip~\cite{cai.hpca15, cai.date13, fukami.di17}.\footnote{Due to space
limitations, we refer the reader
to prior works~\cite{luo.jsac16, parnell.globecom14}
for a detailed methodology on how to obtain the threshold voltage
distribution.} Read-retry
allows us to fine-tune the read reference voltage used for each read operation.
The smallest amount by which we can change the read reference voltage is called
a \emph{voltage step}.
\chI{We normalize each threshold voltage \chI{value} to the number of voltage steps needed
to reach that particular voltage \chI{value}.}$^{\ref{fnt2}}$

\chI{%
\paratitle{Limitations} In our experiments, we characterize 
\chI{3D NAND flash memory chips of the same model} from one major vendor.
Our approach ensures that any variation that we observe in
our characterization is the result of only manufacturing process variation, and
not a result of differences in flash chip architecture or different manufacturing
techniques used by different vendors.  While we do not take vendor-related
variation into account, we believe that our general \chI{qualitative findings on
\chI{the effects of self-recovery and temperature}}
can be generalized to 3D charge trap NAND flash memory manufactured by
other vendors. This is because the underlying structure of \chI{3D charge trap
cells} (see Section~\ref{sec:background:3d}) is similar across 
different vendors~\cite{park.jssc15,mizoguchi.imw17,choi.vlsit16}.
Thus, while the exact numbers reported in this \chIII{chapter} may differ from vendor to
vendor, our \chI{qualitative} findings\chI{, which are a result of charge
detrapping from the tunnel oxide (see Section~\ref{sec:background:recovery}),}
should be similar across vendors.


We are unable to perform repeated runs of our test procedures on the \chI{\emph{same}} block,
as each run of a test procedure induces further wearout on a block.  The amount
of wearout affects the error rate of NAND flash
memory~\cite{cai.procieee17, cai.arxiv17, mielke.irps08, parnell.globecom14,
cai.date13, luo.jsac16}.
\chI{To ensure an accurate comparison between multiple test procedure runs,
we use eight \emph{target} wordlines in the same stack layer from eight
randomly-selected flash blocks that are set to the \emph{same} level of wearout
for the same test procedure.  By selecting wordlines in the same layer, we 
eliminate the potential impact of cross-layer process variation.}
Note that we do not characterize \emph{chip-to-chip}
process variation, as an accurate study of such variation requires a large-scale
study of \chI{a large number (e.g., hundreds)} of 3D NAND flash memory chips, which
we do not have access to.
Hence, we leave such a \chI{large-scale} study for future work.}

\subsection{\chI{Characterizing the} Dwell Time Effect}
\label{sec:heatwatch:characterization:dwell}

To measure the \chI{effect} of dwell time on flash reliability (see Section~\ref{sec:background:recovery}), we
characterize the RBER and the threshold voltage distribution.
\chI{We wear out each of our target blocks} by repeatedly writing pseudorandom data
until the block reaches 3,000 P/E cycles. For the last 300 P/E cycles, 
we use a different dwell time for each block, spanning a range of \SIrange{64}{8192}{\second}.
Prior work \chI{shows} that the magnitude of the self-recovery effect is 
correlated with the dwell time for only the last 10\% of P/E cycles performed
on a block~\cite{mielke.irps06}.  We show in Section~\ref{sec:heatwatch:characterization:cycle}
that the dwell time used during
only the \emph{last 20 P/E cycles} affects self-recovery.

\chI{We measure how the dwell time affects retention loss speed and
program variation,
by performing a \emph{retention test} on each target wordline
\emph{immediately} after the block containing the wordline
reaches 3,000 P/E cycles.}
In this test,
we program
pseudorandom data to the target wordline, and repeatedly \chI{measure} the
threshold voltage distribution using the methodology described in
Section~\ref{sec:heatwatch:characterization:methodology} for up to
\SI{71}{\minute} (i.e., \SI{4260}{\second}) after the data was written.
We conduct this experiment at an environmental temperature of \SI{70}{\celsius},
\chI{which} accelerates \chI{both} self-recovery and retention loss to
reduce the
experiment duration to a reasonable length~\cite{mielke.irps06}.\footnote{\chI{Based
on Arrhenius' Law~\cite{arrhenius.zpc1889},} the same
experiment would take more than 11 years to finish had we performed it at room
temperature (\SI{20}{\celsius}).} We later repeat a small portion of the test
under room temperature (\SI{20}{\celsius}), and verify
that all of our observed trends remain the same.

\paratitle{\chI{Effect on RBER}}
\chI{First, we study how self-recovery affects
the raw bit error rate.}
Figure~\ref{fig:dwell-vs-error} \chI{shows the} \chI{RBER} \chI{as retention} time 
($t_s$) increases, \chI{for different} dwell times ($t_d$) used for the last \chI{300} P/E cycles. 
We use a color gradient for the curves, where the reddest (topmost) curve
has the shortest dwell time, and the \chI{blackest} (bottommost) curve has the longest
dwell time.
Note that the x-axis and y-axis are both in log scale.

\begin{figure}[h]
\centering
\ifeps
\includegraphics[trim=10 10 10 10,clip,width=.7\linewidth]
{figs/myerror-dwell-hynixDwellTest2-Temp-70.eps}
\else
\includegraphics[trim=10 10 10 10,clip,width=.7\linewidth]
{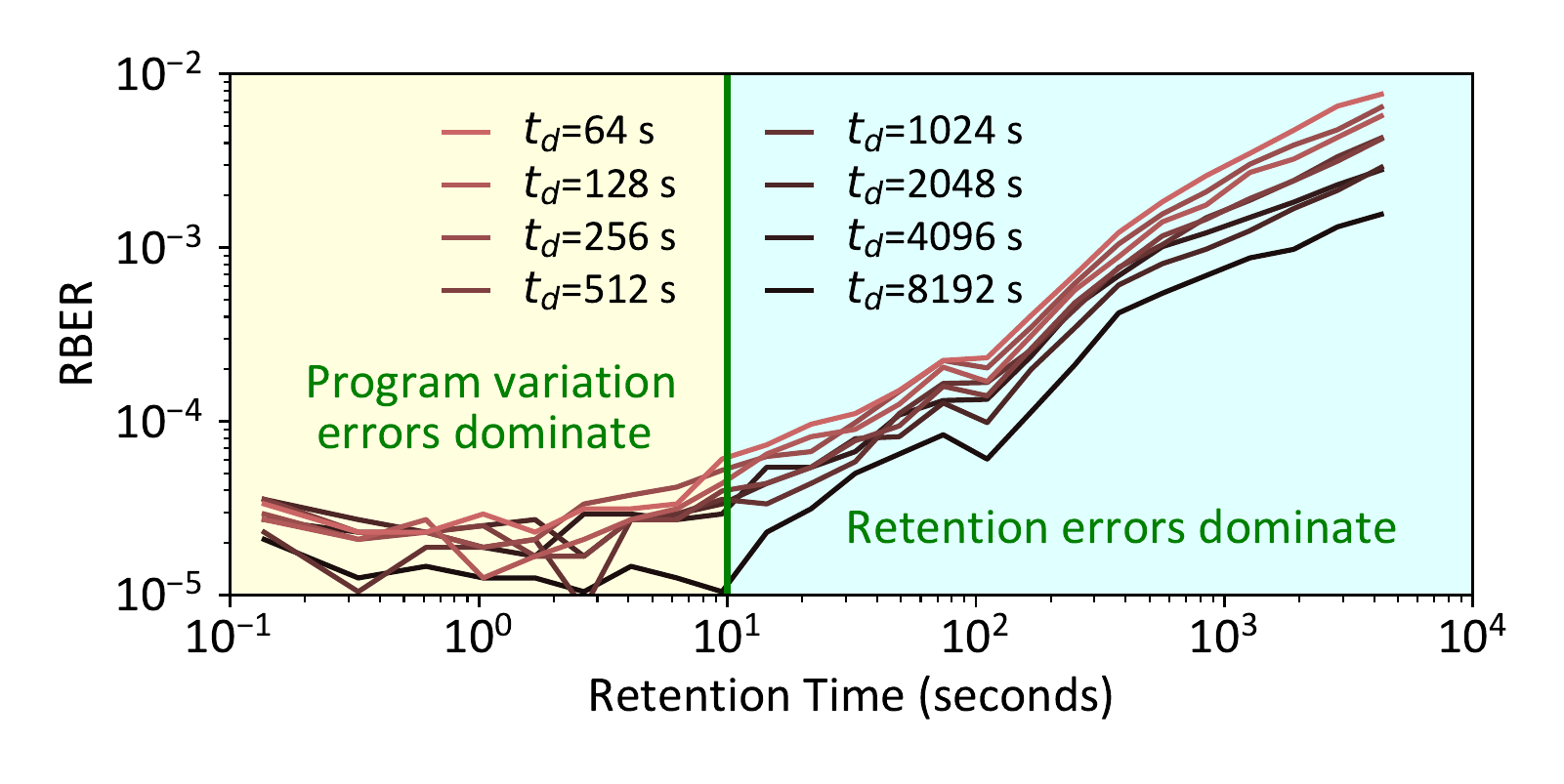}
\fi
\caption{Change in RBER over retention time for flash pages that were programmed using different dwell
times.}
\label{fig:dwell-vs-error}
\end{figure}

\chI{We make two observations from Figure~\ref{fig:dwell-vs-error}.} First, when the retention time is short (i.e.,
$t_r <$~\SI{10}{\second}),
the RBER is \emph{similar} across different dwell times.
\chI{During this} time, \chI{the RBER} is
dominated by program variation \chI{errors~\cite{cai.date12, mielke.irps08, luo.jsac16}}.
Recall that a longer dwell time \chI{increases} the amount of detrapped charge
during self-recovery.
However, since \chI{the RBER} is similar \chI{across different curves} regardless of the dwell time,
this means that self-recovery does \emph{not} mitigate program variation errors.
Therefore, unlike previous \chI{works}~\cite{mohan.hotstorage10, chen.codes13, wu.hotstorage11},
we conclude that self-recovery does \emph{not} repair \emph{all} of the errors that
occur due to wearout in 3D NAND flash memory.
Second, when the retention time is large (i.e., $t_r >$~\SI{10}{\second}), there is a strong
correlation between a longer dwell time and a lower RBER. 
\chI{During this time}, the RBER is dominated by retention errors
(hence the growth in RBER as the retention time increases).
\chI{We conclude that a longer dwell time after an erase operation
\chI{mitigates} retention errors, but not program variation errors, in 3D NAND flash 
memory.}

\paratitle{\chI{Effect on Threshold Voltage}}
Next, we study the threshold voltage distribution of the flash pages under test, to
understand how self-recovery affects the threshold voltage shift (and thus the
\chI{RBER}) due to retention loss.
Figure~\ref{fig:real-shifted-distribution} shows the threshold
voltage distribution \chI{\emph{before}} (black dots, $t_r =$~\SI{1}{\minute}) and \chI{\emph{after}} (red dots,
$t_r =$~\SI{71}{\minute}) a large retention time elapses, for a flash page 
programmed using a \SI{64}{\second} dwell time (top plot), and for a flash
page programmed using a \SI{8192}{\second} dwell time (bottom plot).
We observe from the figure that when the dwell time is higher,
the threshold voltage distribution shift due to retention loss is significantly
smaller.

\begin{figure}[h]
\centering
\ifeps
\includegraphics[trim=10 10 10 10,clip,width=.7\linewidth]
{figs/distribution-comparison.eps}
\else
\includegraphics[trim=10 10 10 10,clip,width=.7\linewidth]
{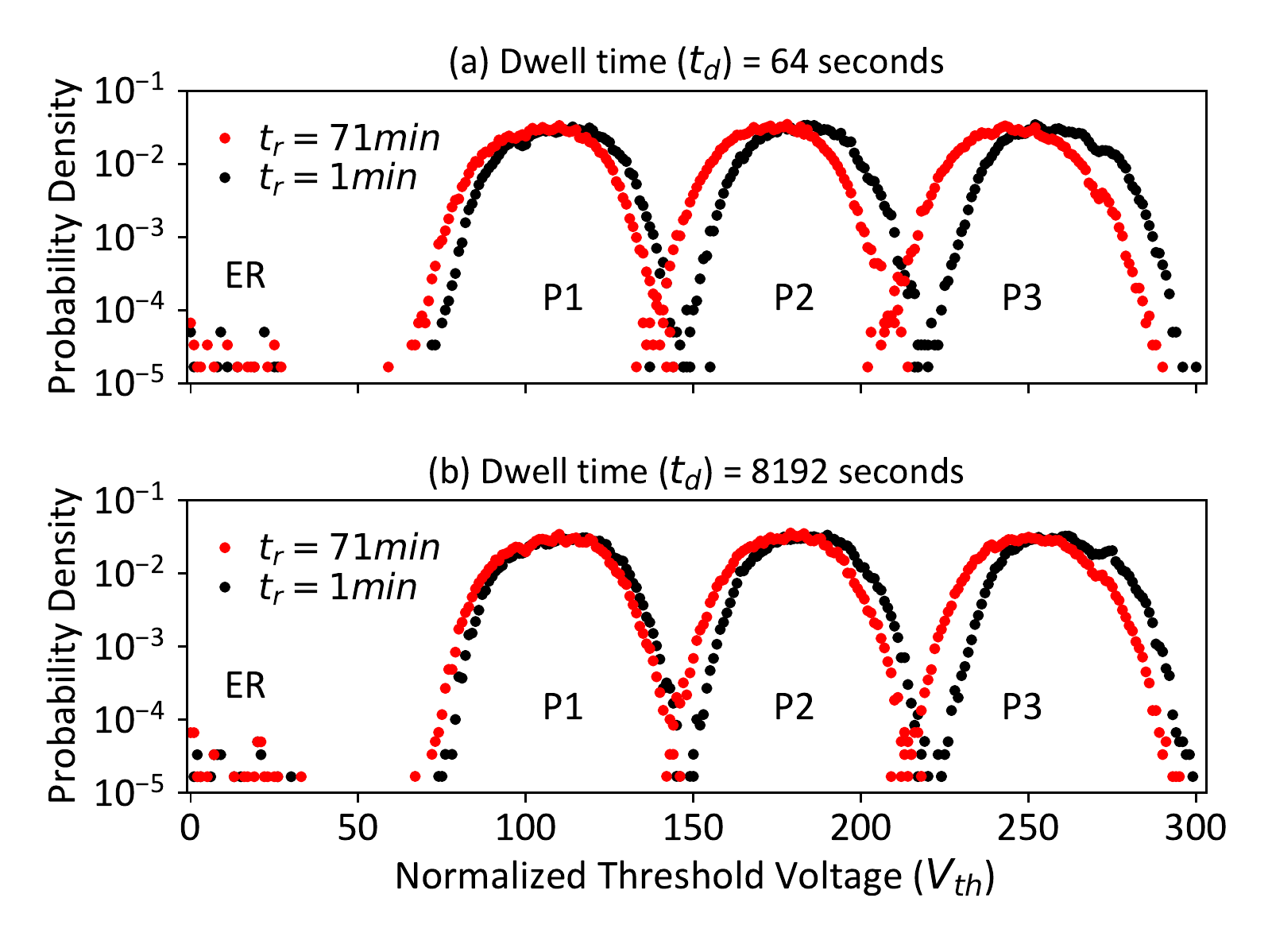}
\fi
\caption{Threshold voltage distribution before and after a long retention time,
\chI{for different dwell times.}}
\label{fig:real-shifted-distribution}
\end{figure}


To quantify \chI{the threshold voltage shift} as a function of the dwell time,
we plot the statistical means of the
threshold \chI{voltage distributions} of cells programmed to the P1, P2, and P3 states in
Figure~\ref{fig:dwell-vs-mean}. We use the same color gradient that we 
used in Figure~\ref{fig:dwell-vs-error} to represent the different dwell times. 
Note that for these experiments, the smallest retention time that we show on
the x-axis ($t_r = $~\SI{64}{\second}) is much larger than the smallest
retention time shown in Figure~\ref{fig:dwell-vs-error} ($t_r = $~\SI{0.5}{\second}),
because it takes significantly longer for us to sweep the threshold voltage of the
cells in a wordline \chI{(as in Figure~\ref{fig:dwell-vs-mean})}, compared to simply measuring the RBER
of the wordline \chI{(as in Figure~\ref{fig:dwell-vs-error})}.
We make two observations from the figure.
First, for all three states, the mean threshold voltage
changes by a smaller amount when the dwell time is higher, corroborating
the threshold voltage distribution shifts shown in Figure~\ref{fig:real-shifted-distribution}.
Second, for a fixed dwell time, the change in voltage is linearly related with
the log of retention time.\chI{\footnote{\chI{A similar \chI{linear} relationship 
\chI{between the change in threshold voltage and the log of the retention time} is observed 
for planar NAND flash memory in past works~\cite{jesd218.jedec10,
mielke.irps06}}.}}
We use this relationship to develop a \chI{simple model that} can quantify how
retention loss speed changes with dwell time \chI{(see below)}.

\begin{figure}[h]
\centering
\ifeps
\includegraphics[trim=10 10 10 10,clip,width=.7\linewidth]
{figs/mean-dwell-hynixDwellTest2-Temp-70.eps}
\else
\includegraphics[trim=10 10 10 10,clip,width=.7\linewidth]
{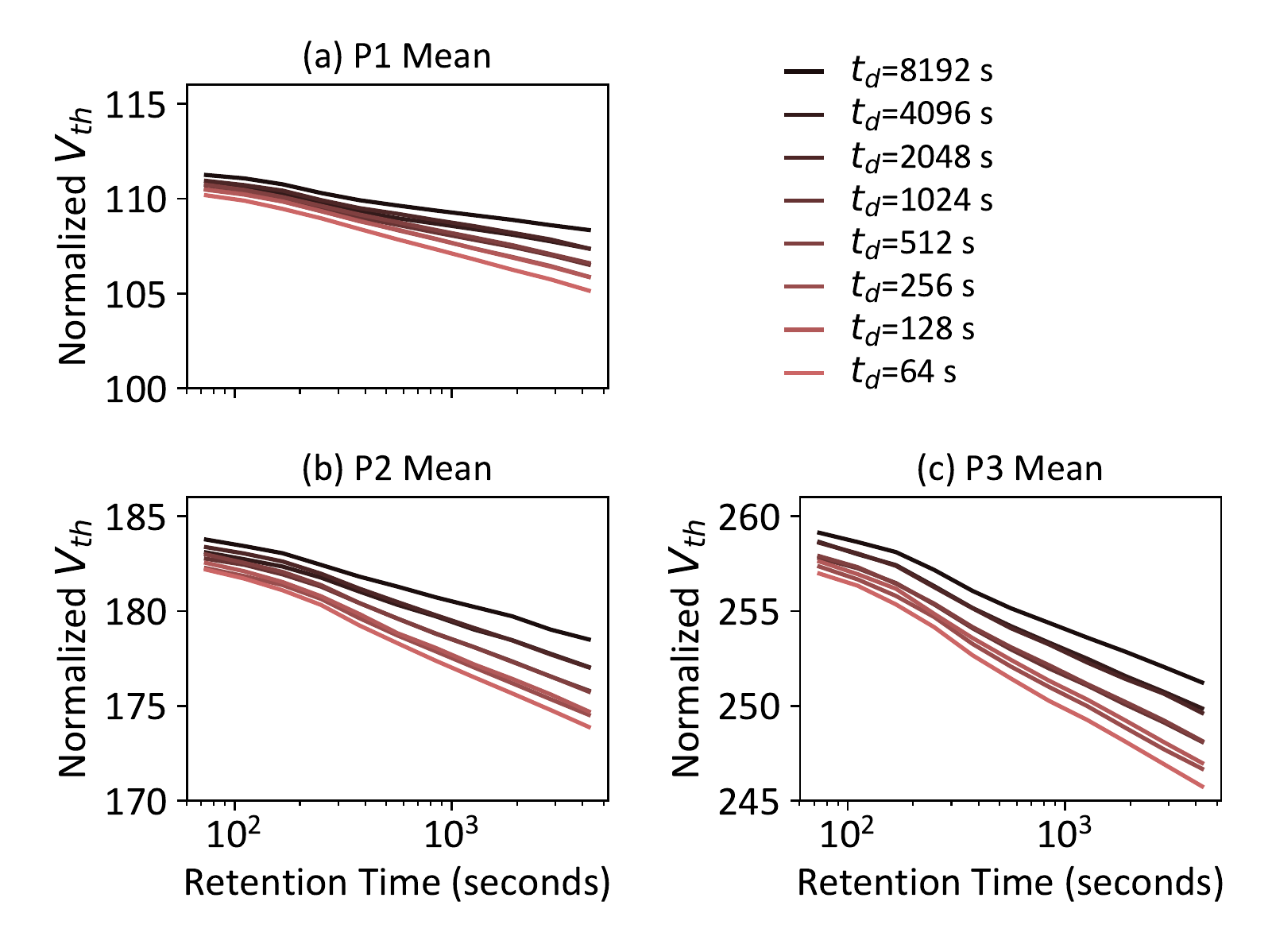}
\fi
\caption{Threshold voltage distribution mean \chI{vs.}\ retention time for 
different dwell times.}
\label{fig:dwell-vs-mean}
\end{figure}

We also calculate how the width of the distribution changes due to retention loss
for different dwell times (not plotted).  We observe that the change in the 
distribution width is relatively small, and thus choose to ignore it to simplify
\chI{the analysis.}


\paratitle{\chI{Effect on Retention Loss Speed and Program Variation}}
To quantify how \chI{self-recovery} changes 
(1)~retention loss speed and
(2)~program variation,
we construct a simple model of how the threshold voltage \chI{and RBER change} due to 
these two factors.
\chI{As we observe in Figure~\ref{fig:dwell-vs-mean}, the threshold voltage
distribution mean is linearly correlated with the logarithm of the
retention time ($t_r$).  Thus,}
we fit our measurements to the following linear model, \chI{for a given
dwell time}:
\begin{align}
    \chI{Y(t_r)} = \alpha \cdot \ln(t_r) + \beta
\label{eqn:simple-retention}
\end{align}
In this model, 
\chI{$Y$ can represent either
(1)~the mean of the threshold voltage distribution of each high-voltage state
(i.e., P1/P2/P3); or
(2)~the logarithm of the RBER, i.e., $\log(RBER)$;\chI{\footnote{\chI{We 
model the \emph{logarithm} of the RBER, because \chI{when retention loss
linearly shifts the threshold
voltage distribution, which roughly follows a Gaussian distribution~\cite{luo.jsac16}, 
the linear distribution shift results} in
a logarithmic change in \chI{RBER, which is quantified as}
the overlapping area between two neighboring distributions.}}}
based on the values chosen for $\alpha$ and $\beta$.
$\alpha$ represents the retention loss speed.
\chI{$\beta$ represents the \emph{program
offset}, which is the initial value of $Y$
immediately after programming.}}

\chI{We use the \emph{absolute value} of the program offset 
(i.e., $|\beta|$) to quantify the impact of program variation.
For the threshold
voltage distribution mean of each high-voltage state, $Y$ and $\beta$ are positive, and
a \emph{more positive} program offset results in a \emph{higher} initial mean.
\chI{As we observe under \emph{Effect on Threshold Voltage} in 
Section~\ref{sec:heatwatch:characterization:temperature}, when the mean voltages of
neighboring distributions increase, the overlap between the distributions
decreases,}
which in turn reduces the 
number of program variation errors.
For $\log(RBER)$, $Y$ and $\beta$ are negative, because the RBER is always less
than 1. A \emph{more negative} program offset (i.e., a greater $|\beta|$) 
corresponds to a \emph{more negative} initial value of $\log(RBER)$
(i.e., \emph{fewer} errors).}
\chI{For each dwell time, we fit Equation~\ref{eqn:simple-retention} to our
experimental characterization data in order to calculate the values of 
$\alpha$ and $|\beta|$ when $Y$ represents (1)~the mean voltage of each
higher-voltage state, or (2)~$\log(RBER)$.}
Figure~\ref{fig:normalized-dwell} (left) illustrates the relation between
dwell time and retention loss speed ($\alpha$), normalized to the greatest observed
retention loss speed.
Figure~\ref{fig:normalized-dwell} (right) illustrates the relation between 
dwell time and program \chI{offset ($|\beta|$),
normalized} to the \chI{\chI{greatest} observed program offset}.
Note that the x-axis (i.e., the dwell time) is in log scale.
In both figures, the markers represent our measured data points from real NAND flash
memory chips, and the dashed lines show a linear trend line for the \chI{data.}


\begin{figure}[h]
\centering
\ifeps
\includegraphics[trim=10 10 10 10,clip,width=.7\linewidth]
{figs/normalized-dwell.eps}
\else
\includegraphics[trim=10 10 10 10,clip,width=.7\linewidth]
{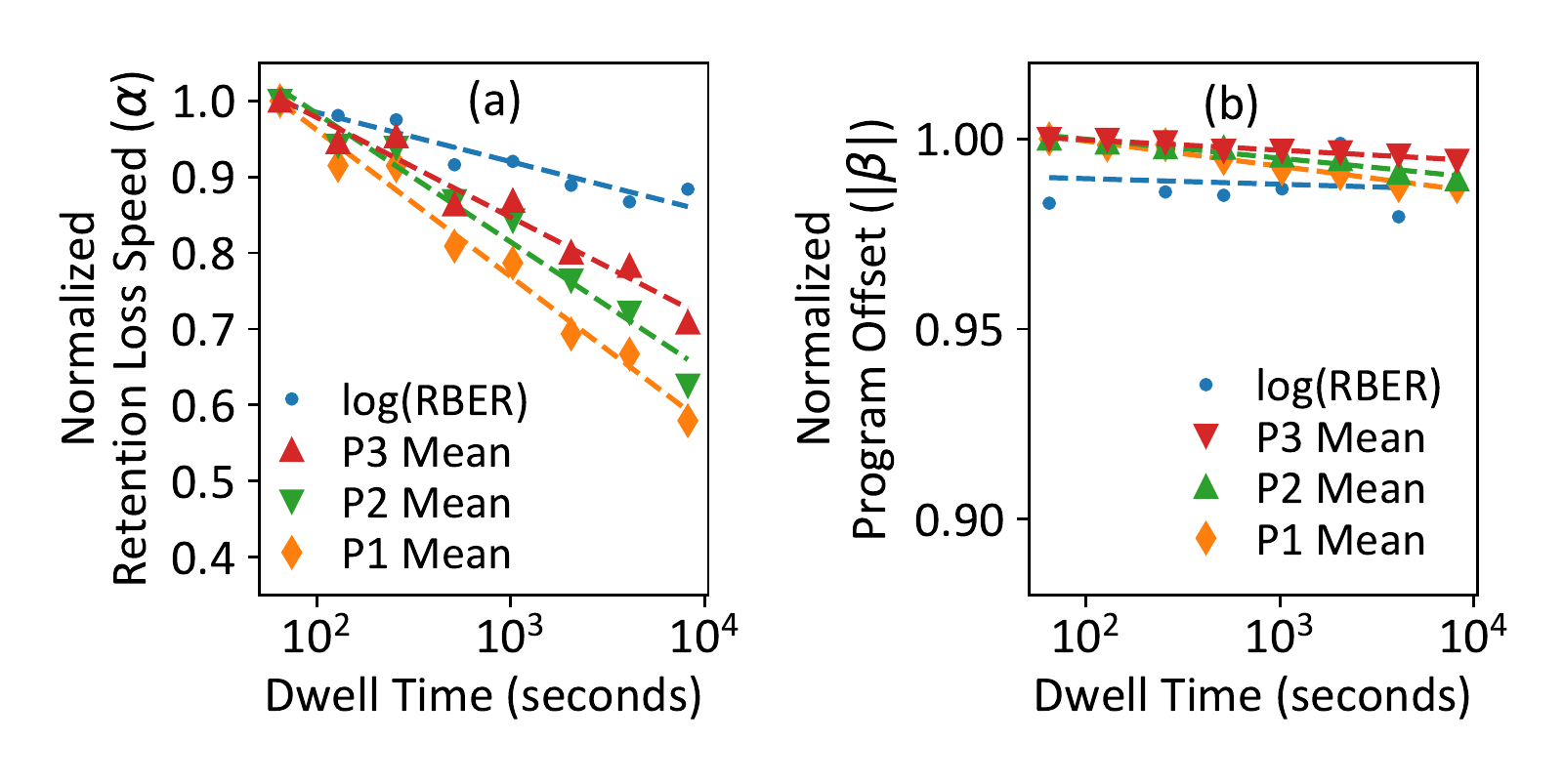}
\fi
\caption{Retention loss speed (left) and program \chI{offset} \chI{(right), for}
different dwell \chI{times}.}
\label{fig:normalized-dwell}
\end{figure}

We make two key \chI{observations from \chI{Figure~\ref{fig:normalized-dwell}}.} 
First, the self-recovery
effect reduces \chI{the} retention loss speed linearly with the logarithm of dwell time.
We observe, however, that the change in retention loss speed is \emph{different for
each state}.
As Figure~\ref{fig:normalized-dwell}~(left) shows, our data points follow the
linear trend line closely (with an $R^2$ value larger than 90\% for \chI{each state}
and for the RBER). 
Second, as we concluded from Figure~\ref{fig:dwell-vs-error},
self-recovery has little effect on program \chI{variation}
within the tested dwell time range. As
Figure~\ref{fig:normalized-dwell}~(right) shows, 
the maximum difference in program \chI{offset} for any of our threshold
voltage states is less than 2.1\%.
Note that our second finding is \emph{new}, and it shows that, unlike
previous \chI{findings} for planar NAND flash
memory~\cite{chen.codes13,wu.hotstorage11,wu.imw11,mohan.hotstorage10},
self-recovery does \emph{not} reduce the number of program variation \chI{errors,
and hence the amount of wear,}
in 3D NAND flash memory.

\subsection{\chI{Characterizing the} Temperature Effect}
\label{sec:heatwatch:characterization:temperature}

Next, we measure the \chI{effect} of temperature on self-recovery and flash
reliability (see Section~\ref{sec:background:recovery}). 
Similar to the
experiment in Section~\ref{sec:heatwatch:characterization:dwell}, we use eight target
wordlines in the same stack layer from randomly-selected flash blocks. First, for
each block, we wear out the block in 1,000 P/E cycle intervals up to a total of
10,000 P/E cycles, writing pseudorandom data and using a fixed dwell time of
\SI{0.5}{\second}.
We then put the test chip in a
temperature-controlled box, and set a target temperature. 
After the temperature of the test chip \chI{settles} to the target temperature,
we perform 20 more P/E cycles to each target wordline at the target temperature, 
using a \SI{0.5}{\second} dwell time.
Though these P/E cycles are performed at different temperatures for
each test, the dwell time we use is small, and thus we assume that the
difference between the equivalent dwell times at room temperature
are small across our tests.
Then, we perform the retention test described in
Section~\ref{sec:heatwatch:characterization:dwell} for all target wordlines up to
a retention time of \SI{71}{\minute}. We repeat the retention test under a range of target
temperatures in each round: 0, 10, 20, 28, 35, 50, 60, and \SI{70}{\celsius}.
During the retention test, data is both programmed and read under
the target temperature.

\paratitle{\chI{Effect on RBER}}
\chI{First, we study} how the RBER changes with retention time
under different temperatures, \chI{as shown in 
Figure~\ref{fig:temperature-vs-error}} for a wordline with 10,000 P/E cycles. Each curve
represents the RBER under a different programming temperature.
We use a color gradient to indicate the \chI{temperature:
the} reddest color represents the highest temperature
(\SI{70}{\celsius}) and \chI{the blackest} represents the lowest temperature
(\SI{0}{\celsius}).

\begin{figure}[h]
\centering
\ifeps
\includegraphics[trim=10 10 10 10,clip,width=.7\linewidth]
{figs/myerr-pec-10K.eps}
\else
\includegraphics[trim=10 10 10 10,clip,width=.7\linewidth]
{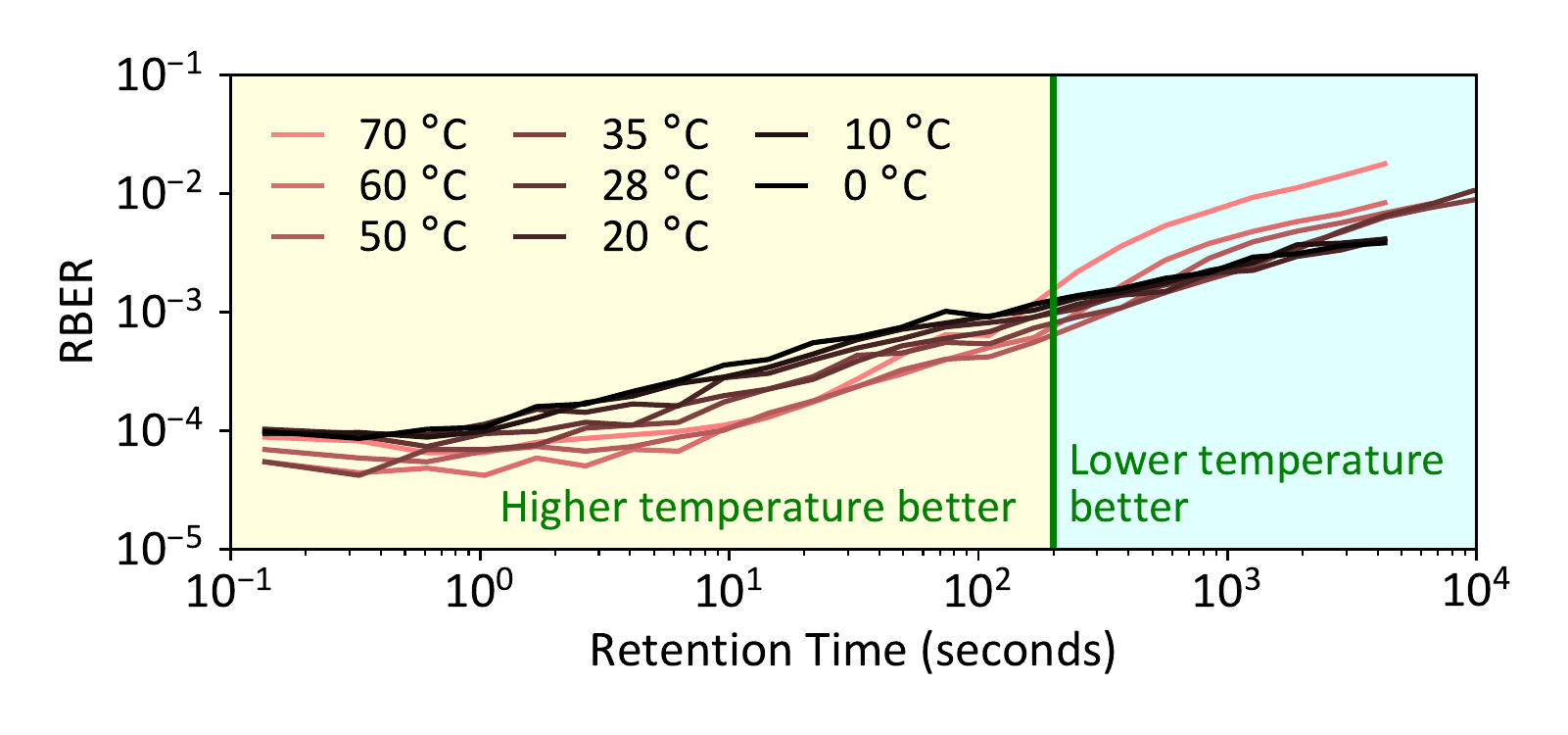}
\fi
\caption{RBER over retention time at 10,000 P/E cycles under different
programming temperatures.}
\label{fig:temperature-vs-error}
\end{figure}

We make two key observations from the figure. First, when the retention time is
small (\chI{i.e.,} \chI{$t_r < 2 \cdot 10^2$}), the RBER is lower for higher temperatures (i.e., the red
curves). 
\chI{Recall that when the retention time is small, the RBER is
dominated by program variation errors~\cite{cai.date12, mielke.irps08, luo.jsac16}.
Thus, we expect that the RBER decreases \chI{with higher temperatures} because
higher programming temperatures \emph{decrease} the number of program variation errors
(we discuss this in more detail under \emph{Effect on Threshold Voltage} below).}
Second, when the retention time
is larger (\chI{i.e.,} \chI{$t_r > 2 \cdot 10^2$}), the RBER becomes higher for higher programming \chI{temperatures}.
This is because 
\chI{as the temperature increases, the \chI{retention errors increase} at a faster rate.  Due to 
this faster growth,
the RBER for a higher temperature overtakes the RBER for a
lower temperature} at a retention time between \SIrange{e2}{e3}{\second}. This \chI{indicates} that
the threshold voltage shift due to high-temperature retention is
faster than that for low-temperature retention, which \chI{is in line} with Arrhenius'
Law~\cite{arrhenius.zpc1889} (see Section~\ref{sec:background:recovery}).

\paratitle{\chI{Effect on Threshold Voltage}}
\chI{Next, we study how the programming temperature affects the threshold 
voltage of a flash cell.  We begin by studying how the initial threshold voltage
(i.e., \chI{the threshold voltage} immediately after programming) changes
with temperature.
\chI{We} measure the threshold voltage distribution under
different programming \chI{temperatures,}
and then fit our data to Equation~\ref{eqn:simple-retention} to compensate for
any retention loss that \chI{occurs} during the measurements.
Figure~\ref{fig:predicted-shifted-distribution} shows the resulting threshold 
voltage distribution for each state, at \SI{0}{\celsius} (the black dotted curves)
and at \SI{70}{\celsius} (the red curves).  Note that the ER state distribution at 
\SI{70}{\celsius} completely falls below the lowest voltage that we can measure,
and hence is not shown.}


\begin{figure}[h]
\centering
\ifeps
\includegraphics[trim=10 10 10 10,clip,width=.7\linewidth]
{figs/distribution-comparison-temperature.eps}
\else
\includegraphics[trim=10 10 10 10,clip,width=.7\linewidth]
{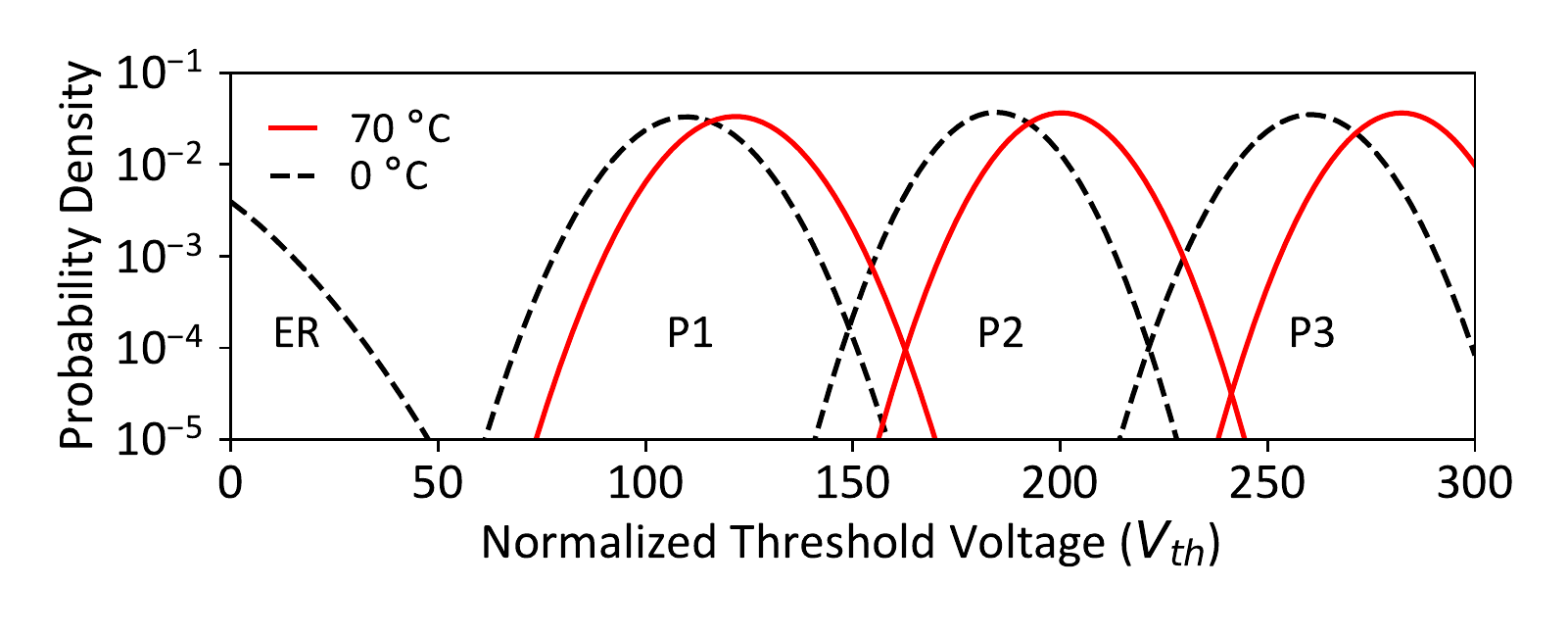}
\fi
\caption{Threshold voltage distribution right after programming at different
programming temperatures, predicted by \chI{our} retention loss model \chI{(Equation~\ref{eqn:simple-retention})}.}
\label{fig:predicted-shifted-distribution}
\end{figure}

\chI{We make two observations from \chI{Figure~\ref{fig:predicted-shifted-distribution}}. First, the higher programming
temperature shifts the P1, P2, and P3 states
to higher \chI{threshold} \chI{voltages,}
\chI{and the}
ER state to lower \chI{threshold} voltages.
\chI{The threshold voltage shifts are}
likely due to the increased electron
mobility under high temperature, which improves the speed of the
program operation (and likely the erase operation as
\chI{well~\cite{mielke.irps06, wilson.mascots14}}).
As a result, each programming pulse adds a greater amount of charge.
Second, due to the threshold voltage distribution shifts,
the amount of overlap between the P1 and P2 distributions, and between the P2 and
P3 distributions, decreases \chI{at a higher programming temperature}.  This is because while the distribution means shift 
to higher voltages at a higher programming temperature, the distribution widths do not 
change.  Because of the smaller amount of overlap \chI{between two neighboring
distributions}, there are fewer program variation errors \chI{at higher temperatures},
as we \chI{have shown in}
Figure~\ref{fig:temperature-vs-error}.}


\chI{Next, we study how threshold voltage shifts due to retention loss change 
with the programming temperature.
For brevity, we do not plot these results.
We observe that when the retention time is large ($t_r > 2 \cdot 10^2$),
the retention loss speed increases due to high temperature, which is similar in nature to
the effect of programming temperature on RBER.}

\paratitle{\chI{Effect on Retention Loss Speed and Program \chI{Variation}}}
\chI{We use} our \chI{model from Equation~\ref{eqn:simple-retention}}
to calculate the retention loss speed and program \chI{offset} for each programming
temperature, based on our characterization data.
\chI{Figure~\ref{fig:normalized-temperature} illustrates how the programming 
temperature affects retention loss speed (left) and program \chI{offset} 
(right).}
We \chI{fit} a quadratic trend line for retention loss speed, and a linear trend line for
program \chI{offset} \chI{(shown as dashed} \chI{lines).}

\begin{figure}[h]
\centering
\ifeps
\includegraphics[trim=10 10 10 10,clip,width=.7\linewidth]
{figs/normalized-temperature.eps}
\else
\includegraphics[trim=10 10 10 10,clip,width=.7\linewidth]
{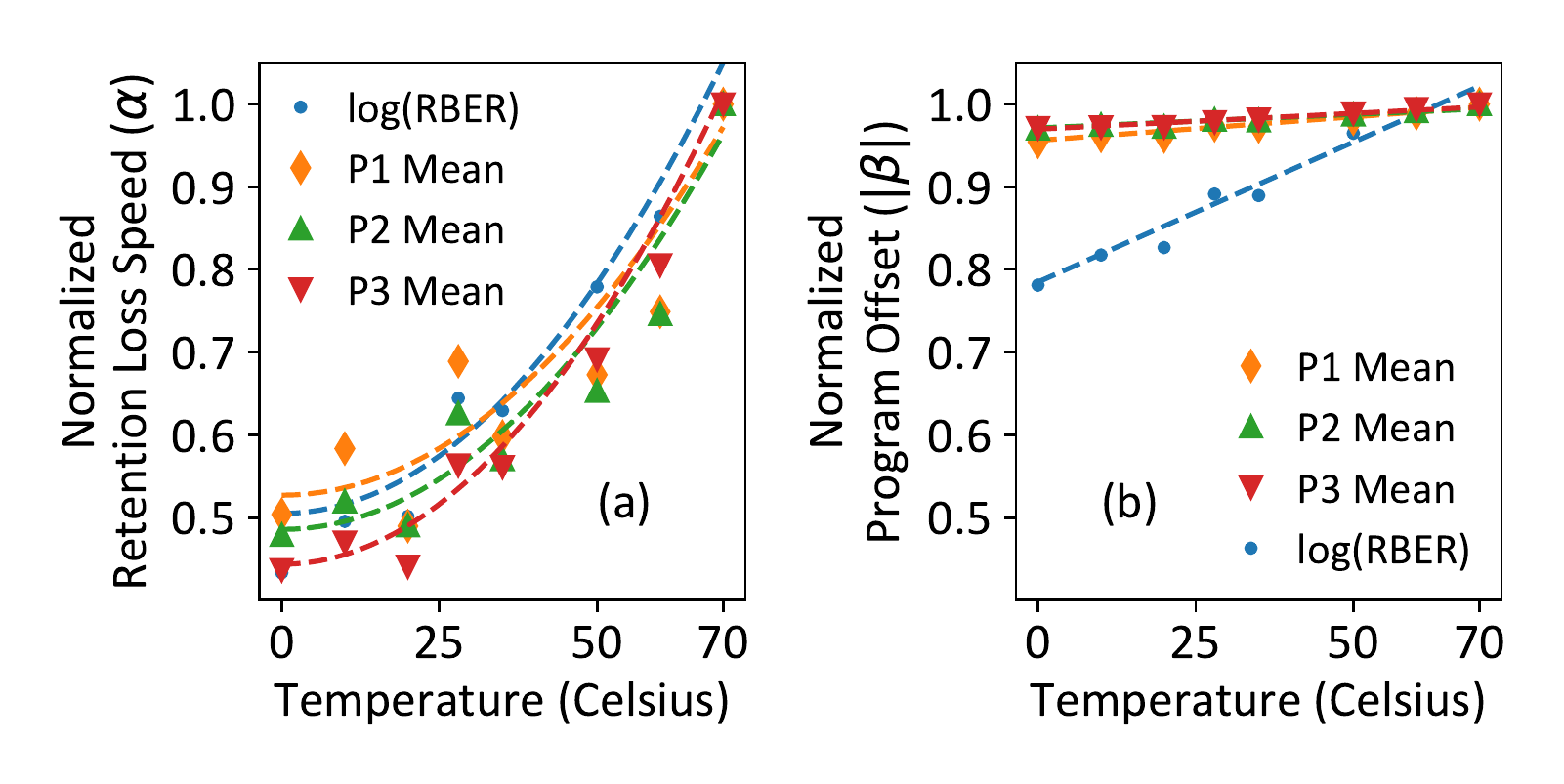}
\fi
\caption{Retention loss speed (left) and program \chI{offset} (right) across
different programming temperatures.}
\label{fig:normalized-temperature}
\end{figure}

We make \chI{two observations} from \chI{Figure~\ref{fig:normalized-temperature}}.
First, \chI{a} \chI{higher
temperature} accelerates retention \chI{loss at} a superlinear rate. We show in
Section~\ref{sec:heatwatch:modeling} that this trend complies with Arrhenius'
Law~\cite{arrhenius.zpc1889}.
Second, we find that \chI{the} programming temperature \chI{affects
program} \chI{variation}. This
effect has \emph{not} been accounted for in prior work, which usually assumes that
\chI{program operations are \chI{performed}} at room temperature~\cite{jesd218.jedec10}. 
In Figure~\ref{fig:normalized-temperature} (right), we find
that \chI{the program offset is higher}
at higher programming
\chI{\chI{temperatures.} As \chI{already shown} in Figure~\ref{fig:predicted-shifted-distribution}, 
the \chI{\chI{higher} initial threshold voltage} \chI{at higher programming
temperatures} reduces the amount of overlap between neighboring
\chI{threshold voltage distributions}, which in turn \chI{\emph{reduces}} the number of 
program variation errors.}
\chI{We conclude that higher temperature \chI{\emph{increases}} retention errors \chI{but}
\chI{\emph{reduces}} program variation errors.}

\subsection{\chI{Characterizing the} Recovery Cycle Effect}
\label{sec:heatwatch:characterization:cycle}

\chI{We} \chI{measure the \chI{effect} of \emph{recovery cycles}, \chI{i.e.,}
\chI{P/E cycles performed with a long dwell time,}
on self-recovery and flash reliability}.
\chI{We measure how the \emph{number} of recovery cycles
affects retention loss speed.}
We focus on retention loss speed
in this \chI{experiment because,} as we saw in Section~\ref{sec:heatwatch:characterization:dwell},
the dwell time does \emph{not} affect program \chI{variation}.
\chI{We first wear out each block by repeatedly writing pseudorandom data
with a dwell time of \SI{0.5}{\second}, until the block reaches 3,000 P/E cycles.}
Then, we \chI{start self-recovery, performing recovery cycles using} a 6-hour dwell time.
During the idle time of each recovery cycle,
we perform our 71-minute retention test at an operating temperature of
\SI{70}{\celsius} \chI{to measure the current retention loss speed}.
We keep performing recovery cycles until the change in retention
loss speed is less than 1\%, which indicates that an additional
recovery cycle does \emph{not} \chI{significantly} increase the \chI{effect} of
self-recovery.

\paratitle{\chI{Effect on Retention Loss Speed}}
\chI{Based on our characterization data, we calculate the retention loss speed 
($\alpha$) after each recovery cycle.
We use \chI{Equation~\ref{eqn:simple-retention}} to calculate $\alpha$,
as we did in Figures~\ref{fig:normalized-dwell} and \ref{fig:normalized-temperature}.}
\chI{Figure~\ref{fig:normalized-cycles} shows how the \chI{retention loss speed}
changes as a function of the number of recovery cycles performed.}
We fit \chI{power law trend lines} to the data, shown as a dashed line
\chI{in the figure}.

\begin{figure}[h]
\centering
\ifeps
\includegraphics[trim=10 10 10 10,clip,width=.7\linewidth]
{figs/normalized-cycles.eps}
\else
\includegraphics[trim=10 10 10 10,clip,width=.7\linewidth]
{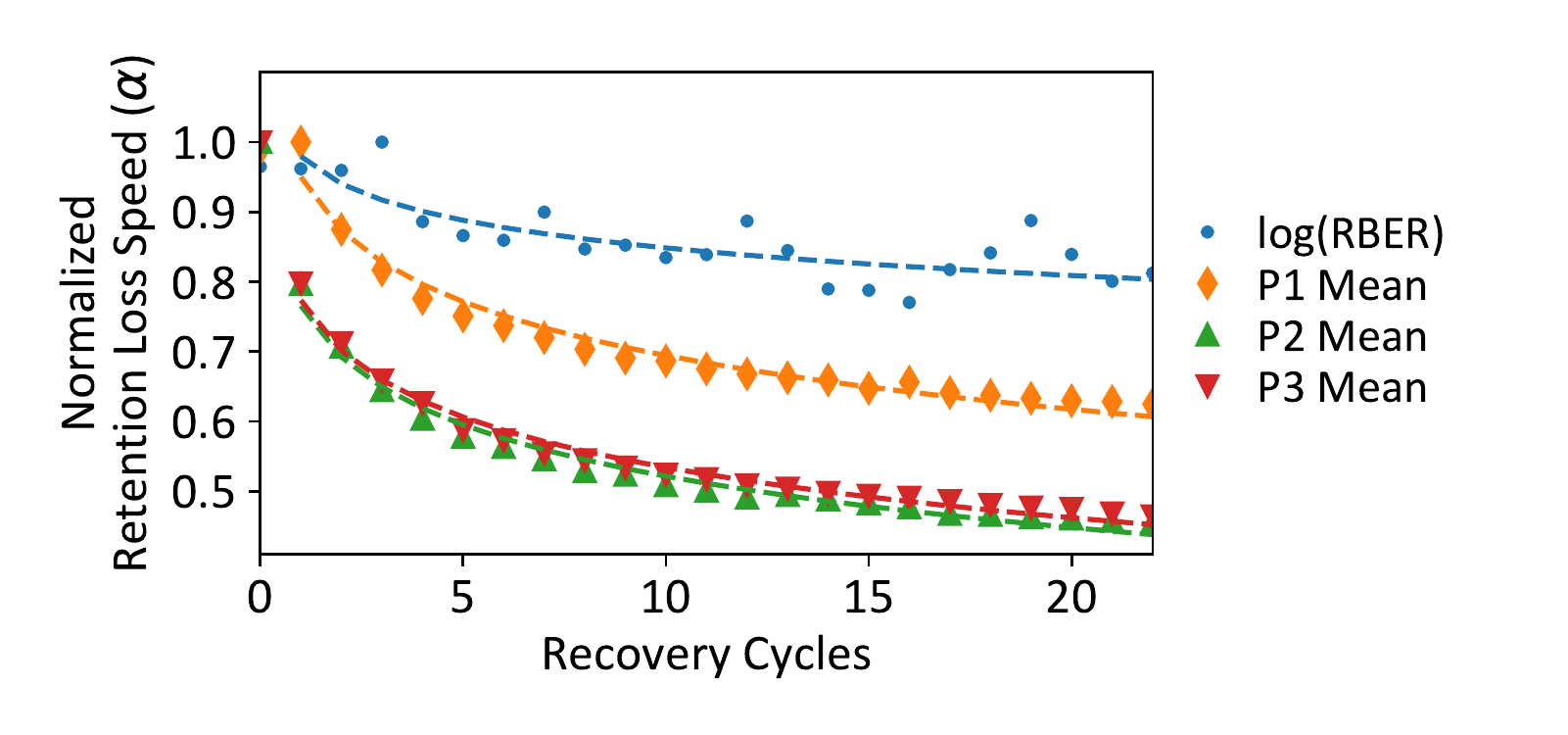}
\fi
\caption{Retention loss speed \chI{vs.}\ recovery cycles.}
\label{fig:normalized-cycles}
\end{figure}

We make two key observations from Figure~\ref{fig:normalized-cycles}.
First, to our surprise, the reduction in retention loss speed due to 
self-recovery becomes insignificant very quickly.
We find \chI{that, for} wordlines that have endured 3,000 P/E cycles, 
most of the benefits of self-recovery occur within 22 recovery
cycles for 3D NAND flash memory.  This is much smaller than the number of
cycles required according to prior work~\cite{mielke.irps06}, which \chI{finds}
that most of the benefits of self-recovery occur only when the number of
recovery cycles is 10\% of the total P/E cycle count.  In other words,
according to prior work, it should have taken \chI{at least} 300 recovery cycles to
achieve
most of the benefits of self-recovery.
Importantly, this implies that we can reap the benefits of self-recovery with
a much lower overhead \chI{(i.e., significantly fewer recovery cycles)} than previously-proposed mechanisms~\cite{mielke.irps06,
jesd22a117c.jedec11, compagnoni.irps10}.
\chI{Thus, we can improve the overall performance of NAND flash memory devices
that perform self-recovery.}
Second, \chI{the RBER does \chI{\emph{not}} change significantly \chI{\emph{until after}} the
first three recovery cycles.}
\chI{To reduce the latency of self-recovery, prior works~\cite{chen.codes13,
wu.imw11, mohan.hotstorage10, wu.hotstorage11} distribute recovery cycles
throughout the flash lifetime, and periodically perform only a single recovery
cycle.  Based on our observation, performing only one recovery cycle may
\emph{not} change the RBER significantly, and these mechanisms may \chI{\emph{not} significantly} improve
the flash lifetime as currently designed.}

\subsection{Summary of Key Observations}
\label{sec:heatwatch:characterization:summary}

We conclude that (1)~\chI{the} self-recovery effect reduces retention loss speed
linearly
with the logarithm of dwell time, and has little effect on program \chI{variation};
(2)~\chI{the} temperature effect increases retention loss speed at a superlinear
rate,
and increases program \chI{variation}; and (3)~the reduction in retention
loss speed due
to self-recovery becomes insignificant after 20 recovery cycles.


\section{Self-Recovery Effect Modeling}
\label{sec:heatwatch:modeling}
\label{sec:heatwatch:modeling:summary}

\chI{\chI{We} use our observations from Section~\ref{sec:heatwatch:characterization}
to construct \chI{\emph{Unified Recovery and Temperature} (URT),} 
a comprehensive analytical model of the impact of 
\chI{retention, wearout, self-recovery, and temperature}
on two \chI{output} parameters: 
\chI{(1)~the threshold voltage of a flash cell, and 
(2)~the raw bit error rate (RBER) of a flash page}.
\chI{URT calculates each output parameter $Y$ as:}
\begin{equation}
Y = Y_0 + \Delta Y(t_r \cdot AF, t_d \cdot AF, PEC)
\label{eqn:urt}
\end{equation}
\chI{In the equation, $Y_0$ is the initial value of the output parameter 
immediately after a cell is programmed, and $\Delta Y$ is the change in the 
output parameter due to retention loss. $\Delta Y$ is a function of the
\chI{(1)}~retention time ($t_r$) and \chI{(2)}~the dwell time ($t_d$), both of which are
scaled by an \emph{acceleration factor} $AF$ (see below),
and \chI{(3)}~the P/E cycle count ($PEC$).}}

\chI{URT consists of three components.
The \emph{program variation component} (PVM; 
Section~\ref{sec:heatwatch:modeling:linear})
predicts $Y_0$ based on the amount of program 
variation that took place.
The \emph{effective retention/dwell time component} (RDTM; 
Section~\ref{sec:heatwatch:modeling:arrhenius}) computes $AF$, which scales
\chI{the retention or dwell time at the} current temperature of the NAND flash memory to \chI{the}
\emph{effective time} at room temperature that has the same impact on $Y$.
The \emph{self-recovery and retention component} (SRRM; 
Section~\ref{sec:heatwatch:modeling:urm}), predicts $\Delta Y$ based on}
\chI{the effective retention/dwell time and the P/E cycle count.}
We show how \chI{URT can be used to} improve
flash reliability in Section~\ref{sec:heatwatch:mechanism}.

\subsection{Program Variation Component}
\label{sec:heatwatch:modeling:linear}
\label{sec:heatwatch:modeling:pvm}

\chI{First, we build a \emph{program variation model} (PVM) to predict
the \emph{initial values} ($Y_0$ in Equation~\ref{eqn:urt}) of the threshold
voltage and RBER immediately after a cell is programmed. The initial values
are determined by
(1)~\chI{the} target programming voltage, which is fixed for each state, and 
(2)~the \chI{program offset}
(Section~\ref{sec:heatwatch:characterization}).}
Recall that program \chI{offset}
is affected by the programming temperature
\chI{(Section~\ref{sec:heatwatch:characterization:temperature})}. Prior work \chI
{shows} that the P/E cycle count significantly affects program \chI{offset} as
well~\cite{luo.jsac16, mielke.irps08, cai.date13, cai.iccd13}.


\chI{To} account for both variables (i.e., programming temperature and P/E cycle count), we
use a multivariate linear regression model to model program variation:

\mbox{}\par
\kern-2\baselineskip
\begin{equation}
    Y_0 = A \cdot T_p \cdot PEC + B \cdot T_p + C \cdot PEC + D
\label{eqn:pvm}
\end{equation}
In \chI{PVM}, 
$Y_0$ is a function of the P/E cycle count of the flash cell
($PEC$) and the programming temperature ($T_p$), which are input parameters. $A$, $B$, $C$, and $D$ are
model constants that change based on which value we model (e.g., \chI{initial threshold voltage,
initial log of RBER}).
We fit PVM to our characterization data using the ordinary least squares
implemented in Statsmodels~\cite{seabold.statsmodels10}, and conclude that the model fits well,
with an $R^2$ value of 91.7\%. \chI{We provide the values of all the fitted model
parameters online~\cite{luo.tr18}.}





\subsection{Effective Retention/Dwell Time Component}
\label{sec:heatwatch:modeling:arrhenius}
\label{sec:heatwatch:modeling:etm}
\label{sec:heatwatch:modeling:rdtm}

Next, we build an \emph{effective retention/dwell time model} (RDTM) to \chI{calculate
the \emph{acceleration factor} ($AF$ in Equation~\ref{eqn:urt}), which scales the
retention time or dwell time under \emph{any} temperature ($t_{real}$) to the effective time
under room temperature ($t_{room}$)}.
Arrhenius' Law~\cite{arrhenius.zpc1889} (see Section~\ref{sec:background:recovery})
is commonly used by prior works to scale the retention time and dwell
time of flash memory across different temperatures (e.g.,~\cite{jesd91a.jedec03,
cai.hpca15, mielke.irps06, cai.iccd12, cai.itj13}).
RDTM \chI{uses the same general equation as} Arrhenius' Law
(Equation~\ref{eqn:arrhenius}):
\begin{align}
    AF = \chI{\frac{t_{real}}{t_{room}} = }
    \exp \left( \frac{E_a}{k_B} \cdot \left( \frac{1}{T_{real}} - \frac{1}
    {T_{room}} \right) \right)
\label{eqn:rdtm}
\end{align}
\chI{In RDTM, $AF$ is a function of  
the room temperature ($T_{room}$), 
the current temperature of the NAND flash memory
($T_{real}$), and
the activation energy ($E_a$).
$k_B$ is the Boltzmann constant.}
\chI{\chI{To accurately model the amplification factor}, we
(1)~\chI{experimentally} calculate a new value of $E_a$ that we can use for 3D NAND flash memory; and
(2)~verify the accuracy of Arrhenius' Law through experimental characterization, 
which \chI{no previous work has done for 3D NAND flash memory}.}



While $E_a$ is commonly accepted to be \SI{1.1}{\electronvolt} for \chI{\emph{planar}} NAND 
flash memory~\cite{jesd91a.jedec03,
cai.hpca15}, we cannot use the same value of $E_a$ for \chI{\emph{3D}} NAND
flash memory, due to changes in materials and manufacturing process.
\chI{Fortunately, we have extensive \chI{real device} characterization data on retention loss
at different temperatures
(Section~\ref{sec:heatwatch:characterization:temperature}), which enables us to} determine the
correct $E_a$ for 3D NAND flash memory.
%
We define $t_1$ as the time required for a 3D NAND flash memory device to
experience a fixed amount of retention loss, $\Delta Y$, at temperature $T_1$.
We define $t_2$ as the time required for the \emph{same} amount of retention loss to occur
at temperature $T_2$.
\chI{Using Equation~\ref{eqn:arrhenius},} the activation energy ($E_a$) can be calculated as:
\begin{align}
    E_a = \frac{k_B \cdot \ln \left( \frac{t_1}{t_2} \right) \cdot T_1T_2}
               {T_2 - T_1}
\label{eqn:activation}
\end{align}
We define $t_1$ as the time required for 3D NAND flash memory to
experience a fixed amount of retention loss, $\Delta Y$, at temperature $T_1$.
We define $t_2$ as the time required for the \emph{same} amount of retention loss to occur
at temperature $T_2$.

\chI{We} choose $T_2 =$~\SI{343.15}{\kelvin} \chI{(\SI{70}{\celsius})} as the reference temperature, and $t_2 =$~\SI{3600}{\second}
as the reference retention time, as our model is more resilient to
noise at a larger retention time.
Using our characterization data from
Section~\ref{sec:heatwatch:characterization:temperature}, we find the equivalent $t_1$
for different temperature values of $T_1$, spanning \SIrange{20}{70}{\celsius}, and
for different P/E cycle counts, spanning 1,000--10,000 P/E cycles. We use
ordinary least squares \chI{estimates} to fit Equation~\ref{eqn:activation} to these
data points.
From the fit, we determine that for the best fit, $E_a =$~\SI{1.04}{\electronvolt} for the
3D NAND flash memory chips that we test.
The 95\% confidence interval for $E_a$ is \SIrange{1.01}{1.08}{\electronvolt}. 
\chI{\chI{The value of $E_a$ is
based on the materials used for the cell, and should be similar for 3D charge
trap cells manufactured by other vendors~\cite{jeong.jssc16, mizoguchi.imw17, choi.vlsit16}.}} 

\chI{Next, we verify that Arrhenius' Law \chI{holds} for 3D NAND flash memory, 
by calculating} the coefficient of determination ($R^2$)
of the fit to the equation for Arrhenius' Law.  \chI{We} find that $R^2 = 0.76$. 
This means that Arrhenius' Law
explains 76\% of the variations due to temperature. This is a good fit given that
we use a single value for $E_a$ (best fit) across \emph{all} of our data points, because it is 
known that activation energy
changes across different temperatures and P/E cycle counts~\cite{marquart.fms15,
belgal.irps02}. We use a single value of $E_a$ regardless of temperature and
P/E cycle count to simplify RDTM. We leave more accurate
activation energy modeling for future work.





\subsection{Self-Recovery and Retention Component}
\label{sec:heatwatch:modeling:urm}
\label{sec:heatwatch:modeling:srrm}

\chI{We build a \emph{self-recovery and retention model} (SRRM) to 
accurately predict the \emph{threshold voltage shift and change in RBER}
($\Delta Y$ in Equation~\ref{eqn:urt}) due to retention loss.
SRRM predicts the effect of (1)~effective retention time,
(2)~effective dwell time, and (3)~P/E cycle count, which all
directly affect retention loss speed \chI{(see Section~\ref{sec:heatwatch:characterization:dwell})}.}

To construct SRRM, we repeat our dwell time
experiments from Section~\ref{sec:heatwatch:characterization:dwell} at \emph{room temperature}.
We cover a slightly larger dwell time range than our previous experiments,
testing from \SI{32}{\second} to \SI{4.6}{\hour}. To include the effect of the P/E cycle count,
we perform the retention test described in Section~\ref{sec:heatwatch:characterization:dwell}
for up to a 24-day retention time under room
temperature, using eight randomly-selected flash blocks,
and spanning a range between 1,000 and 10,000 P/E
cycles at \chI{1,000-P/E-cycle} intervals.  
We observe similar trends in terms of retention time, dwell time, and
temperature sensitivity as the findings we observe at a higher temperature
in Section~\ref{sec:heatwatch:characterization}.
\chI{For brevity, we do not plot these results, \chI{but we provide them online~\cite{luo.tr18}}.}

\chI{From an analysis of} the results of these experiments, we find that the threshold voltage shift
in 3D NAND flash memory is much less sensitive to the P/E cycle count than
planar NAND flash memory. Thus, we develop a new model \chI{that \chI{predicts} the
impact of retention and self-recovery on} 3D NAND flash memory, instead of relying on prior models
for planar NAND flash memory. Our SRRM model is as follows:
\begin{align}
    \Delta Y(t_{er}, t_{ed}, PEC) =
        b \cdot (PEC + c) \cdot 
        \ln \left( 1 + \frac{t_{er}}{t_0 + a \cdot t_{ed}} \right)
\label{eqn:urm}
\end{align}
\chI{In SRRM, $\Delta Y$ is} a function of three \chI{input parameters}:
(1)~\chI{the effective} retention time of the data stored in the cell (\chI{$t_{er}$}),
(2)~\chI{the effective} dwell time (\chI{$t_{ed}$}), and
(3)~\chI{the} P/E cycle count for a flash cell ($PEC$).
The model has four \chI{constants, whose values} change depending on
which \chI{output parameter ($\Delta Y$)} we are evaluating:
$b$ and $c$ control the impact of P/E cycle count on
retention loss speed, and $t_0$ and $a$ control the impact of dwell time on
retention loss speed. \chI{To determine the values for these constants, we use
the non-linear least squares algorithm implemented in
SciPy~\cite{jones.scipy14, levenberg.qam44} to fit SRRM to the
characterization data we collected.}

Figure~\ref{fig:urm-all-fit} illustrates how predictions from SRRM compare with our
measured characterization data.
\chI{Figure~\ref{fig:urm-all-fit}a compares the SRRM predictions and measured values
of the threshold voltage shift for cells in state P1, P2, or \chI{P3.
Figure}~\ref{fig:urm-all-fit}b compares the SRRM
predictions and measured values of the change in the log of RBER.}
Each cross (`x') \chI{in} the figure
represents a data point,
where the x-coordinate of each data point is the \chI{value predicted} by SRRM, and
the y-coordinate of each data point is the value measured during our characterization.
The dashed line shows the perfect fit (i.e., where the predicted and measured values
are equal).
We observe that \chI{for both the threshold voltage shift and the change in RBER,}
all of the data points are very close to the perfect fit line. 
Thus, SRRM accurately predicts \chI{both output parameters}. 
\chI{Overall, the} percentage root mean square error
($\%RMSE$)
for SRRM is 4.9\%. As a comparison, \chI{if we were to \chI{predict} these two
output parameters using UDM~\cite{mielke.irps06},
a previously-proposed model for retention loss in planar NAND flash memory,
the average $\%RMSE$ of the predictions would be 17.8\%,}
which is much less accurate than \chI{the predictions from} our model.
\chI{We} conclude that SRRM is highly accurate for \chI{predicting the
change in RBER and the threshold voltage shift at room temperature in}
3D NAND flash memory.

\begin{figure}[h]
\centering
\ifeps
\includegraphics[trim=10 10 10 10,clip,width=.7\linewidth]
{figs/UDM-all-fit.eps}
\else
\includegraphics[trim=10 10 10 10,clip,width=.7\linewidth]
{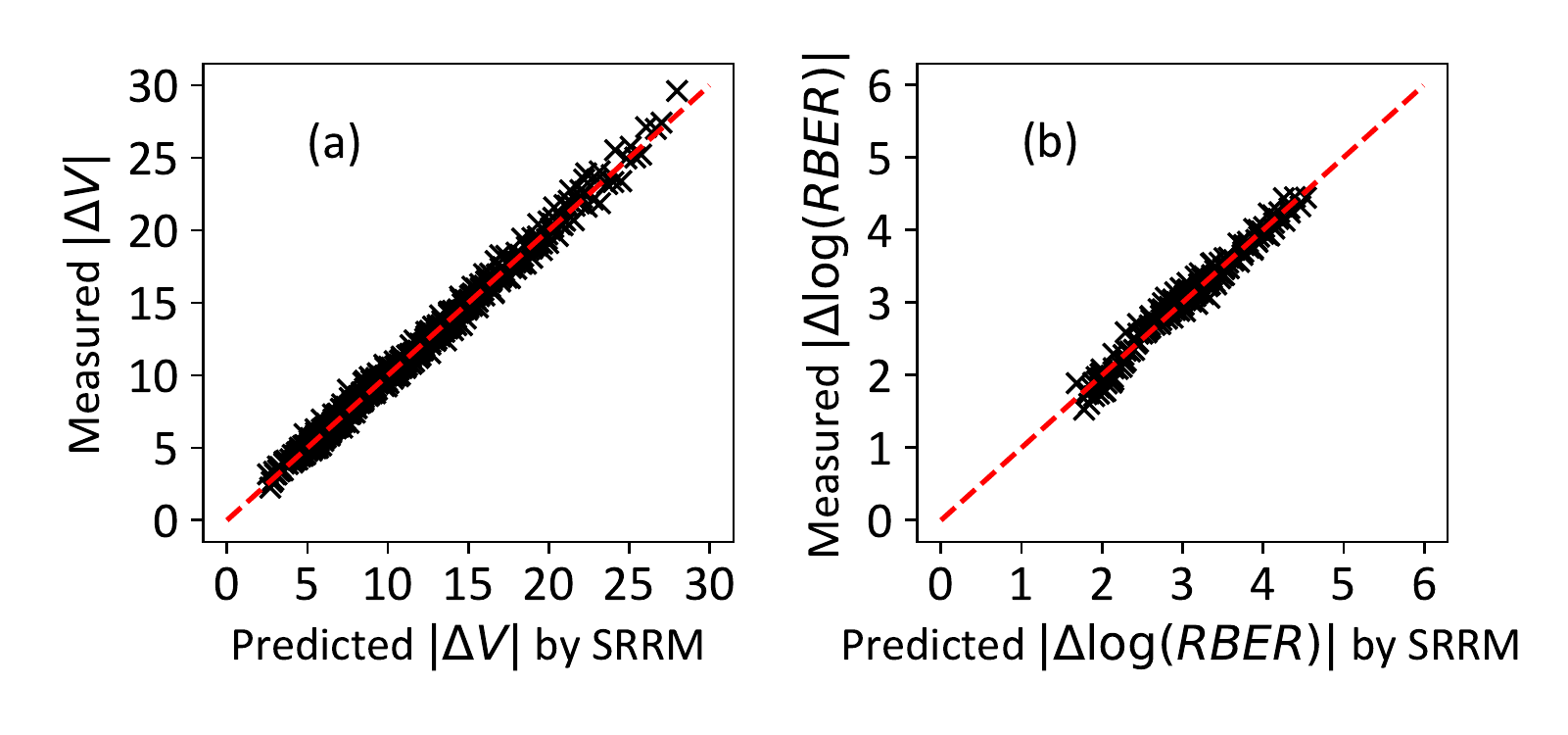}
\fi
\caption{\chI{SRRM prediction accuracy.}}
\label{fig:urm-all-fit}
\end{figure}

\section{Improving 3D NAND Reliability}
\label{sec:heatwatch:mechanism}

Our \emph{goal} in this section is to improve flash reliability and
lifetime by developing a new \chI{mechanism that} makes
use of our findings (Section~\ref{sec:heatwatch:characterization}) and our new model
(Section~\ref{sec:heatwatch:modeling}). \chI{Our new mechanism is called \emph{HeatWatch}}.

\subsection{Observations}
\label{sec:heatwatch:mechanism:observation}

We make three key observations \chI{from the following three experiments} that
lead to the design of \chI{HeatWatch}.
First,
we observe that \emph{SSD write intensity and the SSD environmental temperature significantly
impact flash lifetime}.
The write intensity of an SSD is measured as the number of full drive writes per day.
Given a fixed SSD size, the write intensity is inversely proportional to the 
average dwell time, \chI{thus \chI{affecting flash} lifetime (Section~\ref{sec:heatwatch:characterization:dwell})}.
This is because modern SSDs \chI{use} \emph{wear-leveling} techniques
to keep all flash blocks in the SSD at a similar P/E cycle
\chI{count~\cite{cai.procieee17, cai.arxiv17, gal.cs05, chang.sac07}}.
The environmental
temperature affects \chI{the} \chI{flash} \chI{lifetime}
\chI{(Section~\ref{sec:heatwatch:characterization:temperature})},
\chI{because it changes} the temperature of the underlying NAND flash
memory.

Figure~\ref{fig:dwell-vs-pec} shows \chI{the flash} lifetime under
different write intensities and environmental temperatures, assuming 
that the vendor guarantees a three-month retention time for the data,
which is typical for enterprise
\chI{SSDs~\cite{cai.procieee17, cai.arxiv17, cai.iccd12, cai.hpca15, pan.hpca12}}. 
The SSD write intensity is shown on the x-axis in log scale. We plot the results 
by using separate curves for each temperature that we evaluate, 
which ranges from \SI{0}{\celsius} to \SI{70}{\celsius}.

\begin{figure}[h]
\centering
\ifeps
\includegraphics[trim=10 10 10 10,clip,width=.7\linewidth]
{figs/dwell-vs-pec.eps}
\else
\includegraphics[trim=10 10 10 10,clip,width=.7\linewidth]
{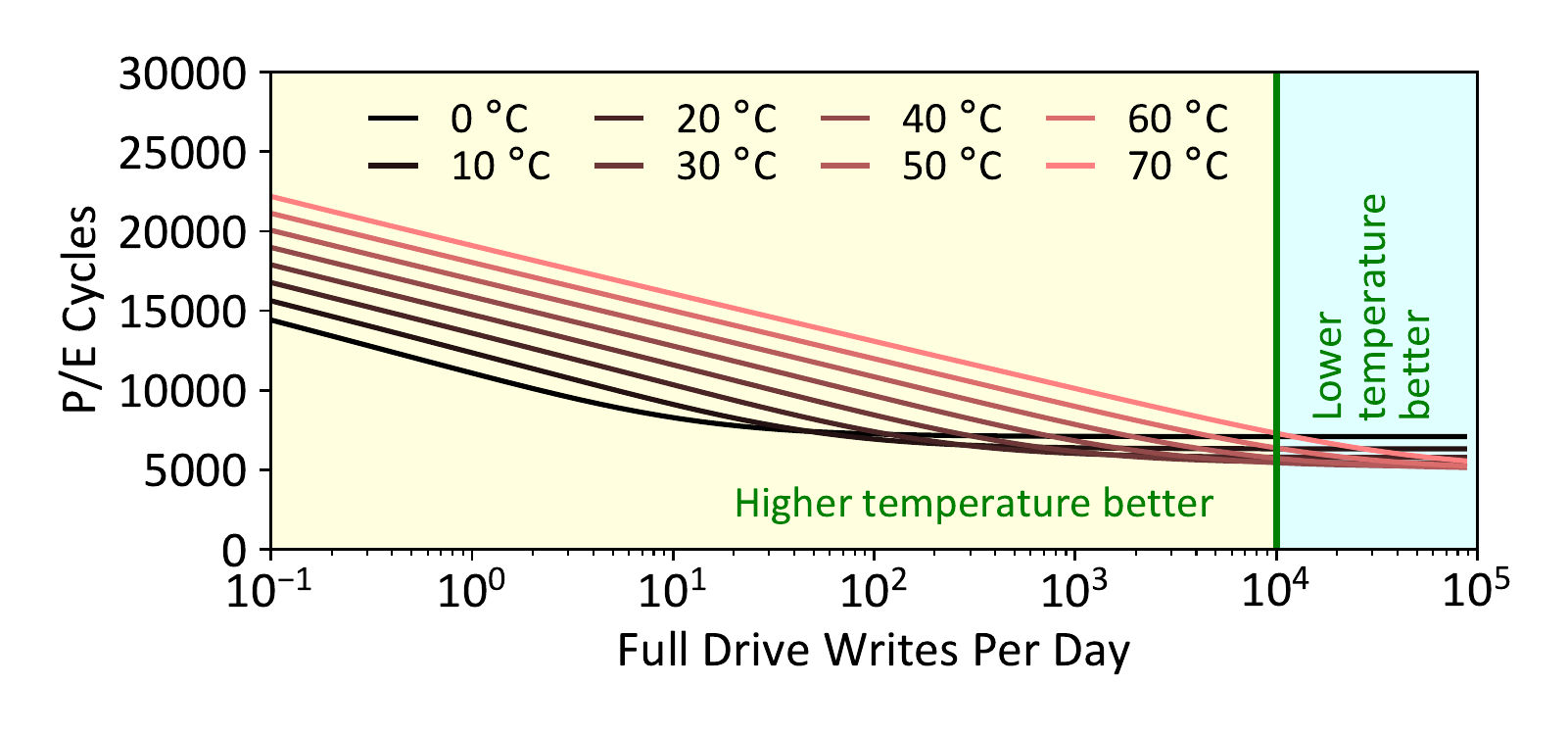}
\fi
\caption{Change in \chI{flash} lifetime due to write intensity and
environmental temperature \chI{($t_r =$ 3~months)}.}
\label{fig:dwell-vs-pec}
\end{figure}

From the figure, we see that \chI{the flash} lifetime initially decreases as SSD write intensity
\chI{increases,} but stops decreasing at around 5,000 P/E cycles.
\chI{\chI{When the write intensity is low ($<10^4$ drive writes/day,
\chI{which covers the range of write intensities of most contemporary workloads\chI{~\cite{cai.iccd12}}}),
a higher temperature is desirable,
as it improves both the effective dwell time and program variation
and thus \chI{leads to} a \chI{longer lifetime}. In contrast, when \chI{the}
write intensity is high, a lower
temperature is better \chI{due to an improved effective retention time}.
Note that these curves drastically shift \chI{(not shown)}
(1)~with different assumptions
about retention time, or (2)~when the temperature is not constant. Thus, \chI{we find that} no
single temperature or temperature range is ideal.}}

Second, we observe that \emph{the choice of the read reference voltage ($V_{ref}$)
significantly affects RBER and flash lifetime}. The voltage at which the lowest
RBER can be achieved is called the \emph{optimal read reference voltage} ($V_{opt}$).
$V_{opt}$ changes based on conditions such as retention time and P/E cycle 
\chI{count. Adapting} the optimal read reference voltage to these
changing conditions significantly \chI{increases} flash
lifetime~\cite{cai.iccd13, luo.jsac16, cai.hpca15, cai.procieee17, cai.arxiv17,
cai.date13, parnell.globecom14}.
\chI{Based on our} \chI{experiments under room temperature,}
Figure~\ref{fig:vref-vs-rber} shows how the RBER increases as the \chI{applied read
reference voltage} \chI{becomes} further away from the optimal read reference voltage
(which we refer to as the \chI{$|V_{ref} - V_{opt}|$} \emph{distance}).
We find that the RBER increases exponentially as the \chI{$|V_{ref} - V_{opt}|$}
distance increases. We conclude, as prior works \chI{have~\cite{cai.procieee17,
cai.arxiv17, cai.hpca15, luo.jsac16, cai.iccd13}},
that it is important to accurately predict the
optimal read reference voltage, as even \chI{a small $|V_{ref} - V_{opt}|$ distance}
can greatly increase the error rate (and thus hurt the flash lifetime).

\begin{figure}[h]
\centering
\ifeps
\includegraphics[trim=10 10 10 10,clip,width=.7\linewidth]
{figs/vref-vs-rber.eps}
\else
\includegraphics[trim=10 10 10 10,clip,width=.7\linewidth]
{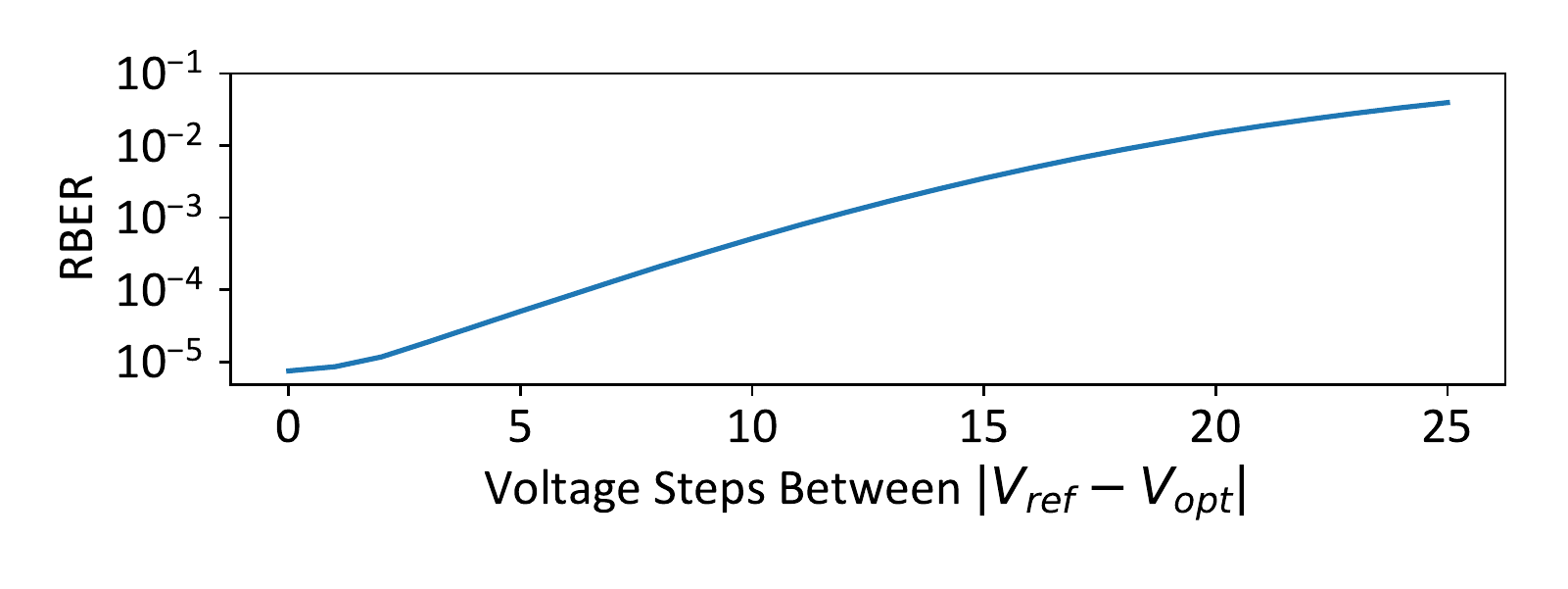}
\fi
\caption{RBER \chI{vs.\ $|V_{ref} - V_{opt}|$ distance}.}
\label{fig:vref-vs-rber}
\end{figure}

Third, we observe that \emph{the optimal read reference voltage in 3D
NAND flash memory changes over time, and can be accurately predicted as the value that falls in the middle of two
neighboring threshold voltage distributions}. Figure~\ref{fig:optvrefs} shows
the \chI{\emph{measured}} $V_{opt}$ from our 
characterization (blue dots in the figure),
and the value of $V_{opt}$ calculated by using \chI{our URT model} to predict the threshold
voltage distributions of each state (orange curve),
for read reference voltages $V_b$ and $V_c$ (see Section~\ref{sec:background:3d}).
The x-axis shows the retention time of the data in log scale. 
We see that after \SI{4000}{\second} of retention time, the \chI{measured} optimal
read reference voltages for $V_b$ and $V_c$ change by 8 and 11 voltage steps,
respectively.\footnote{Our characterization shows that $V_a$ does \emph{not} change
\chI{significantly based on} retention time, so we do not plot it. We find that URT accurately
predicts the lack of change in $V_a$.} Comparing the blue dots with the orange curves, we
find that our URT-based $V_{opt}$ prediction is \emph{always} within 1 voltage step of
the empirical optimal read reference voltage.
\chI{We} conclude that URT \chI{accurately predicts} the optimal
read reference voltage.

\begin{figure}[h]
\centering
\ifeps
\includegraphics[trim=0 10 10 10,clip,width=.7\linewidth]
{figs/myoptvrefs.eps}
\else
\includegraphics[trim=0 10 10 10,clip,width=.7\linewidth]
{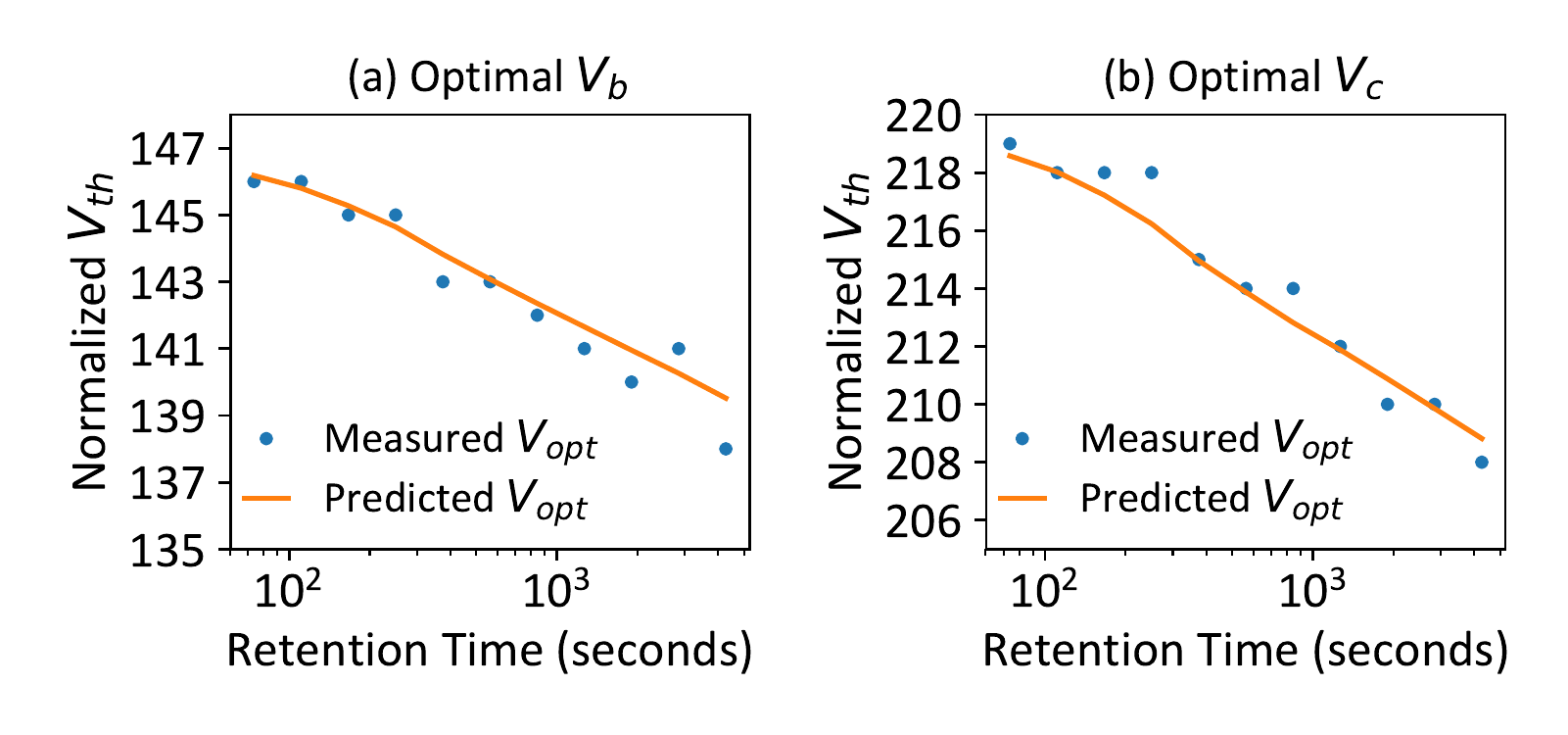}
\fi
\caption{\chI{Measured} and URT-predicted $V_{opt}$.}
\label{fig:optvrefs}
\end{figure}

\subsection{HeatWatch Mechanism}
\label{sec:heatwatch:mechanism:mechanism}

Motivated by our observations in Section~\ref{sec:heatwatch:mechanism:observation}, we
propose \emph{HeatWatch}, a mechanism that improves flash reliability
and lifetime by predicting and applying the optimal read reference voltage
using our URT model from Section~\ref{sec:heatwatch:modeling}.
\chI{HeatWatch consists of (1)~three \emph{tracking components}, which
monitor and efficiently record the SSD temperature, dwell time, retention time, 
and P/E cycle count; and
(2)~two \emph{prediction components}, which compute
the URT model using this tracked information} to accurately predict the optimal read reference
voltage for each read operation.

\paratitle{\chI{Tracking Component 1: SSD Temperature}}
Modern SSDs already contain multiple
temperature sensors to prevent overheating~\cite{lenny.fms14, meza.sigmetrics15}. 
\chI{HeatWatch \emph{efficiently} logs the readings from these sensors,}
which the RDTM component of \chI{URT} (see Section~\ref{sec:heatwatch:modeling:etm})
uses to scale the dwell time and retention time.
\chI{HeatWatch records the temperature every second, and}
\chI{\emph{precomputes} the
acceleration factors ($AF_i$) for every \emph{logarithmic time interval} $i$.
HeatWatch uses logarithmic intervals because the effects of dwell time and
retention time are logarithmic with respect to their duration (see
Section~\ref{sec:heatwatch:characterization:dwell}).
HeatWatch uses RDTM to calculate $AF_0$.
For all other intervals, $AF_i$ is calculated as a piecewise integral, by
summing up the \emph{two} most recent
values of $AF_{i-1}$, since interval~$i$ covers \emph{double} the amount of
time as interval~$i-1$.  Therefore, for each interval, HeatWatch stores the
current \emph{and} previous values of $AF_i$,
in an \emph{acceleration factor log}.}

\paratitle{\chI{Tracking Component 2: Dwell Time}}
\chI{The} self-recovery effect is dependent on the
average dwell time across \emph{multiple} P/E cycles (Section~\ref{sec:heatwatch:mechanism:observation}).  The average dwell time is determined by the
workload write intensity. Thus, we use the SSD controller to (1)~monitor the
workload write intensity online, and (2)~calculate the average dwell time for each
flash block. \chI{HeatWatch does \chI{\emph{not}} need to track the variation in dwell 
time across different flash pages within the \emph{same} block, as we find from
our \chI{experimental} characterization that
the effect of page-to-page dwell time variation is negligible \chI{(Section~\ref{sec:heatwatch:mechanism:observation})}.}

HeatWatch approximates the effective dwell time by taking the
average unscaled dwell time across the last 20 P/E cycles, and scaling it using
\chI{RDTM}. The SSD controller keeps a counter that
tracks the amount of data
written to the SSD, and logs the timestamps of the last 20 full drive writes. When a
flash block is programmed during drive write $n$, the SSD controller calculates the average
unscaled dwell time as the difference between the current time and the timestamp of
drive write $n-20$. Then, the SSD controller computes and stores the
effective room temperature dwell time by scaling it using the
aforementioned acceleration factor log.

\paratitle{\chI{Tracking Component 3: P/E Cycles and Retention Time}}
The SSD controller already records the P/E cycle count 
of each block to use in wear-leveling
algorithms~\cite{cai.procieee17, cai.arxiv17, gal.cs05, chang.sac07}. 
To track the
retention time of each flash block, HeatWatch simply logs the
timestamp when the block is selected for programming. Then,
HeatWatch calculates the effective
retention time for each read operation as the difference
between the program time and read time, scaled by \chI{RDTM}.

\paratitle{\chI{Prediction Component 1: Optimizing the} Read Reference Voltage}
\chI{The} optimal read reference voltage between
two states can
be predicted accurately by averaging the means of the threshold
voltage distributions for each state (Section~\ref{sec:heatwatch:mechanism:observation}).
As HeatWatch knows the P/E cycle count, programming temperature, effective dwell time,
and effective retention time, it can use the URT model from Section~\ref{sec:heatwatch:modeling} to 
predict the \chI{means of the threshold voltage distributions for each state,
and thus the} optimal read reference voltage.
For each read operation, the SSD controller simply gathers all
\chI{the metadata} for the flash block \chI{that is to be} read, and then predicts and applies the optimal
read reference voltage. \chI{The information gathering and
prediction happen \chI{\emph{after}} the FTL translates the \chI{logical address of the read to a}
physical address \chI{(see~\cite{cai.procieee17, cai.arxiv17} for detail)}, since the 
information for the flash block
is indexed using the physical address}.

\chI{
\paratitle{\chI{Prediction Component 2:} Fine-Tuning URT Model Parameters Online} 
To accommodate for chip-to-chip variation, URT learns its model parameters
\emph{online}.  We initialize the URT model parameters using a set of parameters 
that have been learned \chI{offline,
which} the vendor can provide at manufacture time.
Over time, URT fine-tunes its model parameters by
(1)~sampling a number of flash wordlines in the chip \chI{(10 in our evaluations)}, which are selected at random from 
blocks that span a range of different
P/E cycles, effective dwell/retention time, and programming temperatures;
(2)~learning the optimal read reference voltages for the sampled flash
wordlines online, using a technique similar to Retention Optimized Reading (ROR)~\cite{cai.hpca15}; and 
(3)~using the sampled data to
fit the fine-tuned URT model parameters for each chip, which can be done \chI{relatively} easily
in the firmware with little overhead~\cite{luo.jsac16}.
The overall latency for online training is dominated by the latency to \chI{predict}
the optimal read reference voltage for each wordline.
In the worst case, ROR performs 300 read operations per wordline, 
using a different voltage step for each read.  For the 10~wordlines sampled by
\chI{URT model tuning}, assuming a read latency of \SI{100}{\micro\second}, 
the total worst-case latency of \chI{URT model tuning} is \SI{0.3}{\second}.
Note that this tuning
procedure needs to be performed \chI{only} infrequently (e.g., every 1,000 P/E cycles), and can be performed in the 
background (i.e., when the chip is idle), thus incurring negligible performance
overhead.
}

\paratitle{\chI{Storage} Overhead}
\chI{HeatWatch needs to store three sets of information.}
(1)~\chI{HeatWatch stores the acceleration factor for only logarithmic \chI{time} intervals from
\SI{0.5}{\second} to \SI{1}{\year} (26 intervals in total).}
\chI{HeatWatch stores the current and previous value of each acceleration factor,
\chI{as described in \emph{Tracking SSD Temperature} above}.}
Assuming that each acceleration factor is stored as a \SI{4}{\byte} floating-point
number, \chI{the total log requires \SI{208}{\byte} of storage}.
(2)~\chI{HeatWatch stores} the programming temperature, dwell time,
and program time for each flash block. Assuming that each piece of information
uses \SI{4}{\byte}, \chI{for a \SI{1}{\tera\byte}} SSD with an \SI{8}{\mega\byte} flash block size, 
\chI{HeatWatch uses \SI{1.5}{\mega\byte} of
storage in total to store this information}. (3)~\chI{HeatWatch needs} a 32-bit counter, and must store the 
timestamps for the last 20 full disk writes
\chI{(Section~\ref{sec:heatwatch:characterization:cycle})}, which requires \SI{84}{\byte} of storage. 
\chI{In total, the three sets of information require \chI{less than} \SI{1.6}{\mega\byte} of 
storage.  To minimize the performance overhead of accessing this data,}
HeatWatch buffers the data in the on-board DRAM \chI{in the SSD controller~\cite{cai.procieee17,cai.arxiv17}}.
The storage overhead is very small,
as a \SI{1}{\tera\byte} SSD \emph{typically} contains \SI{1}{\giga\byte} of DRAM for
caching~\cite{cai.procieee17, cai.arxiv17}.

\paratitle{\chI{Latency Overhead}} 
\chI{HeatWatch performs two operations that contribute to its latency.
(1)~\chI{Every} second, HeatWatch updates the acceleration factor log with the latest temperature
reading. This update can be done in the
background, and, thus, its performance overhead is negligible. 
(2)~HeatWatch computes the URT model during each read operation, 
which involves performing \chI{\emph{only}} 16 arithmetic computations in the 
SSD controller (Section~\ref{sec:heatwatch:modeling}).
The model computation latency is negligible
compared to \chI{the flash} read latency (\chI{$>$\SI{25}{\micro\second}}\chI{~\cite{grupp.fast12}}).}


\subsection{Evaluation}
\label{sec:heatwatch:mechanism:evaluation}

To \chI{evaluate} \chI{HeatWatch}, we \chI{examine} the raw bit error rate (RBER)
and lifetime \chI{of} four different configurations:

\begin{itemize}[topsep=0pt,partopsep=0pt,noitemsep,leftmargin=10pt]

\item \emph{Fixed $V_{ref}$}, which always uses the default read reference voltage to
read the data.

\item \chI{\chI{\emph{Retention-Only}}, which predicts the optimal read reference voltage
based on a model that considers only P/E cycle count and retention 
time\chI{~\cite{parnell.globecom14, papandreou.glsvlsi14, cai.hpca15, cai.date13, cai.iccd13,
luo.jsac16}}.}
Note that this model always assumes a
\emph{fixed} retention loss speed, regardless of the dwell time and temperature.

\item \emph{HeatWatch}, 
our proposed mechanism to \chI{accurately predict the optimal read reference voltage
by tracking dwell time and temperature and using our URT model.}

\item \emph{Oracle}, which always \chI{\emph{ideally}} uses the \chI{measured} optimal read reference voltage,
and does \emph{not} incur any performance overhead. \chI{Note that this is
\chI{\emph{not}} implementable.}

\end{itemize}

We evaluate these four configurations using 28 \chI{commonly-used} real \chI{storage}
traces~\cite{narayanan.tos08}, which have varying write intensities.
Each trace represents seven days of workload data, and contains
timestamps we can use to calculate the dwell time and retention time of each access. We
simulate temperature variation over the course of a day as the superposition of a
sinusoidal function and some Gaussian noise. The sinusoidal model has a mean of
\SI{35}{\celsius}, an amplitude of \SI{15}{\celsius}, and a 1-day period,
representing how the temperature changes during a \chI{daily} cycle. 
The Gaussian noise model that we use has a standard deviation
of \SI{3}{\celsius}.

Figure~\ref{fig:rber-vs-pec} shows how the RBER increases \chI{with P/}E cycle count for
our four evaluated configurations, using a workload that appends \emph{all} 28 disk traces
together to mimic an SSD that runs multiple workloads back-to-back. The
dotted line shows an error correction capability (see Section~\ref{sec:errors})
of $2 \cdot 10^{-3}$ errors per bit~\cite{cai.procieee17, cai.arxiv17}. 
We determine the lifetime for each configuration using the point at which the 
RBER intersects the error correction capability.
We use \emph{Fixed $V_{ref}$}, which has the highest RBER, as our baseline. 
From the figure, we see that \chI{\emph{Retention-Only}} reduces the RBER by 83.1\%,
on average across all P/E cycles, compared to the baseline,
\chI{\emph{HeatWatch} \chI{reduces} the RBER by 93.5\% compared to the baseline.}
This is very close to the average RBER improvement under \emph{Oracle} (93.9\%).
\chI{\emph{HeatWatch}} \chI{significantly improves}
lifetime \chI{due to its RBER improvement}.
The lifetime \chI{with} \emph{HeatWatch} is
21,065 P/E cycles, which is 3.19$\times$ and
1.29$\times$ the lifetime \chI{with} \emph{Fixed $V_{ref}$} and \chI{\emph{Retention-Only}}, respectively.
\emph{HeatWatch} comes within \chI{only} 200 P/E cycles of the \emph{Oracle}
lifetime.

\begin{figure}[h]
\centering
\ifeps
\includegraphics[trim=10 10 10 10,clip,width=.7\linewidth]
{figs/rber-vs-pec-mixed.eps}
\else
\includegraphics[trim=10 10 10 10,clip,width=.7\linewidth]
{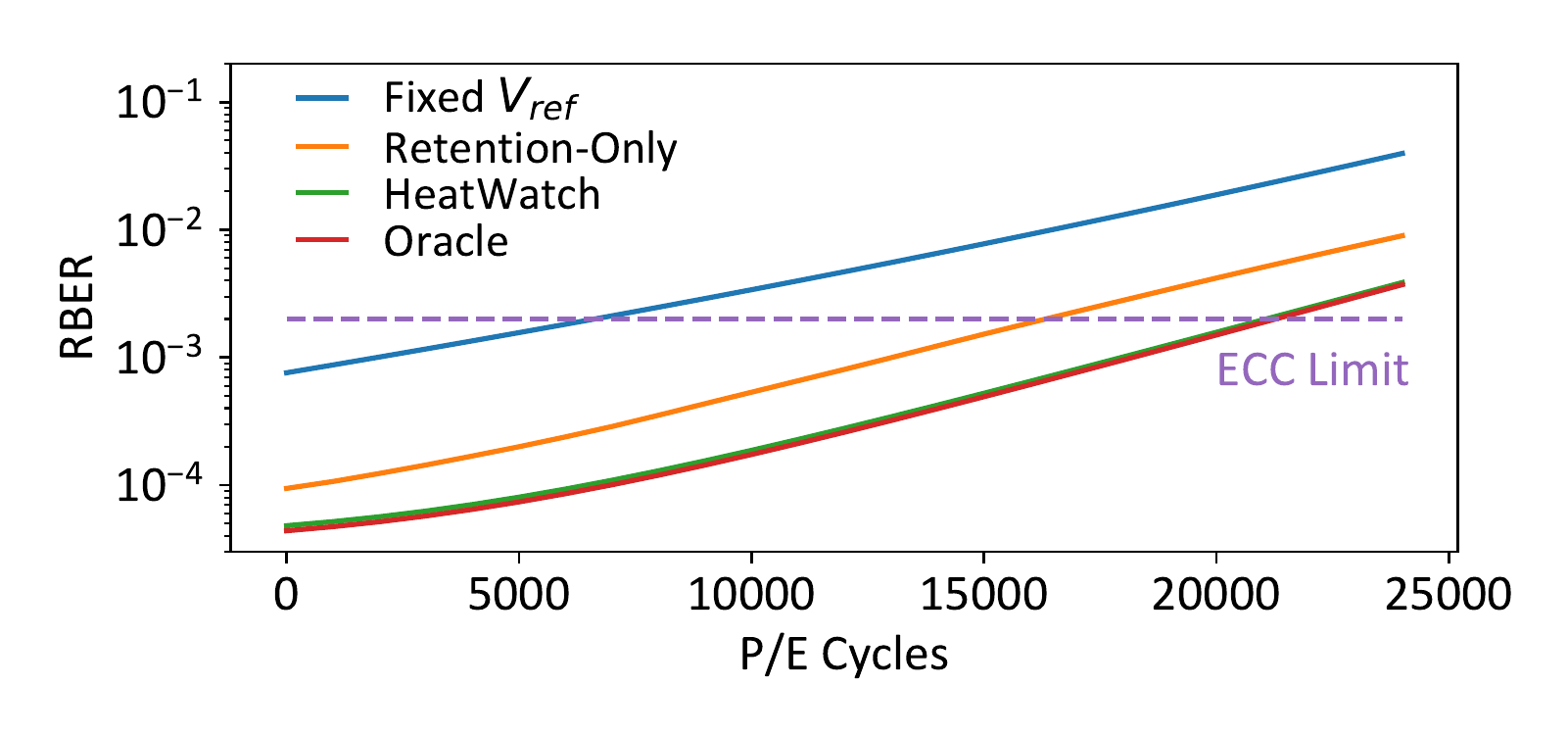}
\fi
\caption{RBER \chI{vs.}\ P/E cycle count.}
\label{fig:rber-vs-pec}
\end{figure}

We repeat the same experiment for each workload individually, and determine
the lifetime for each workload under \chI{the} four configurations, as shown
in Figure~\ref{fig:workload-lifetime}. On average across all of our workloads, 
The lifetime under \chI{\emph{Retention-Only}} is $3.11\times$ the lifetime of the \emph{Fixed $V_{ref}$} baseline.
\emph{HeatWatch} improves the lifetime further, with a lifetime that is $3.85\times$ the baseline
lifetime.  Again, this is very close to the lifetime improvement \chI{of} \emph{Oracle} 
($3.89\times$).

\begin{figure}[h]
\centering
\includegraphics[trim=7 275 235 1,clip,width=.7\linewidth]
{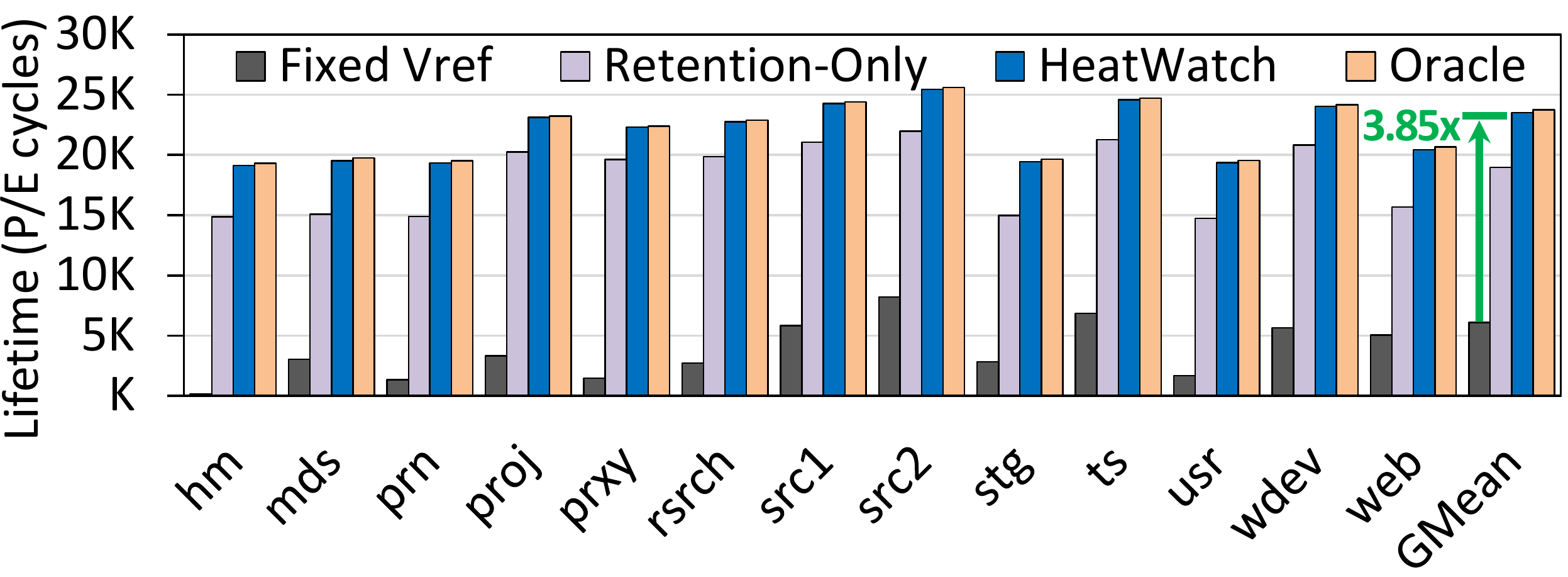}
\caption{P/E cycle lifetime \chI{for each workload}.}
\label{fig:workload-lifetime}
\end{figure}

We conclude that by incorporating dwell time and temperature information to
predict the optimal read reference voltage, HeatWatch improves the lifetime of
3D NAND flash memory devices over a state-of-the-art 
mechanism~\cite{papandreou.glsvlsi14}, and approaches the lifetime of
\chI{an ideal} mechanism \chI{that has perfect} knowledge of the optimal read
reference voltage.

\section{Related Work}
\label{sec:heatwatch:related}

To our knowledge, this \chV{dissertation} is the first to 
\chI{(1)~\chI{experimentally characterize} and accurately model the self-recovery and temperature 
effects in 3D NAND flash memory; and
(2)~devise a mechanism that improves 3D NAND flash memory lifetime by 
comprehensively taking into account retention time, wearout, self-recovery, and
temperature.} \chI{We discuss the \chI{closely-related} works.}

\paratitle{Data Retention in NAND Flash Memory}
Many prior works focus on flash \chI{memory} retention loss and retention errors, and
show that retention loss is the most dominant source of errors in modern NAND flash
memory~\cite{cai.procieee17, cai.arxiv17, cai.hpca15, cai.iccd12, papandreou.glsvlsi14, mielke.irps08,
pan.hpca12, choi.sigmetrics17}. 
These works \chI{do \chI{\emph{not}} consider} the effects of self-recovery and temperature on retention loss.
Our work in this chapter investigates these effects through an extensive
characterization of state-of-the-art \chI{3D} NAND flash memory chips.

\paratitle{3D NAND Flash Memory Characterization}
Recent works study the
error characteristics of 3D NAND flash memory, and identify differences 
between 3D and planar NAND
flash memory due to memory cell design and architectural
changes~\cite{cai.procieee17, cai.arxiv17, mizoguchi.imw17, park.jssc15,
choi.vlsit16}. None of these works
provide a detailed characterization of the impact of self-recovery, retention, P/E cycle count, and
temperature on \chI{real} 3D NAND flash memory.

\paratitle{Retention Loss Models}
\chI{Our URT model is inspired by and improves upon the Unified Detrapping Metric
(UDM) \chI{model~\cite{mielke.irps06}}.}
There are three reasons why \chI{prior models developed for planar NAND flash memory,
such as UDM, are} insufficient for 3D NAND \chI{flash memory}. 
First, \chI{3D charge trap cells are} more resilient to P/E
cycling \chI{than the floating-gate cells used by planar NAND flash memory}~\cite{park.jssc15,mizoguchi.imw17,choi.vlsit16}. \chI{Thus, the PEC
component in our model (Equation~\ref{eqn:urm}) is different from the equivalent component
in UDM.}  Second, 3D NAND \chI{flash memory} has a
different activation energy than planar NAND \chI{flash memory}, as we \chI{experimentally} \chI{show} in
Section~\ref{sec:heatwatch:modeling:arrhenius}. Third, 3D NAND \chI{flash memory} reliability is affected
by the \chI{programming} temperature, as we \chI{show} in
Section~\ref{sec:heatwatch:modeling:linear}.  
\chI{Because UDM does \emph{not} accurately capture these changes,
its error rate for 3D NAND flash memory is}
\chI{3.6$\times$ greater than the error rate for the SRRM
component of our \chI{new} URT model,
\chI{as we show in Section~\ref{sec:heatwatch:modeling:urm}}.}

\paratitle{Improving Flash Reliability}
Many prior works propose mechanisms to improve flash lifetime and reduce
raw bit errors (see~\cite{cai.procieee17, cai.arxiv17} for a detailed survey).
For example, flash refresh techniques limit the number of
retention errors to achieve higher P/E cycle lifetime\chI{~\cite{cai.iccd12,
pan.hpca12, cai.itj13, luo.msst15}}. Prior work also adjusts the read reference voltage
according to P/E cycle \chI{count} and retention time to reduce \chI{the}
RBER\chI{~\cite{parnell.globecom14, papandreou.glsvlsi14, cai.hpca15, cai.date13, cai.iccd13,
luo.jsac16}}.
Different from prior work, we develop a new mechanism that tracks workload write
intensity and SSD temperature online \chI{and} adjusts read reference voltage
accordingly
to improve flash lifetime. We compare our mechanism to prior \chI{mechanisms} \chI{that} are
agnostic to these factors and show that \chI{it} can \chI{significantly} reduce RBER
\chI{and improve} flash lifetime.

\section{\chIV{Limitations}}

\chIV{Neither our URT model nor the HeatWatch mechanism consider other error
sources in 3D NAND flash memory such as retention interference, process
variation\chV{,} and read disturb, but they can
be augmented in the future to consider all possible error types to further
improve 3D NAND reliability. The activation energy may change under different
temperatures and P/E cycles. To simplify our URT model, we use a single static
value for activation energy, which is good enough for our use case.
When the temperature variation is larger, we may need a dynamic model to
estimate the activation energy to more accurately predict the temperature
effect.
We do not consider temperature variation across layers, which can happen if
the NAND flash chip heats up from inside during program/erase operations. We
expect future work to improve our model and technique to consider temperature
variation.
It is possible that certain FTL algorithms, such
as WARM, may write to certain flash blocks more frequently, which makes the
average dwell time of these blocks significantly shorter than the average.
This limits the accuracy of the HeatWatch dwell time tracking component. For
these FTL algorithms, we may need to track dwell time differently for these
frequently-written blocks.}


\section{Conclusion}
\label{sec:heatwatch:conclusion}

\chI{We} perform the first detailed \emph{experimental} characterization of the impact of
self-recovery and temperature on the reliability of 3D NAND flash memory.
We find that due to significant changes in the
memory design and manufacturing process,
prior findings and models for planar NAND flash memory \chI{are \chI{\emph{not}} accurate for} 
3D NAND flash memory.
We use key findings from our characterization to 
develop URT, a unified \chI{and accurate} \chI{cell threshold voltage and
raw bit error rate} model \chI{that takes into account the} combined effects of self-recovery, temperature,
retention loss, and wearout.
\chI{We develop a new mechanism, HeatWatch, that uses URT to}
dynamically \chI{adapt} the read reference voltage to the
\chI{data retention time, dwell time,}
\chI{SSD temperature, and wearout}.
We show that HeatWatch \chI{greatly reduces \chI{the raw bit error rate} and improves flash lifetime}.
\chI{We conclude that the effects of self-recovery and temperature in 3D NAND
flash memory can be \chI{accurately modeled and} \chI{successfully used} to improve flash reliability.
We hope that our \chI{data, model, and new mechanism} inspire others to develop \chI{other} new mechanisms that
take advantage of the self-recovery \chI{and temperature effects in 3D} NAND flash memory.}



\chapter{\chIV{System-Level Implications and Lessons Learned}}
\label{sec:systems}

\chIV{This dissertation proposes several mechanisms that improve flash reliability
by mitigating raw bit errors that occur on flash devices. In this chapter, we
discuss the system-level implications of these mechanisms and a summary of
lessons learned.}

\section{System-Level Implications}
\label{sec:systems:implications}

\chIV{
Throughout this dissertation, we have focused on evaluating the traditional
metrics
of flash reliability, P/E cycle \chV{lifetime.
This P/E cycle} metric assumes that the same workload \chV{runs on a
each individual SSD (i.e., the workload cannot be redistributed
across multiple SSDs).} This metric also assumes the NAND flash memory always
deploys a fixed amount of ECC that has a fixed RBER limit
(as illustrated in Figure~\ref{fig:this-work-rber}
and~\ref{fig:rber-vs-pec} as the \emph{ECC limit}). However, these assumptions
may change in a real
system. For example, \chV{in a distributed storage system, certain SSDs may contain
more write-hot data than the others due to uneven workload
distribution across server machines~\cite{huang.fast17}},
\chV{and} the flash controller may use a simpler ECC with
lower storage overhead. In this section, we discuss the system-level
implications or the reliability impact of our mechanisms under different
assumptions.
}

\subsection{Impact on Tolerable Write Frequency}

\chIV{As we have mentioned in
Section~\ref{sec:application:lifetime} and~\ref{sec:3derror:mitigation:all},
server machines in enterprises and data centers, along with the flash memory
devices used in those machines, are replaced after a predefined period,
typically several years. \chV{Since a flash memory device can tolerate a
limited total number of writes before having to be replaced, it can
only tolerate a certain number of writes \emph{per day} on average (i.e., 
\emph{tolerable write frequency}). To ensure that each flash memory device
does not exceed its P/E cycle lifetime prematurely before being replaced,}
the system software throttles the write frequency of each flash memory device
to only a few Drive Writes Per Day (DWPD)\chV{~\cite{lee.fast12}}. Instead of
increasing the flash device lifetime, we can use our mechanisms to improve
the tolerable write frequency.
When the P/E cycle lifetime improves after deploying the mechanisms
proposed in this dissertation, the tolerable write frequency by each flash
memory device increases proportionally, assuming the ECC limit remains the
same. \chV{Thus, the increase in tolerable write frequency is the same as the
increase in lifetime:} WARM improves tolerable write frequency (or DRPD) by
3.24$\times$; Two applications of our online flash channel model improve
DWPD by 48.9\% and 69.9\%; Combining LaVAR, LI-RAID and ReMAR improves
3D NAND DWPD by 85.0\% (Section~\ref{sec:3derror:mitigation:all}); HeatWatch
improves DWPD by 3.85$\times$.
}

\subsection{Impact on ECC Cost}

\chIV{As we have shown in
Figure~\ref{fig:technology-trend-2} and in
Section~\ref{sec:3derror:mitigation:all}, sustaining a high ECC limit incurs
a high storage overhead \chV{of the} ECC bits (i.e., \emph{ECC redundancy}).
Using BCH code~\cite{bose.ic60} as an example, assuming that the
codeword length is fixed (i.e., \SI{8}{\kilo\byte}), the ECC limit is linearly
correlated with ECC redundancy~\cite{deal.whitepaper09}. \chV{Instead of
increasing the lifetime, we can use our mechanisms to reduce ECC redundancy,
since we need to tolerate fewer raw bit errors by the end of the flash
lifetime.}}

\chIV{
\chV{To calculate the reduction in the required ECC cost, we use the same
method and assumptions as we use in Section~\ref{sec:3derror:mitigation:all}.
First, we obtain the P/E cycle lifetime of the baseline
SSD using state-of-the-art error correction mechanisms that can tolerate up
to $3\cdot10^{-3}$ raw bit error rate. This requires to a 12.8\% ECC
redundancy using a BCH code~\cite{deal.whitepaper09}. Second, we obtain the
worst raw bit error rate of our mechanism at the end of the same P/E cycle
lifetime of the baseline SSD. Third, we calculate the ECC redundancy required
by our mechanism to achieve the same data reliability in terms of the
error correction failure rate ($P_{ECFR}$ in Equation~\ref{eq:E1}). The
required ECC redundancy is obtained by varying the number of bits within a
codeword correctable by the ECC ($t$ in Equation~\ref{eq:E1}) until
$P_{ECFR}$ reaches $10^{-15}$ required by the JEDEC
standard~\cite{jep122h.jedec16}.} Our online
flash channel model reduces ECC redundancy by 37.5\%; Combining LaVAR,
LI-RAID and ReMAR reduce ECC redundancy by 78.9\%
(Section~\ref{sec:3derror:mitigation:all}); HeatWatch reduces ECC redundancy
by 78.2\%. ECC redundancy savings when using LDPC code should be similar, but
are much more difficult to evaluate because LDPC codes do not have a fixed ECC
limit~\cite{cai.arxiv17,cai.procieee17}.
}

\subsection{Impact on Performance and Flash Management Policies}

\chIV{We have
already shown that the performance overhead of each proposed mechanism is
small. The amortized cost for training and predicting using an online model is
negligible (i.e., $<$1\%) compared to flash read latency. In the meantime, our
mechanisms \chV{reduce} raw bit error rate at \chV{any P/E cycles}, which leads to
reductions in ECC decoding latency and in read-retry counts, and thus reduces
the overall flash read latency, \chV{comparing to a baseline SSD with the same
amount of P/E cycles}. The exact reduction in flash read latency
depends on several design choices of the ECC code: (1)~a shorter codeword
length or fewer ECC bits \chV{simplify decoder logic thus \chV{reducing} the ECC
decoding latency. However, this reduces the error correction strength of the
ECC, leading to more read-retries as the weaker ECC is more likely to fail,} 
(2)~a stronger ECC code may increase ECC decoding latency, but \chV{reduces}
read-retry counts. A similar
tradeoff exists when designing a soft-decoding LDPC code to balance the ECC
latency and soft-decoding sensing levels. WARM affects existing flash
management policies by dividing flash blocks into write-hot and write-cold
block pools, which increases the average response time by 1.3\% across all
workloads; All of our online model-based techniques, including online flash
channel modeling, LaVAR, ReMAR and HeatWatch do not change the data layout
or mapping, thus does not increase FTL or GC overhead; LI-RAID
only changes the data layout within a flash block, and only reduces the
storage capacity by less than 0.4\%, thus has negligible overhead to the FTL
and GC.
}

\section{Lessons Learned}
\label{sec:systems:lessons}

\chIV{
This dissertation provides several new analyses \chV{and techniques} to improve
flash reliability efficiently. When doing these analyses and developing these
new techniques, we learned two important lessons that applies for future
research in this direction. In this section, we summarize these two lessons.
}

\subsection{Combining Large-Scale and Small-Scale Characterization Studies}

\chIV{
This dissertation focuses on small-scale studies that perform extensive
characterization of only a few NAND flash memory devices. While our
observations should apply to any device that uses similar manufacturing
technologies (Section~\ref{sec:vthmodel:overview:methodology},
\ref{sec:3derror:summary}, and~\ref{sec:heatwatch:characterization:summary}), and
our online modeling techniques can adapt to any variation across chips 
(Section~\ref{sec:static} and~\ref{sec:heatwatch:mechanism}),
\chV{we have not been able to verify this behavior across a large number of
devices}. Ideally, we would
like to combine an in-depth small-scale study with a lightweight large-scale
study \chV{that does not require characterizing the full threshold voltage
distribution}, which complement each other to provide a stronger result.
Our small-scale study provides a deeper understanding of the
characteristics of different error types. A large-scale study can show the
effectiveness of our proposed mechanisms in real deployment.
We are unable to do this now due to limitations in the number of chips
available.
}

\subsection{Improve Systems Reliability Rather Than Device Reliability Alone}

\chIV{
This dissertation focuses on reducing raw bit errors within the flash chips.
These errors are usually invisible to application and system
designers because a vast majority of them are already contained within the SSD
using strong but expensive ECCs. While containing all raw bit errors within
the SSD provides a strong device reliability and reduces the complexity of
system design, it leads to a suboptimal design that uses less cost-efficient
techniques to improve reliability. \chV{For example, the ECC currently designed for
the least-reliable flash chip may consume more storage overhead than needed
on an average flash chip; and a very low uncorrectable error rate may not be
necessary in a distributed storage system because the data is already
replicated in other servers to tolerate more catastrophic server failures.} An
ideal design should
combine device-level techniques that mitigates raw bit errors cost-effectively
and provides reasonable device reliability, and system-level techniques that
utilize many devices to tolerate device failures. This design could lead to
significant cost savings while achieving higher overall system reliability.
}

\chapter{Conclusions}
\label{sec:conclusion}

In this dissertation, we present a multitude of low-cost architectural
techniques to improve NAND flash memory reliability. Following our thesis
statement, all of our proposed techniques (1)~take advantage of device-level
error characteristics or workload characteristics, (2)~can \chIII{be} implemented with
low overhead in the flash controller as a part of the firmware, and
(3)~\chIII{improve NAND} flash reliability at \chIII{low cost.}

First, we propose a new technique \chV{that} exploits workload
\emph{write-hotness} and device \emph{data retention} characteristics to
improve flash lifetime. We observe that pages with
different degrees of \emph{write-hotness} have \chIII{widely-ranging} retention time
requirements. WARM aims to eliminate redundant refreshes for
write-hot pages with minimal storage and performance overhead. The first key
idea of WARM is to effectively partition pages stored in flash into two groups
based on the write frequency of the pages. The second key idea of WARM is to
apply the most suitable flash management policies to the two different groups
of flash pages/blocks. \chV{Our evaluations show that WARM eliminates a significant
amount of unnecessary refreshes and improves flash lifetime significantly, and
that WARM can also be combined with refresh techniques to provide larger
benefits in flash lifetime.}

Second, we propose a new framework for \emph{online flash channel modeling},
which learns and exploits an online threshold voltage distribution model in
the flash controller to improve flash reliability. We observe from our
characterization of \chIII{real, state-of-the-art planar NAND flash memory}
chips that (1)~the threshold voltage
distribution can be approximated using \chIII{a} modified version of the Student's
t-distribution, and that (2)~the amount by which the distribution shifts as
the P/E cycle count increases is governed by the power law. Using \chIII{our}
characterization results, we build an \emph{accurate} and
\emph{easy-to-compute} model of the threshold voltage distribution of modern MLC NAND flash
memory. We demonstrate various applications of our model in a flash
controller. \chV{Our evaluations show that these applications improve flash
lifetime significantly.}

Third, \chV{this dissertation provides} the \emph{first comprehensive} experimental
characterization and modeling
of 3D NAND device characteristics using real, state-of-the-art MLC 3D NAND flash
memory chips, and proposes four mechanisms to exploit these device
characteristics in the flash controller for improving flash reliability. We
find several differences in the device characteristics
between 3D and planar NAND devices.
Three of the key differences we find are inherent to the internal architecture
of 3D NAND flash memory:
layer-to-layer process variation, early retention loss, and retention
interference. We develop models and techniques to exploit these 3D NAND flash
device characteristics. \chV{Our evaluations show that, when combined
together, our techniques improve flash lifetime significantly over
state-of-the-art error mitigation
techniques developed for planar NAND flash memory.}

\chIII{Fourth, \chV{this dissertation is the first to} experimentally characterize
and model the self-recovery and
temperature effect in 3D NAND flash memory using real 3D NAND devices, \chV{and to}
propose a new technique called \emph{HeatWatch},
which exploits the awareness of the workload's write-intensity and the SSD
temperature in the flash controller to improve flash reliability.} Through the
characterization, we observe that (1)~a longer dwell time slows down flash
retention loss speed, and (2)~high temperature accelerates flash retention
loss but improves program accuracy. We develop URT, a unified model for the
combined effects of self-recovery, temperature, retention loss, and wearout.
Our evaluations show that URT accurately predicts 3D NAND device
characteristics. HeatWatch efficiently tracks the dwell time of the
workload and the temperature of the SSD, and uses URT to dynamically adjust
the read reference voltage to the workload write-intensity and SSD
temperature. Our evaluations on 28 real workload traces show that, by
accurately predicting and applying the optimal read reference voltage,
HeatWatch improves flash lifetime by $3.85\times$, over a baseline that uses a
fixed read reference voltage.

\chV{Finally, this dissertation provides an analysis of the system-level
implications of our proposed techniques. We show that, aside from improving
flash lifetime, our techniques can also improve tolerable write frequency
and reduce ECC cost when a longer flash lifetime is not needed. In addition,
we show that our techniques may reduce read latency and add very little
overhead to the existing flash management policies.}





\chapter{Future Research Directions}
\label{sec:future}

This dissertation has shown several example approaches that significantly
\chIII{improve NAND} flash reliability at a low cost by making the flash controller \chIII{device-
and workload-aware}.
\chIII{We believe} that it is promising to continue exploring along
this direction, \chIII{because (1)~NAND flash memory continues to grow in
population as it becomes a cheaper and more accessible technology \chV{and} (2)~flash
reliability issue will grow as flash vendors more aggressively increase the density of NAND
flash memory to satisfy the demand}. In this chapter, we
describe potential research directions
to further improve the reliability and efficiency of NAND flash memory in a
system.

\section{Temperature Effects on Read Operations}
\label{sec:future:temperature}

\chV{In Chapter~\ref{sec:heatwatch}, we have shown that SSD temperature
significantly affects retention loss speed and program variation of NAND
flash memory. Recent work shows that temperature could also affect read
operations significantly (i.e., \emph{read temperature variation})~\cite{li.fms12}, and
proposes circuit-level techniques to compensate
for this type of temperature variation~\cite{tanzawa.patent07, cho.patent05}.
In reality, read temperatures can change significantly in a very
short time, especially on mobile devices~\cite{cai.hpca15}, leading to
a significant increase in raw bit error rate under extreme temperatures.}

\chV{To the best of our knowledge, there is no characterization data or model
available for read temperature effect in open literature. So we believe that
it is valuable to study
this phenomenon and propose flash controller techniques to tolerate the read
temperature variation. We believe that the flash controller can learn an
accurate online model, like the ones we use in
Chapter~\ref{sec:vthmodel}, \ref{sec:3derror}, and~\ref{sec:heatwatch}, to
compensate for the read temperature variation better than existing
circuit-level techniques. The key challenge in
understanding the read temperature effect is to design a rigorous testing
procedure that eliminates the potential noise caused by the
unknown amount of retention errors introduced
when the read temperature slowly converges to the target level. It is
difficult to accurately quantify the number of retention errors during this
time because the temperature\chV{,} and hence the retention loss speed\chV{,} change at a
variable speed. Hence, one potential way to rigorously characterize read
temperature effect is to perform the characterization under low temperature
such that the number of newly introduced retention errors is minimal. Another
potential way is to
constantly monitor the temperature and use integration to compute the
retention loss scaled by the varying temperature, like HeatWatch does when
precomputing the temperature amplification factor
(Section~\ref{sec:heatwatch:mechanism}), and account for the retention
errors in the characterization.}

\chV{To conclude, we believe it is interesting to (1)~design a rigorous testing
procedure to characterize read temperature variation, (2)~understand the cause
of read temperature variation, e.g., it might be caused by the temperature
sensitivity of the sense
amplifier, (3)~develop new models for read temperature variation, and
(4)~design new flash controller techniques to
mitigate and compensate for read temperature variation.}

\section{SSD Errors At Scale}
\label{sec:future:scale}

Today's data center servers already use SSDs as a high-performance alternative
to hard disk drives to store frequently-accessed persistent data. As the
storage density of \chIII{NAND-}flash-based SSDs \chIII{continues} to increase, and
\chIII{the price of SSDs continues}
to decrease, data center servers are \chIII{increasingly} likely to deploy
SSDs instead of hard disk drives as the primary storage \chIII{medium}. This creates
both opportunities and challenges for \chIII{improving} SSD reliability. In this
section, we discuss \chIII{these opportunities and challenges,} and \chIII
{discuss several} potential \chIII{research} directions for improving SSD
reliability at scale.

\subsection{3D NAND Errors In the Field}
\label{sec:future:scale:errors}

As we have discussed in Section~\ref{sec:errors:largescale}, recent works have
analyzed the reliability of hundreds of thousands of SSDs in production data
centers~\cite{meza.dsn15, schroeder.fast16, narayanan.systor16,
narayanan.sigmetrics16}. However, none of these works include any \chIII
{analysis of SSDs using
3D NAND} because 3D NAND technology \chIII{has only been} recently introduced
and has
not \chIII{accumulated enough} device hours \chIII{in the field to perform
long-term studies}. As we have discussed in
Section~\ref{sec:background:3d}, 3D NAND devices have \chIII{a} different
flash cell
design, \chIII{a different} flash chip organization, \chIII{and use a} larger
manufacturing process
technology than planar NAND devices. Thereby, as we have shown in
Section~\ref{sec:3derror:errors} using \chIII{a} controlled error study, 3D NAND
devices
have different error characteristics from planar NAND\@. Hence, \chIII{we
expect that} 3D NAND
devices in the field \chIII{will also demonstrate} different reliability
characteristics.

In addition, the density of future 3D NAND devices is increased using
different methods than for planar NAND\@. For planar NAND, manufacturers
\chIII{increased} its density in each product generation using aggressive process
technology shrinking. This, unfortunately, decreases \chIII{the} flash cell size as
well
as the distance between cells, thus significantly \chIII{decreasing} flash
reliability.
For 3D NAND, in the foreseeable future, manufacturers can increase storage
density by increasing the number of stacked layers instead of using
aggressive process technology \chIII{scaling}. Hence, future generation 3D NAND
devices are
more likely have larger layer-to-layer process variation. Thus, using a
large-scale field study could allow us to observe new challenges for scaling
the 3D NAND devices as data centers will deploy multiple generations of 3D
NAND chips over time.

To conclude, we believe it is interesting to (1)~investigate 3D NAND error
characteristics in the field, and compare the \chIII{characteristics to those of}
planar \chIII{NAND,} (2)~compare 3D NAND error characteristics across
multiple generations to identify new challenges in scaling flash density, such
as layer-to-layer process variation, (3)~investigate SSD failure due to other
components than flash chips, such as the controller and the command or data
bus, (4)~investigate 3D NAND chip-to-chip process variation by comparing error
characteristics across many devices, and (5)~investigate and compare the
effectiveness of various state-of-the-art error mitigation or error recovery
techniques in the field, such as WARM (\chIII{Chapter}~\ref{sec:warm}), online flash
channel modeling (Chapter~\ref{sec:vthmodel}), HeatWatch
(Chapter~\ref{sec:heatwatch}), RFR~\cite{cai.hpca15}, \chIII{and
RDR~\cite{cai.dsn15}.}
We believe that the insights derived from these \chIII{investigations} can
enable and inspire new techniques, especially new system-level
techniques~\cite{huang.fast17}, to tolerate SSD errors more efficiently.

\subsection{Predicting and Preventing SSD Failures}
\label{sec:future:scale:uber}

SSD failures caused by uncorrectable errors on a single SSD are designed to be
very infrequent (e.g., typically less than $10^{-15}$), \chIII{due to the use
of} strong ECC
within the controller. In a large-scale data center, however, \chIII{the
occurrence of these infrequent
uncorrectable errors is} amplified \chIII{due to the use of} \emph{hundreds of
thousands} of SSDs at the same time. As a result, uncorrectable \chIII{errors} become a
relatively frequent event \chIII{in a} data center, which can result in
\chIII{frequent} machine failures and data \chIII{loss}.

SSD failures caused by component failures are also infrequent on a single SSD,
because each SSD contains only a few components in total~\cite{mielke.irps08,
grupp.fast12, cai.procieee17, cai.arxiv17}. In a data center,
however, SSD component failures are frequent because the data center consists
of thousands of SSD controllers, millions of flash chips, \chIII{and millions of data
buses~\cite{meza.sigmetrics15,schroeder.fast16,narayanan.systor16}}. Such component \chIII{failures} can be catastrophic, because
\chIII{they} can lead
to many uncorrectable errors at the same \chIII{time,} and \chIII{to} immediate loss
of large
chunks of data.

Fortunately, these SSD failures are typically preceded by early \chIII{warning} signs,
such as correctable errors, and have strong spatial
locality~\cite{meza.sigmetrics15,schroeder.fast16,narayanan.systor16}.
Thus, we believe it is
possible to prevent a majority of these failures by predicting them in
advance. We believe that by investigating SSD failures in the field
(Section~\ref{sec:future:scale:errors}), we can identify \chIII{predictable}
patterns of these SSD
failures. Based on these patterns, we can (1)~develop models of SSD failures,
including models of uncorrectable errors and models of SSD component failures,
(2)~predict SSD failures before they happen, \chIII{and} (3)~design mechanisms
to prevent
a majority of SSD failures from affecting system and data reliability, e.g.,
taking SSDs offline before failure happens, or duplicating data to more
reliable SSDs in advance.

\subsection{Tolerating Reliability Variation Across SSDs}
\label{sec:future:scale:variation}

Improving and managing SSD reliability at a \chIII{larger scale introduces} new
challenges,
\chIII{as reliability} can vary significantly across different SSDs within a data
center. \chV{There are \emph{four} major reasons for the} variation in SSD reliability. First, the SSDs within a
data center might have been deployed at
different times. Thus, different batches of SSDs may use different \chIII{generations}
of flash technology, which have very different reliabilities. They may also
have different \chIII{deployment dates}, leading to different amounts of
wearout on the SSD\@.
Second, even for SSDs within the same batch, SSD reliability can vary due to
the process variation across different flash chips. For example, some SSDs may
consist of more reliable flash chips which have longer endurance and can store
data for a longer retention time. Third, \chIII{even} in an ideal \chIII{world
without} any
process variation, each SSD within a data center \chIII{may run} a different workload
throughout its lifetime. For example, write intensity can vary by as much as
5,500$\times$ across different workloads~\cite{huang.fast17, cai.iccd12},
leading to drastically different P/E cycle lifetimes for different SSDs.
The reliability variation can cause failures to happen sooner on some SSDs
\chIII{than the designed lifetime of the SSD}, requiring \chIII{more frequent}
SSD replacement in a data center.

We believe \chIII{that} by tolerating reliability variation across SSDs in a
data center, \chIII{the}
overall storage reliability can be improved, and the cost for managing SSD
reliability can be reduced. There can be many ways to tolerate this variation.
\chV{For example, (1)~we} can apply RAID across different SSDs and tolerate SSD failures using
erasure \chIII{codes~\cite{shu.book04}}. To minimize \chIII{the} RAID failure rate,
less reliable SSDs should be
evenly distributed across different RAID groups instead of being arbitrarily
grouped together. \chV{As another example, (2)~we} can either apply global \chIII{wear-leveling} across all SSDs
within a data center to mitigate such variation~\cite{huang.fast17}, or move
the data (e.g., write-hot vs.\ write-cold data) to its most suitable SSD
(e.g., move write-hot data to less reliable SSDs)~\cite{luo.msst15,luo.hpca18}.
We believe it is interesting to compare the benefit \chIII{of} these two solutions for
different use cases.

\section{Enabling Cold Storage in \chIII{SSDs}}
\label{sec:future:cold}

SSDs are already popular \chIII{for storing frequently-accessed} data in data
centers.
The key challenges for the larger-scale deployment of SSDs are (1)~higher cost
compared to hard disk drives, and (2)~limited retention time guarantee. The
advent of 3D NAND technology increases the density potential of NAND flash
memory, but increases \chIII{manufacturing costs} compared to planar NAND
\chIII{due to the} more
complex manufacturing \chIII{process required}. \chIII{Furthermore}, as we
have shown in
\chIII{Chapter}~\ref{sec:3derror}, due to early retention loss, 3D NAND also has
\chIII{greater} retention errors than planar NAND\@. Thus, we believe it is important
to reduce SSD cost by increasing \chIII{both} storage density \chIII{and
SSD retention time}. In this section, we discuss the problems and
potential directions associated with each potential solution.

\subsection{Identifying Suitable Data for SSD Cold Storage}
\label{sec:future:cold:data}

\chIII{Similar to our approach for} WARM \chIII{(Chapter~\ref{sec:warm})}, we would like
to identify
suitable data for cold storage in SSDs
and manage them in a more efficient way. For flash-based SSDs, write
operations not only consume P/E cycle lifetime but also reduce the dwell time
of the SSD, which accelerates retention loss
(Section~\ref{sec:heatwatch:characterization}). Thus, write-cold data is more
suitable for SSD cold storage. However, \chIII{infrequently-accessed} data (i.e.,
read-cold and write-cold data) should be stored in the cheapest \chIII
{possible} storage such
as tape. Thus, we believe write-cold, read-hot data can benefit the most
from the fast random access performance of SSD cold storage.

We believe it is interesting to investigate the best technique to identify
write-cold, read-hot data, such as: (1)~identifying data used by read-only
applications, which requires frequent accesses to large amounts of static data
that do not change, (2)~identifying write-cold data within each SSD using
multiple queues like WARM, \chIII{Bloom} filters~\cite{liu.isca12}, log-structured
merge trees~\cite{oneil.ai96}, or any other write-cold data identification
techniques, or \chIII{(3)~using} programmer
annotations~\cite{kang.hotstorage14}. After we identify such data, the SSD
controller or the storage manager can decide to aggregate \chIII{the
write-cold, write-hot} data and
manage \chIII{it} in a more efficient way.

\subsection{Increasing SSD Retention Time}
\label{sec:future:cold:retention}

To use \chIII{SSDs for} cold \chIII{storage, the SSDs must provide a} long
enough
retention time. Otherwise, as we have shown in WARM, frequent refreshes will
eat away the majority of the P/E cycle lifetime and \chIII{potentially} reduce
SSD
performance. SSD cold storage has several properties that \chIII{could}
benefit SSD reliability and extend \chIII{the} retention time. First, write-cold data
is seldom updated, \chIII{and} thus consumes fewer P/E \chIII{cycles}. \chIII
{As a result, SSDs used for cold data have a longer lifetime.} Second,
as we have
shown in Section~\ref{sec:heatwatch:characterization}, thanks to flash memory
self-recovery, a longer dwell time slows down flash retention loss. Thus, if
an SSD contains only write-cold data, the dwell time of all flash blocks
within the SSD will be long, increasing the retention time of the SSD\@.

We believe it is interesting to investigate techniques to further improve the
retention time for SSD cold storage to make it more appealing
\chIII{than cold data storage using hard disk drives}.
First, we can investigate techniques
that \chIII{partition} the data identified for SSD cold storage (e.g., write-cold,
read-hot data) to a separate pool of SSDs or flash chips. This will reduce the
dwell time for cold data, and increase the dwell time for other data.
\chIII{This can lead to more efficient flash policies for all data, as it
reduces the retention loss speed for cold data
(Chapter~\ref{sec:heatwatch}), and reduces the required retention time
guarantee for the other data (Chapter~\ref{sec:warm}).}
Second, we can investigate suitable flash management policies for write-cold
data. Note that, according to our findings in
Section~\ref{sec:3derror:comprehensive}, read disturb errors are minimal in 3D
NAND\@. Also note that, for write-cold \chIII{data, the vast} majority of writes are
\chIII{for refresh operations, and} thus its dwell time is approximately equal
to its retention time.
This could allow for much higher \chIII{lifetimes for SSDs used for} cold
storage.
Third, once write-cold data is stored on separate SSDs, we can even manage
SSD cooling \chIII{by using} the best storage temperature that maximizes SSD
retention time.
Since writes are infrequent in cold storage, we can schedule them when the
temperature is high, which reduces SSD programming errors. Fourth, we can
apply retention error recovery techniques to further relax retention time
constraints for SSD cold storage. Since the data in cold storage will remain
static for a long time, the SSD errors in cold storage \chIII{are} dominated by
retention errors. Thus, retention error recovery can be very effective at
correcting errors in SSD cold storage even when the raw bit error rate
exceeds the ECC \chIII{correction capability}.

\subsection{Increasing SSD Capacity}
\label{sec:future:cold:capacity}

Scaling NAND flash density is hard because it often trades off flash
reliability. In SSD cold storage, however, we can give up certain unused SSD
reliability and performance to increase SSD capacity. This lowers the
cost-per-bit of SSD cold storage, which makes it more appealing to use. We
believe there are two reasons we can \chIII{trade off} some aspects of SSD reliability
and performance for cold storage. First, since SSD cold storage does not
require \chIII{a} high P/E cycle lifetime, we can limit SSD lifetime to only a few
hundred P/E cycles. \chIII{This allows for much more aggressive scaling to
increase the density of SSD cold storage, leading to a higher
cost-efficiency.} Second,
since the write performance is less important for
cold storage, we can use SSD write modes with long delay. By trading off P/E
cycle lifetime and write performance, we believe we can significantly improve
SSD capacity, and hence reduce the cost for SSD cold storage through many
ways. For example, we can configure the flash chips to TLC or QLC mode for SSD
cold storage, which increases the number of bits stored on each flash cell,
\chIII{in turn reducing} the error margin and \chIII{increasing} SSD write
latency. \chIII{To compensate for the potentially increased raw bit error
rate in TLC or QLC NAND flash memory, we can develop more aggressive error
mitigation techniques tailored for SSD cold storage to tolerate these errors.}
\chIII{For example, we} can
configure SSD cold storage to use \chIII{a} smaller program step size to
\chIII{trade off} write
performance. This could allow us to use weaker ECC on SSD cold storage, which
frees up SSD capacity \chIII{that} originally used for ECC redundancy.


\chapter*{Other Works of This Author}
\addcontentsline{toc}{chapter}{Other Works of This Author}
\label{sec:other}

During the course of my Ph.D., I \chV{have} had the opportunity to collaborate with many
of my fellow graduate students. These projects not only help me to learn
about NAND flash memory, DRAM, and cost-reliability trade-offs, which \chV{is useful}
for this dissertation, but also help me gain insights, ideas,
and skills to \chV{conduct} good research, which \chV{is useful} when writing this
dissertation. In this chapter, I would like to acknowledge these projects and
my other works related to this dissertation.

In collaboration with Justin Meza, I have worked on single-level storage
systems. We \chV{rethought} the interface for efficiently managing a heterogeneous
memory and storage system which consists of DRAM, NVM, \chV{NAND} flash memory, and hard
disk drive. We explored the design of a \emph{Persistent Memory Manager} that
coordinates the management of memory and storage under a single hardware unit
in a single address space~\cite{meza.weed13}. Efficient management of NVM and
flash reliability is one of the concerns for designing the persistent memory
manager.

In collaboration with Vivek Seshadri, I have worked on processing using
DRAM\@. We propose RowClone, a new and simple mechanism to perform bulk copy
and initialization completely within DRAM using the internal row
buffers~\cite{seshadri.micro13}. \chV{This gave me the opportunity} to learn DRAM architecture,
which has a lot in common with NAND flash memory.

During my PhD, I have also worked on improving the cost-reliability trade-offs
of DRAMs in data centers. First, we propose the idea of
\emph{heterogeneous-reliability memory}, which is a hardware/software
cooperative system design that chooses the best memory reliability
provisioning for each chunk of data to lower data center cost while achieving
high reliability~\cite{luo.dsn14}. Second, we propose \emph{Capacity- and
Reliability-Adaptive Memory} (CREAM), a hardware mechanism that adapts
error-correcting DRAM modules to offer multiple levels of error protection,
and provides the capacity saved from using weaker protection to
applications~\cite{luo.arxiv17}.

In collaboration with Yu Cai, I have worked on many ideas to improve flash
reliability aside from my dissertation work. Over the years, we propose
several online or offline techniques to mitigate data retention
errors~\cite{cai.hpca15}, read disturb errors~\cite{cai.dsn15}, and program
errors~\cite{cai.hpca17}. We have also worked on a survey and tutorial of the
best practices for error characterization, mitigation, and recovery in
flash-based SSDs~\cite{cai.procieee17}. In collaboration with Aya Fukami, I
have worked on improving the reliability of chip-off forensic analysis of NAND
flash memory devices~\cite{fukami.di17}. \chV{In collaboration with Arash Tavakkol, we have
worked on improving the performance and fairness of the I/O request scheduler
in the SSD controller~\cite{tavakkol.isca18}.} All of these works are closely
related to this dissertation. During these collaborations, we all gained
valuable expertise on improving flash reliability by learning from each other.


\backmatter%


\renewcommand{\bibsection}{\chapter{\bibname}}
\bibliographystyle{plainnat}
\bibliography{refs} 

\end{document}